\documentclass[10pt]{article}
\pdfoutput=1

\usepackage{amsmath}
\usepackage{mathtools}
\usepackage{stackrel}
\usepackage{scalerel}
\usepackage{leftidx}
\usepackage{xfrac}
\usepackage{stmaryrd}

\usepackage{authblk}

\usepackage[charter]{mathdesign}
\usepackage{euscript}
\let\mathbb\undefined
\usepackage{bbold}
\usepackage{bbm}
\usepackage{yfonts}
\usepackage{dsfont}
\DeclareSymbolFont{usualmathcal}{OMS}{cmsy}{m}{n}
\DeclareSymbolFontAlphabet{\mathcal}{usualmathcal}

\usepackage{varwidth}

\usepackage{authblk}

\usepackage[shortlabels]{enumitem}

\usepackage{hyperref}
\usepackage{cleveref}
\hypersetup{colorlinks=true, linkcolor=red!50!black, citecolor=green!50!black, urlcolor=blue!80!black}

\usepackage{xcolor}
\usepackage{tcolorbox}
\usepackage{graphicx}
\usepackage{subfigure}

\usepackage[all,2cell,arrow,matrix]{xy} \UseAllTwocells \SilentMatrices
\usepackage{tikz-cd}
\usepackage{tikz}
\usetikzlibrary{arrows,arrows.meta,decorations.markings,decorations.pathreplacing,
decorations.pathmorphing,calc}

\usepackage{verbatim}
\usepackage{comment}

\usepackage[nottoc]{tocbibind} 
\usepackage{appendix}

\usepackage{geometry}
\tikzset{->-/.style={decoration={markings,mark=at position #1 with {\arrow{stealth}}},postaction={decorate}},->-/.default=0.55}
\colorlet{e_ext}{red}
\colorlet{m_ext}{blue!20}
\tikzset{e_str/.style={very thick,red!80}}
\tikzset{m_str/.style={very thick,blue!80}}
\tikzset{m_dual_str/.style={thick,dashed,blue}}
\tikzset{link_label/.style={scale=0.8,black}}

\tikzcdset{scale cd/.style={every label/.append style={scale=#1},
    cells={nodes={scale=#1}}}}

\usepackage[amsmath,amsthm,thmmarks]{ntheorem}
\theoremstyle{definition}

\newtheorem{thm}{Theorem}[subsection]

\newtheorem{prop}[thm]{Proposition}
{
\theoremsymbol{\mbox{\footnotesize ${\square}$}}
\newtheorem{pthm}[thm]{Theorem$^{\text{ph}}$}
}
\newtheorem{pprop}[thm]{Proposition$^{\text{ph}}$}
\newtheorem{pcor}[thm]{Corollary$^{\text{ph}}$}
\newtheorem{plem}[thm]{Lemma$^{\text{ph}}$}
\newtheorem{cor}[thm]{Corollary}
\newtheorem{lem}[thm]{Lemma}

\newtheorem{conv}[thm]{Convention}
\newtheorem{conj}[thm]{Conjecture}
\newtheorem{notation}[thm]{Notation}

{
\theoremsymbol{\mbox{\footnotesize ${\square}$}}
\newtheorem{defn}[thm]{Definition}
}
{
\theoremsymbol{\mbox{\footnotesize $\heartsuit$}}
\newtheorem{expl}[thm]{Example}
}
{
\theoremsymbol{\mbox{\footnotesize $\diamondsuit$}}
\newtheorem{rem}[thm]{Remark}
}
\qedsymbol{\mbox{\footnotesize $\square$}}

\numberwithin{equation}{subsection}

\newcommand\nn             {\nonumber \\}
\newcommand\be            {\begin{equation}}
\newcommand\ee            {\end{equation}}
\newcommand\bea           {\begin{eqnarray}}
\newcommand\eea         {\end{eqnarray}}
\newcommand\bnu          {\begin{enumerate}}
\newcommand\enu          {\end{enumerate}}
\newcommand\bit          {\begin{itemize}}
\newcommand\eit          {\end{itemize}}

\newcommand{\pf}{\begin{proof}}
\newcommand{\epf}{\qed\end{proof}}

\makeatletter
\providecommand{\leftsquigarrow}{%
  \mathrel{\mathpalette\reflect@squig\relax}%
}
\newcommand{\reflect@squig}[2]{%
  \reflectbox{$\m@th#1\rightsquigarrow$}%
}
\makeatother

\newcommand\Cb			{\mathbb{C}}
\newcommand\Nb			{\mathbb{N}}

\newcommand\Rb			{\mathbb{R}}
\newcommand\Zb			{\mathbb{Z}}
\newcommand\bk			{\mathbb{k}}

\newcommand\CA			{\EuScript{A}}
\newcommand\CB			{\EuScript{B}}
\newcommand\CC			{\EuScript{C}}
\newcommand\CD			{\EuScript{D}}
\newcommand\CE			{\EuScript{E}}
\newcommand\CF			{\EuScript{F}}
\newcommand\CG			{\EuScript{G}}

\newcommand\CK			{\EuScript{K}}
\newcommand\CL			{\EuScript{L}}
\newcommand\CM			{\EuScript{M}}
\newcommand\CN			{\EuScript{N}}
\newcommand\CO			{\EuScript{O}}
\newcommand\CP			{\EuScript{P}}
\newcommand\CQ			{\EuScript{Q}}

\newcommand\CS			{\EuScript{S}}
\newcommand\CT			{\EuScript{T}}
\newcommand\CU			{\EuScript{U}}
\newcommand\CV			{\EuScript{V}}

\newcommand\CX			{\EuScript{X}}
\newcommand\CY			{\EuScript{Y}}

\newcommand{\FZ}			{\text{\usefont{U}{euf}{m}{n}Z}}

\newcommand\SA			{\mathsf{A}}
\newcommand\SB			{\mathsf{B}}
\newcommand\SC			{\mathsf{C}}
\newcommand\SD			{\mathsf{D}}
\newcommand\SE			{\mathsf{E}}
\newcommand\SF			{\mathsf{F}}
\newcommand\SG			{\mathsf{G}}

\newcommand\SK			{\mathsf{K}}
\newcommand\SL			{\mathsf{L}}
\newcommand\SM			{\mathsf{M}}
\newcommand\SN			{\mathsf{N}}
\newcommand\SO			{\mathsf{O}}
\newcommand\SP			{\mathsf{P}}
\newcommand\SQ			{\mathsf{Q}}

\newcommand\ST			{\mathsf{T}}

\newcommand\SX			{\mathsf{X}}
\newcommand\SY			{\mathsf{Y}}
\newcommand\SZ			{\mathsf{Z}}

\newcommand{\Ising}{\mathrm{Is}}

\DeclareMathOperator{\Hom}{Hom}
\DeclareMathOperator{\End}{End}
\DeclareMathOperator{\Aut}{Aut}

\DeclareMathOperator{\Id}{Id}
\DeclareMathOperator{\id}{id}

\DeclareMathOperator{\Ind}{Ind}

\DeclareMathOperator{\ev}{ev}

\DeclareMathOperator{\Fun}{\mathsf{Fun}}

\DeclareMathOperator{\Vect}{Vect}

\DeclareMathOperator{\Alg}{\mathsf{Alg}}

\DeclareMathOperator{\Mod}{\mathsf{Mod}}
\DeclareMathOperator{\LMod}{\mathsf{LMod}}
\DeclareMathOperator{\RMod}{\mathsf{RMod}}
\DeclareMathOperator{\BMod}{\mathsf{BMod}}

\newcommand{\op}			{\mathrm{op}}
\newcommand{\rev}			{\mathrm{op}}

\newcommand{\one}			{\mathbb{1}}

\newcommand{\ob}          {\mathrm{ob}}

\newcommand\Cat			{\mathsf{Cat}}

\newcommand\vect			{\mathrm{Vec}}

\newcommand\Rep			{\mathrm{Rep}}
\newcommand\rep			{\mathsf{Rep}}

\newcommand\Algc     {\mathsf{Alg}^{\mathrm{c}}}

\newcommand\Pic			{\mathrm{Pic}}
\newcommand\Irr			{\mathrm{Irr}}

\newcommand{\LC} {\hat{\EuScript{C}}}
\newcommand{\LD} {\hat{\EuScript{D}}}
\newcommand{\LM} {\hat{\EuScript{M}}}

\newcommand{\TC}{\mathsf{TC}}

\newcommand\EE{\mathrm{E}}

\newcommand{\TME}{\mathsf{MC}}

\newcommand\Kar{\mathrm{Kar}}

\newcommand\forget  {\mathsf{f}}
\newcommand\Obs{\mathrm{Obs}}
\newcommand{\ising} {\EuScript{I}s}
\newcommand{\Isphase} {\mathsf{Is}}
\newcommand{\KarCat}{\mathsf{KarCat}}

\newcommand\condense{\mathrel{\,\hspace{.75ex}\joinrel\rhook\joinrel\hspace{-.75ex}\joinrel\rightarrow}}

\begin{document}

\title{Higher condensation theory}
\author[a,b]{Liang Kong \thanks{Email: \href{mailto:kongl@sustech.edu.cn}{\tt kongustc@outlook.com}}}
\author[c,d,a]{Zhi-Hao Zhang \thanks{Email: \href{mailto:zhangzhihao@bimsa.cn}{\tt zhangzhihao@bimsa.cn}}}
\author[e,f]{Jiaheng Zhao \thanks{Email: \href{mailto:zhaojiaheng171@mails.ucas.ac.cn}{\tt zhaojiaheng171@mails.ucas.ac.cn}}}
\author[g,c]{Hao Zheng \thanks{Email: \href{mailto:haozheng@mail.tsinghua.edu.cn}{\tt haozheng@mail.tsinghua.edu.cn}}}
\affil[a]{International Quantum Academy, Shenzhen 518048, China}
\affil[b]{Shenzhen Institute for Quantum Science and Engineering, \authorcr Southern University of Science and Technology, Shenzhen, 518055, China}
\affil[c]{Beijing Institute of Mathematical Sciences and Applications, Beijing 101408, China}
\affil[d]{Wu Wen-Tsun Key Laboratory of Mathematics of Chinese Academy of Sciences, \authorcr
School of Mathematical Sciences, \authorcr
University of Science and Technology of China, Hefei, 230026, China}
\affil[e]{Academy of Mathematics and Systems Science, 
Chinese Academy of Sciences, Beijing 100190, China}
\affil[f]{University of Chinese Academy of Sciences, Beijing 100049, China}
\affil[g]{Institute for Applied Mathematics, Tsinghua University, Beijing, 100084, China}
\date{\vspace{-5ex}}

\maketitle

\vspace{1cm}
\begin{abstract}
We develop a unified mathematical theory of defect condensations for topological orders in all dimensions based on higher categories, higher algebras and higher representations. A $k$-codimensional topological defect $A$ in an $n+$1D (potentially anomalous) topological order $\SC^{n+1}$ is condensable if it is equipped with the structure of a condensable $\EE_k$-algebra. Condensing such a defect $A$ amounts to a $k$-step process. In the first step, we condense the defect $A$ along one of its transversal directions, thus obtaining a $(k-1)$-codimensional defect $\Sigma A$, which is naturally equipped with the structure of a condensable $\EE_{k-1}$-algebra. In the second step, we condense the defect $\Sigma A$ along one of the remaining transversal directions, thus obtaining a $(k-2)$-codimensional defect $\Sigma^2 A$, so on and so forth. In the $k$-th step, we condense the 1-codimensional defect $\Sigma^{k-1}A$ along the only transversal direction, thus defining a phase transition from $\SC^{n+1}$ to a new $n+$1D topological order $\SD^{n+1}$. We give precise mathematical descriptions of each step in above process, including the precise mathematical characterization of $i$-codimensional topological defects in the condensed phase $\SD^{n+1}$ for $1\leq i \leq k$ in terms of $\EE_i$-modules over the $\EE_i$-algebra $\Sigma^{k-i}A$. 
When $\SC^{n+1}$ is anomaly-free, the same phase transition, as a $k$-step process, can be alternatively defined by replacing the last two steps by a single step of condensing the $\EE_2$-algebra $\Sigma^{k-2}A$ directly along the remaining two transversal directions. 
When $n=2$, this modified last step is precisely a usual anyon condensation in a 2+1D topological order. In addition to above algebraic approach, we also provide a geometric theory of defect condensations. We derive many new mathematical results physically along the way, 
and develop some powerful tools for the construction and classification of condensable $\EE_1$-algebras and Lagrangian $\EE_2$-algebras from the precise characterization of the non-surjectivity and non-fully-faithfulness of the center functors. We also establish the connections among various notions of `gauging' symmetries. At the end of this work, we briefly discuss questions, generalizations and applications that naturally arise from our theory, including higher Morita theory, a theory of integrals and the condensations of liquid-like gapless defects in topological orders. 

\end{abstract}

\newpage

\tableofcontents

\newpage

\section{Introduction} \label{sec:intro}
In this work, we develop a unified mathematical theory of defect condensations for topological orders in all dimensions based on higher categories, higher algebras and higher representations, and explore its consequences and applications. In this section, we explain the motivations, summarize the main results, and fix some conventions and notations. The symbol ``$n$D'' represents the spacetime dimension.

\subsection{Motivations and historical remarks}

One of fundamental themes in modern theoretical physics is to establish a new paradigm of phases and phase transitions that can unify traditional spontaneous symmetry-breaking (SSB) orders with exotic new phases discovered since 1980's, including topological orders \cite{Wen90}, symmetry protected/enriched topological (SPT/SET) orders \cite{GW09,CGW10} and gapless quantum liquids \cite{KZ22b}. This yet-to-be-established new paradigm should be a theoretical framework that can catch the universal and model independent features of all these quantum phases and phase transitions among them. It was known that many phase transitions among topological orders are driven by the condensations of topological defects. In 2+1D, a condensation of 0+1D defects was known as an anyon condensation \cite{MS89a,BSS02,BSS03,BS09,BSH09,BSS11,KS11a,Lev13,BJQ13,Kon14e}. A condensation of 1+1D topological defects was known as a string condensation \cite{ZLZHKT23}. A defect condensation can also be viewed as the (higher) gauging of generalized or categorical symmetries (see for example \cite{FFRS10,GKSW15,BT18,TW24,RSS23,CH25,HPRW25})\footnote{Within the framework of TQFT, there is a program, which is parallel to defect condensations,  of `orbifold TQFTs' developed by Carqueville, Runkel, Schaumann and Mulevi\v{c}ius in a series of works \cite{CR16,CRS18,CRS19,CRS20,CMRSS21,Mul24,MR23,CM23,CMRSS24} (see \cite{Car23} for a survey, see also Remark\,\ref{rem:Runkel_works}) based on state-sum constructions. It was preceded by some earlier works on QFTs with defects \cite{FFRS07,FFRS10,DKR11}. See Remark\,\ref{rem:FQFT_AQFT} and Remark\,\ref{rem:ingo} for further comments.}. Defect condensations can also be used to control or define phase transitions among gapless quantum liquids \cite{KZ20,KZ21,CJKYZ20,CW23a,LY23,CJW25,CAW24}. Therefore, the mathematical theory of defect condensations, which is indeed universal and model independent, is an indispensable ingredient of the new paradigm of phase transitions.


\medskip
Since both the experimental discovery of topological orders \cite{TSG82,Wen90} (see \cite{Wen17,Wen19} for reviews) and the theoretical discovery of its macroscopic descriptions occurred in 2+1D \cite{MS89,Wit89,FRS89,FG90,Reh90,MR91,Wen91,Kit06}, the first thoroughly studied defect condensation is the boson condensation (somehow misnamed as anyon condensation) in 2+1D topological orders. It was originated in the study of  the extensions of chiral algebras \cite{MS89b} and was first studied by Moore and Seiberg in a special case in \cite{MS89a}. It was later systematically developed and named as `boson condensation' by Bais, Schroers, Slingerland in \cite{BSS02,BSS03} as a generalization of Landau's symmetry-breaking theory, and was further developed in \cite{BS09,BSH09,BSS11}. The precise mathematical description of an anyon condensation was first obtained for 2+1D finite abelian gauge theories \cite{KS11,Lev13,BJQ13}, and was fully established in a categorical language for all 2+1D topological orders in \cite{Kon14e} based on many earlier mathematical works on algebras in braided fusion categories or subfactors \cite{Par95,BEK99,BEK00,BEK01,KO02,FFRS06,DMNO13}. More works on this categorical description followed \cite{ERB14,HW15a,NHKSB16,NHKSB16a,Bur18}. We provide more references on recent developments in Remark\,\ref{rem:recent_references_anyon_condensation}. 

However, the anyon condensation theory do not provide the complete model-independent story of a phase transition between two 2+1D topological orders. A manifestation of this fact is that an anyon condensation from one 2+1D topological order to another is not reversible. On the other hand, an experimental or lattice model realization of a phase transition is always reversible. Another manifestation of this fact is that two 2+1D topological orders connected by a gapped domain wall, in general, cannot be obtained from each other via anyon condensations. The reason behind both phenomena is that anyons do not cover all topological defects in a 2+1D topological order. Indeed, anyons are topological defects of codimension 2 (or 2-codimensional topological defects). There are non-trivial 1-codimensional topological defects in a 2+1D topological order \cite{KS11a,KK12,FSV13,CR16,Kon14e}. These topological defects form a monoidal 2-category \cite{KK12,KW14}. It turns out that all of these 1-codimensional topological defects can be obtained from anyons via condensations \cite{FSV13,CR16,Kon14e}, thus are called condensation descendants or condensed defects \cite{KW14}. Mathematically, the process of finding all condensation descendants from a subset of defects (such as anyons) is called condensation completion. It was first introduced by Carqueville and Runkel in \cite{CR16} for 2-categories under the name of ``orbifold completion'', and was proposed as a fundamental question for higher dimensional topological orders in \cite{KW14}. It was later thoroughly developed by Douglas and Reutter in \cite{DR18} under the name of ``idempotent completion'', and was used there to give a mathematical definition of multi-fusion 2-category and a state-sum construction of 3+1D TQFT's. Combining it with other known mathematical results, it becomes clear that the monoidal 2-category of topological defects of codimension 1 or higher in a 2+1D topological order \cite{KK12,Kon14e} is actually a fusion 2-category (see \cite{KLWZZ20} for a detailed explanation). The theory of condensation completion was generalized to higher categories by Gaiotto and Johnson-Freyd in \cite{GJF19} under the name of ``Karoubi completion'' or ``condensation completion'' (see also \cite{CRS19} and a rigorous treatment for 3-categories in \cite{CM23}). This theory leads Johnson-Freyd to a mathematical theory of a (multi-)fusion $n$-category in \cite{JF22}. The compatibility between condensation completion and boundary-bulk relation \cite{KWZ15,KWZ17} was first shown in \cite{KTZ20} for 3+1D Dijkgraaf-Witten theories, and was proved in general context by Johnson-Freyd in \cite{JF22} (see \cite{KZ22b} for a slightly different approach). This compatibility was also proposed in \cite{KLWZZ20} as general physical principle beyond topological orders (see also \cite{KZ22b}). The theory of condensation completion was also generalized to gapless quantum liquids \cite{KZ22b,KZ22}.

\medskip
Although the theory of defect condensations was significantly advanced by that of condensation completion, it is still largely incomplete. The condensation of a 1-codimensional topological defect was briefly studied for 1+1D anomalous topological orders in \cite{KS11a,FSV13,CR16,Kon14e,DR18} and for higher dimensional cases in the condensation completion theory \cite{GJF19}. 
The condensation theory of $k$-codimensional topological defects was known for $k=2$ in 2+1D \cite{Kon14e} and 3+1D \cite{ZLZHKT23,DX24,Xu24}, and is widely open for $k>2$ in higher dimension theories. Most of these results are too brief and formal, and are lack of physical explanations and systematic constructions of examples. The relation between the condensations of the defects of different codimensions have not been studied. Moreover, there are serious gaps in turning the theory into a computable one, i.e., expressing the condensation data and results in explicit coordinate systems (see Theorem$^{\mathrm{ph}}$\,\ref{thm:condensable_E1Alg_in_coordinates_classification} (2),(4),(5)). 
Recently, the category theory of anyon condensations has inspired a lot of works on topological phase transitions in 1+1D (see for example \cite{CJKYZ20,CW23a,LY23,CJW25,CAW24} and references therein). As we are entering a new era of systematically studying topological phase transitions, there are increasing demands to have a rather complete theory of defect condensations in all dimensions. 

The main goal of this work is to develop such a theory, i.e., a theory of the condensations of $k$-codimensional topological defects in a (potentially anomalous) $n+$1D topological order for $1\leq k \leq n$. We show that this theory naturally leads us to the mathematical theory of higher algebras in higher monoidal categories and their higher representations. We also provide some systematic constructions of examples. This theory can be naturally generalized to the condensations of liquid-like gapless defects in topological orders or in gapped/gapless quantum liquids. We provide a glimpse of this generalized theory at the end of this work, together with other questions or generalizations naturally arising from our theory.

\subsection{Main results}

Since invertible topological orders do not play any role in our defect condensation theory, throughout this work, a topological order is defined up to invertible ones. 

\medskip
In order to state the main results, we need first introduce some mathematical notions. We want to keep it brief because a detailed review of these notions is in later sections. 

Consider a potentially anomalous $n+$1D topological order $\SC^{n+1}$ (see Definition\,\ref{def:anomaly_TO}). We want to define what we mean by condensing a $k$-codimensional topological defect $A$ in $\SC^{n+1}$. This condensation produces a condensed phase denoted by $\SD^{n+1}$ and a gapped domain wall denoted by $\SM^n$. We denote the category of topological defects of codimension 1 and higher in $\SC^{n+1}$ (resp. $\SD^{n+1}$) by $\CC$ (resp. $\CD$), which is a multi-fusion $n$-category \cite{JF22} with a tensor unit $\one_\CC$ (resp. $\one_\CD$). If $\SC^{n+1}$ is simple (see Definition\,\ref{def:simple_TO}), then $\CC$ is a fusion $n$-category. The gapped domain wall $\SM^n$ can be described by a pair $(\CM,m)$, where $\CM$ is the category of all wall conditions (as objects) and topological defects (as morphisms) and $m$ is a distinguished wall condition (i.e., $m\in \CM$). 
We set $\Omega\CC:=\hom_\CC(\one_\CC,\one_\CC)$, $\Omega_m\CM := \hom_\CM(m,m)$ and $\Omega^k\CC = \Omega(\Omega^{k-1}\CC)$. It is known that $\Omega^{k-1}\CC$ is an $\EE_k$-multi-fusion $(n-k+1)$-category \cite{KZ22b}, which is a direct sum of indecomposable separable $(n-k+1)$-categories. We denote the direct summand of $\Omega^{k-1}\CC$ consisting of the tensor unit by $\Sigma\Omega^k\CC$, which is called the delooping of $\Omega^k\CC$. Mathematically, the delooping of an $\EE_1$-multi-fusion $l$-category $\CA$, denoted by $\Sigma\CA$, can be defined by a one-point delooping of $\CA$ followed by condensation completion \cite{DR18,GJF19} (see a physical review in Section\,\ref{sec:cc}), and can be mathematically identified with the category $\RMod_{\CA}((l+1)\vect)$ of right separable $\CA$-modules, or equivalently, right $\CA$-modules in $(l+1)\vect$ \cite{DR18,GJF19}, where $n\vect := \Sigma^n\Cb$ \cite{GJF19} and $\vect$ is the 1-category of finite dimensional vector spaces over $\Cb$. 

\medskip
Now we summarize the main results of this work. 
\begin{pthm} \label{pthm:main_result_intro}
Consider a (potentially anomalous) $n+$1D simple topological order $\SC^{n+1}$, i.e., $\CC$ is a fusion $n$-category. A $k$-codimensional topological defect $A$ in $\SC^{n+1}$ is fully condensable if $A$ is a condensable $\EE_k$-algebra in $\Omega^{k-1}\CC$ (see Definition\,\ref{def:condensable_algebra}). By condensing $A$, we mean a $k$-step process explained below. 
\bnu

\item In the first step, we condense $A$ along one of the transversal directions $x^1,\cdots, x^k$ (see Convention\,\ref{rem:convention_op},\ \ref{conv:xi-direction}) of $A$, thus obtaining a $(k-1)$-codimensional defect $\Sigma A$ in $\Sigma\Omega^{k-1}\CC \hookrightarrow \Omega^{k-2}\CC$. 
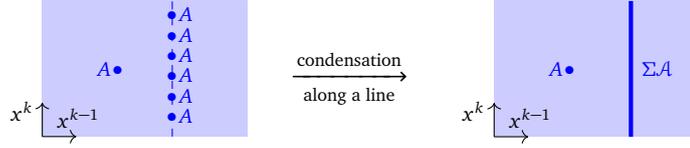
$$
\begin{array}{c}
\begin{tikzpicture}[scale=0.9]
\fill[blue!20] (0,0) rectangle (3,2) ;
\draw[->] (0,0) -- (0.5,0) node [near end,above] {\footnotesize $\quad x^{k-1}$};
\draw[->] (0,0) -- (0,0.5) node [near end,left] {\footnotesize $x^k$};
\node at (1,1) {\footnotesize \textcolor{blue}{$A\, \bullet$}} ;
\node at (2,0.3) {\footnotesize \textcolor{blue}{$\bullet\, A$}} ;
\node at (2,0.6) {\footnotesize \textcolor{blue}{$\bullet\, A$}} ;
\node at (2,0.9) {\footnotesize \textcolor{blue}{$\bullet\, A$}} ;
\node at (2,1.2) {\footnotesize \textcolor{blue}{$\bullet\, A$}} ;
\node at (2,1.5) {\footnotesize \textcolor{blue}{$\bullet\, A$}} ;
\node at (2,1.8) {\footnotesize \textcolor{blue}{$\bullet\, A$}} ;
\draw[blue, dashed] (1.9,0) -- (1.9,2) ;
\end{tikzpicture}
\end{array}
\quad
\xrightarrow[\text{\scriptsize along a line}]{\text{\scriptsize condensation}}
\quad
\begin{array}{c}
\begin{tikzpicture}[scale=0.9]
\fill[blue!20] (0,0) rectangle (3,2) ;
\draw[->] (0,0) -- (0.5,0) node [near end,above] {\footnotesize $\quad x^{k-1}$};
\draw[->] (0,0) -- (0,0.5) node [near end,left] {\footnotesize $x^k$};
\node at (1,1) {\footnotesize \textcolor{blue}{$A\, \bullet$}} ;
\draw[blue, ultra thick] (2,0) -- (2,2) node [midway,right] {\footnotesize $\Sigma A$}; 
\end{tikzpicture}
\end{array}
$$
It turns out that $\Sigma\Omega^{k-1}\CC$ can be identified with $\RMod_{\Omega^{k-1}\CC}((n-k+2)\vect)$, which can be viewed as a coordinate system of $\Sigma\Omega^{k-1}\CC$. In this coordinate system, $\Sigma A$ can be identified with the category $\RMod_A(\Omega^{k-1}\CC)$ of right $A$-modules in $\Omega^{k-1}\CC$, i.e., $\Sigma A = \RMod_A(\Omega^{k-1}\CC) \in \Sigma\Omega^{k-1}\CC\hookrightarrow \Omega^{k-2}\CC$. 

\item It turns out that $\Sigma A$ is automatically a condensable $\EE_{k-1}$-algebra in $\Sigma\Omega^{k-1}\CC$ and, therefore, a condensable $\EE_{k-1}$-algebra in $\Omega^{k-2}\CC$, which can be further condensed along one of the remaining transversal directions as the second step. So on and so forth. 
In the $(k-1)$-th step, we obtain a 1-codimensional topological defect $\Sigma^{k-1}A$, which is automatically a condensable $\EE_1$-algebra in $\CC$. 

\item In the $k$-th step, we condense $\Sigma^{k-1}A$ in $\CC$ along the unique remaining transversal direction. We obtain a new topological order $\SD^{n+1}$ and a gapped domain wall $\SM^n$ such that 
\be \label{pic:C-M-D}
\begin{array}{c}
\begin{tikzpicture}[scale=1.8]
\fill[blue!20] (1,0) rectangle (3,1.2) ;
\node at (1.2,0.5) {\scriptsize $\SC^{n+1}$} ;
\node at (2,-0.1) {\scriptsize $\SM^n=(\CM,m)$} ;
\fill[teal!20] (2.02,0) rectangle (3,1.2) ;
\draw[blue, ultra thick,->-] (2,0) -- (2,1.2) node [midway,left] {\scriptsize $\SM^n$}; 
\node at (2.8,0.5) {\scriptsize $\SD^{n+1}$} ;
\node at (1.5,1) {\scriptsize $\Omega^i\CC \xrightarrow{L_{i+1}} \Omega_m^i\CM$} ;
\node at (2.55,1) {\scriptsize $\Omega_m^i\CM \xleftarrow{R_{i+1}} \Omega^i\CD$} ;
\draw[->] (1,0) -- (1.25,0) node [near end,above] {\footnotesize $\quad x^1$};
\draw[->] (1,0) -- (1,0.25) node [near end,left] {\footnotesize $x^2$};
\end{tikzpicture}
\end{array}
\quad\quad
\begin{array}{l}
\CD \simeq \Mod_{\Sigma^{k-1}A}^{\EE_1}(\CC); \\
\\
\CM \simeq \RMod_{\Sigma^{k-1}A}(\CC)=\Sigma^k A,  \\
  m=\Sigma^{k-1}A \in \CM, 
\end{array}  
\ee
where $\Mod_{\Sigma^{k-1}A}^{\EE_1}(\CC)$ denotes the category of $\EE_1$-modules (i.e., bimodules) over $\Sigma^{k-1}A$ in $\CC$ and is an indecomposable multi-fusion $n$-category; and $\RMod_{\Sigma^{k-1}A}(\CC)$ denotes the category of right $\Sigma^{k-1}A$-modules in $\CC$ and is a separable $n$-category. Moreover, $\CD$ is Morita equivalent to $\CC$ with the Morita equivalence defined by the invertible $\CC$-$\CD$-bimodule $\CM$. When the algebra $A$ is indecomposable, $\CD$ is a fusion $n$-category and $\CM$ is an indecomposable left $\CC$-module. 

\item For $1\leq i \leq k$, the category of topological defects of codimension $i$ and higher in $\SD^{n+1}$ is determined by the following canonical $\EE_i$-monoidal equivalence:
\[
\Omega^{i-1}\CD = \Omega^{i-1} \Mod_{\Sigma^{k-1}A}^{\EE_1}(\CC) \simeq \Mod_{\Sigma^{k-i}A}^{\EE_i}(\Omega^{i-1}\CC),
\] 
where $\Mod_{\Sigma^{k-i}A}^{\EE_i}(\Omega^{i-1}\CC)$ denotes the category of $\EE_i$-modules over $\Sigma^{k-i}A$ in $\Omega^{i-1}\CC$. Roughly speaking, an $\EE_i$-module over $\Sigma^{k-i}A$ in $\Omega^{i-1}\CC$ is an object in $\Omega^{i-1}\CC$ equipped with an $i$-dimensional $A$-action. 

For $1\leq i \leq k$, the category of topological defects of codimension $(i-1)$ and higher in $\SM^n$ is determined by the following canonical equivalences: 
\[
\Omega_m^{i-1}\CM = \Omega^{i-1} \Sigma^k A = \Omega^{i-1} \RMod_{\Sigma^{k-1}A}(\CC) 
\simeq
\RMod_{\Sigma^{k-i}A}(\Omega^{i-1}\CC). 
\]

\item For $1\leq i \leq k$, $i$-codimensional topological defects in $\Omega^{i-1}\CC$ and those in $\Omega^{i-1}\CD$ move onto the wall according to the following functors, respectively, as illustrated in (\ref{pic:C-M-D}),
\[
L_i \coloneqq -\otimes \Sigma^{k-i}A \colon \Omega^{i-1}\CC \to \Omega_m^{i-1}\CM ;
\quad\quad
R_i \coloneqq \Sigma^{k-i}A \otimes_{\Sigma^{k-i}A} - \colon \Omega^{i-1}\CD \to \Omega_m^{i-1}\CM .
\]

\item $\SC^{n+1}$ and $\SD^{n+1}$ share the same gravitational anomaly. In other words, they are gapped boundaries of the same anomaly-free $n+$2D topological order as illustrated below,
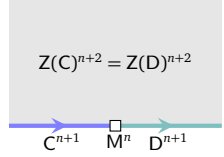
\[
\begin{array}{c}
\begin{tikzpicture}[scale=1.4]
\fill[gray!20] (0,0) rectangle (2,1.2) ;
\node at (1,0.6) {\scriptsize $\SZ(\SC)^{n+2}=\SZ(\SD)^{n+2}$} ;
\draw[blue!50, ultra thick,->-] (0,0) -- (1,0) ; 
\draw[teal!50, ultra thick,->-] (1,0) -- (2,0) ; 
\draw[fill=white] (0.95,-0.05) rectangle (1.05,0.05) node[midway,below,scale=1] {\scriptsize $\hspace{1mm}\SM^n$} ;
\node at (0.5,-0.15) {\scriptsize $\SC^{n+1}$} ;
\node at (1.5,-0.15) {\scriptsize $\SD^{n+1}$} ;
\end{tikzpicture}
\end{array}
\]
where $\SZ(\SC)^{n+2}$ and $\SZ(\SD)^{n+2}$ denote the bulk of $\SC^{n+1}$ and $\SD^{n+1}$, respectively. Mathematically, it implies the following results. 
\begin{align}
\FZ_1(\CC) &\simeq \FZ_1(\Mod_{\Sigma^{k-1}A}^{\EE_1}(\CC)), \\
\FZ_0(\RMod_{\Sigma^{k-1}A}(\CC)) &\simeq \Mod_{\Sigma^{k-1}A}^{\EE_1}(\CC)^\rev \boxtimes_{\FZ_1(\CC)} \CC, \\
\CD^\rev \simeq \Mod_{\Sigma^{k-1}A}^{\EE_1}(\CC)^\rev 
&\simeq \FZ_1(\CC, \FZ_0(\RMod_{\Sigma^{k-1}A}(\CC))),
\end{align}
where $\FZ_0(-),\FZ_1(-),\FZ_1(-,-)$ denote the $\EE_0$-center, the $\EE_1$-center and $\EE_1$-centralizer (see Definition\,\ref{def:centralizer+center}) and the superscript $^\rev$ is defined in Definition\,\ref{def:convention_op}. Our convention of left and right is explained below Figure\,\ref{fig:functoriality_bbr}. 
\enu 
\end{pthm}

\begin{rem} \label{rem:SET}
In most physical literature, physicists study condensations in an anomaly-free topological orders (see Definition\,\ref{def:anomaly_TO}). In this work, we provide a condensation theory for potentially anomalous topological orders (e.g.,  the boundary phase $\SC^{n+1}$ of a non-trivial bulk phase $\SZ(\SC)^{n+2}$). It is not only because our theory naturally covers all the anomalous cases, but also because a defect condensation in an anomalous $n+$1D topological orders, according to the so-called topological Wick rotation \cite{KZ18b,KZ20,KZ21,KLWZZ20,KLWZZ20a,KZ22b}, automatically defines a phase transition between two anomaly-free $n+$1D gapped quantum liquids (including SPT/SET/SSB orders) with a finite internal symmetry, and automatically defines the gauging of categorical symmetries in a gapless QFT. In other words, our main results 
automatically give a mathematical theory of condensations in SPT/SET/SSB orders 
(see some concrete examples in 1+1D in  \cite{KWZ22,XZ24}), and a theory of the gauging of categorical symmetries in gapless QFT's (see Section\,\ref{sec:general_gauging} for more discussions on the gauging of symmetries in QFT's). 
\end{rem}

\begin{rem}
We state the main results for simple topological orders. In the later sections, we slightly generalize them to composite topological orders (see Definition\,\ref{def:simple_TO}) for the sake of completeness. Physics oriented readers can ignore the composite cases at least in their first reading. 
\end{rem}

In the special case when $\SC^{n+1}$ is anomaly-free, we can define the same condensation alternatively by combining the last two-steps into one as stated below (see Section\,\ref{sec:2codim_nd_general_theory} for more related results). 
\begin{pthm}
When $\SC^{n+1}$ is anomaly-free and simple, if $A$ is a simple condensable $\EE_k$-algebra in $\Omega^{k-1}\CC$ for $k\geq 2$, then we can modify the $k$-step process of condensing $A$ by combining the last two steps into a single step of condensing $\Sigma^{k-2}A$ in $\Omega\CC$ directly in the remaining two transversal directions. Note that $\Sigma^{k-2}A$ is automatically a condensable $\EE_2$-algebra in $\Omega\CC$. This condensation reproduces the same condensed phase $\SD^{n+1}$ and the same gapped domain wall $\SM^n$ as explained below. 
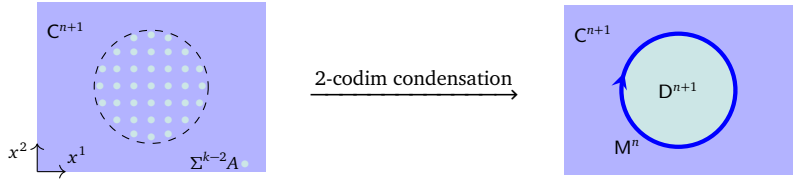
\[
\begin{array}{c}
\begin{tikzpicture}[scale=0.75]
\fill[blue!30] (-2,0) rectangle (2,3) ;
\node at (-1.5,2.5) {\scriptsize $\SC^{n+1}$} ;
\node at (0,0.9) {\scriptsize \textcolor{teal!20}{$\bullet$}} ;
\node at (0,1.2) {\scriptsize \textcolor{teal!20}{$\bullet$}} ;
\node at (0,1.5) {\scriptsize \textcolor{teal!20}{$\bullet$}} ;
\node at (0,1.8) {\scriptsize \textcolor{teal!20}{$\bullet$}} ;
\node at (0,2.1) {\scriptsize \textcolor{teal!20}{$\bullet$}} ;
\node at (0,2.4) {\scriptsize \textcolor{teal!20}{$\bullet$}} ;
\node at (0,0.6) {\scriptsize \textcolor{teal!20}{$\bullet$}} ;
\node at (0.3,1.5) {\scriptsize \textcolor{teal!20}{$\bullet$}} ;
\node at (-0.3,1.5) {\scriptsize \textcolor{teal!20}{$\bullet$}} ;
\node at (0.6,1.5) {\scriptsize \textcolor{teal!20}{$\bullet$}} ;
\node at (-0.6,1.5) {\scriptsize \textcolor{teal!20}{$\bullet$}} ;
\node at (-0.6,1.8) {\scriptsize \textcolor{teal!20}{$\bullet$}} ;
\node at (-0.6,2.1) {\scriptsize \textcolor{teal!20}{$\bullet$}} ;
\node at (-0.6,1.2) {\scriptsize \textcolor{teal!20}{$\bullet$}} ;
\node at (-0.6,0.9) {\scriptsize \textcolor{teal!20}{$\bullet$}} ;
\node at (-0.9,1.5) {\scriptsize \textcolor{teal!20}{$\bullet$}} ;
\node at (0.9,1.5) {\scriptsize \textcolor{teal!20}{$\bullet$}} ;
\node at (0.3,1.2) {\scriptsize \textcolor{teal!20}{$\bullet$}} ;
\node at (-0.3,1.2) {\scriptsize \textcolor{teal!20}{$\bullet$}} ;
\node at (0.3,1.8) {\scriptsize \textcolor{teal!20}{$\bullet$}} ;
\node at (-0.3,1.8) {\scriptsize \textcolor{teal!20}{$\bullet$}} ;
\node at (-0.85,1.8) {\scriptsize \textcolor{teal!20}{$\bullet$}} ;
\node at (-0.85,1.2) {\scriptsize \textcolor{teal!20}{$\bullet$}} ;
\node at (-0.3,2.1) {\scriptsize \textcolor{teal!20}{$\bullet$}} ;
\node at (-0.3,2.35) {\scriptsize \textcolor{teal!20}{$\bullet$}} ;
\node at (-0.3,0.9) {\scriptsize \textcolor{teal!20}{$\bullet$}} ;
\node at (-0.3,0.65) {\scriptsize \textcolor{teal!20}{$\bullet$}} ;
\node at (0.6,1.8) {\scriptsize \textcolor{teal!20}{$\bullet$}} ;
\node at (0.6,2.1) {\scriptsize \textcolor{teal!20}{$\bullet$}} ;
\node at (0.6,1.2) {\scriptsize \textcolor{teal!20}{$\bullet$}} ;
\node at (0.6,0.9) {\scriptsize \textcolor{teal!20}{$\bullet$}} ;
\node at (0.3,2.1) {\scriptsize \textcolor{teal!20}{$\bullet$}} ;
\node at (0.3,2.35) {\scriptsize \textcolor{teal!20}{$\bullet$}} ;
\node at (0.3,0.9) {\scriptsize \textcolor{teal!20}{$\bullet$}} ;
\node at (0.3,0.65) {\scriptsize \textcolor{teal!20}{$\bullet$}} ;
\node at (0.85,1.8) {\scriptsize \textcolor{teal!20}{$\bullet$}} ;
\node at (0.85,1.2) {\scriptsize \textcolor{teal!20}{$\bullet$}} ;

\node at (1.2,0.2) {\scriptsize $\Sigma^{k-2}A\,$\textcolor{teal!20}{$\bullet$}} ;

\draw[dashed] (1,1.5) arc (360:0:1) ;
\draw[->] (-2,0) -- (-1.5,0) ; 
\node at (-1.3,0.3) {\scriptsize $x^1$};
\draw[->] (-2,0) -- (-2,0.5) node [near end,left] {\scriptsize $x^2$};
\end{tikzpicture}
\end{array}
\quad\xrightarrow{\mbox{\footnotesize 2-codim condensation}} \quad
\begin{array}{c}
\begin{tikzpicture}[scale=0.75]
\fill[blue!30] (-2,0) rectangle (2,3) ;
\fill[teal!20] (0,1.5) circle (1);
\node at (-1.5,2.5) {\scriptsize $\SC^{n+1}$} ;
\node at (0,1.5) {\scriptsize $\SD^{n+1}$} ;
\node at (-0.9,0.5) {\scriptsize $\SM^n$} ;
\draw[blue, ultra thick,->-] (1,1.5) arc (360:0:1) ;
\end{tikzpicture}
\end{array}
\]
\bnu
\item We have the following monoidal equivalences: 
\be
\Omega\CD \simeq \Mod_{\Sigma^{k-2} A}^{\EE_2}(\Omega\CC) \quad \mbox{and} \quad
\Omega_m\CM \simeq \RMod_{\Sigma^{k-2} A}(\Omega\CC), 
\ee
where the first one is a braided equivalence or an $\EE_2$-monoidal equivalence.

\item The 2-codimensional topological defects in $\SC^{n+1}$ and $\SD^{n+1}$ move onto the wall according to the following functors: 
\[
L_2 \coloneqq -\otimes \Sigma^{k-2} A \colon \Omega\CC \to \Omega_m\CM, \quad\quad
R_2 \coloneqq \Sigma^{k-2} A \otimes_{\Sigma^{k-2} A} - \colon \Omega\CD \to \Omega_m\CM. 
\]
both of which are central functors that can be lifted to an $\EE_2$-monoidal (or braided) equivalence: 
\[
\FZ_1(\RMod_{\Sigma^{k-2} A}(\Omega\CC)) \simeq \Omega\CC \boxtimes \Mod_{\Sigma^{k-2} A}^{\EE_2}(\Omega\CC)^\rev, 
\]
or equivalently, an $\EE_2$-monoidal (or braided) equivalence: 
\[
\Mod_{\Sigma^{k-2} A}^{\EE_2}(\Omega\CC)^\rev \simeq \FZ_2(\Omega\CC, \FZ_1(\RMod_{\Sigma^{k-2} A}(\Omega\CC)). 
\]

\item Its compatibility with the $k$-step condensation leads to the following equivalences: 
\[
\CD \simeq \Mod_{\Sigma^{k-1}A}^{\EE_1}(\CC) \simeq \Sigma \Mod_{\Sigma^{k-2} A}^{\EE_2}(\Omega\CC), \quad\quad \RMod_{\Sigma^{k-1}A}(\CC) \simeq \Sigma\RMod_{\Sigma^{k-2} A}(\Omega\CC). 
\]

\enu
\end{pthm}

Some other related results are better summarized as pure mathematical results.  
\begin{thm}
Let $\CB$ be an $\EE_k$-fusion $n$-category and $A$ a condensable $\EE_k$-algebra in $\CB$. We set $\Sigma^0 A=A$ and $\Sigma^i A=\RMod_{\Sigma^{i-1}A}(\Sigma^{i-1}\CB)$  for $0\leq i\leq k$. We have a natural $\EE_{k-i}$-monoidal equivalence: 
\be 
\Mod_{\Sigma^i A}^{\EE_{k-i}}(\Sigma^i\CB) \simeq \Omega^{k-i}\Mod_{\Sigma^k A}^{\EE_0}(\Sigma^k\CB), 
\ee
a natural $\EE_{k-1-i}$-monoidal equivalence $\Sigma^i\RMod_A(\CB) \simeq \RMod_{\Sigma^i A}(\Sigma^i\CB)$ and an $\EE_{k-1}$-monoidal equivalence $
\RMod_A(\CB) \simeq \Omega^i\RMod_{\Sigma^i A}(\Sigma^i\CB)$ (see Theorem\,\ref{thm:Ek-fusion_nCat_EkAlg_A_condense}). We have an $\EE_k$-monoidal equivalence: 
\be 
\Mod_A^{\EE_k}(\CB)^\rev \simeq \FZ_k(\CB, \FZ_{k-1}(\RMod_A(\CB))), 
\ee
where $\FZ_k(-,-)$ is the $\EE_k$-centralizer and $\FZ_{k-1}(-)$ is the $\EE_{k-1}$-center (see Definition\,\ref{def:centralizer+center}). When $\CB=n\vect$, we obtain $\Mod_A^{\EE_k}(n\vect)^\rev \simeq \FZ_{k-1}(\Sigma A)$ (see Theorem\,\ref{cor:EkMod_Zk_centralizer}). 
\end{thm}

In addition to the general theory summarized above, we also provide some general constructions of $k$-codimensional defect condensations. These results are summarized below. 
\begin{pthm} \label{thm:condensable_E1Alg_in_coordinates_classification}
When $\CC$ is equipped with an explicit coordinate system, we can give a classification or a mathematical characterization of all possible condensations. 
\bnu 

\item[(1)] When $\SC^{n+1}$ is the trivial phase, i.e., $\SC^{n+1}=\mathbf{1}^{n+1}$, we have $\CC=n\vect$. An indecomposable condensable $\EE_k$-algebra $\CA$ in $\Omega^{k-1}\CC=(n-k+1)\vect$ is precisely an indecomposable $\EE_k$-multi-fusion $(n-k)$-categories. By condensing $\CA$, we obtain a new phase $\SD^{n+1}$ and a gapped domain wall $\SM^n$ such that (see Section\,\ref{sec:general_example_k-codim})
\begin{align*}
\CD &\simeq \Mod_{\Sigma^{k-1}\CA}^{\EE_1}(n\vect) \simeq \FZ_0(\Sigma^k\CA)^\rev, \quad\quad
(\CM,m) \simeq (\RMod_{\Sigma^{k-1}\CA}(n\vect), \Sigma^{k-1}\CA), \\
\Omega^{k-1}\CD &\simeq \Mod_\CA^{\EE_k}((n-k+1)\vect) \simeq \FZ_{k-1}(\Sigma\CA)^\op, \quad\quad
\Omega_m^{k-1}\CM \simeq \RMod_\CA((n-k+1)\vect).
\end{align*}
When $k=1$, we can obtain all non-chiral $n+$1D topological orders $\SD^{n+1}$ in this way by choosing a proper $\CA$, and $\CA$ is `Lagrangian' (in the sense that $\SD^{n+1}=\mathbf{1}^{n+1}$) if and only if $\CA$ is a non-degenerate indecomposable multi-fusion $(n-1)$-category (see Definition\,\ref{def:non-degenerate_Lagrangian}). 

When $k=2$, $\CD \simeq \Sigma\Mod_\CA^{\EE_2}((n-1)\vect)$, and 
$\CA$ is `Lagrangian' (in the sense that $\SD^{n+1}=\mathbf{1}^{n+1}$) if and only if $\CA$ is a non-degenerate braided fusion $(n-2)$-category (see Definition\,\ref{def:non-degenerate_Lagrangian}).

\item[(2)] When $\SC^{n+1}$ is anomaly-free and simple, we have $\CC=\Sigma\Omega\CC=\RMod_{\Omega\CC}(n\vect)$ as fusion $n$-categories. 
An indecomposable condensable $\EE_1$-algebra in $\CC$ is precisely an indecomposable multi-fusion $(n-1)$-category $\CA$ equipped with a braided monoidal functor $\phi: \Omega\CC \to \FZ_1(\CA)$. 
By condensing $\CA$ in $\CC$, we obtain a condensed phase $\SD^{n+1}$ and a gapped domain wall $\SM^n$ such that (see Theorem$^{\mathrm{ph}}$\,\ref{thm:condensable_E1Alg_af_C})
\begin{align*}
&\CD \simeq \Mod_\CA^{\EE_1}(\CC) \simeq \Sigma \FZ_2(\Omega\CC, \FZ_1(\CA))^\op, \quad\quad
\Omega\CD \simeq  \FZ_2(\Omega\CC, \FZ_1(\CA)); \\
&\CM \simeq \RMod_\CA(\CC) \simeq \RMod_\CA(n\vect), \quad\quad  m=\CA \in \RMod_\CA(\CC).
\end{align*}
Moreover, $\CA$ is Lagrangian in the sense that $\SD^{n+1}=\mathbf{1}^{n+1}$ if and only if $\phi$ is an equivalence. 

\item[(3)] Given two Morita equivalent anomaly-free simple topological orders $\SC^{n+1}$ and $\SD^{n+1}$ connected by a gapped domain wall $\SK^n$ as illustrated in (\ref{pic:chiral_TO_1codim_condensations}), there is a one-to-one correspondence between 
\bnu
\item[-] $(\CK\boxtimes \CT)$-modules modulo the equivalence relation: $\CX \sim \CY$ if $\CY \simeq \CX \boxtimes \CV$ for $\CV \in (n+1)\vect^\times$; 
\item[-] indecomposable condensable $\EE_1$-algebras in $\CC$ that define 1-codimensional defect condensations from $\SC^{n+2}$ to $\SD^{n+2}$ in the Morita class $[\SK^{n+1} \boxtimes \ST^{n+1}]$, 
\enu
defined by $\CX \mapsto A_\CX=\Fun_{\CK \boxtimes \CT}(\CX,\CX)^\op$ (in the coordinate $\CC=\RMod_{\Omega\CC}(n\vect)$). The condensable $\EE_1$-algebra $A_\CX$ is simple if and only if the module $\CX$ is indecomposable (see Theorem$^{\mathrm{ph}}$\,\ref{pcor:1codim_condensation_bw_two_chiral_TO}).

\item[(4)] When $\SC^{n+1}$ is anomaly-free, simple and non-chiral, i.e., $\SC^{n+1}$ admitting a gapped boundary $\SB^n$, in this case, we have a new coordinate system $\CC=\BMod_{\CB|\CB}(n\vect)^\op$ for an indecomposable multi-fusion $(n-1)$-category $\CB$. In this case, an indecomposable condensable $\EE_1$-algebra in $\CC$ is precisely an indecomposable multi-fusion $(n-1)$-category $\CA^\op$ equipped with a monoidal functor $\psi: \CB \to \CA$. By condensing $\CA^\op$ in $\CC$, we obtain a condensed phase $\SD^{n+1}$ and a gapped domain wall $\SM^n$ such that (see Theorem$^{\mathrm{ph}}$\,\ref{pthm:construct_cond_E1_algebras})
\begin{align*}
\CD &\simeq \Mod_{\CA^\rev}^{\EE_1}(\CC) \simeq \Mod_{\CA^\rev}^{\EE_1}(n\vect) \simeq \Mod_{\CA}^{\EE_1}(n\vect)^\rev \simeq \Sigma\FZ_1(\CA), 
\\
(\CM,m) &= (\LMod_{\CA}(\CC),\CA) = (\BMod_{\CA|\CB}(n\vect), \CA). 
\end{align*}
Moreover, $\CA^\rev$ is Lagrangian if and only if $\CA$ is a non-degenerate multi-fusion $(n-1)$-category. 

\item[(5)] When $\SC^{n+1}$ is anomaly-free, simple and non-chiral, there are two coordinate systems, i.e., $\CC=\RMod_{\Omega\CC}(n\vect)$ and $\CC=\BMod_{\CB|\CB}(n\vect)^\op$. The same condensable $\EE_1$-algebra in $\CC$ can be expressed in two coordinates. The coordinate transformation for condensable $\EE_1$-algebras in $\CC$ is illustrated below (see Theorem$^{\mathrm{ph}}$\,\ref{pthm:construct_cond_E1_algebras_II}). 
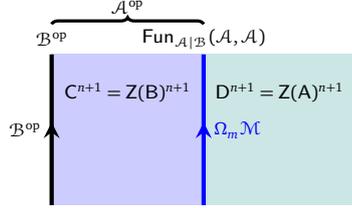
$$
\begin{array}{c}
\begin{tikzpicture}[scale=1]
\fill[blue!20] (-2,0) rectangle (0,2) ;
\fill[teal!20] (0,0) rectangle (2,2) ;
%
\draw[->-,ultra thick] (-2,0)--(-2,2) node[midway,left] {\scriptsize $\CB^\op$} ;

\draw[blue,->-,ultra thick] (0,0)--(0,2) node[midway,right] {\scriptsize $\Omega_m\CM$} ; 
%
%
%
\node at (-1,1.5) {\scriptsize $\SC^{n+1}=\SZ(\SB)^{n+1}$} ;
\node at (1,1.5) {\scriptsize $\SD^{n+1}=\SZ(\SA)^{n+1}$} ;
\node at (0,2.2) {\scriptsize $\Fun_{\CA|\CB}(\CA,\CA)$} ;
\node at (-2,2.2) {\scriptsize $\CB^\op$} ;

\draw[decorate,decoration=brace,very thick] (-2,2.4)--(0,2.4) node[midway, above] {\scriptsize $\CA^\op$};

\end{tikzpicture}
\end{array}
\quad\quad
\begin{array}{l}
\mbox{\small Two definitions of the same algebra in $\CC$}:  \\
\mbox{\small (2) the central functor $\FZ_1(\CB) \to \Fun_{\CA|\CB}(\CA,\CA)$} \\
\mbox{\small \quad\,\, defines $\Fun_{\CA|\CB}(\CA,\CA) \in \Algc_{\EE_1}(\RMod_{\FZ_1(\CB)}(n\vect))$}; \\
\mbox{\small (1) the monoidal functor $\CB \to \CA$} \\
\mbox{\small \quad\,\, defines $\CA^\op \in \Algc_{\EE_1}(\BMod_{\CB|\CB}(n\vect)^\op)$}. 
\end{array}
$$


\item[(6)] Let $\SD^{n+1}$ be the anomaly-free, simple and non-chiral phase obtained in the case (1) by condensing an $\EE_k$-fusion $(n-k)$-category $\CA$ as a $k$-codimensional defect in $\mathbf{1}^{n+1}$. A simple condensable $\EE_k$-algebra $B$ in $\Omega^{k-1}\CD \simeq \Mod_\CA^{\EE_k}((n-k+1)\vect) \simeq \FZ_{k-1}(\Sigma\CA)^\op$ is precisely defined by an $\EE_k$-monoidal functor $\CA\to \CB$ for an $\EE_k$-fusion $(n-k)$-category $\CB$. By condensing $B\in \Omega^{k-1}\CD$ in $\SD^{n+1}$ in $k$ steps, we obtain a new condensed phase $\SE^{n+1}$ such that 
$$
\begin{array}{c}
\begin{tikzpicture}[scale=0.6]
\fill[blue!20] (-2,0) rectangle (2,3) ;
\fill[teal!20] (0,0) rectangle (2,3) ;
\draw[blue, ultra thick, ->-] (0,0) -- (0,3) node[near start, left] {\footnotesize $\SM^n$};
\node at (-1,1.5) {\footnotesize $\mathbf{1}^{n+1}$} ;
\node at (1,1.5) {\footnotesize $\SD^{n+1}$} ;
\fill[red!20] (2,0) rectangle (4,3) ;
\draw[purple, ultra thick, ->-] (2,0) -- (2,3) node[near start, right] {\footnotesize $\SN^n$};
\node at (3.3,1.5) {\footnotesize $\SE^{n+1}$} ;
\end{tikzpicture}
\end{array}
\quad\quad
\begin{array}{c}
\CE=\Mod_{\Sigma^{k-1}B}^{\EE_1}(\CD) \simeq \Mod_{\Sigma^{k-1}\CB}^{\EE_1}(n\vect), \\
\\
\Omega^{k-1}\CE \simeq \Mod_\CB^{\EE_k}((n-k+1)\vect) \simeq \FZ_{k-1}(\Sigma\CB)^\op,
\end{array} 
$$ 
where $\Sigma^{k-1}\CB$ is equipped with an $\EE_1$-monoidal functor $\Sigma^{k-1} \CA \to \Sigma^{k-1}\CB$ and should be viewed as a condensable $\EE_1$-algebra in $\CD$. Moreover, the gapped domain wall $\SN^n=(\CN,n)$ produced by this condensation is given by 
\begin{align*}
&(\CN,n) = (\BMod_{\Sigma^{k-1}\CA|\Sigma^{k-1}\CB}(\Sigma^{k-1}\CB,\Sigma^{k-1}\CB), \Sigma^{k-1}\CB), \\
&\Omega_n^{k-1}\CN = \Omega^{k-2}\Fun_{\Sigma^{k-1}\CA|\Sigma^{k-1}\CB}(\Sigma^{k-1}\CB,\Sigma^{k-1}\CB) \simeq \RMod_\CB(\Mod_\CA^{\EE_k}((m+1)\vect)). 
\end{align*}
Note that this result is an $\EE_k$-analogue of (4) and is explained in details in Theorem$^{\mathrm{ph}}$\,\ref{pthm:iterate_Ek_condensation}, and the $k=2$ case is explained in Theorem$^{\mathrm{ph}}$\,\ref{pthm:iterate_E2_condensation}. 
\enu
\end{pthm}

Using (2) in above theorem and Theorem$^{\mathrm{ph}}$\,\ref{pthm:Z1map-made-1to1} and \ref{pcor:chiral_TO_wall_1to1}, we obtain an efficient way to construct and classify all 1-codimensional defect condensations between any two $n+$2D topological orders (see Corollary$^{\mathrm{ph}}$\ \ref{pcor:n=3_classification}), and a one-to-one correspondence between Lagrangian $\EE_1$-algebras (or Lagrangian $\EE_2$-algebras) in a non-chiral $n+$2D topological order and certain module $n$-categories (see Theorem$^{\mathrm{ph}}$\,\ref{pcor:4D_Lagrangian_1to1_modules}, \ref{pthm:iterate_E2_condensation} and Corollary$^{\mathrm{ph}}$\ \ref{cor:A_connected_classification_bdy_Z(A)}). We summarize the key results below. 

\begin{pthm} \label{pthm:intro_gapped boundary_classification}
For a non-chiral simple topological order $\SZ(\SA)^{n+2}$ with a simple gapped boundary $\SA^{n+1}$ (i.e., $\CA$ is a fusion $n$-category) and a non-degenerate fusion $n$-category $\CT$, there are one-to-one correspondences among the following sets: (the idea is illustrated below)
$$
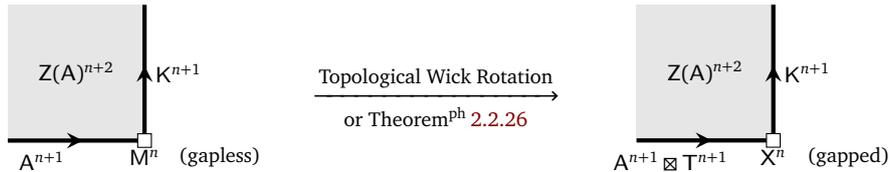
\begin{array}{c}
\begin{tikzpicture}[scale=0.9]
\fill[gray!20] (-3,1) rectangle (-1,3) ;
\draw[ultra thick,->-] (-1,1)--(-1,3) node [midway, right] {\footnotesize $\SK^{n+1}$} ;
\draw[ultra thick,->-] (-3,1)--(-1,1) ;
\draw[fill=white] (-1.1,0.9) rectangle (-0.9,1.1) node[midway,below] {\footnotesize $\SM^n$} ;
\node at (-2.5,0.7) {\footnotesize $\SA^{n+1}$} ;
\node at (0.1,0.75) {\footnotesize (gapless)} ;
\node at (-2,2) {\footnotesize $\SZ(\SA)^{n+2}$} ;
\end{tikzpicture} 
\end{array}
\quad \xrightarrow[\mbox{\footnotesize or Theorem$^{\mathrm{ph}}$\,\ref{pthm:TME-action_transitive}}]{\mbox{\footnotesize Topological Wick Rotation}} \quad
\begin{array}{c}
\begin{tikzpicture}[scale=0.9]
\fill[gray!20] (-3,1) rectangle (-1,3) ;
\draw[ultra thick,->-] (-1,1)--(-1,3) node [midway, right] {\footnotesize $\SK^{n+1}$} ;
\draw[ultra thick,->-] (-3,1)--(-1,1) ;
\draw[fill=white] (-1.1,0.9) rectangle (-0.9,1.1) node[midway,below] {\footnotesize $\SX^n$} ;
\node at (-2.5,0.7) {\footnotesize $\SA^{n+1} \boxtimes \ST^{n+1}$} ;
\node at (0.1,0.75) {\footnotesize (gapped)} ;
\node at (-2,2) {\footnotesize $\SZ(\SA)^{n+2}$} ;
\end{tikzpicture} 
\end{array}
$$
\begin{itemize}

\item[(1)] left indecomposable $\CA\boxtimes \CT$-modules modulo the equivalence relation: $\CX\sim \CY$ if $\CY \simeq \CX \boxtimes \CV$ for an invertible separable $n$-category $\CV$; 

\item[(2)] simple gapped boundaries of $\SZ(\SA)^{n+2}$ in the Morita class $[\SA^{n+1}\boxtimes \ST^{n+1}]$ (see Definition\,\ref{def:TME}); 

\item[(3)] Lagrangian $\EE_1$-algebras in $\Sigma\FZ_1(\CA)$ in the Morita class $[\CA\boxtimes \CT]$ (see Remark\,\ref{rem:MorEq_condensable_E1Alg});

\item[(4)] Lagrangian $\EE_2$-algebras in $\FZ_1(\CA)$ in the 2-Morita class $[\CA\boxtimes \CT]$ (see Theorem$^{\mathrm{ph}}$\,\ref{conj:Lagrangian=boundary}). 

\end{itemize} 
By allowing $\CT$ to run through all Morita classes, we obtain a classification of all gapped boundaries of $\SZ(\SA)^{n+2}$, all Lagrangian $\EE_1$-algebras in $\Sigma\FZ_1(\CA)$ and all Lagrangian $\EE_2$-algebra in $\FZ_1(\CA)$. When $n=1$, we recover an old result in 1-categories \cite{KR08,DMNO13}. 

Moreover, when the fusion $n$-category $\CA$ is indecomposable as a separable $n$-category, i.e., $\CA\simeq \Sigma\Omega\CA$, there is a one-to-one correspondence between the set of braided monoidal functors $\Omega\CA\to \CO$ (up to isomorphisms) for a non-degenerate braided fusion $(n-1)$-category $\CO$ and that of Lagrangian $\EE_2$-algebras in $\FZ_1(\CA)$. This result generalizes an earlier result for the $n=2$ case \cite{JFR23,DX24}. 
\end{pthm}

We also obtain many other interesting new results along the way as we develop our theory. We list a few important examples and highlight other important contents of this work. 
\bnu

\item[(1)]  The theory of the categories of topological defects in topological orders in all dimensions was initiated in \cite{KW14,KWZ15}, and was fully established in \cite{GJF19,JF22} (see also \cite{KLWZZ20}), and was further developed in \cite{KZ22,KZ24, KZ22b}. We give a comprehensive review of the historical developments of this theory with an emphasis on the physical intuitions and physical principles instead of mathematical rigor. Although the main results of this theory have already appeared in literature \cite{KWZ15,JF22,KLWZZ20,KZ22}, the complete physical reasoning behind this theory, including its limitations, is told for the first time in Section\,\ref{sec:topological_defects}. 

\item[(2)] We review the complete mathematical formulation of boundary-bulk relation as the center functor in Section\,\ref{sec:center_functor}. In Section\,\ref{sec:measure_non-fully-faithfulness}, we point out that the proof of Proposition$^*$ 3.13 in \cite{KZ24} actually provide a precise mathematical characterization of the non-surjectivity and the non-fully-faithfulness of the center functor, which is stated explicitly in Theorem\,\ref{thm:center-functor_surjective_full_faithful} or \ref{thm:center-functor_surjective_full_faithful_2}. This further leads to interesting result such as Theorem$^{\mathrm{ph}}$\,\ref{pthm:Z1map-made-1to1}, Theorem$^{\mathrm{ph}}$\,\ref{pcor:chiral_TO_wall_1to1} and Corollary$^{\mathrm{ph}}$\,\ref{cor:A_connected_classification_bdy_Z(A)}, Theorem$^{\mathrm{ph}}$\,\ref{pthm:TME-action_transitive}, Theorem$^{\mathrm{ph}}$\,\ref{pthm:recover_all_bdy_from_one} and results summarized in Theorem$^{\mathrm{ph}}$\,\ref{pthm:intro_gapped boundary_classification}. 

\item[(3)] In addition to the general constructions of condensable $\EE_k$-algebras (see  Section\,\ref{sec:general_example_1-codim}, \ref{sec:general_example_2-codim}, \ref{sec:general_example_k-codim}), we also provide some concrete examples of 1-codimensional, 2-codimensional and $k$-codimensional condensations in Section\,\ref{sec:1codim_2d}, \ref{sec:example_1codim_4D}, \ref{sec:example_2codim_3D}, \ref{sec:example_2codim_4D}, \ref{sec:example_12codim_3D} and \ref{sec:example_kcodim}. Even for the well-known theory of anyon condensations in 2+1D, we retell the story from a new perspective, emphasizing the role of higher algebras and higher representations in the condensation theory. This perspective makes the later generalizations to condensations in higher (co)-dimensions very natural. We also provide many examples of anyon condensations in Section\,\ref{sec:example_2codim_3D}. Readers can use it as a guide to literature. 

\item[(4)] We develop a geometric approach towards defect condensations. In particular, we develop the geometric approach towards the particle condensations in 1+1D in Section\,\ref{sec:geometric_theory_1codim_2} and the anyon condensations in 2+1D in Section\,\ref{sec:2codim_3D_geometric_approach} (see the key result Theorem\,\ref{thm:geo_A=internal_hom}). In this geometric approach, we emphasize the construction of condensable algebras via internal homs, a mathematical notion which is reviewed in Section\,\ref{sec:internal_hom}. The complete geometric theory is quite involved. In order to control the complexity of mathematics, we move some mathematically technical results to Appendix\,\ref{sec:proof_A=internal_hom} and \ref{Appendix^EMcal}. In particular, in Appendix\,\ref{Appendix^EMcal}, we develop a relative version of classical Eilenberg-Watts calculus that has independent values. These geometric theories can be generalized to higher dimensions and lead to a theory of integrals, which is briefly outlined in Section\,\ref{sec:FH}. It seems that it includes but generalizes the so-called $\alpha$-version of factorization homology \cite{Lur17,AFT17,AFT16,AF20} and should be related (or even equivalent) to the so-called $\beta$-version of factorization homology \cite{AFR18,AF24} (see Remark\,\ref{rem:alpha_version_FH}).

\item[(5)] In Section\,\ref{sec:gauging_G-symmetry}, we establish the connections or equivalences among various notions of gauging 0-form $G$-symmetries in a 2+1D topological order, including $G$-crossed extension approach \cite{BBCW19}, minimal nondegenerate extension approach \cite{LKW16a,KLWZZ20} and the 1-codimensional defect condensation approach (or via a monoidal functor \cite{LYW24}). 

\item[(6)] Although we mainly focus on the condensations of topological defects in topological orders, our theory naturally generalizes to situations beyond gapped defects. In Section\,\ref{sec:gapless}, we briefly discuss how the theory generalizes to the condensations of liquid-like gapless defects in topological orders and the condensations of quantum liquids by focusing only on the topological data of gapless defects or phases. We also provide an illustrating example in 2+1D (see the key result in Theorem$^{\mathrm{ph}}$\,\ref{pthm:condensing_cylinder_3D}). More detailed study of the condensations of gapless defects will appear elsewhere. 

\item[(7)] The higher representation theory of higher algebras is much richer than that of an $\EE_1$-algebra. In addition to results summarized in this section, we obtain some other interesting results along the way. In Section\,\ref{sec:Higher_Morita_Theory}, we briefly discuss some basic ingredients in the higher Morita theory for condensable $\EE_k$-algebras, including introducing the notions of various higher Morita equivalences and proving some basic results. We also introduce the notion of the enveloping algebra of a condensable $\EE_k$-algebra, which is an important tool in the study of higher representation theory. 

\enu

In additional to many important algebraic constructions, we emphasize a geometric way to understand higher algebras and higher representations (see Section\,\ref{sec:E1_algebra}, \ref{sec:E2=E1+E1} and \ref{sec:anyon_cond_algebra}) and almost all mathematical results appeared in this work.  Moreover, we emphasize a geometric way to construct these condensable $\EE_k$-algebras as a theory of integrals (i.e., integrating local observables to a global ones). We show that some computations can be carried out by purely geometrical and physical intuitions (see Section\,\ref{sec:geometric_intuition_1codim}, \ref{sec:geometric_intuition_1codim}, \ref{sec:2codim_3D_geometric_approach}, \ref{sec:2codim_nd_general_theory}, \ref{sec:FH} and \ref{sec:gapless_condensation_3D}). One of the main goals of this work is to convince readers that almost all ingredients in our theory of topological orders and defect condensations, based on higher categories, higher algebras and higher representations, have geometric meanings. Moreover, many deep and new algebraic results can be derived from the geometric intuitions with the help from only a handful physical principles.

\begin{rem} \label{rem:math_assumptions_intro}
Our theory shows that defect condensations is essentially a theory of higher algebras (e.g., $\EE_k$-algebras) and higher representations (e.g., $\EE_k$-modules). Although we try to explain many of these mathematical notions in details, a mathematically rigorous treatment of this subject is far beyond this paper, which originally aimed at physics oriented readers. A more mathematical treatment of this subject will appear in a mathematical companion of this paper \cite{KZZZ25}. In this work, we provide an almost self-contained story, in which most mathematical complexity is replaced by physical principles and self-evident geometric or physical intuitions. We show that these geometric and physical intuitions, together with necessary physical principles, are powerful enough for us to build a self consistent and rather complete theoretical framework. This framework allows us to establish a precise correspondence between geometric or physical manipulations and algebraic computations. Once this correspondence is established, we are able to do precise computations based on purely geometric intuitions. This leads us to many new results that are not available in mathematical literature. In order to make this framework to work, we make a technical assumption that we always work within separable higher categories \cite{KZ22}, and (higher) algebras are always separable (also called condensable in this work), and their higher centers or higher centralizers, internal homs and their categories of $\EE_k$-modules are always separable (as higher algebras or higher categories). Some parts of this assumption were known as mathematical theorems for 1-categories and 2-categories (see for example \cite{Ost03,Dec25,Dec23a,Dec23b,DX24,Xu24}).
\end{rem}

\begin{rem}
All the condensations studied in this work are bosonic in nature. We leave a systematic study of fermionic condensations to the future. See \cite{ALW19,WW17,LSCH21,GK21,ZWG22} and references therein for existing works on this topics. Although we restrict our study in the setting of $\EE_k$-multi-fusion higher categories, we believe that the defect condensations in QFT's with continuous symmetries are similar. Namely, they should also be defined by higher algebras and higher representations in more general $\EE_k$-monoidal higher categories. 
\end{rem}


\begin{rem} \label{rem:FQFT_AQFT}
On the one hand, for high energy theorists or the mathematical physicists following the program of functorial quantum field theories (FQFT), the notion of a QFT is often defined via path integrals over all spacetime manifolds. In this setup, gauging a $q$-form symmetry in an $n$D QFT was defined roughly by a process of ``summing over insertions of topological defects of codimension $(q+1)$ in an $n$-dimensional spacetime manifold'' \cite{RSS23}. It turns out that, long before the rise of higher form symmetries, there was a precisely and rigorous mathematical theory of this gauging process under the name of `orbifold TQFT' \cite{FFRS07,FFRS10,CR16,CRS18,CRS19,CRS20,CMRSS21,Mul24,MR23,CM23,CMRSS24}. 
On the other hand, for condensed matter theorists, the notion of a topological order (or a SPT/SET/SSB order) is naturally defined on a spatial open disk (often by a lattice model). Therefore, it is natural to view a topological order as a `fully extended TQFT defined on a spatial open disk' or an algebraic quantum field theory (AQFT) defined by local observables (which often form a higher categorical structure \cite{KW14,KWZ15,JF22,KZ22b}). The relation between a FQFT and an AQFT was known in principle. At least for fully dualizable QFT's, in principle or as a folklore, the partition function on a non-trivial spacetime manifold $M$ in a FQFT can be obtained from an AQFT by integrating observables on an open disk over $M$. For TQFT's, the mathematical foundation of this integration is an on-going program of the so-called factorization homology \cite{Lur17,AF20,AFT17,AFT16,AFR18,AF24} (see Remark\,\ref{rem:alpha_version_FH}). 

The manifest difference between the mathematical formulation of `gaugings' in FQFT's and that of `condensations' in AQFT's is due to the fact that an integration over a spacetime manifold changes the mathematical structure of the observables. For example, integrating a system of anyons (i.e., a modular tensor category) over a closed spatial surface $M$ produces a pair $(\vect, u_M)$, where $u_M$ is the space of ground states on $M$ \cite{AKZ17}. Note that the data $u_M$ is the only global observable on $M$. All the $\EE_2$-monoidal structures (i.e., the fusions and braidings of anyons) that are well-defined on an open 2-disk are lost after the integration. In other words, anyons, together with their fusions and braidings, are not global observables (see \cite{AKZ17,KZ22a} for more discussion). Therefore, the explicitly structured mathematical theory of defect condensations based on higher categories, higher algebras and higher representations is a distinct feature of an AQFT\footnote{In orbiford TQFT's \cite{CRS20}, it is unavoidable to introduce something similar or equivalent to a higher algebra as the initial data in their construction (see Remark\,\ref{rem:ingo}). 
}. It remains an important open problem to compare the partition functions obtained by integrating our condensation results with those obtained by gauging $q$-form symmetries and those in orbiford TQFT's. 
\end{rem}

\subsection{Layout, conventions and notations}
The discouraging length of this paper is an avoidable consequence of the incredible richness of the theory of defect condensations and our goal of providing a comprehensive story of it. For readers convenience, we summarize the layout of this work, the conventions and the notations used in this work. 

\bigskip
\noindent{\bf Layout}: In Section\,\ref{sec:topological_defects}, we review the mathematical theory of topological defects in an $n+$1D topological orders based on the remote detectable principle, boundary-bulk relation and condensation complete principle. We also review the notions of $\EE_m$-multi-fusion $n$-categories and explain that an $\EE_m$-monoidal $n$-category can be viewed as an $n$-category equipped with a tensor product in each of $n$ independent dimensions. We also introduce other useful notations, new notions and new results along the way. Importantly, in Section\,\ref{sec:center_functor}, we derive some new results from some earlier results on the center functor in \cite{KZ24}. These results are very important for later constructions and classifications of the condensations of defects of codimension 1 or 2. In Section\,\ref{sec:condense_1-codim_defect}, we show that a phase transition from a topological order $\SC^{n+1}$ to another $\SD^{n+1}$ can be defined by the condensation of a 1-codimensional topological defects. In Section\,\ref{sec:condense_2-codim_defect}, we show that, when $\SC^{n+1}$ is anomaly-free, the same condensation can be defined by condensing a 2-codimensional topological defect directly. When $n=2$, this condensation is precisely the usual anyon condensation in 2+1D topological orders. We start from a review of anyon condensation in 2+1D but from a new perspective in Section\,\ref{sec:anyon_cond_algebra} and provide a large amount of examples in Section\,\ref{sec:example_2codim_3D}. We also develop a geometric anyon condensation theory in Section\,\ref{sec:2codim_3D_geometric_approach}. Then in Section\,\ref{sec:2codim_condensation_n+1D}, we generalize it to higher dimension. In Section\,\ref{sec:condense_k-codim_defect}, we study the condensation of $k$-codimensional topological defects and give the main results of this work as summarized in this Section. In Section\,\ref{sec:app}, we outline a few directions that this theory should be further developed or generalized, including a under-developed higher Morita theory, a under-developed theory of integrals (which might be related 
to the theory of factorization homology), a condensation theory of (potentially gapless) liquid-like defects or quantum liquids and some applications. 

\bigskip
\noindent {\bf Conventions}: Throughout this work, a topological order is defined up to invertible ones. We use ``Theorem$^{\mathrm{ph}}$'' to highlight a physical result and use ``Theorem'' to represent a mathematical result. 

We use $n+$1D to denote the spacetime dimension. We mainly use spacetime dimension in this work because the $n+$1D spacetime dimension matches with the $n+$1-levels of hierarchical structure of an $n$-category. However, almost all the illustrating pictures are drawn in the spatial dimensions because $\EE_k$-monoidal structure matches with the spatial dimension. For example, particles on a line in the spatial dimension can be fused in one spatial dimension, thus forming an $\EE_1$-fusion 1-category or simply a fusion 1-category. 


\bigskip
\noindent {\bf Notations}: In this work, we systematically introduce a lot of formal notations. For example, we denote a simple statement like: ``$\CA$ is an indecomposable multi-fusion $n$-category'' by ``$\CA \in \Algc_{\EE_1}((n+1)\vect)$''. Although we are aware of the danger of turning physics oriented readers away, we do it for an important reason. We slowly introduce the formal language when we review the more familiar situations in lower dimensions (but often from a new perspective). Once the readers get used to the new formal language, they will start to appreciate the power of categorical language when we move on to higher dimensional theories. For example, a simple notation $A \in \Algc_{\EE_1}(\Mod_A^{\EE_k}(\CC))$ means a thousand words. Indeed, it summarizes incredible amount of structures, which were completely out of reach before the advent of this categorical language. Without such a powerful language, the higher condensation theory is simply impossible.

Most of the notations are introduced carefully when they first appear. Here, we highlight a few important conventions on notations. 
\bnu

\item We denote $n+$1D (potentially anomalous) topological orders by $\SA^{n+1},\SB^{n+1},\SC^{n+1},\SD^{n+1},\cdots$, and denote their categories of (`intrinsic' topological defects see Section\,\ref{sec:RDP}): 
\bnu
\item[-] topological defects of codimension 1 and higher by $\CA, \CB, \CC, \CD,\cdots$, respectively; 
\item[-] topological defects of codimension 2 and higher by $\Omega\CA,\Omega\CB, \Omega\CC, \Omega\CD, \cdots$, respectively; 
\item[-] topological defects of codimension $(k-1)$ and higher by $\Omega^k\CA, \Omega^k\CB, \Omega^k\CC, \Omega^k\CD, \cdots$, respectively. 
\enu
Note that we carefully distinguish $\SA^{n+1}$ with $\CA$ simply because they have different meanings and play different roles in the category of topological orders. 

\item If a topological order $\SA^{n+1}$ is anomalous, it has a unique $n+$2D bulk phase denoted by $\SZ(\SA)^{n+2}$ such that $\SA^{n+1}$ becomes a gapped boundary of $\SZ(\SA)^{n+2}$ \cite{KW14}. Recall that $\SZ(\SA)^{n+2}$ is also called {\it gravitational anomaly} of $\SA^{n+1}$ in \cite{KW14}. 

\item We use $\ST\SC^3$ and $\Isphase^3$ to denote 2+1D toric code phase and 2+1D Ising topological order, respectively. For a finite group $G$ and $\omega\in Z^{n+1}(G, U(1))$, we use $\SG\ST_{(G,\omega)}^{n+1}$ to denote the $n+$1D twisted finite gauge theory (or Dijkgraaf-Witten theory). When $\omega$ is trivial, it is called a finite gauge theory and is denoted by $\SG\ST_G^{n+1}$. Note that $\ST\SC^3=\SG\ST_{\Zb_2}^3$. Following the first convention, we denote the category of topological defects in $\ST\SC^3$, $\Isphase^3$, $\SG\ST_{(G,\omega)}^{n+1}$ and $\SG\ST_G^{n+1}$ by $\CT\CC$, $\ising$, $\CG\CT_{(G,\omega)}^{n+1}$ and $\CG\CT_G^{n+1}$, respectively. The usual Ising modular tensor category is denoted by $\Omega\ising$. Restricted in Example\,\ref{expl_Ising_G_crossed}, the notion $\Ising_\pm$ is used to denote two Ising type fusion 1-categories. 

\item We use $\vect$ to denote the category of finite dimensional vector spaces over the complex number field $\Cb$. It can be treated as the definition of the delooping of $\Cb$, i.e., $\Sigma\Cb :=\vect$ \cite{GJF19}. 

\item We use $(n+1)\vect$, which is defined by iterated deloopings $\Sigma^{n+1}\Cb$ \cite{GJF19} (see Section\,\ref{sec:cc} for more details), to denote the category of separable $n$-categories. Intuitively, one can view a separable $n$-category $\CS$ as a ``finite dimensional $n$-vector space''. We set $\CS^\op:=\CS^{\op 1}$, i.e., flipping all 1-morphisms. 

\item We denote the full subcategory of $(n+1)\vect$ consisting of invertible objects by $(n+1)\vect^\times$. 

\item $\CB \in \Algc_{\EE_k}((n+1)\vect)$ means that $\CB$ is an $\EE_k$-multi-fusion $n$-category, and, at the same time, a condensable $\EE_k$-algebra in $(n+1)\vect$. 

\item $A, B\in \Algc_{\EE_k}(\CB)$ means that $A$ and $B$ are condensable $\EE_k$-algebras in $\CB$. 

\item $\LMod_A(\CB), \RMod_A(\CB), \BMod_{A|B}(\CB)$ represent the categories of left $A$-modules, right $A$-modules, $A$-$B$-bimodules in $\CB$, respectively. 

\item $\Mod_A^{\EE_k}(\CB)$ denotes the categories of $\EE_k$-$A$-modules (or $\EE_k$-modules over $A$) in $\CB$. See Section\,\ref{sec:anyon_cond_algebra} and Section\,\ref{sec:main_theorem} for more explanation of the notion of an $\EE_k$-$A$-module. 

\item $B \in \Alg_{\EE_k}(\Mod_A^{\EE_k}(\CB))$ means that $B$ is an $\EE_k$-$A$-module in $\CB$ and, at the same time, an $\EE_k$-algebra in $\CB$ such that the defining data of the $\EE_k$-algebra structure on $B$ are all (higher) morphisms in $\Mod_A^{\EE_k}(\CB)$.

\item For $\CA,\CB\in \Algc_{\EE_1}((n+1)\vect)$ and $\CM,\CN \in \BMod_{\CA|\CB}((n+1)\vect)$, we use $\Fun_{\CA|\CB}(\CM,\CN)$ to denote the category of bimodule functors   from $\CM$ to $\CN$. 

\item In this work, $[x,y]$ denotes the internal hom which is briefly reviewed in Section\,\ref{sec:internal_hom}. 

\item For an $\EE_k$-monoidal functor $F: \CA \to \CB$, we use $\FZ_k(F)$ or $\FZ_k(\CA,\CB)$ to  denote the $\EE_k$-centralizer of $F$ (see Definition\,\ref{def:centralizer+center}), and use $\FZ_k(\CA)=\FZ_k(\id_\CA)$ to denote $\EE_k$-center of $\CA$. 

\item For a monoidal $n$-category $\CC$, the meaning of the notation $\CC^\op$ is explained in Convention\,\ref{rem:convention_op} and Definition\,\ref{def:convention_op}. 

\enu

\bigskip
\noindent {\bf Acknowledgements}: We would like to thank David Ayala, Ansi Bai, Nils Carqueville, John Francis, Ling-Yan Hung, Tian Lan, Gregory Moore, Ingo Runkel, Nathan Seiberg, Rui Wen, Xiao-Gang Wen and Hao Xu for useful discussion and comments. LK and ZHZ are supported by NSFC (Grant No. 11971219) and by Guangdong Provincial Key Laboratory (Grant No. 2019B121203002) and by Guangdong Basic and Applied Basic Research Foundation (Grant No. 2020B1515120100). ZHZ is also supported by Wu Wen-Tsun Key Laboratory of Mathematics at USTC of Chinese Academy of Sciences. JHZ is partially supported by NSFC (Grant No.11571343). HZ is supported by NSFC under Grant No. 11871078 and by Startup Grant of Tsinghua University and BIMSA.

\newpage

\section{Categories of topological defects} \label{sec:topological_defects}

Based on three important guiding principles: Remote Detectable Principle \cite{Lev13,KW14}, Boundary-Bulk Relation \cite{KK12,KWZ15,KWZ17} and Condensation Completion Principle \cite{KS11a,KK12,FSV13,CR16,Kon14e,KW14,DR18,KTZ20,GJF19,JF22,KLWZZ20}, the mathematical theory of topological defects in topological orders was initiated and developed in some earlier works \cite{KW14,KWZ15,LKW18,LW19,KTZ20}, and was fully established in \cite{GJF19,JF22} (also in \cite{KLWZZ20}), and was further developed and generalized to SPT/SET orders and gapped/gapless quantum liquids in \cite{KLWZZ20,KLWZZ20a,KZ22b,KZ24,KZ22}. It forms the theoretical foundation of our higher condensation theory and are used throughout this work. In this section, we review this theory in details and introduce some important notions and notations along the way. Although the main results of this theory have already appeared in literature \cite{KWZ15,JF22,KLWZZ20,KZ22b}, we believe that the complete physical reasoning behind this theory, including its limitations, is told here for the first time. In Section\,\ref{sec:center_functor}, we use center functor to derive some new results that become important to the classification of 1-codimensional defect condensations and that of Lagrangian $\EE_2$-algebras in later sections.

\subsection{Categories of topological defects} \label{sec:Cat_top_defects}

\subsubsection{2+1D topological orders} \label{sec:Cat_top_defects_3D}

The systematic study or classification of topological defects in TQFT's or topological orders was initiated from that of the Wilson lines in 2+1D TQFT's \cite{Wit89,RT91} and, around the same time, that of the anyons in 2+1D anomaly-free topological orders \cite{MS89a,FFK89,FRS89,FG90,Reh90,MR91,Wen91} (preceded by many earlier works on fractional statistics of fields and anyons \cite{Fro76,LM77,GMS80,GMS81,Wil82,Wu84,ASW84,Fro88}). We start from a review of the theory of anyons in a 2+1D anomaly-free topological order before we move onto the higher dimensional theories. 

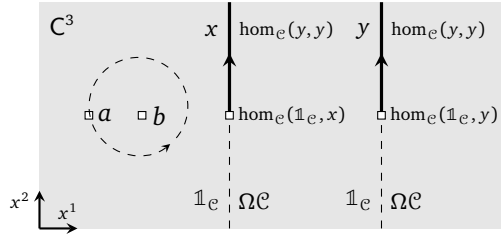
\begin{figure}
\centering
\begin{tikzpicture}
\fill[gray!20] (-2,0) rectangle (4.2,3) ;
\draw[very thick,->-] (0.5,1.5)--(0.5,3) node[near end,left] {\small $x$} 
node[near end,right] {\scriptsize $\hom_\CC(y,y)$} ;
\draw[dashed] (0.5,0)--(0.5,1.5) node[near start,left] {\small $\one_\CC$} node[near start,right] {\small $\Omega\CC$} ;
\draw[very thick,->-] (2.5,1.5)--(2.5,3) node[near end,left] {\small $y$} node[near end,right] {\scriptsize $\hom_\CC(y,y)$};
\draw[dashed] (2.5,0)--(2.5,1.5) node[near start,left] {\small $\one_\CC$} node[near start,right] {\small $\Omega\CC$}; 
\draw[fill=white] (2.45,1.45) rectangle (2.55,1.55) node[midway,right,scale=1] {\scriptsize $\hom_\CC(\one_\CC,y)$} ; 
\draw[fill=white] (0.45,1.45) rectangle (0.55,1.55) node[midway,right,scale=1] {\scriptsize $\hom_\CC(\one_\CC,x)$} ; 
\draw[fill=white] (-1.4,1.45) rectangle (-1.3,1.55) node[midway,right,scale=1] {$a$} ;
\draw[fill=white] (-0.7,1.45) rectangle (-0.6,1.55) node[midway,right,scale=1] {$b$} ; 
\node at (-1.7,2.7) {\small $\SC^3$} ;
\draw[thick,-stealth] (-2,0) -- (-1.5,0) node[near end,above] {\scriptsize $x^1$} ;
\draw[thick,-stealth] (-2,0) -- (-2,0.5) node[near end,left] {\scriptsize $x^2$} ;
\draw[-stealth,dashed] (-0.2,1.2) arc (-40:310:0.65) ;
\end{tikzpicture}
\caption{This picture illustrate some defects in a 2+1D topological order $\SC^3$ in spatial dimensions: two 1-codimensional defect $x,y\in \CC$, a domain wall between $\one_\CC$ and $x$, a domain wall between $\one_\CC$ and $y$ and two anyons $a,b\in \Omega\CC$.}
\label{fig:defects_in_2d-TO}
\end{figure}

\medskip
Consider a 2+1D anomaly-free topological order $\SC^3$ (see Remark\,\ref{rem:simple_TO}). It can have 2-codimensional topological defects, which are particle-like topological excitations in spatial dimension picture and are also called anyons or particles. In the spacetime picture, an anyon is also called a Wilson line, which is the world line of the anyon. It is possible to have $0$D defects connecting two potentially different Wilson lines. Such a $0$D defect, or a 3-codimensional defect, is also called an instanton. Anyons can be fused. More precisely, for two anyons $a$ and $b$, their fusion is denoted by $a\otimes b$. Moreover, anyons can be braided. This amounts to a braiding isomorphism 
\[
c_{a,b}: a\otimes b \xrightarrow{\simeq} b\otimes a. 
\]
The composed isomorphism $(a\otimes b \xrightarrow{c_{a,b}} b\otimes a \xrightarrow{c_{b,a}} a\otimes b)$ is called the double braiding, which amounts to an adiabatic move of the $a$-particle along a circle around the $b$-particle as illustrated in Figure\,\ref{fig:defects_in_2d-TO}. The complete set of anyons and instantons, together with the fusions and braidings among anyons, form a mathematical structure called a braided fusion 1-category\footnote{In physics, $\Omega\CC$ is necessarily a modular tensor category, which is a non-degenerate braided fusion 1-category equipped with a ribbon structure \cite{MS89,Tur20}. Moreover, unitarity is often assumed. We prefer to ignore both structures for the reason that the roles played by them in condensation theory are not essential. But we secretly assume the unitarity in all pictures.} \cite{Wit89,MS89,FRS89, ZHK89,FG90, Reh90, MR91,Wen91,Kit06} (see \cite{KZ22a} for a recent review). We denoted this braided fusion 1-category by $\Omega\CC$, a notation which represents the looping of $\CC$ and is explained later in (\ref{eq:omega_C}). Moreover, the braidings of $\Omega\CC$ are required to be {\it non-degenerate} in the sense that the $S$-matrix, whose entries are the trace of double braidings of two simple objects, is non-degenerate \cite{MS89,Tur20} (see also \cite{BK01,Kit06} for reviews). The simplest non-degenerate braided fusion 1-category is the category $\vect$ of finite dimensional vector spaces over $\Cb$. It is precisely the category of anyons in the 2+1D trivial topological order, which is denoted by $\mathbf{1}^3$.  

Concrete lattice models realizing non-chiral 2+1D topological orders (i.e. those admitting gapped boundaries), such as Kitaev's quantum double model \cite{Kit03} and Levin-Wen models \cite{LW05}, were constructed. Anyons in these models were intensively studied and were shown to form a non-degenerate braided fusion 1-category \cite{Kit03,KK12}. We want to emphasize that not only simple anyons (i.e., simple objects in $\Omega\CC$) are physical, but also all the composite anyons (i.e., direct sums of simple objects in $\Omega\CC$), which can be realized in lattice models (see \cite{KZ22a} for a more detailed explanation). It was proposed in \cite{Kit06} that a 2+1D topological order $\SC^3$ can be completely characterized by the data $\Omega\CC$ (up to invertible topological orders). For example, a non-chiral 2+1D topological order (see Definition\,\ref{def:anomaly_TO}) can be described by the Drinfeld center $\FZ_1(\CA)$ \cite{LW05,KK12} of a fusion 1-category $\CA$, where $\FZ_1(\CA)$ is automatically a non-degenerate braided fusion 1-category \cite{Mueg03a}. This proposal was soon accepted by the community. 

\begin{rem} \label{rem:simple_TO}
The 2+1D anomaly-free topological orders in above discussion are assumed to satisfy a {\it simpleness} (or a stability) condition: the ground state degeneracy of this topological order defined on a 2-sphere is trivial. This condition is automatically assumed in most physical literature. We proceed our story with this simpleness assumption until Section\,\ref{sec:cc}, where we introduce a slightly more general notion of a composite topological order (see Definition\,\ref{def:simple_TO}). It is natural and indispensable in the study of the fusions among topological orders and condensations \cite{KWZ15,KZ22b}. 
\end{rem}

However, this proposal became puzzling and questionable after a new discovery. In 
\cite{KS11a,KK12,FSV13,CR16,Kon14e}, it was shown that a 2+1D topological order $\SC^3$ can have non-trivial 1-codimensional topological defects as depicted in Figure\,\ref{fig:defects_in_2d-TO}, and these defects, together with all higher codimensional topological defects form a monoidal 2-category $\CC$. We explain the ingredients of this 2-category $\CC$ below.  
\begin{itemize}
\item $0$-morphisms (or objects) are 1-codimensional topological defects (i.e., string-like defects in the spatial dimension, or strings for simplicity); 

\item 1-morphisms are 2-codimensional topological defects, including those 2-codimensional topological defects that are domain walls between two (potentially non-trivial) 1-codimensional topological defects. 

\item 2-morphisms are 3-codimensional topological defects (i.e., instantons or 0D operators in spacetime). 
\end{itemize}
The tensor product of $\CC$ is given by the horizontal fusion of two 1-codimensional topological defects (e.g., $x\otimes y$ in Figure\,\ref{fig:defects_in_2d-TO}), and the tensor unit of $\CC$ is the trivial 1-codimensional topological defect, denoted by $\one_\CC$. Since a particle can be viewed as a domain wall between two trivial 1-codimensional defects, the category of particles in $\SC^3$ is precisely given by the following 1-category:  
\be \label{eq:omega_C}
\Omega\CC:=\hom_\CC(\one_\CC,\one_\CC), 
\ee 
which is called the looping of $\CC$. We denote the full sub-2-category of $\CC$ consisting of a single object $\one_\CC$ by $\mathrm{B}\Omega\CC$, which is called one-point delooping of $\Omega\CC$, i.e.
\be
\mathrm{B}\Omega\CC := 
\begin{array}{c}
\xymatrix{
\one_\CC \ar@(ul,ur)[]^{\Omega\CC}
}
\end{array}
\, .
\ee
Notice that $\mathrm{B}\Omega\CC$ does not contain more information than $\Omega\CC$.

\begin{rem} \label{rem:liquid-like_defect}
We do not give a precise definition of a topological defect. Instead, we give a brief clarification. In this work, by a $k$-codimensional topological defect in $\SC^{n+1}$, we mean an anomalous $(n-k+1)$D topological order (i.e., a gapped quantum liquid without symmetry). In other words, we assume a liquid-like property \cite{KZ22b,KZ22}. Physically, by `liquid-like' we mean that the defect can be bent freely without any change\footnote{This condition amounts to say that a liquid-like defect commutes with the energy-momentum tensor, and, therefore, can be viewed as a generalized symmetry \cite{FFRS10,GKSW15} (see \cite{SN23} for a recent review).}, and the fusions among such defects are well-defined. Mathematically, it means that the category of all such gapped liquid-like defects are well-defined and fully dualizable \cite{Lur09}. Note that it was known that there are non-liquid quantum phases \cite{Cha05,Haa11}, which can be viewed as non-liquid-like defects in the higher dimensional trivial phase. In general, if we stack a $k$D topological defect with a gapped $k$D non-liquid, we obtain a gapped non-liquid defect. Non-liquid-like defects are not dualizable in general. In this work, we ignore non-liquid gapped defects or domain walls completely. In particular, by a `gapped defect', we always mean a topological defect unless we specified otherwise. 
\end{rem}

When $\SC^3$ admits gapped boundaries, we have $\Omega\CC=\FZ_1(\CA)$ for a fusion 1-category $\CA$ \cite{KK12,FSV13,Kon14e}. In this case, it was shown in \cite{KK12} that a 1-codimensional topological defect in $\SC^3$ can be constructed in a generalized Levin-Wen models \cite{LW05} based on the defining data of a finite semisimple $\CA$-$\CA$-bimodule $\CM$. By ``finite semisimple'', we mean a 1-category that is equivalent to a finite direct sum of $\vect$. We denote the 2-category of finite semisimple 1-categories by $2\vect$. Then a finite semisimple $\CA$-$\CA$-bimodule simply means an $\CA$-$\CA$-bimodule in $2\vect$. We denote by $\BMod_{\CA|\CA}(2\vect)$ the 2-category of 
$\CA$-$\CA$-bimodules in $2\vect$ (as objects), bimodule functors (as 1-morphisms) and bimodule natural transformations (as 2-morphisms). Then the main results in \cite{KK12} says that 
\be \label{eq:KK12}
\CC =\BMod_{\CA|\CA}(2\vect).
\ee
The 2-category $\BMod_{\CA|\CA}(2\vect)$ has a natural monoidal structure defined by the relative tensor product $\boxtimes_\CA$ (see \cite{EGNO15} for a review) and the tensor unit $\CA$. In this case, we have 
\be \label{eq:non-chiral_OmegaC}
\Omega\CC = \Omega\BMod_{\CA|\CA}(2\vect)=\Fun_{\CA|\CA}(\CA,\CA) \simeq \FZ_1(\CA),
\ee
which is exactly our original setup.

\begin{rem} \label{rem:domain_wall=boundary}
There is another way to look at the 1-codimensional topological defects in $\SC^3$. By the folding trick, a 1-codimensional topological defect in $\SC^3$ can be viewed as a gapped boundary of the double-layered system $\SC^3\boxtimes \overline{\SC^3}$ (see Example\,\ref{expl:folding}). As a consequence, there are one-to-one correspondences among simple 1-codimensional topological defects in $\SC^3$,  indecomposable $\Omega\CC$-modules (up to equivalences) and Lagrangian algebras in $\Omega\CC \boxtimes \Omega\CC^\rev$ \cite{KR09,DMNO13,KK12}. 
\end{rem}

We give some examples of $\CC$ that are frequently used in this work. 
\begin{expl} \label{expl:2Vec}
2+1D trivial topological order $\mathbf{1}^3$: The 1-category of anyons is given by $\vect$. Since $\mathbf{1}^3$ can be realized by Levin-Wen model defined by the trivial fusion 1-category $\vect$. The monoidal 2-category of all defects in $\mathbf{1}^3$ can be identified with $\BMod_{\vect|\vect}(2\vect) \simeq 2\vect$. 
The objects in $2\vect$ are precisely the labels of 1+1D anomaly-free topological orders. 
\bnu
\item the tensor unit $\one=\vect$ in $2\vect$ labels the trivial 1-codimensional topological defects in $\mathbf{1}^3$, which is, at the same time, the trivial 1+1D topological order $\mathbf{1}^2$. Note that $\hom_{2\vect}(\one,\one)\simeq \vect$ is the precisely the category of particles in $\mathbf{1}^2$. 

\item What about a generic object $\one^{\oplus k}$ in $2\vect$? It is a composite 1-codimensional topological defect in $\mathbf{1}^3$ and, at the same time, a composite 1+1D topological order $(\mathbf{1}^2)^{\oplus k}$, which naturally occurs in the process of dimensional reduction. For example, if we define a toric code model on a narrow strap, viewed as a quasi-1+1D system, with both boundaries chosen to be the smooth boundary, we obtain the composite 1+1D topological order $(\mathbf{1}^2)^{\oplus 2}$ (see \cite{KZ22a} for a more detailed explanation). Note that $\hom_{2\vect}(\one^{\oplus k}, \one^{\oplus k})$ is precisely the multi-fusion 1-category of particles living on the composite 1+1D topological order $(\mathbf{1}^2)^{\oplus k}$ \cite{KWZ15,AKZ17}. 

\enu
\end{expl}

\begin{expl} \label{expl:3D_toric_code}
2+1D $\Zb_2$ topological order $\TC^3$: Since it is realizable by the 2+1D toric code model \cite{Kit03}, we denote it by $\TC^3$. We denote the 2-category of topological defects in $\TC^3$ by $\CT\CC$. In this case, $\Omega\CT\CC$ has four simple objects $1,e,m,f$ and the following fusion rules: 
\[
e\otimes e \simeq m\otimes m \simeq f\otimes f \simeq 1, \quad \quad f\simeq e\otimes m \simeq m \otimes e.
\]
The double braiding between $e$ and $m$ is $-1$. The self-double-braidings of $e$ and $m$ are both trivial, and the self-double-braiding of $f$ is $-1$. Mathematically, $\Omega\CT\CC$ can be identified with the Drinfeld center $\FZ_1(\Rep(\Zb_2))$ of $\Rep(\Zb_2)$, where $\Rep(\Zb_2)$ denote the category of finite dimensional representations of the $\Zb_2$ group. The monoidal 2-category $\CT\CC$ can be obtained either by the following monoidal equivalence \cite{KK12}: 
$$
\CT\CC \simeq \BMod_{\Rep(\Zb_2)|\Rep(\Zb_2)}(2\vect)
$$ 
or by explicit constructions in the toric code model \cite{KZ22a}. 
\bnu

\item There are six simple 1-codimensional topological defects $\one, \vartheta, \mathrm{ss}, \mathrm{sr}, \mathrm{rs}, \mathrm{rr}$ in $\TC^3$, where $\one$ denotes the trivial 1-codimensional defect; $\vartheta$ denotes the invertible defect that exchanges the $e$-particles with the $m$-particles thus defines the $e$-$m$-duality \cite{Bom10,KK12}; the remaining four defects $\mathrm{ss},\mathrm{sr}, \mathrm{rr}, \mathrm{rs}$ denote the defects obtained by stacking of two smooth/rough gapped boundaries of the toric code model as illustrated below: 
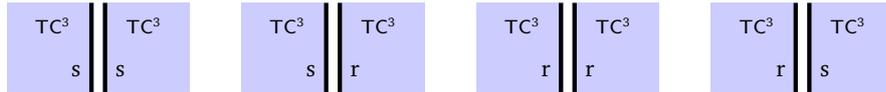
\[
\begin{array}{c}
\begin{tikzpicture}[scale=0.6]
\fill[blue!20] (-2,0) rectangle (-0.1,2) ;
\fill[blue!20] (0.1,0) rectangle (2,2) ;
\draw[ultra thick] (-0.15,0)--(-0.15,2) node [near start, left] {\small $\mathrm{s}$} ;
\draw[ultra thick] (0.15,0)--(0.15,2) node [near start, right] {\small $\mathrm{s}$} ;
\node at (-1,1.5) {\scriptsize $\ST\SC^3$} ;
\node at (1,1.5) {\scriptsize $\ST\SC^3$} ;
\end{tikzpicture}
\end{array}
\quad
\begin{array}{c}
\begin{tikzpicture}[scale=0.6]
\fill[blue!20] (-2,0) rectangle (-0.1,2) ;
\fill[blue!20] (0.1,0) rectangle (2,2) ;
\draw[ultra thick] (-0.15,0)--(-0.15,2) node [near start, left] {\small $\mathrm{s}$} ;
\draw[ultra thick] (0.15,0)--(0.15,2) node [near start, right] {\small $\mathrm{r}$} ;
\node at (-1,1.5) {\scriptsize $\ST\SC^3$} ;
\node at (1,1.5) {\scriptsize $\ST\SC^3$} ;
\end{tikzpicture}
\end{array}
\quad
\begin{array}{c}
\begin{tikzpicture}[scale=0.6]
\fill[blue!20] (-2,0) rectangle (-0.1,2) ;
\fill[blue!20] (0.1,0) rectangle (2,2) ;
\draw[ultra thick] (-0.15,0)--(-0.15,2) node [near start, left] {\small $\mathrm{r}$} ;
\draw[ultra thick] (0.15,0)--(0.15,2) node [near start, right] {\small $\mathrm{r}$} ;
\node at (-1,1.5) {\scriptsize $\ST\SC^3$} ;
\node at (1,1.5) {\scriptsize $\ST\SC^3$} ;
\end{tikzpicture}
\end{array}
\quad
\begin{array}{c}
\begin{tikzpicture}[scale=0.6]
\fill[blue!20] (-2,0) rectangle (-0.1,2) ;
\fill[blue!20] (0.1,0) rectangle (2,2) ;
\draw[ultra thick] (-0.15,0)--(-0.15,2) node [near start, left] {\small $\mathrm{r}$} ;
\draw[ultra thick] (0.15,0)--(0.15,2) node [near start, right] {\small $\mathrm{s}$} ;
\node at (-1,1.5) {\scriptsize $\ST\SC^3$} ;
\node at (1,1.5) {\scriptsize $\ST\SC^3$} ;
\end{tikzpicture}
\end{array}
\]
where the letter `$\mathrm{s}$' represents the smooth boundary and the letter `$\mathrm{r}$' represents the rough boundary. The fusion rules given in the following table \cite[Table\,1]{KZ22a}: 
\be \label{table:fusion_rule_1d_domain_wall_toric_code}
\begin{tabular}{c|c|c|c|c|c|c}
$\otimes$ & $\one$ & $\vartheta$ & \mbox{ss} & \mbox{sr} & \mbox{rs} & \mbox{rr} \\
\hline
$\one$ & $\one$ & $\vartheta$ & \mbox{ss} & \mbox{sr} & \mbox{rs} & \mbox{rr} \\
\hline
$\vartheta$ & $\vartheta$ & $\one$ & \mbox{rs} & \mbox{rr} & \mbox{ss} & \mbox{sr} \\
\hline
\mbox{ss} & \mbox{ss} & \mbox{sr} & 2\mbox{ss} & 2\mbox{sr} & \mbox{ss} & \mbox{sr} \\
\hline
\mbox{sr} & \mbox{sr} & \mbox{ss} & \mbox{ss} & \mbox{sr} & 2\mbox{ss} & 2\mbox{sr} \\
\hline
\mbox{rs} & \mbox{rs} & \mbox{rr} & 2\mbox{rs} & 2\mbox{rr} & \mbox{rs} & \mbox{rr} \\
\hline
\mbox{rr} & \mbox{rr} & \mbox{rs} & \mbox{rs} & \mbox{rr} & 2\mbox{rs} & 2\mbox{rr}
\end{tabular}
\ee
where  `$2\mathrm{ss}$' represents $\mathrm{ss} \oplus \mathrm{ss}$ (see \cite{KZ22a} for more explanation of this fusion rule). All six defects $\one, \vartheta, \mathrm{ss}, \mathrm{sr}, \mathrm{rs}, \mathrm{rr}$ can be obtained by condensing certain (potentially composite) anyons as we show explicitly in Example\,\ref{expl:toric_code_SigmaA} (see also the second table in this example). 

\item There are six simple $\Rep(\Zb_2)$-$\Rep(\Zb_2)$-bimodules: $\Rep(\Zb_2), \vect_-, \Rep(\Zb_2)\boxtimes \Rep(\Zb_2), \Rep(\Zb_2)\boxtimes \vect, \vect \boxtimes \Rep(\Zb_2), \vect \boxtimes \vect$, where $\vect_-$ is an invertible bimodule defined by the associator $(a\odot x)\odot a \xrightarrow{-1} a\odot (x\odot a)$ when $a\in \Rep(\Zb_2)$ is the non-trivial simple object and $x\in \vect_-$. They correspond to six simple 1-codimensional topological defects $\one, \vartheta, \mathrm{ss}, \mathrm{sr}, \mathrm{rs}, \mathrm{rr}$ in $\TC^3$, respectively. 

\enu

Above two ways of representing $\CT\CC^3$ provide two coordinate systems of $\CT\CC^3$. Similar to differential geometry, computations are often carried out in a concrete coordinate systems of a chart, and the coordinate transformations allow us to relate different charts and play an important role in the theory. It turns out that $\CT\CC^3$ has a few more natural coordinate systems. 
\bnu
\item The monoidal equivalence $\CT\CC^3 \simeq \RMod_{\FZ_1(\Rep(\Zb_2))}(2\vect)$ \cite{KLWZZ20}, explained in later subsections (see Eq.\,(\ref{eq:SigmaZA=BMod_AA}) and Theorem\,\ref{thm:SigmaA=RMod}), provides another important coordinate system. 
\item Since $\vect_{\Zb_2}$ and $\Rep(\Zb_2)$ are Morita equivalent, we have $\CT\CC^3 \simeq \BMod_{\vect_{\Zb_2}|\vect_{\Zb_2}}(2\vect)$. 
\enu
The coordinate transformations among different coordinate systems are summarized in the following table. 
\begin{table}[htbp] \centering \renewcommand\arraystretch{1.3}
\scriptsize{
\begin{tabular}{c|c|c|c|c|c|c}
Coordinates of $\CT\CC^3$ & $\one$ & $\vartheta$ & \mbox{ss} & \mbox{sr} & \mbox{rs} & \mbox{rr} \\
\hline
$\RMod_{\FZ_1(\Rep(\Zb_2))}(2\vect)$ & $\FZ_1(\Rep(\Zb_2))$ & $\vect \oplus \vect$ & $\Rep(\Zb_2)$ & $\vect$ & $\vect$ & $\vect_{\Zb_2}$ \\
\hline
$\BMod_{\Rep(\Zb_2)|\Rep(\Zb_2)}(2\vect)$ & $\Rep(\Zb_2)$ & $\vect_-$ & $\Rep(\Zb_2)\boxtimes\Rep(\Zb_2)$ & $\Rep(\Zb_2)\boxtimes\vect$ & $\vect\boxtimes\Rep(\Zb_2)$ & $\vect\boxtimes\vect$ \\
\hline
$\BMod_{\vect_{\Zb_2}|\vect_{\Zb_2}}(2\vect)$ & $\vect_{\Zb_2}$ & $\vect_-$ & $\vect\boxtimes \vect$ & $\vect\boxtimes \vect_{\Zb_2}$ & $\vect_{\Zb_2}\boxtimes \vect$ & $\vect_{\Zb_2}\boxtimes \vect_{\Zb_2}$ \\
\hline
$(H,\alpha)$ & $(\{ 1\},0)$ & $(\{ 1,f \},0)$ & $(\{ 1,m \},0)$ & $(\{ 1,e,m,f \},\alpha_0)$ & 
$(\{ 1,e,m,f \},\alpha_1)$ & $(\{ 1,e \},0)$ \\
\hline
$\Sigma A$ for $A \in \Algc_{\EE_1}(\Omega\CT\CC)$ & $1$ & $A_f=1\oplus f$ & $A_m=1\oplus m$ & $A_4=A_m\otimes A_e$ & $A_4^\op=A_e\otimes A_m$ & $A_e=1\oplus e$ \\
\end{tabular}
}
\caption{Coordinates in $\CT\CC^3$ and coordinate transformations}\label{table:coordinates_in_TC} 
\end{table}

\noindent In the second arrow, two simple objects in $\vect \oplus \vect$ are precisely the two twist defects $\chi_+, \chi_-$ constructed in \cite{Bom10,KK12} and the right $\FZ_1(\Rep(\Zb_2))$-module structure is defined by $\chi_\pm\otimes e=\chi_\pm\otimes m =\chi_\mp$; the fifth row is explained in Example\,\ref{expl:3D_GT_G} (see (\ref{eq:six_(H,a)})); the sixth row is explained in Example\,\ref{expl:toric_code_SigmaA}. 
\end{expl}

\begin{expl} \label{expl:3D_GT_G}
2+1D finite gauge theories $\SG\ST_G^3$ for a finite gauge group $G$: 
Note that $\SG\ST_{\Zb_2}^3=\ST\SC^3$. The category $\Omega\CG\CT_G^3$ of topological defects of codimension 2 (or anyons) and higher in $\SG\ST_G^3$ form the non-degenerate braided fusion 1-category $\FZ_1(\Rep(G))$, which is the Drinfeld center of $\Rep(G)$. The category $\CG\CT_G^3$ of all topological defects form the monoidal 2-category: 
\[
\CG\CT_G^3 \simeq \BMod_{\Rep(G)|\Rep(G)}(2\vect). 
\]
A simple object in $\CG\CT_G^3$ is an indecomposable $\Rep(G)$-$\Rep(G)$-bimodules in $2\vect$. Since $\Rep(G) \boxtimes \Rep(G)^\rev \simeq \Rep(G) \boxtimes \Rep(G) \simeq \Rep(G \times G)$, it is also an indecomposable left $\Rep(G\times G)$-module in $2\vect$, which had been classified in \cite{BO04,Ost03}. More precisely, an indecomposable left $\Rep(G \times G)$-module in $2\vect$ is given by $\Rep^\prime(\tilde{H})$, where $\tilde{H}$ is the central extension of a subgroup $H$ of $G \times G$ determined by a 2-cohomology class of $\alpha \in Z^2(H, \Cb^\times)$:
\[
1 \to \Cb^\times \to \tilde{H} \to H \to 1
\]
and $\Rep^\prime(\tilde{H})$ is the category of $\tilde{H}$-representations such that $\Cb^\times$ acts as identity. The category $\Rep^\prime(\tilde{H})$ is also denoted by $\Rep(H,\alpha)$, in which an object  is a vector space $V$ equipped with a map $\rho \colon H \to \mathrm{GL}(V)$ such that $\rho(g) \rho(h) = \alpha(g,h) \rho(gh)$ for all $g,h \in H$.
The $\Rep(G\times G)$-action on $\Rep^\prime(\tilde{H})$ or $\Rep(H,\alpha)$ is defined by the usual vector space tensor product. 

The fusion product in $\CG\CT_G^3$ is defined by $\boxtimes_{\Rep(G)}$, where the two-side $\Rep(G)$-actions on $\Rep^\prime(\tilde{H})$ are induced from the two natural monoidal functors $-\boxtimes \one_{\Rep(G)}, \one_{\Rep(G)} \boxtimes -:  \Rep(G) \to \Rep(G) \boxtimes \Rep(G)$. It is possible to work out the fusion rules, but we do not need them in this work. We give a complete classification of condensable $\EE_1$-algebras in $\CG\CT_G^3$ in later sections without using the fusion rules. 

We give only one example, when $G=\Zb_2$, by abusing the notation, we denote the elements in $\Zb_2 \times \Zb_2$ by $1,e,m,f$ with the obvious multiplication rules. The six simple $\Rep(\Zb_2)$-$\Rep(\Zb_2)$-bimodules given above are precisely $\Rep(H,\alpha)$ for the following six pairs $(H,\alpha)$: 
\be \label{eq:six_(H,a)}
(\{ 1\},0), \quad (\{ 1,f \},0), \quad (\{1, m\},0), \quad (\{ 1,e,m,f \}, \alpha_0), \quad
(\{ 1,e,m,f \}, \alpha_1), \quad (\{ 1, e\},0),
\ee
where $0\in H^3(Z_2,U(1))=0$ and $\alpha_0, \alpha_1\in H^3(\Zb_2\times \Zb_2, U(1))\simeq \Zb_2$, respectively (see Table\,\ref{table:coordinates_in_TC} for a more complete dictionary). The fusion rules are given in the table (\ref{table:fusion_rule_1d_domain_wall_toric_code}). 
\end{expl}


\begin{expl} \label{expl:ising}
2+1D Ising topological order $\Isphase^3$: 
We denote the 2-category of topological defects in $\Isphase^3$ by $\ising$. 
In this case, the 1-category $\Omega\ising$ of anyons is the well known Ising modular tensor category. It consists of three simple anyons $1, \psi, \sigma$ with the commutative fusion rules $\psi\otimes \psi =1, \psi \otimes \sigma \simeq \sigma, \sigma\otimes \sigma \simeq 1 \oplus \psi$. All its non-trivial braidings are given as follows: 
\begin{align*}
\psi \otimes \psi = 1 \xrightarrow{c_{\psi,\psi}=-1} 1 =\psi \otimes \psi; 
\quad
&\psi \otimes \sigma = \sigma \xrightarrow{c_{\psi,\psi}=e^{-\frac{\pi i}{2}}}  \sigma = \sigma \otimes \psi; \\
\sigma \otimes \psi = \sigma \xrightarrow{c_{\psi,\sigma}=e^{-\frac{\pi i}{2}}}  \sigma = \psi \otimes \sigma; \quad
&\sigma \otimes \sigma = 1 \oplus \psi \xrightarrow{c_{\sigma,\sigma}=e^{-\frac{\pi i}{8}}\oplus e^{\frac{3\pi i}{8}}} 1 \oplus \psi = \sigma \otimes \sigma. 
\end{align*}
According to \cite{FRS02}, all condensable $\EE_1$-algebras in $\Omega\ising$ are Morita equivalent to $1$. This result implies that $\Isphase^3$ has no non-trivial 1-codimensional topological defect. Therefore, we obtain that $\ising \simeq \mathrm{B}\Omega\ising$ as monoidal 2-categories. 
\end{expl}

\begin{rem}
While we are preparing the paper, a few more examples of $\CC$ in 2+1D have been worked out explicitly in \cite{YWL25}.   
\end{rem}

The construction of 1-codimensional topological defects in \cite{KS11a,KK12,FSV13,CR16,Kon14e} makes the proposal in \cite{Kit06} questionable because the characterization by the pair $(\Omega\CC,c)$ does not contain any 1-codimensional topological defects except the trivial one. 
\be \label{eq:question}
\mbox{Does it mean that the characterization $(\Omega\CC,c)$ is incomplete? }
\ee
The work \cite{KW14} was devoted to answer this question. In order to proceed, one needs guidances from new physical principles.

\subsubsection{Remote detectable principle and `intrinsic' topological defects} \label{sec:RDP}

Since these new physics principles apply to topological orders in all dimensions, we explain them in an arbitrary dimension. We start from recalling some basic notions that are used throughout this work. 

\begin{defn} \label{def:anomaly_TO}
Let $\SC^{n+1}$ be an $n+$1D topological order. 
\bnu
\item $\SC^{n+1}$ is called {\it anomaly-free} if it can be realized by an $n$D gapped lattice model; it is called {\it anomalous} otherwise \cite{KW14}. 
\item An anomaly-free $\SC^{n+1}$ is called {\it non-chiral} if it admits a gapped boundary; and is called {\it chiral} otherwise (i.e., all its boundaries are gapless). 
\enu
We denote the trivial $n+$1D topological order by $\mathbf{1}^{n+1}$. 
\end{defn}

Since an anomalous topological order $\SC^{n+1}$ should be always realizable as a topological defect in a higher (but still finite) dimensional lattice model, by a dimensional reduction argument \cite{KW14}, $\SC^{n+1}$ is always realizable as a gapped boundary of an $n+$2D anomaly-free topological order. Moreover, 
it was shown that this $n+$2D anomaly-free topological order is necessarily unique \cite{KW14}, and is called the {\it gravitational anomaly} or the {\it bulk} of $\SC^{n+1}$. We denoted it by $\SZ(\SC)^{n+2}$. Then the condition that $\SC^{n+1}$ is anomaly-free can be mathematically expressed as 
\be \label{eq:anomaly-free_SC}
\SZ(\SC)^{n+2}=\mathbf{1}^{n+2}. 
\ee
For an anomalous topological order $\SC^{n+1}$, we also have an interesting identity\footnote{This identity says that the bulk of a bulk is trivial. It is somewhat dual to the well-known statement in topology: the boundary of a boundary is empty, which leads to a homology theory. Therefore, we expect that the identity (\ref{eq:Z2=1}) should lead us to a non-trivial but yet-unknown cohomology theory.} \cite{KW14}: 
\be \label{eq:Z2=1}
\SZ(\SZ(\SC))^{n+3} = \mathbf{1}^{n+3}. 
\ee

\begin{expl} \label{expl:folding}
When $\SC^{n+1}$ is anomaly-free, then the trivial 1-codimensional topological defect in $\SC^{n+1}$ can be viewed as an anomalous $n$D topological order, denoted by $\SC^n$, and is precisely the trivial domain wall between $\SC^{n+1}$ and $\SC^{n+1}$. We apply the folding trick such that $\SC^n$ becomes a gapped boundary of a double-layered system as illustrated below: 
\be \label{eq:folding_trick}
\begin{array}{c}
\begin{tikzpicture}[scale=1.5]
\draw[black,thick,-stealth] (-1,0) -- (-0.5,0) node[very near end, below] {\scriptsize $x^1$}; 
\draw[black, ultra thick,->-] (0,0) -- (1,0) ;
 \draw[black, ultra thick,->-] (1,0) -- (2,0) ;
\draw[fill=black] (0.97,-0.03) rectangle (1.03,0.03) node[midway,above,scale=1] {\scriptsize $\hspace{1mm}\SC^n$} ;
\node at (0.5,0.15) {\scriptsize $\SC^{n+1}$} ;
\node at (1.5,0.15) {\scriptsize $\SC^{n+1}$} ;
\end{tikzpicture}
\end{array}
\quad \xrightarrow{\mbox{\scriptsize folding trick}} \quad
\begin{array}{c}
\begin{tikzpicture}[scale=1.5]
\draw[black, ultra thick,->-] (0,0.2) -- (1,0) node[midway,above] {\scriptsize $\overline{\SC^{n+1}}$} ;
\draw[fill=black] (0.97,-0.03) rectangle (1.03,0.03) node[midway,below,scale=1] {\scriptsize $\hspace{1mm}\SC^n$} ;
\draw[black, ultra thick,->-] (0,-0.2) -- (1,0) node[midway,below] {\scriptsize $\SC^{n+1}$} ;
\end{tikzpicture}
\end{array}
\ee
where $\overline{\SC^{n+1}}$ represents the $n+$1D topological order obtained from $\SC^{n+1}$ by folding and $\boxtimes$ represents the stacking of two $n+$1D topological orders (without introducing interactions between them). 
We obtain the following identity: 
\be \label{eq:folding}
\SZ(\SC^n)^{n+1} = \SC^{n+1} \boxtimes \overline{\SC^{n+1}}. 
\ee
In this work, we assume that a folding is done in the $x^1$-th spatial dimension, which is transversal to the domain wall and is depicted as the horizontal direction in pictures (see Convention\,\ref{rem:convention_op}). 
\end{expl}

Since a topological order is a macroscopic notion, in principle, it should be characterized by the complete set of macroscopic observables, which mainly (but not completely) consists of topological defects. Therefore, the anomaly-free condition of a topological order should be translated to a condition on its macroscopic observables or topological defects. 

\medskip
Now we are ready to state a guiding principle: Remote Detectable Principle, which was proposed in \cite{KW14} (preceded by a physical discussion of $\Omega\CC$ in $n=2$ cases in \cite{Lev13}). 
\be
\mbox{\fbox{\begin{varwidth}{\textwidth}
{\bf Remote Detectable Principle}: In an anomaly-free topological order, \\
a topological defect should be detectable up to condensation descendants \\
by some topological defects via double braidings.  
\end{varwidth}}}  \label{Remote_Detectable_Principle}
\ee

This principle immediately provides an answer to the question \eqref{eq:question}. Consider an anomaly-free 2+1D topological order $\SC^3$. All topological defects in $\SC^3$ form a fusion 2-category $\CC$. 
\bnu
\item[(1)] In this case, only defects of codimension 2 (i.e., anyons) can be braided. If a simple anyon $x$ has trivial double braidings with all anyons, by the Remote Detectable Principle, $x$ must be the trivial particle, i.e. $x=1_{\one_\CC}$. We can restate this result mathematically. For a braided fusion 1-category $\CB$, we define its $\EE_2$-center $\FZ_2(\CB)$ (also called M\"{u}ger center) by the full subcategory consisting of objects with trivial double braidings with all objects in $\CB$, i.e., 
$$
\FZ_2(\CB) := \{ z\in \CB | c_{z,b}\circ c_{b,z} = 1_{z\otimes b}, \forall b\in \CB \}. 
$$
Using this language, the anomaly-free condition of $\SC^3$ can be translated into the following condition:
\be \label{eq:Z2_trivial_condition}
\FZ_2(\Omega\CC) \simeq \vect.
\ee
It turns out that this equivalence \eqref{eq:Z2_trivial_condition}, viewed as an anomaly-free condition, is equivalent to the non-degeneracy condition of the $S$-matrix \cite{Mueg03a} (see also \cite{Lev13} for a physical discussion). 

\item[(2)] Note that 1-codimensional defects in $\SC^3$ cannot be braided at all. In this case, the Remote Detectable Principle implies that all 1-codimensional topological defects in $\CC$ are necessarily the condensation descendants (or condensed defects) of anyons or the trivial 1-codimensional topological defect $\one_\CC$. Indeed, it was shown in \cite{Kon14e} that all 1-codimensional topological defects and those higher codimensional topological defects in $\SC^3$ that are not in $\Omega\CC$ are the condensation descendants of those in $\Omega\CC$. This result implies that all 1-codimensional defects can be obtained from the trivial 1-codimensional defect $\one$ or $\mathrm{B}\Omega\CC$ via condensations. The process of including all condensation descendants of $\mathrm{B}\Omega\CC$ is called the condensation completion (or Karoubi completion) of $\mathrm{B}\Omega\CC$, and is denoted by $\Kar(\mathrm{B}\Omega\CC)$ \cite{GJF19}, and can be mathematically defined \cite{DR18,GJF19} (see Section\,\ref{sec:cc}). Following \cite{DR18,GJF19}, we set 
$$
\Sigma(-):=\Kar(\mathrm{B}(-)).  
$$
In this new notation, we have $\CC=\Sigma\Omega\CC := \Kar(\mathrm{B}\Omega\CC)$. 
Note that all objects in $\CC$ are connected by 2-codimensional topological defects because they are all condensation descendants of the trivial 1-codimensional topological defect $\one_\CC$. 
\enu
In summary, we have answered the question (\ref{eq:question}). Namely, the pair $(\Omega\CC,c)$ indeed provides a complete mathematical characterization of the anomaly-free 2+1D topological order $\SC^3$, and all the missing topological defects can be recovered from those in $\Omega\CC$ via condensations, i.e., $\CC=\Sigma\Omega\CC$. 

\begin{rem}
When $\SC^3$ is non-chiral, we have $\CC \simeq \BMod_{\CA|\CA}(2\vect)$ for a fusion 1-category $\CA$ (recall (\ref{eq:KK12})) \cite{KK12}. In this case, Remote Detectable Principle implies the following monoidal equivalence: 
\be \label{eq:SigmaZA=BMod_AA}
\Sigma\FZ_1(\CA) \simeq \BMod_{\CA|\CA}(2\vect). 
\ee
In Section\,\ref{sec:general_example_1-codim}, we rederive this result as a consequence of a 1-codimensional defect condensation in $\mathbf{1}^3$. 
\end{rem}

\medskip
Before we discuss how Remote Detectable Principle applies to $n+$1D anomaly-free topological orders for $n\geq 3$, we first review some useful notions and notations in higher category theory. An $n$-category $\CS$ consists of objects, which are also called $0$-morphisms, and $k$-morphisms between two $(k-1)$-morphisms for $1\leq k\leq n$. For a $k$-morphism $f$ in $\CS$, we denote the identity morphism on $f$ by $1_f$. We simplify the notation $1_{1_f}$ to $1_f^2$ and define $1_f^k:=1_{1_f^{k-1}}$ inductively. We define $\Omega^0\CS:=\CS$ and $\Omega_f\CS:=\hom_{\CS}(f,f)$. If an $n$-category $\CA$ is monoidal, it is equipped with a canonical object the tensor unit $\one_\CA$. In this case, we abbreviate $\Omega_{\one_\CA}\CA$ to $\Omega\CA$. For example, the category $\Omega_f\CS:=\hom_{\CS}(f,f)$ is monoidal. Therefore, we can define: 
$$
\Omega_f^k\CS:=\Omega(\Omega_f^{k-1}\CS)=\Omega^{k-1}(\Omega_f\CS) = \hom_{\CS}(1_f^{k-1},1_f^{k-1}). 
$$
In particular, $\Omega^{n-k}\CS$ is a $k$-category for $k\leq n$. For a monoidal $n$-category $\CA$, we define the looping of $\CA$, denoted by $\Omega\CA$, by $\hom_\CA(\one_\CA,\one_\CA)$; and define the one-point delooping of $\CA$, denoted by $\mathrm{B}\CA$, to be the $(n+1)$-category consisting of a single object $\bullet$ such that $\Omega_\bullet \mathrm{B}\CA:=\hom_{\mathrm{B}\CA}(\bullet, \bullet)=\CA$ as illustrated below. 
\be \label{eq:def_1-pt-delooping}
\mathrm{B}\CA := 
\begin{array}{c}
\xymatrix{
\bullet \ar@(ul,ur)[]^{\CA=\hom_{\mathrm{B}\CA}(\bullet, \bullet)}
}
\end{array}
\ee
Note that the monoidal structure of $\CA$ is necessary for $\mathrm{B}\CA$ to be well-defined because $\hom_{\mathrm{B}\CA}(\bullet, \bullet)$ has a monoidal structure defined by the composition of 1-morphisms.

Then we state a result which is a consequence of a stronger result: Theorem$^{\mathrm{ph}}$\,\ref{pthm:1codim_condensation_nd_II}. We simply take it for grant in this section (see Remark\,\ref{rem:infinite-type_condensation} for a discussion on the limitation of this result). 
\begin{pthm}[\cite{KW14,GJF19}] \label{pthm:condensed=connected}
Given two $k$D topological defects $\SA^k$ and $\SB^k$ (or anomalous topological orders),  $\SB^k$ can be obtained from $\SA^k$ via a condensation of higher codimensional topological defects in $\SA^k$ if and only if they can be connected by a $(k-1)$D gapped domain wall\footnote{This condition is also the defining axiom of the notion of Morita equivalence introduced in Definition\,\ref{def:TME}.}. 
\end{pthm}

Now we consider an anomaly-free $n+$1D topological order $\SC^{n+1}$ for $n\geq 3$. It can have topological defects of all codimensions. All possible topological defects in $\SC^{n+1}$ form a monoidal $n$-category $\tilde{\CC}$, in which $k$-morphisms are $(k+1)$-codimensional topological defects for $0\leq k\leq n$; when $k=n$, they are instantons (i.e., $0$D defects or $(n+1)$-codimensional defects). We reserve the notation $\CC$ for a subcategory of $\tilde{\CC}$ described later. The monoidal structure on $\tilde{\CC}$, i.e., a fusion product $\otimes$ in $\tilde{\CC}$, is defined by the fusion of two 1-codimensional topological defects. The tensor unit is the trivial 1-codimensional topological defect, denoted by $\one$. 
The category $\tilde{\SC}^{n+1}$ is very large. In particular, for $k\leq n$, by stacking a $k$D anomaly-free topological order to an $(n-k)$-morphism in $\tilde{\CC}$, we obtain a new $(n-k)$-morphism in $\tilde{\CC}$ \cite{KW14,KLWZZ20a}. When $\SC^{n+1}=\mathbf{1}^{n+1}$, an anomaly-free $k$D topological order for $k\leq n$ should be viewed as $(n-k+1)$-codimensional topological defects in $\mathbf{1}^{n+1}$, thus defines an $(n-k)$-morphism in $\tilde{\CC}$ \cite{KLWZZ20a}. Since we have excluded gapless defects, a $k$D chiral topological order is not connected to the trivial $k$D topological order $\mathbf{1}^k$ in $\SC^{n+1}=\mathbf{1}^{n+1}$ by any $(n-k+1)$-morphisms. Moreover, for $n\geq 3$, in general, two 1-codimensional topological defects are not connected by gapped domain walls. When $\SC^{n+1}=\mathbf{1}^{n+1}$, we denote the $\tilde{\CC}$ in this case by $n\widetilde{\vect}$. 


Now we discuss how Remote Detectable Principle affects the categorical description of the anomaly-free $n+$1D topological order $\SC^{n+1}$.

\bnu
\item[(1)] Topological defects of codimension 2 are braidable and should be detectable via double braidings. As a consequence, if a simple $k$-codimensional defect for $k\geq 2$ has trivial double braidings with all topological defects of codimension 2, it must be the trivial $k$-codimensional topological defect $1_\one^k$ or its condensation descendants. Note that the double braiding should be insensitive to condensation descendants as illustrated below. 
$$
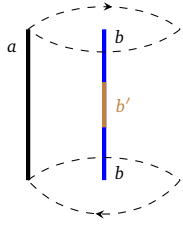
\begin{array}{c}
\begin{tikzpicture}
\draw[black,ultra thick] (0,0) -- (0,2) node[very near end, left] {\scriptsize $a$}; 
\draw[blue,ultra thick] (1,0) -- (1,2) ; 
\draw[brown,ultra thick] (1,0.7) -- (1,1.3) node[midway, right] {\scriptsize $b'$}; 
\draw[dashed,->-] (0,2) .. controls (0.5,2.4) and (1.5,2.4) .. (2,2);
\draw[dashed] (2,2) .. controls (1.5,1.6) and (0.5,1.6) .. (0,2);
\node at (1.2,1.9) {\scriptsize $b$} ;
\node at (1.2,0.1) {\scriptsize $b$} ;
\draw[dashed] (0,0) .. controls (0.5,0.4) and (1.5,0.4) .. (2,0);
\draw[dashed,->-] (2,0) .. controls (1.5,-0.6) and (0.5,-0.6) .. (0,0);
\end{tikzpicture}
\end{array}
\quad\quad
\begin{array}{l}
\mbox{During the adiabatic move of a double braiding}, \\
\mbox{a creation followed by an annihilation of} \\
\mbox{a condensation descendant $b'$ of $b$} \\
\mbox{should contribute no detectable difference}. 
\end{array}
$$
In higher dimensional cases, the `trivial double braiding' does not require the double braiding to be trivial on the nose, instead, it only requires the double braiding to be trivial up to higher isomorphisms. More explicitly, for $a,b \in \Omega\CC$, the double braiding being trivial means that there is a higher isomorphism $c_{b,a}\circ c_{a,b} \to 1_{a\otimes b}$ satisfying natural physical conditions. When $n=3$, these conditions have been explicitly spelled out in \cite[Definition 3.10]{KTZ20} (preceded by a definition in the semistrict\footnote{Physical applications demand us to consider non-semistrict monoidal 2-categories \cite{KTZ20}.} cases in \cite{Cra98}) as the defining properties of the so-called sylleptic center, which is an $\EE_2$-center\footnote{The notion of $\EE_2$-center is a well-defined mathematical notion \cite{Lur17} (see Definition\,\ref{def:centralizer+center}). Actually, all types of centers, such as usual center of an algebra, Drinfeld center, M\"{u}ger center and left/right or full centers in 2D conformal field theories, are all universally defined by the same universal property but in different (higher) categories \cite{Lur17,Ost03,Dav10,KYZ21}.} and a 2-categorical generalization of the M\"{u}ger center.

\item[(2)] The category $n\widetilde{\vect}$ of topological defects in $\mathbf{1}^{n+1}$ is not compatible with Remote Detectable Principle. 
\bnu
\item Lower dimensional anomaly-free $k$D topological orders for $k\leq n$ can not be detected by double braidings because braidings among lower dimensional anomaly-free $k$D topological orders for $k\leq n$ are completely symmetric. Mathematically, it means that the category $n\widetilde{\vect}$ is symmetric monoidal. As a consequence, Remote Detectable Principle requires that we either modify the categorical description $n\widetilde{\vect}$ or demands that all lower dimensional anomaly-free $k$D topological orders are condensation descendants of the trivial $k$D topological orders for $k\leq n$.

\item However, by Theorem$^{\mathrm{ph}}$\,\ref{pthm:condensed=connected}, lower dimensional chiral topological orders are not condensation descendants of the trivial $k$D topological defect. As pointed out in \cite{KW14}, if the domain wall between two topological defects are gapless and not gappable, it is still possible to relate these two topological defects via a `generalized condensation'. However, such a `generalized condensation' is necessarily beyond the condensation theory discussed in this work, which focus only on a finite type of condensations in the sense that neither infinite direct sums nor infinite dimensional vector spaces are allowed in the definition of a condensable algebra or in that of a (multi-)fusion $n$-category. It was suggested in \cite{KW14} that we should exclude all lower dimensional chiral topological orders in the categorical description of $\mathbf{1}^{n+1}$ (or $\SC^{n+1}$ in general). See Remark\,\ref{rem:infinite-type_condensation} for more discussion. 
\enu
Therefore, by restricting to only the finite type of condensations studied in this work, Remote Detectable Principle demands us to replace $n\widetilde{\vect}$ by a subcategory $n\vect$, which consists of only trivial $(k-1)$-codimensional topological defects $1_\one^k$ for $k=0,1, \cdots, n$ and their condensation descendants. As a consequence, we obtain $n\vect = \mathrm{Kar}(\mathrm{B}\,(n-1)\vect)  = \Sigma(n-1)\vect = \Sigma^{n-1}\vect$. By setting $\Sigma\Cb :=\vect$, we obtain $n\vect = \Sigma^n\Cb$. We define $n\vect$ more carefully in Section\,\ref{sec:nVec}.

\item[(3)] Similarly, in the categorical description of $\SC^{n+1}$, we would like to replace $\tilde{\CC}$ by a subcategory $\CC$, which contains only a set of `elementary' (or `intrinsic') topological defects \cite{KW14}. In other words, we should exclude all `extrinsic' topological defects, which include all lower dimensional chiral topological orders and all condensation descendants obtained from stacking a lower dimensional chiral topological order with an `intrinsic' topological defect. In spite of many attempts to make the meaning of `intrinsic' (or `elementary') precise in \cite[Section\ XI]{KW14}, it was still not very successful because a full fledged mathematical theory of condensations in higher categories is missing as it was explicitly pointed out in \cite[Conjecture\ 10 \& 11]{KW14}. This missing theory was developed later and is explained in details in Section\,\ref{sec:cc}. 
Since $\tilde{\CC}$ is never used in this work, by abusing language, we abbreviate the category $\CC$ of all `intrinsic' topological defects to the category $\CC$ of all topological defects. We provide more precise definition of $\CC$ after we introduce the notion of a separable $n$-category and that of a multi-fusion $n$-category. Now we derive a few important properties of $\CC$ via Remote Detectable Principle. 
\bnu
\item $\CC$ consists of the trivial 1-codimensional topological defect $\one$ (also denoted by $\one_\CC$) and is necessarily closed under fusion. Therefore, $\CC$ is a monoidal $n$-category. Since condensation descendants of `intrinsic' topological defects are necessarily `intrinsic', $\CC$ is necessarily condensation complete (or Karoubi complete) (see Section\,\ref{sec:cc} for more discussion).

\item Since topological defects of codimension 2 and higher are braidable, Remote Detectable Principle demands that only topological defects that have the trivial double braidings with all topological defects of codimension 2 and higher are the trivial topological defects $1_\one^k$ for $k=0,1, \cdots, n$ and their condensation descendants. Mathematically, the category of all defects of codimension 2 or higher that have trivial double braidings\footnote{In higher categories, a `trivial double braiding' does not mean that the double braiding is trivial on the nose. Instead, it means that it is `trivial' up to higher isomorphisms, which enter the definition of the `trivialness' as defining data. It means that $\FZ_2(\Omega\CC)$ is not a subcategory of $\Omega\CC$ in general.} to all defects is called $\EE_2$-center of $\Omega\CC$, denoted by $\FZ_2(\Omega\CC)$. Therefore, we obtain the following anomaly-free condition, for $n\geq 2$, 
\be \label{eq:anomaly-free_condition_OmegaC}
\FZ_2(\Omega\CC) \simeq (n-1)\vect. 
\ee


\item Since all the other defects not in $\Omega\CC$, including all non-trivial 1-codimensional defects, cannot be braided, they are necessarily the condensation descendants of $\Omega\CC$ or better $\mathrm{B}\Omega\CC$ \cite{KW14,JF22,KLWZZ20}. Namely, those defects not in $\mathrm{B}\Omega\CC$ can all be obtained from those in $\mathrm{B}\Omega\CC$ via condensations. In other words, we have 
$$
\CC=\Sigma\Omega\CC := \Kar(\mathrm{B}\Omega\CC),
$$
where $\Kar(-)$ denotes the condensation completion or Karoubi completion. As a consequence, all objects in $\CC$ are connected by 1-morphisms (see Remark\,\ref{rem:anomalous_cases}).  

\enu

\enu

\begin{rem} \label{rem:infinite-type_condensation}
Strictly speaking, replacing $n\widetilde{\vect}$ by its subcategory $n\vect$ of `intrinsic topological defects' is more of a mathematical convenience than a demand of a physical principle because the mathematical theory of `generalized condensations' (of infinite type) is simply not available. However, a complete condensation theory should include the process of approaching the critical point of the phase transition, where the system closes the energy gap and becomes gapless. This process is not a finite process in nature. Therefore, a complete condensation theory might demand us to include certain `generalized condensations' of infinite type and `generalized condensation descendants' in the theory. As an example of thinking along this line, $n\widetilde{\vect}$ was used in \cite{KLWZZ20a}. Moreover, we provides a glimpse of such a `generalized condensation' in Section\,\ref{sec:gapless}, where some hom spaces become infinitely dimensional (see Remark\,\ref{rem:infinite-type_condensation_gapless}). 
\end{rem}

\begin{rem} \label{rem:anomalous_cases}
Above discussion applies to the case when $\SC^{n+1}$ is anomaly-free. If $\SC^{n+1}$ is anomalous, in general, 1-codimensional topological defects in $\SC^{n+1}$ are not necessarily the condensation descendants of those in $\Omega\CC$. It means that $\CC$ is disconnected in general when $\SC^{n+1}$ is anomalous. 
\end{rem}

\begin{rem}
Although $\Omega\CC$ has already encoded all the data in $\CC$ when $\SC^{n+1}$ is anomaly-free, as we show later, $\Omega\CC$ does not directly provide a complete story of defect condensations. For certain problems in condensation theory, it is necessary to consider $\CC=\Sigma\Omega\CC$ instead (see Section\,\ref{sec:condense_2-codim_defect}). 
\end{rem}

\begin{conv} \label{rem:convention_op}
In this work, a monoidal $n$-category $\CS$ is always interpreted physically as the category of topological defects. Therefore, we always endows $\CS$ with a local spacetime coordinate system 
$$
(x^1, x^2, \cdots, x^n,x^{n+1}=t)
$$ 
such that the fusion product in $\Omega^{i-1}\CS$, denoted by $\otimes^i$, is the fusion of defects along the positive $x^i$-th direction (see Figure\,\ref{fig:defects_in_2d-TO} and (\ref{eq:folding_trick})). In pictures, we always depicted the fusion of the lower-codimensional topological defects in the horizontal direction. According to this convention, the $n$-category of all topological defects of $\overline{\SC^{n+1}}$ is given by $\CC^\op$, which is the same category as $\CC$ but equipped with a new fusion product $(\otimes^1)^\op$, which is defined by $a(\otimes^1)^\op b :=b\otimes^1 a$ for $a,b\in \CC$ (see also Definition\,\ref{def:convention_op}). 
\end{conv}


\begin{defn} \label{def:op-Cat}
Given an $n$-category $\CS$, we denote the category obtained from flipping all $k$-morphisms in $\CS$ by $\CS^{\op k}$ and that obtained from flipping both $k$-morphisms and $l$-morphisms by $\CS^{\op k, \op l}$. For convenience, 
we abbreviate $\CS^{\op 1}$ to $\CS^\op$. 
\end{defn}

\subsubsection{Boundary-bulk relation} \label{sec:bbr}

There is an obvious question: can we use $\CC$ (instead of $\Omega\CC$) to give a mathematically equivalent  characterization of the anomaly-free condition on $\SC^{n+1}$? We show below how this question naturally leads us to a new guiding principle. 

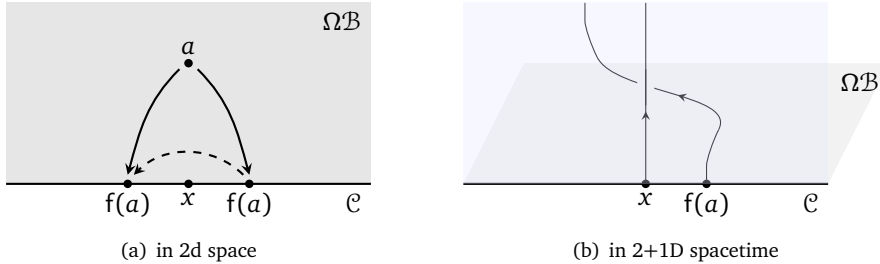
\begin{figure}[htbp]
\centering
\subfigure[in 2d space]{
\begin{tikzpicture}[scale=0.8]
\fill[gray!20] (-3,0) rectangle (3,3) node[at end,below left,black] {$\Omega\CB$} ;
\draw[thick] (-3,0)--(3,0) node[at end,below left] {$\CC$} ;

\fill (0,0) circle (0.07) node[below] {$x$} ;
\fill (0,2) circle (0.07) node[above] {$a$} ;
\fill (-1,0) circle (0.07) node[below] {$\forget(a)$} ;
\fill (1,0) circle (0.07) node[below] {$\forget(a)$} ;

\draw[-stealth,dashed,thick] (0.9,0.15) to [out=135,in=45] (-0.9,0.15) ;
\draw[-stealth,thick] (-0.15,1.9) to [out=225,in=75] (-1,0.15) ;
\draw[-stealth,thick] (0.15,1.9) to [out=-45,in=105] (1,0.15) ;
\end{tikzpicture}
}
\hspace{5ex}
\subfigure[in 2+1D spacetime]{
\begin{tikzpicture}[scale=0.8]
\fill[gray!10] (-1,0)--(5,0)--(6,2)--(0,2)--cycle ;

\draw (3,0)--(3,0.3) .. controls (3,0.4) and (3.1,0.7) .. (3.2,0.9) ; 
\draw[->-=0.4] (3.2,0.9) .. controls (3.5,1.5) and (1.5,1.5) .. (1.2,2.1) ;
\draw (1.2,2.1) .. controls (1.1,2.3) and (1,2.6) .. (1,2.7)--(1,3) ;
\begin{scope}
\clip (2,1.6) circle (0.3) ;
\draw[gray!10,double=black,double distance=0.4pt,line width=3pt] (2,0)--(2,3) ;
\end{scope}
\draw[->-=0.4] (2,0)--(2,3) ;

\node[below left] at (6,2) {$\Omega\CB$} ;
\draw[thick] (-1,0)--(5,0) node[at end,below left] {$\CC$} ;
\fill (2,0) circle (0.07) node[below] {$x$} ;
\fill (3,0) circle (0.07) node[below] {$\forget(a)$} ;

\fill[blue!10,opacity=0.3] (-1,0) rectangle (5,3) ;
\end{tikzpicture}
}
\caption{the half-braiding of $\forget(a)$ with $x$}
\label{fig:central_structure}
\end{figure}

\medskip
This question is already non-trivial when $n=1$. In this case, only topological defects in $\SC^2$ are particles and instantons, and they form a 1-category $\CC$. Since particles can be fused along the 1-dimensional space, $\CC$ is a monoidal 1-category. By Remote Detectable Principle, if $\SC^2$ is anomaly-free, then $\CC$ can only contain the trivial particles. In other words, $\CC$ must be trivial, i.e. $\CC\simeq \vect$. This is compatible with the fact that there is no non-trivial 1+1D anomaly-free topological order, proved first in a microscopic approach \cite{CGW10}. In order to formulate the anomaly-free condition on $\CC$, we need consider the case that $\SC^2$ is anomalous, i.e., $\SB^3:=\SZ(\SC)^3\neq \mathbf{1}^3$. This case can be physically realized in lattice model \cite{LW05,KK12}. 
In this case, $\CC$ have non-trivial particles. The bulk particles move onto the boundary and become a boundary particle. This process defines a functor $\forget: \Omega\CB \to \CC$, which clearly preserves the fusion products, i.e., $\forget(a) \otimes \forget(b) \simeq \forget(a\otimes b)$ \cite{KK12,FSV13}. Mathematically, such a functor is called a monoidal functor. Moreover, for $a\in \Omega\CB$, $\forget(a)$ can be half-braided with $x\in \CC$ as illustrated in Figure\,\ref{fig:central_structure}. 
This half-braiding structure endows the functor $\forget$ with a structure called a central functor \cite{Bez04} (see \cite{FSV13} and a recent review \cite{KZ22a} for its physical meanings). This notion is defined by the property that the functor $\forget$ can be factorized as the composition of two functors $\Omega\CB \to \FZ_1(\CC) \to \CC$, where $\FZ_1(\CC)$ is the $\EE_1$-center (or Drinfeld center) of $\CC$ and the second functor is the so-called forgetful functor. Then the condition that particles in $\CC$ can be detected by bulk particles via the half braidings \cite{Lev13} can be mathematically reformulated precisely as the following condition \cite{KK12,Kon14e}: 
\be \label{eq:bbr_2d}
\Omega\CB\simeq \FZ_1(\CC),
\ee
which is also called boundary-bulk relation. 
In other words, the functor $\forget: \Omega\CB \to \CC$ can be identified with the forgetful functor $\forget: \FZ_1(\CC) \to \CC$. As a consequence, the anomaly-free condition on $\SC^2$ can be translated to a condition on $\CC$ expressed as follows: 
\be \label{eq:E1-center=1Vec}
\FZ_1(\CC) \simeq \vect. 
\ee
Note that the only fusion 1-category\footnote{There are more multi-fusion 1-category solutions discussed later.} solution of above equation is $\CC \simeq \vect$  \cite{EGNO15}. This mathematical fact provides a categorical proof of the physical fact that there is no non-trivial 1+1D topological order \cite{KW14,KWZ15}. 

The boundary-bulk relation was generalized to formulations other than (\ref{eq:bbr_2d}) (see (\ref{eq:bbr_Z0})) and to all dimensions and codimensions as the center functor (see Section\,\ref{sec:center_functor}) and to quantum phases far beyond topological orders \cite{KWZ15,KWZ17,KZ22b}. In summary, boundary-bulk relation, in its simplest form, can be stated as follows. 
\be 
\mbox{\fbox{\begin{varwidth}{\textwidth}
{\bf Boundary-Bulk Relation}: The bulk $\SZ(\SC)^{n+2}$ is the center
 of a boundary $\SC^{n+1}$.
\end{varwidth}}}  \label{eq:bbr}
\ee
This result\footnote{This result has a precise meaning in the symmetric monoidal higher category $\ST\SO^n$ of potentially anomalous $n$D topological orders. A morphism in $\ST\SO^n$ is defined in \cite{KWZ15,KWZ17} (see also Figure\,\ref{fig:domain_wall=functor}), 
its symmetric fusion product is defined by the stacking of two topological orders, and its tensor unit is $\mathbf{1}^n$. The notion of the center in $\ST\SO^n$ is well defined by its universal property (see Definition\,\ref{def:centralizer+center}).} was proved via an important notion of a morphism between topological orders \cite{KWZ15,KWZ17}, a notion which can be generalized to gapped/gapless quantum liquids \cite{KWZ17,KZ22b}. We should regard (\ref{eq:bbr}) as a formal guiding principle, which leads to concrete predictions when we replace $\SZ(\SC)^{n+2}$ and $\SC^{n+1}$ by the concrete categories of topological defects. For example, we should still have (\ref{eq:bbr_2d}) for $n>1$, i.e.,
\be \label{eq:bbr_nd}
\Omega\CB\simeq \FZ_1(\CC).
\ee
It is important to note that this replacement is guaranteed by the fact that the notion of a monoidal functor is compatible with that of a morphism between two topological orders \cite[Theorem\, 3.2.3]{KZ18} (see also \cite{KWZ15}) and \cite[Remark\ 3.30]{KZ24}.

\begin{rem} \label{rem:bbr}
One should use the boundary-bulk relation with caution. Notice that both $\Omega\CB$ and $\CC$ give completable descriptions of defects in $\SB^{n+2}$ and $\SC^{n+1}$, respectively. This is a necessary condition in the applications of the boundary-bulk relation. If we do not respect this rule and apply it naively, we could have, for example, $\Omega^2\CB \simeq \FZ_2(\Omega\CC)$, which is not true in general (unless $\CC=\Sigma\Omega\CC$). For example, when $\CC=2\vect_G$ for a finite group $G$, $\Omega^2\CB=\Rep(G)$ \cite{LKW18,KTZ20} but $\FZ_2(\Omega\CC)=\vect$. Therefore, although the boundary-bulk relation (\ref{eq:bbr}) is clearly compatible with the anomaly-free condition on $\Omega\CC$ given in (\ref{eq:anomaly-free_condition_OmegaC}), 
we should not view (\ref{eq:anomaly-free_condition_OmegaC}) as a direct application of the boundary-bulk relation, but instead, a consequence of the condition (\ref{eq:anomaly-free_condition_C}). 
\end{rem}

The boundary-bulk relation (\ref{eq:bbr_nd}) immediately implies the following anomaly-free condition on $\CC$ generalizing (\ref{eq:E1-center=1Vec}) \cite{KWZ15,JF22,KLWZZ20} (see Remark\,\ref{rem:compatibility_center_Kar}): 
\be \label{eq:anomaly-free_condition_C}
\FZ_1(\CC) \simeq n\vect, 
\ee
where $n\vect$ is the category of `intrinsic' topological defects in $\mathbf{1}^{n+1}$ (recall Section\,\ref{sec:RDP}) and its precisely definition is given in Section\,\ref{sec:cc}. Since $\Omega\CC$ encodes all the information of  $\CC$ up to condensation descendants, we expect that the two anomaly-free condition (\ref{eq:anomaly-free_condition_OmegaC}) and (\ref{eq:anomaly-free_condition_C}) are mathematically equivalent \cite{KWZ15,KLWZZ20}. Based on a corrected definition of a multi-fusion $n$-category, this conjecture was proved (at a physical level of rigor) by Johnson-Freyd in \cite{JF22} (see also \cite{KZ24} for a different approach).

\begin{rem} \label{rem:compatibility_center_Kar}
We give a historical remark on the proposal or conjecture (\ref{eq:anomaly-free_condition_C}). Although, in retrospect, (\ref{eq:anomaly-free_condition_C}) seems a natural guess,  there is no mathematical evidence to support this guess beyond the known result (\ref{eq:E1-center=1Vec}) in 2014-2015\footnote{Although an obvious thing to do is check if it is true for $n=2$ when $\CC=\BMod_{\CA|\CA}(2\vect)$, how to compute the Drinfeld center of $\CC$ was not known until years later.} when the program was initiated in \cite{KW14,KWZ15}. There are some natural questions that made the conjecture (\ref{eq:anomaly-free_condition_C}) questionable. The bulk $\SZ(\CC)^{n+2}$ is needed simply because it fix the braiding non-detectability of $\SC^{n+1}$ by the half-braidings between the bulk defects and the boundary defects.  Therefore, one could view (\ref{eq:anomaly-free_condition_C}) as a reformulation of Remote Detectable Principle. On the one hand, condensation descendants do not play an essential role in braidings. It is possible that condensation descendants are not needed for boundary-bulk relation to hold. For example, replacing $\CC$ by $\mathrm{B}\Omega\CC$ in (\ref{eq:anomaly-free_condition_C}) could be another possibility. Moreover, `extrinsic' topological defects should be excluded \cite{KW14}. But how to distinguish `intrinsic' and `extrinsic' defects contributed additional complication to the theory \cite[Section\ XI]{KW14}. On the other hand, the mathematical notion of center is universal (or maximal in some sense). It suggests that it is reasonable to include all condensation descendants for boundary-bulk relation to hold. Since $\CC$ is the maximal possible choice (for all `intrinsic' topological defects), it is called a maximal $\mathrm{BF}^{pre}$-category instead of a (multi-)fusion higher category in \cite{KW14}. The later term was reserved for the one that is compatible with the notion of center (see  \cite[Definition\ 2.21]{KWZ15}). The key question is whether the notion of center is compatible with condensation completion. In mathematics, the notion of a fusion 2-category was officially introduced by Douglas and Reutter in \cite{DR18} in 2018 without knowing the physically required compatibility. It was based on a detailed study of the idempotent completion of 2-categories previously introduced by Carqueville and Runkel in \cite{CR16}, and was also supported by their success in the state-sum construction of 3+1D TQFT. However, it is unclear if this mathematical notion of a fusion 2-category is compatible with the notion of center as required by physics. The first evidence of its compatibility with center appeared in 2019 in \cite{KTZ20}. The original goal of this work was to find the correct mathematical definition of a (modular) fusion 2-category. By explicitly computing the $\EE_1$-center $\FZ_1(2\vect_G^\omega)$ of $2\vect_G^\omega$, it was shown that $\FZ_1(2\vect_G^\omega)$ contains all 2-codimensional condensation descendants in 3+1D Dijkgraaf-Witten theories. Then it becomes clear that the conjecture (\ref{eq:anomaly-free_condition_C}) is reasonable\footnote{Based on the new results in \cite{DR18,KTZ20}, the first author of this paper explained to Johnson-Freyd in the spring of 2019 the conjectured equivalence between two anomaly-free conditions (\ref{eq:anomaly-free_condition_OmegaC}) and (\ref{eq:anomaly-free_condition_C}) in its correct form instead of the incorrect one (missing condensation completion) in \cite{KWZ15}, but did not correct this mistake before \cite{JF22,KLWZZ20} due to an ambitious but unfinished plan of rewriting the whole paper \cite{KWZ15}.}. In 2020, two independent works \cite{JF22,KLWZZ20} on this conjecture appeared within the same week on arXiv. Based on Gaiotto and Johnson-Freyd's work \cite{GJF19} on condensation completion in higher categories, in the beautiful work \cite{JF22}, Johnson-Freyd introduced the mathematical definition of a (multi-)fusion $n$-category and established the category theory of `intrinsic' topological defects, including a proof of the equivalence between two anomaly-free conditions (\ref{eq:anomaly-free_condition_OmegaC}) and (\ref{eq:anomaly-free_condition_C}) (first conjectured in \cite[Definition\ 2.21]{KWZ15}). Somewhat orthogonal to \cite{JF22}, the work \cite{KLWZZ20} was devoted to developing a categorical theory of SPT/SET orders, which automatically includes that of topological orders as a special case, and the categorical theory of topological orders and (multi-)fusion $n$-categories were largely taken for granted. In \cite{KLWZZ20}, the compatibility between center and condensation completion was proposed as a general principle called {\it Condensation Completion Principle}\footnote{There is no mathematical proof of the necessity of the condensation completion of the category of boundary defects. However, it is possible to argue from the Naturality Principle that things should appear naturally unless they are forbidden by a physical law.}, which applies to all gapped/gapless quantum liquids (including topological orders). 
A more systematic approach based on separable $n$-categories was developed later in \cite{KZ22b} and is reviewed in Section\,\ref{sec:separable_nCat_composite_TO}. 
\end{rem}

Notice that the boundary-bulk relation (\ref{eq:bbr_nd}) only involves $\Omega\CB$. It is natural to ask if it is possible to have a formulation of boundary-bulk relation in terms of $\CB$? 
Observe that if we include all 1-codimensional topological defects in $\SB^{n+2}$, then fusing a non-trivial 1-codimensional topological defect to the boundary $\SC^{n+1}$ changes the boundary condition. Since $\SB^{n+2}$ is anomaly-free, an 1-codimensional topological defects in $\SB^{n+2}$ is necessarily a condensation descendant of the trivial 1-codimensional topological defect $\one_\CB$. It means that a 1-codimensional topological defect in $\SB^{n+2}$ is always connected to $\one_\CB$ by a $n$D gapped domain wall. Therefore, new boundary conditions obtained by fusing 1-codimensional topological defects in $\SB^{n+2}$ with $\SC^{n+1}$ are necessarily obtained from $\SC^{n+1}$ via condensations and connected to $\SC^{n+1}$ by a $n$D gapped domain wall\footnote{However, it is not true that all gapped boundaries of $\SB^{n+2}$ can be connected to $\SC^{n+1}$ by an $n$D gapped domain wall. For example, when $\SB^{n+2}=\mathbf{1}^4$, a chiral anomaly-free 2+1D topological order $\SC^3$ cannot connect to a non-chiral anomaly-free 2+1D topological order via a gapped domain wall, or equivalently, such a domain wall is necessarily gapless.}. We denote by $\Sigma\CC$ the $(n+1)$-category of all gapped boundary conditions of $\SB^{n+2}$ that can be obtained from $\SC^{n+1}$ via condensations. The notation $\Sigma\CC$ represents the condensation completion of $\mathrm{B}\CC$ and is explained in Section\,\ref{sec:cc}. Applying the boundary-bulk relation again, we should expect $\CB \simeq \FZ_0(\Sigma\CC)$, where $\FZ_0(\Sigma\CC)$ is the $\EE_0$-center 
of $\Sigma\CC$ (explained in Section\,\ref{sec:center}) and can be defined explicitly by $\Fun(\Sigma\CC,\Sigma\CC)$, i.e., the category of $\Cb$-linear functors from $\Sigma\CC$ to $\Sigma\CC$. We summarize boundary-bulk relation below.

\begin{pthm}[Boundary-Bulk Relation]
Let $\SB^{n+2}$ be an anomaly-free  topological order with a gapped boundary $\SC^{n+1}$. We have
\begin{align}
\Omega\CB &\simeq \FZ_1(\CC), \label{eq:bbr_Z1} \\
\CB &\simeq \FZ_0(\Sigma\CC) = \Fun(\Sigma\CC,\Sigma\CC). \label{eq:bbr_Z0}
\end{align}
Moreover, these two formulations of boundary-bulk relation are necessarily equivalent. This equivalence immediately implies the following braided equivalence: 
\be \label{eq:Z1_Z0}
\FZ_1(\CC) \simeq \Omega\FZ_0(\Sigma\CC). 
\ee
\end{pthm}

\begin{rem}
The result (\ref{eq:Z1_Z0}) is a mathematical fact proved in \cite{Fra12,JF22,KZ22b}. 
\end{rem}

In summary, we have given a description of the category of topological defects in an $n+$1D anomaly-free (simple) topological order $\SC^{n+1}$, and we show that the anomaly-free condition on $\SC^{n+1}$ can be translated into a condition on $\Omega\CC$ (\ref{eq:anomaly-free_condition_OmegaC}) or a condition on $\CC$ (\ref{eq:anomaly-free_condition_C}). It is clear that, in order to move further, we need make the $\Omega\CC, \CC, n\vect$ in these conditions (\ref{eq:anomaly-free_condition_OmegaC}) and (\ref{eq:anomaly-free_condition_C}) mathematically precise. This is the subject of the next subsection.

\subsection{Condensation completion} \label{sec:cc}

For a topological order $\SC^{n+1}$, the process of completing $\mathrm{B}\Omega\CC$ by its condensation descendants is called the {\it condensation completion} (or {\it Karoubi completion}) of $\mathrm{B}\Omega\CC$ \cite{CR16,DR18,GJF19}. More generally, for an arbitrary $k$-category $\CA$, it is possible to define the Karoubi completion of $\CA$, denoted by $\Kar(\CA)$. This notion was first introduced in the $k=2$ cases by Carqueville and Runkel in \cite{CR16}, and was later thoroughly developed by Douglas and Reutter in \cite{DR18}, and was later generalized to all $k\geq 2$ by Gaiotto and Johnson-Freyd in \cite{GJF19}. In this subsection, we briefly review the basic ideas of condensation completion \cite{GJF19} via a physical approach along the line of \cite{KW14,KWZ15,KLWZZ20a}, and introduce the notions of separable $n$-categories \cite{KZ22b} and multi-fusion $n$-categories \cite{JF22,KZ22b}. 

\subsubsection{Category \texorpdfstring{$n\vect$}{nVec}} \label{sec:nVec}
In the previous subsection, $\CC$ is assumed a priori and automatically condensation complete because it contains all (`intrinsic') topological defects in $\SC^{n+1}$, including all condensation descendants, by definition. We want to take a closer look at topological defects in $\SC^{n+1}$, especially when $\SC^{n+1}=\mathbf{1}^{n+1}$. 

\medskip
Recall Theorem$^{\mathrm{ph}}$\,\ref{pthm:condensed=connected}. We show in Theorem$^{\mathrm{ph}}$\,\ref{pthm:1codim_condensation_nd_II} that the condensation in Theorem$^{\mathrm{ph}}$\,\ref{pthm:condensed=connected} can be defined by the gapped domain wall directly. Now we simply take it for granted. One immediate corollary is that a $k$D topological defect $\SB^k$ can be obtained from another one $\SA^k$ via a condensation of higher codimensional defects living in $\SA^k$ if and only if $\SA^k$ can be obtained from $\SB^k$ via a condensation of higher codimensional defects living in $\SB^k$. Therefore, any two $k$-morphisms $f,g \in \CC$, the $(n-k)$-category $\hom_\CC(f,g)$ is a direct sum of connected full sub-categories. When $\SC^{n+1}$ is anomaly-free, the monoidal $n$-category $\CC$ is connected because every 1-codimensional defect is a condensation descendant of the trivial one $\one_\CC$ and, therefore, is connected to $\one_\CC$ via a gapped domain wall (recall discussion in Section\,\ref{sec:RDP}). We used the following connected diagram to give an intuitive description of $\CC$: for $x\in \CC$, 
\[
\xymatrix{
\one_\CC \ar@(ul,ur)[]^{\Omega\CC} \ar@/^/[rr]^{\hom_\CC(\one_\CC,x)} & & x \ar@(ul,ur)[]^{\hom_\CC(x,x)} \ar@/^/[ll]^{\hom_\CC(x,\one_\CC)}
}
\]
Since all objects in $\CC$ are condensation descendants of $\one_\CC$, $\CC$ is the condensation completion (or Karoubi completion) of $\mathrm{B}\Omega\CC$, i.e. $\CC = \Kar(\mathrm{B}\Omega\CC)$. The notion of a condensation completion can be mathematically defined for any higher category \cite{GJF19}. 
Recall that the one-point delooping is defined for any monoidal higher category $\CA$. We define the delooping $\Sigma\CA$ of a condensation complete $\CA$ by: 
\be \label{eq:def_delooping}
\Sigma\CA := \Kar(\mathrm{B}\CA) = \Kar( \begin{array}{c}
\xymatrix{
\bullet \ar@(ul,ur)[]^{\CA}
}
\end{array} ) 
=
\begin{array}{c}
\xymatrix{
\bullet \ar@(ul,ur)[]^{\CA} \ar@/^/[rr]^{\LMod_A(\CA)} & & A \ar@(ul,ur)[]^{\BMod_{A|A}(\CA)} \ar@/^/[ll]^{\RMod_A(\CA)}
}
\end{array}
\ee
where an object $A$ in $\Sigma\CA$ is a `nice' algebra (called condensation monad) in $\CA$ as illustrated in the last `$=$' in (\ref{eq:def_delooping}) \cite{CR16,DR18,GJF19}. 
Then we have $\Sigma\Omega\CC = \CC$. We set $\Sigma\Cb:=\vect$. 
\begin{pthm}[\cite{GJF19}] \label{pthm:GJF_nVec}
$n\vect = \Sigma^n \Cb$. Physically, $n\vect$ is the category of `intrinsic' topological defects in $\mathbf{1}^{n+1}$, consisting of `roughly' all lower dimensional non-chiral topological orders and gapped domain walls among them (see Remark\,\ref{rem:meaning_of_an_object_in_nVec} and Theorem$^{\mathrm{ph}}$\,\ref{pthm:physical_meaning_separable_n-category} for more precise statement).  
\end{pthm}
\pf
This follows immediately from the fact that $n\vect$ contains only the trivial $k$-morphisms for $k\geq 0$ and their condensation descendants, and, as we have shown earlier, this fact is a consequence of Remote Detectable Principle and the limitation of our condensation theory (see Remark\,\ref{rem:infinite-type_condensation}). 
\epf

We can gain a better understanding of $n\vect$ by working out $n\vect$ for $n=2,3$ explicitly. 
\begin{expl} \label{expl:2Vec_condensation}
$2\vect=\Sigma\vect$ is the 2-category of topological defects in $\mathbf{1}^3$. In Example\,\ref{expl:2Vec}, we have shown that it is precisely that 2-category of finite semisimple 1-categories or finite direct sums of $\vect$. Now we work it out from the point of view of condensation completion. The trivial defect in $2\vect$ is precisely the trivial 1+1D topological order $\mathbf{1}^2$, particles in which form the trivial fusion 1-category $\vect$. Condensation completion demands us to add all the condensation descendants of the trivial 1+1D topological order. We can proceed in two slightly different but equivalent ways. 
\be \label{diag:2Vec}
\xymatrix{
\mathbf{1}^2 \ar@(ul,ur)[]^{\vect} \ar@/^/[rrr]^{\hom_{2\vect}(\vect,\CX)} & &  & \CX \ar@(ul,ur)[]^{\Fun(\CX,\CX)} \ar@/^/[lll]^{\hom_{2\vect}(\CX,\vect)} 
}
\quad\quad\quad\quad
\xymatrix{
\mathbf{1}^2 \ar@(ul,ur)[]^{\vect} \ar@/^/[rrr]^{\BMod_{\Cb|A}(\vect)} & & &  A \ar@(ul,ur)[]^{\BMod_{A|A}(\vect)} \ar@/^/[lll]^{\BMod_{A|\Cb}(\vect)} 
}
\ee
\bnu
\item The category of particles of each of these condensation descendants is necessarily a solution to the anomaly-free condition (\ref{eq:E1-center=1Vec}). All solutions are given by the indecomposable multi-fusion 1-category $\Fun(\CX,\CX)$ for a finite semisimple 1-category $\CX$. We see immediately that these condensation descendants form the 2-category $2\vect$ because objects in $2\vect$ are precisely finite semisimple 1-category $\CX$ and hom 1-category in $2\vect$ is precisely $\Fun(\CX,\CX)$ as illustrated in the first diagram in (\ref{diag:2Vec})\footnote{For readers who are familiar with Levin-Wen models with gapped boundaries \cite{KK12}, the 2-category $2\vect$ is just a special case of the 2-category of gapped boundary conditions of a fixed 2+1D Levin-Wen model based on a a fusion 1-category $\CA$. The objects in this 2-category are labeled by finite semisimple right $\CA$-modules and hom 1-categories are $\Fun_{\CC^\op}(\CM,\CN)$. This 2-category is precisely the 2-category of right $\CA$-modules in $2\vect$, denoted by $\RMod_\CA(2\vect)$. As a special case, $2\vect=\RMod_{\vect}(2\vect)$.}. 

Note that $\CX$ labels a 2D anomaly-free topological order, the category of topological defects in which is precisely the indecomposable multi-fusion 1-category $\Fun(\CX,\CX)$. When $\CX \simeq \vect^{\oplus k}$ for $k>1$, however, the associated 2D topological order does not satisfy the simpleness condition (recall Remark\,\ref{rem:simple_TO}) and can be identified with the direct sum of $k$ trivial phases: $(\mathbf{1}^2)^{\oplus k}=\mathbf{1}^2 \oplus \cdots \oplus \mathbf{1}^2$. Such a topological order is called  a composite topological order (see Definition\,\ref{def:simple_TO}) and naturally appear in dimensional reduction processes and condensations. For example, a 2+1D toric code model defined on a narrow strap with the smooth boundary condition on two sides can be viewed as a quasi-1+1D system that realizes the composite 1+1D anomaly-free topological order $\mathbf{1}^2 \oplus \mathbf{1}^2$ \cite{KWZ15,KZ22a}. More discussion and examples of composite 1+1D topological orders can be found in \cite{KZ22a}.

\item According \cite{CR16, Kon14e, DR18}, a condensation in $\mathbf{1}^2$ is defined by a separable algebra $A$ in $\vect$. The category of particles in the condensed 1+1D phase is given by the 1-category of $A$-$A$-bimodules in $\vect$, denoted by $\BMod_{A|A}(\vect)$ as illustrated in the second diagram in (\ref{diag:2Vec}). 
Then the condensation completion of the trivial 1+1D topological order can be identified with the 2-category of separable algebras in $\vect$ (as objects), bimodules (as 1-morphisms) and bimodule maps (as 2-morphisms). 
Since a separable algebra $A$ in $\vect$ is nothing but a finite direct sum of matrix algebras, $\BMod_{A|A}(\vect) \simeq \Fun(\CX,\CX)$ for some $\CX\in 2\vect$. Moreover, $\CX$ can be identified with $\RMod_A(\vect)$, which is the category of right $A$-modules in $\vect$ and is invariant up to the Morita equivalence of $A$. For this reason, one can relabel $A$ by its complete Morita invariant $\CX=\RMod_A(\vect)$. Therefore, this 2-category is precisely $2\vect$. 

\enu
In both cases, the finite semisimple 1-category $\CX$ (i.e. $\CX\in 2\vect$) pops out. Its physical means is the category of particles on the domain wall between $\mathbf{1}^2$ and one of its condensation descendants (say $x$), i.e., $\CX=\hom_{2\vect}(\vect, x)$.  
It is easy to see that $2\vect=\RMod_{\vect}(2\vect)$, which is a special case of a more general fact (see Theorem\,\ref{thm:SigmaA=RMod}). 
\end{expl}

\begin{expl} \label{expl:3Vec}
$3\vect=\Sigma 2\vect$ is the monoidal 2-category of topological defects in $\mathbf{1}^4$. The tensor unit $\bullet=\one$ is the trivial 1-codimensional topological defect, which is precisely the trivial 2+1D topological order $\mathbf{1}^3$, i.e., $\bullet=\one=\mathbf{1}^3$. A simple object $X$ in $3\vect$ is a condensation descendant of $\mathbf{1}^3$, and, in this case only, can be conveniently\footnote{This convenient interpretation of an object in $3\vect$ needs to be slightly modified for that of an object in $n\vect$ for $n>3$ (see Remark\,\ref{rem:meaning_of_an_object_in_nVec}, Theorem$^{\mathrm{ph}}$\,\ref{pthm:physical_meaning_separable_n-category} and Example\,\ref{expl:4Vec}).} interpreted as a 2+1D non-chiral topological orders $\SX^3$ (recall Definition\,\ref{def:anomaly_TO}). 
The 3-category $3\vect$ is connected as shown in the following diagram. 
$$
\xymatrix{
\mathbf{1}^3 \ar@(ul,ur)[]^{2\vect} \ar@/^/[rrr]^{\hom_{3\vect}(\mathbf{1}^3,\SX^3)} & & & \SX^3 \ar@(ul,ur)[]^{\CX} \ar@/^/[lll]^{\hom_{3\vect}(\SX^3,\mathbf{1}^3)} 
\ar@/^/[rrr]^{\hom_{3\vect}(\SX^3,\SY^3)} & & & 
\SY^3 \ar@(ul,ur)[]^{\CY} \ar@/^/[lll]^{\hom_{3\vect}(\SY^3,\SX^3)} 
}
$$
The 2-category $\hom_{3\vect}(\mathbf{1}^3,\SX^3)$ is the 2-category of gapped boundary conditions of $\SX^3$. Its objects $a,b,c,\cdots$ are the labels for gapped boundary conditions of $\SX^3$, and the 1-category of morphisms $\hom_{3\vect}(a,a)$ is the category of particles on the boundary $a$ and is an indecomposable multi-fusion 1-category, and $\hom_{3\vect}(a,b)=\hom_{3\vect}(b,a)^{\op}$ is the category of particles on the 0+1D gapped domain wall connecting $a$ and $b$. By Theorem$^{\mathrm{ph}}$\,\ref{pthm:condensed=connected}, the 2-category $\hom_{3\vect}(\mathbf{1}^3,\SX^3)$ is connected by non-zero 1-morphisms as illustrated below: 
$$
\hom_{3\vect}(\mathbf{1}^3,\SX^3)
\quad = \quad
\begin{array}{c}
\xymatrix{
a \ar@(ul,ur)[]^{\hom_{3\vect}(a,a)} \ar@/^/[rrr]^{\hom_{3\vect}(a,b)} & & & b \ar@(ul,ur)[]^{\hom_{3\vect}(b,b)} \ar@/^/[lll]^{\hom_{3\vect}(b,a)}
}
\end{array}
\quad = \quad 
\Sigma \hom_{3\vect}(a,a). 
$$
Since $\CX:=\hom_{3\vect}(\SX^3,\SX^3)$ is the precisely the monoidal 2-category of topological defects in $\SX^3$, by (\ref{eq:bbr_Z0}), we obtain
\be \label{eq:3vect_Fun_1}
\CX \simeq \FZ_0(\Sigma\hom_{3\vect}(a,a)) = \Fun(\hom_{3\vect}(\mathbf{1}^3,\SX^3), \hom_{3\vect}(\mathbf{1}^3,\SX^3)). 
\ee
The $\Omega \CX$ is the category of particles in the non-chiral 2+1D topological order $\SX^3$. By the boundary-bulk relation \cite{KK12,Kon14e}, it must be the Drinfeld center of the category of particles on each boundary $a \in \hom_{3\vect}(\mathbf{1}^3, \SX^3)$, i.e., 
$$
\Omega \CX \simeq \FZ_1(\hom_{3\vect}(a,a)),
$$ 
which a mathematical fact following the equivalence (\ref{eq:Z1_Z0}). Note that $\hom_{3\vect}(a,a)$ and $\hom_{3\vect}(b,b)$ are Morita equivalent. By the Morita theory of multi-fusion 1-categories \cite{Mueg03,Ost03}, we obtain the following equivalence of 2-categories: 
$$
\Sigma \hom_{3\vect}(a,a) \simeq \RMod_{\hom_{3\vect}(a,a)}(2\vect). 
$$
As a mathematical result, it was first obtained by Douglas and Reutter in \cite{DR18}. 

The direct sum of two non-chiral topological orders $\SX^3 \oplus \SY^3$ also lives in $3\vect$ and we have 
$$
\hom_{3\vect}(\mathbf{1}^3,\SX^3 \oplus \SY^3) \simeq 
\hom_{3\vect}(\mathbf{1}^3,\SX^3) \oplus \hom_{3\vect}(\mathbf{1}^3,\SY^3). 
$$
Such a generalized `topological order' is called a composite topological order (see Definition\,\ref{def:simple_TO}). It is not only a mathematical convenience but also a natural physical requirement when we study the dimensional reductions and the condensation theory of topological orders. In particular, we show in Section\,\ref{sec:general_example_1-codim} that this generalized `topological order' $\SX^3 \oplus \SY^3$ can be obtained from $\mathbf{1}^3$ by condensing the algebra $A=\hom_{3\vect}(x,x) \oplus \hom_{3\vect}(y,y)$ in $2\vect$ for $x\in \hom_{3\vect}(\mathbf{1}^3,\SX^3)$ and $y\in \hom_{3\vect}(\mathbf{1}^3,\SY^3)$. Therefore, from now on, we use $\SX^3$ and $\SY^3$ to denote generic objects in $3\vect$ which can be a direct sum. 

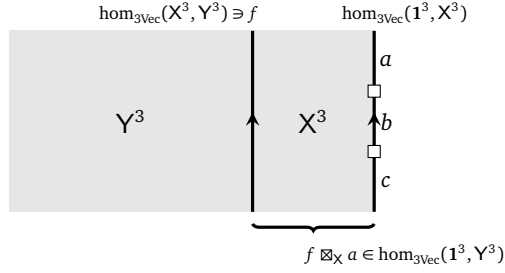
\begin{figure}
$$
\begin{array}{c}
\begin{tikzpicture}[scale=0.8]
\fill[gray!20] (-3,0) rectangle (3,3) ;
\draw[very thick,->-] (3,0)--(3,3) ; 
\draw[very thick,->-] (1,0)--(1,3) ;
\draw[fill=white] (2.9,0.9) rectangle (3.1,1.1) ; 
\node at (3.2,2.5) {\small $a$} ;
\node at (3.2,1.5) {\small $b$} ;
\node at (3.2,0.5) {\small $c$} ;
\draw[fill=white] (2.9,1.9) rectangle (3.1,2.1) ; 
\node at (3.5,3.3) {\scriptsize $\hom_{3\vect}(\mathbf{1}^3,\SX^3)$} ;
\node at (-0.2,3.3) {\scriptsize $\hom_{3\vect}(\SX^3,\SY^3)\ni f$} ;
\node at (2,1.5) { $\SX^3$} ;
\node at (-1,1.5) {$\SY^3$} ;
\draw[decorate,decoration=brace,very thick] (3,-0.2)--(1,-0.2) ;
\node at (3.5,-0.7) {\scriptsize $f\boxtimes_{\SX} a \in \hom_{3\vect}(\mathbf{1}^3,\SY^3)$} ;
\end{tikzpicture}
\end{array}
$$
\caption{a domain wall $f$ defines a functor $f\boxtimes_\SX -$ 
}
\label{fig:domain_wall=functor}
\end{figure}

Note that a gapped domain wall $f\in \hom_{3\vect}(\SX^3,\SY^3)$ naturally 
defines a functor 
$$
\hom_{3\vect}(\mathbf{1}^3,\SX^3)\xrightarrow{f\boxtimes_\SX -} \hom_{3\vect}(\mathbf{1}^3,\SY^3).
$$
The physical meaning of this functor is illustrated in Figure\,\ref{fig:domain_wall=functor}. 
Conversely, we claim that any functor from $\hom_{3\vect}(\mathbf{1}^3,\SX^3)$ to $\hom_{3\vect}(\mathbf{1}^3,\SY^3)$ should be realizable by such a domain wall, i.e.,  
\be \label{eq:3vect_Fun_2}
\hom_{3\vect}(\SX^3,\SY^3) = \Fun(\hom_{3\vect}(\mathbf{1}^3,\SX^3), \hom_{3\vect}(\mathbf{1}^3,\SY^3)).
\ee
There are two ways to prove this claim. 
\bnu
\item This claim follows from the so-called {\it Naturality Principle}, which says that if there is no physical law to forbid certain functors to be physically realizable (or occur physically), then all of them should be physically realizable. This principle is often used in bootstrap analysis, such as the bootstrap theory of rational CFT's \cite{MS89} and that of anyon condensations \cite{Kon14e}.  

\item In this approach, the precise mathematical definitions of the categories of boundary conditions of $\SX^3$ and $\SY^3$ were known \cite{JF22,KLWZZ20,KZ22b} (or see \cite{KZ22a} for a review). Then this claim is simply a mathematical fact, which follows immediately from \cite[Theorem\, 3.2.3]{KZ18} (see also \cite{KWZ15}). 

\enu
These two results (\ref{eq:3vect_Fun_1}) and (\ref{eq:3vect_Fun_2}) simply says that the same category $3\vect$ can be equivalently defined by the following replacement:
\begin{align*}
\SX^3 &\mapsto \hom_{3\vect}(\mathbf{1}^3,\SX^3), \\
\hom_{3\vect}(\SX^3,\SY^3) &\mapsto \Fun(\hom_{3\vect}(\mathbf{1}^3,\SX^3), \hom_{3\vect}(\mathbf{1}^3,\SY^3)).
\end{align*}
This fact can be reformulated in more mathematical language. Let $\Cat_2^{\Cb}$ be the 3-category of $\Cb$-linear 2-categories \cite{DR18}. This fact means that the following functor is fully faithful. 
\be \label{eq:yoneda_functor_3Vec}
\hom_{3\vect}(\bullet, -): 3\vect \to \Cat_2^{\Cb}
\ee
It turns out that the fully faithfulness of the functor (\ref{eq:yoneda_functor_3Vec}) can be generalized to $(n+1)\vect$. The generalization provides a natural way to define the notion of a separable $n$-category. 
\end{expl}

\subsubsection{Separable \texorpdfstring{$n$}{n}-categories} \label{sec:separable_nCat_composite_TO}

Let $\Cat_n^{\Cb}$ be the $(n+1)$-category of $\Cb$-linear $n$-categories. The following result is taken from \cite[Cor.\,3.2]{KZ22b} based on the works \cite{GJF19,JF22}. 
\begin{thm} \label{pthm:nVec_embedding}
The functor $\hom_{(n+1)\vect}(\bullet, -): (n+1)\vect \to \Cat_n^{\Cb}$ is fully faithful. 
\end{thm}
\pf
All the arguments in Example \ref{expl:3Vec} automatically generalizes to $(n+1)\vect$ to provide a physical proof of this theorem except that the reference \cite[Theorem 3.2.3]{KZ18} should be replaced by \cite[Theorem\ 3.26, Remark 3.27]{KZ24}. 
\epf

\begin{defn} \label{def:separable_ncat}
A {\it separable $n$-category} is a $\Cb$-linear $n$-category that lies in the essential image of $\hom_{(n+1)\vect}(\bullet, -)$. A separable $n$-category is condensation complete by definition. If it is not a direct sum of two non-zero ones, it is called {\it indecomposable} (or {\it connected}). We call an indecomposable summand of a separable $n$-category $\CA$ as an indecomposable (or connected) component of $\CA$. 
\end{defn}

\begin{expl} \label{expl:separable_01cat}
Separable 0-categories are finite dimensional vector spaces over $\Cb$, i.e., objects in $\vect$. Separable 1-categories are finite semisimple $\Cb$-linear 1-categories. 
\end{expl}

\begin{expl} \label{expl:separable_2cat}
We gives a few examples of separable 2-categories. 
\bnu

\item For a finite group $G$, $2\vect_G$ is the 2-category of $G$-graded separable 1-categories. A generic object is a direct sum $X= \oplus_{g\in G} X_g$, where the subscript of $X_g$ is the grading and $X_g$ is a separable 1-category, which is equivalent to $\vect^{\oplus n_g}$ for $n_g\in \Nb$. The only simple objects are $\vect_g$ for $g\in G$. We illustrate the hom spaces associated to simple objects by the following diagram:
$$
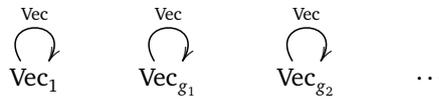
\xymatrix{
\vect_1 \ar@(ul,ur)[]^{\vect} &  \vect_{g_1} \ar@(ul,ur)[]^{\vect} & 
\vect_{g_2} \ar@(ul,ur)[]^{\vect} & \cdots
}
$$

\item $2\Rep(\Zb_2)=\Sigma\Rep(\Zb_2)$ has two simple objects $\one$ and $\one_c$ \cite{DR18}. We illustrate the hom spaces associated to simple objects by the following diagram: 
$$
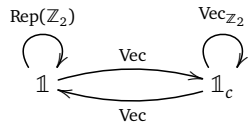
\xymatrix{
\one \ar@(ul,ur)[]^{\Rep(\Zb_2)} \ar@/^/[rr]^{\vect} & &  \one_c \ar@(ul,ur)[]^{\vect_{\Zb_2}} \ar@/^/[ll]^{\vect}
}
$$
Note that $2\Rep(\Zb_2)$ has a unique connected component.

\item The monoidal center $\FZ_1(2\vect_{\Zb_2})$ of $2\vect_{\Zb_2}$ as a separable 2-category is equivalent to a direct sum of two $2\Rep(\Zb_2)$ \cite{KTZ20}
as illustrated below.  
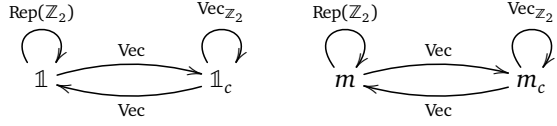
\be \label{eq:quiver_4D_toric_code}
\xymatrix{
\one \ar@(ul,ur)[]^{\Rep(\Zb_2)} \ar@/^/[rr]^{\vect} & &  \one_c \ar@(ul,ur)[]^{\vect_{\Zb_2}} \ar@/^/[ll]^{\vect}
}
\quad\quad
\xymatrix{
m \ar@(ul,ur)[]^{\Rep(\Zb_2)} \ar@/^/[rr]^{\vect} & &  m_c \ar@(ul,ur)[]^{\vect_{\Zb_2}} \ar@/^/[ll]^{\vect}
}
\ee
Since $\FZ_1(2\vect_{\Zb_2})$ is the category of topological defects of codimension 2  and higher, each objects represents a string-like defect. In particular, $\one$ is the trivial string; $\one_c$ is the condensation descendant of the $\one$-string \cite{KW14,EN17,KTZ20a} and is sometimes called the Cheshire string \cite{EN17}; $m$ is the $m$-string or magnetic string \cite{HZW05}; and $m_c$ is the condensation descendant of the $m$-string. The only non-trivial simple object in $\hom(\one,\one)=\Rep(\Zb_2)$ is the $e$-particle \cite{HZW05}. 

\item The monoidal center $\FZ_1(2\vect_G)$ of $2\vect_G$ as a separable 2-category is equivalent to a direct sum \cite{KTZ20}: 
$$
\FZ_1(2\vect_G) \simeq \oplus_{[h]\in \mathrm{Cl}} \,\, 2\Rep(C_G(h)),
$$
where $\mathrm{Cl}$ is the set of conjugacy classes of $G$ and $C_G(h)$ is the centralizer of $h\in G$. 
\enu
\end{expl}

\begin{expl} \label{expl:separable_ncat}
We give a few useful separable $n$-categories 
\bnu

\item $n\Rep(G):=\Sigma^{n-1}\Rep(G)$. It can also be defined by the category of functors $\Fun(\mathrm{B}G, n\vect)$ \cite{KZ24}. It has a unique connected component. 

\item $n\vect_G$ is the category of $G$-graded separable $(n-1)$-categories, defined by the condensation complete $\Cb$-linear monoidal $n$-category $\oplus_{g\in G}n\vect$ \cite{KZ24}. Its number of  connected components is $|G|$. It has a monoidal structure with the tensor product induced by the multiplication of $G$. For $\omega\in Z^{n+2}(G, \Cb^\times)$, there is a way to `twist' the monoidal structure on $n\vect_G$ by $\omega$ to give a new monoidal  $n$-category denoted by $n\vect_G^\omega$. It turns out that $\Sigma n\vect_G^\omega$ is a separable $(n+1)$-category (see Definition\,\ref{def:multi-fusion_ncat}).

\item The monoidal center $\FZ_1(n\vect_G^\omega)$ is a separable $n$-category\footnote{For $\omega_1\neq \omega_2$, in general, $\FZ_1(n\vect_G^{\omega_1})$ and $\FZ_1(n\vect_G^{\omega_2})$ are not equivalent even as separable $n$-categories. Example\ 3.4 in \cite{KTZ20} provided an example of this non-equivalence for $n=2$.}. Its physical meaning is the category of topological defects of codimension 2 and higher in $n+$2D twisted $G$-gauge theory\footnote{Twisted finite gauge theories are also called Dijkgraaf-Witten theories.} $\SG\ST_{(G,\omega)}^{n+2}$, i.e., $\Omega\CG\CT_{(G,\omega)}^{n+2}=\FZ_1(n\vect_G^\omega)$. If the twist $\omega$ is trivial, then $\SG\ST_{(G,\omega)}^{n+2}$ is called a finite gauge theory and is denoted by $\SG\ST_{G}^{n+2}$ for simplicity. It was conjectured in \cite{KTZ20} that we have the following equivalence of separable $n$-categories:
$$
\Omega\CG\CT_{(G,\omega)}^{n+2} = \FZ_1(n\vect_G^\omega) \simeq \oplus_{[h]\in\mathrm{Cl}} \,\, n\Rep(C_G(h), \tau_h(\omega)),
$$
where $\mathrm{Cl}$ is the set of conjugacy classes of $G$, $C_G(h)$ is the centralizer of $h\in G$, and $\tau_h$ is the transgression map $C^{k+1}(G,\Cb^\times) \to C^k(C_G(h), \Cb^\times)$, and $n\Rep(C_G(h), \tau_h(\omega))$ is the category of right separable module categories over the fusion $n$-category $n\vect_{C_G(h)}^{\tau_h(\omega)}$. When $\omega$ is trivial, we simply have 
$$
\Omega\CG\CT_{G}^{n+2} = \FZ_1(n\vect_G) \simeq \oplus_{[h]\in\mathrm{Cl}} \,\, n\Rep(C_G(h)). 
$$
When $n=1$, this conjecture was rigorously proved in \cite{Wil08}, and the $n=2$ cases were rigorously proved in \cite{KTZ20}. In this work, we simply takes above conjectured equivalences for granted for all $n$. Moreover, we expect $\FZ_1(n\vect_G^\omega)$ to be $\EE_2$-fusion. It turns out that $\Sigma^2\FZ_1(n\vect_G^\omega)$ is a separable $(n+2)$-category (see Definition\,\ref{def:Em-MultiFusion-nCat}). 

\item Let $z\in G$ be the fermion parity symmetry and lives in the center of $G$. The category $\Rep(G,z)$ is a symmetric fusion 1-category, then $n\Rep(G,z) := \Sigma^{n-1}\Rep(G,z)$ is a separable $n$-category (actually a symmetric fusion $n$-category). 
\enu
\end{expl}

As a consequence, we can identify the category $(n+1)\vect$ as the $(n+1)$-category of separable $n$-categories (as objects), $\Cb$-linear $n$-functors (as 1-morphisms), $\Cb$-linear $n$-natural transformations (as 2-morphisms), so on and so forth. When $n=0$, a separable $0$-category is a finite dimensional vector space over $\Cb$; when $n=1$, a separable $1$-category is a finite semisimple 1-categories; when $n=2$, a separable $1$-category is a finite semisimple 2-category \cite{DR18}. 

By the folding trick, a gapped domain wall between two non-chiral topological order $\SX^n$ and $\SY^n$ is automatically a gapped boundary of $\SX^n \boxtimes \overline{\SY^n}$. For $x,y\in n\vect$, this means that $\hom_{n\vect}(x,y) \simeq \hom_{n\vect}(\one, y\otimes x^R)$, where $x^R$ is the right dual of $x$. By Definition\,\ref{def:separable_ncat}, $\hom_{n\vect}(x,y)$ is a separable $(n-1)$-category. It further implies that every hom space in a separable higher category is automatically separable.

\begin{defn} 
A $0$-morphism $f$ in a separable $1$-category $\CA$ is called {\it simple} if $\hom_\CA(f,f)\simeq \Cb$. By induction, a $k$-morphism $g$ in a separable $n$-category $\CB$ is called {\it simple} if $1_g$ is simple in $\hom_\CB(g,g)$.   
\end{defn}

\begin{rem}
An indecomposable separable $n$-category is precisely a simple object in $(n+1)\vect$. 
\end{rem}

\begin{rem} \label{rem:meaning_of_an_object_in_nVec}
In Example\,\ref{expl:3Vec}, for the sake of convenience, we have used the non-chiral 2+1D topological order $\SX^3$ to label a simple object $X$ in $3\vect$. There is nothing wrong in this case. For $n\vect$ and $n>3$, however, we need slightly modify the physical meaning of a simple object $X \in n\vect$. Note that $X$ as a label for an object in $n\vect$ is not physically detectable. What is physically detectable is the category $\hom_{n\vect}(\bullet,X)$, which can be identified with $X$ via the fully faithful functor $\hom_{(n+1)\vect}(\bullet, -): (n+1)\vect \to \Cat_n^{\Cb}$. It suggests that the physical meaning of the object $X$ should come from the category $\hom_{n\vect}(\bullet,X)$. Therefore, as an indecomposable separable $(n-1)$-category in $n\vect$, $X$ is precisely a connected component of the 2-category of all gapped boundary conditions of a non-chiral topological order. This slight modification is necessary because the higher category of gapped boundary conditions of a higher dimensional non-chiral topological order is in general disconnected, i.e., splitting into a direct sum of connected full sub-categories (see Example\,\ref{expl:4Vec}). We summarize this discussion into the following physical theorem. 
\end{rem}

\begin{pthm} \label{pthm:physical_meaning_separable_n-category}
Physically, a simple object in $(n+1)\vect$, or equivalently, an indecomposable separable $n$-category $\CS$ is a connected component of the category of all gapped boundary conditions (up to invertible $n$D topological orders\footnote{Invertible $n$D topological orders are also gapped boundaries of $\mathbf{1}^{n+1}$. However, they are invisible in the theory of separable higher categories.}) of a non-chiral $n+$1D topological order $\SX^{n+1}$, i.e., a pair $(\SX^{n+1},x)$ where $x$ labels the connected component. An object $s\in \CS$ is a gapped boundary condition (up to invertible $n$D topological orders) of $\SX^{n+1}$, and $\Omega_s\CS:=\hom_\CS(s,s)$ is the category of all topological defects on the boundary $s$ of $\SX^{n+1}$. A 1-morphism in $\CS$ is a domain wall between two gapped boundaries of $\SX^{n+1}$ and a $k$-morphism for $1\leq k\leq n$ is a $k$-codimensional topological defect on the boundaries of $\SX^{n+1}$. 
\end{pthm}

\begin{expl} \label{expl:4Vec}
We explain Remark\,\ref{rem:meaning_of_an_object_in_nVec} and Theorem$^{\mathrm{ph}}$\,\ref{pthm:physical_meaning_separable_n-category} via the example $4\vect$. Note that $4\vect$ is the precisely the category of all topological defects in $\mathbf{1}^5$. 
\bnu
\item The tensor unit $\bullet=\one$ is precisely the trivial 4D topological order $\mathbf{1}^4$. 

\item Let $\SP^3$ be a 2+1D chiral topological order and $\CP$ the category of all topological defects in $\SP^3$. As we show in Section\,\ref{sec:general_example_1-codim}, by condensing $A=\CP$ in $3\vect$, we obtain an indecomposable separable $3$-category $X=\Sigma\CP$, which is an invertible object in $4\vect$ \cite{JF22,KZ22b} and is precisely a connected component of the category of all gapped boundary conditions (up to invertible $3$D topological orders) of $\mathbf{1}^4$, The simple objects in $X=\Sigma\CP$ labels all 2+1D chiral topological orders that are connected to $\SP^3$ via 1+1D gapped domain walls. 

\item A simple 1-morphism $f\in \hom_{4\vect}(\bullet, X) \simeq \hom_{4\vect}(3\vect, \Sigma\CP)$ can be identified with a simple object in $\Sigma\CP$. It is precisely the label of a simple gapped boundary of $\mathbf{1}^4$, or equivalently, a 2+1D chiral topological order (up to invertible ones), whose category of particles is Witt equivalent to the non-degenerate braided fusion 1-category $\Omega\CP$ \cite{DMNO13}. 

\item A generic simple object $X \in 4\vect$ is a condensation descendant of $\mathbf{1}^4$. As an indecomposable separable 3-category, $X$ is precisely a connected component of the category of all gapped boundaries of a non-chiral 4D topological order, whose category of topological defects is given by $\FZ_0(X)=\Fun(X,X)$. 

\enu
Notice that, although $3\vect$ consists of only 2+1D non-chiral topological orders, $4\vect$ consists of all chiral or non-chiral 2+1D topological orders (up to chiral central charges) as simple 1-morphisms from $\bullet$ to invertible objects. When the invertible object is also the tensor unit $\bullet$, then such simple 1-morphisms are precisely those non-chiral 2+1D topological orders. It is worth reminding readers that the group of all invertible objects in $4\vect$ are precisely the Witt group of non-degenerate braided fusion 1-categories \cite{JF22,KZ22b}. 
\end{expl}

\subsubsection{Composite topological orders} \label{sec:composite_TO}
The physical meanings of the morphisms in $(n+1)\vect$ provides a way to generalize the notion of a topological order to composite ones. 
This generalization is not necessary if we limit ourselves to those $n$D topological orders that are simple in the sense that they have a unique ground state on an $(n-1)$-sphere (or stable under local perturbations) \cite{KWZ15}. However, the generalization is necessary for a complete theory as illustrated in the following picture. 
\be \label{pic:Morita_equivalent}
\begin{array}{c}
\begin{tikzpicture}[scale=1.5]
\fill[gray!20] (2,0) arc (0:180:1) -- cycle ;
\node at (1,0.5) {\footnotesize $\SZ(\SA)^{n+1}=\SZ(\SB)^{n+1}$} ;
\draw[black, ultra thick,->-] (0,0) -- (1,0) ; 
\draw[black, ultra thick,->-] (1,0) -- (2,0) ; 
\draw[fill=white] (0.95,-0.05) rectangle (1.05,0.05) node[midway,below,scale=1] {\scriptsize $\hspace{1mm}\SM^{n-1}$} ;
\node at (0.5,-0.15) {$\SA^n$} ;
\node at (1.5,-0.15) {$\SB^n$} ;
\end{tikzpicture}
\end{array} 
\quad\quad \xrightarrow{\mbox{closing the fan}} \quad\quad
\begin{array}{c}
\begin{tikzpicture}[scale=1.5]
\draw[black, ultra thick,->-] (0,1) -- (0,0) ; 
\draw[fill=white] (-0.05,-0.05) rectangle (0.05,0.05) node[midway,below,scale=1] {\scriptsize $\hspace{1mm}\SM^{n-1}$} ;
\node at (1,0.8) {\small $\SX^n:=\overline{\SB^n} \boxtimes_{\SZ(\SA)} \SA^n$} ;
\end{tikzpicture}
\end{array} 
\ee
The topological orders $\SA^n$ and $\SB^n$ in (\ref{pic:Morita_equivalent}), i.e., sharing the same bulk and connected by a gapped domain wall $\SM^{n-1}$, are called Morita equivalent (see also Definition\,\ref{def:TME}). Note that, even if we choose $\SA^n$ and $\SB^n$ to be simple, the domain wall $\SM^{n-1}$, as an $(n-1)$D topological order, is not simple in general. Moreover, by closing the fan, we obtain an $n$D topological order $\SX^n$ having $\SM^{n-1}$ as its gapped boundary, and $\SX^n$ is not simple in general. For example, when $n=2$ and $\CA=\CM=\CB=\vect_G$ for a non-trivial finite group $G$, $\SA^2$ is a simple topological order but $\SM^1$ and $\SX^2$ are not simple. In this case, it turns out that $\SM^1$ and $\SX^2$ are non-simple or composite in two different ways. More precisely, $\SX^2$ is of type-I and $\SM^1$ is of type-II as defined below. 
\begin{defn} \label{def:physics_TO}
Roughly speaking, a composite topological order $\SC^{n+1}$ is a `direct sum' of simple ones. Moreover, $\SC^{n+1}$ is called {\it type-I} if all its simple direct summands are Morita equivalent, and is called {\it type-II} otherwise.  (see Definition\,\ref{def:simple_TO} for a more complete definition). 
\end{defn}

In order to give a precise meaning of the `direct sum', we need work within a precise category. It is possible to do it for topological orders (up to invertible ones) as we show in Definition\,\ref{def:simple_TO}, which was already implicitly used in \cite{KZ22b}. However, physically oriented readers are recommended to understand the notion of a composite topological order at above intuitive level. 

\begin{defn} \label{def:simple_TO}
We can define a generalized notion of an $n$D (potentially anomalous) topological order (up to invertible ones) $\SC^n$ by a 1-morphism $x: \bullet \to \CS$ for an object $\CS \in (n+1)\vect$. Since $\CS$ can be viewed as a separable $n$-category, a 1-morphism $x: \bullet \to \CS$ can be identified with an object $x\in \CS$.  
The category of topological defects in the topological order $x$ is defined by the category $\Omega_x\CS:=\hom_\CS(x,x)$. 
\bnu

\item If $x=0$, then $\SC^n$ is called the zero $n$D topological order denoted by $0$. We identify all zero $n$D topological orders and identify the topological order $y: \bullet \to \CT$ with the topological order $0\oplus y: \bullet \to \CS \oplus \CT$, i.e., $y=0\oplus y$. 

\item If $x$ lives in a single connected component of $\CS$, then $\SC^n$ is called {\it type-I}.  In this case, if $x$ is simple, then the topological order $\SC^n$ is called {\it simple}; if $x$ is not simple, then $\SC^n$ is called {\it type-I composite}. We denote the number of simple direct summands of $x$ by $\mathrm{gsd}(x)$ or $\mathrm{gsd}(\SC^n)$ because it coincides with the ground state degeneracy of $\SC^n$ defined on an $(n-1)$-sphere. 

\item If $\CS$ is not connected and $x$ does not live in a single connected component of $\CS$, then the topological order $\SC^n$ is called {\it type-II composite} or {\it type-II} for simplicity. We denote the number of simple direct summands of $x$ by $\mathrm{gsd}(x)$. We call the minimal number of non-zero type-I direct summands of $x$ the multiplicity of $\SC^n$ and denote it by $\mathrm{Mult}(\SC^n)$. 

\enu
$\SC^n$ is called {\it composite} if it is either type-I composite or type-II composite. 
When $\SC^n$ is a boundary/wall, the boundary/wall is called simple, type-I/type-II composite if $\SC^n$ is simple, type-I/type-II composite, respectively. 
\end{defn}


\begin{rem}
We provide some examples of type-I and type-II topological orders later in Example\,\ref{expl:composite_TO}. Strictly speaking the definition of a mathematical object is incomplete if we do not define the (iso)morphisms between this notion. We do not want to do it here. One can consult with the definition of the slice category $\bullet/(n+1)\vect$ in \cite{KZ22b}. In this work, we simply identify all isomorphic objects in $(n+1)\vect$, i.e., identify all equivalent separable $n$-categories, for all physical discussion. 
\end{rem}

\begin{rem} \label{rem:two_simpleness}
Both type-I and type-II topological orders are necessary for a complete condensation theory. When $\SC^n$ is anomaly-free, the simpleness of $\SC^n$ defined in Definition\,\ref{def:simple_TO} coincides with the following simpleness condition: the ground state degeneracy of $\SC^n$ defined on an $(n-1)$-sphere is trivial. 
This simpleness condition is essentially a stability condition. As we shown in \cite{KWZ15}, a type-I topological order is not stable unless it is simple (see Remark\,\ref{rem:simple=stable}), so is a type-II topological order, which is necessarily a gapped boundary of a type-I topological order. It means that if we restricted ourselves to only stable phases, we can restricted our theory to only simple topological orders. However, this restriction is not very natural if we work with a condensation complete theory. 
\end{rem}

\begin{rem} \label{rem:ignore_composite}
The notion of a composite topological order is necessary for the study of the categories of topological orders and the condensation theory (see Remark\,\ref{rem:composite_complete_condensation_theory}). However, for most physicists who are only interested in the simple condensations of anomaly-free simple topological orders, it is beneficial to ignore anomalous and composite cases and skip relevant discussion at least in the first reading. 
\end{rem}

\subsubsection{Morita equivalence between topological orders}

\begin{defn} \label{def:TME}
Two $n+$1D topological orders $\SA^{n+1}, \SB^{n+1}$ are called {\it Morita equivalent}\footnote{This notion was called Witt equivalence in \cite{KW14}.} if they share the same bulk (or gravitational anomaly), i.e., $\SZ(\SA)^{n+2}=\SZ(\SB)^{n+2}$, and are connected by a gapped domain wall $\SM^n$ as illustrated below. We use $[\SA^{n+1}]$ to denote the Morita class of $\SA^{n+1}$. 
\be \label{eq:TO_equivalences}
\begin{array}{c}
\begin{tikzpicture}[scale=1.5]
\fill[gray!20] (0,0) rectangle (2,0.8) ;
\node at (1,0.4) {\footnotesize $\SZ(\SA)^{n+2}=\SZ(\SB)^{n+2}$} ;
\draw[black, ultra thick,->-] (0,0) -- (1,0) ; 
\draw[black, ultra thick,->-] (1,0) -- (2,0) ; 
\draw[fill=white] (0.95,-0.05) rectangle (1.05,0.05) node[midway,below,scale=1] {\scriptsize $\hspace{1mm}\SM^n$} ;
\node at (0.5,-0.15) {$\SA^{n+1}$} ;
\node at (1.5,-0.15) {$\SB^{n+1}$} ;
\end{tikzpicture}
\end{array}
\ee
(This terminology comes from the fact that the associated multi-fusion $n$-categories $\CA$ and $\CB$ are Morita equivalent. We also use $[\CA]$ to denote the Morita class of $\CA$. If $\SA^{n+1}$ and $\SB^{n+1}$ share the same bulk but are not Morita equivalent, then a domain wall $\SM^n$ between them is necessarily gapless.) 
\end{defn}


\begin{expl}
Any two 2+1D anomaly-free simple topological orders $\SA^3$ and $\SB^3$ share the same trivial 3+1D bulk.  In this case, $\SA^3$ and $\SB^3$ are Morita equivalent if and only if two fusion 2-categories $\CA$ and $\CB$ are Morita equivalent, or equivalently, 
$\Omega\CA$ and $\Omega\CB$ are Witt equivalent \cite{DMNO13}. If $\SA^3$ and $\SB^3$ are not Morita equivalent, then a domain wall between them is necessarily gapless. 
\end{expl}

\begin{rem}
By the structure theorem of multi-fusion categories \cite{KZ22b} (reviewed in Proposition\,\ref{prop:structure_theorem_MF}), a type-I topological order is always Morita equivalent to a simple one. Two type-II topological orders are Morita equivalent if and only if they share the same multiplicity and there is one-to-one pairing among their type-I direct summands such that each pair are Morita equivalent. 
\end{rem}

\begin{rem}
Two (potentially anomalous) topological orders $\SA^{n+1}$ and $\SB^{n+1}$ are Morita equivalent if and only if they share the same bulk and $\SA^{n+1}\boxtimes_{\SZ(\SA)^{n+2}} \SB^{n+1} \simeq \SZ(\SM)^{n+1}$. There is another equivalent way to characterize the Morita equivalence as a centralizer $\SZ(\SA,\SZ(\SM))^{n+1} \simeq \overline{\SB^{n+1}}$ (explained in Example\,\ref{expl:MoritaEq=centralizer}). 
\end{rem}

We recall a result from \cite[Proposition\ 4]{KW14} on Morita equivalence (recall Definition\,\ref{def:TME}) between two anomaly-free topological orders. 
\begin{pprop}[\cite{KW14}] \label{plem:2way_Witt_eq_phases}
Two anomaly-free $n+$1D topological order $\SC^{n+1}$ and $\SD^{n+1}$ are Morita equivalent if and only if there exist two $n$D topological orders $\SP^n$ and $\SQ^n$ such that 
\be \label{eq:Witt_eq}
\SC^{n+1} \boxtimes \SZ(\SP)^{n+1} = \SD^{n+1} \boxtimes \SZ(\SQ)^{n+1}. 
\ee
\end{pprop}
\pf
($\Rightarrow$). If $\SC^{n+1}$ and $\SD^{n+1}$ are Morita equivalent, then they can be connected by a gapped anomaly-free domain wall $\SM^n$. By the folding trick, we have $\SZ(\SM)^{n+1}=\SC^{n+1}\boxtimes \overline{\SD^{n+1}}$. Then we obtain 
$$
\SC^{n+1} \boxtimes \SZ(\SD^n)^{n+1} = \SC^{ n+1} \boxtimes \overline{\SD^{n+1}} \boxtimes \SD^{n+1} = \SZ(\SM)^{n+1} \boxtimes \SD^{n+1}. 
$$

($\Leftarrow$). It is quite obvious because $\SC^{n+1}$ is clearly Morita equivalent to the left side of (\ref{eq:Witt_eq}), and $\SD^{n+1}$ is Morita equivalent to the right side of (\ref{eq:Witt_eq}), as illustrated in the following picture. 
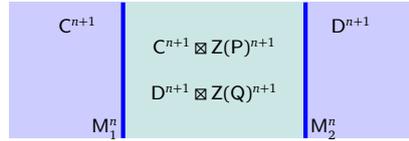
\[
\begin{array}{c}
\begin{tikzpicture}[scale=0.6]
\fill[blue!20] (-4.5,0) rectangle (4.5,3) ;
\fill[teal!20] (-2,0) rectangle (2,3) ;
\draw[blue, ultra thick] (-2,3) -- (-2,0) ; 
\draw[blue, ultra thick] (2,3) -- (2,0) ; 
\node at (-2.4,0.2) {\scriptsize  $\SM_1^n$} ;
\node at (2.4,0.2) {\scriptsize  $\SM_2^n$} ;
\node at (-3,2.5) {\scriptsize $\SC^{n+1}$} ;
\node at (3,2.5) {\scriptsize $\SD^{n+1}$} ;
\node at (0,2) {\scriptsize $\SC^{n+1} \boxtimes \SZ(\SP)^{n+1}$} ; 
\node at (0,1) {\scriptsize $\SD^{n+1} \boxtimes \SZ(\SQ)^{n+1}$} ; 
\end{tikzpicture}
\end{array}
\]
\epf


We denote by $\TME^n(\SC^{n+1})$ the set of Morita classes of gapped (necessarily type-I) boundaries of a non-chiral simple topological order $\SC^{n+1}$. For simplicity, we set $\TME^n:=\TME^n(\mathbf{1}^{n+1})$. Let $\SM^n$ be a gapped boundary of $\SC^{n+1}$. We denote the associated equivalence class by $[\SM^n]$. Stacking topological orders gives a well-defined map: 
\begin{align*}
\TME^n(\SB^{n+1}) \times \TME^n(\SC^{n+1}) &\xrightarrow{\boxtimes} \TME^n(\SB^{n+1} \boxtimes \SC^{n+1}) \\
([\SX^n], [\SM^n]) \quad &\mapsto \quad [\SX^n \boxtimes \SM^n]. 
\end{align*}
When $\SB^{n+1}=\mathbf{1}^{n+1}$, this map defines a $\TME^n$-action on $\TME^n(\SC^{n+1})$. When $\SB^{n+1}=\SC^{n+1}=\mathbf{1}^{n+1}$, this map defines an associative and commutative multiplication on $\TME^n$. 
\begin{pthm}[\cite{KW14}] \label{pthm:TME_group}
$\TME^n=(\TME^n, \boxtimes, [\mathbf{1}^n])$ is an abelian group and $[\SX^n]^{-1}=[\overline{\SX^n}]$ for any anomaly-free $\SX^n$.
\end{pthm}

\begin{rem} \label{rem:TM-group_Witt-group}
When $n=3$, $\TME^3$ is precisely the usual Witt group \cite{DMNO13}. Therefore, $\TME^n$ for $n>3$ can be viewed as higher Witt groups. 
\end{rem}

The following result is essentially a slightly refined reformulation of (1)$\Leftrightarrow$(3) part of Proposition 3.13 in [KZ24] (see Theorem\,\ref{thm:center-functor_surjective_full_faithful} and \ref{thm:center-functor_surjective_full_faithful_2} and also Theorem$^{\mathrm{ph}}$\,\ref{pthm:recover_all_bdy_from_one}), and first appeared (but formulated slightly differently but equivalently) in Theorem$^{\mathrm{ph}}$\, 3.2.13 in the first version of this paper, and was independently obtained a few months later in the third version of \cite{LYW24}. 
\begin{pthm} \label{pthm:TME-action_transitive}
For a non-chiral simple topological order $\SC^{n+1}$, the $\TME^n$-action on $\TME^n(\SC^{n+1})$ is free and transitive, i.e., $\TME^n(\SC^{n+1})$ is a $\TME^n$-torsor. 
(A mathematical reformulation of this result can be found in Theorem\,\ref{thm:same_bulk=Morita+invertible}.)
\end{pthm}
\pf
Let $\SM^n$ and $\SN^n$ be two gapped boundaries of $\SC^{n+1}$. We set $\SX^n:= \overline{\SM^n} \boxtimes_{\SC^{n+1}} \SN^n$. Since $\SM^n \boxtimes \overline{\SM^n}$ is connected to $\SC^n$ (i.e. the trivial 1-codimensional topological defect in $\SC^{n+1}$) by the gapped domain wall $\SM^{n-1}$ (recall Example\,\ref{expl:folding}: $\SM^{n-1}$ is the trivial 1-codimensional topological defect in $\SM^n$), then  
$$
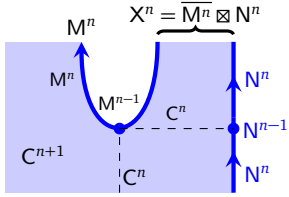
\begin{array}{c}
\begin{tikzpicture}[>=Stealth]
\fill[blue!20] (-4,0) rectangle (-1,2) ;
\fill[white] (-3,2) .. controls (-3,0.5) and (-2,0.5) .. (-2,2)--(-3,2)..controls (-2.5,2) .. (-2,2)--cycle;
\draw[blue,ultra thick,->-] (-1,0) -- (-1,1) node[near start, right] {\footnotesize $\SN^n$} ; 
\draw[blue,ultra thick,->-] (-1,1) -- (-1,2) node[midway, right] {\footnotesize $\SN^n$} ; 
\draw[blue,ultra thick,-stealth] (-2,2) .. controls (-2,0.5) and (-3,0.5) .. (-3,2) ;
\draw[dashed] (-2.5,0) -- (-2.5,0.8) ;
\draw[decorate,decoration=brace,very thick] (-2,2.1)--(-1,2.1) ;
\node at (-3.25,1.5) {\scriptsize $\SM^n$} ;
\filldraw [blue] (-2.5,0.85) circle [radius=2pt] ;
\filldraw [blue] (-1,0.85) circle [radius=2pt] node[right] {\footnotesize $\SN^{n-1}$};
\node at (-2.5,1.2) {\scriptsize $\SM^{n-1}$} ;
\draw[dashed] (-2.5,0.85) -- (-1,0.85) node[midway, above] {\scriptsize $\SC^n$} ; 
\node at (-3,2.2) {\footnotesize $\SM^n$} ;
\node at (-1.5,2.4) {\footnotesize $\SX^n=\overline{\SM^n}\boxtimes \SN^n $};
\node at (-3.5,0.5) {\footnotesize $\SC^{n+1}$} ;
\node at (-2.3,0.2) {\footnotesize $\SC^n$} ;
\end{tikzpicture}
\end{array}
\quad\quad
\begin{array}{l}
\mbox{$\SM^{n-1}\boxtimes_{\SC^n} \SN^{n-1}$ defines a gapped domain wall between} \\
\SM^n \boxtimes \SX^n=\SM^n \boxtimes (\overline{\SM^n} \boxtimes_{\SC^{n+1}} \SN^n) = (\SM^n \boxtimes \overline{\SM^n}) \boxtimes_{\SC^{n+1}} \SN^n \\
\mbox{and $\SC^n \boxtimes_{\SC^{n+1}} \SN^n = \SN^n$.}
\end{array}
$$ 
Therefore, $\SN^n$ and $\SM^n \boxtimes \SX^n$ are Morita equivalent. 
\epf

\begin{rem}
Once we give a coordinate system $\SC^{n+1} = \SZ(\SB)^{n+1}$, then $\TME^n(\SC^{n+1})$ is equipped with a natural coordinate given by the bijection $\TME^n \to \TME^n(\SC^{n+1})$ defined by $[\SX^n] \mapsto [\SX^n\boxtimes \SB^n]$. 
\end{rem}

Once we adapt the new notion of a topological order, certain things discussed in Section\,\ref{sec:Cat_top_defects} need slight modifications.  We postpone the discussion to the next subsubsection after we introduce the notion of a multi-fusion $n$-category.

\subsubsection{Multi-fusion \texorpdfstring{$n$}{n}-categories} \label{sec:MF_ncat}

\begin{defn} \label{def:multi-fusion_ncat}
A {\it multi-fusion $n$-category}, or an {\it $\EE_1$-multi-fusion $n$-category}, is a condensation complete monoidal $n$-category $\CA$ such that $\Sigma\CA$ is a separable $(n+1)$-category. It is called {\it indecomposable} if it is not a direct sum of two multi-fusion $n$-categories. It is called a {\it fusion $n$-category} if $\one_\CA$ is simple. 
\end{defn}

\begin{rem}
The notion of a (multi-)fusion $2$-category was first introduced by Douglas and Reutter in \cite{DR18}. That of a (multi-)fusion $n$-category was first introduced by Johnson-Freyd in \cite{JF22}. Here we follow an approach given in \cite{KZ22b}. 
\end{rem}

\begin{expl}
We provide a few examples of fusion 2-categories below. 
\bnu
\item The separable 2-category $2\vect_G$ introduced in Example\,\ref{expl:separable_2cat} has an obvious structure of fusion 2-category with the tensor product defined by $\vect_g \otimes \vect_h = \vect_{gh}$ for $g,h\in G$ and the tensor unit defined by $\vect_1$. 

\item The separable 2-category $2\Rep(\Zb_2)$ is also equipped with a structure of fusion 2-category with the (non-trivial) fusion rule given by 
$\one_c\otimes \one_c = \one_c \oplus \one_c$.

\item The separable 2-category $\FZ_1(2\Rep(\Zb_2)) \simeq 2\Rep(\Zb_2) \oplus 2\Rep(\Zb_2)$ (recall (\ref{eq:quiver_4D_toric_code})) is equipped with a structure of a fusion 2-category. The fusion rules are given below. 
$$
\quad 
\one_c\otimes \one_c = \one_c \oplus \one_c, \quad m\otimes m=\one, \quad m_c = m\otimes \one_c = \one_c \otimes m, \quad
m_c \otimes m_c \simeq m_c \oplus m_c. 
$$
\enu
\end{expl}

\begin{expl}
The separable $n$-categories $n\Rep(G), n\vect_G, \FZ_1(n\vect_G)$ introduced in Example\,\ref{expl:separable_ncat} are examples of fusion $n$-categories. An indecomposable multi-fusion $n$-category $\CA$ can be easily constructed from two separable $n$-categories $\CS_1$ and $\CS_2$ as 
$$
\CA = \Fun(\CS_1\oplus \CS_2, \CS_1\oplus \CS_2) =
\left( \begin{array}{cc}
\Fun(\CS_1, \CS_1) & \Fun(\CS_1, \CS_2)  \\
\Fun(\CS_2, \CS_1)  & \Fun(\CS_2, \CS_2) 
\end{array} \right).
$$
\end{expl}

A separable $n$-category is a finite direct sum of indecomposable ones. An indecomposable separable $n$-category $\CS$ is illustrated by the following connected diagram (connected by 1-morphisms). 
\be \label{diag:Morita_pair}
\begin{array}{c}
\xymatrix{
a \ar@(ul,ur)[]^{\hom_{\CS}(a,a)} \ar@/^/[rrr]^{\hom_{\CS}(a,b)} & & & b \ar@(ul,ur)[]^{\hom_{\CS}(b,b)} \ar@/^/[lll]^{\hom_{\CS}(b,a)}
}
\end{array} \quad\quad\quad \forall a,b\in \CS. 
\ee
Note that $\hom_\CS(a,a)$ and $\hom_\CS(a,a)$ are indecomposable multi-fusion $(n-1)$-categories, which define physically the categories of topological defects on the gapped boundaries labeled by $a$ and $b$, respectively. Since $a$ and $b$ are connected, two multi-fusion $(n-1)$-categories are necessarily Morita equivalent. Moreover, $\hom_\CS(a,b)$ and $\hom_\CS(b,a)$ are precisely the invertible bimodules that define the Morita equivalence. Physically, $\hom_\CS(a,b)$ and $\hom_\CS(b,a)$ are the category of wall conditions of gapped domain walls between $a$ and $b$. In particular, we have 
the following monoidal equivalence:  
\be \label{eq:hombb=ednofun_homab}
\hom_\CS(b,b) \simeq \Fun_{\hom_\CS(a,a)^\rev}(\hom_\CS(a,b), \hom_\CS(a,b)). 
\ee
The physical meaning of this fact is illustrated in Figure\,\ref{fig:functoriality_bbr}, in which two gapped boundaries $\SA^{n+1}$ and $\SZ^{(1)}(\SL)$ of the same non-chiral topological order $\SZ(\SA)^{n+2}$ are connected by a gapped domain wall $\SL^n$. By the functoriality of the boundary-bulk relation, we have $\FZ_1^{(1)}(\CL):=\Fun_{\CA^\rev}(\CL,\CL)$. By replacing $b$ by $b\oplus c$ in (\ref{eq:hombb=ednofun_homab}), we obtain a new equivalence: 
\be \label{eq:hombc=fun_homab_homac}
\hom_\CS(b,c) \simeq \Fun_{\hom_\CS(a,a)^\rev}(\hom_\CS(a,b), \hom_\CS(a,c)), 
\ee
which simply says that, $\forall a\in \CS$, we have
\be
\CS = \Sigma\hom_{\CS}(a,a) \simeq \RMod_{\hom_\CS(a,a)}(n\vect).
\ee
This result provides a convenient and useful characterization of the delooping. 
\begin{thm}[\cite{GJF19,KZ22b}] \label{thm:SigmaA=RMod}
For a multi-fusion $n$-category $\CA$, we have $\Sigma\CA \simeq \RMod_\CA((n+1)\vect)$ and the equivalence is defined by the functor $\hom_{\Sigma\CA}(\bullet,-): x\mapsto \hom_{\Sigma\CA}(\bullet,x)$. 
\end{thm}

\begin{rem} \label{rem:Yoneda_functor=id}
If we identify $\Sigma\CA$ with $\RMod_\CA((n+1)\vect)$, i.e., $\Sigma\CA = \RMod_\CA((n+1)\vect)$, then we can also identify $\hom_{\Sigma\CA}(\bullet,-)$ with $\id_{\Sigma\CA}$. This naive fact turns out to be very useful in the study of topological orders because all the physical observables are encoded directly in the hom spaces in $\Sigma\CA$ instead of its objects, which are, in priori, only abstract labels of topological defects and are not directly observable. Then the functor $\hom_{\Sigma\CA}(\bullet,-)=\id_{\Sigma\CA}$ allows us to translate abstract data of the objects in $\Sigma\CA$ into physically observable data. One physical application of this result is when $\CA$ is a non-degenerate braided fusion $(n-1)$-category. In this case, $\Sigma\CA=\RMod_\CA(n\vect)$ can be viewed as the category of topological defects of an anomaly-free topological order as illustrated below. 
\be \label{pic:objects_in_SigmaA}
\begin{array}{c}
\begin{tikzpicture}
\fill[gray!20] (-2,0) rectangle (4.2,3) ;
\draw[very thick,->-] (0.5,1)--(0.5,3) node[near start,left] {\small $x$} 
node[near start,right] {\scriptsize $\hom_{\Sigma\CA}(x,x)$} ;
\draw[dashed] (0.5,0)--(0.5,1) node[near start,left] {\small $\one$} node[near start,right] {\small $\CA$} ;
\draw[very thick,->-] (2.5,1)--(2.5,3) node[near start,left] {\small $y$} node[near start,right] {\scriptsize $\hom_{\Sigma\CA}(y,y)$};
\draw[dashed] (2.5,0)--(2.5,1) node[near start,left] {\small $\one$} node[near start,right] {\small $\CA$}; 
\draw[fill=white] (2.45,0.95) rectangle (2.55,1.05) node[midway,right,scale=1] {\scriptsize $\hom_{\Sigma\CA}(\one,y)=y$} ; 
\draw[fill=white] (0.45,0.95) rectangle (0.55,1.05) node[midway,left,scale=1] {\scriptsize $x=\hom_{\Sigma\CA}(\one,x)$} ; 
\draw[fill=white] (0.45,2.45) rectangle (0.55,2.55) node[midway,right,scale=1] {\scriptsize $\hom_{\Sigma\CA}(x,x')$} ; 
\node at (-1.7,2.7) {\footnotesize $\Sigma\CA=\RMod_\CA(n\vect)$} ;
\draw[thick,-stealth] (-2,0) -- (-1.5,0) node[near end,above] {\scriptsize $x^1$} ;
\draw[thick,-stealth] (-2,0) -- (-2,0.5) node[near end,left] {\scriptsize $x^2$} ;
\end{tikzpicture}
\end{array}
\ee
In the coordinate system $\Sigma\CA=\RMod_\CA(n\vect)$, an object $x\in \Sigma\CA$ (the label of a 1-codimensional topological defect) can be identified with $\hom_{\Sigma\CA}(\one, x)$, i.e., the defect junction between two 1-codimensional defects $\one\in \Sigma\CA$ and $x$. The horizontal tensor product $x\otimes^1 y$ in $\Sigma\CA$ is precisely the horizontal fusion of two defect junctions $\hom_{\Sigma\CA}(\one,x) \boxtimes_\CA \hom_{\Sigma\CA}(\one,y)$. Moreover, a morphism $f: x\to x'$ induces a morphism $f\circ -: \hom_{\Sigma\CA}(\one,x) \to \hom_{\Sigma\CA}(\one,x')$. Therefore, $f$ is faithfully and physically realized by vertically fusing the defect junction $\hom_{\Sigma\CA}(\one,x')$ with $\hom_{\Sigma\CA}(\one,x)$. We use these facts in many physical discussions in later sections. 
\end{rem}

In a separable $n$-category, an object $x \in \CS$ is simple if and only if its identity 1-morphism $1_x$ is a simple object in $\hom_\CS(x,x)$. In this case, $\hom_\CS(x,x)$ is a fusion $(n-1)$-category. When $x$ is not simple, it represents a composite defect or a composite boundary condition. In this case, $\hom_\CS(x,x)$ is a multi-fusion $n$-category. If all direct summands in $x$ all live in the same connected component of $\CS$, then $\hom_\CS(x,x)$ is an indecomposable multi-fusion $n$-category; otherwise, $\hom_\CS(x,x)$ is a direct sum of indecomposable multi-fusion $n$-category. As an illustrating example, when $x=a\oplus b$ and $a,b$ are simple and connected, then we have the following decomposition (see also (\ref{diag:Morita_pair})):
$$
\hom_\CS(x,x) = \left( \begin{array}{cc}
\hom_\CS(a,a) & \hom_\CS(a,b) \\
\hom_\CS(b,a) & \hom_\CS(b,b)
\end{array} \right).
$$
All indecomposable multi-fusion $n$-category is of such a matrix-type \cite{KZ22b} as stated in details in the following proposition. 
\begin{prop}[\cite{KZ22b}] \label{prop:structure_theorem_MF}
If $\CA$ is an indecomposable multi-fusion $n$-category, then we have
\bnu

\item $\CA \simeq \oplus_{i,j=1}^p \CA_{i,j}$, where $\CA_{ij}:=e_i \otimes \CA \otimes e_j$ and $e_i$ are simple summands of $\one_\CA$, i.e. $\one_\CA = \oplus_{i=1}^p e_i$; 

\item $\Sigma\CA = \Sigma\CA_{ii}$ and $\CA_{ii}$ is a fusion $n$-category; 

\item the canonical $\CA_{ii}$-$\CA_{kk}$-bimodule map $\CA_{ij} \boxtimes_{\CA_{jj}} \CA_{jk} \to \CA_{ik}$ induced from the tensor product functor is an equivalence; 

\item the $\CA_{ii}$-$\CA_{jj}$-bimodule $\CA_{ij}$ is the inverse of the $\CA_{jj}$-$\CA_{ii}$-bimodule $\CA_{ji}$, i.e., $[\CA_{ii}]=[\CA_{jj}]$. 

\enu
\end{prop}

\begin{rem} \label{rem:simple=stable}
The physical meaning of an indecomposable multi-fusion $n$-category $\CA$ is the category of topological defects in an $n+$1D type-I topological order $\SA^{n+1}$ \cite{KWZ15}. The physical meaning of the number of simple direct summands of the tensor unit $\one_\CA$ (i.e., the number $p$) is the ground state degeneracy of the associated topological order $\SA^{n+1}$ defined on an $n$-sphere $S^n$. In other words, if $\one_\CA$ is not simple, then $\SA^{n+1}$ is unstable, which means that by adding local perturbations the phase $\SA^{n+1}$ can flow to a stable one associated to $\CA_{ii}$ for an $i=1,\cdots, p$ \cite{KWZ15}. Some 1D and 2D examples have been explicitly constructed in lattice models (see for example \cite{KK12,KZ22a}). A decomposable multi-fusion $n$-category is the category of all topological defects in an $n+$1D type-II topological order. 
\end{rem}

\begin{rem} \label{rem:MFC=MFC}
This definition of a multi-fusion 1-category is equivalent to the usual definition, i.e., a $\Cb$-linear monoidal 1-category satisfying the following properties. 
\bnu
\item[(1)] {\it Rigidity}: Every particle $x$ has a right dual particle $x^R$, together with a creation morphism $b_x^R: \one \to x^R \otimes x$ and an annihilation morphism $d_x: x\otimes x^R \to \one$, satisfying some natural conditions; and a left dual particle $x^L$, together with a creation morphism $b_x^L: \one \to x\otimes x^L$ and an 
annihilation morphism $d_x^L: x^L\otimes x \to x$, satisfying some natural conditions (see \cite{EGNO15} for a review). 

\item[(2)] {\it Separability}: $\CA$ is a separable 1-category, i.e., a finite direct sum of $\vect$, or equivalently, an object in $2\vect$ (recall Example\,\ref{expl:2Vec_condensation}). 
\enu
Both properties follow immediately from the fact $\Sigma\CA=\RMod_\CA(2\vect)$ is a separable 2-category and the fully-dualizability of $3\vect$. Conversely, a multi-fusion 1-category $\CA$ in the usual sense satisfies the condition that $\Sigma\CA\simeq \RMod_\CA(2\vect)$ is a separable 2-category \cite{DR18}. 
\end{rem}

Recall that the entire Section\,\ref{sec:Cat_top_defects} is based on the assumption of the simpleness of the topological order (recall Remark\,\ref{rem:simple_TO}). 
Once we adapt the generalized notion of a topological order in Definition\,\ref{def:simple_TO}, one result in Section\,\ref{sec:Cat_top_defects} needs to be modified slightly. A topological order $\SC^n$ (in the sense of Definition\,\ref{def:simple_TO}) is called {\it anomaly-free} if $\SZ(\SC)^{n+1}=\mathbf{1}^{n+1}$. 
We encounter a new phenomenon. 
\bnu

\item[(1)] If $\SC^n$ is anomaly-free, then $\SC^n$ is necessarily type-I. 
If, in addition, $\SC^n$ is type-I composite, then $\CC \neq \Sigma\Omega\CC$ and $\FZ_2(\Omega\CC)$ is non-trivial. 

\item[(2)] Both boundary-bulk relations (\ref{eq:bbr_Z1}) and (\ref{eq:bbr_Z0}) still hold for the generalized notion of a topological order. But we encounter new phenomena worth discussion. If $\SC^n$ is type-II composite, then $\SB^{n+1}=\SZ(\SC)^{n+1}$ is type-I composite. More explicitly, assume $\SC^n=\SC_1^n \oplus \SC_2^n$ is a 1-morphism in $(n+1)\vect$: 
$$
\bullet \xrightarrow{\SC_1^n \oplus \SC_2^n} \CS_1 \oplus \CS_2
$$
for two indecomposable separable $n$-categories $\CS_1$ and $\CS_2$. In this case, we have 
\begin{align*}
\SB^{n+1} &= \SZ(\SC)^{n+1}=\SZ(\SC_1)^{n+1}\oplus \SZ(\SC_2)^{n+1}, \\
\CB &\simeq \FZ_0(\CS_1 \oplus \CS_2) = \left( \begin{array}{cc}
\Fun(\CS_1,\CS_1) & \Fun(\CS_1,\CS_2) \\
\Fun(\CS_2,\CS_1) & \Fun(\CS_2,\CS_2)
\end{array} \right), \\
\Omega\CB &\simeq \Omega\FZ_0(\CS_1) \oplus \Omega\FZ_0(\CS_2) \simeq \Omega\FZ_0(\Sigma\CC_1) \oplus \Omega\FZ_0(\Sigma\CC_2) \simeq \FZ_1(\CC_1) \oplus \FZ_1(\CC_2).
\end{align*}
Note that the key point here is that even though $\SC_1^n$ and $\SC_2^n$ are not connected by higher morphisms in $(n+1)\vect$, their bulks $\SZ(\SC_1)^{n+1}$ and $\SZ(\SC_2)^{n+1}$ are both non-chiral, and, therefore, as 1-morphisms in $(n+2)\vect$, are connected by higher morphisms. In this case, we still have the bulk of a bulk is trivial, i.e., $\SZ(\SB)^{n+2}=\mathbf{1}^{n+2}$. In particular, we have $\FZ_1(\CB) \simeq n\vect$ \cite{JF22,KZ22b}. Therefore, $\SB^{n+1}$ is type-I composite and anomaly-free. 
\enu

Now we summarize and update results in Section\,\ref{sec:Cat_top_defects} and \ref{sec:cc} in the following physical theorem, which is essentially a reformulation (with a minor refinement) of \cite[Definition\,2.21]{KWZ15}. 
\begin{pthm} \label{pthm:TO_multi-fusion_ncat}
Let $\SC^{n+1}$ be an $n+$1D topological order (in the sense of Definition\,\ref{def:simple_TO}). The category $\CC$ of topological defects in $\SC^{n+1}$ is a multi-fusion $n$-category. 
\bnu
\item If $\SC^{n+1}$ is type-I, then $\CC$ is an indecomposable multi-fusion $n$-category, and its bulk $\SZ(\SC)^{n+2}$ is simple. If $\SC^{n+1}$ is simple, then $\CC$ is a fusion $n$-category. 

\item If $\SC^{n+1}$ is anomaly-free (i.e., $\SZ(\SC)^{n+2}=\mathbf{1}^{n+2}$), then $\SC^{n+1}$ is necessarily type-I, $\FZ_1(\CC)\simeq n\vect$ and $\CC$ is an indecomposable multi-fusion $n$-category. 
\bnu
\item If, in addition, $\SC^{n+1}$ is simple, then $\CC$ is a simple separable $n$-category, i.e. $\CC=\Sigma\Omega\CC$, and we have $\FZ_2(\Omega\CC) \simeq (n-1)\vect$. 

\item If, in addition, $\SC^{n+1}$ is type-I composite, then $\CC\neq \Sigma\Omega\CC$, and $\Omega\CC$ is non-trivial and splits into a direct sum. 
\enu

\item If $\SC^{n+1}$ is type-II, then $\SB^{n+2}=\SZ(\SC)^{n+2}$ is type-I composite  and anomaly-free, and we have 
$$
\mathrm{gsd}(\SZ(\SC)^{n+2}) = \mathrm{Mult}(\SC^{n+1}).
$$ 
In this case, $\CC=\oplus_{i=1}^{\mathrm{Mult}(\SC^{n+1})}\CC_i$ is a direct sum of indecomposable multi-fusion $n$-categories $\CC_i$, and $\CB=\FZ_0(\Sigma\CC)$ is an indecomposable multi-fusion $(n+1)$-category, and we have $\Omega\CB \simeq \FZ_1(\CC) \simeq \oplus_i \FZ_1(\CC_i)$. 

\enu
\end{pthm}

In the rest of this work, by a (simple/composite) topological order we always means the one defined in Definition\,\ref{def:simple_TO}.

\begin{expl} \label{expl:composite_TO}
We provide some examples of type-I and type-II 2+1D topological orders. 
\bnu

\item A 2+1D anomaly-free simple topological order is the one that satisfies the simpleness condition: the ground state degeneracy on $S^1$ is trivial. As we have reviewed in Section\,\ref{sec:Cat_top_defects_3D}, an anomaly-free simple topological order $\SC^3$ can be described by a pair $(\CC,c)$ or $(\Omega\CC,c)$, where $\Omega\CC$ is a non-degenerate braided fusion 1-category and $\CC=\Sigma\Omega\CC$ is a fusion 2-category. It can be identified with the simple object $\bullet \in \Sigma\CC$ (or equivalently, a 1-morphism $\bullet \to \Sigma\CC$ in $4\vect$), where $\Sigma\CC$ is a separable 3-category and an object in $4\vect$.

\item For each simple object $x\in \Sigma\CC$, $\Omega_x^2\Sigma\CC$ is a non-degenerate braided fusion 1-category that is Witt equivalent to $\Omega\CC$. Moreover, $\Sigma\CC$ is an invertible object in $4\vect$. For example, if $\SC^3$ is non-chiral, then $\Sigma\CC \simeq 3\vect=\bullet \in 4\vect$. All invertible objects arise in this way and form a abelian group which is isomorphic to the usual Witt group \cite{JF22}. 

\item Let $x,y\in \Sigma\CC$ be two simple objects. We denote the 2+1D anomaly-free topological orders defined by $x,y$ by $\SX^3$ and $\SY^3$, respectively. Both $\SX^3$ and $\SY^3$ are Morita equivalent to $\SC^3$. Now we consider 
The direct sum $x\oplus y: \bullet \to \Sigma\CC$, as a 1-morphism in $4\vect$, defines a type-I composite anomaly-free 2+1D topological order, denoted by $(\Sigma\CC,x\oplus y)^3$, i.e., $\SZ(\Sigma\CC,x\oplus y)^4=\mathbf{1}^4$. Its category of topological defects is given by the following indecomposable multi-fusion 2-category:
$$
\hom_{\Sigma\CC}(x\oplus y, x\oplus y) = \left( \begin{array}{cc}
\hom_{\Sigma\CC}(x,x) & \hom_{\Sigma\CC}(x,y) \\
\hom_{\Sigma\CC}(y,x) & \hom_{\Sigma\CC}(y,y)
\end{array} \right).
$$
Note that $\hom_\CS(x,x)=\CX$ and $\hom_\CS(y,y)=\CY$ are the categories of defects in $\SX^3$ and $\SY^3$, respectively. We have the following monoidal equivalences: 
$$
\FZ_1(\hom_{\Sigma\CC}(x\oplus y, x\oplus y)) \simeq \FZ_1(\CX) \simeq \FZ_1(\CY) \simeq 2\vect,
$$
where the first and second $\simeq$ follows from Example\,\ref{expl:Z1_ind_MF}, justifying the boundary-bulk relation. However, the category of particles in $(x\oplus y)^3$ is given by $\Omega\CX \oplus \Omega\CY$, whose M\"{u}ger center is $\vect \oplus \vect$ and non-trivial.

The separable 2-category $\hom_{\Sigma\CC}(x,y)=\hom_{\Sigma\CC}(y,x)^\op$ is precisely the category of gapped walls between $\SX^3$ and $\SY^3$, and we have a monoidal equivalence
$$
\CX^\rev \boxtimes \CY \simeq \FZ_0(\hom_{\Sigma\CC}(x,y)). 
$$

Moreover, for two gapped walls $p,q\in \hom_{\Sigma\CC}(x,y)$ between $\SX^3$ and $\SY^3$, $\CP=\Omega_p\hom_{\Sigma\CC}(x,y)$ and $\CQ=\Omega_q\hom_{\Sigma\CC}(x,y)$ are the categories of topological defects on the walls $a$ and $b$, respectively, as illustrated in the following picture and diagram.
$$
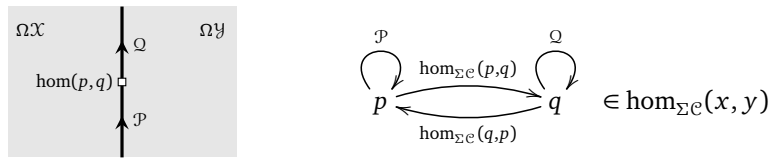
\begin{array}{c}
\begin{tikzpicture}[scale=1]
\fill[gray!20] (-1.5,0) rectangle (1.5,2) ;
%
\draw[->-,very thick] (0,0)--(0,1) node[midway, right] {\scriptsize $\CP$} ; 
\draw[->-,very thick] (0,1)--(0,2) node[midway, right] {\scriptsize $\CQ$} ;
%
\draw[fill=white] (-0.05,0.95) rectangle (0.05,1.05) node[midway,left] {\scriptsize $\hom(p,q)$} ;
%
%
\node at (-1.2,1.7) {\scriptsize $\Omega\CX$} ;
\node at (1.2,1.7) {\scriptsize $\Omega\CY$} ;
\end{tikzpicture}
\end{array}
\quad\quad\quad\quad
\begin{array}{c}
\xymatrix{
p \ar@(ul,ur)[]^{\CP} \ar@/^/[rr]^{\hom_{\Sigma\CC}(p,q)} & &  q \ar@(ul,ur)[]^{\CQ} \ar@/^/[ll]^{\hom_{\Sigma\CC}(q,p)}
} \quad \in \hom_{\Sigma\CC}(x,y)
\end{array}
$$
They are Morita equivalent indecomposable multi-fusion 1-categories both equipped with a braided equivalence 
$$
\Omega\CX \boxtimes \Omega\CY^\op \simeq \FZ_1(\CP), \quad\quad
\Omega\CX \boxtimes \Omega\CY^\op \simeq \FZ_1(\CQ),
$$
respectively. The 0+1D domain wall $\hom_{\Sigma\CC}(p,q)=\hom_{\Sigma\CC}(q,p)^\op$ is the separable 1-category uniquely fixed by $\CP, \CQ$ and $\Omega\CX, \Omega\CY$ (see \cite{AKZ17}).

\item Let $\CS$ and $\CT$ be two simple separable 3-categories. Two objects $x\in \CS$ and $y\in \CT$ define two 2+1D type-I topological orders denoted by $(\CS,x)^3$ and $(\CT,y)^3$. Their category of topological defects are $\Omega_x\CS$ and $\Omega_y\CT$, respectively. Now we consider the direct sum of 1-morphisms $x\oplus y \in \CS \oplus \CT$. It defines a 2+1D type-II topological orders, denoted by $(\CS\oplus \CT, x\oplus y)^3$. In this case, we have $(\CS\oplus \CT, x\oplus y)^3= (\CS,x)^3 \oplus (\CT,y)^3$. The category of topological defects in $(\CS\oplus \CT, x\oplus y)^3$ is given by the following (decomposable) multi-fusion 2-category
$$
\hom_{\CS\oplus \CT}(x\oplus y, x\oplus y) \simeq  \hom_\CS (x,x) \oplus 
\hom_\CT(y,y) = \Omega_x\CS \oplus \Omega_y\CT. 
$$
The bulk of $(\CS\oplus \CT, x\oplus y)^3$ is non-trivial and type-I composite, i.e.,
$$
\SB^4=\SZ(\CS\oplus \CT,x\oplus y)^4 = \SZ(\CS,x)^4 \oplus \SZ(\CT,y)^4 \neq \mathbf{1}^4. 
$$
We have the following (braided) monoidal equivalences: 
\begin{align*}
\CB &\simeq \FZ_0(\CS \oplus \CT) = \left( \begin{array}{cc} 
\Fun(\CS,\CS) & \Fun(\CS,\CT) \\
\Fun(\CT,\CS) & \Fun(\CT,\CT)
\end{array} \right), \\
\Omega\CB &= \Omega\Fun(\CS,\CS) \oplus \Omega\Fun(\CT,\CT) \simeq \FZ_1(\Omega_x\CS) \oplus \FZ_1(\Omega_y\CT), 
\end{align*}
where we have used (\ref{eq:Z1_Z0}), justifying the boundary-bulk relation. Note that $\SB^4$ is anomaly-free in that $\FZ_1(\CB) \simeq 3\vect$. But we have 
$$
\FZ_2(\Omega\CB) \simeq \FZ_2(\FZ_1(\Omega_x\CS)) \oplus \FZ_2(\FZ_1(\Omega_y\CT)) \simeq 2\vect \oplus 2\vect \neq 2\vect. 
$$

Note that, when $\CS=\CT=\Sigma\CC$, the bulk 
$$
\SB^4=\SZ(\Sigma\CC \oplus \Sigma\CC, x\oplus y)^4 = \SZ(\Sigma\CC,x)^4 \oplus \SZ(\Sigma\CC,y)^4 = \mathbf{1}^4 \oplus \mathbf{1}^4 \neq \mathbf{1}^4
$$ 
is still non-trivial and type-I composite. In this case, one can see a fundamental difference between $(\Sigma\CC, x\oplus y)^3$ and $(\Sigma\CC \oplus \Sigma\CC, x\oplus y)^3$, where the first one is type-I composite and the second one is type-II composite. 


\enu
All the constructions in this example generalize tautologically to higher dimensions. 
\end{expl}

\begin{rem} \label{rem:composite_complete_condensation_theory}
As we show in later sections, 
the condensations in a simple topological order $\SC^{n+1}$ can produce type-I composite topological orders $\SD^{n+1}$ as the condensed phase and type-II composite topological orders $\SM^n$ as the gapped domain wall between $\SC^{n+1}$ and $\SD^{n+1}$. Therefore, composite topological orders should be included in order to have a complete condensation theory (recall (\ref{pic:Morita_equivalent})). 
\end{rem}

In an anomaly-free simple $n+$1D topological order $\SC^{n+1}$, the category $\CC$ of all topological defects is a fusion $n$-category. It is the condensation completion of the braided fusion $(n-1)$-category $\Omega\CC$ of topological defect of codimension 2 and higher, i.e., $\CC \simeq \Sigma\Omega\CC$. 
When $\SC^{n+1}$ is non-chiral, $\Omega\CC \simeq \FZ_1(\CA)$ for a fusion $(n-1)$-category $\CA$. In this case, there is another way to represent the fusion $n$-category $\CC$ (recall (\ref{eq:SigmaZA=BMod_AA})), i.e., 
\[
\CC = \Sigma \Omega\CC \xrightarrow[\simeq]{\phi} \BMod_{\CA|\CA}(2\vect). 
\]
These two representations provide two coordinate systems of $\CC$. The coordinate transformations between them are given by the monoidal equivalence $\phi$ and $\phi^{-1}$ defined below \cite[Theorem$^{\mathrm{ph}}$\,3.28]{KLWZZ20}:
\[
\phi: X \mapsto X \boxtimes_{\FZ_1(\CA)} \CA; \quad\quad\quad 
\phi^{-1}: Y \mapsto Y \boxtimes_{\CA^\op\boxtimes \CA} \CA. 
\]
The coordinate transformation between different coordinate systems are very useful to concrete computations (see Table\,\ref{table:coordinates_in_TC}, Theorem${}^{\mathrm{ph}}$\,\ref{pthm:construct_cond_E1_algebras_II} and Example\,\ref{expl:GT_G4_GtoH_II}).

\begin{rem} \label{rem:GJF_assumptions}
We have reviewed the theory of condensation completion physically in the spirits of \cite{KW14,KWZ15,KLWZZ20} instead of the more rigorous theory by Gaiotto and Johnson-Freyd \cite{GJF19,JF22}, which is not yet completely mathematically rigorous in the sense that no concrete model of weak $n$-categories is chosen and certain theory of colimits in higher categories was assumed. However, Gaiotto and Johnson-Freyd's theory has a lot of advantages, which allow us to proceed to answer many questions that are essentially orthogonal to higher coherence data at least at the physical level of rigor as shown in \cite{JF22,KZ22b,KZ24}. 
\end{rem}

\begin{rem}
A few physical and computational works on condensation completion in 2+1D appeared recently \cite{KZ22a,XLWWC25,YWL25}. They contain some explicit calculations of the condensation completion for a few 2+1D topological orders. 
\end{rem}

Recall that if two topological orders are Morita equivalent, then they share the same bulk. Conversely, if they share the same bulk, then it is not necessarily true that they are Morita equivalent because they might be connected by a gapless domain wall. Mathematically, it is possible to find an equivalent characterization of the sharing-the-same-bulk condition. The following result is a mathematical reformulation of Theorem$^{\mathrm{ph}}$\,\ref{pthm:TME-action_transitive}, and is a direct generalization of similar results for fusion 1-categories \cite{ENO11} and fusion 2-categories \cite{Dec25}. 
\begin{thm} \label{thm:same_bulk=Morita+invertible}
Two indecomposable multi-fusion $n$-categories $\CA$ and $\CB$ share the same $\EE_1$-center, i.e., $\FZ_1(\CA) \simeq \FZ_1(\CB)$, if and only if $\Sigma\CA \simeq \Sigma\CB \boxtimes \CS$ for an invertible separable $n$-category $\CS$. 
\end{thm}
\pf
It follows from the same proof of `(1)$\Leftrightarrow$(3)' part of \cite[Proposition* 3.13]{KZ24}. 
\epf

\begin{rem}
This result is essentially a special case of \cite[Proposition* 3.13]{KZ24}, which is reformulated in a slightly stronger way in Theorem$^{\mathrm{ph}}$\, \ref{thm:center-functor_surjective_full_faithful_2} but followed from the same proof. It first appeared (but formulated differently but equivalently) in Theorem$^{\mathrm{ph}}$\, 3.2.13 in the first arXiv version of this paper, which is now Theorem$^{\mathrm{ph}}$\,\ref{pthm:recover_all_bdy_from_one} (see also Remark\,\ref{rem:center=Morita+invertible}). 
The current reformulation in Theorem\,\ref{thm:same_bulk=Morita+invertible} first appeared  in the third arXiv version of \cite{LYW24}, which appeared a few months later than the first arXiv version of this paper. 
\end{rem}

\subsection{Higher algebras} \label{sec:higher_alg}
In this subsection, we explain the main idea of higher algebras in $(n+1)\vect$ based on physical or geometric intuitions. 

\subsubsection{Monoidal 1-categories as \texorpdfstring{$\EE_1$}{E1}-algebras} \label{sec:E1_algebra}

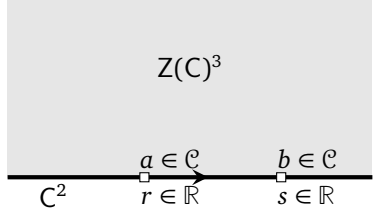
\begin{figure}
$$
\begin{tikzpicture}[scale =1.2]
\fill[gray!20] (-2,0) rectangle (2,2) ;
\draw[color = black, ultra thick,->-](-2,0)--(2,0);
\draw[fill=white] (-0.55,-0.05) rectangle (-0.45,0.05) node[midway,above,scale=1] {$\quad\quad a\in \CC$} node[midway,below,scale=1] {\quad\quad $r \in \Rb$} ;
\draw[fill=white] (0.95,-0.05) rectangle (1.05,0.05) node[midway,above,scale=1] {$\quad\quad b\in \CC$} node[midway,below,scale=1] {\quad\quad $s \in \Rb$};
\node[black] at(-1.5,-0.2) {$\SC^2$};
\node[black] at(0,1.2) {$\SZ(\SC)^3$};
\end{tikzpicture}
$$
\caption{An anomalous topological order $\SC^2$ and its bulk $\SZ(\SC)^3$}
\label{fig:anomalous_1d_TO}
\end{figure}

A potentially anomalous 1+1D topological order $\SC^2$ can be realized as a gapped boundary of a 2+1D topological order $\SZ(\SC)^3$ as depicted in Figure\,\ref{fig:anomalous_1d_TO}. It can have particle-like topological defects or particles for simplicity. These particles can be fused along the unique spatial dimension. Therefore, these particles, together with their fusion, form a fusion 1-category $\CC$ \cite{KK12}, in which 1-morphisms are instantons, i.e. 0D defect in spacetime. This result is well known. It is, however, only a part of a more complete story. We would like to sketch some key points in this more complete story (see also \cite[Remark\,3.4.19 \& 3.4.66]{KZ22a}) in order to introduce some important notions. 

\medskip
Consider a lattice model realization of the 2+1D topological order $\SZ(\SC)^3$ with the gapped boundary $\SC^2$. Suppose two anyons $a,b\in \CC$ are realized in the model at two different sites $r,s \in \Rb$, respectively, as depicted Figure\,\ref{fig:anomalous_1d_TO}. We chose the boundary line to be the real line. Recall that a particle is invariant under the action of local operators. Therefore, it represents a subspace of the total Hilbert space that are invariant under the net of local operators, and is more precisely defined as a superselection sector \cite{HK64,Haa96}. The physical configuration depicted in Figure\,\ref{fig:anomalous_1d_TO} already defines a fusion product $a \otimes_{(r,s)} b$, which can be viewed a subspace of the total Hilbert space. If we realize $a,b$ in the same model but at different sites $r',s'$, then a different fusion product $a \otimes_{(r',s')} b$ is realized as a different subspace of a potentially different total Hilbert space\footnote{ It depends on how one realize the defects. If we allow to introduce auxillary space localized around a site $r$ for the realization of a particle at $r$, then total Hilbert spaces are different for anyon at different sites but isomorphic via adiabatic moves.}. It means that there are infinitely many fusion products parameterized by the elements in the configuration space $\{ (r,s) \in \Rb^2 \mid r \neq s \}$. An adiabatic move of $(a,b)$ from $(r,s)$ to $(r',s')$ along a path $\gamma$ in the configuration space defines a linear isomorphism: 
$$
T^\gamma_{a,b} \colon a\otimes_{(r,s)} b \xrightarrow{\simeq} a\otimes_{(r',s')} b. 
$$ 
In mathematical language, $T^\gamma \coloneqq \{ T^\gamma_{a,b} \}_{a,b\in\CC} \colon \otimes_{(r,s)} \rightarrow \otimes_{(r',s')}$ defines a natural isomorphism. One of the defining properties of a topological order is the following principle:
\be
\mbox{\fbox{\begin{varwidth}{\textwidth}
{\bf Adiabatic Principle}: if two paths $\gamma_1$ and $\gamma_2$ in the configuration space  \\
$\{ (r,s) \in \Rb^2 \mid r \neq s \}$ are homotopy equivalent, then $T^{\gamma_1}=T^{\gamma_2}$.  
\end{varwidth}}}  \label{Adiabatic_Principle}
\ee
This property allows us to reduce the infinitely many fusion products $\otimes_{(r,s)}$ to a two homotopically inequivalent fusion products $\otimes \coloneqq \otimes_{(-1,1)}$ and $\otimes^\op \coloneqq \otimes_{(1,-1)}$. It is clear that we have $a \otimes^\op b = b\otimes a$. 

\medskip
We have sketched only some key ideas of a complete story, which is essentially the same as the mathematical proof of the fact that an $\EE_1$-algebra in the symmetric monoidal 2-category $\Cat_1$ is precisely a monoidal 1-category \cite{SW03,Lur17,Fre17}. We have used the notation $\Cat_1$ to denote the category of 1-categories (as objects), functors (as 1-morphisms) and natural transformations (as 2-morphisms). $\Cat_1$ is symmetric monoidal with the tensor product given by the Cartesian product $\times$ and the tensor unit given by the category with a single object $\bullet$ and a single 1-morphism $1_\bullet$. The $\EE_1$-algebra (or an algebra over the $\EE_1$-operad) structure on $\CA$ consists of the infinitely many fusion products $\otimes_{(r,s)}$ that are defined in 1-dimensional space, satisfying some natural conditions. It is different from that of a monoidal structure by definition. In $\Cat_1$, they are equivalent due to the fact that there is no higher morphisms in $\Cat_1$ to catch the information of higher homotopy data in the $\EE_1$-operad \cite{SW03,Lur17,Fre17}. In this work, we do not distinguish the 2-category of $\EE_1$-algebras, $\EE_1$-algebra homomorphisms and $\EE_1$-algebra 2-homomorphisms with that of monoidal categories, monoidal functors and monoidal natural transformations, and denote both categories by $\Alg_{\EE_1}(\Cat_1)$. Then the fact that $\CA$ is a monoidal 1-category can be easily summarized by a compact notation $\CA \in \Alg_{\EE_1}(\Cat_1)$. From now on, we use the physical intuition of infinitely many fusion products freely whenever we talk about a 1+1D topological order $\SA^2$ or a monoidal 1-category $\CA$.

For $\CA \in \Alg_{\EE_1}(\Cat_1)$, we can define the notion of a left/right $A$-module in $\Cat_1$, that of a module functor and that of a module natural transformation \cite{Ben65,Ost03} (see \cite{EGNO15} for a review). We denote the 2-category of left, right, bi- $\CA$-modules in $\Cat_1$ by $\LMod_\CA(\Cat_1)$, $\RMod_\CA(\Cat_1)$), $\BMod_{\CA|\CA}(\Cat_1)$, respectively. 

\subsubsection{Braided monoidal 1-categories as \texorpdfstring{$\EE_2$}{E2}-algebras} \label{sec:E2=E1+E1}

For 2+1D topological order $\SC^3$, the category $\CC$ of topological defects in $\SC^3$ is monoidal with a tensor unit $\one_\CC=\one$. Its anyons form a braided monoidal category $\Omega\CC$, which is usually defined by a tensor product $\otimes$, a tensor unit $1_\one$, the associators $\alpha_{a,b,c}: a\otimes (b\otimes c) \to (a\otimes b) \otimes c$, the left unitors $l_a: 1_\one \otimes a \to a$, the right unitors $r_a: a\otimes 1_\one\to a$ and the braidings isomorphisms $c_{a,b}: a\otimes b \to b\otimes a$ satisfying some natural conditions. In this subsection and the next, we reinterpret this structure in two seemingly different but equivalent ways that are ready to be generalized to higher dimensions. 

\medskip
In the first way, we go back to the physical intuition provided by the lattice model realizations of a 2+1D topological order. We mainly repeat the discussion in \cite[Remark\,3.4.19 \& 3.4.66]{KZ22a}.

\begin{figure}
$$
\begin{tikzpicture}[>=Stealth]
\fill[gray!20] (-2.5,0) rectangle (2.5,3) ;
\draw[->] (-2.5,0)--(-2,0) node[near end,above] {\small $x^1$};
\draw[->] (-2.5,0)--(-2.5,0.5) node[near end,left] {\small $x^2$};
\draw[fill=white] (-1.4,1.45) rectangle (-1.3,1.55) node[midway,right,scale=1] {$a$} node[midway,left,scale=1] {\scriptsize $\xi$};
\draw[fill=white] (-0.05,1.55) rectangle (0.05,1.65) node[midway,right,scale=1] {$b$} node[midway,left,scale=1] {\scriptsize $\eta$} ; 
\node at (-2.1,2.7) {\small $\Omega\CC$} ;
\draw[->] (-1.35,1.55) .. controls (0,3) and (1,2.5) .. (1.5,1.8) node[midway, above,scale=1] {\footnotesize $\gamma_3$};
\draw[->] (-1.35,1.45) .. controls (0,0.5) and (1,1) .. (1.5,1.7) node[midway, above,scale=1] {\footnotesize $\gamma_2$};
\draw[->] (-1.35,1.45) .. controls (0,-0.5) and (1,0.2) .. (1.5,1.7) node[midway,above,scale=1] {\footnotesize $\gamma_1$};
\draw[fill=white] (1.45,1.7) rectangle (1.55,1.8) node[midway,right,scale=1] {$a$} node[midway,left,scale=1] {\scriptsize $\xi'$};
\end{tikzpicture}
$$
\caption{three paths in configuration space}
\label{fig:AP}
\end{figure}
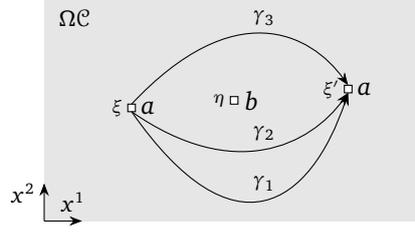

Suppose we have a lattice model realization of the 2+1D topological order $\SC^3$, and suppose two anyons $a,b\in \Omega\CC$ are realized in the model at two different sites $\xi,\eta \in \Rb^2=\Cb$, respectively, as depicted in Figure\,\ref{fig:AP}. Recall that an anyon is invariant under the action of local operators. Therefore, it represents a subspace of the total Hilbert space that are invariant under the net of local operators, which can be mathematically defined as superselection sectors \cite{HK64,Haa96}. This physical configuration already defines a fusion product $a \otimes_{(\xi,\eta)} b$, which can be viewed a subspace of the total Hilbert space. If we realize $a,b\in \Omega\CC$ in the same model but at different sites $\xi',\eta'$, then a different fusion product $a \otimes_{(\xi',\eta')} b$ is realized as a different subspace of a potentially different total Hilbert space. Similar to 1+1D topological orders, it means that there are infinitely many fusion products parameterized by the elements in the configuration space $\{ (\xi,\eta) \in \Cb^2 \mid \xi \neq \eta \}$. An adiabatic move of $(a,b)$ from $(\xi,\eta)$ to $(\xi',\eta')$ along a path $\gamma$ in the configuration space defines an isomorphism: 
$$
T^\gamma_{a,b} \colon a\otimes_{(\xi,\eta)} b \xrightarrow{\simeq} a\otimes_{(\xi',\eta')} b. 
$$ 
In mathematical language, $T^\gamma \coloneqq \{ T^\gamma_{a,b} \}_{a,b\in\Omega\CC} \colon \otimes_{(\xi,\eta)} \rightarrow \otimes_{(\xi',\eta')}$ defines a natural isomorphism. Similar to 1+1D topological order, we have the following defining property of a 2+1D topological order:
\be
\mbox{\fbox{\begin{varwidth}{\textwidth}
{\bf Adiabatic Principle}: if two paths $\gamma_1$ and $\gamma_2$ in the configuration space  \\
$\{ (\xi,\eta) \in \Cb^2 \mid \xi \neq \eta \}$ are homotopy equivalent, then $T^{\gamma_1}=T^{\gamma_2}$.  
\end{varwidth}}}  \label{cond:homotopy_trivialness}
\ee
Again Adiabatic Principle allows us to reduce the infinitely many fusion products $\otimes_{(\xi,\eta)}$ to a single fusion product $\otimes \coloneqq \otimes_{(-1,1)}$ and a braiding isomorphisms $c_{a,b}$ defined by
$$
a\otimes b \xrightarrow{c_{a,b} :=T_{a,b}^{\gamma_+}} b \otimes a, \quad\quad \gamma_+:=\{ (e^{i(1-t)\pi}, e^{i(2-t)\pi}) \in \Cb^2 | t\in [0,1] \}. 
$$
A physical work on adiabatic moves and braided monoidal structures can also be found in \cite{KL20}.

The point of this discussion is that, instead of strictly following the matmatical definition of a braided monoidal category, i.e. one tensor product plus braidings, we can return freely to above physical intuition of the infinitely many fusion products $\otimes_{(\xi,\eta)}$ whenever we discuss the anyons in a 2+1D topological order. Importantly, we have only sketched some key points of a complete story, which is essentially equivalent to the proof of the well-known mathematical result that an $E_2$-algebra in the 2-category $\Cat_1$ is equivalent to a braided monoidal 1-category \cite{SW03,Lur17,Fre17}. We denote the category of $E_2$-algebras in $\Cat_1$ (or equivalently, braided monoidal 1-categories) by $\Alg_{\EE_2}(\Cat_1)$. Its 1-morphisms are $E_2$-algebra homomorphisms (or braided monoidal functors), and 2-morphisms are morphisms between two $E_2$-algebra homomorphisms (or braided monoidal natural transformations). Then the fact that $\Omega\CC$ is a braided 1-category can be simply represented by a compact notation $\Omega\CC \in \Alg_{\EE_2}(\Cat_1)$.

\begin{figure}[htbp]
$$
\begin{tikzpicture}[>=Stealth]
\fill[gray!20] (-2,0) rectangle (3,2) ;
\draw[->] (-2,0)--(-1.5,0) node[near end,above] {\small $x^1$};
\draw[->] (-2,0)--(-2,0.5) node[near end,left] {\small $x^2$};
\draw[dashed] (1,0)--(1,2) ; 
\draw[dashed] (2,0)--(2,2) ; 
\draw[fill=white] (1.95,1.45) rectangle (2.05,1.55) node[midway,right,scale=1] {$d$} ; 
\draw[fill=white] (0.95,1.45) rectangle (1.05,1.55) node[midway,right,scale=1] {$c$} ;
\draw[fill=white] (1.95,0.65) rectangle (2.05,0.75) node[midway,right,scale=1] {$b$} ; 
\draw[fill=white] (0.95,0.65) rectangle (1.05,0.75) node[midway,right,scale=1] {$a$} ;  
\draw[fill=white] (-1.4,0.95) rectangle (-1.3,1.05) node[midway,right,scale=1] {$a$} node[midway,left,scale=1] {\scriptsize $\xi$};
\draw[fill=white] (-0.7,0.95) rectangle (-0.6,1.05) node[midway,above,scale=1] {$b$} node[midway,below,scale=1] {\scriptsize $\eta$} ; 
\node at (-1.7,1.8) {\small $\Omega\CC$} ;
\node at (1.2,0.2) {\small $\one$} ;
\node at (2.2,0.2) {\small $\one$} ;
\draw[-stealth,dashed] (-0.2,0.6) arc (-40:310:0.65) ;
\end{tikzpicture}
$$
\caption{the idea of braiding and 2-tuply monoidal 1-category}
\label{fig:braiding_E2}
\end{figure}
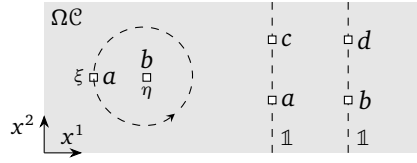

\medskip
In the second way, we rederive the braiding structure from the monoidal structure of $\mathrm{B}\Omega\CC$. Since $\mathrm{B}\Omega\CC$ is monoidal, its one point delooping $\mathrm{B}^2 \Omega\CC$ is a well defined 3-category, a fact which is equivalent to $\Omega\CC$ being braided monoidal. The 3-category $\mathrm{B}^2 \Omega\CC$ is somtimes called a 2-tuply monoidal 1-category (a list of proposed definitions of a weak $n$-category can be found in \cite{Lei02}; see also the list in \href{https://ncatlab.org/nlab/show/n-category}{nLab}). This result simply says that the braiding structure can be replaced by the fusion product in two independent directions as illustrated in the right half of Figure\,\ref{fig:braiding_E2}. 
\bnu

\item[(1)] Anyon $a, b\in \Omega\CC$ can be fused along the trivial string $\one$ (i.e., vertically in Figure\,\ref{fig:braiding_E2}). We denote this fusion product in $x^2$-direction by $a \otimes^2 b$. Mathematically, this fusion is precisely the composition of two 1-morphisms $a, c \in \hom_\CC(\one,\one)$:  
\begin{align*}
\hom_\CC(\one,\one) \times \hom_\CC(\one,\one) &\xrightarrow{\circ} \hom_\CC(\one,\one) \\
(c, a) &\mapsto c\circ a = a\otimes^2 c. 
\end{align*}
$1_\one$ is the unit of this funsion product, i.e. $1_\one \otimes^2 a \simeq a \simeq a\otimes^2 1_\one$. 

\item[(2)] There is another fusion product $\otimes^1$ on $\Omega\CC$ defined by horizontally fusion two $\one$-strings in Figure\,\ref{fig:braiding_E2}. Mathematically, it comes from the fusion product of $\mathrm{B}\Omega\CC$.
\begin{align*}
\hom_\CC(\one, \one) \times \hom_\CC(\one,\one) &\xrightarrow{\otimes} \hom_\CC(\one\otimes \one, \one\otimes\one) \simeq \hom_\CC(\one,\one) \\
(a,b) &\mapsto a \otimes^1 b. 
\end{align*}
It is obvious that $1_\one$ is also the unit of the fusion product $\otimes^1$. 
\enu

Physically, it is obvious that these two fusion products are compatible with each other. It means that the following diagram
\be \label{diag:otimes1_otimes2}
\xymatrix{
\Omega\CC \times \Omega\CC \times \Omega\CC \times \Omega\CC\ar[d]_{\otimes^1 \times \otimes^1}   &  & 
\Omega\CC \times \Omega\CC \times \Omega\CC \times \Omega\CC \ar[d]^{\otimes^2 \times \otimes^2 }  \ar[ll]_{1\times \tau \times 1} \\
\Omega\CC \times \Omega\CC \ar[dr]_{\otimes^2} & \overset{\delta}{\Longrightarrow} & 
\Omega\CC \times \Omega\CC 
\ar[dl]^{\otimes^1} \\
& \Omega\CC &
}
\ee
where $\tau$ is defined by $(b,c) \mapsto (c,b)$ for $c,b\in\Omega\CC$, 
is commutative up to a natural isomorphism $\delta$. Equivalently, there is a natural isomorphism, for $a,b,c,d\in \Omega\CC$, 
\be \label{eq:define_delta}
\delta_{a,b,c,d}: (a\otimes^1 b) \otimes^2 (c \otimes^1 d) \xrightarrow{\simeq} (a\otimes^2 c) \otimes^1 (b\otimes^1 d). 
\ee
\begin{itemize}
\item By restricting to a special case, we obtain 
\be \label{eq:delta_a11d}
a\otimes^2 d \simeq (a\otimes^1 1_\one) \otimes^2 (1_\one \otimes^1 d) \xrightarrow{\delta_{a,1_\one,1_\one,d}} (a\otimes^2 1_\one) \otimes^1 (1_\one \otimes^2 d) \simeq a\otimes^1 d. 
\ee
which says that two fusion products are isomorphic. If we use the physical intuition discussed in the first way. Without loss of generality, we can set $\otimes^1=\otimes_{(0,1)}$ and $\otimes^2 = \otimes_{(0,i)}$, then $\delta_{a,1_\one,1_\one,d}$ can be defined physically by the isomorphism $T_{a,d}^\gamma$, where $\gamma:=\{ (0, e^{i\frac{\pi}{2}(1-t)}) \in \Cb^2 | t\in [0,1] \}$. 

\item By restricting to another special case, we recovers the braiding isomorphism 
\begin{align}
c_{b,c}^1: \,\, b \otimes^1 c &\simeq 
(1_\one \otimes^2 b) \otimes^1 (c \otimes^2 1_\one) \xrightarrow{\delta_{1_\one,b,c,1_\one}^{-1}} (1_\one \otimes^1 c) \otimes^2 (b \otimes^1 1_\one) \nn
&\simeq (c\otimes^1 1_\one) \otimes^2 (1_\one \otimes^1 b) \xrightarrow{\delta_{c,1_\one,1_\one,d}} (c\otimes^2 1_\one) \otimes^1 (1_\one \otimes^1 b) \simeq
c\otimes^1 b, \label{eq:braiding_from_2fusion_1}
\end{align}
\begin{align}
c_{b,c}^2: \,\, b \otimes^2 c &\simeq 
(1_\one \otimes^1 b) \otimes^2 (c \otimes^1 1_\one) \xrightarrow{\delta_{1_\one,b,c,1_\one}} (1_\one \otimes^2 c) \otimes^1 (b \otimes^2 1_\one) \nn
&\simeq (c\otimes^2 1_\one) \otimes^1 (1_\one \otimes^2 b) \xrightarrow{\delta_{c,1_\one,1_\one,d}^{-1}} (c\otimes^1 1_\one) \otimes^2 (1_\one \otimes^1 b) \simeq
c\otimes^2 b, \label{eq:braiding_from_2fusion_2}
\end{align}
where we have used the fact that $1_\one$ is the tensor unit of both $\otimes^1$ and $\otimes^2$. We leave it as an exercise to prove that the isomorphism defined in (\ref{eq:braiding_from_2fusion_1}) and (\ref{eq:braiding_from_2fusion_2}) satisfies the axioms of a braiding. 

\end{itemize}

In summary, we have sketched a proof of the fact that a 1-category $\CA$ being braided monoidal if and only if $\mathrm{B}\CA$ is well defined and monoidal, or equivalently, $\mathrm{B}^2\CA$ is a well-define 3-category. In more concrete terms, we have shown that the braiding monoidal structure of $\Omega\CC$ can be equivalently encoded by two fusions in two independent directions and their compatibilities. Now we translate this compatibility into more formal mathematical language.

Recall that the category of $\EE_1$-algebras in $\Cat_1$ (i.e., monoidal 1-categories) is denoted by $\Alg_{\EE_1}(\Cat_1)$. Its 1-morphisms are $\EE_1$-algebra homomorphisms (i.e., monoidal functors), and 2-morphisms are morphisms between two $\EE_1$-algebra homomorphisms (i.e., monoidal natural transformations). Now the fact that the pair $(\otimes^2,1_\one)$ endows $\Omega\CC$ with a monoidal structure can be represented by the following notation: 
$$
(\Omega\CC, \otimes^2, 1_\one) \in \Alg_{\EE_1}(\Cat_1) \quad \mbox{or} \quad \Omega\CC \in \Alg_{\EE_1}(\Cat_1) \,\,\, \mbox{for simplicity}. 
$$ 
Moreover, $\Alg_{\EE_1}(\Cat_1)$ is again symmetric monoidal with the same Cartesian product $\times$ and the same tensor unit $\bullet$ as in $\Cat_1$. In particular, we obtain $\Omega\CC \times \Omega\CC \in \Alg_{\EE_1}(\Cat_1)$. The monoidal structure on $\Omega\CC \times \Omega\CC$ is defined by the composed functor $(\otimes^2 \times \otimes^2) \circ (1\times \tau \times 1)$. Then the compatibility condition (\ref{diag:otimes1_otimes2}) simply says that $\otimes^1:\Omega\CC \times \Omega\CC \to \Omega\CC$ is an $E_1$-algebra homomorphism (or a monoidal functor). It means that $\otimes^1:\Omega\CC \times \Omega\CC \to \Omega\CC$ is a 1-morphism in $\Alg_{\EE_1}(\Cat_1)$. Therefore, $\otimes^1$ endows $\Omega\CC$ with a structure of an $\EE_1$-algebra in $\Alg_{\EE_1}(\Cat_1)$. In other words, we obtain
\be \label{eq:Dun88_v1}
\Omega\CC \in \Alg_{\EE_1}(\Alg_{\EE_1}(\Cat)).  
\ee

This notation is heavily loaded. It means that $\Omega\CC$ has two multiplication maps $\otimes^1$ and $\otimes^2$, and $\otimes^1$ is a 1-morphism in $\Alg_{\EE_1}(\Cat)$. 
What we have explained in this second way is a key fact
$$
\Alg_{\EE_2}(\Cat_1) \simeq \Alg_{\EE_1}(\Alg_{\EE_1}(\Cat_1)). 
$$
It is a well-known mathematical result that holds in more general context \cite{Dun88,Lur17}. 

\subsubsection{\texorpdfstring{$\EE_m$}{Em}-multi-fusion \texorpdfstring{$n$}{n}-categories} \label{sec:Em-fusion-n_cat}

We briefly review  the notion of an $\EE_m$-multi-fusion $n$-category and some basic facts associated to it from \cite[Section\ 3]{KZ22b} and \cite{KZ24}. 

\begin{defn}
An {\it $\EE_1$-monoidal $n$-category} $\CA$ is a pair $(\CA, \mathrm{B}\CA)$, where $\mathrm{B}\CA$ is the one-point delooping of $\CA$. The identity 1-morphism $1_\bullet$ is the tensor unit of $\CA$. By induction, an {\it $\EE_m$-monoidal $n$-category} is a pair $(\CA,\mathrm{B}\CA)$ such that $\mathrm{B}\CA$ is an $\EE_{m-1}$-monoidal $n$-category. By abusing notation, we simply denote such an $\EE_{m-1}$-monoidal $n$-category by $\CA$, i.e. 
$$
\CA = (\CA, \mathrm{B}\CA, \mathrm{B}^2\CA, \cdots, \mathrm{B}^m\CA).
$$ 
When $m=1$, $\CA$ is also called a {\it monoidal $n$-category}; when $m=2$, $\CA$ is also called a {\it braided monoidal $n$-category}. 
\end{defn}

\begin{conv} \label{conv:xi-direction}
We adapt the following convention that is compatible with Convention\,\ref{rem:convention_op}. Note that $\mathrm{B}^{m-1}\CA$ is $\EE_1$-monoidal and is equipped with a tensor product $\otimes^1$; $\mathrm{B}^{m-2}\CA$ is $\EE_2$-monoidal and is equipped with two tensor products $\otimes^1,\otimes^2$; $\CA$ is $\EE_m$-monoidal and is equipped with $m$ tensor products $\otimes^1, \cdots, \otimes^m$. We always use a coordinate system $(x^1,x^2, \cdots, x^n)$ for the $n$-spatial dimensions such that $\otimes^i$ is the tensor product in the positive $x^i$-direction for $i=1,\cdots, n$. 
\end{conv}

\begin{expl} \label{expl:En+2_ncat=Einfty}
A symmetric monoidal 1-category is an $\EE_3$-algebra in $\Cat_1$. It turns out that an $\EE_3$-algebra in $\Cat_1$ is automatically $\EE_\infty$-monoidal because $\Cat_1$ does not have higher morphisms to distinguish higher commutativities. If one consider symmetric monoidal 3-category $\Cat_2$ (i.e., the 3-category of 2-categories), then $\EE_3$-algebras in $\Cat_2$ that are not $\EE_\infty$. In this case, a monoidal 2-category is an $\EE_1$-algebra in $\Cat_2$; a braided monoidal 2-category is an $\EE_2$-algebra in $\Cat_2$; a syllepitc monoidal 2-category is an $\EE_3$-algebra in $\Cat_2$; a symmetric monoidal 2-category is an $\EE_4$-algebra in $\Cat_2$, which is automatically an $\EE_\infty$-algebra in $\Cat_2$. In general, an $E_{n+2}$-monoidal $n$-category in $\Cat_n$ is automatically an $\EE_\infty$-algebra in $\Cat_n$. 
\end{expl}

\begin{defn}
An $\EE_0$-monoidal $n$-category $\CA$ is a pair $(\CA,\one_\CA)$, where $\CA$ is an $n$-category and $\one_\CA\in \CA$ is a distinguished object. An $\EE_0$-monoidal functor $(\CA,\one_\CA) \to (\CB,\one_\CB)$ is a functor $f: \CA \to \CB$ such that $f(\one_\CA)=\one_\CB$. An $\EE_0$-monoidal (higher) natural transformation is a (higher) natural transformation that is trivial on the distinguished object. We use $\EE_0\Cat_n$ to denote the $(n+1)$-category of $\EE_0$-monoidal categories, $\EE_0$-monoidal functors and $\EE_0$-monoidal (higher) natural transformations. We set 
$$
\Fun^{\EE_0}((\CA,\one_\CA), (\CB,\one_\CB)):=\hom_{\EE_0\Cat_n}((\CA,\one_\CA), (\CB,\one_\CB)), 
$$
which is (not full) subcategory of $\Fun(\CA,\CB)$. We use $\EE_m\Cat_n$ to denote the full $(n+1)$-subcategory $\EE_0\Cat_n$ consisting of all the iterated 1-point deloopings $\mathrm{B}^m\CC$. We set
$$
\Fun^{\EE_m}(\CA,\CB) := \Fun^{\EE_0}(\mathrm{B}^m\CA,\mathrm{B}^m\CB). 
$$
\end{defn}

\begin{defn} \label{def:convention_op}
For $k>-m$, we use $\CA^{\op k}$ to denote the $\EE_m$-monoidal $n$-category obtained by reversing all the $k$-morphisms, i.e.,
$\mathrm{B}^m(\CA^{\op k}) = (\mathrm{B}^m\CA)^{\op (k+m)}$.
We set $\CA^\rev:=(\mathrm{B}^m\CA)^{\op 1}=\CA^{\rev (1-m)}$. 
\end{defn}

\begin{rem}
According to Convention\,\ref{rem:convention_op} and \ref{conv:xi-direction}, $\CA^\rev$ flips the fusion product $\otimes^1$ (in the $x^1$-direction) to the opposite fusion product $(\otimes^1)^\op$ defined by $a(\otimes^1)^\op b := b\otimes^1 a$. 
\end{rem}

We denote by $\KarCat_n^\Cb$ the $(n+1)$-category of condensation complete (or Karoubi complete) $\Cb$-linear $n$-categories ($\Cb$-linear functors, etc) \cite{JF22,KZ22b}. It is also a symmetric monoidal $(n+1)$-category \cite{JF22}. Its tensor product, denoted by $\boxtimes$, is defined by $$\Fun_\Cb(\CA\boxtimes \CB, -) \simeq \Fun_\Cb(\CA, \Fun_\Cb(\CB,-)),$$ where $\Fun_\Cb(,)$ denotes the category of $\Cb$-linear $n$-functors. Its tensor unit is $n\vect:=\Sigma^n\Cb$. We denote by $\EE_m\KarCat_n^\Cb$ the $(n+1)$-category of $\EE_m$-monoidal $n$-categories, $\EE_m$-monoidal functors, $\EE_m$-monoidal (higher) natural transformations, etc. The notion of an $\EE_m$-multi-fusion $n$-category was introduced in \cite{JF22}. We take the definition from \cite{KZ22b}. 

\begin{defn} \label{def:Em-MultiFusion-nCat}
An {\it $\EE_m$-multi-fusion $n$-category} $\CA$ is a condensation-complete $\Cb$-linear $\EE_m$-monoidal $n$-category, i.e., $\CA\in \EE_m\KarCat_n^\Cb$, such that $\Sigma^m \CA$ is a separable $(n+m)$-category, i.e., $\Sigma^m \CA \in (n+m+1)\vect$. It is called {\it indecomposable} if $\CA$ is not a direct sum of two non-zero separable $n$-categories. If the tensor unit $\one_\CA$ is simple, then $\CA$ is called an {\it $\EE_m$-fusion $n$-category}. An $\EE_m$-fusion $0$-category is just $\Cb$. 
\end{defn}

\begin{expl} \label{expl:def_nRep(G)}
We provide some useful examples. 
\bnu
\item If $\CA$ is also $\EE_\infty$-monoidal, it means that $\Sigma^n\CA$ can be defined for all $n$. In this case, $\Sigma^n\CA$ is also $\EE_\infty$-monoidal. For example, symmetric fusion 1-categories are $\Rep(G)$ or $\Rep(G,z)$ for a finite group $G$. Then $(n+1)\Rep(G)=\Sigma^n\Rep(G)$ and $(n+1)\Rep(G,z)=\Sigma^n\Rep(G,z)$ are $\EE_\infty$-monoidal for all $n$. One can equivalently define $n\Rep(G)$ by $n\Rep(G):=\Fun(\mathrm{B}G, n\vect)$ and prove that $n\Rep(G) \simeq \Sigma (n-1)\Rep(G)$ as a theorem \cite{KZ24}. 
We treat different but equivalent definitions of $n\Rep(G)$ as different coordinate systems of $n\Rep(G)$.

\item For $\omega\in Z^{n+2}(G,U(1))$, $n\vect_G^\omega$ is an $\EE_1$-multi-fusion $n$-category \cite{KZ24}, and $\FZ_1(n\vect_G^\omega)$ is an $\EE_2$-multi-fusion $n$-category. When $\omega$ is trivial and $G$ is abelian, $n\vect_G$ is clearly a symmetric or $\EE_\infty$-fusion $n$-category.  

\item An $n$-groupoid is an $n$-category such that all morphisms are invertible. An $n$-group is a monoidal $(n-1)$-category $\CG$ such that $\mathrm{B}\CG$ is an $n$-groupoid. It is called finite if all homotopy groups $\pi_i(\CG)$ are finite. For a finite $n$-group $\CG$, we recall some results in \cite{KZ24}. 
\bnu
\item[-] $n\Rep(\CG):=\Fun(\mathrm{B}G, n\vect)$ is a symmetric fusion $n$-category and $(n+1)\Rep(\CG) \simeq \Sigma n\Rep(\CG)$.
\item[-] $n\vect_\CG:=\Omega L(\mathrm{B}\CG)$ is a fusion $n$-category, where $L$ is the left adjoint functor of the forgetful functor $\forget: \KarCat_{n+1}^\Cb \to \KarCat_{n+1}$.  
\enu
When $G$ is a finite abelian group. Let $\chi(G):=\hom(G,\Cb^\times)$ be the character group. We have $n\vect_{\mathrm{B}^{n-1}G} \simeq n\Rep(\chi(G))$ and $n\Rep(\mathrm{B}^{n-1}G) \simeq n\vect_{\chi(G)}$ \cite{KZ24}.

\item Using the $\EE_k$-center for $k\geq 0$ (see Definition\,\ref{def:centralizer+center}), we can construct an $\EE_{k+1}$-multi-fusion $n$-category from an $\EE_k$-multi-fusion $n$-category. For example, $\FZ_k(n\Rep(G))$ is an $\EE_{k+1}$-multi-fusion $n$-category. In this case, unfortunately, it does not generate new examples when $k\geq 3$ (see (\ref{eq:Zk-1_nRepG})). 

\item Another way to generate a new $\EE_k$-multi-fusion $n$-category from an old one $\CC$ is by condensations. More precisely, the new category is the category of $\EE_k$-modules over a condensable $\EE_k$-algebra in $\CC$ explained Section\,\ref{sec:condense_k-codim_defects}.  

\enu
\end{expl}


The following results are useful tools to determine or find new $\EE_k$-multi-fusion higher categories. 
\begin{thm}[\cite{KZ22b}]
Let $\CA$ be a $\Cb$-linear condensation-complete $\EE_m$-monoidal $n$-category for $m\geq 1$. If $\Sigma\CA$ is a separable $(n+1)$-category, then $\CA$ is an $\EE_m$-multi-fusion $n$-category. 
\end{thm}

\begin{cor}[\cite{KZ22b}]
If $\CA$ is an $\EE_m$-multi-fusion $n$-category for $n\geq 1$, then $\Omega\CA$ is an $\EE_{m+1}$-multi-fusion $(n-1)$-category. 
\end{cor}

\subsubsection{Condensable higher algebras} \label{sec:condensation_monad}

We have seen that an $\EE_m$-multi-fusion $n$-category is an $\EE_m$-algebra in $(n+1)\vect$. It turns out that they are also condensable (or separable) in a sense that we want to explain now. However, the mathematical theory of condensable (or separable) higher algebras in higher categories is very rich and highly non-trivial. It is beyond this work. We will develop such a theory in \cite{KZZZ25}. In this work, we only briefly sketch some facts that are useful to later sections and provide some intuitions. 


\medskip
A 0-category is defined by a set. A 0-condensation is an equality between objects of a 0-category.
\begin{defn}[\cite{GJF19}]
For two objects $X$ and $Y$ in a weak $n$-category $\CC$, an $n$-{\it condensation} of $X$ onto $Y$, denoted by $X\condense Y$, is a pair of 1-morphism $f: X \to Y$ and $g: Y \to X$, together with an $(n-1)$-condensation $fg\condense \id_Y$, i.e. 
$$
X\condense Y := \{  f: X \rightleftarrows Y: g, \quad fg\condense \id_Y \}
$$

The object $Y$ is called a condensate of $X$. A condensation is an $n$-condensation for some $n$. 
\end{defn}

\begin{expl} \label{expl:separable-algebra_1}
Let us unravel the definition in a few lower dimensional cases. 
\bnu
\item When $n=1$, $\CC$ is a 1-category. In this case, a 1-condensation is a pair of 1-morphisms $f: X\to Y$ and $g: Y\to X$ such that $fg=\id_Y$. As a consequence, $p:=gf$ is a idempotent, i.e. $p^2=p$. 

\item When $n=2$, $\CC$ is a 2-category. A 2-condensation in $\CC$ is a pair of 1-morphisms $f: X\to Y$ and $g: Y\to X$ such that $fg\condense \id_Y$ is a 1-condensation, i.e. there is a pair of 2-morphism $i: fg\to \id_Y$ and $j: \id_Y \to fg$ such that $ij=\id_{\id_Y}$.  
\bnu
\item Let $\CA$ be a multi-fusion category, and let $\LMod_\CA(2\vect)$ be the 2-category of left finite semisimple $\CA$-modules. Note that 
\be \label{eq:OmegaCA=CAop}
\Omega_\CA(\LMod_\CA(2\vect)) =\Fun_\CA(\CA,\CA) \simeq \CA^\rev
\ee
Let $A=(A,\mu_A,\eta_A)$ be an algebra in $\CA$, where $\mu_A: A\otimes A\to A$ is the multiplication and $\eta_A: \one_\CA \to A$ is the unit. Then $\CM:=\RMod_A(\CA) \in \LMod_\CA(2\vect)$. Now we would like to construct a 2-condensation $\CA \condense \CM$ when the algebra $A$ is separable. It means that we should construct 
$$
f:\CA \to \CM, \quad g: \CM \to \CA, \quad i: fg \to \id_\CM, \quad j: \id_\CM \to fg
$$ 
\begin{itemize}
\item $f=-\otimes A: \CA \to \CM$ defines a 1-morphism in $\LMod_\CA(2\vect)$. 

\item $g=[A,-]: \CM \to \CA$ is precisely the forgetful functor and is automatically a left $\CA$-module functor, and, at the same time, the right adjoint of $f$. 

\item $fg=[A,-]\otimes A$ is equipped with an evaluation 2-morphism 
$$
fg=[A,-]\otimes A \xrightarrow{i=\ev} \id_\CM, 
$$
i.e. $\ev_M=\mu_M: M\otimes A \to M$, which is nothing but the right $A$-action on $M$. Note that $i$ is a 1-morphism in $\Fun_\CA(\CM,\CM)$. 

\item We need a natural transformation $j: \id_\CM \to [A,-]\otimes A$ splitting $i$ in $\Fun_\CA(\CM,\CM)$. By \cite[Theorem\ 4.10]{KZ17}, this condition is equivalent to the condition that $A$ is separable algebra in $\CA$ in the usual sense, i.e., $\mu_A: A\otimes A \to A$ splits as $A$-$A$-bimodules.

\end{itemize}

\item Using \cite[Proposition\ 3.1.5]{GJF19}, a separable algebra $A$ in $\CA^\rev$ can be defined as $A=u^R\circ u$, where $u: \CA \to \CM \in \LMod_\CA(2\vect)$ extends to a 2-condensation in $2\vect$ and $u^R$ is the right adjoint of $u$. This idea can be used to define a separable higher algebra in a higher multi-fusion categories. We will show that elsewhere. 

\enu
\enu
\end{expl}


The precise definition of a condensable $\EE_m$-algebra in an $\EE_m$-multi-fusion $n$-category will be given elsewhere. We only provide a working definition that provides some physical intuitions here. For $m=0$, a condensable $\EE_0$-algebra in an $\EE_0$-multi-fusion $n$-category $(\CA,\one_\CA)$ is a pair $(A,\eta_A)$, where $A$ in an object in $\CA$ equipped with a unit morphism $\one_\CA \xrightarrow{\eta_A} A$. For $m\leq 1$, roughly speaking, there is a morphism $\mu_A^i: A \otimes A \to A$ of an $\EE_m$-algebra that defines the multiplication in the $i$-th independent direction for $i=1,2,\cdots,m$. All of them are compatible in the sense that $\mu_A := \mu_A^1 \simeq \mu_A^i$ (see Section\,\ref{sec:anyon_cond_algebra} for more discussion). It has a right adjoint $\mu_A^R$, and both $\mu_A$ and $\mu_A^R$ intertwine the $m$-dimensional $A$-actions. If $\mu_A$ is an $n$-morphism, its right adjoint is not defined. In this case, for conveniences and physical applications, we assume that top two morphisms form a $\ast$-category. In particular, the space of $n$-morphisms are Hilbert space and $f^\ast$ is defined for each $n$-morphisms $f$.\footnote{In this case, we have $g^R=g^L=g^\ast$ for all $k$-morphisms $g$ \cite[Proposition\ A.7]{KWZ15}.} We set $\mu_A^R=\mu_A^\ast$. Since $A$ plays the role of the vacuum in the condensed phase, we need impose a stability condition of the vacuum as explained in \cite{Kon14e} for the $m=1,2$ and $n=1$ cases. More explicitly, it means that $\mu_A \circ \mu_A^R=\id_A$ when $m=1,2$ and $n=1$, and it can be generalized to higher dimensions.  
\begin{defn} \label{def:condensable_algebra}
In an $\EE_m$-multi-fusion $n$-category $\CA$, a condensable (or separable) $\EE_m$-algebra $A$ is an $\EE_m$-algebra $A$ equipped with an $(n-1)$-condensation $\mu_A \circ \mu_A^R \condense \id_A$ in $\Mod_A^{\EE_m}(\CA)$ (i.e., modules equipped with $m$-dimensional $A$-actions explained later), where $\mu_A^R$ is the right adjoint of the multiplication 1-morphism $\mu_A: A\otimes A \to A$. It is called {\it simple} if the morphism $\eta_A: \one_\CA \to A$ is simple for $n>1$ and $\hom(\one_\CA,A)\simeq \Cb$ for $n=1$. It is called {\it indecomposable} if $A$ is not a direct sum of two algebras. 
\end{defn}

\begin{expl}
When $m=1$ and $n=2$, above definition is the same as the notion of a separable algebra in \cite{Dec23a,DX24}. According to \cite{Dec23a,DX24}, the following statements are true. 
\bnu
\item Condensable $\EE_1$-algebras in $2\vect$ are precisely multi-fusion 1-categories, and simple ones are fusion 1-categories, and indecomposable ones are indecomposable multi-fusion 1-categories. 

\item Condensable $\EE_2$-algebras in $2\vect$ are precisely $\EE_2$-multi-fusion 1-categories or braided multi-fusion 1-categories. Indecomposable ones are automatically simple and are $\EE_2$-fusion 1-categories (or braided fusion 1-categories). 

\item Condensable $\EE_3$-algebras in $2\vect$ are symmetric multi-fusion 1-categories, which is automatically $\EE_\infty$-monoidal. 

\item Condensable $\EE_2$-algebras in $\FZ_1(2\vect_G)$ for a finite group $G$ are exactly $G$-crossed braided multi-fusion 1-categories. 

\item Let $\CB$ be an $\EE_2$-fusion 1-category. A braided multi-fusion 1-category $\CA$ equipped with a braided functor $\CB \to \CA$ is a condensable $\EE_2$-algebra in the $\EE_2$-fusion 2-category $\FZ_1(\Sigma\CB)$. 
\enu
\end{expl}

\begin{expl}
We give a few examples of condensable $\EE_k$-algebras in higher categories.  
\bnu

\item Condensable $\EE_k$-algebras in $(n+1)\vect$ are precisely $\EE_k$-multi-fusion categories. 

\item When $G$ is abelian, condensable $\EE_k$-algebras in $(n+1)\vect_G$ are precisely $G$-graded $\EE_k$-multi-fusion categories. 

\item Condensable $\EE_k$-algebras in $(n+1)\Rep(G)$ in the coordinate $(n+1)\Rep(G) = \Fun(\mathrm{B}G, (n+1)\vect)$ are precisely $G$-equivariant $\EE_k$-multi-fusion categories, a notion which means an $\EE_k$-multi-fusion category $\CA$ equipped with a monoidal functor $G \to \mathrm{Aut}^{\EE_k}(\CA)$ and $\mathrm{Aut}^{\EE_k}(\CA)$ is the category of $\EE_k$-monoidal auto-equivalences of $\CA$. 

\item A condensable $\EE_2$-algebras in $\FZ_1((n+1)\vect_G)$ is precisely a $G$-crossed braided multi-fusion $n$-category (see Example\,\ref{expl:G_crossed_BF_nCat} (1) for a physical explanation).

\enu
\end{expl}

We will develop the theory of condensable $\EE_m$-algebras in an $\EE_m$-multi-fusion category elsewhere. In this work, we simply take it for granted that, for $m\geq 1$, condensable $\EE_m$-algebras in $(n+1)\vect$ are precisely $\EE_m$-multi-fusion $n$-categories.  Indecomposable condensable $\EE_m$-algebras in $(n+1)\vect$ are $\EE_m$-fusion $n$-categories for $m>1$ and are indecomposable multi-fusion $n$-categories for $m=1$; 
simples ones are $\EE_m$-fusion $n$-categories for all $m\geq 1$. For an $\EE_m$-multi-fusion $n$-category $\CA$, we denote the category of condensable $\EE_m$-algebras in $\CA$ by $\Algc_{\EE_m}(\CA)$. We denote the statement that $A$ is a condensable $\EE_m$-algebra $A$ in $\CA$ by $A\in \Algc_{\EE_m}(\CA)$. Then we also have $\CA \in \Algc_{\EE_m}((n+1)\vect)$. As we have explained in Section\,\ref{sec:E2=E1+E1}, by iterating the same argument, we obtain 
$$
\Algc_{\EE_m}(\CA) \simeq \Algc_{\EE_1}(\Algc_{\EE_{m-1}}(\CA)). 
$$
Moreover, from Example\,\ref{expl:separable-algebra_1}, we see that a condensable $\EE_1$-algebra $A$ in a multi-fusion 1-category can always be realized by an internal hom $[A,A]$ (see Section\ \ref{sec:internal_hom} for a brief introduction of this notion). This fact can be generalized to all condensable $\EE_m$-algebras in any $\EE_m$-multi-fusion $n$-categories. We will develop the theory elsewhere. In this work, we simply take it for granted that all internal homs naturally constructed are condensable higher algebras, and vice versa. A condensable $\EE_1$-algebra is a direct sum of indecomposable ones\footnote{See the structure theorem of a condensable $\EE_1$-algebra in a multi-fusion 1-category in Theorem\,\ref{thm:structure_theorem_separable_algebra} and a proposed structure theorem for such algebras in a multi-fusion $n$-category in Remark\,\ref{rem:structure_theorem}.}; and a condensable $\EE_m$-algebra for $m>1$ is direct sum of simple ones. We also take it for granted that the higher representation theory of condensable $\EE_m$-algebras in an $\EE_m$-multi-fusion $n$-category is compatible with condensation completion. In particular, all the categories of $\EE_k$-modules appear in this work are assumed to be separable. The special case for $m=n=2$ was proved in \cite{DX24}. We also want to remind readers about Remark\,\ref{rem:math_assumptions_intro} for our assumptions on condensability or separability used throughout this work.

\begin{rem} \label{rem:unitarity}
For physics oriented readers, it is helpful and harmless to ignore the subtleness in the definition of separability and take it for granted that all physically natural algebras or higher monoidal categories are automatically condensable or separable and unitary. Indeed, in all pictures, when we illustrate the physical intuitions, the unitarity is automatically assumed to avoid the framing issues. Under the unitary assumption, the left and right duals of a topological defect $x$ are the same, i.e., $x^R=x^L=x^\ast$, which should be viewed as the anti-defect of $x$. 
\end{rem}

\subsubsection{\texorpdfstring{$\EE_m$}{Em}-center and \texorpdfstring{$\EE_m$}{Em}-centralizer} \label{sec:center}

The notion of the center of an algebraic structure is very common in both physics and mathematics. The familiar notions of center include the center of a group $G$, defined by $Z(G):=\{ z\in G | zg=gz, \forall g\in G\}$; the center of a $\Cb$-algebra $A$, defined by $Z(A):=\{ z\in A | za=az, \forall a\in A\}$; the Drinfeld center of a fusion category; the M\"{u}ger center of a braided fusion category; the full center in 1+1D rational CFTs. The reason that they are all called `centers' because they satisfy the same universal properties (defined in different categories). 

In this subsubsection, we review the definitions of and some useful results on $\EE_m$-centralizer and $\EE_m$-centers from \cite{KZ24}. For physics oriented readers, the definition is quite abstract. We suggest readers to first focus on the results instead of the definitions. 

\begin{defn} \label{def:centralizer+center}
For $\CA, \CB\in \EE_m\KarCat_n^\Cb$ and a $\Cb$-linear $\EE_m$-monoidal functor $F: \CA \to \CB$, the $\EE_m$-centralizer of $F$ is the universal condensation-complete $\Cb$-linear $\EE_m$-monoidal $n$-category $\FZ_m(F)$ equipped with a unital action $G: \FZ_m(F) \boxtimes \CA \to \CB$, i.e. 
a $\Cb$-linear $\EE_m$-monoidal functor exhibiting the following diagram commutative (up to equivalences).
$$
\xymatrix{
& \FZ_m(F) \boxtimes \CA \ar[rd]^G & \\
\CA \ar[rr]^{F} \ar[ur]^{\one_{\FZ_m(F)} \boxtimes \id_\CA} & & \CB 
}
$$
For convenience, when $F$ is clear from the context, we also denote $\FZ_m(F)$ by $\FZ_m(\CA,\CB)$. When $F=\id_\CA$, then $\FZ_m(\CA):=\FZ_m(\id_\CA)=\FZ_m(\CA,\CA)$ defines the $\EE_m$-center of $\CA$. 
\end{defn}

\begin{rem}
We spell out the universal property of $\FZ_m(F)$ more explicitly. If $\CX\in \EE_m\KarCat_n^\Cb$ is equipped with a unital action $H: \CX \boxtimes \CA \to \CB$ (i.e., exhibiting the bottom triangle in the following diagram commutative), 
\be \label{diag:universal_property_center}
\begin{array}{c}
\xymatrix{
& \FZ_m(F) \boxtimes \CA \ar@/^2pc/[ddr]^G   &    \\
& \CX \boxtimes \CA \ar[dr]^{H} \ar[u]^{\exists ! \,f \boxtimes \id_\CA}   & \\
\CA \ar@/^2pc/[uur]^{\one_{\FZ_m(F)}\boxtimes \id_\CA} \ar[ur]^{\one_\CA \boxtimes \id_\CM}  \ar[rr]^{F} & & \CB \, 
}
\end{array}
\ee
then there is a unique $\Cb$-linear $\EE_m$-monoidal functor $f: \CX \to \FZ_m(F)$ rendering above diagram commutative.
\end{rem}

\begin{expl}
 We want to emphasize that the diagram  (\ref{diag:universal_property_center}) defines the notion of a centralizer or center in any category. In other words, any thing satisfies the universal property defined by the diagram (\ref{diag:universal_property_center}) in any (higher) categories is qualified to be called a `centralizer' (or a `center' for $F=\id_\CA$). We explain a few well known centers in terms of $\EE_m$-centers. 
\bnu

\item The center of a group is an $\EE_1$-center in the category of sets. Note that a group can be viewed as an $\EE_1$-algebra in the category of sets.  

\item The center of a $\Cb$-algebra $A$ is an $\EE_1$-center in the category of vector spaces over $\Cb$. 

\item  The Drinfeld center of a monoidal 1-category is an $\EE_1$-center in the 2-category $\mathsf{Cat}_1$ of 1-categories (as objects), functors (as 1-morphisms) and natural transformations (as 2-morphisms). An $\EE_1$-algebra in $\mathsf{Cat}_1$ is precisely a monoidal 1-category. 

\item The M\"{u}ger center of a braided monoidal 1-categories is an $\EE_2$-center in the 2-category $\mathsf{Cat}_1$. An $\EE_2$-algebra in $\mathsf{Cat}_1$ is precisely a braided monoidal 1-category. 

\enu
\end{expl}

\begin{expl}
In the category of potentially anomalous topological orders (or quantum liquids or QFT's), a morphism $f: \SA^n \to \SB^n$ between $\SA^n$ and $\SB^n$ was introduced in \cite{KWZ15,KWZ17} and is defined by a gapped domain wall $f^n$ between $\SZ(\SA)^{n+1}$ and $\SZ(\SB)^{n+1}$ such that $f^n\boxtimes_{\SZ(\SA)^{n+1}}\SA^n = \SB^n$ as illustrated below. 
\[
\begin{tikzpicture}[scale=0.8]
\draw[line width=2pt] (-3,0) -- (0,0) node[midway,above] {\footnotesize $\SZ(\SB)^{n+1}$} ;
\draw[line width=2pt] (0,0) -- (3,0) node[midway,above] {\footnotesize $\SZ(\SA)^{n+1}$} ;
\fill[black] (3,0) circle (0.12) node[above] {\footnotesize $\SA^n$} ;
\fill[black] (0,0) circle (0.12) node[above] {\footnotesize $f^n$} ;
\draw[decorate,decoration=brace,very thick] (3,-0.3)--(0,-0.3) ;
\node at (1.5,-0.8) {\footnotesize $f^n\boxtimes_{\SZ(\SA)^{n+1}}\SA^n = \SB^n$} ;
\end{tikzpicture}
\]
By checking the universal property of the centralizer (\ref{diag:universal_property_center}), it is an easy exercise to show that $f^n=\FZ(f)$ and $\SZ(\SA)^n=\FZ(\id_{\SA^n})$ (see \cite{KWZ15,KWZ17} for a proof of the later case). Similarly, in the category of fusion $n$-categories, for a monoidal functor $f: \CA \to \CB$, we have $\FZ_1(f) \simeq \Fun_{\CA|\CB}(\CB,\CB)$, where the left $\CA$-module structure on $\CB$ is defined by the monoidal functor $f$.  
\end{expl}

\begin{rem}
There are more general notions of centers. For example, consider an algebra $A$ in a monoidal 1-category $\CA$. It is impossible to define the center of $A$ within $\CA$ because there is no structure of commutativity in $\CA$. However, it is possible to define the notion of the center of $A$ in the Drinfeld center of $\CA$ \cite{Dav10,KYZ21}. 
\end{rem}

\begin{expl}
For $F: \CA \to \CB$, $\FZ_0(F)=\FZ_0(\CA,\CB) = (\Fun(\CA,\CB), F)$ because giving a unital action $\CC\boxtimes\CA \to \CB$ is equivalent to giving a $\Cb$-linear functor $\CC\to \Fun(\CA,\CB)$ that maps $\one_\CC$ to $F$. 
\end{expl}

\begin{expl} \label{expl:Z1Z0_centralizer}
For $\CA \in \Algc_{\EE_1}((n+1)\vect$ and $\CM \in \LMod_\CA((n+1)\vect)$,  by definition, $\CM$ is equipped with a monoidal functor $F: \CA \to \FZ_0(\CM)$. We have
\be \label{eq:Z1_Z0_centralizer}
\FZ_1(F) = \FZ_1(\CA,\FZ_0(\CM)) \simeq \Fun_\CA(\CM,\CM). 
\ee 
In fact, given a unital action $\CX \boxtimes \CA \to \FZ_0(\CM)$ is equivalent to giving a $\Cb$-linear monoidal functor $\CX \to \Fun_\CA(\CM,\CM)$. The physical meaning of (\ref{eq:Z1_Z0_centralizer}) is illustrated below. 
\be \label{eq:Z1Z0_centralizer}
\begin{array}{c}
\begin{tikzpicture}[scale=1.5]
\fill[gray!20] (0,0) rectangle (2,1.2) ;
\node at (1,0.6) {\small $\SZ(\SA)^{n+2}=\SZ(\SB)^{n+2}$} ;
\draw[black, ultra thick,->-] (0,0) -- (1,0) ; 
\draw[black, ultra thick,->-] (1,0) -- (2,0) ; 
\draw[fill=white] (0.95,-0.05) rectangle (1.05,0.05) node[midway,below,scale=1] {\scriptsize $\hspace{1mm}\SM^n$} ;
\node at (0.5,-0.15) {\small $\SA^{n+1}$} ;
\node at (1.5,-0.15) {\small $\SB^{n+1}$} ;
\end{tikzpicture}
\end{array}
\quad\quad
\begin{array}{l}
\SM^n=(\CM,m), \quad m\in \CM,  \\
\CB^\rev = \Fun_\CA(\CM,\CM) \simeq \FZ_1(\CA,\FZ_0(\CM)), \\
\CA = \Fun_{\CB^\rev}(\CM,\CM) \simeq \FZ_1(\CB^\op, \FZ_0(\CM)),
\end{array}
\ee
where the categories of all topological defects in two simple $n+$1D topological order $\SA^{n+1}$ and $\SB^{n+1}$ are the fusion $n$-category $\CA$ and $\CB$, respectively, and the gapped domain wall $\SM^n$, as we show later, can be described by a pair $(\CM,m)$, where $\CM \in \BMod_{\CA|\CB}((n+1)\vect)$ is the category of wall conditions and $m$ is an object in $\CM$ specifying a wall condition. For example, when $\CA=n\vect$, $\CM\in (n+1)\vect$, if $F: n\vect \to \FZ_0(\CM)$ is the tensor unit of $\FZ_0(\CM)$, we obtain $\FZ_1(F) \simeq \FZ_0(\CM)$. 
\end{expl}

\begin{expl} \label{expl:MoritaEq=centralizer}
The notion of a morphism between two (potentially anomalous) topological orders is defined by a gapped domain wall. For example, $\SM^n$ in Figure\,\ref{fig:domain_wall=functor} defines a morphism between $\SA^{n+1}$ and $\SB^{n+1}$.\footnote{For intrigued readers, we leave it as an exercise to compute the centralizer of this morphism $\SM^n$. We will discuss it elsewhere.} However, this notion only distinguish topological orders up to Morita equivalence. A more fundamental notion of a morphism between two topological orders should distinguish them up to isomorphisms. Such a notion was introduced \cite{KWZ15} (see also \cite{KWZ17} and Figure\,\ref{fig:domain_wall=functor}). For example, roughly speaking, $\SB^{n+1}$ in the picture in (\ref{eq:Z1Z0_centralizer}) defines a morphism $F: \SA^{n+1} \to \SZ(\SM)^{n+1}$. This notion leads to a new symmetric monoidal higher category of $n+$1D topological orders \cite{KWZ15}. In this category, the bulk $\SZ(\SA)^{n+2}$ of $\SA^{n+1}$ is indeed the center of $\SA^{n+1}$. This justifies the notation of the bulk. Moreover, one can also define the notion of centralizer in this category as in Definition\,\ref{def:centralizer+center}. For example, we have $\SZ(F)^{n+1} = \SZ(\SA,\SZ(\SM))^{n+1} = \overline{\SB^{n+1}}$. 
\end{expl}

\begin{prop}
Let $F: \CA \to \Omega\CB$ be a $\Cb$-linear $\EE_m$-monoidal functor for $\CA\in \EE_m\KarCat_n^\Cb$ and $\CB \in \EE_{m-1}\KarCat_n^\Cb$. We have 
\be \label{eq:FZm-FZm-1}
\FZ_m(\CA,\Omega\CB) \simeq \Omega \FZ_{m-1}(\Sigma\CA,\CB). 
\ee
\end{prop}

\begin{thm}[\cite{KZ24}]
Let $F:\CA\to \CB$ be a $\Cb$-linear $\EE_m$-monoidal functor for $\CA,\CB \in \EE_m\KarCat_n^\Cb$. For $0\leq k\leq m$, $F$ induces canonically a functor $\Sigma^k F: \Sigma^k\CA \to \Sigma^k\CB$. We have
\be \label{eq:Zm_centralizer}
\FZ_m(F) = \FZ_m(\CA,\CB) = \Omega^k \FZ_{m-k}(\Sigma^k\CA,\Sigma^k\CB) = \Omega^m(\Fun(\Sigma^m\CA, \Sigma^m\CB), \Sigma^mF). 
\ee
As a special case, we have 
\be \label{eq:Zm_center}
\FZ_m(\CA) \simeq \Omega\FZ_{m-1}(\Sigma \CA) \simeq \Omega^m\FZ_0(\Sigma^m\CA) \simeq \Omega^m \Fun(\Sigma^m\CA, \Sigma^m\CA). 
\ee

\end{thm}

\begin{expl} \label{expl:Z1_ind_MF}
When $m=1$, we obtain 
$$
\FZ_1(\CA,\CB) = \Omega(\Fun(\Sigma\CA,\Sigma\CB), F) \simeq \Fun_{\CA|\CB}(\CB,\CB) \simeq \Fun_{\CA|\CB}(\CA,\CB). 
$$
Let $\CA$ be an indecomposable multi-fusion $n$-category. By Proposition\,\ref{prop:structure_theorem_MF}, we have the decomposition $\CA \simeq \oplus_{i,j} \CA_{ij}$ as separable $n$-categories. We obtain a useful result:
$$
\FZ_1(\CA) \simeq \Omega\FZ_0(\Sigma\CA) \simeq \Omega\FZ_0(\Sigma\CA_{ii}) \simeq \FZ_1(\CA_{ii}). 
$$ 
\end{expl}

\begin{cor}[\cite{KZ24}] \label{cor:Zm_Ek-fusion}
If $\CA, \CB \in \Algc_{\EE_m}((n+1)\vect)$ and $F: \CA\to \CB$ is $\Cb$-linear $\EE_m$-monoidal, then $\FZ_m(F)=\FZ_m(\CA,\CB) \in \Algc_{\EE_m}((n+1)\vect)$. 
\end{cor}

\begin{rem}
The special cases of $n=2$ and $m=0,1,2,3$ of Corollary\,\ref{cor:Zm_Ek-fusion} are rigorously proved recently in \cite{Xu24}. 
\end{rem}

\begin{defn} \label{def:non-degenerate_Lagrangian}
We introduce the following notions (as working definitions for convenience\footnote{We will take a more systematic approach in \cite{KZZZ25} to define the notion of a Lagrangian $\EE_m$-algebra in a non-degenerate $\EE_m$-multi-fusion $n$-category.}). 
\bnu

\item An indecomposable $\EE_1$-multi-fusion $n$-category $\CA$ is called {\it non-degenerate} if $\FZ_1(\CA) \simeq n\vect$. 
In this case, for $A\in \Algc_{\EE_1}(\CA)$, $A$ is called {\it Lagrangian} if $\Mod_A^{\EE_1}(\CA) \simeq n\vect$. 

\item An $\EE_2$-fusion $n$-category $\CB$ is called {\it non-degenerate} if $\FZ_2(\CB) \simeq n\vect$. In this case, for $B\in \Algc_{\EE_2}(\CB)$, $B$ is called {\it Lagrangian} if $\Mod_B^{\EE_2}(\CB) \simeq n\vect$. 

\enu
\end{defn}

\begin{rem}
An multi-fusion $n$-category $\CC$ is non-degenerate if and only if $\CC$ as a $\CC$-$\CC$-bimodule is closed, i.e., the canonical monoidal functor $\CC\boxtimes \CC^\rev \to \Fun(\CC,\CC)$ is an equivalence \cite{KZ24}. The $\EE_1$-center of a direct sum of two multi-fusion $n$-categories is the direct sum of the $\EE_1$-centers of the summands. Moreover, the $\EE_1$-center of an indecomposable multi-fusion $n$-category is an $\EE_2$-fusion $n$-category. 
Therefore, a non-degenerate multi-fusion $n$-category is automatically an indecomposable multi-fusion $n$-category. 
\end{rem}

\subsection{Center functor} \label{sec:center_functor}
The previous boundary-bulk relation can be generalized to include gapped domain walls. This more complete version of boundary-bulk relation can be formulated precisely in terms of the so-called center functor \cite{KWZ15,KZ18,KZ24}. In this subsection, we review the center functor and related results that are useful in later sections. 

\subsubsection{Complete boundary-bulk relation as a center functor}
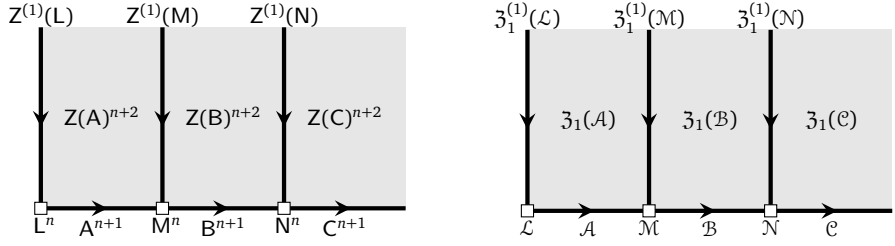
\begin{figure}[htbp]
$$
\begin{array}{c}
\begin{tikzpicture}[scale=0.8]
\fill[gray!20] (-3,0) rectangle (3,3) ;
\draw[ultra thick,->-] (-3,3)--(-3,0) ; 
\draw[ultra thick,->-] (1,3)--(1,0) ;
\draw[ultra thick,->-] (-1,3)--(-1,0) ;
\draw[ultra thick,->-] (-3,0)--(-1,0) ;
\draw[ultra thick,->-] (-1,0)--(1,0) ;
\draw[ultra thick,->-] (1,0)--(3,0) ;
\draw[fill=white] (-3.1,-0.1) rectangle (-2.9,0.1) node[midway,below] {\footnotesize $\hspace{1mm}\SL^n$}; 
\draw[fill=white] (-1.1,-0.1) rectangle (-0.9,0.1) node[midway,below] {\footnotesize $\hspace{1mm}\SM^n$} ;
\node at (-2,-0.3) {\footnotesize $\SA^{n+1}$} ;
\node at (0,-0.3) {\footnotesize $\SB^{n+1}$} ;
\node at (2,-0.3) {\footnotesize $\SC^{n+1}$} ;
\draw[fill=white] (0.9,-0.1) rectangle (1.1,0.1) node[midway,below] {\footnotesize $\hspace{1mm}\SN^n$} ;
\node at (-2,1.5) {\footnotesize $\SZ(\SA)^{n+2}$} ;
\node at (0,1.5) {\footnotesize $\SZ(\SB)^{n+2}$} ;
\node at (2,1.5) {\footnotesize $\SZ(\SC)^{n+2}$} ;
\node at (-3,3.2) {\footnotesize $\SZ^{(1)}(\SL)$} ;
\node at (-1,3.2) {\footnotesize $\SZ^{(1)}(\SM)$} ;
\node at (1,3.2) {\footnotesize $\SZ^{(1)}(\SN)$} ;
\end{tikzpicture}
\end{array}
\quad\quad
\begin{array}{c}
\begin{tikzpicture}[scale=0.8]
\fill[gray!20] (-3,0) rectangle (3,3) ;
\draw[ultra thick,->-] (-3,3)--(-3,0) ; 
\draw[ultra thick,->-] (1,3)--(1,0) ;
\draw[ultra thick,->-] (-1,3)--(-1,0) ;
\draw[ultra thick,->-] (-3,0)--(-1,0) ;
\draw[ultra thick,->-] (-1,0)--(1,0) ;
\draw[ultra thick,->-] (1,0)--(3,0) ;
\draw[fill=white] (-3.1,-0.1) rectangle (-2.9,0.1) node[midway,below] {\footnotesize $\CL$}; 
\draw[fill=white] (-1.1,-0.1) rectangle (-0.9,0.1) node[midway,below] {\footnotesize $\CM$} ;
\node at (-2,-0.3) {\footnotesize $\CA$} ;
\node at (0,-0.3) {\footnotesize $\CB$} ;
\node at (2,-0.3) {\footnotesize $\CC$} ;
\draw[fill=white] (0.9,-0.1) rectangle (1.1,0.1) node[midway,below] {\footnotesize $\CN$} ;
\node at (-2,1.5) {\footnotesize $\FZ_1(\CA)$} ;
\node at (0,1.5) {\footnotesize $\FZ_1(\CB)$} ;
\node at (2,1.5) {\footnotesize $\FZ_1(\CC)$} ;
\node at (-3,3.2) {\footnotesize $\FZ_1^{(1)}(\CL)$} ;
\node at (-1,3.2) {\footnotesize $\FZ_1^{(1)}(\CM)$} ;
\node at (1,3.2) {\footnotesize $\FZ_1^{(1)}(\CN)$} ;
\end{tikzpicture}
\end{array}
$$
\caption{the functoriality of the boundary-bulk relation
}
\label{fig:functoriality_bbr}
\end{figure}

We have discussed the boundary-bulk relation in Section\,\ref{sec:bbr}. 
It turns out that (\ref{eq:bbr}) is only a part of more complete boundary-bulk relation \cite{KWZ15,KZ24}, which is also very useful to our higher condensation theory. We review it now. Consider the physical configuration depicted in Figure\,\ref{fig:functoriality_bbr}, in which $\SA^{n+1},\SB^{n+1}, \SC^{n+1}$ are potentially anomalous type-I topological orders (see Remark\,\ref{rem:bbr_type-I} for this `type-I' assumption); $\SM^n$ and $\SN^n$ are gapped domain walls. 

The orientation of the $\SZ(\SA)^{n+1}$ follows that in Figure\,\ref{fig:defects_in_2d-TO} and that of $\SA^{n+1}$ is the one induced from the bulk, and $\SZ(\SA)^{n+2}$ is on the left side of $\SA^{n+1}$ if one follows the arrow on $\SA^{n+1}$, which is also the direction we fuse 1-codimensional defects in $\SA^{n+1}$. There are two different ways to look at $\SM^n$. 
\bnu

\item $\SM^n$ is a gapped defect junction of $\SA^{n+1}$, $\SB^{n+1}$ and $\SZ^{(1)}(\SM)^{n+1}$. 

\item When $\SM^n$ is viewed as gapped domain wall between $\SA^{n+1}$ and $\SB^{n+1}$, it is potentially anomalous in the sense that the $n+$1D topological order $\SZ^{(1)}(\SM)^{n+1}$, which can be viewed as the anomaly of $\SM^{n+1}$ as a domain wall of $\SA^{n+1}$ and $\SB^{n+1}$, can be non-trivial. According to \cite{KW14,KWZ15}, $\SM^{n+1}$ uniquely determines $\SZ^{(1)}(\SM)^{n+1}$, called the {\it relative bulk} of $\SM^n$. This fact also explains the notation. Moreover, $\SZ^{(1)}(\SM)^{n+1}$ is also type-I. 
Note that the bulk $\SZ(\SM)^{n+1}$ of $\SM^n$ is different from the relative bulk $\SZ^{(1)}(\SM)^{n+1}$ of $\SM^n$ but can be expressed in terms of the later as follows: 
$$
\SZ(\SM)^{n+1} = \overline{\SB^{n+1}} \boxtimes_{\SZ(\SB)} \SZ^{(1)}(\SM)^{n+1} \boxtimes_{\SZ(\SA)} \SA^{n+1},
$$
where $\SZ^{(1)}(\SM)^{n+1} \boxtimes_{\SZ(\SA)} \SA^{n+1}$ represents the fusion of $\SZ^{(1)}(\SM)^{n+1}$ and $\SA^{n+1}$ along $\SZ(\SA)^{n+2}$. 
\enu

\begin{rem}
In Figure\,\ref{fig:functoriality_bbr}, even when the defect junction $\SN^n$ is gapless, its relative bulk $\SZ(\SN)^{n+1}$ is always gapped because any domain walls between two non-chiral topological orders $\SZ(\SA)^{n+2}$ and $\SZ(\SC)^{n+2}$ are necessarily gappable. 
\end{rem}

\begin{pthm}[Functoriality of Boundary-Bulk Relation \cite{KWZ15}]  \label{pthm:bbr}
The boundary-bulk relation, i.e., the assignment: 
$$
\SA^{n+1} \mapsto \SZ(\SA)^{n+1}, \quad\quad 
\SM^n \mapsto \SZ^{(1)}(\SM)^{n+1},
$$
gives a well-defined functor called {\it center functor} $\SZ: \ST\SO_{\mathrm{gp-wall}}^{n+1} \to {}^{\mathrm{af}}\ST\SO_{\mathrm{af-gp-wall}}^{n+2}$, where
\bnu
\item[-] $\ST\SO_{\mathrm{gp-wall}}^{n+1}$ is the category of potentially anomalous $n+$1D type-I topological orders as objects and potentially anomalous gapped domain walls as 1-morphisms; 

\item[-] ${}^{\mathrm{af}}\ST\SO_{\mathrm{af-gp-walll}}^{n+2}$ is the category of anomaly-free   simple $n+$2D topological orders as objects and anomaly-free type-I gapped domain walls as 1-morphisms. 
\enu
More precisely, in the physical situation depicted in Figure\,\ref{fig:functoriality_bbr}, we have an equivalence of type-I topological orders: 
\be \label{eq:functorialty_bbr_phases}
\SZ^{(1)}(\SN)^{n+1} \boxtimes_{\SZ(\SB)} \SZ^{(1)}(\SM)^{n+1} = \SZ^{(1)}(\SM\boxtimes_\SB \SN)^{n+1},
\ee
where $\boxtimes_{\SZ(\SB)}$ means horizontally fuse $\SZ^{(1)}(\SM)^{n+1}$ with $\SZ^{(1)}(\SN)^{n+1}$ along $\SZ(\SB)^{n+2}$. 
\end{pthm}

\begin{rem} \label{rem:bbr_type-I}
The functoriality of boundary-bulk relation for type-II topological orders is not natural (but possible by imposing some technical conditions). We do not need it. 
\end{rem}

Above result can be translated into a mathematical one. Note that the categories of topological defects $\CA,\CB,\CC$ in the type-I topological orders $\SA^{n+1},\SB^{n+1}, \SC^{n+1}$, respectively, are indecomposable multi-fusion $n$-categories. Let $\CM$  be a $\CA$-$\CB$-bimodule and $\CN$ a $\CB$-$\CC$-bimodule. 
\begin{thm}[\cite{KWZ15,KZ18,KZ24}] \label{thm:bbr_functor}
The following assignment (see Figure\,\ref{fig:functoriality_bbr}): 
\be \label{def:Z1_functor}
\CA \mapsto \FZ_1(\CA), \quad \quad {}_\CA\CM_\CB \mapsto \FZ_1^{(1)}(\CM) := \Fun_{\CA|\CB}(\CM,\CM)
\ee
defines the so-called {\it center functor} $\FZ_1: \CM\CF us_n^{\mathrm{ind}} \to {}^{\mathrm{ndg}}\CB\CF us_n^{\mathrm{cl}}$, where 
\bnu
\item[-] $\CM\CF us_n^{\mathrm{ind}}$ is the symmetric monoidal 1-category of indecomposable multi-fusion $n$-categories as objects and the equivalence classes of non-zero separable bimodules as 1-morphisms; 
\item[-] ${}^{\mathrm{ndg}}\CB\CF us_n^{\mathrm{cl}}$ is the symmetric monoidal 1-category of non-degenerate braided fusion $n$-categories as objects and the equivalence classes of closed multi-fusion $\CB$-$\CA$-modules as morphisms from $\CA$ to $\CB$. 
\enu
In particular, this functoriality\footnote{This functoriality is stronger than the weak notion of a lax functor. This is a manifestation of an interesting physical principle, which will be discussed elsewhere.} means that it respects the horizontal fusion: 
\begin{align}
\Fun_{\CB|\CC}(\CN,\CN) \boxtimes_{\FZ_1(\CB)} \Fun_{\CA|\CB}(\CM,\CM)
&\xrightarrow{\simeq}
\Fun_{\CA|\CC}(\CM\boxtimes_\CB\CN,\CM\boxtimes_\CB\CN) \label{eq:funtoriality_bbr} \\
g\boxtimes_{\FZ_1(\CB)} f \quad\quad &\mapsto \quad\quad 
f\boxtimes_\CB g. \nonumber
\end{align}
\end{thm}

It turns out that this center functor $\FZ_1(-)$ is a consequence of a simpler center functor $\FZ_0: \CS ep_{n+1}^{\mathrm{ind}} \to {}^{\mathrm{ndg}}\CF us_{n+1}^{\mathrm{cl}}$ \cite{KZ24}. 
\bnu
\item[-] The domain category $\CS ep_{n+1}^{\mathrm{ind}}$ is the symmetric monoidal 1-category of indecomposable separable $(n+1)$-categories, i.e., $\Sigma\CA$ for an indecomposable multi-fusion $n$-category, as objects and the equivalence classes of non-zero $\Cb$-linear functors as 1-morphisms. 
\item[-] The codomain category ${}^{\mathrm{ndg}}\CF us_{n+1}^{\mathrm{cl}}$ is the symmetric monoidal 1-category of non-degenerate fusion $(n+1)$-categories as objects and the equivalence classes of pairs $(\CM,X)$ as morphisms from $\CA$ to $\CB$, where $\CM$ is a closed $\CA$-$\CB$-bimodule and $X$ is a non-zero object of $\CM$. Two pairs $(\CM,X)$ and $(\CN,Y)$ are equivalent if there is a $\Cb$-linear bimodule equivalence $F: \CM \to \CN$ such that $F(X) \simeq Y$. 
\enu
The tensor product is given by $\boxtimes$. The center functor $\FZ_0: \CS ep_{n+1}^{\mathrm{ind}} \to {}^{\mathrm{ndg}}\CF us_{n+1}^{\mathrm{cl}}$ is defined by 
\be \label{def:Z0_functor}
\Sigma \CA \mapsto \FZ_0(\Sigma\CA)=\Fun(\Sigma\CA, \Sigma\CA), \quad\quad
(\Sigma\CA \xrightarrow{F} \Sigma\CB) \mapsto \FZ_0^{(1)}(F):=(\Fun(\Sigma\CA,\Sigma\CB), F), 
\ee
and is symmetric monoidal. The center functor $\FZ_1$ is simply a consequence of the center functor $\FZ_0$ because the delooping functors 
\be \label{eq:Sigma=equivalence}
\Sigma: \CM\CF us_n^{\mathrm{ind}} \to \CS ep_n^{\mathrm{ind}}\quad\quad 
\mbox{and} \quad\quad
\Sigma: {}^{\mathrm{ndg}}\CB\CF us_n^{\mathrm{cl}} \to {}^{\mathrm{ndg}}\CF us_n^{\mathrm{cl}}
\ee 
are symmetric monoidal equivalences. More explicitly, we illustrate the relation between the $\FZ_0$-functor and the $\FZ_1$-functor in the following diagrams: 
\be \label{diag:relation_Z0_Z1-functors}
\xymatrix{
\CA \ar@{|->}[r]^-{\FZ_1}  \ar[d]^\Sigma_\simeq & \FZ_1(\CA) \ar[d]^\Sigma_\simeq \\
\Sigma\CA \ar@{|->}[r]^-{\FZ_0} & \FZ_0(\Sigma\CA), 
}
\quad\quad\quad\quad
\xymatrix{
{}_\CA\CM_\CB \ar@{|->}[r]^-{\FZ_1^{(1)}} \ar@{|->}[d]  & \Fun_{\CA|\CB}(\CM, \CM) \ar@{|->}[d] \\
F_\CM \ar@{|->}[r]^-{\FZ_0^{(1)}} & (\Fun(\Sigma\CA,\Sigma\CB), F_\CM), 
}
\ee
where $F_\CM:=-\boxtimes_\CA \CM \in \Fun(\Sigma\CA,\Sigma\CB)$.

\subsubsection{Measure non-surjectivity and non-fully-faithfulness}
\label{sec:measure_non-fully-faithfulness}
It turns out that the deviations of the $\FZ_0$-functor (or the $\FZ_1$-functor) from being essentially subjective, full and faithful are very important in the study of topological orders. In the proof of Proposition* 3.13 in \cite{KZ24}, we actually proved that invertible separable higher categories precisely measure the non-surjectivity and non-fully-faithfulness of the center functor, a result which is a slightly refinement of the original statement of Proposition* 3.13 in \cite{KZ24}. We state it more explicitly below. 
\begin{thm}[\cite{KZ24}] \label{thm:center-functor_surjective_full_faithful}
We denote the full subcategory of $(n+1)\vect$ consisting of invertible objects (i.e., invertible separable $n$-categories) by $(n+1)\vect^\times$ and denote its underlying group by $\mathrm{BrPic}(n\vect)$. 
\bnu
\item[(1)] $\FZ_0: \CS ep_{n-1}^{\mathrm{ind}} \to {}^{\mathrm{ndg}}\CF us_{n-1}^{\mathrm{cl}}$ is `essentially surjective' up to invertible separable $n$-categories in the sense that   
\bnu
\item $\CX \in {}^{\mathrm{ndg}}\CF us_{n-1}^{\mathrm{cl}}$ if and only if $\Sigma\CX \in (n+1)\vect^\times$;
\item $\CX\in \FZ_0(\CS ep_{n-1}^{\mathrm{ind}})$ if and only if $\Sigma\CX \simeq n\vect \in (n+1)\vect^\times$. 
\enu

\item[(2)] $\FZ_0: \CS ep_n^{\mathrm{ind}} \to {}^{\mathrm{ndg}}\CF us_n^{\mathrm{cl}}$ is `full' up to invertible separable $n$-categories in the sense that, for any 1-morphism 
$\FZ_0(\CS) \xrightarrow{(\CX,x)} \FZ_0(\CS')$ in ${}^{\mathrm{ndg}}\CF us_n^{\mathrm{cl}}$ and $\CS, \CS' \in \CS ep_n^{\mathrm{ind}}$, 
there exists a $\CU \in (n+1)\vect^\times$ and a $\Cb$-linear functor $\CS \xrightarrow{F} \CS'$ such that $\CX \simeq \FZ_0(F) \boxtimes \CU$.  

\item[(3)] $\FZ_0: \CS ep_{n+1}^{\mathrm{ind}} \to {}^{\mathrm{ndg}}\CF us_{n+1}^{\mathrm{cl}}$ is `faithful' up to invertible separable $n$-categories in the sense that
any two 1-morphisms $F, G \in \Fun(\CS,\CS')$ for $\CS, \CS'\in \CS ep_{n+1}^{\mathrm{ind}}$ share the `same' image, i.e., $\FZ_0^{(1)}(F) \simeq \FZ_0^{(1)}(G)$, if and only if there exists a $\CV \in (n+1)\vect^\times$ such that $G(-) \simeq F(-) \boxtimes \CV$. 

\enu
\end{thm}

The proof of Theorem\,\ref{thm:center-functor_surjective_full_faithful} is identical to that of Proposition* 3.13 in \cite{KZ24}. We do not repeat it here. Instead, we provide a physical proof of this result reformulated in terms of the $\FZ_1$-functor below. 
\begin{thm}[\cite{KZ24}] \label{thm:center-functor_surjective_full_faithful_2}
For readers convenience, we rewrite above theorem in terms of the $\FZ_1$-functor based on the relation (\ref{diag:relation_Z0_Z1-functors}). See Figure\,\ref{fig:functoriality_bbr} for the associated physical configuration. 
\bnu
\item[(1)] $\FZ_1: \CM\CF us_{n-2}^{\mathrm{ind}} \to {}^{\mathrm{ndg}}\CB\CF us_{n-2}^{\mathrm{cl}}$ is `essentially surjective' up to invertible separable $n$-categories, i.e.,  
\bnu
\item $\CC \in {}^{\mathrm{ndg}}\CB\CF us_{n-2}^{\mathrm{cl}}$ if and only if $\Sigma^2\CC \in (n+1)\vect^\times$; 
\item $\CC\in \FZ_1(\CM\CF us_{n-2}^{\mathrm{ind}})$ if and only if $\Sigma^2\CC \simeq n\vect \in (n+1)\vect^\times$.
\enu

\item[(2)] $\FZ_1: \CM\CF us_{n-1}^{\mathrm{ind}} \to {}^{\mathrm{ndg}}\CB\CF us_{n-1}^{\mathrm{cl}}$ is `full' up to invertible separable $n$-categories, i.e., for any 1-morphism $\CK: \FZ_1(\CA) \to \FZ_1(\CB)$ in ${}^{\mathrm{ndg}}\CB\CF us_{n-1}^{\mathrm{cl}}$ and $\CA,\CB \in \CM\CF us_{n-1}^{\mathrm{ind}}$, there exists a $\CU \in (n+1)\vect^\times$ and a non-zero $\CM \in \BMod_{\CA|\CB}((n+1)\vect)$ such that $\Sigma\CK \simeq \Sigma\FZ_1^{(1)}(\CM) \boxtimes \CU$. 

\item[(3)] $\FZ_1: \CM\CF us_n^{\mathrm{ind}} \to {}^{\mathrm{ndg}}\CB\CF us_n^{\mathrm{cl}}$ is `faithful' up to invertible separable $n$-categories. That is, two 1-morphisms ${}_\CA\CM_\CB, {}_\CA\CN_\CB$ for $\CA,\CB \in \CM\CF us_n^{\mathrm{ind}}$ share the `same' image, i.e., $\FZ_1^{(1)}(\CM) \simeq \FZ_1^{(1)}(\CN)$, if and only if there exists an $\CV \in (n+1)\vect^\times$ such that $\CN \simeq \CM \boxtimes \CV$. 

In the special case $\CA=\CB=n\vect$, two separable $n$-categories $\CM$ and $\CN$ share the same center, i.e., $\FZ_0(\CM) \simeq \FZ_0(\CN)$, if and only if there exists a $\CV \in (n+1)\vect^\times$ such that $\CN \simeq \CM \boxtimes \CV$. 
\enu
\end{thm}
\pf
We provide a physical proof here. The proof of (1) is obvious. The proof of (2) is precisely given by that of Theorem$^{\mathrm{ph}}$\,\ref{pthm:TME-action_transitive}. Now we give a proof of (3). Look at the first picture in Figure\,\ref{fig:functoriality_bbr}. Note that stacking $\SM^n$ with an anomaly-free $n$D topological order $\ST^n$ followed by a condensation within $\SM^n\boxtimes\ST^n$ do not change anything around the defect junction. In particular, $\SZ^{(1)}(\SM)^{n+1}$ remains intact. Moreover, if a replacement of the junction $\SM^n$ by a new junction $\SN^n$ leave its neighborhood intact, then, by Theorem$^{\mathrm{ph}}$\,\ref{pthm:TME-action_transitive}, it is necessary that $\CN \simeq \CM \boxtimes \CV$ for $\CV \in (n+1)\vect^\times$. 
\epf

\begin{rem} The (2)-part of Theorem\,\ref{thm:center-functor_surjective_full_faithful_2} first explicitly appeared (but formulated differently but equivalently) in Theorem$^{\mathrm{ph}}$\, 3.2.13 in the first version of this paper (see also Theorem$^{\mathrm{ph}}$\,\ref{pthm:TME-action_transitive} and \ref{pthm:recover_all_bdy_from_one} in this version). It also appeared later in the third version of \cite{LYW24} and was formulated 
as in Theorem\,\ref{thm:same_bulk=Morita+invertible}. 
\end{rem}

\begin{cor} \label{cor:center-functor_surjective_full_faithful}
This corollary is essentially the Example 3.14 in \cite{KZ24}. There is no non-trivial invertible separable $n$-category only $n=0,1,2$. For $n=0,1$, $\FZ_0: \CS ep_n^{\mathrm{ind}} \to {}^{\mathrm{ndg}}\CF us_n^{\mathrm{cl}}$ is a symmetric monoidal equivalence; 
for $n=2$, it is fully faithful; for $n=3$, it is faithful but neither full nor essentially surjective; for $n>3$, $\FZ_0$ is neither essentially surjective nor full nor faithful. 
\end{cor}

\begin{rem} \label{cor:BPic-Aut}
Theorem\,\ref{thm:bbr_functor}, \ref{thm:center-functor_surjective_full_faithful} or \ref{thm:center-functor_surjective_full_faithful_2} are important in that it can lead us to many new results. For example, for $\CA\in \CM\CF us_{n-1}^{\mathrm{ind}}$, we have an exact sequence of group homomorphisms:
$$
1 \rightarrow \mathrm{BrPic}(n\vect)_\CA \hookrightarrow \mathrm{BrPic}(n\vect) \xrightarrow{\CA\boxtimes -} \mathrm{BrPic}(\CA) \xrightarrow{\FZ_1^{(1)}} 
\mathrm{\Aut}_0^{\EE_2}(\FZ_1(\CA)) \to 1. 
$$
\bnu
\item[-] We denote the group of the invertible $A$-$A$-bimodules by $\mathrm{BrPic}(\CA)$. We define $\mathrm{BrPic}(n\vect)_\CA :=\{ \CV \in \mathrm{BrPic}(n\vect) | \CA \boxtimes \CV \simeq \CA \}$ to be the stabilizer subgroup of $\mathrm{BrPic}(n\vect)$. 

\item[-] Note that the group of the invertible objects in $\Sigma\FZ_1(\CA)$ is naturally embedded in that of the braided auto-equivalences of $\FZ_1(\CA)$. We denote its image by $\mathrm{\Aut}_0^{\EE_2}(\FZ_1(\CA))$. 

\enu
A detailed study of this topics and its relations to minimal modular extension (see \cite[Section\ 4.3]{KZ24}) will be given elsewhere. 
\end{rem}


The center functor $\SZ: \ST\SO_{\mathrm{gp-wall}}^{n+1} \to {}^{\mathrm{af}}\ST\SO_{\mathrm{af-gp-wall}}^{n+2}$ is neither full nor faithful in general. 
However, the center functor $\SZ$ can be modified to give a full functor if we enlarge the domain category $\ST\SO_{\mathrm{gp-wall}}^{n+1}$ to $\ST\SO_{\mathrm{wall}}^{n+1}$ by allowing the 1-morphisms $\SM^n$ and $\SN^n$ to be gapless. Indeed, for a generic type-I gapped domain wall $\SK^{n+1}$ between $\SZ(\SA)^{n+2}$ and $\SZ(\SB)^{n+2}$, the associated defect junction $\SM^n$ is gapless in general (see the first picture in (\ref{pic:TWR})). 
\bnu
\item[-] According to \cite{KZ18b,KZ20,KZ21}, $\SM^n$ is holographically dual (via topological Wick rotation) to a pair $(\ST^{n+1},\SX^n)$, where $\ST^{n+1}$ is an anomaly-free $n+$1D simple topological order and $\SX^n$ is a gapped defect junction among $\SA^{n+1}\boxtimes \ST^{n+1}$, $\SK^{n+1}$ and $\SB^{n+1}$, as illustrated below. 
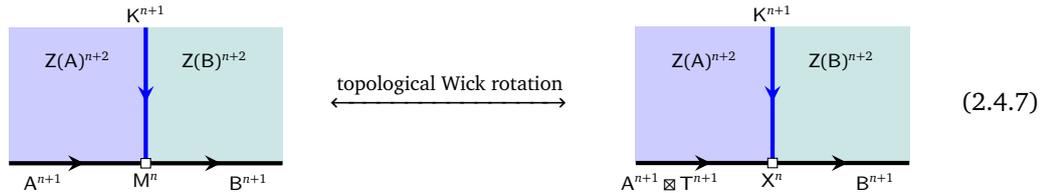
\be \label{pic:TWR}
\begin{array}{c}
\begin{tikzpicture}[scale=0.9]
\fill[blue!20] (-2,0) rectangle (0,2) ;
\fill[teal!20] (0,0) rectangle (2,2) ;
\draw[->-,ultra thick] (-2,0)--(0,0) node[near start,below] {\scriptsize $\SA^{n+1}$} ;
\draw[->-,ultra thick] (0,0)--(2,0) node[near end,below] {\scriptsize $\SB^{n+1}$} ;

\draw[blue,->-,ultra thick] (0,2)--(0,0) ; 
\draw[fill=white] (-0.07,-0.07) rectangle (0.07,0.07) ; 
%
%
\node at (-1,1.5) {\scriptsize $\SZ(\SA)^{n+2}$} ;
\node at (1,1.5) {\scriptsize $\SZ(\SB)^{n+2}$} ;
\node at (0,-0.25) {\scriptsize $\SM^n$} ;
\node at (0,2.2) {\scriptsize $\SK^{n+1}$} ;


\end{tikzpicture}
\end{array}
\quad \xleftrightarrow{\mbox{\footnotesize topological Wick rotation}} \quad
\begin{array}{c}
\begin{tikzpicture}[scale=0.9]
\fill[blue!20] (-2,0) rectangle (0,2) ;
\fill[teal!20] (0,0) rectangle (2,2) ;
\draw[->-,ultra thick] (-2,0)--(0,0) node[near start,below] {\scriptsize $\SA^{n+1}\boxtimes \ST^{n+1}$} ;
\draw[->-,ultra thick] (0,0)--(2,0) node[near end,below] {\scriptsize $\SB^{n+1}$} ;

\draw[blue,->-,ultra thick] (0,2)--(0,0) ; 
\draw[fill=white] (-0.07,-0.07) rectangle (0.07,0.07) ; 
%
%
\node at (-1,1.5) {\scriptsize $\SZ(\SA)^{n+2}$} ;
\node at (1,1.5) {\scriptsize $\SZ(\SB)^{n+2}$} ;
\node at (0,-0.25) {\scriptsize $\SX^n$} ;
\node at (0,2.2) {\scriptsize $\SK^{n+1}$} ;


\end{tikzpicture}
\end{array}
\ee
Note that the gapped wall $\SK^{n+1}$ is necessarily Morita equivalent to $\overline{\SA^{n+1}}\boxtimes \overline{\ST^{n+1}} \boxtimes \SB^{n+1}$ for an anomaly-free $\ST^{n+1}$. 

\item[-] Alternatively, using Theorem$^{\mathrm{ph}}$\,\ref{pthm:TME-action_transitive}, one can always find the anomaly-free topological order $\ST^{n+1}$, which is unique up to Morita equivalences, such that $\SK^{n+1}$ is Morita equivalent to $\overline{\SA^{n+1}}\boxtimes \overline{\ST^{n+1}} \boxtimes \SB^{n+1}$.

\enu
This result, together with Theorem\,\ref{thm:center-functor_surjective_full_faithful} and \ref{thm:center-functor_surjective_full_faithful_2}, leads us to the following result, which is useful in later sections. 
\begin{pthm} \label{pthm:Z1map-made-1to1}
The map $\SZ^{(1)}: \SX^n \mapsto \SZ^{(1)}(\SX)^{n+1}$ (or equivalently, ${}_{\CA \boxtimes \CT}\CX_\CB \mapsto \Fun_{\CA\boxtimes \CT|\CB}(\CX,\CX)$) defines a one-to-one correspondence between: 
\bnu

\item[(1)] $(\CA \boxtimes \CT)$-$\CB$-bimodules modulo the equivalence relation: $\CX\sim \CY$ if $\CY \simeq \CX \boxtimes \CV$ for $\CV \in (n+1)\vect^\times$; 

\item[(2)] type-I gapped walls between $\SZ(\SA)^{n+2}$ and $\SZ(\SB)^{n+2}$ in the Morita class $[\overline{\SA^{n+1}} \boxtimes \overline{\ST^{n+1}} \boxtimes \SB^{n+1}]$. 

\enu
The wall is simple if and only if the associated $(\CA \boxtimes \CT)$-$\CB$-bimodule is indecomposable. When $\SB^{n+1}=\mathbf{1}^{n+1}$, the slightly modified map\footnote{Flipping the orientation of $\SZ^{(1)}(\SX)^{n+1}$ is necessary for the boundary to have the orientation induced from that of the bulk.}
\be \label{eq:SX-CX-mapsto-a-boundary}
\SX^n \mapsto \overline{\SZ^{(1)}(\SX)^{n+1}} \quad\quad \mbox{or equivalently} \quad\quad
{}_{\CA \boxtimes \CT}\CX \mapsto \Fun_{\CA\boxtimes \CT}(\CX,\CX)^\op
\ee
defines a one-to-one correspondence between left $(\CA \boxtimes \CT)$-modules modulo the equivalence relation $\sim$ and gapped boundaries of $\SZ(\SA)^{n+2}$ in the Morita class $[\SA^{n+1} \boxtimes \ST^{n+1}]$. 
\end{pthm}

\begin{rem}
Note that the $n=1$ case of Theorem$^{\mathrm{ph}}$\,\ref{pthm:Z1map-made-1to1} recovers the well-known one-to-one correspondence between left indecomposable $\CA$-modules and simple gapped boundaries of the 3D non-chiral simple topological order $\SZ(\SA)^3$ \cite{KK12,DMNO13,Kon14e}. For $n\geq 1$, it provides a powerful tool to classify gapped boundaries of $\SZ(\SA)^{n+2}$ (or Lagrangian $\EE_1$- or $\EE_2$-algebras, see Corollary\,\ref{pcor:4D_gapped boundary_classification} and Corollary\,\ref{pcor:4D_Lagrangian_1to1_modules}). 
\end{rem}

\begin{rem}
Note that it is possible that $\CX \simeq \CX \boxtimes \CV$ for $\CV \in (n+1)\vect^\times$. For example, when $\CA=\CT=\CB=n\vect$, consider $\CV \in (n+1)\vect^\times$ such that $\CV^{\boxtimes m}\simeq n\vect$ (e.g., $(\Sigma\ising)^{\boxtimes 16} \simeq 3\vect$). We set $\CX := \oplus_{i=0}^{m-1} \CV^{\boxtimes i}$. Then we have $\CX \boxtimes \CV \simeq \CX$.  
\end{rem}


\begin{cor} \label{cor:A_connected_classification_bdy_Z(A)}
When $\CA$ is a fusion $n$-category and is connected as a separable $n$-category, i.e., $\CA=\Sigma\Omega\CA$, there is a one-to-one correspondence between 
\bnu
\item[(1)] braided monoidal functors $\Omega\CA\to \CO$ for a non-degenerate braided fusion $(n-1)$-category $\CO$; 

\item[(2)] simple gapped boundaries of $\SZ(\SA)^{n+2}$. 

\enu
Moreover, the Morita class of the boundary is precisely the Morita class $[\CA\boxtimes \Sigma\CO^\op]$. 
\end{cor}
\pf
Let $\CT$ be a non-degenerate fusion $n$-category $\CT$. An indecomposable left $\CA \boxtimes \CT$-module is again indecomposable as a separable $n$-category, i.e., $\Sigma \CQ$ for a random choice of fusion $(n-1)$-category $\CQ$ within the same Morita class. 
Therefore, an indecomposable left $\CA \boxtimes \CT$-module $\Sigma\CQ$, which is defined by a monoidal functor $\CA\boxtimes\CT \to \FZ_0(\Sigma\CQ)\simeq \Sigma\FZ_1(\CQ)$, determines a braided monoidal functor $\Omega\CA \boxtimes \Omega\CT \to \FZ_1(\CQ)$, which further induces a braided monoidal functor $\Omega\CA \to \FZ_2(\Omega\CT, \FZ_1(\CQ))$ by the universal property of the $\EE_2$-centralizer, where $\FZ_2(\Omega\CT, \FZ_1(\CQ))$ is non-degenerate as a braided fusion $(n-1)$-category. 

Conversely, given a braided monoidal functor $f: \Omega\CA\to \CO$ for a non-degenerate braided fusion $(n-1)$-category $\CO$ within the Witt class $[\Omega\CT]^{-1}$, we obtain a monoidal functor 
$$
\CA \boxtimes \CT \xrightarrow{\Sigma f \boxtimes 1} \Sigma\CO \boxtimes \CT \simeq  \FZ_0(\CM), \quad\quad \mbox{for some $\CM \in (n+1)\vect$}, 
$$
which further defines an indecomposable $\CA\boxtimes\CT$-module $\CM$ precisely up to the equivalence relation $\sim$ by Theorem\,\ref{thm:center-functor_surjective_full_faithful_2} (3). Moreover, we have an equality of the Morita class $[\CA\boxtimes\CT]=[\CA\boxtimes \Sigma\CO^\op]$. 
\epf

\begin{rem}
Although we do have an explicit formula (recall (\ref{eq:SX-CX-mapsto-a-boundary})) for the gapped boundary associated to the braided monoidal functor $f: \Omega\CA \to \CO$, it does not reveal the meaning of $f$ directly. When we identify each simple gapped boundary with a Lagrangian $\EE_2$-algebra (see Corollary$^{\mathrm{ph}}$\,\ref{cor:LA_n+2D_finite_gauge_theory}), we see that the braided monoidal functor $f: \Omega\CA \to \CO$ precisely defines a Lagrangian $\EE_2$-algebra in $\FZ_1(\CA)$ because we have $\FZ_1(\CA) \simeq \FZ_1(\Sigma\Omega\CA) \simeq \Mod_{\Omega\CA}^{\EE_2}(n\vect)^\op$ (see (\ref{eq:E2_Z1_LMod})). 
\end{rem}

\subsubsection{Generalized center functors}

In the center functor defined in Theorem$^{\mathrm{ph}}$\,\ref{pthm:bbr}, the image of the functor are all non-chiral topological orders in the bulk. In order to include all chiral topological orders, we can either allow the boundary to be gapless, or somewhat `equivalently', consider a relative version of the center functor. 

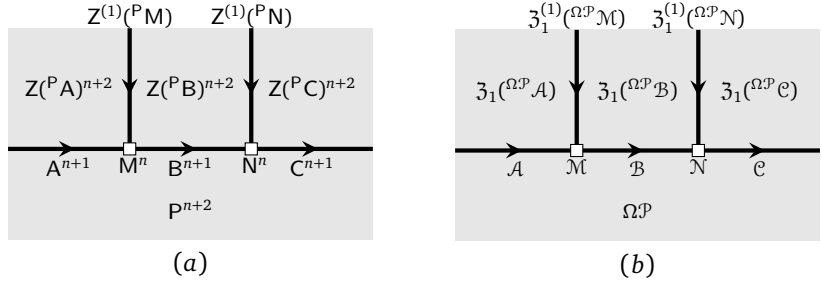
\begin{figure}[htbp]
$$
\begin{array}{c}
\begin{tikzpicture}[scale=0.8]
\fill[gray!20] (-3,-0.5) rectangle (3,3) ;
\draw[ultra thick,->-] (1,3)--(1,1) ;
\draw[ultra thick,->-] (-1,3)--(-1,1) ;
\draw[ultra thick,->-] (-3,1)--(-1,1) ;
\draw[ultra thick,->-] (-1,1)--(1,1) ;
\draw[ultra thick,->-] (1,1)--(3,1) ;
\draw[fill=white] (-1.1,0.9) rectangle (-0.9,1.1) node[midway,below] {\footnotesize $\hspace{1mm}\SM^n$} ;
\node at (-2,0.7) {\footnotesize $\SA^{n+1}$} ;
\node at (0,0.7) {\footnotesize $\SB^{n+1}$} ;
\node at (2,0.7) {\footnotesize $\SC^{n+1}$} ;
\draw[fill=white] (0.9,0.9) rectangle (1.1,1.1) node[midway,below] {\footnotesize $\hspace{1mm}\SN^n$} ;
\node at (-2,2) {\footnotesize $\SZ({}^\SP\SA)^{n+2}$} ;
\node at (0,2) {\footnotesize $\SZ({}^\SP\SB)^{n+2}$} ;
\node at (2,2) {\footnotesize $\SZ({}^\SP\SC)^{n+2}$} ;
\node at (-1,3.2) {\footnotesize $\SZ^{(1)}({}^\SP\SM)$} ;
\node at (1,3.2) {\footnotesize $\SZ^{(1)}({}^\SP\SN)$} ;
\node at (0,0) {\footnotesize $\SP^{n+2}$} ;
\end{tikzpicture} \\
(a)
\end{array}
\quad\quad
\begin{array}{c}
\begin{tikzpicture}[scale=0.8]
\fill[gray!20] (-3,-0.5) rectangle (3,3) ;
\draw[ultra thick,->-] (1,3)--(1,1) ;
\draw[ultra thick,->-] (-1,3)--(-1,1) ;
\draw[ultra thick,->-] (-3,1)--(-1,1) ;
\draw[ultra thick,->-] (-1,1)--(1,1) ;
\draw[ultra thick,->-] (1,1)--(3,1) ;
\draw[fill=white] (-1.1,0.9) rectangle (-0.9,1.1) node[midway,below] {\footnotesize $\CM$} ;
\node at (-2,0.7) {\footnotesize $\CA$} ;
\node at (0,0.7) {\footnotesize $\CB$} ;
\node at (2,0.7) {\footnotesize $\CC$} ;
\draw[fill=white] (0.9,0.9) rectangle (1.1,1.1) node[midway,below] {\footnotesize $\CN$} ;
\node at (-2,2) {\footnotesize $\FZ_1({}^{\Omega\CP}\CA)$} ;
\node at (0,2) {\footnotesize $\FZ_1({}^{\Omega\CP}\CB)$} ;
\node at (2,2) {\footnotesize $\FZ_1({}^{\Omega\CP}\CC)$} ;
\node at (-1,3.2) {\footnotesize $\FZ_1^{(1)}({}^{\Omega\CP}\CM)$} ;
\node at (1,3.2) {\footnotesize $\FZ_1^{(1)}({}^{\Omega\CP}\CN)$} ;
\node at (0,0) {\footnotesize $\Omega\CP$} ;
\end{tikzpicture} \\
(b)
\end{array}
$$
\caption{a relative version of boundary-bulk relation
}
\label{fig:generalized_functoriality_bbr}
\end{figure}

\medskip
Consider the physical configuration depicted in Figure\,\ref{fig:generalized_functoriality_bbr} (a). We use $\SZ({}^\SP\SA)^{n+2}$ to denote the bulk of $\SA^{n+1}$ relative to $\SP^{n+2}$. 
The notation ${}^\SP\SA$ does not have any extra meaning other than the pair $(\SP,\SA)$. 
However, this notation has some advantages in addition to being short. It reminds us that $\SP^{n+2}$ and $\SA^{n+1}$ have different dimensions and $\SP^{n+2}$ acts on $\SA^{n+1}$ in some sense. Similarly, the notation ${}^{\Omega\CP}\CA$ also does not have any extra meaning other than the pair $(\Omega\CP,\CA)$. Although it might cause confusion because we also use ${}^{\Omega\CP}\CA$ to represent an $\Omega\CP$-enriched category only in Section\,\ref{sec:gapless_condensation_3D}, this confusion is harmless and is guaranteed to give the same result by the idea of topological Wick rotation \cite{KZ18b,KZ20,KZ21,KZ22b}.  

\begin{pthm} \label{pthm:generalized_bbr}
For type-I topological orders $\SA^{n+1}$, the assignment $\SA^{n+1} \mapsto \SZ({}^\SP\SA)^{n+2}$ and $\SM^n \mapsto \SZ^{(1)}({}^\SP\SM)$ is functorial in the sense that the following identity holds. 
$$
\SZ^{(1)}({}^\SP\SN)^{n+1} \boxtimes_{\SZ({}^\SP\SB)} \SZ^{(1)}({}^\SP\SM)^{n+1} =
\SZ^{(1)}({}^\SP(\SM\boxtimes_\SB\SN))^{n+1}. 
$$
Mathematically, this amount to the following assignment (see Figure\,\ref{fig:generalized_functoriality_bbr} (b)): 
$$
\CA \mapsto \FZ_1({}^{\Omega\CP}\CA) := \FZ_2(\Omega\CP^\op, \FZ_1(\CA)), \quad\quad
\CM \mapsto \FZ_1^{(1)}({}^{\Omega\CP}\CM) := \Fun_{\CA\boxtimes_{\Omega\CP}\CB^\op}(\CM,\CM), 
$$
which defines a relative version of the center functor, well-defined due to the monoidal equivalence: 
$$
\Fun_{\CB\boxtimes_{\Omega\CP}\CC^\op}(\CN,\CN) \boxtimes_{\FZ_2(\Omega\CP^\op, \FZ_1(\CB))}\Fun_{\CA\boxtimes_{\Omega\CP}\CB^\op}(\CM,\CM) \simeq 
\Fun_{\CA\boxtimes_{\Omega\CP}\CC^\op}(\CM\boxtimes_\CB\CN,\CM\boxtimes_\CB\CN).  
$$
We have not specified the (co-)domain category of this functor. We leave it to intrigued readers. 
\end{pthm}

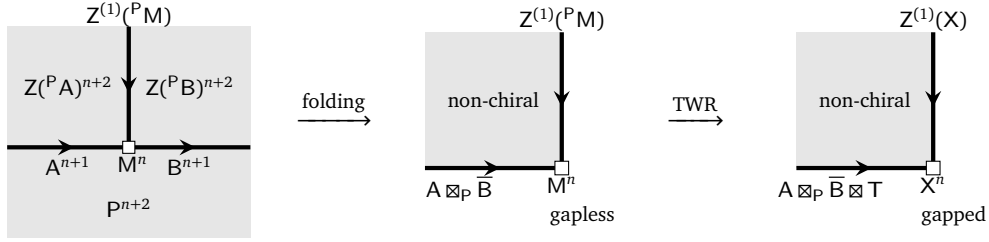
\begin{figure}[htbp]
$$
\begin{array}{c}
\begin{tikzpicture}[scale=0.8]
\fill[gray!20] (-3,-0.5) rectangle (1,3) ;
\draw[ultra thick,->-] (-1,3)--(-1,1) ;
\draw[ultra thick,->-] (-3,1)--(-1,1) ;
\draw[ultra thick,->-] (-1,1)--(1,1) ;
\draw[fill=white] (-1.1,0.9) rectangle (-0.9,1.1) node[midway,below] {\footnotesize $\hspace{1mm}\SM^n$} ;
\node at (-2,0.7) {\footnotesize $\SA^{n+1}$} ;
\node at (0,0.7) {\footnotesize $\SB^{n+1}$} ;
\node at (-2,2) {\footnotesize $\SZ({}^\SP\SA)^{n+2}$} ;
\node at (0,2) {\footnotesize $\SZ({}^\SP\SB)^{n+2}$} ;
\node at (-1,3.2) {\footnotesize $\SZ^{(1)}({}^\SP\SM)$} ;
\node at (-1,0) {\footnotesize $\SP^{n+2}$} ;
\end{tikzpicture} 
\end{array}
\quad\xrightarrow{\mbox{\footnotesize folding}} \quad
\begin{array}{c}
\begin{tikzpicture}[scale=0.9]
\fill[gray!20] (-3,1) rectangle (-1,3) ;
\draw[ultra thick,->-] (-1,3)--(-1,1) ;
\draw[ultra thick,->-] (-3,1)--(-1,1) ;
\draw[fill=white] (-1.1,0.9) rectangle (-0.9,1.1) node[midway,below] {\footnotesize $\SM^n$} ;
\node at (-2.5,0.7) {\footnotesize $\SA\boxtimes_\SP\overline{\SB}$} ;
\node at (-0.7,0.3) {\footnotesize gapless} ;
\node at (-2,2) {\footnotesize non-chiral} ;
\node at (-1,3.2) {\footnotesize $\SZ^{(1)}({}^\SP\SM)$} ;
\end{tikzpicture} 
\end{array}
\quad \xrightarrow{\mbox{\footnotesize TWR}} \quad
\begin{array}{c}
\begin{tikzpicture}[scale=0.9]
\fill[gray!20] (-3,1) rectangle (-1,3) ;
\draw[ultra thick,->-] (-1,3)--(-1,1) ;
\draw[ultra thick,->-] (-3,1)--(-1,1) ;
\draw[fill=white] (-1.1,0.9) rectangle (-0.9,1.1) node[midway,below] {\footnotesize $\SX^n$} ;
\node at (-2.5,0.7) {\footnotesize $\SA\boxtimes_\SP\overline{\SB} \boxtimes \ST$} ;
\node at (-0.7,0.3) {\footnotesize gapped} ;
\node at (-2,2) {\footnotesize non-chiral} ;
\node at (-1,3.2) {\footnotesize $\SZ^{(1)}(\SX)$} ;
\end{tikzpicture} 
\end{array}
$$
\caption{reduce a wall to a boundary by the folding trick followed by a TWR
}
\label{fig:chiral_TO_folding_TWR}
\end{figure}

It is possible to generalize Theorem\,\ref{thm:center-functor_surjective_full_faithful} or Theorem\,\ref{thm:center-functor_surjective_full_faithful_2} to the generalized center functor defined in Theorem$^{\mathrm{ph}}$\,\ref{pthm:generalized_bbr}. For the purpose of studying condensations in this work, however, we do not gain any truely new result because the folding trick reduces the problem of constructing or classifying gapped domain walls (or 1-codimensional defect condensations in later sections) between two Morita equivalent chiral topological orders to the problem of constructing or classifying gapped boundaries of a non-chiral topological order. More precisely, applying the folding trick to the physical configuration in the first picture in Figure\,\ref{fig:chiral_TO_folding_TWR}, followed by a topological Wick rotation (TWR), we obtain a reformulation of Theorem$^{\mathrm{ph}}$\,\ref{pthm:Z1map-made-1to1} for chiral topological orders as summarized below. 

\begin{pthm} \label{pcor:chiral_TO_wall_1to1}
For an anomaly-free simple $n+$1D topological order $\ST^{n+1}$, there is a one-to-one correspondence between 
\bnu
\item[(1)] left $((\CA\boxtimes_{\Omega\CP}\CB^\op) \boxtimes \CT)$-modules modulo the equivalence relation: $\CX \sim \CY$ if $\CY \simeq \CX \boxtimes \CV$ for $\CV \in (n+1)\vect^\times$; 
\item[(2)] type-I gapped walls between two (potentially chiral) simple topological orders $\SZ({}^\SP\SA)^{n+2}$ and $\SZ({}^\SP\SB)^{n+2}$ in the Morita class $[(\SB \boxtimes_{\SP} \overline{\SA})^{n+1} \boxtimes \overline{\ST^{n+1}}]$
\enu 
defined by $\CX \mapsto \Fun_{\CA\boxtimes_{\Omega\CP}\CB^\op \boxtimes \CT}(\CX,\CX)$. The gapped wall is simple, i.e., $\Fun_{\CA\boxtimes_{\Omega\CP}\CB^\op}(\CX,\CX)$ is a fusion $n$-category, if and only if the left $((\CA\boxtimes_{\Omega\CP}\CB^\op) \boxtimes \CT)$-module $\CX$ is indecomposable. 
\end{pthm}

\begin{rem} \label{rem:classify_wall=1codim_condensation_chiral}
We show in Section\,\ref{sec:general_example_1-codim} that a type-I gapped domain wall between $\SC^{n+2}$ and $\SD^{n+2}$ (more precisely, $\Fun_{\CA\boxtimes_{\Omega\CP}\CB^\op \boxtimes \CT}(\CX,\CX)^\op$) is precisely an indecomposable condensable $\EE_1$-algebra in $\CC$, which defines a 1-codimensional defect condensation from $\SC^{n+2}$ to $\SD^{n+2}$. Therefore, Theorem$^{\mathrm{ph}}$\,\ref{pcor:chiral_TO_wall_1to1} is also a classification of 1-codimensional defect condensations from $\SC^{n+2}$ to $\SD^{n+2}$. 
\end{rem}

\begin{rem} \label{rem:P=C=A_B=K}
In applications, one can choose a convenient $\SP^{n+2}$ in order to simplify the calculation. For example, if we already know an example of gapped domain wall $\SK^{n+1}$ between $\SC^{n+2}=\SZ({}^\SP\SA)^{n+2}$ and $\SD^{n+2}=\SZ({}^\SP\SB)^{n+2}$. We can choose $\SP^{n+2}=\SC^{n+2}$, $\SA^{n+1}=\SC^{n+1}$ and $\SB^{n+1}=\SK^{n+1}$. Then we obtain a one-to-one correspondence, defined by $\CX \mapsto \Fun_{\CK^\op \boxtimes \CT}(\CX,\CX)$, between 
\bnu
\item[(1)] left $(\CK^\op\boxtimes \CT)$-modules modulo the equivalence relation: $\CX \sim \CY$ if $\CY \simeq \CX \boxtimes \CV$ for $\CV \in (n+1)\vect^\times$; 
\item[(2)] type-I gapped walls between two (potentially chiral) simple $\SC^{n+2}$ and $\SD^{n+2}$ in the Morita class $[\SK^{n+1} \boxtimes \overline{\ST^{n+1}}]$. 
\enu 
\end{rem}

\begin{rem}
Theorem$^{\mathrm{ph}}$\,\ref{pcor:chiral_TO_wall_1to1} and Remark\,\ref{rem:P=C=A_B=K} provide a powerful tool to construct and classify of 1-codimensional defect condensations between two chiral topological orders in higher dimensions. Unfortunately, we are not able to provide any concrete applications of this result in $n+$1D for $n>3$ in this work because we do not know the categorical description of any non-trivial chiral topological orders in $n+$1D for $n>3$. 
\end{rem}

\begin{rem}
Theorem$^{\mathrm{ph}}$\,\ref{pthm:bbr} is enough for the purpose of including chiral topological orders in the image of the center functor. However, it is even possible to further enlarge both the domain category and the codomain category to obtain an even more general center functor by allowing $\SP^{n+1}$ to vary for different $\SA^{n+1}$, $\SB^{n+1}$ and $\SM^n$ or even slightly beyond that. This further generalization is essentially equivalent to allowing $\SA^{n+1}$, $\SB^{n+1}$ and $\SM^n$ to be gapless. Indeed, in \cite[Theorem\ 4.15]{KZ21}, we introduced this generalized functor for $n=2$, which can be generalized to higher dimensions tautologically by restricting to only the topological skeletons \cite{KZ22b}. In particular, the illustrating picture of this functor in Figure 10 in \cite{KZ22} works for all dimensions tautologically. Moreover, it is even possible to further generalize the center functor such that the bulk phases $\SZ(\SA)^{n+2}$ and $\SZ(\SB)^{n+2}$ are also gapless \cite{KZ22b}. We do not need these generalized center functors in this work. We will come back to this issue elsewhere. 
\end{rem}

\newpage


\section{Condensations of 1-codimensional topological defects} \label{sec:condense_1-codim_defect}

In this section, we study the condensations of 1-codimensional topological defects. We start from reviewing the theory of the particle condensations in 1+1D (potentially anomalous) topological orders via an algebraic bootstrap approach \cite{Kon14e} in Section\,\ref{sec:1d_cond_alg}; then we redo it via a geometric approach \cite{KS11a,FSV13} which is ready to be generalized to higher dimensional topological orders in Section\,\ref{sec:geometric_intuition_1codim}; then we provide the general theory of the condensations of 1-codimensional topological defects for topological orders in all dimensions in Section\,\ref{sec:1codim_nd}.

\subsection{Particle condensations in 1+1D: algebraic approach} \label{sec:1d_cond_alg}
In this subsection, we briefly review the bootstrap arguments in \cite{Kon14e} that lead to the particle condensation theory for potentially anomalous 1+1D topological orders.

\subsubsection{Algebraic bootstrap} 
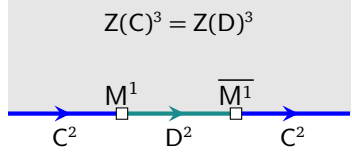
\begin{figure}[htbp]
$$
\begin{tikzpicture}[scale = 1.5]
\fill[gray!20] (-1,0) rectangle (2,1) ;
\draw[color = blue, ultra thick,->-](-1,0)--(0,0);
\draw[color = teal!90, ultra thick,->-](0,0)--(1,0);
\draw[color = blue, ultra thick,->-](1,0)--(2,0);
\draw[fill=white] (-0.05,-0.05) rectangle (0.05,0.05) node[midway,above,scale=1] {$\SM^1$} ;
\draw[fill=white] (0.95,-0.05) rectangle (1.05,0.05) node[midway,above,scale=1] {$\overline{\SM^1}$} ;
\node[black] at(0.5,-0.2) {\small $\SD^2$};
\node[black] at(-0.5,-0.2) {\small $\SC^2$};
\node[black] at(1.5,-0.2) {\small $\SC^2$};
\node[black] at (0.5,0.8) {\small $\SZ(\SC)^3=\SZ(\SD)^3$}; 
\end{tikzpicture}
$$
\caption{a condensation in the topological order $\SC^2$}
\label{fig:1d-cond}
\end{figure}

Consider a 1+1D (potentially anomalous) simple topological order $\SC^2$ as depicted in only the spatial dimension in Figure\,\ref{fig:1d-cond}. Its particles form a fusion 1-category $\CC$ (recall Theorem$^{\mathrm{ph}}$\,\ref{pthm:TO_multi-fusion_ncat}). Now we consider a particle condensation happening in a large but connected  region and producing a new 1+1D topological order $\SD^2$, whose particles form a new fusion 1-category $\CD$ (see Figure\,\ref{fig:1d-cond}), and two gapped domain wall $\SM^1$ and $\overline{\SM^1}$, where $\overline{\SM^1}$ can be viewed as the time reversal of $\SM$. The bulk $\SZ(\SC)^2$ remains unchanged. 

Since $\SD^2$ is obtained from a condensation of $\SC^2$, particles in $\CD$ necessarily come from those in $\CC$, i.e. $\ob(\CD) \subset \ob(\CC)$. The hom space $\hom_\CC(a,b)$ for $a,b\in \CC$ encodes the information of how many independent channels of fusing or splitting $a$ into $b$. It is clear that the fusion-splitting channels between particles in $\CD$ necessarily come from those between particles in $\CC$, i.e. $\hom_\CD(a,b) \subset \hom_\CC(a,b)$. As a consequence, $\CD$ is necessarily a sub-category of $\CC$, i.e. $\CD\subset \CC$. In particular, the vacuum or the trivial particle in $\CD$ is necessarily a (composite) particle $A \in \CC$.

A particle condensation is triggered by introducing interactions among particles. For example, for $a,b\in \CD \subset \CC$, the consequence of introducing interactions between particles is that a sub-Hilbert space of the Hilbert space associated to $a\otimes_\CC b$ becomes energy favorable. This sub-Hilbert space gives a physical definition of the fusion product $a\otimes_\CD b$ in the condensed phase $\SD$. In other words, the condensation 
process produces a family of projections (called condensation maps): 
$$
a \otimes_\CC b \xrightarrow{p_{a,b}} a\otimes_\CD b, \quad\quad \forall a,b\in \CD. 
$$
In particular, we have the following condensation maps: 
\begin{align*}
&\mu_A: A\otimes_\CC A \xrightarrow{p_{A,A}} A\otimes_\CD A \simeq A.  \\
&\forall x\in \CD, \quad \quad \mu_x^L: A\otimes_\CC x \xrightarrow{p_{A,x}} A\otimes_\CD x \simeq x, \quad\quad \mu_x^R: x\otimes_\CC A \xrightarrow{p_{x,A}} x\otimes_\CD A \simeq x. 
\end{align*}
Moreover, since the vacuum in $\CC$ should condense into the vacuum in $\CD$, we expect to have a morphism $\eta_A: \one_\CC \to A$. The triple $(A,\mu_A,\eta_A)$ defines an algebra or an $\EE_1$-algebra in $\CC$ and the triple $(x,\mu_x^L,\mu_x^R)$ defines an $A$-$A$-bimodule in $\CC$. Moreover, $\mu_A$ splits as an $A$-$A$-bimodule map due to a natural condition on the stability of the vacuum of the condensed phase $\SD^2$ \cite{Kon14e}. Namely, there exists an $A$-$A$-bimodule map $e_A: A \to A\otimes A$ such that $\mu_A \circ e_A = \id_A$, which is the defining condition of a condensable (or separable) $\EE_1$-algebra in $\CC$. In the unitary case, $e_A$ can be chosen to be $\mu_A^\dagger$ and $A$ is naturally a special $\dagger$-Frobenius algebra. If $\dim \hom_\CC(\one_\CC, A)=1$, the condensable $\EE_1$-algebra $A$ is called simple. If a condensable $\EE_1$-algebra $A$ can not be decomposed into a direct sum of algebras, then $A$ is called indecomposable. In general, a condensable $\EE_1$-algebra $A$ in $\CC$ is a direct sum of indecomposable ones. We postpone the structure theorem of a condensable $\EE_1$-algebra to Section\,\ref{sec:internal_hom} after we review the notion of an internal hom. 

\begin{rem} \label{rem:E1-algebras_in_C}
As we explained in Section\,\ref{sec:E1_algebra}, $\CC \in \Alg_{\EE_1}(2\vect)$ and there are infinite number of tensor products $\otimes_{(r,s)}$ on $\CC$, which can be reduced to a single fusion product $\otimes$ (and $\otimes^\op$) due to the Adiabatic Principle (\ref{Adiabatic_Principle}). Therefore, there is also infinitely many multiplication maps $A\otimes_{(r,s)} A \to A$, which endows $A$ with the structure of an $\EE_1$-algebra in $\CC$. In this case, an $\EE_1$-algebra in $\CC$ is precisely an associatvie algebra in $\CC$. 
We use $\Algc_{\EE_1}(\CC)$ to denote the category of condensable $\EE_1$-algebras in $\CC$. 

Similarly, the infinitely many $A$-action on $x\in \CD$: $x\otimes_{(r,s)} A \to x$ endows $A$ with a structure of an $\EE_1$-$A$-module. In this case, an $\EE_1$-$A$-module in $\CC$ is precisely an $A$-$A$-bimodule in $\CC$. We denote the 1-category of $\EE_1$-$A$-modules (as objects) and $\EE_1$-$A$-modules maps (as 1-morphisms) by $\Mod_A^{\EE_1}(\CC)$, which can be identified with $\BMod_{A|A}(\CC)$, i.e. $\Mod_A^{\EE_1}(\CC)=\BMod_{A|A}(\CC)$. 
\end{rem}

The following bootstrap result was obtained in \cite{Kon14e} under the name of `1-d condensation'.  
\begin{pthm} \label{pthm:1d_cond_part1}
Using some obvious physical intuitions and certain stability arguements, one can easily derive the following results. 
\bnu
\item[(1)] The triple $(A,\mu_A,\eta_A)$ defines a condensable $\EE_1$-algebra in $\CC$. 

\item[(2)] For $x\in \CD$, the triple $(x,\mu_x^L,\mu_x^R)$ defines an $A$-$A$-bimodule in $\CC$. 

\item[(3)] $\CD = \Mod_A^{\EE_1}(\CC)$, where $\Mod_A^{\EE_1}(\CC)=\BMod_{A|A}(\CC)$ is the category of $A$-$A$-bimodules or $\EE_1$-$A$-modules in $\CC$  (as objects) and intertwiners (as 1-morphisms). The fusion product $\otimes_\CD$ can be identified with the relative tensor product $\otimes_A$, i.e. $\otimes_\CD = \otimes_A$, and $\one_\CD=A$. If $A=\one_\CC$, then $\CD=\CC$ and the condensation is called a trivial condensation. 

\enu
\end{pthm}

The gapped domain wall $\SM^1$ can also support particles. These particles form a separable 1-category $\CM$. Since particles in $\CC$ can fuse onto the wall from left and particles in $\CD$ can fuse onto the wall from right, $\CM$ is necessarily a left $\CC$-module and a right $\CD$-module. More precisely, a left $\CC$-module is a category $\CM$ equipped with a left $\CC$-action functor $\odot: \CC \times \CM \to \CM$ that is unital and associative, i.e. there exist the following natural isomorphisms:
$$
\one_\CC \odot x \simeq x, \quad\quad a\odot (b \odot x) \simeq (a\otimes_\CC b) \odot x, \quad \forall a,b\in \CC, x\in \CM,
$$
where we have used the following notation $\odot(a,x):= a\odot x$, satsifying some physically natural conditions (see \cite{EGNO15} for a review). A right $\CD$-module is similar. Moreover, the two side actions on $\CM$ are obviously commutive. This means that there are natural isomorphisms 
$$
a\odot (x \odot d) \simeq (a\odot x)\odot d, \quad\quad \forall a\in \CC, x\in \CM, d\in \CD, 
$$
satisfying some natural conditions. These data and conditions are the defining data and axioms of a $\CC$-$\CD$-bimodule category. In summary, $\CM$ is a $\CC$-$\CD$-bimodule in $2\vect$. However, $\CM$ does not catch the complete information of $\SM$. The physical domain wall $\SM$ must specify a particle $m\in\CM$ (as a wall condition) such that fusing particles in $\CC$ (or $\CD$) onto the wall gives a well defined map $\CC \to \CM$ defined by $a \mapsto a\odot m$ for $a\in\CC$, and a well define map $\CD \to \CM$ defined by $d\mapsto m\odot d$ for $d\in \CD$. Therefore, the physical domain wall $\SM$ can be mathematically described by a pair $(\CM, m)$. Such a pair defines a so-called $\EE_0$-algebra in the symmetric monoidal 2-category $2\vect$ \cite{Lur17}. A morphsm between two $\EE_0$-algebras $(\CM,m)$ and $(\CM',m')$ is a functor $f: \CM \to \CM$, together with a morphism $g: f(m) \to m'$. We denote the category of $\EE_0$-algebras in $2\vect$ by $\Alg_{\EE_0}(2\vect)$. 

\begin{rem} \label{rem:E1_module_category}
Similar to Remark\,\ref{rem:E1-algebras_in_C}, there are infinitely many $\CC$-action on $\CM$ given by $\odot_{(r,s)}: \CC \times \CM \to \CM$ for $r,s\in \Rb$ and $r\neq s$. When $\CC=\CD$, the category $\CM$ is naturally equipped with the structure of an $\EE_1$-$\CC$-module, i.e. $\CM \in \Mod_\CC^{\EE_1}(2\vect)=\BMod_{\CC|\CC}(2\vect)$.  
\end{rem}

Since the domain wall $\SM^1$ is also created from a particle condensation of $\SC^2$, the particles and the fusion-splitting chanels in $\CM$ should all come from $\CC$. In other words, $\CM$ must be a subcategory of $\CC$. Moreover, the condensation  again produces a condensation map $x\otimes A \to x\odot A \simeq x$, which clearly defines a right $A$-module structure on $x$. Note that an $A$-$A$-module in $\CC$ is clearly a deconfined particle living in $\CD$, while a right $A$-module in $\CC$ defines a particle that is necessarily confined to the domain wall. Therefore, we obtain $\CM=\RMod_A(\CC)$, where $\RMod_A(\CC)$ denotes the category of right $A$-modules in $\CC$ as objects and intertwiners (of the $A$-action) as morphisms. As a consequence, we have $a\odot x := a\otimes x$ and $x\odot d := x\otimes_A d$ for $x\in\CM$, $a\in \CC$ and $d\in \CD$.
If we do not specify any particle in the condensed phase in $\CD$, any site in the $\SD^2$-phase naturally has the trivial particle $\one_\CD=A$ on it. Similarly, if we do not specify a particle on the wall $\SM^1$, then the trivial particle $A$ naturally lives on it. Therefore, we should have $\SM^1=(\RMod_A(\CC),A)$.

For the other wall $\overline{\SM^1}$, by the similar reason, we should expect that the category of particles on this wall is given by $\CM^\op=\LMod_A(\CC)$. As a $\CD$-$\CC$-bimodule $\CM^\op$, the action is defined by $d\odot m \odot c := c^\ast \odot m \odot d^\ast$, where we have  used convention $d^\ast=d^L=d^R$ as in the unitary case to avoid framing issues. 

\medskip
We summarize the result below. 

\begin{pthm}[\cite{Kon14e}] \label{pthm:1d_cond_part2}
The gapped domain wall $\SM$ can be mathematically characterized by the pair $(\CM,A)$, where $\CM=\RMod_A(\CC)$ and $m\in \CM$. 
\bnu
\item[(1)] Particles in $\CC$ move to the wall $\SM$ according to the map $a \mapsto a\otimes A, \forall a\in \CC$; 
\item[(2)] particles in $\CD$ move to the wall $\SM$ according to the map $d \mapsto A\otimes_A d=d, \forall d\in \CD$. 
\enu
We also have $\overline{\SM} = (\CM^\op,A)$, where $\CM^\op \simeq \LMod_A(\CC)$. 
\end{pthm}

\subsubsection{Internal homs and structure theorem} \label{sec:internal_hom}
These mathematical precise results rely heavily on an important categorical notion called an {\it internal hom}, which is very important to the condensation theory and to all QFTs as well (see \cite{Dav10,DKR15,KYZ21} and Remark\,\ref{rem:RCFT_internal_hom}). For this reason, we first review this notion (see more details in the book \cite{EGNO15}).\footnote{Although it is very tempting to suggest physics oriented readers to skip this subsubsection for their first reading, we want to emphasize that the notion of internal hom is so important and natural that it is precisely the mathematical structure needed to catch many physical intuitions. Since it is so basic, natural and important to QFT's, we are sure that it will become one of the most basic concepts in physics and a powerful tool for the physicists in the next generation.}

\begin{defn}[\cite{Ost03}] \label{def:internal_hom}
Let $\CC$ be a multi-fusion 1-category and $\CM$ be a separable left $\CC$-module\footnote{Our assumptions on $\CC$ and $\CM$ guarantee the existence of the internal homs.}, i.e. a category equipped with a (unital and associative) $\CC$-action $\odot: \CC \times \CM \to \CM$. For $x,y\in \CM$, the internal hom $[x,y]_\CC$ is an object uniquely determined by
the following isomorphisms:  
\be \label{eq:internal_hom_def1}
\hom_\CM(a\odot x, y) \simeq \hom_\CC(a, [x,y]_\CC), \quad\quad \forall a\in \CC, x,y\in \CM
\ee
that are natural in the variable $a$, i.e. the following diagram:
$$
\xymatrix{
\hom_\CM(a'\odot x, y) \ar[r]^{\simeq} \ar[d]_{-\ \circ \ (h\odot \id_x)} & \hom_\CC(a', [x,y]_\CC) \ar[d]^{-\ \circ\ h}  \\
\hom_\CM(a\odot x, y) \ar[r]^{\simeq}  & \hom_\CC(a, [x,y]_\CC)
}
$$
is commutative for all $h: a\to a'$. We sometimes abbreviate $[x,y]_\CC$ to $[x,y]$ for simplicity.
\end{defn}

There is a distinguished morphism $\ev: [x,y]_\CC \odot x \to y$ in $\CM$ defined by the preimage of $\id_{[x,y]_\CC}$ under the isomorphism in (\ref{eq:internal_hom_def1}). 
It allows us to give an equivalent definition of the internal hom via its universal property. Both definitions are useful and important in this work. 
\begin{defn}
The internal hom can be equivalently defined by a pair $([x,y]_\CC, \ev)$, where $\ev: [x,y]_\CC \odot x \to y$ is a morphism in $\CM$, such that it is terminal among all such pairs. That is, given another pair $(F, f)$, where $F\in \CC$ and $f: F\odot x \to y$ is a morphism in $\CM$, there exists a unique morphism $g: F \to [x,y]_\CC$ such that $\ev\circ (g\odot \id_x) = f$. This universal property is often expressed by the following commutative diagram: 
\be \label{diag:internal_hom_universal_property}
\xymatrix{
& [x,y]_\CC \odot x \ar[rd]^{\ev}  & \\
F \odot x \ar@{-->}[ur]^{\exists !\, g\odot \id_x} \ar[rr]^f & & y
}
\ee
where the notation `$\exists$' represents `exists' and the notation `$!$' represents `unique'. 
\end{defn}

\begin{expl}
We give two most useful examples, which also tell us how to compute internal homs.  
\bnu
\item when $\CM=\CC$ and $\odot=\otimes_\CC$, we have $[x,y]_\CC = y\otimes_\CC x^L$; 
\item when $\CM=\RMod_A(\CC)$ and $\odot=\otimes_\CC$, we have 
$[x,y]_\CC = (x\otimes_A y^R)^L$ \cite{Ost03}. It includes the first case as a special case when $A=\one_\CC$. 
\enu
\end{expl}

\begin{rem} \label{rem:internal_hom_functor}
The internal homs define an internal hom functor $[-,-]_\CC: \CM^\op \otimes \CM \to \CC$, i.e. $(x,y) \mapsto [x,y]_\CC$. As a consequence, the isomorphisms in (\ref{eq:internal_hom_def1}) are natural in all three variables $a,x,y$. Moreover, the internal hom can be equivalently defined by stating that $[x,-]: \CM \to \CC$ is the right adjoint of the functor $-\odot x: \CC \to \CM$. 
\end{rem}

What is remarkable about the internal hom is that, by the universal property of the internal hom $[x,z]$, there is a distinguished morphism 
\be \label{eq:ev}
[y,z] \otimes_\CC [x,y] \xrightarrow{\ev} [x,z]
\ee
determined by the composed morphism: $[y,z] \odot ([x,y] \odot x) \xrightarrow{\id_{[y,z]} \otimes_\CC \ev} [y,z] \odot y \xrightarrow{\ev} z$ and the universal property. Moreover, we have a canonical morphism $\eta: \one_\CC \to [x,x]$ defined by $\one_\CC \odot x \simeq x$. 
\begin{lem}
The triple $([x,x],\ev,\eta)$ defines an $\EE_1$-algebra in $\CC$; the pair $([x,y], \ev)$, where $\ev: [x,y]\otimes_\CC [x,x]\to [x,y]$, defines a right $[x,x]$-module structure on $[x,y]$; and $[x,y]$ is naturally a $[y,y]$-$[x,x]$-bimodule. 
\end{lem}

\begin{rem} \label{rem:RCFT_internal_hom}
Intenal homs play very important role in QFT's. For example, in 1+1D rational CFT's, all modular-invariant bulk CFT's, boundary CFT's and defect or wall CFT's are internal homs.   More precisely, given a rational vertex operator algebra (VOA) $V$, i.e. $\CC=\Mod_V$ is a modular tensor category (MTC) \cite{MS89,Hua08}. A category of boundary conditions preserving the chiral symmetry $V$ is defined by an indecomposable separable left $\CC$-module $\CM$, i.e. each object in $\CM$ represents a boundary condition preserving the chiral symmetry $V$. 
\bnu
\item The 0+1D boundary CFT associated to the boundary condition $x\in \CM$ is precisely the internal hom $[x,x]_\CC$. The algebraic structure defined by the triple $([x,x],\ev, \eta)$ precisely encodes the mathematical structure of OPE among boundary fields \cite{FRS02,HK04,KR09,DKR15}. 

\item The 0D domain wall between two boundary CFT's associated to the boundary conditions $x$ and $y$ is given by the internal hom $[x,y]_\CC$ \cite{FFRS07,DKR15}. The morphism defined in  (\ref{eq:ev}) precisely encode the most general OPE between two fields living in two domain walls.

\item The modular-invariant bulk CFT is again the internal hom $[\id_\CM,\id_\CM]_{\FZ_1(\CC)}$ but living in $\FZ_1(\CC)$ (i.e. the Drinfeld center of $\CC$) \cite{FRS02,KR09,Dav10}, where $\id_\CM \in \Fun_\CC(\CM,\CM)$ is the identity functor, and $\Fun_\CC(\CM,\CM)$ is the category of $\CC$-module functors from $\CM$ to $\CM$ and, at the same time, a left $\FZ_1(\CC)$-module defined by $(z\odot F)(-):=z\odot F(-)$ for $z\in \FZ_1(\CC)$ and $F\in  \Fun_\CC(\CM,\CM)$. This internal hom $[\id_\CM,\id_\CM]_{\FZ_1(\CC)}$ is automatically a Lagrangian algebra in $\FZ_1(\CC)$ \cite{KR09,DMNO13}. 

\enu
\end{rem}

The following mathematical theorem tells us the basic structure of a condensable $\EE_1$-algebra in a multi-fusion 1-category $\CA$. 
\begin{thm}[\cite{KZ17}]  \label{thm:structure_theorem_separable_algebra}
A condensable $\EE_1$-algebra $A$ in a multi-fusion 1-category $\CA$ is a direct sum of indecomposable condensable $\EE_1$-algebras, each of which is a matrix algebra. More explicitly, assume $A$ is indecomposable, let $A=\oplus_{i=1}^m x_i$ be the decomposition of the right A-module into simple ones. We have the following results. 
\bnu

\item $A\simeq \oplus_{i,j}^m [x_i, x_j]$; 

\item the functor $[x_i, -]: \RMod_A(\CA) \to \RMod_{[x_i,x_i]}(\CA)$ is an equivalence; 

\item $[x_i, x_i]$ is a simple condensable $\EE_1$-algebra; 

\item $[x_i, x_j]$ is an invertible $[x_j,x_j]$-$[x_i,x_i]$-bimodule; 

\item $[x_j, x_l] \otimes_{[x_j, x_j]} [x_i, x_j] \simeq [x_i, x_l]$ as $[x_l,x_l]$-$[x_i, x_i]$-bimodules. 

\enu
\end{thm}

\begin{rem} \label{rem:structure_theorem}
Up to the standard assumptions about higher $n$-categories and their (co)-limit theory (see \cite{GJF19}), the proof of this theorem in \cite{KZ17} can be generalized to condensable $\EE_1$-algebras in a multi-fusion $n$-category. In this work, we simply take it for granted that condensable $\EE_1$-algebras in a multi-fusion $n$-category as the same structure theorem as the $n=1$ case. 
\end{rem}

Now we slightly generalize the bootstrap results in Theorem$^{\mathrm{ph}}$\,\ref{pthm:1d_cond_part1}  and Theorem$^{\mathrm{ph}}$\,\ref{pthm:1d_cond_part2} for a potentially composite topological order $\SC^2$ (i.e., $\CC$ is a multi-fusion 1-category). 
\begin{pthm} \label{pthm:condensation_type12_1d}
A condensable algebra $A$ in a multi-fusion 1-category $\CC$ is a direct sum of non-zero condensable algebras in each indecomposable summand of $\CC$. 
\bnu
\item $\SC^2$ is type-I (i.e., $\CC$ is an indecomposable multi-fusion 1-category): 
\bnu
\item If $A$ is indecomposable, then the condensed phase $\SD^2$ is type-I and simple (i.e., $\CD$ is a fusion 1-category), and $\CM=\RMod_A(\CC)$ is an indecomposable left $\CC$-module. 
\item If $A$ is a direct sum of indecomposable ones, i.e., $A=\oplus_{i=1}^l A_i$, then the condensed phase $\SD^2$ is type-I, $\mathrm{gsd}(\SD^2)=l$, and $\CD$ is an indecomposable multi-fusion 1-category, which can be decomposed as a direct sum of separable 1-categories: $\CD=\oplus_{i,j=1}^l \BMod_{A_i|A_j}(\CC)$. The domain wall $\SM^1=(\CM,A)$ is of type-II and we have the decomposition of left $\CC$-modules: $\CM = \RMod_A(\CC) = \oplus_{i=1}^l \RMod_{A_i}(\CC)$. 
\enu
\item $\SC^2$ is type-II (i.e., $\CC=\oplus_{i=1}^k \CC_i$ and $\CC_i$ are indecomposable multi-fusion 1-categories): 
\bnu
\item In this case, we have $A=\oplus_{i=1}^k A_i$, where $A_i=A\cap \CC_i \neq 0$. 

\item The condensed phase $\SD^2$ is also type-II, $\mathrm{mult}(\SC^2) \simeq \mathrm{mult}(\SD^2)$, 
and $\CD \simeq \oplus_{i=1}^k \BMod_{A_i|A_i}(\CC_i)$. 
\item The domain wall $\SM^1$ is type-II,  
and we have the decomposition of left $\CC$-modules: $\CM = \RMod_A(\CC) = \oplus_{i=1}^k \RMod_{A_i}(\CC_i)$. 
\enu

\enu
\end{pthm}

\begin{rem}
Note that the number $\mathrm{gsd}(\SC^2)$ is not preserved in a condensation but the number $\mathrm{mult}(\SC^2)$ is preserved. More precisely, when $\SC^2$ is type-I, $\mathrm{gsd}(\SD^2)$ only depends on the number of direct summands of algebras in $A$.  For example, by Proposition\,\ref{prop:structure_theorem_MF}, we have $\CC=\oplus_{i,j}^m \CC_{i,j}$ as separable 1-categories and $\one_\CC=\oplus_{i=1}^m e_i$. If $A=e_1$, then $\CD\simeq \CC_{11}$ and $\mathrm{gsd}(\SD^2)=1$; if $A=e_1\oplus e_2$, then $\CD\simeq \oplus_{i,j=1}^2 \CC_{ij}$ and $\mathrm{gsd}(\SD^2)=2$. 
\end{rem}

\subsubsection{Examples of particle condensations in 1+1D}


Since the only simple 1+1D anomaly-free topological order is the trivial one, we provide examples of condensations in an anomalous 1+1D simple topological order $\SC^2$. In this case, $\CC$ is a fusion 1-category. Physically, such a condensation can be viewed either as a phase transition occurred on the gapped boundary of $\SZ(\SC)^3$ or, by topological Wick rotation or topological holography \cite{KZ18b,KZ20,KZ21,KLWZZ20a,KLWZZ20,KZ22b}, as condensation among 1+1D gapped quantum liquids.

\begin{expl}
The topological defects on the trivial 1-codimensional topological defect in the 2+1D $\Zb_2$-gauge theory form a fusion category $\FZ_1(\Rep(\Zb_2))$. By condensing simple condensable $\EE_1$-algebras in $\FZ_1(\Rep(\Zb_2))$, we obtain all simple 1-codimensional topological defects, which are simple objects in the fusion 2-category $\Sigma \FZ_1(\Rep(\Zb_2))$. This fusion 2-category has six simple objects, denoted by $\one, \vartheta, \mathrm{rr}, \mathrm{rs}, \mathrm{sr}, \mathrm{ss}$ \cite{KZ22a} (recall Example\,\ref{expl:3D_toric_code}).

Since $\FZ_1(\Rep(\Zb_2)) \simeq \vect_{\Zb_2 \times \Zb_2}$, the condensable $\EE_1$-algebras in $\FZ_1(\Rep(\Zb_2))$ are classified by subgroups of $\Zb_2 \times \Zb_2$ and the corresponding cohomology groups. There are four simple objects in $\FZ_1(\Rep(\Zb_2))$ denoted by $1,e,m,f$.
\bnu
\item The tensor unit $1$ is a condensable $\EE_1$-algebra. By condensing $1$ we obtain the trivial 1-codimensional topological defect $\one$.
\item There are three subgroups of $\Zb_2 \times \Zb_2$ which are isomorphic to $\Zb_2$. Since $H^2(\Zb_2;U(1)) = 0$ is trivial, there are three corresponding condensable $\EE_1$-algebra $1 \oplus e, 1 \oplus m,1 \oplus f$. By condensing them we obtain the 1-codimensional topological defects $\mathrm{rr},\mathrm{ss},\vartheta$, respectively.
\item Since $H^2(\Zb_2 \times \Zb_2;U(1)) \simeq \Zb_2$, there are two condensable $\EE_1$-algebra structures on the object $1 \oplus e \oplus m \oplus f$. By condensing them we obtain the 1-codimensional topological defects $\mathrm{rs}$ and $\mathrm{sr}$, respectively.
\enu
\end{expl}

\begin{expl} \label{expl:condensations_in_ising}
Recall the Ising fusion category $\ising$ in Example\,\ref{expl:ising}. According to \cite{FRS02}, all condensable $\EE_1$-algebras $A$ in $\ising$ are Morita equivalent to the trivial $\EE_1$-algebras $\one$. More explicitly, one can show that they are the internal hom algebras: $[x,x]=x\otimes x^\ast, \forall x\in \ising$ (see Section\ \ref{sec:internal_hom} for a brief introduction of this notion). As a consequence, we have $\BMod_{A|A}(\ising) \simeq \ising$ for all condensable $\EE_1$-algebras $A$. In other words, all particle condensations in $\ising$ along a line, which can be viewed as a boundary phase of the 2+1D double Ising topological order, only produces the same and unique boundary phase of the 2+1D double Ising topological order. This fact is compatible with the fact that there is only one Lagrangian algebra in $\FZ_1(\ising)$ \cite{KR08,DMNO13}. 
\end{expl}

\begin{expl} \label{expl:algebras_in_RepG}
Let $G$ be a finite group. Consider the 2+1D finite gauge theory $\SG\ST_G^3$.  
\bnu
\item[(1)] It has a gapped boundary whose topological defects form the fusion 1-category $\Rep(G)$ (when $G = \Zb_2$, this is the smooth boundary of the toric code model). The classification of condensable $\EE_1$-algebras in $\Rep(G)$ is given by Ostrik \cite{Ost03}. A condensable $\EE_1$-algebra in $\Rep(G)$ is a separable algebra $A$ in $\vect$ equipped with a $G$-action (i.e., a group homomorphism $G \to \Aut(A)$), and it is simple if and only if $G$ acts on the set of primitive (minimal) central idempotents of $A$ transitively. We give some simple examples. 
\bnu
\item When we condense $A=\mathrm{Fun}(G)$ in the gapped boundary associated to $\Rep(G)$, we obtain the condensed boundary associated to $\BMod_{A|A}(\Rep(G))\simeq \vect_G$, and the gapped domain wall $\CM=\RMod_A(\Rep(G)) \simeq \vect$. When $G=\Zb_2$, this condensation defines a boundary phase transition from the smooth boundary to the rough boundary.\footnote{According to topological Wick rotation \cite{KZ22b}, this condensation also defines the spontaneous symmetry-breaking from the symmetric phase to the symmetry-broken phase in the 1+1D Ising chain \cite{KWZ22}.} 

\item When we condense $A=\mathrm{Fun}(G/H)$ for a subgroup $H\leq G$ in the gapped boundary associated to $\Rep(G)$, we obtain a condensed boundary whose boundary particles form the fusion category:  
\[
\BMod_{A|A}(\Rep(G))\simeq \Fun(\vect_{G/H}, \vect_{G/H})^G,
\] 
where $\Fun(\vect_{G/H}, \vect_{G/H})^G$ denotes the $G$-equivariantization of $\Fun(\vect_{G/H}, \vect_{G/H})$ \cite{XZ24}.\footnote{According to topological Wick rotation, this condensation defines the spontaneous symmetry-breaking from the symmetric phase to the symmetry-partially-broken (from $G$-symmetry to $H$-symmetry) phase in a 1+1D spin chain \cite{XZ24}.}  


\enu

\hspace{0.5cm} Now we give a general construction of simple condensable $\EE_1$-algebras in $\Rep(G)$ via internal homs in (\ref{eq:AVH-alpha}). By \cite{BO04,Ost03}, an indecomposable finite semisimple left $\Rep(G)$-module is necessarily equivalent to $\Rep^\prime(\tilde{H})$, where $\tilde{H}$ is the central extension of a subgroup $H \leq G$ determined by a 2-cohomology class $[\alpha] \in H^2(H, \Cb^\times)$:
\[
1\to \Cb^\times \to \tilde{H} \to H \to 1
\]
and $\Rep^\prime(\tilde{H})$ is the category of $\tilde{H}$-representations such that $\Cb^\times$ acts via the identity character. Equivalently, $\Rep^\prime(\tilde H)$ is equivalent to the category $\Rep(H,\alpha)$ of $\alpha$-twisted projective $H$-representations, where an object in $\Rep(H,\alpha)$ is a vector space $V$ equipped with a map $\rho \colon H \to \mathrm{GL}(V)$ such that $\rho(g) \rho(h) = \alpha(g,h) \rho(gh)$ for all $g,h \in H$. The $\Rep(G)$-action on $\Rep(H,\alpha)$ is defined by the usual vector space tensor product. For $V \in \Rep(H,\alpha)$, the internal hom
\be \label{eq:AVH-alpha}
A_V^{(H,\alpha)}:=[V,V]_{\Rep(G)} = \Ind^G_H \mathrm{End}(V)
\ee
defines a condensable $\EE_1$-algebra in $\Rep(G)$ such that $\Mod_{A_V^{(H,\alpha)}}(\Rep(G)) \simeq \Rep(H,\alpha)$. In particular, it means that different choices of $V$ define algebras in the same Morita class. Importantly, $\Rep(H,\alpha)$ and $\Rep(K,\beta)$ are equivalent as $\Rep(G)$-modules if and only if these two pairs $(H,\alpha)$ and $(K,\beta)$ are conjugate in the sense that there exists an element $g \in G$ such that $gHg^{-1} = K$ and $[\alpha] = [g^* \beta] \in H^2(H;U(1))$, where  $(g^\ast\beta)(x,y):=\beta(gxg^{-1}, gyg^{-1})$ for all $x,y\in H$. By condensing $A_V^{(H,\alpha)}$, we obtain a condensed phase, particles on which form the fusion 1-category: 
$$
\Mod_{A_V^{(H,\alpha)}}^{\EE_1}(\Rep(G)) \simeq \Fun_{\Rep(G)}(\Rep(H,\alpha),\Rep(H,\alpha))^\op \simeq 
\FZ_1(\Rep(G), \FZ_0(\Rep(H,\alpha)))^\op. 
$$
Note that $\Rep(G,\alpha)$ is an invertible $\Rep(G)$-$\Rep(G)$-bimodule. Therefore, by condensing $A_V^{(G,\alpha)}$, we obtain the condensed phase $\Mod_{A_V^{(G,\alpha)}}^{\EE_1}(\Rep(G)) \simeq \Rep(G)$. 

\item[(2)] It has another gapped boundary of $\SG\ST_G^3$ whose topological defects form a fusion category $\vect_G$ (when $G = \Zb_2$, this is the rough boundary of the toric code model). A condensable $\EE_1$-algebra in $\vect_G$ is a separable algebra $A$ in $\vect$ equipped with a $G$-grading, and it is simple if the trivial component $A_e = \Cb$ is trivial. The simple condensable $\EE_1$-algebras in $\vect_G$ are given by the twisted group algebras $\Cb[H,\alpha]$ equipped with the obvious $G$-grading, where $H \leq G$ is a subgroup and $\alpha \in Z^2(H;U(1))$ is a 2-cocycle. For $\chi \in C^1(H,\Cb^\times)$, $\Cb[H,\alpha]$ and $\Cb[H,\alpha+d\chi]$ are isomorphic as algebras. Two $\EE_1$-condensable algebras $\Cb[H,\alpha]$ and $\Cb[K,\beta]$ are Morita equivalent if and only if the pairs $(H,\alpha)$ and $(K,\beta)$ are conjugate. 
\enu
By applying topological Wick rotation, condensations on the gapped boundaries of $\SG\ST_G^3$ can also be viewed as phase transitions, which are beyond Landau's spontaneous symmetry-breaking theory, among 1+1D gapped quantum liquids with the same onsite symmetry $G$ \cite{KZ22b}. 
\end{expl}

\begin{expl}[\cite{FS03,Ost03,Ost03a}]
Consider the fusion 1-category $\vect_G^\omega$ for a finite group $G$ and $\omega\in H^3(G,\Cb^\times)$. For each subgroup $H\leq G$ such that $\omega|_H =1$ and $\psi \in H^2(H,\Cb^\times)$, the twist group algebra $\Cb[H,\psi]$ defines a simple condensable $\EE_1$-algebra in $\vect_G^\omega$ with the $G$-grading supported on $H$. Conversely, any simple condensable $\EE_1$-algebra in $\vect_G^\omega$ is isomorphic to $\Cb[H,\psi]$ for some $H$ and $\psi$. Two algebras $\Cb[H,\psi]$ and $\Cb[H,\phi]$ are isomorphic if and only if $\phi=\psi + d\chi$ for $\chi\in C^1(H,\Cb^\times)$. Two algebras $\Cb[H,\psi]$ and $\Cb[K,\phi]$ are Morita equivalent if two pairs $(H,\psi)$ and $(K,\phi)$ belong to the same conjugacy class. In other words, the indecomposable left $\vect_G^\omega$-modules are classified by the conjugacy classes of the pairs $(H,\psi)$. Moreover, there is a one-to-one correspondence between indecomposable $\vect_G^\omega$-modules (or equivalently, the Morita classes of simple condensable $\EE_1$-algebras $A$ in $\vect_G^\omega$) and Lagrangian $\EE_2$-algebras in $\FZ_1(\vect_G^\omega)$ \cite{DMNO13} defined by $A \mapsto Z(A)$, where $Z(A)$ is the full center of $A$ \cite{FFRS07,KR08,Dav10}. This recovers the classification of the Lagrangian $\EE_2$-algebras in $\FZ_1(\vect_G^\omega)$ in \cite{DS17}. 
\end{expl}

\subsection{Particle condensations in 1+1D: geometric approach} \label{sec:geometric_intuition_1codim}

In this subsection, we provide a geometric approach towards particle condensations. The key points in this approach had already appeared in \cite{KS11a,FSV13}. Our presentation uses
a generalization of the classical Eilenberg-Watts calculus, which is covered in Appendix\,\ref{Appendix^EMcal} and is also used in the geometric approach towards anyon condensation theory in 2+1D. 

\subsubsection{Physical intuitions} \label{sec:1d_geo_physical}
We provide a physical intuition of a particle condensation. Let $\SC^2$ and $\SD^2$ be two Morita equivalent 1+1D potentially anomalous type-I topological orders, i.e., $\CC$ and $\CD$ are indecomposable multi-fusion 1-categories. A condensation from $\SC^2$ to $\SD^2$ can be achieved by 
\bnu
\item[(1)] first proliferating the $\SD^2$-phases inside the $\SC^2$-phase 
as illustrated below in spatial dimensions;  
\be \label{pic:proliferate}
\begin{array}{c}
\begin{tikzpicture}[scale = 1.5]
\draw[blue!60, ultra thick,->-](-3,0)--(-2,0) node[midway,below,scale=1] {$\SC^2$};
\draw[color = teal!70, ultra thick,->-](-2,0)--(-1,0) node[midway,below,scale=1] {$\SD^2$};
\draw[color = blue!60, ultra thick,->-](-1,0)--(0,0) node[midway,below,scale=1] {$\SC^2$};
\draw[color = teal!70, ultra thick,->-](0,0)--(1,0) node[midway,below,scale=1] {$\SD^2$};
\draw[color = blue!60, ultra thick,->-](1,0)--(2,0) node[midway,below,scale=1] {$\SC^2$};
\draw[color = teal!70, ultra thick,->-](2,0)--(3,0) node[midway,below,scale=1] {$\SD^2$};
\draw[color = blue!60, ultra thick,->-](3,0)--(4,0) node[midway,below,scale=1] {$\SC^2$};
\draw[fill=blue] (-2.05,-0.05) rectangle (-1.95,0.05) node[midway,above,scale=1] {$\SM^1$} ;
\draw[fill=blue] (-1.05,-0.05) rectangle (-0.95,0.05) node[midway,above,scale=1] {$\overline{\SM^1}$} ;
\draw[fill=blue] (-0.05,-0.05) rectangle (0.05,0.05) node[midway,above,scale=1] {$\SM^1$} ;
\draw[fill=blue] (0.95,-0.05) rectangle (1.05,0.05) node[midway,above,scale=1] {$\overline{\SM^1}$} ;
\draw[fill=blue] (1.95,-0.05) rectangle (2.05,0.05) node[midway,above,scale=1] {$\SM^1$} ;
\draw[fill=blue] (2.95,-0.05) rectangle (3.05,0.05) node[midway,above,scale=1] {$\overline{\SM^1}$} ;
\end{tikzpicture}
\end{array}
\ee
\item[(2)] then shrinking each line segment $\overline{\SM^1} \boxtimes_\SC \SM^1$ produces a particle in $\SD^2$; 

\item[(3)] then annihilating this particle, i.e. mapping it to the trivial particle $\one_\CD$ in $\SD^2$. 

\enu
As a consequence, the entire 1d (spatial dimension) line turn into a green line.

\medskip
Notice that shrinking a green line segment in (\ref{pic:proliferate}), i.e. $\SM\boxtimes_\SD \overline{\SM}$ defines a particle $A\in \CC$. The step (2) and (3) simply produces a morphism $\mu_A: A\otimes_\CC A \to A$. Moreover, we should have a canonical map $\eta_A: \one_\CC \to A$, which creates the green line out of nothing. This map makes the proliferating possible in the first place. 
Since the physical meanings of 1-morphisms $\mu_A$ and $\eta_A$ are instantons, we drew these instantons in the following spacetime pictures. 
\be \label{pic:condensable_E1_algebra}
\begin{array}{c}
\begin{tikzpicture}[scale=0.9]
\draw[black,thick, ->] (-5,0) -- (-5,0.5) node[very near end, left] {\small $t$} ;
\draw[black,thick, ->] (-5,0) -- (-4.5,0) node[very near end, below] {\small $x_1$} ;
\fill[blue!20] (-4,0) rectangle (-1,2) ;
\fill[teal!20] (-3,2) .. controls (-3,1) and (-2,1) .. (-2,2)--(-3,2)..controls (-2.5,2) .. (-2,2)--cycle;
\draw[blue,ultra thick] (-3,2) .. controls (-3,1) and (-2,1) .. (-2,2) ;
\node at (-3,2.2) {\small $\SM$} ;
\node at (-2,2.2) {\small $\overline{\SM}$} ;
\quad\quad\quad
\fill[blue!20] (0,0) rectangle (5,2) ;
\fill[teal!20] (1,0) .. controls (1,1) and (2,1) .. (2,2)--(3,2) .. controls (3,1) and (4,1) .. (4,0)--(3,0) .. controls (3,1) and (2,1) .. (2,0)--cycle ;
\draw[blue,ultra thick] (1,0) .. controls (1,1) and (2,1) .. (2,2) ;
\node at (2,-0.2) {\small $\overline{\SM}$} ;
\draw[blue,ultra thick]  (3,2) .. controls (3,1) and (4,1) .. (4,0) ;
\node at (3,-0.2) {\small $\SM$} ;
\draw[blue,ultra thick] (2,0) .. controls (2,1) and (3,1) .. (3,0) ;
\node at (2,2.2) {\small $\SM$} ;
\node at (3,2.2) {\small $\overline{\SM}$} ;
\draw[dashed] (-3.2,1) rectangle (-1.8,1.6) ;
\draw[dashed] (0.8,0.4) rectangle (4.2,1) ;
\end{tikzpicture}
\end{array}
\ee
Then the following identities of 1-morphisms in $\CC$:
$$
\mu_A \circ (\mu_A \otimes \id_A) = \mu_A \circ (\id_A \otimes \mu_A), \quad \mu_A \circ (\eta_A \otimes \id_A) =\id_A = \mu_A \circ (\id_A \otimes \eta_A)
$$
holds by the obvious physical intuitions. For example, the first identity simply says that first running the procedure (2) and (3) to a blue line the first, then running the same procedures to the blue line next to it makes no difference to running in the opposite order. As a consequence, the triple $(A,\mu_A,\eta_A)$ should define 
a condensable $\EE_1$-algebra in $\CC$, which defines the fusion of two green lines into one.

\begin{rem} \label{rem:coalgebra_structure}
In the unitary case, by turning the pictures in (\ref{pic:condensable_E1_algebra}) upside down, we obtain the counit $\epsilon_A$ and comultiplication $\Delta_A=\mu_A^\dagger$: 
\[
\begin{array}{c}
\begin{tikzpicture}[scale=0.9]
\draw[black,thick, ->] (-5,0) -- (-5,0.5) node[very near end, left] {\small $t$} ;
\draw[black,thick, ->] (-5,0) -- (-4.5,0) node[very near end, below] {\small $x_1$} ;
\fill[blue!20] (-4,0) rectangle (-1,2) ;
\fill[teal!20] (-3,0) .. controls (-3,1) and (-2,1) .. (-2,0)--(-3,0)..controls (-2.5,0) .. (-2,0)--cycle;
\draw[blue,ultra thick] (-3,0) .. controls (-3,1) and (-2,1) .. (-2,0) ;
\node at (-3,-0.2) {\small $\SM$} ;
\node at (-2,-0.2) {\small $\overline{\SM}$} ;
\quad\quad\quad
\fill[blue!20] (0,0) rectangle (5,2) ;
\fill[teal!20] (1,2) .. controls (1,1) and (2,1) .. (2,0)--(3,0) .. controls (3,1) and (4,1) .. (4,2)--(3,2) .. controls (3,1) and (2,1) .. (2,2)--cycle ;
\draw[blue,ultra thick] (1,2) .. controls (1,1) and (2,1) .. (2,0) ;
\node at (2,2.2) {\small $\overline{\SM}$} ;
\draw[blue,ultra thick]  (3,0) .. controls (3,1) and (4,1) .. (4,2) ;
\node at (3,2.2) {\small $\SM$} ;
\draw[blue,ultra thick] (2,2) .. controls (2,1) and (3,1) .. (3,2) ;
\node at (1,2.2) {\small $\SM$} ;
\node at (4,2.2) {\small $\overline{\SM}$} ;
\draw[dashed] (-3.2,0.4) rectangle (-1.8,1) ;
\draw[dashed] (0.8,1) rectangle (4.2,1.6) ;
\end{tikzpicture}
\end{array}
\]
which is co-unital and co-associative. Moreover, $(A,\mu_A,\eta_A,\Delta_A,\epsilon_A)$ defines a Frobenius algebra \cite{KS11a,FSV13}. Moreover, we have $m_A\circ \Delta_A=\id_A$, i.e., $A$ is a special Frobenius algebra in $\CC$ \cite{FRS02}. 
\end{rem}

A particle $d\in \CD$ in the condensed phase $\SD$ produces a particle in $\SC$ by viewing the green line segment decorated by $d\in \CD$ as a particle in $\CC$ as illustrated below. 
\be \label{pic:d-particle_forget}
\begin{tikzpicture}[scale = 1.5]
\draw[color = blue!60, ultra thick](-3,0)--(-2,0) node[midway,below,scale=1] {$\SC$};
\draw[color = teal!70, ultra thick](-2,0)--(-1,0) node[midway,below,scale=1] {$\SD$};
\draw[color = blue!60, ultra thick](-1,0)--(0,0) node[midway,below,scale=1] {$\SC$};
\draw[color = teal!70, ultra thick](0,0)--(1,0) node[midway,below,scale=1] {$\SD$};
\draw[color = blue!60, ultra thick](1,0)--(2,0) node[midway,below,scale=1] {$\SC$};
\draw[color = teal!70, ultra thick](2,0)--(3,0) node[midway,below,scale=1] {$\SD$};
\draw[color = blue!60, ultra thick](3,0)--(4,0) node[midway,below,scale=1] {$\SC$};
\draw[fill=blue] (-2.05,-0.05) rectangle (-1.95,0.05) node[midway,above,scale=1] {$\SM$} ;
\draw[fill=blue] (-1.05,-0.05) rectangle (-0.95,0.05) node[midway,above,scale=1] {$\overline{\SM}$} ;
\draw[fill=blue] (-0.05,-0.05) rectangle (0.05,0.05) node[midway,above,scale=1] {$\SM$} ;
\draw[fill=blue] (0.95,-0.05) rectangle (1.05,0.05) node[midway,above,scale=1] {$\overline{\SM}$} ;
\draw[fill=blue] (1.95,-0.05) rectangle (2.05,0.05) node[midway,above,scale=1] {$\SM$} ;
\draw[fill=blue] (2.95,-0.05) rectangle (3.05,0.05) node[midway,above,scale=1] {$\overline{\SM}$} ;
\fill[black] (0.5,0) circle (0.04) node[black] at(0.5,0.2) {$d$};
\draw[decorate,decoration=brace,thick] (1,-0.3)--(0,-0.3) ;
\node at (0.5,-0.5) {\small $d\in \CC$} ;
\end{tikzpicture}
\ee
When $d=\one_\CD$, this particle is $\one_\CD=A\in \CC$. Note that $A$ is precisely $\one_\CD$ viewed as an object in $\CC$ because $\CD \subset \CC$. Therefore, for general $d\in \CD$, we expect that this particle is precisely the same particle $d$ viewed as an object in $\CC$. Then it is clear that procedures (2) and (3) define an $A$-$A$-bimodule structure on the particle $d\in \CC$. Therefore, we have recovered the same result $\CD=\BMod_{A|A}(\CC)$.

As we have shown in Section\,\ref{sec:1d_cond_alg}, the gapped domain wall $\SM^1$ can also support particles, which form a separable 1-category $\CM$. First, $m\in \CM$, $m$ is necessarily an object in $\CC$. Secondly, $m$ is necessarily a right $A$-module. This suggest that $\CM=\RMod_A(\CC)$. The fact that the green line shrink to the particle in $\CC$ and the fact that the blue line shrink to a particle in $\CD$ simply suggests that 
$$
\CM^\op \boxtimes_\CC \CM \simeq \CD, \quad\quad \CM \boxtimes_\CD \CM^\op \simeq \CC
$$
as $\CD$-$\CD$-bimodule categories and $\CC$-$\CC$-bimodule categories, respectively. It means that $\CM=\RMod_A(\CC)$ should be an invertible $\CC$-$\CD$-bimodule that defines the Morita equivalence between $\CC$ and $\CD=\BMod_{A|A}(\CC)$. It is indeed true as a well-known mathematical fact. It is also clear that particles from two sides of $\SM^1$ move onto the domain wall $\SM$ according to $a\mapsto a\otimes A$ and $d\mapsto A\otimes_A d$ for $a\in \CC, d\in \CD$. Therefore, we have recovered the results in Theorem$^{\mathrm{ph}}$ \ref{pthm:1d_cond_part1} and \ref{pthm:1d_cond_part2} via a geometric approach.

\medskip
In physics, a phase transition can always go backwards by tuning the coupling constant of certain interactions in the opposite direction. Particle condensations in 1+1D topological orders are also reversible. More precisely, 
\bnu
\item[(1)] one can view Figure\,\ref{pic:proliferate} as a process of proliferating $\SC^2$-phase segments inside a large $\SD^2$-phase; 
\item[(2)] then shrink the green line segment to a particle in $\SC^2$-phase; 
\item[(3)] then annihilating this particle, i.e. mapping it to the trivial particle $\one_\CC$ in $\SC^2$. 
\enu
In this way, we obtain a particle condensation in $\SD^2$ that reproduces the $\SC^2$-phase. In other words, by interchanging the role played by $\SC^2$ and $\SD^2$, we obtain the reversed process. More precisely, by shrinking each blue line segment $\overline{\SM} \boxtimes_\SC \SM$ to a particle in $\SD^2$, we obtain a condensable particle in $\SD^2$. More precisely, this particle is automatically equipped with a structure of condensable $\EE_1$-algebra $B$ in $\CD$, thus defines a particle condensation in $\CD$. Moreover, we must have $\CC=\BMod_{B|B}(\CD)$ and $\CM=\LMod_B(\CD)$.

\subsubsection{Precise geometric theory} \label{sec:geometric_theory_1codim_2}
The physical intuitions described in Section\,\ref{sec:1d_geo_physical} can be made mathematically precise. Now we provide the precise mathematical foundation and calculation behind the physical intuitions discussed in Section\,\ref{sec:1d_geo_physical}. Physically oriented readers are recommended to skip this subsubsection at least in the first reading. 

\medskip
Agian, the condensation from the 1+1D anomalous type-I topological order $\SC^2$ to $\SD^2$ can be achieved by the following procedures. 
\bnu
\item[(1)] First, proliferate the $\SD^2$-phases inside the $\SC^2$-phase with $\SM=(\CM,m)$ and $\overline{\SM}=(\CM^\op,m)$ for some $m\in \CM$ as illustrated in the picture (\ref{pic:proliferate}). 

\item[(2)] Secondly, shrink the black line $\overline{\SM} \boxtimes_\SC \SM$ to a particle in $\SD^2$. By \cite[Proposition\ 2.2.7]{KZ18}, this particle can be explicitly computed as the image of $m\boxtimes_\CC m$ the following equivalences: 
\begin{align} 
\CM^\op \boxtimes_\CC \CM &\simeq \Fun_\CC(\CM,\CM) \simeq \CD^\rev \nn
m\boxtimes_\CC m &\mapsto [m,-]_\CC \odot m \label{eq:shrink-1} 
\end{align}
where $[m,-]_\CC: \CM \to \CC$ is the functor defined by $[m,-]_\CC(m'):=[m,m']_\CC \in \CC$ for $m'\in \CM$ and $[m,m']_\CC$ is the internal hom (recall Remark\,\ref{rem:internal_hom_functor}).

\item[(3)] Thirdly, map this particle in $\CD$ to the trivial particle $\one_\CD\in \CD$. Since $\CD^\rev \simeq \Fun_\CC(\CM,\CM)$ is defined by $d\mapsto -\odot d$, $\one_\CD$ is mapped to $-\odot \one_\CD \simeq \id_\CM\in\Fun_\CC(\CM,\CM)$. Therefore, mapping it to $\one_\CD$ is just the canonical natural transformation between two $\CC$-module functors:  
\be \label{eq:shrink-2}
\ev: [m,-] \odot m \to \id_\CM
\ee

\enu
After these three procedures, the entire 1d line turn into a green line, thus defines the condensation.

Again shrinking a green line segment in (\ref{pic:proliferate}), i.e. $\SM\boxtimes_\SD \overline{\SM}$, defines a particle $A\in \CC$, which can explicitly computed as in the following picture. 
\be  \label{fig:geo-1d-cond-alg}
\begin{array}{c}
\begin{tikzpicture}[scale = 1.5]
\draw[color = blue!60, ultra thick](-1,0)--(0,0);
\draw[color = teal!70, ultra thick](0,0)--(1,0);
\draw[color = blue!60, ultra thick](1,0)--(2,0);
\draw[fill=blue] (-0.05,-0.05) rectangle (0.05,0.05) node[midway,above,scale=1] {$m\in \CM$} ;
\draw[fill=blue] (0.95,-0.05) rectangle (1.05,0.05) node[midway,above,scale=1] {$m\in \CM^\op$} ;
\node[teal!70] at(0.5,-0.2) {$\CD$};
\node[blue!60] at(-0.5,-0.2) {$\CC$};
\end{tikzpicture}
\end{array}
= 
\begin{array}{c}
\begin{tikzpicture}[scale = 1.5]
\draw[color = blue!60, ultra thick](-1.5,0)--(1.5,0);
\fill[blue!60] (0,0) circle (0.06) node[black] at(0.2,0.2) {$A=[m,m]\in \CC$};
\node[blue!60] at(-0.5,-0.2) {$\CC$};
\end{tikzpicture} 
\end{array} 
\ee
Mathematically, it is given by the canonical equivalence $\CM \boxtimes_\CD \CM^\op \simeq \CC$ defined by $x\boxtimes_\CD y \mapsto [x,y]_\CC^\ast$.

Moreover, according to (\ref{eq:shrink-1}) and (\ref{eq:shrink-2}), the procedure (2) and (3) defines a morphism $\mu_A: A\otimes A\to A$ as the following composed map: 
\be \label{eq:multiplication}
[m,m] \otimes [m,m] \simeq [m, [m,m]\odot m] \xrightarrow{[m,\ev_m]} [m,m],
\ee
where the first `$\simeq$' is due to the canonical isomorphism $a\otimes [x,y] \simeq [x,a\otimes y]$ \cite{Ost03}. It turns out that the composed map (\ref{eq:multiplication}) coincides with the canonical morphism $[m,m] \otimes [m,m] \to [m,m]$ defined by the universal property of the internal hom.

We have a canonical morphism from $\eta: \one_\CC \to [m,m]$, which can be defined physically in two steps. 
\bnu
\item In the first step, we simply factorizes $\one_\CC$ as follows: 
\begin{align*}
\begin{array}{c}
\begin{tikzpicture}[scale = 1.5]
\draw[color = blue!60, ultra thick](-1.5,0)--(1.5,0);
\fill[blue!60] (0,0) circle (0.06) node[blue!60] at(0.2,0.2) {$\one_{\CC}$};
\node[blue!60] at(-0.5,-0.2) {$\CC$};
\end{tikzpicture}
\end{array}
&= \int_{x \in \CM}^\CD
\begin{array}{c}
\begin{tikzpicture}[scale = 1.5]
\draw[color = blue!60, ultra thick](-1,0)--(0,0);
\draw[color = teal!70, ultra thick](0,0)--(1,0);
\draw[color = blue!60,ultra thick](1,0)--(2,0);
\draw[fill=blue] (-0.05,-0.05) rectangle (0.05,0.05) node[midway,blue,above,scale=1] {$x\in \CM$} ;
\draw[fill=blue] (0.95,-0.05) rectangle (1.05,0.05) node[midway,blue,above,scale=1] {$x\in \CM^\op$} ;
\node[teal!70] at(0.5,-0.2) {$\CD$};
\node[blue!60] at(-0.5,-0.2) {$\CC$};
\end{tikzpicture}
\end{array}
\end{align*}
That is, the canonical equivalence $\CC \simeq \CM\boxtimes_\CD \CM^\op$ maps $\one_\CC$ to an object $\int_{x\in \CM}^\CD x\boxtimes_\CD x$ in $\CM\boxtimes_\CD \CM^\op$, where the integral defines a limit (called `end') in the category $\CC$. In Appendix\,\ref{Appendix^EMcal}, we explain this notion as a generalization of the classical Eilenberg-Watts calculus. That $\one_\CC \mapsto \int_{x\in \CM}^\CD x\boxtimes_\CD x$ in $\CM\boxtimes_\CD \CM^\op$ is explained in Example\,\ref{expl:idM=int_mRm}. Physically, it means that one can always split a black line by inserting a green line segment, then all possible particles $x\in \CM, x\in \CM^\op$ are accumulating at two ends in pairs. Such a pair of particles can be viewed as a pair of boundary conditions at the two ends of the green line segment. By integrating all these possible boundary conditions, we simply recover the trivial particle in $\CC$. This step is completely invertible. 

\item In the second step, we define $\eta$ by ``projecting out'' all the $x$-components for $x\neq m$. Mathematically, this ``projection'' is precisely the defining morphism $\int_{x\in \CM}^\CD x\boxtimes_\CD x \to m\boxtimes_\CD m$ of the limit. 
It turns out that such physically defined morphism $\eta$ is precisely the one defined by the univeresal property of the internal hom $[m,m]$. 

\enu

\begin{expl}
It is helpful to look at a special case $\CD=\CC=\CM$. In this case, we have (see also Example\,\ref{expl:1xRx})
$$
\int_{x\in \CM}^\CD x\boxtimes_\CD x = \int_{x\in \CC}^\CC x\otimes x^\ast\simeq \one_\CC. 
$$  
\end{expl}

If we already know that $\SD^2$ and $\SM^1$ are obtained from $\SC^2$ by condensing a condensable $\EE_1$-algebra $A\in \CC$. In this case, we have $\CD=\BMod_{A|A}(\CC)$, $\CM=\RMod_A(\CC)$, $\CM^\op=\LMod_A(\CC)$ and $\SM=(\CM,A)$. 
\bnu

\item The bulk-to-wall map $\CC \to \CM$ is $a\mapsto a\otimes A, \forall a\in \CC$. It right adjoint functor is $[A,-]: \CM \to \CC$ is precisely the forgetful functor $\forget: \RMod_A(\CC) \to \CC$ defined by $x\mapsto x$ by forgetting the $A$-module structure on $x$. Therefore, $[A,A]_\CC=A$. 

\item The $d$-particle in $\SD^2$, as depicted in (\ref{pic:d-particle_forget}), becomes the particle $[A,d\otimes_A A]=d$ in $\CC$. This assignment $\CD \ni d\mapsto d \in \CC$ is precisely the forgetful functor $\forget: \RMod_A(\CC) \to \CC$. 

\enu

Very importantly, what we have shown is that the data on the wall $\SM^1=(\CM,m)$, which should be viewed as a boundary condition, uniquely determines a condensation. This is compatible with the usual physical intuition that a condensation can be constructed by proliferating the green line segments with a chosen boundary condition on the wall $\SM^1$. 
Moreover, in (\ref{fig:geo-1d-cond-alg}), the choice of the object $m\in \CM$ is arbitrary. A different choice $m'$ simply produces a different condensable $\EE_1$-algebra $A'=[m',m']^\op$, which defines a `new' condensation. However, this `new' condensation does not produce a new condensed phase. This is because $A$ and $A'$ are Morita equivalent, i.e. $\RMod_A(\CC) \simeq \CM \simeq \RMod_{A'}(\CC)$. As a consequence, we have, in $\Algc_{\EE_1}(2\vect)$,  
$$
\BMod_{A|A}(\CC) \simeq \Fun_\CC(\RMod_A(\CC), \RMod_A(\CC))^\rev \simeq 
\Fun_\CC(\RMod_{A'}(\CC), \RMod_{A'}(\CC))^\rev \simeq
\BMod_{A'|A'}(\CC). 
$$
In other words, $\CD$ is irrelevant to the choice of $m\in \CM$. It only depends on $\CM$ as $\CD\simeq \Fun_\CC(\CM,\CM)^\rev$ in $\Algc_{\EE_1}(2\vect)$. 
Note that an equivalence $\CM \simeq \RMod_A(\CC)$ provides a way of labeling objects in $\CM$, and should be viewed as a coordinate system on $\CM$. An equivalence $\CM \simeq \RMod_{A'}(\CC)$ provides a different coordinate system on $\CM$ (i.e. a relabeling of objects in $\CM$). Similarly, $\BMod_{A|A}(\CC)$ and $\BMod_{A'|A'}(\CC)$ provide two different coordinate systems on $\CD$.

\medskip
The reversed process, a particle condensation in the $\SD^2$-phase that reproduces the $\SC^2$-phase as the condensed phase, can also be made mathematically precise. 
\begin{pthm}\label{pthm:reversed_condensation_1d}
By shrinking the black line segment in the $\SD^2$-phase, we obtain a particle in the $\SD^2$-phase, which is precisely the internal hom algebra $[m,m]_\CD \in \CD$ as illustrated in the following picture. 
\[
\begin{array}{c}
\begin{tikzpicture}[scale = 1.5]
\draw[color = teal!90, ultra thick](-1,0)--(0,0);
\draw[color = blue!60, ultra thick](0,0)--(1,0);
\draw[color = teal!90, ultra thick](1,0)--(2,0);
\draw[fill=blue] (-0.05,-0.05) rectangle (0.05,0.05) node[midway,above,scale=1] {\small $m\in \CM^\op$} ;
\draw[fill=blue] (0.95,-0.05) rectangle (1.05,0.05) node[midway,above,scale=1] {\small $m\in \CM$} ;
\node[black] at(0.5,-0.2) {\small $\CC$};
\node[black] at(-0.5,-0.2) {\small $\CD$};
\node[black] at(1.5,-0.2) {\small $\CD$};
\end{tikzpicture}
\end{array}
= 
\begin{array}{c}
\begin{tikzpicture}[scale = 1.5]
\draw[color = teal!90, ultra thick](-1.5,0)--(1.5,0);
\fill[teal!90] (0,0) circle (0.06) node[black] at(0.2,0.2) {\small $B=[m,m]_\CD\in \CD$};
\node[black] at(-1,-0.2) {\small $\CD$};
\node[black] at(1,-0.2) {\small $\CD$};
\end{tikzpicture} 
\end{array} 
\]
By condensing $B\in \CD$, we obtain $\CC\simeq\BMod_{B|B}(\CD)$ and $\CM\simeq\LMod_B(\CD)$. Particles in $\SD$ move to the wall $\CM$ according to the functor $B\otimes -: \CD \to \CM$, and particles in $\SC$ move to the wall $\CM$ according to the functor $-\otimes_B B: \CC \to \CM$. 
\end{pthm}

\begin{lem}
If we have already known that $\CD=\BMod_{A|A}(\CC)$ and $\SM=(\CM=\RMod_A(\CC), m)$, then 
\bnu
\item[(1)] $B=[m,m]_\CD=m^R\otimes m \in \CD$; 
\item[(2)] $\CC \simeq \BMod_{B|B}(\CD)$ and $\CM \simeq \LMod_B(\CD)$. 
\enu
\end{lem}
\pf
(1). It follows from $\hom_\CM(x\odot d, y) = \hom_\CM(x\otimes_A d, y) \simeq \hom_\CD(d, x^R\otimes y)$. 

(2). Proved in \cite{EGNO15}. The key idea is to check that the functor $m^R \otimes - \otimes m: \CC \to \BMod_{B|B}(\CD)$ defines a monoidal equivalence. 
\epf

In summary, all data appeared in Section\,\ref{sec:geometric_intuition_1codim} become explicit and mathematically precise. It is quite amazing that these precise data automatically make all the physical intuitions works as mathematical facts. What is even more amazing is that this beautiful coincidenece of physical intuitions and precise mathematical results holds in all dimensions.

\subsection{Condensations of 1-codimensional defects in \texorpdfstring{$n$}{n}+1D} \label{sec:1codim_nd}

The mathematical theory of 1-codimensional topological defects in higher dimensions is very similar to that of particle condensations in 1+1D \cite{KS11a,FSV13,CR16,Kon14e}. It was included in \cite{GJF19} more systematically as an ingredient of a mathematical theory of condensation or Karoubi completion of higher categories. Although the physical intuitions behind this mathematical theory was briefly explained in \cite{GJF19}, some interesting ingredients or results and systematic constructions of examples are still lacking. In this subsection, we combine earlier results with other ingredients appeared later \cite{JF22,KZ22b,KZ24} to give a rather complete theory of the condensations of 1-codimensional topological defects in higher dimensional topological orders.

\subsubsection{General theory} \label{sec:general_theory_1codim_nd}
The physical intuition of particle condensations in 1+1D potentially anomalous topological orders automatically carries over to the condensations of 1-codimensional topological defects in potentially anomalous $n+$1D topological orders. 

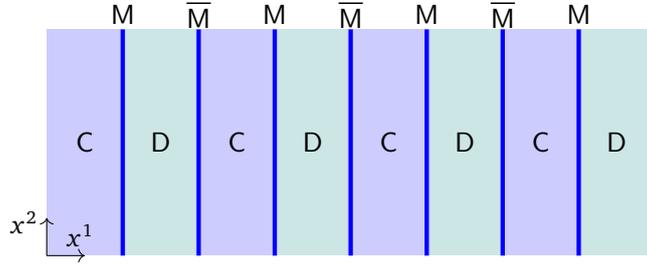
\begin{figure}[htbp]
\[
\begin{tikzpicture}
\fill[blue!20] (-4,0) rectangle (4,3) ;
\draw[->] (-4,0) -- (-4,0.5) node [very near end, left] {$x^2$} ;
\draw[->] (-4,0) -- (-3.5,0) node [very near end, above] {$x^1$} ;
\node at (-3.5,1.5) {$\SC$} ;
\draw[blue, ultra thick] (-3,0) -- (-3,3) ;
\node at (-3,3.2) {$\SM$} ;
\fill[teal!20] (-2.95,0) rectangle (-2.05,3) ;
\node at (-2.5,1.5) {$\SD$} ;
\draw[blue, ultra thick] (-2,0) -- (-2,3) ; 
\node at (-2,3.2) {$\overline{\SM}$} ;
\node at (-1.5,1.5) {$\SC$} ;
\draw[blue, ultra thick] (-1,0) -- (-1,3) ; 
\node at (-1,3.2) {$\SM$} ;
\fill[teal!20] (-0.95,0) rectangle (-0.05,3) ;
\node at (-0.5,1.5) {$\SD$} ;
\draw[blue, ultra thick] (0,0) -- (0,3) ; 
\node at (0,3.2) {$\overline{\SM}$} ;
\node at (0.5,1.5) {$\SC$} ;
\draw[blue, ultra thick] (1,0) -- (1,3) ; 
\node at (1,3.2) {$\SM$} ;
\fill[teal!20] (1.05,0) rectangle (1.95,3) ;
\node at (1.5,1.5) {$\SD$} ;
\draw[blue, ultra thick] (2,0) -- (2,3) ; 
\node at (2,3.2) {$\overline{\SM}$} ;
\node at (2.5,1.5) {$\SC$} ;
\draw[blue, ultra thick] (3,0) -- (3,3) ; 
\node at (3,3.2) {$\SM$} ;
\fill[teal!20] (3.05,0) rectangle (4,3) ;
\node at (3.5,1.5) {$\SD$} ;
\end{tikzpicture}
\]
\caption{Condensation of 1-codimensional topological defects in $\SC^{n+1}$}
\label{fig:1codim_condensation_nd}
\end{figure}

\medskip
Let $\SC^{n+1}$ be a potentially anomalous $n+$1D type-I topological order. The category $\CC$ of topological defects in $\SC^{n+1}$ is an indecomposable multi-fusion $n$-category (recall Theorem$^{\mathrm{ph}}$\,\ref{pthm:TO_multi-fusion_ncat}). More explicitly, a condensation of a (composite) 1-codimensional topological defect in a potentially anomalous $n+$1D topological order $\SC^{n+1}$ that produces a new phase $\SD^{n+1}$ (see Remark\,\ref{rem:boundary_phase_transition}), can be achieved by 
\bnu
\item[(1)] first proliferating the $\SD^{n+1}$-phases inside the $\SC^{n+1}$-phase along the unique transversal direction $x^1$ 
as illustrated Figure\,\ref{fig:1codim_condensation_nd}, where the domain wall $\SM^n$ is assumed to be gapped;  

\item[(2)] then shrinking each blue region $\overline{\SM} \boxtimes_\SC \SM$ to a 1-codimensional topological defect in $\SD^{n+1}$; 

\item[(3)] then annihilating this defect, i.e. mapping it to the trivial 1-codimensional topological defect $\one_\CD$ in $\SD^{n+1}$. 

\enu
As a consequence, the entire phase $\SC^{n+1}$ turn into a new phase $\SD^{n+1}$, which we assume to be indecomposable. 

\begin{rem} \label{rem:boundary_phase_transition}
Such a condensation produces a new phase $\SD^{n+1}$ that share the same gravitational anomaly with $\SC^{n+1}$ as illustrated below. 
$$
\begin{array}{c}
\begin{tikzpicture}[scale=0.8]
\fill[gray!20] (-2,0) rectangle (2,2) ;
\node at (0,1.5) {\scriptsize $\SZ(\SC)^{n+2}=\SZ(\SD)^{n+2}$} ;
\node at (-1.7,-0.3) {\scriptsize $\SC^{n+1}$} ;
\node at (1.7,-0.3) {\scriptsize $\SD^{n+1}$} ;
\draw[blue!50,ultra thick,->-] (-2,0) -- (0,0) ;
\draw[teal!50,ultra thick,->-] (0,0) -- (2,0) ;
\draw[fill=blue] (-0.08,-0.08) rectangle (0.08,0.08) node[midway,below] {\scriptsize $\SM^n$} ;
\end{tikzpicture}
\end{array}
$$
Therefore, one can also view this condensation as a purely boundary phase transition of $\SZ(\SC)^{n+2}$. We assume that the domain wall $\SM^n$ is gapped just for convenience. One can see that if the domain wall $\SM^n$ is gapless, the argument is parallel. This defines a condensation of liquid-like gapless defects in $\SC^{n+1}$, which is beyond the setting of this section. We postpone it to the last section. 
\end{rem}

One can see that condensing a 1-codimensional topological defect in an $n+$1D topological order is completely parallel to that in a 1+1D topological order. We simply summarize the result below.

\begin{pthm} \label{pthm:1codim_condensation_nd}
A condensation of a 1-codimensional topological defect in an $n+$1D type-I topological order $\SC^{n+1}$ can be mathematically described as follows: 
\bnu

\item Since the condensed phase $\SD^{n+1}$ is condensed from $\SC$, the multi-fusion $n$-category $\CD$ is necessarily a sub-2-category of $\CC$. The trivial 1-codimensional topological defect $\one_\CD$ in $\CD$ corresponds to a non-trivial 1-codimensional topological defect $A\in \CC$, which is precisely the defect obtained by shrinking a green strap in Figure\,\ref{fig:1codim_condensation_nd} to a 1-codimensional topological defect in $\SC^{n+1}$. 

\item The $A$-defect is naturally equipped with the structure of an algebra in $\CC$ with the unit 1-morphism $\eta_A$ and the multiplication 1-morphism $\mu_A$ defined by the following pictures in spatial dimensions\footnote{Note that pictures (\ref{pic:condensable_E1_algebra_2}) and (\ref{pic:condensable_E1_algebra}) in order to warn readers that the physical meanings of these two sets of pictures have very different physical meanings.}. 
\be \label{pic:condensable_E1_algebra_2}
\begin{array}{c}
\begin{tikzpicture}
\fill[blue!20] (-4,0) rectangle (-1,2) ;
\fill[teal!20] (-3,2) .. controls (-3,1) and (-2,1) .. (-2,2)--(-3,2)..controls (-2.5,2) .. (-2,2)--cycle;
\draw[blue,ultra thick] (-3,2) .. controls (-3,1) and (-2,1) .. (-2,2) ;
\node at (-3,2.2) {$\SM$} ;
\node at (-2,2.2) {$\overline{\SM}$} ;
\draw[->] (-4,0) -- (-4,0.5) node [very near end, left] {$x^2$} ;
\draw[->] (-4,0) -- (-3.5,0) node [very near end, above] {$x^1$} ;
\quad\quad\quad
\fill[blue!20] (0,0) rectangle (5,2) ;
\fill[teal!20] (1,0) .. controls (1,1) and (2,1) .. (2,2)--(3,2) .. controls (3,1) and (4,1) .. (4,0)--(3,0) .. controls (3,1) and (2,1) .. (2,0)--cycle ;
\draw[blue,ultra thick] (1,0) .. controls (1,1) and (2,1) .. (2,2) ;
\node at (2,-0.2) {$\overline{\SM}$} ;
\draw[blue,ultra thick]  (3,2) .. controls (3,1) and (4,1) .. (4,0) ;
\node at (3,-0.2) {$\SM$} ;
\draw[blue,ultra thick] (2,0) .. controls (2,1) and (3,1) .. (3,0) ;
\node at (2,2.2) {$\SM$} ;
\node at (3,2.2) {$\overline{\SM}$} ;
\draw[dashed] (-3.1,1) rectangle (-1.9,1.6) ; 
\node at (-3.6,1.2) {$\eta_A=$} ;
\draw[dashed] (0.8,0.4) rectangle (4.2,1) ;
\node at (0.3,0.6) {$\mu_A=$} ;
\end{tikzpicture}
\end{array}
\ee
Since the physical meaning of a 1-morphism in $\CC$ is a 2-codimensional topological defect, both $\eta_A$ and $\mu_A$ are defined by the 2-codimensional topological defects encircled by the dashed boxes in above picture. It is clear that we should have the associativity and the unital ``properties'' of $\mu_A$ and $\eta_A$. However, for $n>1$, these ``properties'' are more than properties because higher morphisms (i.e., higher codimensional defects) are needed as defining data. For example, we should have the following 2-isomorphisms
$$
\mu_A \circ (\mu_A \otimes \id_A) \xrightarrow{\simeq} \mu_A \circ (\id_A \otimes \mu_A), \quad \mu_A \circ (\eta_A \otimes \id_A) \xrightarrow{\simeq} \id_A = \mu_A \circ (\id_A \otimes \eta_A).
$$
If $n=2$, these 2-isomorphisms should satisfy a pentagon condition and a triangle condition (see \cite{Dec23a} for a precise mathematical definitions and see also \cite{ZLZHKT23}); if $n>2$, there should be a 3-isomorphism associated to each pentangon and one associated to each triangle, so on and so forth. Our description here is mainly descriptive instead of being precise. A formal theory to take care all the higher coherence data can be found in \cite{Lur17}. The higher coherence data can also be more efficiently encoded by internal homs as we show in Theorem\ \ref{pthm:1codim_condensation_nd_II}. Moreover, we require such algebra to be condensable.

\item Similar to the particle condensations in 1+1D topological orders, 1-codimensional topological defects in $\SD^{n+1}$ are precisely $A$-$A$-modules or $\EE_1$-$A$-modules in $\CC$ (see \cite{Dec23a,ZLZHKT23} for mathematical definitions for $n=2$), 
i.e. $\CD=\Mod_A^{\EE_1}(\CC)$ (recall Remark \ref{rem:E1-algebras_in_C}), which is an indecomposable multi-fusion $n$-category (i.e., $\SD^{n+1}$ is type-I). If $A$ is indecomposable, then $\CD=\Mod_A^{\EE_1}(\CC)$ is a fusion $n$-category. 

\item $\SM^n$ can be described by a pair $(\CM,m)$, where $\CM$ is a separable $n$-category given by $\RMod_A(\CC)$ and $m=A$ is an object in $\CM=\RMod_A(\CC)$ representing a 1-codimensional defect on the wall. The pair $(\CM,m)$ can be viewed as an $\EE_0$-algebra in $(n+1)\vect$.

\item Topological defects in $\SC^{n+1}$ move onto the $\SM^n$-wall according to the left bulk-to-wall map (or functor) 
\be \label{eq:functor_L}
L:=-\otimes A: \CC \to \CM=\RMod_A(\CC) \quad\quad \mbox{defined by $a \mapsto a\otimes A$}. 
\ee
Topological defects in $\SD^{n+1}$ move onto the $\SM^n$-wall according to the right bulk-to-wall map (or functor) 
\be \label{eq:functor_R}
R:= A\otimes_A -: \CD \to \CM=\RMod_A(\CC) \quad\quad \mbox{defined by $d \mapsto A\otimes_A d$}.
\ee

\item Note that the right adjoint $L^R$ of the functor $L=-\otimes A: \CC \to \CM=\RMod_A(\CC)$ is the internal hom functor $[A,-]$ (see Definition\,\ref{defn:internal_hom_ncat}), which, in this case, is precisely the forgetful functor $\forget: \RMod_A(\CC) \to \CC$ by forgeting the right $A$-module structure, i.e. 
$$
L^R = [A,-] = \forget. 
$$
In particular, the condensable $\EE_1$-algebra can be recovered as $L^R(A)=[A,A]=A$. 

\item This condensation preserves the gravitational anomaly (recall Remark\,\ref{rem:boundary_phase_transition}), i.e., $\SZ(\SC)^{n+2}=\SZ(\SD)^{n+2}$. In other words, $\SC^{n+1}$ and $\SD^{n+1}$ are Morita equivalent. Mathematically, it means that the indecomposable multi-fusion $n$-categories $\CC$ and $\CD$ are Morita equivalent, 
and we have the following braided equivalence:
$$
\FZ_1(\CC) \simeq \FZ_1(\Mod_A^{\EE_1}(\CC)). 
$$
\enu  
\end{pthm}

\begin{rem}
When $\SC^{n+1}=\mathbf{1}^{n+1}$, Theorem$^{\mathrm{ph}}$\,\ref{pthm:1codim_condensation_nd} (see also Theorem$^{\mathrm{ph}}$\,\ref{pthm:1codim_condensation_nd_II}) is precisely a reformulation of the layer constructions of the $n+$1D topological order $\SD^{n+1}$ from the trivial $n+$1D topological order $\mathbf{1}^{n+1}$. In this sense, general condensations can be viewed as layer constructions on a non-trivial background phase $\SC^{n+1}$. Notice that such (generalized) layer constructions can only create a new phase $\SD^{n+1}$ that connects to $\SC^{n+1}$ by a gapped domain wall. When $\SC^{n+1}=\mathbf{1}^{n+1}$, all non-chiral $n+$1D topological orders can be obtained from the condensations of 1-codimensional topological defects in $\mathbf{1}^{n+1}$. 
\end{rem}

Then the boundary-bulk relation \cite{KWZ15,KWZ17} immediately gives the following mathematical corollary.
\begin{cor} \label{pcor:E1_Mod_Z0}
Let $\CC$ be an indecomposable multi-fusion $n$-category, $A\in \Algc_{\EE_1}(\CC)$, $\CD=\Mod_A^{\EE_1}(\CC)$ and $\CM=\RMod_A(\CC)$. We have a natural monoidal equivalence (with an illustration of its physical meaning):  
\be \label{eq:1codim_condensation_bbr}
\begin{array}{c}
\begin{tikzpicture}[scale=0.6]
\fill[gray!20] (-2,0) rectangle (2,2) ;
\node at (0,1.5) {\scriptsize $\FZ_1(\CC)\simeq \FZ_1(\CD)$} ;
\node at (-1.7,-0.3) {\scriptsize $\CC$} ;
\node at (1.7,-0.3) {\scriptsize $\CD$} ;
\draw[ultra thick,->-] (-2,0) -- (0,0) ;
\draw[ultra thick,->-] (0,0) -- (2,0) ;
\draw[fill=white] (-0.08,-0.08) rectangle (0.08,0.08) node[midway,below] {\scriptsize $\CM$} ;
\end{tikzpicture}
\end{array}
\quad\quad
\begin{array}{c}
\phi: \Mod_A^{\EE_1}(\CC)^\rev \boxtimes_{\FZ_1(\CC)} \CC \xrightarrow{\simeq} \FZ_0(\RMod_A(\CC)). 
\end{array}
\ee
or equivalently, we have 
\be \label{eq:E1Mod_Z1Z0}
\Mod_A^{\EE_1}(\CC)^\rev \simeq \FZ_1(\CC, \FZ_0(\RMod_A(\CC)). 
\ee
\end{cor}

\begin{rem}
Corollary$^{\mathrm{ph}}$\,\ref{pcor:E1_Mod_Z0} generalizes \cite[Corollary\ 3.30]{DMNO13}. When $n=2$ and $\FZ_1(\CC) \simeq 2\vect$, taking looping on both sides and applying the formula (\ref{eq:FZm-FZm-1}) and Theorem\,\ref{pthm:OmegaSigmaA-AOmega}, we obtain \cite[Corollary\ 3.30]{DMNO13}. 
\end{rem}



\begin{defn} \label{defn:internal_hom_ncat}
Let $\CA$ be a $\Cb$-linear monoidal $n$-category and $\CM$ be a $\Cb$-linear left $\CA$-module. The internal hom $[x,y]$ for $x,y\in \CM$, if exists, is defined to be the object of $\CA$ representing the functor $\hom_\CM(-\odot x, y): \CA^\op \to \Cat_{n-1}^\Cb$. That is, 
\be \label{eq:internal_hom_ncat}
\hom_\CA(-,[x,y]) \simeq \hom_\CM(-\odot x,y). 
\ee
We say that $\CM$ is {\it enriched} in $\CA$ if $[x,y]$ exists for all $x,y\in \CM$. 
\end{defn}

\begin{rem} 
We want to point out that the equivalence (or better adjoint equivalence for physical applications) in (\ref{eq:internal_hom_ncat}) is a defining data of the internal hom. The higher isomorphisms in the definition of the equivalence determine the higher isomorphisms in the definition of the algebraic structure on the internal hom. 
\end{rem}

\begin{prop}[\cite{KZ24}]
For a multi-fusion $n$-category $\CA$, every separable left $\CA$-module (i.e. a left $\CA$-module in $(n+1)\vect$) is enriched in $\CA$. 
\end{prop}

The next result provide explicit constructions of condensable $\EE_1$-algebras in $\CC$ via internal homs. 

\begin{pthm} \label{pthm:1codim_condensation_nd_II}
Given two $n+$1D Morita equivalent type-I topological orders $\SC^{n+1}$ and $\SD^{n+1}$ connected by a gapped domain wall $\SM^n$ with a specified boundary condition $m\in\CM$, i.e. $\SM^n=(\CM,m)$. 
$$
\begin{tikzpicture}[scale=0.7]
\fill[gray!20] (-2,0) rectangle (2,2) ;
\node at (0,1.5) {\scriptsize $\SZ(\SC)^{n+2}=\SZ(\SD)^{n+2}$} ;
\node at (-1.7,-0.3) {\scriptsize $\SC^{n+1}$} ;
\node at (1.7,-0.3) {\scriptsize $\SD^{n+1}$} ;
\draw[ultra thick,->-] (-2,0) -- (0,0) ;
\draw[ultra thick,->-] (0,0) -- (2,0) ;
\draw[fill=white] (-0.08,-0.08) rectangle (0.08,0.08) node[midway,below] {\scriptsize $\SM^n$} ;
\end{tikzpicture}
$$
We have the following results. 
\bnu
\item[(1)] $\CM\in \BMod_{\CC|\CD}((n+1)\vect)$ and the defining bimodule structure on $\CM$ induces a monoidal equivalence: 
\be
\CD^\rev \boxtimes_{\FZ_1(\CC)} \CC \xrightarrow{\simeq} \FZ_0(\CM)=\Fun(\CM,\CM); 
\ee
or equivalently,
\be
\CD^\rev \simeq \Fun_\CC(\CM,\CM) \simeq \FZ_1(\CC,\FZ_0(\CM)). 
\ee

\item[(2)] By shrinking a $\SD^{n+1}$-phase strap in the $\SC^{n+1}$-phase, we obtain a 1-codimensional topological defect in $\SC^{n+1}$, which means, in particular, $\CM\boxtimes_\CD \CM^\op \simeq \CC$ as $\CC$-$\CC$-bimodules. This 1-codimensional topological defect in $\SC^{n+1}$ is given by the internal hom algebra $[m,m]_\CC$ in $\CC$, which is automatically a condensable $\EE_1$-algebra in $\CC$. 
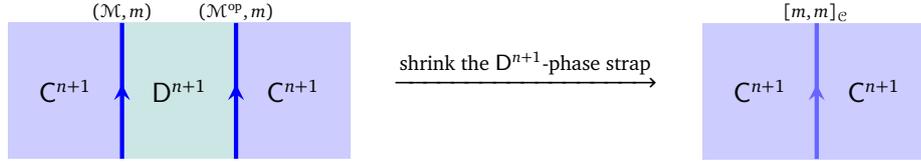
\[
\begin{array}{c}
\begin{tikzpicture}[scale=1.5]
\fill[blue!20] (0,0) rectangle (3,1.2) ;
\node at (0.5,0.6) {$\SC^{n+1}$} ;
\node at (1,1.3) {\scriptsize $(\CM,m)$} ;
\fill[teal!20] (1.02,0) rectangle (1.98,1.2) ;
\draw[blue, ultra thick,->-] (1,0) -- (1,1.2) ; 
\node at (1.5,0.6) {$\SD^{n+1}$} ;
\draw[blue, ultra thick,->-] (2,0) -- (2,1.2) ; 
\node at (2,1.3) {\scriptsize $(\CM^\op,m)$} ;
\node at (2.5,0.6) {$\SC^{n+1}$} ;
\end{tikzpicture}
\end{array}
\quad 
\xrightarrow{\mbox{\footnotesize shrink the $\SD^{n+1}$-phase strap}}
\quad
\begin{array}{c}
\begin{tikzpicture}[scale=1.5]
\fill[blue!20] (0,0) rectangle (2,1.2) ;
\node at (0.5,0.6) {$\SC^{n+1}$} ;
\node at (1,1.3) {\scriptsize $[m,m]_\CC$} ;
\draw[blue!60, ultra thick,->-] (1,0) -- (1,1.2) ; 
\node at (1.5,0.6) {$\SC^{n+1}$} ;
\end{tikzpicture}
\end{array}
\]
We have $\CM \simeq \RMod_{[m,m]_\CC}(\CC)$ and $\CD\simeq\Mod_{[m,m]_\CC}^{\EE_1}(\CC)$. Moreover, all condensable $\EE_1$-algebras in $\CC$ arise as $[m,m]_\CC$ for some $\CD, \CM, m$. .

\item[(3)] We can reverse the process. By shrinking a $\SC^{n+1}$-phase strap in the $\SD^{n+1}$-phase, we obtain a 1-codimensional topological defect in $\SD^{n+1}$, which means, in particular, $\CM^\op\boxtimes_\CC \CM \simeq \CD$ as $\CD$-$\CD$-bimodules. This 1-codimensional topological defect in $\SD$ is given by the internal hom algebra $[m,m]_\CD$ in $\CD$, which is automatically a condensable $\EE_1$-algebra in $\CD$. 
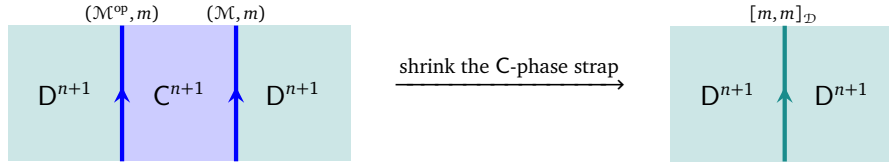
\[
\begin{array}{c}
\begin{tikzpicture}[scale=1.5]
\fill[teal!20] (0,0) rectangle (3,1.2) ;
\fill[blue!20] (1.02,0) rectangle (1.98,1.2) ;
\node at (0.5,0.6) {$\SD^{n+1}$} ;
\node at (1,1.3) {\scriptsize $(\CM^\op,m)$} ;
\draw[blue, ultra thick,->-] (1,0) -- (1,1.2) ; 
\node at (1.5,0.6) {$\SC^{n+1}$} ;
\draw[blue, ultra thick,->-] (2,0) -- (2,1.2) ; 
\node at (2,1.3) {\scriptsize $(\CM,m)$} ;
\node at (2.5,0.6) {$\SD^{n+1}$} ;
\end{tikzpicture}
\end{array}
\quad 
\xrightarrow{\mbox{\footnotesize shrink the $\SC^{n+1}$-phase strap}}
\quad
\begin{array}{c}
\begin{tikzpicture}[scale=1.5]
\fill[teal!20] (0,0) rectangle (2,1.2) ;
\node at (0.5,0.6) {$\SD^{n+1}$} ;
\node at (1,1.3) {\scriptsize $[m,m]_\CD$} ;
\draw[teal!90, ultra thick,->-] (1,0) -- (1,1.2) ; 
\node at (1.5,0.6) {$\SD^{n+1}$} ;
\end{tikzpicture}
\end{array}
\]
We have $\CM \simeq \LMod_{[m,m]_\CD}(\CD)$ and $\CC\simeq\Mod_{[m,m]_\CD}^{\EE_1}(\CD)$. Moreover, all condensable $\EE_1$-algebras in $\CD$ arise as $[m,m]_\CD$ for some $\CC, \CM, m$. 
\enu
In particular, two $n+$1D topological orders are Morita equivalent if and only if they can be obtained from each other by condensing a topological defect of codimension 1 (recall Theorem$^{\mathrm{ph}}$\,\ref{pthm:condensed=connected}).
\end{pthm}

\begin{rem}
When both $\SC^{n+1}$ and $\SD^{n+1}$ are anomaly-free, it is illuminating to rewrite the results in terms of $\Omega\CC$ and $\Omega\CD$ because they determines $\CC$ and $\CD$, respectively, by delooping. Note that $\Omega\CC$ and $\Omega\CD$ are non-degenerate braided fusion $(n-1)$-categories. All topological defects of codimension 1 and higher on a gapped domain wall form a closed multi-fusion $\Omega\CD$-$\Omega\CC$-bimodule $\CX$, i.e. a multi-fusion $(n-1)$-category equipped with a braided equivalence $\phi_\CX:  \Omega\CD^\rev \boxtimes \Omega\CC \to \FZ_1(\CX)$. We obtain the unique closed $\CC$-$\CD$-bimodule $\CM$ as $\Sigma\CX$, i.e. $\CM\simeq \Sigma\CX$. 
Moreover, the condition that $\CX$ is closed multi-fusion $\Omega\CC$-$\Omega\CD$-bimodule is equivalent to the condition that $\Sigma\CX$ is the unique closed $\CC$-$\CD$-bimodule $\CM$ mainly because $\Sigma\FZ_1(\CX) \simeq \FZ_0(\Sigma\CX)$ \cite{JF22,KLWZZ20,KZ22b}. Note that $\Sigma\CX$ is the precisely the separable $n$-category of all possible gapped wall conditions between $\SC^{n+1}$ and $\SD^{n+1}$ within a single Morita class. An object $m\in \CM$ represents a single gapped domain wall and $\hom_\CM(m,m)$ is the multi-fusion $(n-1)$-category of topological defects of codimension 1 and higher on the $m$-wall.  
\end{rem}

\begin{rem}
Similar to the 1+1D cases, once we fixed the domain wall $\SM$ by a pair $(\CM,m)$ (so is its time reversal $\overline{\SM}=(\CM^\op,m)$), then the condensation process is uniquely fixed with the condensable algebra defined by the internal hom algebra\footnote{In 2-categories, the internal hom algebra is automatically equipped with not only the unit and associativity 1-morphisms but also necessary 2-morphisms, which are automatically included as the defining data of the universal property of the internal hom in 2-categories.} $[m,m]_\CC \in \CC$. Different choices of $m$ define different condensations microscopically. But the macroscopic result of the condensation, i.e. the condensed phase, is the same. Different $m$'s give different equivalences $\CD \simeq \Mod_{[m,m]_\CC}^{\EE_1}(\CC)$ and $\CM \simeq \RMod_{[m,m]_\CC}(\CC)$, which provide different relabelings of the objects in $\CD$ and $\CM$. 
\end{rem}

\begin{rem}
A 1-codimensional topological defect $d\in \CD$ in the condensed phase $\SD^{n+1}$ 1-codimensional topological defect in $\SC^{n+1}$ as illustrated below. 
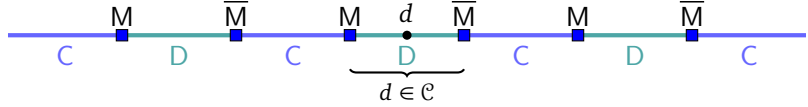
\[
\begin{tikzpicture}[scale = 1.5]
\draw[color = blue!60, ultra thick](-3,0)--(-2,0) node[midway,below,scale=1] {$\SC$};
\draw[color = teal!70, ultra thick](-2,0)--(-1,0) node[midway,below,scale=1] {$\SD$};
\draw[color = blue!60, ultra thick](-1,0)--(0,0) node[midway,below,scale=1] {$\SC$};
\draw[color = teal!70, ultra thick](0,0)--(1,0) node[midway,below,scale=1] {$\SD$};
\draw[color = blue!60, ultra thick](1,0)--(2,0) node[midway,below,scale=1] {$\SC$};
\draw[color = teal!70, ultra thick](2,0)--(3,0) node[midway,below,scale=1] {$\SD$};
\draw[color = blue!60, ultra thick](3,0)--(4,0) node[midway,below,scale=1] {$\SC$};
\draw[fill=blue] (-2.05,-0.05) rectangle (-1.95,0.05) node[midway,above,scale=1] {$\SM$} ;
\draw[fill=blue] (-1.05,-0.05) rectangle (-0.95,0.05) node[midway,above,scale=1] {$\overline{\SM}$} ;
\draw[fill=blue] (-0.05,-0.05) rectangle (0.05,0.05) node[midway,above,scale=1] {$\SM$} ;
\draw[fill=blue] (0.95,-0.05) rectangle (1.05,0.05) node[midway,above,scale=1] {$\overline{\SM}$} ;
\draw[fill=blue] (1.95,-0.05) rectangle (2.05,0.05) node[midway,above,scale=1] {$\SM$} ;
\draw[fill=blue] (2.95,-0.05) rectangle (3.05,0.05) node[midway,above,scale=1] {$\overline{\SM}$} ;
\fill[black] (0.5,0) circle (0.04) node[black] at(0.5,0.2) {$d$};
\draw[decorate,decoration=brace,thick] (1,-0.3)--(0,-0.3) ;
\node at (0.5,-0.5) {\small $d\in \CC$} ;
\end{tikzpicture}
\]
When $d=\one_\CD$, this 1-codimensional topological defect is $\one_\CD=A\in \CC$. Note that $A$ is precisely $\one_\CD$ viewed as an object in $\CC$ because $\CD \subset \CC$. Therefore, for general $d\in \CD$, we expect that this 1-codimensional topological defect is given by $[A,d]_\CC\in \CC$. When $\CD=\Mod_A^{\EE_1}(\CC)$ and $\CM=\RMod_A(\CC)$, $[A,-]: \RMod_A(\CC) \to \CC$ is precisely the forgetful functor. 
\end{rem}

\begin{rem}
Recall that the separable $(n+1)$-category $\Sigma\CC\simeq\Sigma\CD$ is a connected component of the category of gapped boundary conditions of $\SZ(\SC)^{n+2}$. In $\Sigma\CC$, $\SC^{n+1}$ is labeled by the distinguished object $\bullet\in \Sigma\CC$; $\SD^{n+1}$ is labeled by $\CM^\op \in \Sigma\CC=\RMod_\CC((n+1)\vect)$ and $\CD=\Omega_{\CM^\op}(\Sigma\CC)$. In $\Sigma\CD$, $\SD^{n+1}$ is labeled by $\bullet\in \Sigma\CD$; $\SC^{n+1}$ is labeled by $\CM\in \Sigma\CD=\RMod_\CD((n+1)\vect)$ and $\CC=\Omega_{\CM}(\Sigma\CD)$. 
\end{rem}

From a single $\SC^{n+1}$, one can obtain all $\SD^{n+1}$ that are Morita equivalent to $\SC^{n+1}$. In particular, all anomaly-free non-chiral $n+$1D topological orders can be obtained by condensing 1-codimensional topological defects in $\mathbf{1}^{n+1}$, or equivalently, by condensing condensable $\EE_1$-algebras in $n\vect$. We generalize it to a more general situation. This generalized result is essentially a reformulation of a part of Proposition$^{*}$ 3.13 in \cite{KZ24} (see also Remark\,\ref{rem:center=Morita+invertible}), 
and a reformulation of Theorem$^{\mathrm{ph}}$\,\ref{pthm:TME-action_transitive} and Theorem\,\ref{thm:same_bulk=Morita+invertible}, and was obtained independently in the third arXiv version of \cite{LYW24}.

\begin{pthm} \label{pthm:recover_all_bdy_from_one}
Any gapped boundary $\SN^n$ of an $n+$1D anomaly-free simple topological order $\SC^{n+1}$ can be obtained from a single gapped boundary $\SM^n$ by first stacking an anomaly-free $n$D topological order $\SX^n$ to $\SM^n$ then condensing a 1-codimensional topological defect in $\SM^n \boxtimes \SX^n$. 

Mathematically, it means that  two indecomposable multi-fusion $(n-1)$-categories $\CM$ and $\CN$ for $n\geq 2$ share the same $\EE_1$-center, i.e., $\FZ_1(\CM) \simeq \FZ_1(\CN)$ if and only if $\CN$ is Morita equivalent to $\CM \boxtimes \CX$ for a non-degenerate multi-fusion $(n-1)$-category $\CX$, or equivalently, $\Sigma\CN \simeq \Sigma\CM \boxtimes \CS$ for an invertible separable $n$-category $\CS$. 
\end{pthm}

\begin{rem} \label{rem:center=Morita+invertible}
The mathematical meaning of Theorem$^{\mathrm{ph}}$\,\ref{pthm:recover_all_bdy_from_one} can be reformulated slightly differently. Namely, two indecomposable multi-fusion $n$-categories $\CM$ and $\CN$ share the same $\EE_1$-center, i.e., $\FZ_1(\CM)\simeq \FZ_1(\CN)$, if and only if $\Sigma\CN \simeq \Sigma\CM \boxtimes \CS$ for an invertible separable $(n+1)$-category $\CS$. This result follows from the proof of (1)$\Rightarrow$(3) part of Proposition$^{*}$ 3.13 in \cite{KZ24}, which says that 
the center functor $\CS ep_n^{\mathrm{ind}} \to \CF us_n^{\mathrm{cl}}$ is full up to a factor of invertible separable $n$-categories. We can further reformulate it in terms of $\TME^n$ and $\TME^n(\SC^{n+1})$ as in Theorem$^{\mathrm{ph}}$\,\ref{pthm:TME-action_transitive}. 
\end{rem}

\begin{rem} \label{rem:Morita_equivalence_E1_algebras}
Theorem$^{\mathrm{ph}}$\,\ref{pthm:recover_all_bdy_from_one} automatically covers the same result for the gapped domain walls between two topological orders because a wall is a boundary by the folding trick. Since gapped domain walls between two Morita equivalence type-I topological orders $\SA^{n+1}$ and $\SB^{n+1}$ determine condensations from $\SA^{n+1}$ to $\SB^{n+1}$, the notion of Morita equivalence between two gapped domain walls naturally carries over to the notion of Morita equivalence between two condensable $\EE_1$-algebras in $\CA$. This leads to an interesting way to classify condensations from $\SA^{n+1}$ to $\SB^{n+1}$ via the center functor as we show in Theorem$^{\mathrm{ph}}$\,\ref{pthm:construct_cond_E1_algebras_II} and Corollary$^{\mathrm{ph}}$\,\ref{pcor:n=2_classification} and \ref{pcor:n=3_classification}. 
\end{rem}

We have assumed `type-I' condition on $\SC^{n+1}$ in previous discussion. Relaxing this condition gives essentially the same result as Theorem$^{\mathrm{ph}}$\,\ref{pthm:condensation_type12_1d}. For readers convenience, we write it down explicitly. 
\begin{pthm} \label{pthm:condensation_type12_nd}
In general, $\SC^{n+1}$ could be type-I or type-II and $\CC$ is a multi-fusion $n$-category. A condensable algebra $A$ in a multi-fusion $n$-category $\CC$ is a direct sum of non-zero condensable algebras in each indecomposable summand of $\CC$. Condensations preserve the multiplicity. 
\bnu
\item $\SC^{n+1}$ is type-I (i.e., $\CC$ is an indecomposable multi-fusion $n$-category): 
\bnu
\item If $A$ is indecomposable, then the condensed phase $\SD^{n+1}$ is type-I and simple (i.e., $\CD$ is a fusion 1-category), and $\CM=\RMod_A(\CC)$ is an indecomposable left $\CC$-module. 
\item If $A$ is a direct sum of indecomposable ones, i.e., $A=\oplus_{i=1}^l A_i$, then the condensed phase $\SD^{n+1}$ is type-I composite, $\mathrm{gsd}(\SD^{n+1})=l$, and $\CD$ is an indecomposable multi-fusion 1-category, which can be decomposed as a direct sum of separable 1-categories: $\CD=\oplus_{i,j=1}^l \BMod_{A_i|A_j}(\CC)$. The domain wall $\SM^n=(\CM,A)$ is of type-II and we have the decomposition of left $\CC$-modules: $\CM = \RMod_A(\CC) = \oplus_{i=1}^l \RMod_{A_i}(\CC)$. 
\enu
\item $\SC^{n+1}$ is type-II (i.e., $\CC=\oplus_{i=1}^k \CC_i$ and $\CC_i$ are indecomposable multi-fusion 1-categories): 
\bnu
\item In this case, we have $A=\oplus_{i=1}^k A_i$, where $A_i=A\cap \CC_i \neq 0$. 

\item The condensed phase $\SD^{n+1}$ is type-II, $\mathrm{mult}(\SC^2) \simeq \mathrm{mult}(\SD^2)$, 
and $\CD \simeq \oplus_{i=1}^k \BMod_{A_i|A_i}(\CC_i)$. 
\item The domain wall $\SM^n$ is type-II,  
and we have the decomposition of left $\CC$-modules: $\CM = \RMod_A(\CC) = \oplus_{i=1}^k \RMod_{A_i}(\CC_i)$. 
\enu

\enu
\end{pthm}

Since all the subtleness associated to the type-I or type-II composite property are summarized in above theorem, from now on, we mainly focus on the condensations in type-I or simple topological orders.

\subsubsection{General constructions} \label{sec:general_example_1-codim}
In this subsubsection, we provide some general examples, in which we either provide a direct construction of condensable $\EE_1$-algebras in $\CC$ then determine $\CD$ and $\CM$ (as in Theorem$^{\mathrm{ph}}$\,\ref{pthm:1codim_condensation_nd}), or provide $\CC$ and $\CM$ with concrete coordinate systems such that the condensable $\EE_1$-algebra $[m,m]_\CC$ in $\CC$ can be determined explicitly (as in Theorem$^{\mathrm{ph}}$\,\ref{pthm:1codim_condensation_nd_II}). 

\medskip
When $\SC^{n+1}=\mathbf{1}^{n+1}$, a condensable $\EE_1$-algebra in $n\vect$ is precisely a multi-fusion $(n-1)$-category $\CA$. By condensing it, we obtain the condensed phase $\SD^{n+1}$ and a gapped domain wall $\SM^n$ such that 
$$
\CD=\Mod_\CA^{\EE_1}(n\vect), \quad\quad (\CM,m) = (\RMod_\CA(n\vect), \CA) = (\Sigma\CA,\CA), \quad\quad \CD^\op\simeq \FZ_0(\Sigma\CA). 
$$
We want to understand this result from a different perspective in the cases (1) and (2) below. 

\medskip
(1). $\SC^{n+1}=\SD^{n+1}=\mathbf{1}^{n+1}$: It turns out that this seemingly ``trivial'' case can have non-trivial condensations. In this case, a gapped domain wall $\SM^n$ between $\SC^{n+1}$ and $\SD^{n+1}$ is nothing but an anomaly-free $n$D topological order. Again, $\SM^n=(\CM,m)$, where $\CM$ is a separable $n$-category, $m$ is a distinguished object in $\CM$ and labels the $n$D anomaly-free topological order $\SM^n$ as a boundary condition of $\mathbf{1}^{n+1}$. By boundary-bulk relation, we must have
\be \label{eq:Z0_Mm=nVec}
\FZ_0(\CM,m) = \Fun(\CM,\CM) \simeq n\vect. 
\ee
As a consequence, $\CM$ is necessarily indecomposable (actually invertible in $(n+1)\vect$ by \cite[Corollary\ 2.10]{KZ24}). The category of topological defects on $\SM^n$ is $\Omega_m\CM$ is necessarily an indecomposable multi-fusion $(n-1)$-category. Since $\CM$ is indecomposable, we have $\CM = \Sigma\Omega_m\CM$. By boundary-bulk relation, we have $$
\FZ_1(\Omega_m\CM) \simeq (n-1)\vect,
$$ 
which is an immediate consequence of (\ref{eq:Z0_Mm=nVec}), $\CM = \Sigma\Omega_m\CM$ and (\ref{eq:Zm_center}).

Note that the equivalence $\CM = \Sigma\Omega_m\CM \simeq \RMod_{\Omega_m\CM}(n\vect)$ provides a coordinate system that allows us to identify the condensable $\EE_1$-algebra $[m,m]_\CC$ in $\CC$ explicitly. Indeed, in this coordinate system $m=\Omega_m\CM \in \RMod_{\Omega_m\CM}(n\vect)$, and the left bulk-to-wall functor can be rewritten as follows. 
$$
L_1= -\odot m = -\otimes \Omega_m\CM: n\vect \to \RMod_{\Omega_m\CM}(n\vect), 
$$
which is illustrated in the following picture. 
$$
\begin{array}{c}
\begin{tikzpicture}[scale=0.8]
\draw[blue!20,fill=blue!20] (0,0)--(0,2)--(1,3)--(1,1)--cycle ;
\node at (3,1.5) {\footnotesize $\CM=\Sigma\Omega_m\CM=\RMod_{\Omega_m\CM}(n\vect)$} ;
\node at (-1.5,2.5) {\footnotesize $\mathbf{1}^{n+1}$} ;
\node at (3.5,2.5) {\footnotesize $\mathbf{1}^{n+1}$} ;
\node at (-2,1) {\scriptsize $n\vect \xrightarrow{L_1 = -\odot m} \CM$} ;
\node at (-1.5,0.4) {\scriptsize $a \mapsto a\otimes \Omega_m\CM$} ; 
\node at (2.5,0.6) {\scriptsize $m = \Omega_m\CM \in \RMod_{\Omega_m\CM}(n\vect)$} ; 
\end{tikzpicture}
\end{array}
$$
Then its right adjoint functor is precisely the forgetful functor, i.e., 
$$
L_1^R = \forget: \RMod_{\Omega_m\CM}(n\vect) \to n\vect. 
$$ 
Therefore, $L_1^R(\Omega_m\CM)=[\Omega_m\CM,\Omega_m\CM]_{n\vect} =\Omega_m\CM$ is precisely the indecomposable condensable $\EE_1$-algebra in $n\vect$ that defines the condensation. Note that, in the case, the internal hom $[m,m]_\CC$ is no longer an abstract nonsense. It is the indecomposable multi-fusion $(n-1)$-category $\Omega_m\CM$, which is precisely an indecomposable condensable $\EE_1$-algebra in $\CC=n\vect$. 
By Theorem$^{\mathrm{ph}}$\,\ref{pthm:1codim_condensation_nd_II}, we obtain 
$$
\CD = n\vect \simeq \Mod_{\Omega_m\CM}^{\EE_1}(n\vect), \quad\quad
\CM \simeq \RMod_{\Omega_m\CM}(n\vect), 
$$
where the second equivalence is tautological and can be viewed as a consistence check, but the first monoidal equivalence is non-trivial mathematically. We reformulate this result mathematically. 
\begin{prop} \label{cor:condense_nDTO_trivial_phase}
A Lagrangian $\EE_1$-algebra in $n\vect$ (recall Definition\,\ref{def:non-degenerate_Lagrangian}) is precisely a non-degenerate multi-fusion $(n-1)$-category $\CA$. In this case, we have 
\be
\Mod_\CA^{\EE_1}(n\vect) \simeq n\vect. 
\ee
\end{prop}

\begin{rem} \label{rem:bulk-to-wall_map}
The subscript of $L_1$ represents that it is left bulk-to-wall map of 1-codimensional topological defects. We denote the left bulk-to-wall map of $k$-codimensional topological defects by $L_k: \Omega^{k-1}\CC \to \Omega^{k-1}\CM$ and the right bulk-to-wall map of $k$-codimensional topological defect by $R_k: \Omega^{k-1}\CD \to \Omega^{k-1}\CM$. 
\end{rem}

\begin{rem} \label{rem:condensing_afTO}
Proposition\,\ref{cor:condense_nDTO_trivial_phase} is the mathematical formulation of the physical fact that condensing an anomaly-free $n$D topological order $\SM^n$, viewed as 1-codimensional topological defects in the trivial phase $\mathbf{1}^{n+1}$, reproduces the trivial phase $\mathbf{1}^{n+1}$, and gapped domain wall of this condensation is precisely the anomaly-free $n$D topological order $\SM^n$. 
\end{rem}

(2). $\SC^{n+1}=\mathbf{1}^{n+1}\neq \SD^{n+1}$: In this case, the discussion is completely parallel to the previous case. In particular, $\CM$ is a separable $n$-category and $\Omega_m\CM$ is a multi-fusion $(n-1)$-category. If $m$ is non-zero in each connected component of $\CM$, then we obtain 
$$
\CD \simeq \Mod_{\Omega_m\CM}^{\EE_1}(n\vect) \simeq \Fun(\CM,\CM)^\op \simeq \FZ_0(\CM,m)^\op, \quad
\quad \CM = \Sigma\Omega_m\CM = \RMod_{\Omega_m\CM}(n\vect).
$$

We reformulate the associated mathematical result as a generalization of Proposition\,\ref{cor:condense_nDTO_trivial_phase}. 
\begin{prop} \label{cor:condense_nDTO_trivial_phase_2}
A condensable $\EE_1$-algebra in $n\vect$ is precisely a multi-fusion $(n-1)$-category $\CA$. We have 
\be \label{eq:condense_nDTO_trivial_phase_2}
\Mod_\CA^{\EE_1}(n\vect)^\rev \simeq \FZ_0(\Sigma\CA).  
\ee
If $\CA$ is an indecomposable multi-fusion $(n-1)$-category, then we have \cite{KZ22b}
\be
\Mod_\CA^{\EE_1}(n\vect)^\rev \simeq \FZ_0(\Sigma\CA) \simeq \Sigma\FZ_1(\CA). 
\ee
\end{prop}

\begin{rem}
The Levin-Wen model \cite{LW05} can be viewed as a physical way to realize the condensation defined by condensing a simple condensable $\EE_1$-algebra $\CA$ in $2\vect$ (i.e., $\CA$ is a fusion 1-category) with a gapped domain wall $\SM^2=(\Sigma\CA,\CA)$ constructed in \cite{KK12}. Similarly, a higher dimensional generalization of Levin-Wen model based on indecomposable multi-fusion $(n-1)$-category $\CA$ (predicted and briefly described in \cite[Remark\, 5.11]{KZ22b}) should be viewed as a physical way to realize the condensation defined by the indecomposable condensable $\EE_1$-algebra $\CA$ in $n\vect$. 
\end{rem}

\begin{rem}
We expect that the layer construction of 3+1D topological orders via layers of 2+1D topological orders \cite{JQ14} provides a concrete realization of a condensation of 1-codimensional defects in $\mathbf{1}^4$. For example, the layer construction of the 3+1D $\Zb_2$ topological order \cite{ZLZHKT23} should be viewed as a physical realization of the 1-codimensional condensation that produces the 3+1D $\Zb_2$ topological order from $\mathbf{1}^4$.
\end{rem}

\medskip
(3). $\SZ(\SC)^{n+2}=\mathbf{1}^{n+2}$: We further assume that $\SC^{n+1}$ is simple. In this case, $\CC=\Sigma\Omega\CC=\RMod_{\Omega\CC}(n\vect)$ provides a concrete coordinate system to the non-degenerate fusion $n$-category $\CC$. By Remark\,\ref{rem:Yoneda_functor=id}, the fusion product in this coordinate system is defined by $\CK \otimes^1 \CL := \CK \boxtimes_{\Omega\CC} \CL$ as depicted in the following picture.
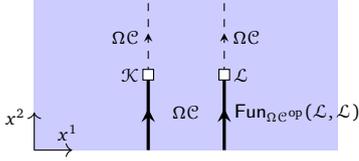
\be \label{pic:x_tensor_y}
\begin{array}{c}
\begin{tikzpicture}[scale=1]
\fill[blue!20] (-2,0) rectangle (2,2) ;
%
\draw[->-,very thick] (-0.5,0)--(-0.5,1) ; 
\draw[->-,very thick] (0.5,0)--(0.5,1) node[midway,right] {\scriptsize $\Fun_{\Omega\CC^\rev}(\CL,\CL)$} ; 
%
\draw[dashed,->-] (-0.5,1)--(-0.5,2) node[midway,left] {\scriptsize $\Omega\CC$} ;
\draw[dashed,->-] (0.5,1)--(0.5,2) node[midway,right] {\scriptsize $\Omega\CC$};
\draw[fill=white] (-0.57,0.93) rectangle (-0.43,1.07) node[midway,left] {\scriptsize $\CK$} ;
\draw[fill=white] (0.43,0.93) rectangle (0.57,1.07) node[midway,right] {\scriptsize $\CL$} ;
%
%
\node at (0,0.5) {\scriptsize $\Omega\CC$} ;
\draw[->] (-2,0) -- (-2,0.5) node [very near end, left] {\scriptsize $x^2$} ;
\draw[->] (-2,0) -- (-1.5,0) node [very near end, above] {\scriptsize $x^1$} ;
\end{tikzpicture}
\end{array}
\quad\quad
\begin{array}{l}
\mbox{\small The right $\Omega\CC$-module structures and tensor product:} \\
\mbox{\small $(1). \quad \CK \times \Omega\CC \xrightarrow{\odot} \CK; \quad (k,a) \mapsto k\otimes^2 a$}; \\  
\mbox{\small $(2). \quad  \CL \times \Omega\CC \xrightarrow{\odot} \CL; \quad (l,a) \mapsto l\otimes^2 a$}; \\
\mbox{\small $(3). \quad \boxtimes_{\Omega\CC}: \CK \times \CL \to \CK \boxtimes_{\Omega\CC} \CL, \quad
(k,l) \to k\boxtimes_{\Omega\CC} l$}. 
\end{array}
\ee

In this case, it is possible to classify all the condensable $\EE_1$-algebras in $\CC$ via the following equivalence\footnote{The proof of this equivalence was known for the $n=2$ case in \cite{BJS21}. The same proof works for general $n$. We provide a physical proof of this result in (\ref{pic:CMD_A}), (\ref{pic:OmegaM=E1_algebra}) and (\ref{eq:OmegaM_two_algebras}).}: 
\be \label{eq:AlgRMod=LModAlg}
\Algc_{\EE_1}(\RMod_{\Omega\CC}(n\vect)) \simeq \LMod_{\Omega\CC}(\Algc_{\EE_1}(n\vect)).
\ee
More explicitly, an indecomposable condensable $\EE_1$-algebra $\CA$ in $\CC$ is precisely an indecomposable multi-fusion $(n-1)$-category $\CA$ equipped with a monoidal action functor $\odot: \Omega\CC \times \CA \to \CA$ that defines the structure of left $\Omega\CC$-module on $\CA$, or equivalently, an indecomposable multi-fusion $(n-1)$-category $\CA$ equipped with a central functor $\Omega\CC \to \CA$ (or equivalently, a braided monoidal functor $\phi: \Omega\CC \to \FZ_1(\CA)$)\footnote{When $n=2$, such $\CA$ is also called a left multi-fusion $\Omega\CC$-module \cite{KZ18} or a module tensor category \cite{HPT16}.}. 

\begin{expl}
For $A\in \Algc_{\EE_2}(\Omega\CC)$ (see Section\,\ref{sec:2codim_2d}), we have $\RMod_A(\Omega\CC) \in \LMod_{\Omega\CC}(\Algc_{\EE_1}(n\vect))$. This example is important later. 
\end{expl}

For an indecomposable condensable $\EE_1$-algebra $\CA$ in $\CC$, we can condense it and obtain a condensed phase $\SD^{n+1}$, which is also simple. We have 
\begin{align*}
&\CD \simeq \Mod_\CA^{\EE_1}(\CC) \simeq \Sigma \FZ_2(\Omega\CC, \FZ_1(\CA)), \quad\quad
\Omega\CD \simeq  \FZ_2(\Omega\CC, \FZ_1(\CA)); \\
&\CM \simeq \RMod_\CA(\CC) \simeq \RMod_\CA(n\vect), \quad\quad  m=\CA \in \RMod_\CA(\CC),
\end{align*}
where the last ``$\simeq$'' is due to the fact that the right $\Omega\CC$-action factors through the right $\CA$-action. Moreover, we have
$$
\Omega_m\CM = \hom_{\RMod_\CA(\CC)}(\CA,\CA) \simeq \Fun_{\CA^\rev}(\CA,\CA) \simeq \CA. 
$$
It means that $\CA$ is precisely the category of topological defects on the gapped wall $(\CM,m)$ between $\SC^{n+1}$ and $\SD^{n+1}$. This fact reveals the geometric relation between $\Omega\CC, \CA, \Omega\CD$ as illustrated in the following picture. 
\be \label{pic:CMD_A}
\begin{array}{c}
\begin{tikzpicture}[scale=1.6]
\fill[blue!20] (1,0) rectangle (3,1.2) ;
\node at (1.2,0.2) {\scriptsize $\Omega\CC$} ;
\node at (2,-0.1) {\scriptsize $\Omega_m\CM=\CA$} ;
\fill[teal!20] (2.02,0) rectangle (3,1.2) ;
\draw[blue, ultra thick,->-] (2,0) -- (2,1.2) node [midway,left] {\scriptsize $m=\CA$}; 
\node at (2.8,0.2) {\scriptsize $\Omega\CD$} ;
\node at (1.5,1) {\scriptsize $\Omega\CC \xrightarrow{- \odot \one_\CA} \CA$} ;
\node at (2.5,1) {\scriptsize $\CA \xleftarrow{\one_\CA \odot -} \Omega\CD$} ;
\end{tikzpicture}
\end{array}
\quad\quad
\begin{array}{c}
\begin{tikzpicture}[>=Stealth]
\fill[blue!20] (-4,0) rectangle (-1,2) ;
\fill[teal!20] (-3,0) .. controls (-3,1.5) and (-2,1.5) .. (-2,0)--(-3,0)..controls (-2.5,0) .. (-2,0)--cycle;
\node at (-2.5,0.5) {\scriptsize $\Omega\CD$} ;
\draw[blue,ultra thick,-stealth] (-3,0) .. controls (-3,1.5) and (-2,1.5) .. (-2,0) ;
\draw[dashed] (-2.5,2) -- (-2.5,1.2) ;
\draw[decorate,decoration=brace,very thick] (-2,-0.1)--(-3,-0.1) ;
\node at (-1.65,0.6) {\scriptsize $\Omega_m\CM$} ;
\node at (-3.4,0.6) {\scriptsize $\Omega_m\CM$} ;
\filldraw [blue] (-2.5,1.15) circle [radius=2pt] ;
\node at (-2.05,1.25) {\scriptsize $\Omega_m\CM$} ;
\node at (-2.5,-0.4) {\scriptsize $\Omega_m\CM \boxtimes_{\Omega\CD} \Omega_m\CM^\rev$};
\draw[->] (-4,0) -- (-4,0.5) node [very near end, left] {\scriptsize $x^2$} ;
\draw[->] (-4,0) -- (-3.5,0) node [very near end, above] {\scriptsize $x^1$} ;
\node at (-3.6,1.5) {\scriptsize $\Omega\CC$} ;
\node at (-2.25,1.8) {\scriptsize $\Omega\CC$} ;
\end{tikzpicture}
\end{array}
\ee
Note that the 1-codimensional defect in $\CC$ obtained by horizontally squeezing the $\Omega\CD$-region in the second picture in (\ref{pic:CMD_A}) is labeled by the object $\Omega_m\CM \in \CC=\RMod_{\Omega\CC}(n\vect)$; and higher codimensional topological defects on it form the following multi-fusion $(n-1)$-category: 
$$
\Omega_{\Omega_m\CM}(\CC) \simeq \Omega_m\CM \boxtimes_{\Omega\CD} \Omega_m\CM^\rev \simeq \Fun_{\Omega\CC^\op}(\Omega_m\CM, \Omega_m\CM). 
$$
In general, when an object $\Omega_m\CM$ is viewed as an algebra in $\CC=\RMod_{\Omega\CC}(n\vect)$, its algebraic structure does not necessarily coincide with the monoidal structure on $\Omega_m\CM^\op$ defined by the particle fusions along the wall $(\CM,m)$. 
However, in this case, two structures indeed coincide as illustrated in the following picture.
\be \label{pic:OmegaM=E1_algebra}
\begin{array}{c}
\begin{tikzpicture}[>=Stealth]
\fill[blue!20] (-4,0) rectangle (1,2) ;
\draw[dashed] (-0.5,2) -- (-0.5,1.25) ;
\draw[dashed] (-2.5,2) -- (-2.5,1.25) ;
\fill[teal!20] (-3,0) .. controls (-3,0.5) and (-3,1.25) .. (-2.5,1.25) .. controls (-2,1.25) and (-2,0.5) .. (-1.5,0.5) .. controls (-1,0.5) and (-1,1.25) .. (-0.5,1.25) .. controls (0,1.25) and (0,0.5) .. (0,0)--cycle ;
\draw[blue,ultra thick,->] (-3,0) .. controls (-3,0.5) and (-3,1.25) .. (-2.5,1.25) .. controls (-2,1.25) and (-2,0.5) .. (-1.5,0.5) .. controls (-1,0.5) and (-1,1.25) .. (-0.5,1.25) .. controls (0,1.25) and (0,0.5) .. (0,0) ;
\filldraw [blue] (-2.5,1.25) circle [radius=2pt] ;
\filldraw [blue] (-0.5,1.25) circle [radius=2pt] ;
\filldraw [blue] (-1.5,0.5) circle [radius=2pt] ;
\filldraw [blue] (-3,0.5) circle [radius=2pt] ;
\filldraw [blue] (0,0.5) circle [radius=2pt] ;
\draw[->] (-4,0) -- (-4,0.5) node [very near end, left] {\scriptsize $x^2$} ;
\draw[->] (-4,0) -- (-3.5,0) node [very near end, above] {\scriptsize $x^1$} ;
\node at (-1.5,0.2) {\scriptsize $\Omega\CD$} ;
\node at (0.4,0.5) {\scriptsize $1_m$} ;
\node at (-3.35,0.5) {\scriptsize $1_m$} ;
\node at (-0.1,1.4) {\scriptsize $\Omega_m\CM$} ;
\node at (-2.1,1.4) {\scriptsize $\Omega_m\CM$} ;
\node at (-1.5,0.75) {\scriptsize $1_m$} ;
\node at (-3.6,1.5) {\scriptsize $\Omega\CC$} ;
\node at (-2.75,1.8) {\scriptsize $\Omega\CC$} ;
\node at (-0.75,1.8) {\scriptsize $\Omega\CC$} ;
\draw[dashed] (-3.15,0.4) rectangle (0.17,0.6) ; 
\draw[dashed] (-2.57,1.15) rectangle (-0.42,1.35) ;
\end{tikzpicture}
\end{array}
\ee
Indeed, the two top blue dots labeled by $\Omega_m\CM$ (or better the top dashed box) represent the object $\Omega_m\CM \boxtimes_{\Omega\CC} \Omega_m\CM$ in $\CC=\RMod_{\Omega\CC}(n\vect)$ because (recall Remark\,\ref{rem:Yoneda_functor=id})
$$
\hom_\CC(\one_\CC, \Omega_m\CM \boxtimes_{\Omega\CC} \Omega_m\CM) \simeq \Fun_{\Omega\CC^\op}(\Omega\CC, \Omega_m\CM \boxtimes_{\Omega\CC} \Omega_m\CM) \simeq \Omega_m\CM \boxtimes_{\Omega\CC} \Omega_m\CM. 
$$
The three lower blue dots labeled by $1_m\in \Omega_m\CM$ (or better the lower dashed box) define a 1-morphism $\Omega_m\CM \boxtimes_{\Omega\CC} \Omega_m\CM \xrightarrow{} \Omega_m\CM$, which defines the algebraic structure on the 1-codimensional defect labeled by $\Omega_m\CM$.  Using Remark\,\ref{rem:Yoneda_functor=id}, we see that it induces on $\hom_\CC(\one_\CC, \Omega_m\CM)=\Omega_m\CM$ an algebraic structure, i.e., $\Omega_m\CM \otimes_{\Omega\CC} \Omega_m\CM \to \Omega_m\CM$, defined by fusing two dashed boxes, which defines the following functor,    
\be \label{eq:OmegaM_two_algebras}
a\boxtimes_{\Omega\CC} b \mapsto 1_m\otimes a \otimes 1_m \otimes b \otimes 1_m \simeq a\otimes b. 
\ee
This means that such defined algebraic structure on $\Omega_m\CM$ in $\CC$ coincides with the monoidal structure (i.e., particle fusions) on $\Omega_m\CM$. This fact also provides a physical explanation or proof of the equivalence (\ref{eq:AlgRMod=LModAlg}). 

By choosing different $\CM,m,\CD$'s, we recover all possible condensations. It means that we can reverse the story. More explicitly, given $\SM^n=(\CM,m)$, where $\CM$ is a separable $n$-category and $\Omega_m\CM$ is a multi-fusion $(n-1)$-category. We assume that $m$ is non-zero in each indecomposable left $\CC$-submodule of $\CM$. 
By moving 2-codimensional topological defects in $\SC^{n+1}$ onto the wall $\SM^n$, we obtain an action $ \Omega\CC \times \Omega_m\CM \to \Omega_m\CM$, which is automatically monoidal, i.e., $\Omega_m\CM$ is monoidal and the monoidal structure is compatible with the $\Omega\CC$-action. As a consequence, we have
$$
\Omega_m\CM \in \LMod_{\Omega\CC}(\Algc_{\EE_1}(n\vect)) \simeq \Algc_{\EE_1}(\RMod_{\Omega\CC}(n\vect)). 
$$
In other words, $\Omega_m\CM$ is a condensable $\EE_1$-algebra in $\CC=\Sigma\Omega\CC$. Moreover, $[m,m]_\CC$ can be identified with $\Omega_m\CM$, and we have
\be \label{eq:D_cond_OmM}
\CD \simeq \Mod_{\Omega_m\CM}^{\EE_1}(\CC), \quad\quad
\CM \simeq \RMod_{\Omega_m\CM}(\CC) \simeq \RMod_{\Omega_m\CM}(n\vect). 
\ee
In this case, we also have $\CD^\rev \simeq \Mod_{\Omega_m\CM}^{\EE_1}(\CC)^\rev \simeq \FZ_1(\CC,\FZ_0(\RMod_{\Omega_m\CM}(\CC)))$. We summarize the key point in this case in the following theorem. 

\begin{pthm} \label{thm:condensable_E1Alg_af_C}
When $\SC^{n+1}$ is anomaly-free and simple, an indecomposable condensable $\EE_1$-algebra in the fusion $n$-category $\CC=\Sigma\Omega\CC$ is precisely given by an indecomposable multi-fusion $(n-1)$-category $\CA$ equipped with a braided monoidal functor $\phi: \Omega\CC \to \FZ_1(\CA)$ (or equivalently, by a type-I gapped domain wall between $\SC^{n+1}$ and the condensed phase obtained by condensing $\CA$). Moreover, we have the following equivalence of categories:  
\be \label{eq:RModAlg=AlgRMod}
\Algc_{\EE_1}(\RMod_{\Omega\CC}(n\vect)) \simeq 
\LMod_{\Omega\CC}(\Algc_{\EE_1}(n\vect)).
\ee
By condensing $\CA$ in $\CC$, we obtain a condensed phase $\SD^{n+1}$ and a gapped domain wall $\SM^n$ such that 
\begin{align*}
&\CD \simeq \Mod_\CA^{\EE_1}(\CC) \simeq \Sigma \FZ_2(\Omega\CC, \FZ_1(\CA)), \quad\quad
\Omega\CD \simeq  \FZ_2(\Omega\CC, \FZ_1(\CA)); \\
&\CM \simeq \RMod_\CA(\CC) \simeq \RMod_\CA(n\vect), \quad\quad  m=\CA \in \RMod_\CA(\CC),
\end{align*}
Moreover, $\CA=\Omega_m\CM$ provides the physical meaning of $\CA$ as the $(n-1)$-category of topological defects on the wall $\SM^n=(\CM,m)$. If $\phi$ is an equivalence, then $\CA$ is ``Lagrangian'' in the sense that $\SD^{n+1}=\mathbf{1}^{n+1}$. 
\end{pthm}

\begin{rem} \label{rem:MorEq_condensable_E1Alg}
When $\SC^{n+1}$ is anomaly-free and simple, above theorem allows us to define the Morita equivalence between two indecomposable $\EE_1$-algebras in $\CC$ according to that between two  type-I gapped domain walls. When $\SC^{n+1}$ is anomalous and simple, in general, (\ref{eq:RModAlg=AlgRMod}) only gives a subset of all condensable $\EE_1$-algebras in $\CC$ because $\Sigma\Omega\CC$ is only a fusion full subcategory of $\CC$. 
\end{rem}

\begin{rem} \label{rem:ingo}
In the orbiford of a 2+1D Reshetikhin-Turaev TQFT \cite{CRS20}, the initial data of this orbiford construction is called `orbiford data', which consists of (1) $\CK$ a modular tensor category that defines the initial 2+1D Reshetikhin-Turaev TQFT; (2)
$A$ is a $\Delta$-separable symmetric Frobenius algebra in $\CK$; (3) $T={}_AT_{AA}$ is an $A$-$A\otimes A$-bimodule in $\CK$; (4) certain choices of morphisms, satisfying conditions in \cite[Eq. (3.16-3.20)]{CRS20}. We were informed by Ingo Runkel that it is possible to show that this data, together with condition (3.16) \& (3.17) in \cite{CRS20}, amounts to say that $\LMod_A(\CK) \in \LMod_{\CK}(\Algc_{\EE_1}(2\vect))$. It suggests that this `orbiford construction' should be viewed as the counterpart (in the TQFT framework) of our 1-codimensional defect condensation theory. We hope to see more detailed clarification of their relation in the future. 
\end{rem}

By Theorem$^{\mathrm{ph}}$\,\ref{pcor:chiral_TO_wall_1to1} and Remark\,\ref{rem:P=C=A_B=K}, we obtain a classification result for 1-codimensional defect condensations between two (potentially chiral) anomaly-free simple topological orders. 

\begin{pthm} \label{pcor:1codim_condensation_bw_two_chiral_TO}
Given two Morita equivalent anomaly-free simple topological orders $\SC^{n+1}$ and $\SD^{n+1}$ connected by a gapped domain wall $\SK^n$ as illustrated in the first picture below. Using the folding trick and topological Wick rotation (TWR) (recall Figure\,\ref{fig:chiral_TO_folding_TWR}), we obtain a one-to-one correspondence between 
\be \label{pic:chiral_TO_1codim_condensations}
\begin{array}{c}
\begin{tikzpicture}[scale=0.9]
\fill[gray!20] (-3,0) rectangle (1,3) ;
\draw[ultra thick,->-] (-1,1)--(-1,3) ; 
\draw[ultra thick,->-] (1,1)--(-1,1) node [midway,below] {\footnotesize $\SK^n$} ;
\draw[fill=white] (-1.1,0.9) rectangle (-0.9,1.1) node[midway,below] {\footnotesize $\SM^n$} ;
\node at (-1,0.3) {\footnotesize gapless} ;

\node at (-2,2) {\footnotesize $\SC^{n+1}$} ;
\node at (0,2) {\footnotesize $\SD^{n+1}$} ;
\end{tikzpicture} 
\end{array}
\quad\quad\xrightarrow{\mbox{\scriptsize folding + TWR}}\quad\quad
\begin{array}{c}
\begin{tikzpicture}[scale=0.9]
\fill[gray!20] (-3,1) rectangle (-1,3) ;
\draw[ultra thick,->-] (-1,1)--(-1,3) ;
\draw[ultra thick,->-] (-3,1)--(-1,1) ;
\draw[fill=white] (-1.1,0.9) rectangle (-0.9,1.1) node[midway,below] {\footnotesize $\SX^{n-1}$} ;
\node at (-2.5,0.7) {\footnotesize $\SK^n \boxtimes \ST^n$} ;
\node at (-0.7,0.3) {\footnotesize gapped} ;
\node at (-2,2) {\footnotesize $(\SC\boxtimes\overline{\SD})^{n+1}$} ;
\end{tikzpicture} 
\end{array}
\ee
\bnu
\item[-] $(\CK\boxtimes \CT)$-modules modulo the equivalence relation: $\CX \sim \CY$ if $\CY \simeq \CX \boxtimes \CV$ for $\CV \in (n+1)\vect^\times$; 
\item[-] indecomposable condensable $\EE_1$-algebras in $\CC$ that define 1-codimensional defect condensations from $\SC^{n+2}$ to $\SD^{n+2}$ in the Morita class $[\SK^{n+1} \boxtimes \ST^{n+1}]$. 
\enu
defined by $\CX \mapsto \Fun_{\CK \boxtimes \CT}(\CX,\CX)^\op$. The condensable $\EE_1$-algebra $A=\Fun_{\CK \boxtimes \CT}(\CX,\CX)^\op$ is simple if and only if the $(\CK\boxtimes \CT)$-module $\CX$ is indecomposable. 
\end{pthm}

\begin{rem}
Theorem$^{\mathrm{ph}}$\,\ref{pcor:1codim_condensation_bw_two_chiral_TO} reduces the problem of classifying 1-codimensional defect condensations between two chiral topological orders to the problem of classifying gapped boundaries of a non-chiral topological order. 
Unfortunately, not much chiral topological orders in higher dimensions are known. We will return to the problem of constructing interesting chiral examples in the future. In this work, we focus on examples from non-chiral topological orders. 
\end{rem}

\medskip
(4). $\SC^{n+1}=\SZ(\SB)^{n+1}$: In this case, $\SC^{n+1}$ is anomaly-free and non-chiral, and admits a gapped boundary $\SB^n$, which is assumed to be simple. Then $\SC^{n+1}$ is also simple. Therefore, this case is just a special case of (3). However, in this case, $\CC$ has another natural coordinate system given by 
$$
\CC = \BMod_{\CB^\op|\CB^\op}(n\vect) \simeq \BMod_{\CB|\CB}(n\vect)^\op,
$$
where $\CB$ is the fusion $(n-1)$-category of all topological defects in $\SB^n$. An indecomposable condensable $\EE_1$-algebra in $\CC$ is precisely an indecomposable multi-fusion $(n-1)$-category $\CA^\op$ equipped with a monoidal functor $\psi:\CB \to \CA$. By condensing the condensable $\EE_1$-algebra $\CA^\op$ in $\CC$, we obtain the condensed phase $\SD^{n+1}$ and $\SM^n$ as follows: 
\begin{align}
\CD  &= \Mod_{\CA^\op}^{\EE_1}(\CC) \simeq \Mod_{\CA^\op}^{\EE_1}(n\vect) \simeq \Sigma \FZ_1(\CA),  \\
(\CM,m) &= (\LMod_{\CA}(\CC),\CA) = (\BMod_{\CA|\CB}(n\vect), \CA), \label{eq:BMod_BA}
\end{align}
where the first ``$\simeq$'' is due to the fact that the $\CB$-module structure factors through the monoidal functor $\psi$. As a consequence, we obtain 
\be \label{eq:Fun_BA_AA}
\Omega\CD \simeq \FZ_1(\CA), \quad\quad
\Omega_m\CM \simeq \Fun_{\CA|\CB}(\CA,\CA). 
\ee
Note that, according to \cite[Theorem\ 3.2.3]{KZ18}\cite{KZ22b}, $\Fun_{\CB|\CA^\op}(\CA,\CA)$ is precisely the closed $\Omega\CD^\op$-$\Omega\CC$-bimodule that defines alternatively the monoidal functor $\psi: \CB^\op \to \CA^\op$ as the following composed functor (with an illustration of its physical meaning). 
\be \label{pic:functor_psi}
\begin{array}{c}
\begin{tikzpicture}[scale=1]
\fill[blue!20] (-2,0) rectangle (0,2) ;
\fill[teal!20] (0,0) rectangle (2,2) ;
%
\draw[->-,ultra thick] (-2,0)--(-2,2) node[midway,left] {\scriptsize $\CB^\op$} ;

\draw[blue,->-,ultra thick] (0,0)--(0,2) node[midway,right] {\scriptsize $\Omega_m\CM$} ; 
%
%
%
\node at (-1,0.5) {\scriptsize $\FZ_1(\CB)$} ;
\node at (1,0.5) {\scriptsize $\Omega\CD$} ;
\node at (-1,1.5) {\scriptsize $\SZ(\SB)^{n+1}$} ;
\node at (1,1.5) {\scriptsize $\SD^{n+1}$} ;
\node at (0,2.2) {\scriptsize $\Fun_{\CA|\CB}(\CA,\CA)$} ;
\node at (-2,2.2) {\scriptsize $\CB^\op$} ;

\draw[decorate,decoration=brace,very thick] (-2,2.4)--(0,2.4) node[midway, above] {\scriptsize $\CA^\op$};

\end{tikzpicture}
\end{array}
\quad\quad
\CB^\op \to  \CB^\op \boxtimes_{\FZ_1(\CB)} \Fun_{\CA|\CB}(\CA,\CA) \simeq \Fun_{\CA}(\CA,\CA) \simeq \CA^\op. 
\ee



\begin{rem}
It is very illuminating to see the connection between the two coordinate systems in this case. By \cite{KLWZZ20}, these 
two coordinate systems $\CC=\RMod_{\Omega\CC}(n\vect)$ and $\CC=\BMod_{\CB^\op|\CB^\op}(n\vect)$ are related by the following monoidal equivalence: 
$$
\begin{array}{c}
\begin{tikzpicture}[scale=0.9]
\fill[blue!20] (-2,0) rectangle (2,2) ;
\draw[->-,very thick] (-2,2)--(0,2) node[near start,above] {\scriptsize $\CB^\op$} ;
\draw[->-,very thick] (0,2)--(2,2) node[near end,above] {\scriptsize $\CB^\op$} ;
\draw[->-,very thick] (0,1.2)--(0,0) ; 
\draw[fill=white] (-0.07,1.93) rectangle (0.07,2.07) ;
\draw[fill=white] (-0.07,1.13) rectangle (0.07,1.27) node[midway,above] {\scriptsize $\CK$} ;
\draw[dotted,thick] (-0.2,1) rectangle (0.2,2.15) ;
\node at (-1.2,1.2) {\scriptsize $\FZ_1(\CB)$} ;
\node at (1.2,1.2) {\scriptsize $\FZ_1(\CB)$} ;
\node[above] at (0,2.12) {\scriptsize $\CK \boxtimes_{\FZ_1(\CB)} \CB$} ;
\end{tikzpicture}
\end{array}
\quad\quad
\begin{array}{c}
\RMod_{\Omega\CC}(n\vect) \xrightarrow{\simeq} \BMod_{\CB^\op|\CB^\op}(n\vect) \\
\quad\quad\,\,\, \CK \mapsto \CK\boxtimes_{\FZ_1(\CB)} \CB.
\end{array}
$$
Using this equivalence, we immediately see that the relation between $\Omega_m\CM \in \Algc_{\EE_1}(\RMod_{\Omega\CC}(n\vect))$ and its corresponding condensable $\EE_1$-algebra $\CA^\op \in \Algc_{\EE_1}(\BMod_{\CB^\op|\CB^\op}(n\vect))$ is precisely the one illustrated in the picture in (\ref{pic:functor_psi}). 
\end{rem}

This result provides a powerful tool to construct condensable 1-codimensional topological defects in a non-chiral topological order, or equivalently, condensable $\EE_1$-algebra in a fusion $n$-category obtained from the delooping of a non-degenerate and non-chiral braided fusion $(n-1)$-category. We summarize it as a theorem. 
\begin{pthm} \label{pthm:construct_cond_E1_algebras}
For a non-chiral simple topological order $\SC^{n+1}=\SZ(\SB)^{n+1}$, all indecomposable condensable $\EE_1$-algebras in the fusion $n$-category $\CC=\BMod_{\CB|\CB}(n\vect)^\op$ are given by an indecomposable multi-fusion $n$-category $\CA^\op$ equipped with a monoidal functor $\CB \to \CA$. By condensing $\CA^\op$ in $\CC$, we obtain the condensed topological order $\SD^{n+1}=\SZ(\SA)^{n+1}$ and a gapped domain $\SM^n$ such that 
\begin{align*} 
\CD &\simeq \Mod_{\CA^\op}^{\EE_1}(\CC) \simeq \Mod_{\CA^\op}^{\EE_1}(n\vect) \simeq \Sigma\FZ_1(\CA), \quad\quad \Omega\CD \simeq \FZ_1(\CA);  \\
\SM^n &= (\CM,m) \simeq (\BMod_{\CA|\CB}(n\vect), \CA), \quad\quad
\Omega_m\CM = \Fun_{\CA|\CB}(\CA,\CA). 
\end{align*}
The relation between this classification and the one in Theorem$^{\mathrm{ph}}$\,\ref{thm:condensable_E1Alg_af_C} is illustrated below. 
$$
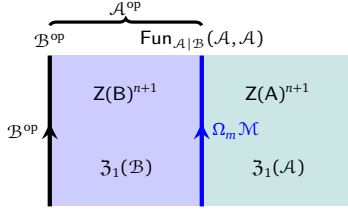
\begin{array}{c}
\begin{tikzpicture}[scale=1]
\fill[blue!20] (-2,0) rectangle (0,2) ;
\fill[teal!20] (0,0) rectangle (2,2) ;
%
\draw[->-,ultra thick] (-2,0)--(-2,2) node[midway,left] {\scriptsize $\CB^\op$} ;

\draw[blue,->-,ultra thick] (0,0)--(0,2) node[midway,right] {\scriptsize $\Omega_m\CM$} ; 
%
%
%
\node at (-1,0.5) {\scriptsize $\FZ_1(\CB)$} ;
\node at (1,0.5) {\scriptsize $\FZ_1(\CA)$} ;
\node at (-1,1.5) {\scriptsize $\SZ(\SB)^{n+1}$} ;
\node at (1,1.5) {\scriptsize $\SZ(\SA)^{n+1}$} ;
\node at (0,2.2) {\scriptsize $\Fun_{\CA|\CB}(\CA,\CA)$} ;
\node at (-2,2.2) {\scriptsize $\CB^\op$} ;

\draw[decorate,decoration=brace,very thick] (-2,2.4)--(0,2.4) node[midway, above] {\scriptsize $\CA^\op$};

\end{tikzpicture}
\end{array}
\quad\quad
\begin{array}{l}
\mbox{\small Two definitions of the same algebra in $\CC$}:  \\
\mbox{\small (1) the monoidal functor $\CB \to \CA$} \\
\mbox{\small \quad\,\, defines $\CA^\op \in \Algc_{\EE_1}(\BMod_{\CB|\CB}(n\vect)^\op)$}; \\
\mbox{\small (2) the central functor $\FZ_1(\CB) \to \Fun_{\CA|\CB}(\CA,\CA)$} \\
\mbox{\small \quad\,\, defines $\Fun_{\CA|\CB}(\CA,\CA) \in \Algc_{\EE_1}(\RMod_{\FZ_1(\CB)}(n\vect))$}. 
\end{array}
$$
For $\CX\in \LMod_\CB(n\vect)$, the canonical monoidal functor $\CB \to \Fun(\CX,\CX)$ defines a Lagrangian $\EE_1$-algebra $\Fun(\CX,\CX)^\op$ in $\CC=\BMod_{\CB|\CB}(n\vect)^\op$. 
\end{pthm}

We provide more concrete constructions in the following theorem. 
\begin{pthm} \label{pthm:construct_cond_E1_algebras_II}
We fixed another non-chiral simple topological order $\SZ(\SP)^{n+1}$ with a simple gapped boundary $\SP^n$. We can construct indecomposable condensable $\EE_1$-algebras in $\CC=\BMod_{\CB|\CB}(n\vect)^\op$ that define condensations from $\SZ(\SB)^{n+1}$ to $\SZ(\SP)^{n+1}$ more concretely in the following way. In this case, $\CP$ is a fusion $(n-1)$-category. Let $\CX \in \BMod_{\CB|\CP}(n\vect)$. Then we have a canonical monoidal functor from $\CB$ to an indecomposable multi-fusion $n$-category $\CA_\CX:=\Fun_{\CP^\op}(\CX,\CX)$: 
$$
\CB \to \Fun_{\CB|\CP}(\CX,\CX) \boxtimes_{\FZ_1(\CB)} \CB \simeq
\Fun_{\CP^\op}(\CX,\CX) =\CA_\CX, 
$$
which defines an indecomposable condensable $\EE_1$-algebra $\CA_\CX^\op$ in $\BMod_{\CB|\CB}(n\vect)^\op$. The physical meaning of each data is illustrated below.  
$$
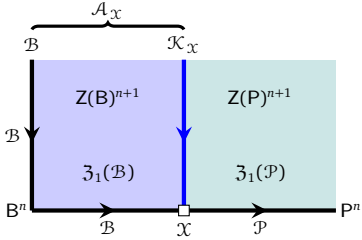
\begin{array}{c}
\begin{tikzpicture}[scale=1]
\fill[blue!20] (-2,0) rectangle (0,2) ;
\fill[teal!20] (0,0) rectangle (2,2) ;
\draw[->-,ultra thick] (-2,0)--(0,0) node[midway,below] {\scriptsize $\CB$} ;
\draw[->-,ultra thick] (-2,2)--(-2,0) node[midway,left] {\scriptsize $\CB$} ;
\draw[->-,ultra thick] (0,0)--(2,0) node[midway,below] {\scriptsize $\CP$} ;

\draw[blue,->-,ultra thick] (0,2)--(0,0) ; 
\draw[fill=white] (-0.07,-0.07) rectangle (0.07,0.07) ; 
%
%
\node at (-1,0.5) {\scriptsize $\FZ_1(\CB)$} ;
\node at (1,0.5) {\scriptsize $\FZ_1(\CP)$} ;
\node at (-1,1.5) {\scriptsize $\SZ(\SB)^{n+1}$} ;
\node at (1,1.5) {\scriptsize $\SZ(\SP)^{n+1}$} ;
\node at (0,-0.25) {\scriptsize $\CX$} ;
\node at (0,2.2) {\scriptsize $\CK_\CX$} ;
\node at (-2.2,0) {\scriptsize $\SB^n$} ;
\node at (2.2,0) {\scriptsize $\SP^n$} ;
\node at (-2,2.2) {\scriptsize $\CB$} ;

\draw[decorate,decoration=brace,very thick] (-2,2.4)--(0,2.4) node[midway, above] {\scriptsize $\CA_\CX$};

\end{tikzpicture}
\end{array}
\quad\quad
\begin{array}{l}
\mbox{\small Two definitions of the same algebra in $\CC$}:  \\
\mbox{\small (1) the monoidal functor $\CB \to \CA_\CX$} \\
\mbox{\small \quad\,\, defines $\CA_\CX^\op \in \Algc_{\EE_1}(\BMod_{\CB|\CB}(n\vect)^\op)$}; \\
\mbox{\small (2) the central functor $\FZ_1(\CB) \to \Fun_{\CB|\CP}(\CX,\CX)^\op$} \\
\mbox{\small \quad\,\, defines $\CK_\CX^\op:=\Fun_{\CB|\CP}(\CX,\CX)^\op \in \Algc_{\EE_1}(\RMod_{\FZ_1(\CB)}(n\vect))$}. 
\end{array}
$$
By condensing $\CA_\CX^\op$ in $\CC$, we obtain the condensed topological order 
$\SD^{n+1}=\SZ(\SP)^{n+1}$ with $\CD \simeq \Mod_{\CA_\CX^\op}^{\EE_1}(\CC)$ 
and a gapped domain wall $\SM^n = (\CM,m) \simeq (\BMod_{\CA_\CX|\CB}(n\vect), \CA_\CX)$. 
\end{pthm}

\begin{rem}
However, this construction $\CX \mapsto \CA_\CX$ (or $\CX \mapsto \CK_\CX$) does not exhaust all possible condensations from $\SZ(\SB)^{n+1}$ to $\SZ(\SP)^{n+1}$ because the domain wall $\SX^{n-1}$ can be gapless in general. If we allow $\SX^{n-1}$ to be gapless\footnote{In this case, $\CX$ is again a $\CB$-$\CP$-bimodule but living in the world of enriched separable higher categories \cite{KZ22b}.}, or equivalently, if we allow to stack anomaly-free $\SY^n$ onto the domain wall $\SZ(\SX)^n$ followed by condensations along the domain wall (recall Theorem\,\ref{pthm:recover_all_bdy_from_one}),  all indecomposable condensable $\EE_1$-algebras in $\CC=\BMod_{\CB|\CB}(n\vect)^\op$ arise in this way (see 
Corollary$^{\mathrm{ph}}$\, \ref{pcor:n=3_classification}). 
\end{rem}

\begin{rem} \label{rem:two_gaugings_condensations}
When $\SP^n=\mathbf{1}^n$, the correspondence between $\CA_\CX$ and $\CK_\CX$ precisely explains the relation between two different but equivalent notions of gauging the fusion category symmetry $\CB$. On the one hand, the gauging of the $\CB$-symmetry in the sense of \cite{FFRS10,BT18,TW24,RSS23,LYW24} produces its dual symmetry $\CK_\CX^\op=\Fun_\CB(\CX,\CX)^\op$ for a left $\CB$-module $\CX$. It amounts to condensing $\CB$ via the gapped domain wall $\CX$, and $\Fun_\CB(\CX,\CX)^\op$ is the fusion category symmetry associated to the condensed phase. 
On the other hand, the gauging process in the sense of \cite{KLWZZ20,KLWZZ20a}  maps the $\CB$-symmetry to an anomaly-free topological order. Mathematically, it amounts to a monoidal functor $\CB \to \Fun(\CX,\CX)=\CA_\CX$ for some separable $(n-1)$-category $\CX$. By \cite[Theorem\ 3.2.3]{KZ18}\cite[Remark\ 3.30]{KZ24}, these two gauging processes are mathematically equivalent. Moreover, Theorem$^{\mathrm{ph}}$\,\ref{pthm:construct_cond_E1_algebras_II} further relates these two gaugings to the condensation of the same 1-codimensional `Lagrangian' topological defect in $\SZ(\SB)^{n+1}$ but in two different coordinate systems, respectively. 
\end{rem}

Using Theorem$^{\mathrm{ph}}$\,\ref{pthm:construct_cond_E1_algebras_II} and \cite[Proposition\,3.13]{KZ21}, we obtain more precise classification results. For $n=2$ (as in Theorem$^{\mathrm{ph}}$\,\ref{pthm:construct_cond_E1_algebras_II}), we recover an old result \cite{KR08,DMNO13} but reformulated in a new context of 1-codimensional condensations; for $n\geq 3$, we obtain a new result.  
\begin{pcor} \label{pcor:n=2_classification}
When $n=2$ (as in Theorem$^{\mathrm{ph}}$\,\ref{pthm:construct_cond_E1_algebras_II}), we obtain a one-to-one correspondence between (the equivalence classes of) $\CB$-$\CP$-bimodules and 1-codimensional defect condensations from $\SZ(\SB)^3$ to $\SZ(\SP)^3$ (i.e., indecomposable condensable $\EE_1$-algebras in $\CC=\Sigma\FZ_1(\CB)$ that produce $\SZ(\SP)^3$, or equivalently, type-I gapped domain walls between $\SZ(\SB)^3$ and $\SZ(\SP)^3$) defined by 
\begin{align*}
{}_\CB\CX_\CP &\mapsto \CK_\CX^\op:=\Fun_{\CB|\CP}(\CX,\CX)^\op \in \Algc_{\EE_1}(\RMod_{\FZ_1(\CB)}(2\vect)), \\
{}_\CB\CX_\CP &\mapsto \CA_\CX^\op:=\Fun_{\CP^\op}(\CX,\CX)^\op \in \Algc_{\EE_1}(\BMod_{\CB|\CB}(2\vect)^\op)
\end{align*}
in two coordinate systems: $\Sigma\FZ_1(\CB)=\RMod_{\FZ_1(\CB)}(2\vect)$ and 
$\Sigma\FZ_1(\CB)=\BMod_{\CB|\CB}(2\vect)^\op$, respectively. 
\end{pcor}
\pf It follows immediately from Theorem$^{\mathrm{ph}}$\,\ref{pthm:construct_cond_E1_algebras_II} and Corollary\,\ref{cor:center-functor_surjective_full_faithful}. 
\epf

Using the idea of topological Wick rotation \cite{KZ18b,KZ20,KZ21,KZ22b}, we obtain a generalization of Corollary$^{\mathrm{ph}}$\,\ref{pcor:n=2_classification} to all dimensions. The following result is an immediate mathematical consequence of Theorem$^{\mathrm{ph}}$\,\ref{pthm:Z1map-made-1to1}. 
\begin{pcor}  \label{pcor:n=3_classification}
For each non-degenerate fusion $(n-1)$-category $\CT$ for $n\geq 2$, there is a one-to-one correspondence (defined in two coordinate systems): 
\begin{align}
{}_{\CB\boxtimes \CT} \CX_\CP &\mapsto \CK_\CX^\op:=\Fun_{\CB\boxtimes \CT|\CP}(\CX,\CX)^\op \in \Algc_{\EE_1}(\RMod_{\FZ_1(\CB)}(n\vect)), \label{map:X_FunBPXX} \\
{}_{\CB\boxtimes \CT}\CX_\CP &\mapsto \CA_\CX^\op:=\Fun_{\CT|\CP}(\CX,\CX)^\op \in \Algc_{\EE_1}(\BMod_{\CB|\CB}(n\vect)^\op) \label{map:X_FunPXX}
\end{align}
between the following two sets: 
\bnu
\item[-] $(\CB\boxtimes \CT)$-$\CP$-bimodules modulo the equivalence relation defined by $\CM \sim \CN$ if $\CN \simeq \CM \boxtimes \CV$ for $\CV \in n\vect^\times$; 
\item[-] 1-codimensional defect condensations from $\SZ(\SB)^{n+1}$ to $\SZ(\SP)^{n+1}$, i.e., indecomposable condensable $\EE_1$-algebras $A$ in $\Sigma\FZ_1(\CB)$, that belong to the same Morita equivalence class as $(\CB\boxtimes\CT) \boxtimes \CP^\op \in \Algc_{\EE_1}(\RMod_{\FZ_1(\CB)}(n\vect))$ (recall Remark\,\ref{rem:Morita_equivalence_E1_algebras}). 
\enu
The condensable $\EE_1$-algebra is simple if and only if the $(\CB\boxtimes \CT)$-$\CP$-bimodule is indecomposable. Note that the $n=2$ case is precisely Corollary$^{\mathrm{ph}}$\,\ref{pcor:n=2_classification} because the only non-degenerate fusion 1-category is the trivial one $\vect$. 
\end{pcor}
\pf
This result follows immediately from Theorem$^{\mathrm{ph}}$\,\ref{pthm:Z1map-made-1to1}. Recall that the key idea of the proof is the holographic dual (via topological Wick rotation) between a potentially gapless defect junction $\SY^{n-1}$ and the pair $(\ST^n, \SX^{n-1})$, where $\ST^n$ is an anomaly-free $n$D topological order and $\SX^{n-1}$ is a gapped defect junction among three anomalous topological orders $\SB^n\boxtimes \ST^n$, $\SK^n$ and $\SP^n$. For readers convenience, we re-illustrate this idea in the pictures below with the corrected labels and dimensions. 
$$
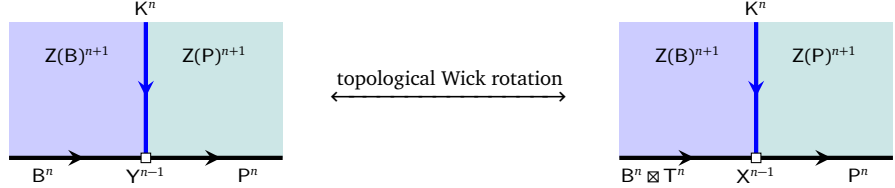
\begin{array}{c}
\begin{tikzpicture}[scale=0.9]
\fill[blue!20] (-2,0) rectangle (0,2) ;
\fill[teal!20] (0,0) rectangle (2,2) ;
\draw[->-,ultra thick] (-2,0)--(0,0) node[near start,below] {\scriptsize $\SB^n$} ;
\draw[->-,ultra thick] (0,0)--(2,0) node[near end,below] {\scriptsize $\SP^n$} ;

\draw[blue,->-,ultra thick] (0,2)--(0,0) ; 
\draw[fill=white] (-0.07,-0.07) rectangle (0.07,0.07) ; 
%
%
\node at (-1,1.5) {\scriptsize $\SZ(\SB)^{n+1}$} ;
\node at (1,1.5) {\scriptsize $\SZ(\SP)^{n+1}$} ;
\node at (0,-0.25) {\scriptsize $\SY^{n-1}$} ;
\node at (0,2.2) {\scriptsize $\SK^n$} ;


\end{tikzpicture}
\end{array}
\quad \xleftrightarrow{\mbox{\footnotesize topological Wick rotation}} \quad
\begin{array}{c}
\begin{tikzpicture}[scale=0.9]
\fill[blue!20] (-2,0) rectangle (0,2) ;
\fill[teal!20] (0,0) rectangle (2,2) ;
\draw[->-,ultra thick] (-2,0)--(0,0) node[near start,below] {\scriptsize $\SB^n\boxtimes \ST^n$} ;
\draw[->-,ultra thick] (0,0)--(2,0) node[near end,below] {\scriptsize $\SP^n$} ;

\draw[blue,->-,ultra thick] (0,2)--(0,0) ; 
\draw[fill=white] (-0.07,-0.07) rectangle (0.07,0.07) ; 
%
%
\node at (-1,1.5) {\scriptsize $\SZ(\SB)^{n+1}$} ;
\node at (1,1.5) {\scriptsize $\SZ(\SP)^{n+1}$} ;
\node at (0,-0.25) {\scriptsize $\SX^{n-1}$} ;
\node at (0,2.2) {\scriptsize $\SK^n$} ;


\end{tikzpicture}
\end{array}
$$
It is worthwhile to note that those $A$'s in $\Sigma\FZ_1(\CB)$ are precisely indecomposable multi-fusion $(n-1)$-categories that are Morita equivalent to $(\CB\boxtimes\CT)\boxtimes \CP^\op$, including those $(\CB\boxtimes\CT)\boxtimes \CP^\op \boxtimes \Sigma\FZ_1(\CQ)$ (for a fusion $(n-2)$-category $\CQ$), which is precisely the image of the $(\CB\boxtimes\CT)$-$\CP$ bimodule $\CX:=\CB\boxtimes\CT\boxtimes\CP\boxtimes \Sigma\CQ$ under the one-to-one correspondence (\ref{map:X_FunBPXX}). 
\epf

\begin{rem}
If we replace $\CT$ by a Morita equivalent one $\CT'$, it simply provides a new coordinate system for the domain of this one-to-one correspondence. 
\end{rem}


The following two results are reformulations of Theorem$^{\mathrm{ph}}$\,\ref{pthm:Z1map-made-1to1} and \ref{cor:A_connected_classification_bdy_Z(A)} in terms of Lagrangian $\EE_1$-algebras. 
\begin{pcor} \label{pcor:4D_gapped boundary_classification}
The map ${}_{\CB\boxtimes \CT} \CX \mapsto \Fun_{\CB\boxtimes \CT}(\CX,\CX)^\op$ defines a one-to-one correspondence between:
\begin{itemize}

\item[-] left indecomposable $\CB\boxtimes \CT$-modules modulo the equivalence relation $\sim$, 

\item[-] Lagrangian $\EE_1$-algebras in $\Sigma\FZ_1(\CB)$ (or simple gapped boundaries of $\SZ(\SB)^{n+1}$) in the Morita class $[\CB\boxtimes \CT]$;

\end{itemize} 
See Theorem$^{\mathrm{ph}}$\,\ref{pthm:Z1map-made-1to1} and Corollary\,\ref{pcor:4D_Lagrangian_1to1_modules} for an equivalent result. 
\end{pcor}

The following result generalizes an earlier result for fusion 2-categories \cite{JFR23,DX24}. 
\begin{cor} \label{cor:LA_n+2D_finite_gauge_theory}
When the fusion $(n-1)$-category $\CB$ is connected as a separable $(n-1)$-category, i.e., $\CB=\Sigma\Omega\CB$, there is a one-to-one correspondence between 
\bnu
\item[-] braided monoidal functors $\Omega\CB\to \CQ$ for a non-degenerate braided fusion $(n-2)$-category $\CQ$; 

\item[-] Lagrangian $\EE_1$-algebra in $\Sigma\FZ_1(\CB)$. 

\enu
Moreover, the Morita class of the Lagrangian $\EE_1$-algebra is determined by the Morita class of $\Sigma\CQ$. 
\end{cor}

\begin{expl} \label{expl:B=n-1RepG_LA_in_GT_G}
A typical example of $\CB=\Sigma\Omega\CB$ is $\CB=(n-1)\Rep(G)$. In this case, Corollary\,\ref{cor:LA_n+2D_finite_gauge_theory} provide a classification of Lagrangian $\EE_1$-algebras in $\CG\CT_G^{n+1}\simeq \Sigma\FZ_1((n-1)\Rep(G))$, or equivalently, that of the gapped boundaries of the finite gauge theory $\SG\ST_G^{n+1}$. 
\end{expl}

\subsubsection{Examples in 2+1D: anomaly-free cases} \label{sec:1codim_2d}

\begin{expl} \label{expl:1codim_cond_TC_1}
The topological defects in the 2+1D $\Zb_2$-gauge theory form the fusion 2-category $\CT\CC$, which can be identified with the 2-category $\RMod_{\FZ_1(\Rep(\Zb_2))}(2\vect)$. By \cite{BJS21,Dec23a} (or Theorem\,\ref{thm:condensable_E1Alg_af_C}), a condensable $\EE_1$-algebra in $\CT\CC$ are precisely multi-fusion left $\FZ_1(\Rep(\Zb_2))$-modules, i.e., a mulit-fusion categories $\CA$ equipped with a braided functors $\FZ_1(\Rep(\Zb_2)) \to \FZ_1(\CA)$. Here we list some examples.
\bnu
\item The fusion category $\FZ_1(\Rep(\Zb_2))$ equipped with the identity functor $\id_{\FZ_1(\Rep(\Zb_2))}$, which should be viewed as a central functor, defines a condensable $\EE_1$-algebra $\FZ_1(\Rep(\Zb_2))$ in the coordinate system $\CT\CC=\RMod_{\FZ_1(\Rep(\Zb_2))}(2\vect)$. It is the tensor unit of $\CT\CC$.

\item The fusion category $\Rep(\Zb_2)$ equipped with the central functor $\forget: \FZ_1(\Rep(\Zb_2)) \to \Rep(\Zb_2)$ defines a condensable $\EE_1$-algebra $\Rep(\Zb_2)^\op=\Rep(\Zb_2)$ in $\RMod_{\FZ_1(\Rep(\Zb_2))}(2\vect)$. It corresponds to the 1-codimensional defect $\mathrm{ss}$ in $\CT\CC$. We have $\Mod_{\Rep(\Zb_2)}^{\EE_1}(\Sigma \FZ_1(\Rep(\Zb_2))) \simeq 2\vect$, which means that the condensed 2+1D topological order is trivial. This condensation defined by $\mathrm{ss}$ is the same as the anyon condensation defined by $\one \oplus m$. Therefore, $\mathrm{ss}$ is a Lagrangian condensable $\EE_1$-algebra in $\CT\CC$. 

\item The fusion category $\vect_{\Zb_2}$ is equipped with the canonical equivalence $\FZ_1(\Rep(\Zb_2)) \to \FZ_1(\vect_{\Zb_2})$. It defines a condensable $\EE_1$-algebra $\vect_{\Zb_2}^\op=\vect_{\Zb_2}$, corresponding to the 1-codimensional defect $\mathrm{rr}$ in $\CT\CC$. We have $\Mod_{\vect_{\Zb_2}}^{\EE_1}(\Sigma \FZ_1(\Rep(\Zb_2))) \simeq 2\vect$, which means that the condensed topological order is trivial. This condensation defined by $\mathrm{rr}$ is the same as the anyon condensation defined by $\one \oplus e$. Therefore, this condensable $\EE_1$-algebra $\vect_{\Zb_2}$ is also Lagrangian.

\item As we review in Example\,\ref{expl:double_Ising_to_toric_code}, the following object
\be \label{eq:11+psipsi}
A:=\one\boxtimes \one \oplus \psi \boxtimes \psi\in \Omega\ising\boxtimes \Omega\ising^\rev \simeq \FZ_1(\Omega\ising)
\ee
has a canonical structure of a simple commutative separable algebra in $\FZ_1(\Omega\ising)$. Then fusion 1-category $\CK:=\RMod_A(\FZ_1(\Omega\ising))^\rev$ is naturally a condensable $\EE_1$-algebra in $\CT\CC$. The category $\CK$ has six simple objects $1,e,m,f,\chi_\pm$ (see Example\,\ref{expl:double_Ising_to_toric_code}). Using the coordinate systems $\CT\CC \simeq \RMod_{\FZ_1(\Rep(\Zb_2))}(2\vect)$ and Table\,\ref{table:coordinates_in_TC}, we see immediately that 
\be \label{eq:K=1+vartheta}
\CK = 1\oplus \vartheta \quad \in \CT\CC^3. 
\ee
The algebraic structure on $1\oplus \vartheta$ can be reconstructed as a special case of a more general construction (see Section\,\ref{sec:gauging_G-symmetry}). Condensing $\CK$ in $\CT\CC$ produces the condensed phase $\SD^3$, which is precisely the 2+1D double Ising topological order, i.e.,
\[
\Mod_\CK^{\EE_1}(\CT\CC) \simeq \ising \boxtimes \ising^\rev
\]
and a gapped domain wall $\SM^2= (\CM,m) =(\RMod_\CK(\CT\CC), \CK)=(\RMod_\CK(2\vect),\CK)$. We explain in Section\,\ref{sec:gauging_G-symmetry} that condensing $\CK$ in $\CT\CC$ can also be understood as the process of gauging the $e$-$m$-duality in $\ST\SC^3$ (or an equivariantization). 
\enu
\end{expl}

\begin{expl} \label{expl:1codim_cond_TC_2}
We can choose a different coordinate system for $\CT\CC$. Indeed, by \cite{KK12}, $\CT\CC$ can be identified with the fusion 2-category $\BMod_{\Rep(\Zb_2)|\Rep(\Zb_2)}(2\vect)^\op$. A condensable $\EE_1$-algebra in $\CT\CC$ is precisely a fusion 1-category $\CB^\op$ equipped with a monoidal functor $\Rep(\Zb_2) \to \CB$. Here we list some examples:
\bnu
\item The fusion category $\Rep(\Zb_2)$ equipped with the identity functor $\Rep(\Zb_2) \to \Rep(\Zb_2)$ is a condensable $\EE_1$-algebra. It is the tensor unit of $\CT\CC$.

\item The fusion category $\vect$ equipped with the forgetful functor $\Rep(\Zb_2) \to \vect$ is a Lagrangian $\EE_1$-algebra, which corresponds to the 1-codimensional topological defect $\mathrm{rr}$ in $\CT\CC$.

\item The multi-fusion category $\Fun(\Rep(\Zb_2), \Rep(\Zb_2))^\op$ equipped with the canonical embedding 
$$
\Rep(\Zb_2) \to \Fun(\Rep(\Zb_2),\Rep(\Zb_2))
$$ 
is a Lagrangian $\EE_1$-algebra, which corresponds to the 1-codimensional topological defect $\mathrm{ss}$ in $\CT\CC$. 

\item We set $\CK:=\RMod_A(\FZ_1(\Omega\ising))$ for $A$ defined in (\ref{eq:11+psipsi}). There is a canonical monoidal functor
$$
\Rep(\Zb_2) \to \CK \boxtimes_{\FZ_1(\Rep(\Zb_2))} \Rep(\Zb_2) \simeq \Omega\ising := \CA,
$$
where the monoidal equivalence `$\simeq$' follows from a general fact\footnote{For a non-degenerate braided fusion category $\CC$ and a condensable $\EE_2$-algebra $A$ and a condensable $\EE_2$-algebra $B$ over $A$ \cite{DMNO13}, we have a monoidal equivalence $ \RMod_A(\CC) \boxtimes_{\Mod_A^{\EE_2}(\CC)}  \RMod_B(\Mod_A^{\EE_2}(\CC)) \simeq \RMod_B(\CC)$.}, 
defines a condensable $\EE_1$-algebra $\CA^\op$ in $\CT\CC=\BMod_{\Rep(\Zb_2)|\Rep(\Zb_2)}(2\vect)^\op$. Condensing $\CA^\op$ produces the condensed topological order $\SZ(\Ising)^3$:
\[
\ising \boxtimes \ising^\rev \simeq \Mod_{\CA^\op}^{\EE_1}(\CT\CC),
\]
and the gapped domain wall 
$\SM^2$ such that (using (\ref{eq:BMod_BA}) and (\ref{eq:Fun_BA_AA}))
\[
 (\CM,m) = (\BMod_{\CA|\CB}(n\vect), \CA) \simeq (\Sigma\CK^\op, \CK).
\]
The physical meaning of above data is depicted in the following picture. 
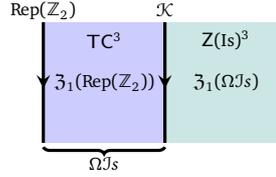
\[
\begin{array}{c}
\begin{tikzpicture}[scale=0.8]
\fill[blue!20] (-2,0) rectangle (0,2) ;
\fill[teal!20] (0,0) rectangle (2,2) ;
\draw[->-,very thick] (-2,2)--(-2,0) ; 
\draw[->-,very thick] (0,2)--(0,0) ; 
%
%
%
\node at (-1,1) {\scriptsize $\FZ_1(\Rep(\Zb_2))$} ;
\node at (1,1) {\scriptsize $\FZ_1(\Omega\ising)$} ;
\node at (-1,1.7) {\scriptsize $\ST\SC^3$} ;
\node at (1,1.7) {\scriptsize $\SZ(\Ising)^3$} ;
\node at (-2,2.2) {\scriptsize $\Rep(\Zb_2)$} ;
\node at (0,2.2) {\scriptsize $\CK$} ;
\draw[decorate,decoration=brace,very thick] (0,-0.07)--(-2,-0.07) node[midway, below] {\scriptsize $\Omega\ising$};
\end{tikzpicture}
\end{array}
\]
\enu
We provide more examples of condensations in $\CT\CC$ as the special cases of condensations in 2+1D finite gauge theories in Example\,\ref{expl:3D_finite_gauge_1codim_condensation}. 
\end{expl}

\begin{expl} \label{expl:3D_finite_gauge_1codim_condensation}
Consider 2+1D finite gauge theory $\SG\ST_G^3$ for a finite group $G$. We provide some examples of condensable $\EE_1$-algebras in $\CG\CT_G^3$ in two coordinate systems. 
\bnu

\item $\CG\CT_G^3 = \RMod_{\FZ_1(\Rep(G))}(2\vect)$: 

\bnu
\item Consider a simple condensable $\EE_2$-algebra $A$ (i.e., \'{e}tale algebra) in $\FZ_1(\Rep(G))$. Such algebras have been classified in \cite{Dav10a} (see Example\,\ref{expl:Davydov}). The central functor 
\be \label{eq:otimesA_ZRep(G)}
- \otimes A: \FZ_1(\Rep(G)) \to \RMod_A(\FZ_1(\Rep(G)))
\ee
defines a condensable $\EE_1$-algebra $B:=\RMod_A(\FZ_1(\Rep(G)))$ in $\RMod_{\FZ_1(\Rep(G))}(2\vect)$. By condensing it, we obtain a new 2+1D topological oder $\SD^3$ and a gapped domain wall $\SM^2$ given by 
\begin{align*}
\CD &\simeq \Mod_{B}^{\EE_1}(\Sigma\FZ_1(\Rep(G))) \simeq \Sigma(\Mod_A^{\EE_2}(\FZ_1(\Rep(G))), \\
(\CM,m) &\simeq (\RMod_{B}(\Sigma\FZ_1(\Rep(G))), B). 
\end{align*}
\bnu

\item When $A=\one_{\FZ_1(\Rep(G))}$, $-\otimes A \simeq \id_{\FZ_1(\Rep(G))}$ defines the trivial condensable $\EE_1$-algebra $\one_{\CG\CT_G^3}$, which defines a trivial condensation. 

\item The group algebra $A=\Cb[G]$ can be viewed as a Lagrangian $\EE_2$-algebra in $\FZ_1(\vect_G) \simeq \FZ_1(\Rep(G))$ (see Example\,\ref{expl:Davydov}). In this case, the central functor $-\otimes A$ can be identified with the canonical central functor: the forgetful functor $\FZ_1(\Rep(G)) \to \Rep(G)$, which defines a Lagrangian $\EE_1$-algebra $\Rep(G)$ in $\CG\CT_G^3$. By condensing it, we obtain $\mathbf{1}^3$ as the condensed phase and a gapped boundary $\SM^2$ such that 
$(\CM,m) \simeq (2\Rep(G),\one)$. 

\item For a subgroup $H\leq G$, the algebra $\mathrm{Fun}(G/H)$ is a condensable $\EE_2$-algebra in $\Rep(G) \hookrightarrow \FZ_1(\Rep(G))$. In this case, the central functor $-\otimes A$ in (\ref{eq:otimesA_ZRep(G)}) defines a condensable $\EE_1$-algebra in $\CG\CT_G^3$. By condensing it, we obtain $\SG\ST_H^3$ as the condensed phase and a gapped boundary $\SM^2$ such that 
$\CM \simeq \Sigma\RMod_A(\FZ_1(\Rep(G)))$. 

\enu

\item We can condense a condensable $\EE_1$-algebra in $\CG\CT_G^3$ to obtain any 2+1D simple non-chiral topological orders as shown in Theorem$^{\mathrm{ph}}$\,\ref{pthm:construct_cond_E1_algebras}. We give some concrete constructions.
\bnu

\item Let $G_1$ and $G_2$ be two finite groups and $H$ a subgroup of both $G_1$ and $G_2$. By restriction, we obtain two monoidal functors $\Rep(G_1) \to \Rep(H)$ and $\Rep(G_2) \to \Rep(H)$. Therefore, $\Rep(H)$ is a $\Rep(G_1)$-$\Rep(G_2)$-bimodule. As a consequence, there is a natural central functor $\FZ_1(\Rep(G_1)) \to \CK:=\Fun_{\Rep(G_1)|\Rep(G_2)}(\Rep(H), \Rep(H))$, which defines a condensable $\EE_1$-algebra $\CK$ in $\CG\CT_{G_1}^3$. By condensing the 1-codimensional topological defect $\CK$ in $\RMod_{\FZ_1(\Rep(G_1))}(2\vect)$, we obtain the $\SG\ST_{G_2}^3$ as the condensed 2+1D phase. 

\smallskip
For example, one can always choose $H$ to be the trivial group. In this case, $\Rep(H)=\vect$ and $\CK\simeq \vect_{G_1 \times G_2}$ as monoidal 1-categories. 

\item In this case, we can obtain a complete classification of all 1-codimensional defect condensations from $\SG\ST_{G_1}^3$ to $\SG\ST_{G_2}^3$ by applying Corollary$^{\mathrm{ph}}$\,\ref{pcor:n=2_classification} as illustrated below. 
\[
\begin{array}{c}
\begin{tikzpicture}[scale=1]
\fill[blue!20] (-2,0) rectangle (0,2) ;
\fill[teal!20] (0,0) rectangle (2,2) ;
\draw[->-,ultra thick] (-2,0)--(0,0) node[midway,below] {\scriptsize $\Rep(G_1)$} ;
\draw[->-,ultra thick] (0,0)--(2,0) node[midway,below] {\scriptsize $\Rep(G_2)$} ;

\draw[blue,->-,ultra thick] (0,2)--(0,0) ; 
\draw[fill=white] (-0.07,-0.07) rectangle (0.07,0.07) ; 
%
%
\node at (-1,1) {\scriptsize $\FZ_1(\Rep(G_1))$} ;
\node at (1,1) {\scriptsize $\FZ_1(\Rep(G_2))$} ;
\node at (0,-0.25) {\scriptsize $\CX$} ;
\node at (0,2.2) {\scriptsize $\CK_\CX=\Fun_{\Rep(G_1)|\Rep(G_2)}(\CX,\CX)$} ;


\end{tikzpicture}
\end{array}
\]
By \cite[Theorem\ 3.3.7]{KZ18}, the functorial assignment 
$\CX \to \CK_\CX$ is fully faithful. It means that there is a one-to-one correspondence between such simple condensable $\EE_1$-algebras $\CK_\CX$ in $\RMod_{\FZ_1(\Rep(G_1))}(2\vect)$ and indecomposable $\Rep(G_1)$-$\Rep(G_2)$-bimodules $\CX$ in $2\vect$, which is equivalent to indecomposable left $\Rep(G_1 \times G_2)$-modules in $2\vect$ and was classified in \cite{BO04,Ost03}. More precisely, an indecomposable left $\Rep(G_1 \times G_2)$-module $\CX$ in $2\vect$ is given by $\Rep^\prime(\tilde{H})$ or $\Rep(H,\alpha)$, where $H \leq G_1 \times G_2$ is a subgroup, $[\alpha]] \in H^2(H,\Cb^\times)$ is a 2-cohomology class and $\tilde{H}$ is the central extension of $H$ determined by $[\alpha]$ (see Example \ref{expl:algebras_in_RepG}). The associated simple condensable $\EE_1$-algebra $\CK_{(H,\omega)}$ in $\RMod_{\FZ_1(\Rep(G_1))}(2\vect)$ is
\[
\CK_{(H,\omega)} = \Fun_{\Rep(G_1)|\Rep(G_2)}(\Rep^\prime(\tilde{H}), \Rep^\prime(\tilde{H})). 
\]
When $G_2$ is the trivial group, all these condensable $\EE_1$-algebras $\CK_{(H,\omega)}^\op$ are Lagrangian. This classification result coincides with the classification of Lagrangian $\EE_2$-algebras in $\FZ_1(\Rep(G_1))$ in \cite{Dav10a} (see also the 5th case in Example\,\ref{expl:Davydov}).

\enu

\enu

\item $\CG\CT_G^3 = \BMod_{\Rep(G)|\Rep(G)}(2\vect)^\op$: 
\bnu

\item The central functor (\ref{eq:otimesA_ZRep(G)}) defines a monoidal functor 
\be \label{eq:RepG_to_RModARepG}
\Rep(G) \to \Rep(G) \boxtimes_{\FZ_1(\Rep(G))} \RMod_A(\FZ_1(\Rep(G))) \simeq \RMod_A(\Rep(G)),
\ee
where the `$\simeq$' is proved in \cite{KZ18,XY25}. By condensing $\RMod_A(\Rep(G))^\op$, we obtain a condensed phase $\SD^3$ such that 
$$
\CD\simeq \Mod_{\RMod_A(\Rep(G))^\op}^{\EE_1}(\CG\CT_G^3) \simeq \Sigma\Mod_A^{\EE_2}(\FZ_1(\Rep(G))),
$$
where the second `$\simeq$' is as a special case of Theorem\,\ref{thm:ModE1SASB=SModE2AB} (see also Example\,\ref{expl:3D_Gauge_condense_E2Alg}).  

When $A=\mathrm{Fun}(G/H)$ for a subgroup $H\leq G$, the monoidal functor (\ref{eq:RepG_to_RModARepG}) is precisely the forgetful functor $\Rep(G) \to \Rep(H)$. By condensing $\Rep(H)^\op$, we obtain the condensed phase $\SD^3$ such that $\CD=\Mod_{\Rep(H)^\op}^{\EE_1}(\CG\CT_G^3)\simeq \Mod_{\Rep(H)^\op}^{\EE_1}(2\vect)=\CG\CT_H^3$.

\item Using Corollary$^{\mathrm{ph}}$\,\ref{pcor:n=2_classification}, we obtain a complete classification of all 1-codimensional defect condensations from $\SG\ST_{G_1}^3$ to $\SG\ST_{G_2}^3$ in the new coordinate. More precisely, required simple condensable $\EE_1$-algebras in $\BMod_{\Rep(G)|\Rep(G)}(2\vect)^\op$ are one-to-one corresponding to indecomposable left $\Rep(G_1 \times G_2)$-modules in $2\vect$, each of which is given by $\Rep^\prime(\tilde{H})$ or $\Rep(H,\alpha)$, where $H \leq G_1 \times G_2$ is a subgroup, $[\alpha]] \in H^2(H,\Cb^\times)$ is a 2-cohomology class and $\tilde{H}$ is the central extension of $H$ determined by $[\alpha]$ (see Example \ref{expl:algebras_in_RepG}) \cite{BO04,Ost03}. The associated simple condensable $\EE_1$-algebra $\CA_{(H,\omega)}^\op$ in $\BMod_{\Rep(G)|\Rep(G)}(2\vect)^\op$ is defined by the monoidal functor 
\[
\Rep(G_1) \to \CK_{(H,\omega)}  \boxtimes_{\FZ_1(\Rep(G_1))}  \Rep(G_1) \simeq 
\Fun_{\Rep(G_2)^\rev}(\Rep^\prime(\tilde{H}), \Rep^\prime(\tilde{H})) \eqqcolon \CA_{(H,\omega)}. 
\]
When $G_2$ is the trivial group, all these condensable $\EE_1$-algebras $\CA_{(H,\omega)}^\op$ are Lagrangian. 

\enu
\enu
\end{expl}

\subsubsection{Examples in 2+1D: anomalous cases}

We provide some concrete examples of 1-codimensional defect condensations in 2+1D anomalous topological orders. By topological Wick rotation (or topological holography) \cite{KZ18b,KZ20,KZ21,KLWZZ20a,KLWZZ20,KZ22b}, a 1-codimensional defect condensation in a 2+1D anomalous topological order can also be understood as a condensation between 2+1D gapped quantum liquids with the same finite onsite symmetry, and as the gauging of 0-form symmetries in a gapless 2+1D QFT.

\begin{expl} \label{expl:E1_algebras_2vectG}
Let $G$ be a finite group. The 3+1D $G$ gauge theory $\SG\ST_G^4$ has a gapped boundary obtained by condensing all particle-like topological defects (symmetry charges).  
The topological defects on this gapped boundary form the fusion 2-category $2\vect_G$.

By \cite{Dec23a}, the condensable $\EE_1$-algebras in $2\vect_G$ are $G$-graded multi-fusion 1-categories, and the simple ones are $G$-graded fusion 1-categories. A $G$-graded (multi-)fusion 1-category is a (multi-)fusion 1-category $\CA$ equipped with a $G$-grading $\CA = \bigoplus_{g \in G} \CA_g$ such that $x \otimes y \in \CA_{gh}$ for any $x \in \CA_g$ and $y \in \CA_h$. In particular, the trivial component $\CA_e$ is also a (multi-)fusion 1-category. We say that the $G$-grading is faithful if $\CA_g \neq 0$ for all $g \in G$. 
\bnu
\item[-] For example, given a subgroup $H \leq G$ and a 3-cocycle $\omega \in Z^3(H,U(1))$, there is a $G$-graded fusion category $\vect_H^\omega$ with the obvious $G$-grading supported on $H$ and the associator given by $\omega$. Two $G$-graded fusion 1-categories $\vect_H^\omega$ and $\vect_K^{\omega'}$ are equivalent as condensable $\EE_1$-algebras in $2\vect_G$ if and only if $H = K$ and $[\omega] = [\omega'] \in H^3(H,U(1))$.
\enu

For a $G$-graded fusion 1-category $\CA$, we give a more explicit mathematical description of the left $2\vect_G$-module $\RMod_\CA(2\vect_G)$. First, we define a 2-category $\mathrm B_G \CA$ as follows.
\bit
\item The objects are elements in $G$.
\item For $g,h \in G$, the hom 1-category $\Hom(g,h)$ is $\CA_{hg^{-1}}$. In particular, the endomorphism 1-category of every object is the fusion 1-category $\CA_e$.
\item The 2-category structure (including the composition, identity, associator, \ldots) is induced by the monoidal structure of $\CA$.
\eit
Secondly, we define $\Sigma_G \CA$ to be the condensation completion of $\mathrm B_G \CA$. Then $\Sigma_G \CA$ is a finite semisimple 2-category (see \cite[Construction 2.1.23]{DR18} and references therein). Note that there is an obvious right translation $G$-action on $\mathrm B_G \CA$:
\[
g \odot h \coloneqq hg^{-1} , \quad \Hom(g \odot h,g \odot k) = \Hom(hg^{-1},kg^{-1}) = \CA_{kh^{-1}} = \Hom(h,k) .
\]
It induces a $G$-action on $\Sigma_G \CA$. Therefore, $\Sigma_G \CA$ is a finite semisimple $2\vect_G$-module.

Now we show that $\Sigma_G \CA \simeq \RMod_\CA(2\vect_G)$ as $2\vect_G$-modules. Denote the (representative of) simple objects in $2\vect_G$ by $\{\delta_g\}_{g \in G}$. There is a 2-functor $\mathrm B_G \CA \to \RMod_\CA(2\vect_G)$ defined by
\bit
\item the map between objects: an object $g \in G$ is mapped to the free right $\CA$-module $\delta_{g^{-1}} \otimes \CA$;
\item the functors between hom 1-categories: the canonical equivalence
\[
\Hom_{\mathrm B_G \CA}(g,h) = \CA_{hg^{-1}} \simeq \Hom_{2\vect_G}(\delta_{g^{-1}},\delta_{h^{-1}} \otimes \CA) \simeq \Hom_{\RMod_\CA(2\vect_G)}(\delta_{g^{-1}} \otimes \CA,\delta_{h^{-1}} \otimes \CA) .
\]
\eit
Moreover, this is a $G$-module 2-functor. Thus it induces a $G$-module 2-functor $\Sigma_G \CA \to \RMod_\CA(2\vect_G)$. Note that every simple module $x \in \RMod_\CA(2\vect_G)$ can be condensed from the free module $x \otimes \CA$. Thus the $G$-module 2-functor $\Sigma_G \CA \to \RMod_\CA(2\vect_G)$ is an equivalence.

There are some immediate corollaries of this result. Let $H$ be the support of the $G$-grading on $\CA$. Then it is easy to see that $\pi_0(\Sigma_G \CA) \simeq G/H$. In particular, $\RMod_\CA(2\vect_G) \simeq \Sigma_G \CA$ is connected if and only if the $G$-grading on $\CA$ is faithful. Moreover, when the $G$-grading on $\CA$ is faithful, there is an equivalence of 2-categories $\Sigma_G \CA \simeq \Sigma \CA_e$. In this case, the $2\vect_G$-action on $\Sigma \CA_e$ is a monoidal 2-functor $2\vect_G \to \BMod_{\CA_e|\CA_e}(2\vect) \simeq \End(\Sigma \CA_e)$, and it induces a 3-group homomorphism $G \to \mathsf{BrPic}(\CA_e)$, where the Brauer-Picard 3-group $\mathsf{BrPic}(\CA_e)$ is the maximal 3-group of $\BMod_{\CA_e|\CA_e}(2\vect)^\times$. By \cite{ENO10}, this 3-group homomorphism is equivalent to a faithful $G$-graded extension of $\CA_e$, which is exactly the $G$-graded fusion 1-category $\CA$.

As an example, for $\omega \in Z^3(G,U(1))$, we have $\RMod_{\vect_G^\omega}(2\vect_G) \simeq 2\vect$, and the pentagonator of the $2\vect_G$-action on $2\vect$ is given by $\omega$. Equivalently, this $2\vect_G$-action on $2\vect$ is induced from the monoidal 2-functor $2\vect_G \to 2\vect$ twisted by $\omega$. Another example is the Ising fusion 1-category $\Omega\ising \simeq \vect_{\Zb_2} \oplus \vect$ as a $\Zb_2$-graded fusion category. The $2\vect_{\Zb_2}$-action on $\RMod_{\Omega\ising}(2\vect_{\Zb_2}) \simeq \Sigma \vect_{\Zb_2} \simeq 2\rep(\Zb_2)$ is given by permuting two simple objects.
\end{expl}

\begin{expl} \label{expl_2RepG_algebra}
The 3+1D $G$ gauge theory $\SG\ST_G^4$ has another gapped boundary obtained by condensing all symmetry fluxes. The fusion 2-category of topological defects on this boundary is equivalent to $2\Rep(G)$. There are two coordinate systems on $2\Rep(G)$.
\bnu
\item[(1)] In the coordinate system, $2\Rep(G) \simeq \Fun(\mathrm{B}G, 2\vect)$, it is clear that objects in $2\Rep(G)$ are separable 1-categories equipped with $G$-actions. A condensable $\EE_1$-algebra in $2\Rep(G)$ is precisely a multi-fusion 1-category $\CA$ equipped with monoidal $G$-action, which is also called a $G$-equivairant multi-fusion 1-category. A condensable $\EE_1$-algebra $\CA$ in $2\Rep(G)$ is simple if $G$ acts transitively on the simple direct summands of the tensor unit of $\CA$.  


\item[(2)] In the coordinate system $2\rep(G) \simeq \Sigma \Rep(G) \simeq \RMod_{\Rep(G)}(2\vect) \simeq \LMod_{\Rep(G)}(2\vect)$, by \cite{BJS21,Dec23a}, a condensable $\EE_1$-algebras in $2\rep(G)$ is precisely a multi-fusion $\Rep(G)$-module, i.e., a multi-fusion 1-category $\CB$ equipped with a braided monoidal functor $\Rep(G) \to \FZ_1(\CB)$, and the simple ones are fusion $\Rep(G)$-modules. Such a multi-fusion $\Rep(G)$-module is also called a multi-fusion 1-category over $\Rep(G)$. By \cite[Lemma 3.5]{DMNO13}, a monoidal functor $\Rep(G) \to \FZ_1(\CB) \to \CB$ factors through $\Rep(G) \to \Rep(H) \to \CB$ for some subgroup $H \leq G$ (up to conjugation) such that $\Rep(H) \to \CB$ is an embedding (fully faithful).
\enu

The coordinate transformations between two coordinate systems of $2\Rep(G)$ are given by 
the equivariantization and de-equivariantization (see \cite[Section 4.1]{DGNO10}). More precisely, 
\begin{align*}
\{\mbox{$\Cb$-linear categories with $G$-action}\} &\longleftrightarrow
\{ \mbox{$\Cb$-linear categories with $\Rep(G)$-action} \} \\
\CA \quad &\mapsto  \quad \CA^G  \\
\CB_G:=\RMod_A(\CB) \quad &\mapsfrom \quad \CB,
\end{align*}
where $\CA^G$ consists of $G$-invariant objects $x$ (i.e., $g(x) \simeq x$ for all $g\in G$) and where $A=\Fun(G)$ is a condensable $\EE_1$-algebra in $\Rep(G)$. These coordinate transformations give one-to-one correspondences between condensable $\EE_1$-algebras in different coordinates \cite[Section 4.2]{DGNO10}: 
$$
\{ \mbox{$G$-equivariant multi-fusion 1-categories} \} \longleftrightarrow
\{ \mbox{multi-fusion 1-categories over $\Rep(G)$} \}
$$

Given a simple condensable $\EE_1$-algebra $\CA \in \LMod_{\vect_G}(2\vect) \simeq \Fun(\mathrm B G,2\vect)$, the set of simple direct summands of the tensor unit is isomorphic to $G/H$ for some subgroup $H$. In other words, its tensor unit spans a multi-fusion subcategory $\Fun(G/H,\vect) \subseteq \CA$. The equivariantization of the monoidal functors $\vect \to \Fun(G/H,\vect) \hookrightarrow \CA$ is 
$$
\vect^G \simeq \rep(G) \to \rep(H) \simeq \Fun(G/H,\vect)^G \hookrightarrow \CA^G,
$$ 
which gives a simple condensable $\EE_1$-algebra in $\LMod_{\rep(G)}(2\vect) \simeq \Sigma \rep(G)$. This is the one-to-one correspondence of simple condensable $\EE_1$-algebras in the two coordinates of $2\rep(G)$. When $H = G$, $\CA$ is a $G$-equivariant fusion 1-category. Under this one-to-one correspondence, $G$-equivariant fusion 1-categories are mapped to fusion 1-categories $\CB$ equipped with braided functor $\Rep(G) \to \FZ_1(\CB)$ such that the composite functor $\Rep(G) \to \FZ_1(\CB) \to \CB$ is fully faithful \cite[Theorem 4.18 (iii)]{DGNO10}.

We give some examples here.
\bnu
\item For every subgroup $H \leq G$, the forgetful functor $\Rep(G) \to \Rep(H)$ is symmetric, hence factors through $\FZ_1(\Rep(H))$. Thus $\Rep(H)$ is a simple condensable $\EE_1$-algebra in $\Sigma \Rep(G)$. We also have $\RMod_{\Rep(H)}(\Sigma \Rep(G)) \simeq \RMod_{\Rep(H)}(2\vect) \simeq \Sigma \Rep(H)$, and the $\Sigma \Rep(G)$-action on $\Sigma \Rep(H)$ is induced by the monoidal functor $\Rep(G) \to \Rep(H)$. After de-equivariantization, the corresponding simple condensable $\EE_1$-algebra in $2\rep(G) \simeq \Fun(\mathrm{B}G, 2\vect)$ is $\Fun(G/H,\vect)$ with the pointwise tensor product and left translation $G$-action.
\item For every subgroup $H \leq G$ and 3-cocycle $\omega \in Z^3(H,U(1))$, there is a braided embedding $\Rep(H) \to \FZ_1(\vect_H^\omega)$. Indeed, these are all the minimal nondegenerate extensions of $\Rep(H)$ \cite{DGNO10,LKW16a}. Thus $\vect_H^\omega$ is a simple condensable $\EE_1$-algebra in $\Sigma \Rep(G)$. 
\enu

For a fusion $\Rep(G)$-module $\CB$, we have $\RMod_\CB(2\rep(G)) \simeq \RMod_\CB(2\vect) \simeq \Sigma \CB$. The left $2\rep(G)$-action on $\Sigma\CB$ is defined by the monoidal 2-functor $\Sigma \Rep(G) \to \Sigma \FZ_1(\CB) \simeq \FZ_0(\Sigma\CB)$, which is induced from the braided monoidal functor $\Rep(G) \to \FZ_1(\CB)$. As discussed above, this braided monoidal functor factors through $\Rep(G) \to \Rep(H) \to \FZ_1(\CB)$ and $\Rep(H) \to \FZ_1(\CB)$ is a braided embedding. It follows that the indecomposable $2\rep(G)$-modules one-to-one correspond to the pairs $(H,\CM)$, where $H \leq G$ is a subgroup of $G$ defined up to conjugation and $\CM$ is a nondegenerate extension of $\Rep(H)$ such that $\CM$ is Witt trivial (i.e., $\CM \simeq \FZ_1(\CB)$ for some fusion category $\CB$). This classification of indecomposable $2\Rep(G)$-module coincides with that of gapped boundaries of $\SG\ST_G^4$ that are Morita equivalent to $2\Rep(G)$ (see also Example\,\ref{expl:E2_algebra_Z12VecG} and \ref{expl:E2_algebra_Z12RepG} and Theorem$^{\mathrm{ph}}$\,\ref{pthm:iterate_E2_condensation}). 
\end{expl}

\begin{expl}
Let $\CT$ be a nondegenerate braided fusion 1-category. We can stack the 2+1D gapped boundary of $\SG\ST_G^4$ associated the fusion 2-category $2\rep(G)$ with an anomaly-free 2+1D topological order corresponding to $\CT$. Then the fusion 2-category of topological defects on this new 2+1D boundary topological order is $2\rep(G) \boxtimes \Sigma \CT \simeq \Sigma (\Rep(G) \boxtimes \CT)$.

Similar to the above example, the condensable $\EE_1$-algebras in $\Sigma(\Rep(G) \boxtimes \CT)$ are multi-fusion $\Rep(G) \boxtimes \CT$-modules, and the simple ones are fusion $\Rep(G) \boxtimes \CT$-modules, that is, a fusion category $\CB$ equipped with a braided functor $\Rep(G) \boxtimes \CT \to \FZ_1(\CB)$. This is equivalent to two braided functors $\Rep(G) \to \FZ_1(\CB)$ and $\CT \to \FZ_1(\CB)$ such that their images in $\FZ_1(\CB)$ are transparent to each other. Since $\CT$ is nondegenerate, the braided functor $\CT \to \FZ_1(\CB)$ is fully faithful \cite[Corollary 3.26]{DMNO13}. Denote the M\"{u}ger centralizer of $\CT$ in $\FZ_1(\CB)$ by $\CM \coloneqq \FZ_2(\CT, \FZ_1(\CB))$. Then $\FZ_1(\CB) \simeq \CM \boxtimes \CT$ by \cite[Theorem 3.13]{DGNO10} (see also \cite[Theorem 4.2]{Mueg03b} for modular case). So the fusion $\Rep(G) \boxtimes \CT$-module structure on $\CB$ is equivalent to a nondegenerate braided fusion category $\CM$, a braided equivalence $\CM \boxtimes \CT \simeq \FZ_1(\CB)$ and a braided functor $\Rep(G) \to \CM$. Similar to the above example, the braided functor $\Rep(G) \to \CM$ factors through $\Rep(G) \to \Rep(H) \to \CM$ such that $\Rep(H) \to \CM$ is fully faithful.

Given a fusion $(\Rep(G) \boxtimes \CT)$-module $\CB$, we have 
$$
\RMod_\CB(2\rep(G) \boxtimes \Sigma \CT) \simeq \RMod_\CB(2\vect) \simeq \Sigma \CB,
$$ 
and the left $\Sigma(\Rep(G) \boxtimes \CT)$-action is induced from the braided monoidal functor 
$$
\Rep(G) \boxtimes \CT \to \Rep(H) \boxtimes \CT \hookrightarrow \CM \boxtimes \CT \simeq \FZ_1(\CB). 
$$
It follows that the indecomposable $(2\rep(G) \boxtimes \Sigma \CT)$-modules are one-to-one corresponding to the pairs $(H,\CM)$, where $H \leq G$ is a subgroup defined up to conjugation and $\CM$ is a nondegenerate extension of $\Rep(H)$ such that $\CM \boxtimes \CT$ is Witt trivial. This classification of indecomposable $2\Rep(G)$-module coincides with that of gapped boundaries of $\SG\ST_G^4$ that are Morita equivalent to $2\Rep(G)\boxtimes \Sigma \CT$ (see also Example\,\ref{expl:E2_algebra_Z12VecG} and \ref{expl:E2_algebra_Z12RepG} and Theorem$^{\mathrm{ph}}$\,\ref{pthm:iterate_E2_condensation}). 
\end{expl}

\begin{expl}
Let $\pi \in Z^4(G,U(1))$. There is a gapped boundary of the 3+1D twisted $G$ gauge theory $\SG\ST_{(G,\pi)}^4$ obtained by condensing all particle-like topological defects (symmetry charges). The fusion 2-category of topological defects on this 2+1D boundary topological order is $2\vect_G^\pi$. It is the same as the $2\vect_G$ as 2-categories but the pentagonator is twisted by $\pi$.

Simple condensable $\EE_1$-algebras in $2\vect_G^\pi$ are `$\pi$-twisted $G$-graded fusion categories': they have the same defining data as $G$-graded fusion categories, but the pentagon equation holds up to $\pi$. For example, given a subgroup $H \leq G$ and a 3-cochain $\pi \in C^3(H,U(1))$ such that $\mathrm d \psi = \pi \vert_H$, there is a simple condensable $\EE_1$-algebra $\vect_H^\psi \in 2\vect_G^\pi$ with the obvious $G$-grading supported on $H$ and the associator twisted by $\psi$. 

The underlying 2-category of $\RMod_{\vect_H^\psi}(2\vect_G^\pi)$ is $2\vect_{G/H} = 2\vect^{\oplus G/H}$, and the left $2\vect_G^\pi$-action is induced by the left translation of $G$ on $G/H$. Its pentagonator is a 3-cocycle in $Z^3(G;\mathrm{Fun}(G/H,U(1)))$, which is determined by $\psi$ under the isomorphism $H^3(H;U(1)) \simeq H^3(G;\mathrm{Fun}(G/H,U(1)))$ by Shapiro's lemma for group cohomology. The underlying 2-category and the (partial) fusion rule of the bimodule 2-category $\BMod_{\vect_H^\psi|\vect_H^\psi}(2\vect_G^\pi)$ can be found in \cite{DY25}.

By Corollary$^{\mathrm{ph}}$\,\ref{pcor:4D_gapped boundary_classification}, the Morita classes of simple condensable $\EE_1$-algebras $A$ in $2\vect_G^\pi$,  or equivalently, indecomposable left $2\vect_G^\pi$-modules, classify Lagrangian $\EE_1$-algebras in $\Sigma\FZ_1(2\vect_G^\pi)$ in the Morita class of $[2\vect_G^\pi]$. We believe that this should lead us to more concrete classification result of Lagrangian $\EE_2$-algebras in $\FZ_1(2\vect_G^\pi)$ (as the full center of $A$) obtained recently in \cite{DHJ+24,Xu24a}. 
\end{expl}

\subsubsection{Examples in 3+1D} \label{sec:example_1codim_4D}

We have already seen condensations of 1-codimensional topological defects in $\mathbf{1}^{n+1}$ in Section\,\ref{sec:general_example_1-codim}. In this subsubsection, we provide some examples in finite gauge theories based on Theorem$^{\mathrm{ph}}$\,\ref{pthm:construct_cond_E1_algebras}. 

\begin{expl}
For any fusion $n$-categories $\CC$, we have $\CC \simeq \CC \boxtimes n\vect$. Therefore, for $n\geq 2$, there are infinitely many condensable $\EE_1$-algebras in any fusion $n$-categories $\CC$ because there are infinitely many multi-fusion $(n-1)$-categories as condensable $\EE_1$-algebras in $n\vect$. 
\end{expl}

\begin{expl}
We denote the 3+1D $G$-gauge theory for a finite group $G$ by $\SG\ST_G^4$. In this case, we have $\Omega\CG\CT_G^4=\FZ_1(2\vect_G)$ \cite{KTZ20}, and there are two coordinate systems of $\CG\CT_G^4$. We provide examples only in the following coordinate systems: 
\[
\CG\CT_G^4=\BMod_{2\Rep(G)|2\Rep(G)}(3\vect)^\op.
\] 
We leave the same constructions in the other coordinate system as exercises. 
\bnu

\item Let $\forget: 2\Rep(G) \to 2\vect$ be the forgetful functor, which is symmetric monoidal. Then $2\vect^\op=2\vect$ is a condensable $\EE_1$-algebra in $\BMod_{2\Rep(G)|2\Rep(G)}(3\vect)^\op$. By condensing it, we obtain the trivial phase $\SD^4=\mathbf{1}^4$ and a gapped boundary $\SM^3$ with 
$$
(\CM,m) = (\RMod_{2\Rep(G)}(3\vect), 2\vect), \quad\quad \Omega_m\CM=\Fun_{2\Rep(G)}(2\vect,2\vect) \simeq 2\vect_G.  
$$
This condensation of 1-codimensional topological defect $2\vect \in \CG\CT_G^4$ is equivalent to a condensation of 3-codimensional topological defect (i.e., a particle) $\Fun(G) \in \Rep(G) \simeq \Omega^2\CG\CT_G^4$. When $G=\Zb_2$, this particle is precisely the $e$-particle in 3+1D toric code model \cite{HZW05,KTZ20a,ZLZHKT23}. When $G=\Zb_2$ the boundary $\SM^3$ is called the rough boundary \cite{ZLZHKT23}, which can also be obtained by condensing the $\one_c$-string \cite{ZLZHKT23}. 

\item Let $\psi: 2\Rep(G) \to \CA=\Fun(2\Rep(G), 2\Rep(G))$ be the canonical monoidal functor defined by the left multiplication $x\otimes -$. Then $\CA^\op$ is a condensable $\EE_1$-algebra in $\BMod_{2\Rep(G)|2\Rep(G)}(3\vect)^\op$. By condensing it, we obtain the trivial phase $\SD^4=\mathbf{1}^4$ and a gapped boundary $\SM^3$ with 
\begin{align*}
(\CM,m) &= (\RMod_{2\Rep(G)|\CA}(3\vect), \CA), \\
\Omega_m\CM &=\Fun_{2\Rep(G)|\CA}(\CA,\CA) \simeq 2\Rep(G).   
\end{align*}
When $G=\Zb_2$, this gapped boundary of $\SG\ST_G^4$ is called the smooth boundary \cite{KTZ20a}, which can also be obtained from condensing $m$-string or the Lagrangian $\EE_2$-algebra $1\oplus m$ in $\FZ_1(2\vect_G)$ \cite{ZLZHKT23}. Note that $\CA^\op$ is a Lagrangian $\EE_1$-algebra in $\CG\CT_G^4$. 

Since $\FZ_1(2\vect_G)$ has a non-trivial braided auto-equivalence $\alpha$ \cite{ZLZHKT23}, we can twist $\psi$ by $\alpha$ as follows: 
\begin{align*}
&\psi^\alpha: 
2\Rep(G) \to 2\Rep(G) \boxtimes_{\FZ_1(2\vect_G)} \FZ_1(2\vect_G) \boxtimes_{\FZ_1(2\vect_G)} 2\Rep(G)  \\
&\hspace{1cm} \xrightarrow{\id \boxtimes_{\FZ_1(2\vect_G)} \varphi \boxtimes_{\FZ_1(2\vect_G)} \id} 
2\Rep(G) \boxtimes_{\FZ_1(2\vect_G)} \FZ_1(2\vect_G) \boxtimes_{\FZ_1(2\vect_G)} 2\Rep(G) \simeq \CA. 
\end{align*}
Then $\psi^\alpha$ endows a different algebraic structure on $\CA$ denoted by $\CA^\alpha$. By condensing $(\CA^\alpha)^\rev$, we obtained the condensed phase $\SD^4=\mathbf{1}^4$ and the so-called twist smooth boundary of $\SG\SG_G^4$ \cite{ZLZHKT23}.

\item Let $H\leq G$ be a subgroup. The forgetful functor $\forget: 2\Rep(G) \to 2\Rep(H)$ is symmetric monoidal \cite{KZ24}, which defines the condensable $\EE_1$-algebra in $\CG\CT_G^4$. By condensing $2\Rep(H)^\rev$, we obtain the finite gauge theory $\SG\ST_H^4$ as the condensed phase, i.e., 
$$
\CG\CT_H^4 \simeq \BMod_{2\Rep(H)|2\Rep(H)}(\CG\CT_G^4)^\op \simeq \Sigma\FZ_1(2\Rep(H)), 
$$
and a gapped domain wall $\SM^3$ such that 
$$
(\CM,m) = (\BMod_{2\Rep(H)|2\Rep(G)}(3\vect), 2\Rep(H)). 
$$

\item By \cite{KTZ20}, there is a canonical braided monoidal embedding $\psi_G: 2\Rep(G) \to \FZ_1(2\vect_G^\omega)$ for $\omega\in Z^4(G,U(1))$. Therefore, $\FZ_1(2\vect_G^\omega) \in \Algc_{\EE_1}(\BMod_{2\Rep(G)|2\Rep(G)}(3\vect))$. More generally, for a subgroup $H\leq G$ and $\sigma\in H^4(H,U(1))$, the following monoidal functor:
$$
2\Rep(G) \xrightarrow{\forget} 2\Rep(H) \xrightarrow{\psi_H} \FZ_1(2\vect_H^\sigma)
$$
endows the category $\FZ_1(2\vect_H^\sigma)$ with a condensable $\EE_1$-algebra structure in $\CG\CT_G^4$. By condensing it, we obtain the double of a 3+1D twisted finite gauge theory as the condensed topological order $\SD^4$, i.e.,
$$
\Omega\CD \simeq \FZ_1(2\vect_H^\sigma) \boxtimes \FZ_1(2\vect_H^\sigma)^\rev
$$
as braided fusion 2-categories, and a gapped domain wall $\SM^3$ such that 
$$
(\CM,m) = (\BMod_{\FZ_1(2\vect_H^\sigma)|2\Rep(G)}(3\vect), \FZ_1(2\vect_H^\sigma)). 
$$
This gapped domain wall was not known before. 

\item We can also construct condensations from $\CG\CT_{G_1}^4$ to any 3+1D twisted finite gauge theory $\CG\CT_{(G_2,\rho)}^4$ for $\rho\in H^4(G_2,U(1))$. We give some examples. 
\bnu
\item For a subgroup $H\leq G$ and $\sigma\in H^4(H,U(1))$, the following composed monoidal functor: 
$$
2\Rep(G) \xrightarrow{\forget} 2\Rep(H) \xrightarrow{\psi_H} \FZ_1(2\vect_H^\sigma) \xrightarrow{\forget} 2\vect_H^\sigma
$$
endows the category $2\vect_H^\sigma$ with a condensable $\EE_1$-algebra structure in $\CG\CT_G^4$. By condensing it, we obtain the 3+1D twist finite gauge theory $\CG\CT_{(H,\omega)}^4$. 

\item The following monoidal functor 
$$
2\Rep(G_1) \xrightarrow{\forget} 2\vect \simeq 2\vect \boxtimes 2\vect \to 2\vect \boxtimes 2\vect_{G_2}^\rho \simeq 2\vect_{G_2}^\rho
$$
defines condensable $\EE_1$-algebra $2\vect_{G_2}^\rho$ in $\CG\CT_{G_1}^4$. By condensing it, we obtain the 3+1D twist finite gauge theory $\CG\CT_{(G_2,\rho)}^4$. This condensation can be viewed as a two-step condensation: first condense $\SG\ST_{G_1}^4$ to $\mathbf{1}^4$ then condense $\mathbf{1}^4$ to $\CG\CT_{(G_2,\rho)}^4$. Of course, this two-step condensation applies to condensations between any two non-chiral topological orders.  

\item One can always use Theorem$^{\mathrm{ph}}$\,\ref{pthm:construct_cond_E1_algebras_II} and Corollary$^{\mathrm{ph}}$\,\ref{pcor:n=3_classification} to construct (or classify) examples systematically. In this way, one can partially reduce the problem to the construction (or classification) of $2\vect_{G_1}$-$2\vect_{G_2}^\rho$-bimodules. 
\enu

\enu
\end{expl}

Using Theorem$^{\mathrm{ph}}$\,\ref{pthm:construct_cond_E1_algebras_II} and Corollary$^{\mathrm{ph}}$\,\ref{pcor:n=3_classification}, we can construct (or even classify to some extent) condensations from $\SG\ST_{G_1}^4$ to $\SG\ST_{G_2}^4$ by specifying a $2\Rep(G_1)$-$2\Rep(G_2)$-bimodule $\CX$ (as illustrated in the following picture, in which we set $\CK_\CX:=\Fun_{2\Rep(G_1)|2\Rep(G_2)}(\CX,\CX)$). 
$$
\begin{array}{c}
\begin{tikzpicture}[scale=1]
\fill[blue!20] (-2,0) rectangle (0,2) ;
\fill[teal!20] (0,0) rectangle (2,2) ;
\draw[->-,ultra thick] (-2,0)--(0,0) node[midway,below] {\scriptsize $2\Rep(G_1)$} ;
\draw[->-,ultra thick] (-2,2)--(-2,0) node[midway,left] {\scriptsize $2\Rep(G_1)$} ;
\draw[->-,ultra thick] (0,0)--(2,0) node[midway,below] {\scriptsize $2\Rep(G_2)$} ;

\draw[blue,->-,ultra thick] (0,2)--(0,0) ; 
\draw[fill=white] (-0.07,-0.07) rectangle (0.07,0.07) ; 
%
%
\node at (-1,0.8) {\scriptsize $\FZ_1(2\Rep(G_1))$} ;
\node at (1,0.8) {\scriptsize $\FZ_1(2\Rep(G_2))$} ;
\node at (-1,1.5) {\scriptsize $\SG\ST_{G_1}^4$} ;
\node at (1,1.5) {\scriptsize $\SG\ST_{G_2}^4$} ;
\node at (0,-0.25) {\scriptsize $\CX$} ;
\node at (0,2.2) {\scriptsize $\CK_\CX$} ;
\node at (-2,2.2) {\scriptsize $2\Rep(G_1)$} ;

\draw[decorate,decoration=brace,very thick] (-2,2.4)--(0,2.4) node[midway, above] {\scriptsize $\CA_\CX$};

\end{tikzpicture}
\end{array}
$$
More precisely, given such a bimodule $\CX$, we obtain a condensable $\EE_1$-algebra $\CA_\CX^\op$ in $\CG\CT_{G_1}^4$ by the following monoidal functor 
$$
2\Rep(G_1) \to 2\Rep(G_1)\boxtimes_{\FZ_1( 2\Rep(G_1))} \CK_\CX \simeq \Fun_{2\Rep(G_2)^\rev}(\CX,\CX) =: \CA_\CX
$$
such that condensing $\CA_\CX^\op$ produces $\SG\ST_{G_2}^4$ as the condensed phase. For example, such $\CX$ can be constructed from a monoidal $\Rep(G_1)$-$\Rep(G_2)$-bimodules, such as $\Rep(G)$ for a subgroup $G \leq G_1 \times G_2$, via delooping, i.e., $\CX = 2\Rep(G)$ is a $2\Rep(G_1)$-$2\Rep(G_2)$-bimodule.

More generally, a $2\Rep(G_1)$-$2\Rep(G_2)$-bimodule $\CX$ is just a right $2\Rep(G_1\times G_2)$-module, i.e., objects in $3\Rep(G_2\times G_2)$. A simple object in $3\Rep(G_2\times G_2)$ is precisely $\Sigma\CL$ for a fusion 1-category $\CL$ over $\Rep(G_2\times G_2)$ (i.e., a fusion 1-category $\CL$ equipped with a braided functor $\phi:\Rep(G_2\times G_2) \to \FZ_1(\CL)$) \cite{KZZZ25}. Note that any indecomposable multi-fusion 1-category $\CQ$ Morita equivalent to $\CL$ produces the same simple $2\Rep(G_1)$-$2\Rep(G_2)$-bimodule $\CX$, i.e., $\Sigma\CL \simeq \Sigma\CQ$.

For example, let $G\leq G_1\times G_2$ be a subgroup. Then $\Rep(G)$ is fusion 1-category over $\Rep(G_2\times G_2)$ and $\CX=\Sigma\Rep(G)$ is a right $2\Rep(G_1\times G_2)$-module\footnote{An indecomposable left $\Rep(G)$-module is $\Rep(H,\alpha)$ for a subgroup $H\leq G$ and $\alpha \in Z^2(H,U(1))$ (Recall Example\,\ref{expl:algebras_in_RepG}). We have the following composed braided monoidal functor 
$\Rep(G_2\times G_2) \xrightarrow{\forget} \Rep(G) \hookrightarrow \FZ_1(\Fun_{\Rep(G)}(\Rep(H,\alpha), \Rep(H,\alpha))^\op)$
which defines the same indecomposable $2\Rep(G_1)$-$2\Rep(G_2)$-bimodule $\Sigma\Rep(G)$.}. 
If we require $\phi$ facts through the embedding $\Rep(G_1\times G_2) \to \FZ_1(\Rep(G_1\times G_2))$, such $\phi$'s can be classified, and it is easy to see that such obtained $\CX$ are all of type $\Sigma\Rep(G)\boxtimes \CS$ for a separable 2-category $\CS$.  

By Corollary$^{\mathrm{ph}}$\,\ref{pcor:n=2_classification} (2), the assignment (\ref{map:X_FunBPXX}) of equivalence classes, i.e., 
$$
\CX \mapsto \Fun_{2\Rep(G_1)|2\Rep(G_2)}(\CX,\CX)^\op \in \Algc_{\EE_1}(\RMod_{\FZ_1(2\Rep(G_1))}(3\vect))
$$
is bijective if we restrict the codomain to the equivalence classes within the Morita class of 
$$
2\Rep(G_1) \boxtimes 2\Rep(G_2)^\op \in \Algc_{\EE_1}(\RMod_{\FZ_1(2\Rep(G_1))}(3\vect))
$$ 
(recall Remark\,\ref{rem:Morita_equivalence_E1_algebras}). 


More generally, an indecomposable condensable $\EE_1$-algebra in $\CG\CT_{G_1}^4=\RMod_{\FZ_1(2\Rep(G_1))}(3\vect)$ that condenses to $\CG\CT_{G_2}^4$ is precisely an indecomposable multi-fusion 2-category $\CK$ that is Morita equivalent to $\CK_\CX\boxtimes \CT \in \Algc_{\EE_1}(\RMod_{\FZ_1(2\Rep(G_1))}(3\vect))$ for a bimodule $\CX$ and a non-degenerate fusion 2-category $\CT$ (or equivalently, using the characterization of $\CK$ given in Corollary$^{\mathrm{ph}}$\,\ref{pcor:n=3_classification}). In the coordinate system $\CG\CT_{G_1}^4=\BMod_{2\Rep(G)|2\Rep(G)}(3\vect)^\op$, this algebra $\CK$ is precisely  the algebra $\CA$ defined by the monoidal functor 
$$
2\Rep(G_1) \to \CK \boxtimes_{\FZ_1(2\Rep(G_1))} 2\Rep(G_1) =:\CA. 
$$ 
In this way, we have covered all condensations from $\CG\CT_{G_1}^4$ to $\CG\CT_{G_2}^4$.


\subsubsection{Examples in higher dimensions} \label{sec:1codim_n+2D}

\begin{expl}
Consider $n+$2D finite gauge theories $\SG\ST_G^{n+2}$. We use the following coordinate. 
$$
\CG\CT_G^{n+2} = \BMod_{n\Rep(G)|n\Rep(G)}((n+1)\vect)^\op.  
$$
An indecomposable condensable $\EE_1$-algebra in $\BMod_{n\Rep(G)|n\Rep(G)}((n+1)\vect)$ is precisely given by an indecomposable multi-fusion $n$-category $\CA^\op$ equipped with a monoidal functor $n\Rep(G) \to \CA$. 
\bnu

\item The canonical monoidal functors $\forget: n\Rep(G) \to n\vect$ and $n\Rep(G) \to \Fun(n\Rep(G), n\Rep(G))$ define two Lagrangian $\EE_1$-algebras in $\CG\CT_G^{n+2}$, which determine the rough boundary (i.e., $\Omega_m\CM=n\vect_G$) and the smooth boundary (i.e., $\Omega_m\CM=n\Rep(G)$), respectively. The canonical forgetful functor $\forget: n\Rep(G) \to n\Rep(H)$ for a subgroup $H\leq G$ \cite{KZ24} defines a condensable $\EE_1$-algebra in $\CG\CT_G^{n+2}$, which produces the finite gauge theory $\SG\ST_H^{n+2}$ as the condensed phase. This example can be recovered as a condensation of higher codimensional defects in Example\,\ref{expl:GT_Gn+1_GtoH}. All above examples can be viewed as special cases of more general constructions explained in Part 4 in this Example.

\item By \cite{KTZ20}, there is a canonical braided monoidal embedding $\psi_G: n\Rep(G) \to \FZ_1(n\vect_G^\omega)$ for $\omega\in Z^{n+2}(G,U(1))$. Therefore, $\FZ_1(n\vect_G^\omega)^\op \in \Algc_{\EE_1}(\BMod_{n\Rep(G)|n\Rep(G)}((n+1)\vect)^\op)$. By condensing it, we obtain the double of twisted finite gauge theory $\SG\ST_{(G,\omega)}^{n+1} \boxtimes \overline{\SG\ST_{(G,\omega)}^{n+1}}$ as the condensed phase $\SD^{n+2}$, i.e.,
$$
\CD\simeq \Mod_{\FZ_1(n\vect_G^\omega)}^{\EE_1}((n+1)\vect)^\op, \quad\quad \Omega\CD \simeq \FZ_1(n\vect_G^\omega) \boxtimes \FZ_1(n\vect_G^\omega)^\rev
$$
where the first equivalence is monoidal and the second one is braided monoidal, and a gapped domain wall $\SM^n$ such that 
$$
(\CM,m) = (\BMod_{\FZ_1(n\vect_G^\omega)|n\Rep(G)}((n+1)\vect), \FZ_1(n\vect_G^\omega)). 
$$

\item For $H\leq G$ and $\sigma\in H^{n+2}(H,U(1))$, the composed braided monoidal functor 
$$
n\Rep(G) \xrightarrow{\forget} n\Rep(H) \xrightarrow{\psi_H} \FZ_1(n\vect_H^\sigma)
$$
defines a condensable $\EE_1$-algebra $\FZ_1(n\vect_H^\sigma)^\op$ in $\BMod_{n\Rep(G)|n\Rep(G)}((n+1)\vect)^\op$. By condensing it, we obtain the double of twisted finite gauge theory $\SG\ST_{(H,\sigma)}^{n+1} \boxtimes \overline{\SG\ST_{(H,\sigma)}^{n+1}}$ as the condensed phase $\SD^{n+2}$, and a gapped domain wall $\SM^n$ such that 
$$
(\CM,m) = (\BMod_{\FZ_1(n\vect_H^\sigma)|n\Rep(G)}((n+1)\vect), \FZ_1(n\vect_H^\sigma)). 
$$
The following composed monoidal functor
$$
n\Rep(G) \xrightarrow{\forget} n\Rep(H) \xrightarrow{\psi_H} \FZ_1(n\vect_H^\sigma) \xrightarrow{\forget} n\vect_H^\sigma
$$
defines a condensable $\EE_1$-algebra $(n\vect_H^\sigma)^\op$, which determines $\SG\ST_{(H,\sigma)}^{n+1}$ as the condensed phase.

\item By Theorem$^{\mathrm{ph}}$\,\ref{pthm:construct_cond_E1_algebras_II}, we can construct a 1-codimensional defect condensation from $\SG\ST_{G_1}^{n+2}$ to $\SG\ST_{G_2}^{n+2}$ by specifying an $n\Rep(G_1)$-$n\Rep(G_2)$-bimodule $\CX$. More precisely, given such a $\CX$, we obtain a condensable $\EE_1$-algebra $\CA_\CX^\op$ in $\CG\CT_{G_1}^{n+1}$ by the following monoidal functor 
$$
n\Rep(G_1) \to \Fun_{n\Rep(G_2)^\rev}(\CX,\CX) =: \CA_\CX
$$
such that condensing $\CA_\CX^\op$ produces $\SG\ST_{G_2}^{n+2}$ as the condensed phase. Examples of such $\CX$ can be constructed as follows. 
\bnu
\item The forgetful functor $\forget: n\Rep(G_1\times G_2)\to n\Rep(H)$ for a subgroup $H\leq G_1 \times G_2$ \cite{KZ24} defines an $n\Rep(G_1)$-$n\Rep(G_2)$-bimodule $\CX=n\Rep(H)$. 

\item A simple $n\Rep(G_1)$-$n\Rep(G_2)$-bimodule $\CX$ is precisely a simple object in $(n+1)\Rep(G_1\times G_2)$, which is of the form $\Sigma\CL$ for a fusion $(n-1)$-category $\CL$ over $(n-1)\Rep(G_1\times G_2)$ \cite{KZZZ25}. If we choose $\CL=(n-1)\Rep(H)$, we recover the construction in (4.a). Note that any indecomposable multi-fusion $(n-1)$-category $\CQ$ that is Morita equivalent to $\CL$ produces the same simple $n\Rep(G_1)$-$n\Rep(G_2)$-bimodule $\CX$, i.e., $\Sigma\CL \simeq \Sigma\CQ$. Note that $\CX=n\Rep(H)\boxtimes \CS$ for any separable $n$-category $\CS$ gives more but obvious examples. 
\enu
When $G_2$ is the trivial group, such constructed condensable $\EE_1$-algebras $\CA_\CX^\op$ are all Lagrangian. 


\enu

\end{expl}

\begin{rem}
From all these examples and from the results of Theorem$^{\mathrm{ph}}$\,\ref{thm:condensable_E1Alg_af_C}, \ref{pthm:construct_cond_E1_algebras} and \ref{pthm:construct_cond_E1_algebras_II} and Corollary$^{\mathrm{ph}}$\,\ref{pcor:n=2_classification} and \ref{pcor:n=3_classification}, we can see that the mathematical theory of the condensations of 1-codimensional defects are much simpler than that of the condensations of anyons. It is also much simpler in constructing examples or sometimes even obtaining a classification. As we show later, it also provides some powerful tools in the study of condensable/Lagrangian $\EE_2$-algebras in (non-degenerate) braided fusion higher categories. 
\end{rem}

\begin{rem}
Since our knowledge of (braided) fusion higher categories are very limited, only explicit examples of condensations we can provide are those in the twisted finite gauge theories in higher dimensions. Actually, there are a few other workable examples from fermionic modular 2-categories \cite{LW19, JF20a}. Examples of condensations can certainly be found there. We encourage intrigued readers to work on these cases. 
\end{rem}

Before we end this section, we give a few examples of 1-codimensional defect condensations in anomalous $n+2$D topological orders. More precisely, we consider condensable $\EE_1$-algebras in $(n+1)\vect_G$, $(n+1)\Rep(G)$ and $(n+1)\vect_G^\pi$.  

\begin{expl}
For a finite group $G$, $(n+1)\vect_G$ is the fusion $(n+1)$-category of topological defects on a gapped boundary of the finite gauge theory $\SG\ST_G^{n+3}$. The condensable $\EE_1$-algebras in $(n+1)\vect_G$ are $G$-graded multi-fusion $n$-categories, and the simple ones are $G$-graded fusion $n$-categories. A $G$-graded (multi-)fusion $n$-category is a (multi-)fusion $n$-category $\CA$ equipped with a $G$-grading, i.e., $\CA = \oplus_{g \in G} \CA_g$, such that $x \otimes y \in \CA_{gh}$ for any $x \in \CA_g$ and $y \in \CA_h$. For example, for a subgroup $H \subseteq G$ and $\omega \in Z^{n+2}(H,U(1))$, the fusion $n$-category $n\vect_H^\omega$ is a $G$-graded fusion $n$-category with the $G$-grading supported on $H$.


For a $G$-graded fusion $n$-category $\CA$, we give a more explicit mathematical description of the left $2\vect_G$-module $\RMod_\CA(2\vect_G)$. First, similar to Example\,\ref{expl:E1_algebras_2vectG}, we define an $(n+1)$-category $\mathrm B_G \CA$ as follows:
(1) the objects are elements in $G$; (2) the hom $n$-categories are $\Hom_{\mathrm B_G \CA}(g,h) \coloneqq \CA_{hg^{-1}}$; (3) the composition $n$-functors
\[
\Hom_{\mathrm B_G \CA}(h,k) \times \Hom_{\mathrm B_G \CA}(g,h) = \CA_{kh^{-1}} \times \CA_{hg^{-1}} \to \CA_{kg^{-1}} = \Hom_{\mathrm B_G \CA}(g,h)
\]
are given by the tensor product functor of $\CA$. We denote the condensation completion of $\mathrm B_G \CA$ by $\Sigma_G \CA$, which is a separable $(n+1)$-category. Similar to Example \ref{expl:E1_algebras_2vectG}, there is an equivalence of $(n+1)\vect_G$-modules $\RMod_\CA((n+1)\vect_G) \simeq \Sigma_G \CA$.
\end{expl}

\begin{expl} \label{expl:E1-algebra_in_n+1RepG}
For a finite group $G$, $(n+1)\Rep(G)$ is the fusion $(n+1)$-category of topological defects on another gapped boundary of the finite gauge theory $\SG\ST_G^{n+3}$. There are two coordinates on $(n+1)\rep(G)$.
\bit
\item $(n+1)\rep(G) \simeq \Fun(\mathrm B G,(n+1)\vect) \simeq \LMod_{n\vect_G}((n+1)\vect)$: The objects are separable $n$-categories equipped with $G$-action. A condensable $\EE_1$-algebra in $(n+1)\rep(G)$ is precisely a multi-fusion $n$-category $\CA$ equipped with a monoidal $G$-action (i.e., a monoidal $n$-functor $G \to \Aut^\otimes(\CA)$). Such $\CA$ is also called a $G$-equivairant multi-fusion $n$-category. 
A condensable $\EE_1$-algebra $\CA$ is simple if and only if $G$ acts transitively on the simple direct summands of the tensor unit of $\CA$.

\item $(n+1)\rep(G) \simeq \Sigma n\rep(G) \simeq \RMod_{n\rep(G)}((n+1)\vect) \simeq \LMod_{n\rep(G)}((n+1)\vect)$: The objects are separable $n$-categories equipped with an $n\rep(G)$-action. The condensable $\EE_1$-algebras are multi-fusion $n\rep(G)$-module $\CB$, i.e., $\CB$ is a multi-fusion $n$-category equipped with a braided monoidal $n$-functor $n\rep(G) \to \FZ_1(\CB)$. A condensable $\EE_1$-algebra $\CB$ is simple if and only if it is fusion.
\eit

These two coordinates should again be related by equivariantization and de-equivariantization. Both notions should be properly generalized to higher categories. For example, the equivariantization of $\CA$, denoted by $\CA^G$, should consist of $G$-equivariant objects of $\CA$. A $G$-equivariant object in $\CA$ is an object $X\in \CA$ equipped with an isomorphisms $u_g: F_g(X) \simeq X$ for $g\in G$ and $F_g$ is the image of $g$ in $\Aut^\otimes(\CA)$ such that the following diagram 
$$
\xymatrix{
F_g(F_h(X)) \ar[rr]^{F_g(u_h)} \ar[d]_\simeq & & F_g(X) \ar[d]^{u_g}\\
F_{gh}(X) \ar[rr]^{u_{gh}} & & X
}
$$ 
commutes up to higher isomorphisms. A 1-morphism should be defined by a 1-morphism in $\CA$ that commutes with $u_g, \forall g\in G$. However, one can see that it is impossible to avoid the higher coherence data and the higher morphisms in $\CA^G$ explicitly. It is easier to define the de-equivariantization: $\CB \mapsto \CB_G$. 
We define $\CB_G:=\RMod_{[\one,\one]}(\CB)$, where $\one$ is the tensor unit of $n\vect$ and $[\one,\one]$ is the internal hom algebra in $n\Rep(G)$ for the $n\Rep(G)$-module $n\vect$. Therefore, the notion of a $G$-equivariant multi-fusion $n$-category can also be understood as the de-equivariantization of a multi-fusion $n\rep(G)$-module $\CB$.

Given a condensable $\EE_1$-algebra $\CB \in \LMod_{n\rep(G)}((n+1)\vect) \simeq (n+1)\rep(G)$, there is an equivalence of $(n+1)$-categories $\RMod_\CB((n+1)\rep(G)) \simeq \RMod_\CB((n+1)\vect) \simeq \Sigma \CB$. The left $(n+1)\rep(G)$-action on $\Sigma \CB$ is defined by a monoidal $(n+1)$-functor
\[
(n+1)\rep(G) \simeq \Sigma n\rep(G) \to \Sigma \FZ_1(\CB) \simeq \FZ_0(\Sigma \CB),
\]
which is induced from the braided monoidal functor $n\Rep(G) \to \FZ_1(\CB)$. Recall that such braided monoidal functors $n\Rep(G) \to \CM$ for a non-degenerate braided fusion $n$-category $\CM$ classify Lagrangian $\EE_1$-algebras in $\CG\CT_G^{n+2}$ by Corollary\,\ref{cor:LA_n+2D_finite_gauge_theory}. 
\end{expl}

\begin{expl}
For a finite group $G$ and $\pi \in Z^{n+3}(G,U(1))$, $(n+1)\vect_G^\pi$ is the fusion $(n+1)$-category of topological defects on a gapped boundary of the twisted finite gauge theory $\SG\ST_{(G,\pi)}^{n+3}$. It is the same $(n+1)$-category as the $(n+1)\vect_G$ but with the associativity twisted by $\pi$ in its highest coherence data. 

A simple condensable $\EE_1$-algebra in $(n+1)\vect_G^\pi$ is precisely a `$\pi$-twisted $G$-graded fusion $n$-category', which has the same defining data as a $G$-graded fusion $n$-category but with the associativity twisted by $\pi$ in its highest coherence data. 
\bnu
\item[-] For example, given a subgroup $H \leq G$ and an $(n+3)$-cochain $\pi \in C^{n+3}(H,U(1))$ such that $\mathrm d \psi = \pi \vert_H$, there is a simple condensable $\EE_1$-algebra $n\vect_H^\psi \in (n+1)\vect_G^\pi$ with the obvious $G$-grading supported on $H$ and the associativity twisted by $\psi$ in its highest coherence data.
\enu

The underlying $(n+1)$-category of $\RMod_{n\vect_H^\psi}((n+1)\vect_G^\pi)$ is $(n+1)\vect_{G/H} \simeq  (n+1)\vect^{\oplus G/H}$. The left $(n+1)\vect_G^\pi$-action on it is defined by the left translation of $G$ on $G/H$. The associativity of the action is twisted by an $(n+3)$-cocycle in $Z^{n+3}(G;\mathrm{Fun}(G/H,U(1)))$, which is determined by $\psi$ under the isomorphism $H^{n+3}(H;U(1)) \simeq H^{n+3}(G;\mathrm{Fun}(G/H,U(1)))$. 

\end{expl}


\newpage


\section{Condensations of 2-codimensional topological defects} \label{sec:condense_2-codim_defect}

In previous section, we have seen that a condensation of 1-codimensional topological defects defines a phase transition from $\SC^{n+1}$ to a new phase $\SD^{n+1}$, where both phase $\SC^{n+1}$ and $\SD^{n+1}$ are potentially anomalous $n+$1D topological orders. When we restrict to the case both $\SC^{n+1}$ and $\SD^{n+1}$ are anomaly-free and simple, both categories of topological defects $\CC$ and $\CD$ are determined by $\Omega\CC$ and $\Omega\CD$, respectively. In this case, it is possible to define a condensation in $\SC^{n+1}$ by condensing a 2-codimensional topological defect in $\Omega\CC$ directly. When $n=2$, such a condensation is precisely what was known as an anyon condensation (or boson condensation) defined in the non-degenerate braided fusion 1-category $\Omega\CC$ \cite{MS89a,BSS02,BSS03,BS09,BSH09,BSS11,KS11a,Lev13,BJQ13,Kon14e}. When $\SC^{n+1}$ is anomalous, it is still possible to define a condensation of 2-codimensional topological defect $A$ by first condensing it to a condensed defect $\Sigma A$ of codimension 1 followed by a condensation of $\Sigma A$. We postpone the study of this two-step condensation to Section\,\ref{sec:condense_k-codim_defect}. 

In this section, we first review the theory of anyon condensations in 2+1D anomaly-free simple topological orders with some new details and from some new perspectives; then we generalize it to higher dimensions.

\subsection{Anyon condensations in 2+1D} \label{sec:2codim_2d}
In Section\,\ref{sec:anyon_cond_algebra}, we review the main results of an algebraic bootstrap approach towards anyon condensation \cite{Kon14e} but from a new perspective of higher algebras and higher representations. We also introduce necessary new notations along the way. In Section\,\ref{sec:example_2codim_3D}, we provide a lot of concrete examples of anyon condensations. In Section\,\ref{sec:2codim_3D_geometric_approach}, we introduce a geometric approach, which can be generalized automatically to higher dimensions.

\subsubsection{Algebraic approach} \label{sec:anyon_cond_algebra}
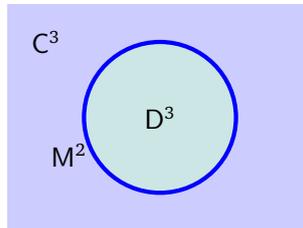
\begin{figure}[htbp] 
\[
\begin{tikzpicture}
\fill[blue!20] (-2,0) rectangle (2,3) ;
\fill[teal!20] (0,1.5) circle (1);
\node at (-1.5,2.5) {$\SC^3$} ;
\node at (0,1.5) {$\SD^3$} ;
\node at (-1.2,1) {$\SM^2$} ;
\draw [blue, ultra thick] (0,1.5) circle [radius=1] ;
\end{tikzpicture}
\]
\caption{an anyon condensation in $\SC^3$ depicted in the spatial dimension}
\label{fig:2d-condensation}
\end{figure}

We sketch a rederivation of the main results in \cite{Kon14e} but from a new perspective explained in Section\,\ref{sec:E2=E1+E1}. Assume that an anyon condensation occurs in a disk-like region within a 2+1D anomaly-free simple topological order $\SC^3$ as depicted in Figure\,\ref{fig:2d-condensation}. We denote the condensed topological order by $\SD^3$. We assume that $\SD^3$ is also simple\footnote{If $\SD^3$ is not simple, it simply means that the associated condensable $\EE_2$-algebra $A$ in $\Omega\CC$ is not simple. The result can be easily derived from the $A$-being-simple case. Note that, in this case, the fact $\CD \neq \Sigma\Omega\CD$ adds some slight inconvenience to state the results. So we simply leave the non-simple cases to intrigued readers.}. We denote the gapped domain wall by $\SM^2$. The categories of all topological defects in $\SC^3$ and $\SD^3$ are $\CC$ and $\CD$, respectively; and that of topological defects of codimension 2 (or anyons) are $\Omega\CC$ and $\Omega\CD$, respectively. Mathematically, $\SM^2=(\CM,m)$, where $\CM$ is the category of walls conditions and $m$ labels a single wall condition and topological particles on $\SM^2$ form a multi-fusion 1-category $\Omega_m\CM$. 

\begin{rem} \label{rem:Runkel_works}
A 2+1D topological order is well-defined on an open 2-disk (of large size) in space dimension. 
Roughly speaking, a 3D topological order can be viewed as a 3D fully dualizable TQFT on an open disk. Therefore, it is natural to ask if an anyon condensation can be defined for a 3D TQFT defined on non-trivial spacetime manifolds. Parallel to anyon condensations in 2+1D topological orders, there is a program of `orbiford construction' (or `gauging defects') in 2+1D Reshetikhin-Turaev TQFT \cite{RT91} developed by Carqueville, Runkel, Schaumann and Mulevicius in a series of works \cite{CR16,CRS18,CRS20,CMRSS21,Mul24,MR23,CM23,CMRSS24} (see \cite{Car23} for a survey). More precisely, they gave a state-sum construction of a new 2+1D Reshetikhin-Turaev TQFT from an old one based on the so-called `an orbifold datum'. This program is not limited to 2+1D \cite{CRS19}. 
\end{rem}

It is possible to derive a precise relation among the categories $\Omega\CC, \Omega\CD, \Omega_m\CM$ via natural physical requirements as shown in \cite{Kon14e}. 
\bnu

\item Anyons in $\SD^3$ come from those in $\SC^3$. In other words, an anyon in $\SD^3$ is automatically an anyon in $\SC^3$, i.e. $\mathrm{ob}(\Omega\CD) \subset \mathrm{ob}(\Omega\CC)$. All fusion-splitting channels among anyons in $\Omega\CD$ come from those in $\Omega\CC$, i.e. $\hom_{\Omega\CD}(x,y) \subset \hom_{\Omega\CC}(x,y)$ for $x,y\in \mathrm{ob}(\Omega\CD)$. As a consequence, $\Omega\CD$ is a sub-category of $\Omega\CC$. 

\item The trivial particle $1_{\one_\CD}$ in $\SD$, i.e., the tensor unit of $\Omega\CD$, is a non-trivial particle $A$ in $\Omega\CC$ unless the condensation is trivial. 

\item Similar to 1+1D cases, particle condensation is triggered by introducing interactions among particles. For $a,b\in \Omega\CD \subset \Omega\CC$, a sub-Hilbert space $a\otimes_{\Omega\CD} b$ of the Hilbert space associated to $a\otimes_{\Omega\CC} b$ becomes energy favorable due to the interaction. In other words, the condensation process produces a family of projections (called condensation maps): 
$$
a \otimes_{\Omega\CC} b \xrightarrow{p_{a,b}} a\otimes_{\Omega\CD} b, \quad\quad \forall a,b\in \Omega\CD. 
$$
In particular, we have the following special condensation maps: 
\begin{align*}
&\mu_A: A\otimes_{\Omega\CC} A \xrightarrow{p_{A,A}} A\otimes_{\Omega\CD} A \simeq A.  \\
&\forall x\in \Omega\CD, \quad \quad \mu_x^L: A\otimes_{\Omega\CC} x \xrightarrow{p_{A,x}} A\otimes_{\Omega\CD} x \simeq x, \quad\quad \mu_x^R: x\otimes_{\Omega\CC} A \xrightarrow{p_{x,A}} x\otimes_{\Omega\CD} A \simeq x. 
\end{align*}
Moreover, since the trivial anyon $1_{\one_\CC}$ in $\Omega\CC$ should condense into the trivial anyon in $\Omega\CD$, we expect to have a morphism $\eta_A: 1_{\one_\CC} \to A$.

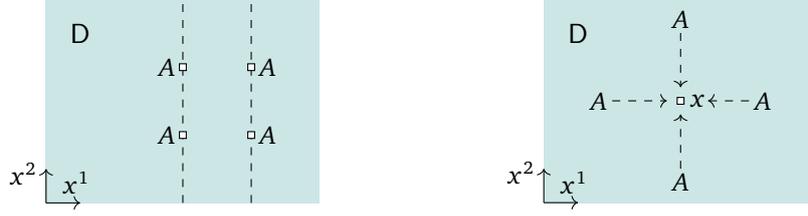
\begin{figure}[t] 
\[
\begin{array}{c}
\begin{tikzpicture}[scale=0.9]
\fill[teal!20] (-2,0) rectangle (2,3) ;
\draw [dashed] (0,0) -- (0,3) ; 
\draw [dashed] (1,0) -- (1,3) ; 
\draw[fill=white] (-0.05,0.95) rectangle (0.05,1.05) node[midway,left] {$A$} ;
\draw[fill=white] (-0.05,1.95) rectangle (0.05,2.05) node[midway,left] {$A$} ;
\draw[fill=white] (0.95,0.95) rectangle (1.05,1.05) node[midway,right] {$A$} ;
\draw[fill=white] (0.95,1.95) rectangle (1.05,2.05) node[midway,right] {$A$} ;
\node at (-1.5,2.5) {$\SD$} ;
\draw[->] (-2,0) -- (-1.5,0) node [very near end, above] {$x^1$} ;
\draw[->] (-2,0) -- (-2,0.5) node [very near end, left] {$x^2$} ;
\end{tikzpicture}
\end{array}
\hspace{2cm}
\begin{array}{c}
\begin{tikzpicture}[scale=0.9]
\fill[teal!20] (-2,0) rectangle (2,3) ;
\draw[fill=white] (-0.05,1.45) rectangle (0.05,1.55) node[midway,right] {$x$} ;
\node at (-1.5,2.5) {$\SD$} ;
\draw [dashed,->] (-1,1.5) -- (-0.2,1.5) ; 
\draw [dashed,->] (1,1.5) -- (0.4,1.5) ; 
\draw [dashed,->] (0,2.5) -- (0,1.7) ; 
\draw [dashed,->] (0,0.5) -- (0,1.3) ; 
\node at (-1.2,1.5) {$A$} ;
\node at (1.2,1.5) {$A$} ;
\node at (0,2.7) {$A$} ;
\node at (0,0.3) {$A$} ;
\draw[->] (-2,0) -- (-1.5,0) node [very near end, above] {$x^1$} ;
\draw[->] (-2,0) -- (-2,0.5) node [very near end, left] {$x^2$} ;
\end{tikzpicture}
\end{array}
\]
\caption{a deconfined particle $x$ as an $\EE_2$-$A$-module  in $\Omega\CC$}
\label{fig:deconfined_particles_E2_module}
\end{figure}

\item Similar to the 1+1D cases, the triple $(A,\mu_A,\eta_A)$ defines a separable algebra in $\Omega\CC$ satisfying a new property: $A$ is commutative. We do not repeat the reason provided for this commutativity in \cite{Kon14e}. Instead, we provide a new proof from the perspective of Section\,\ref{sec:E2=E1+E1}. Consider the physical configuration depicted left picture in Figure\,\ref{fig:deconfined_particles_E2_module}. By the same arguements in 1+1D case, there are two algebraic structures on $A$. One algebraic structure is defined for the horizontal fusion product $\otimes^1$, i.e. $\mu_A^1: A\otimes^1 A \to A$. The other is defined for the vertical fusion product $\otimes^2$, i.e. $\mu_A^2: A\otimes^2 A \to A$. The following diagram should be automatically commutative. 
\be \label{diag:mu1_mu2}
\xymatrix{
(A \otimes^1 A) \otimes^2 (A \otimes^1 A) \ar[d]_{\mu_A^1 \otimes^2 \mu_A^1}  \ar[rr]^{\delta_{A,A,A,A}}_\simeq &  & 
(A \otimes^2 A) \otimes^1 (A \otimes^2 A) \ar[d]^{\mu_A^2 \otimes^1 \mu_A^2 }   \\
A \otimes^2 A \ar[dr]_{\mu_A^2} & & 
A \otimes^1 A
\ar[dl]^{\mu_A^1} \\
& A &
}
\ee
Notice that $(\mu_A^2 \otimes^1 \mu_A^2) \circ \delta_{A,A,A,A}$ defines an algebraic structure on $A\otimes^1 A$ and $(\mu_A^1 \otimes^2 \mu_A^1) \circ \delta_{A,A,A,A}^{-1}$ defines an algebraic structure on $A\otimes^2 A$. Then above commutativity diagram simply says that both morphisms $\mu_A^2: A\otimes^2 A \to A$ and $\mu_A^1: A\otimes^1 A \to A$ are algebra homomorphisms. 

The unit morphism $\eta_A: 1_{\one_\CC} \to A$ is the same for both algebraic structures. We obtain a commutative diagram:
$$
\xymatrix{
1_{\one_\CC} \otimes^1 1_{\one_\CC} \ar[r]^{\eta_A \otimes^1 1} \ar[d]_\simeq & A\otimes^1 1_{\one_\CC} \ar[r]^{1\otimes \eta_A} \ar[d]_\simeq & A\otimes^1 A \ar[ld]^{\mu_A^1} \\
1_{\one_\CC} \ar[r]^{\eta_A} & A
}
$$
where the commutativity of the left square is due to the naturalness of the vertical $\simeq$ and that of the right triangle is due to the unit property of $(A,\mu_A^1,\eta_A)$. Then the commutativity of the outerdiagram means that $\eta_A: 1_{\one_\CC} \to (A,\mu_A^1,\eta_A)$ is an algebraic homomorphism. Similarly, one can show that $\eta_A: 1_{\one_\CC} \to (A,\mu_A^2,\eta_A)$ is also an algebra homomorphism. 

In summary, both $\mu_A^2: A\otimes^2 A \to A$ and $\eta_A$ are morphisms in $\Alg_{\EE_1}(\Omega\CC)$. Therefore, $(A,\mu_A^2,\eta_A)$ defines an algebra in $\Alg_{\EE_1}(\Omega\CC)$. In other words, $A\in \Alg_{\EE_1}(\Alg_{\EE_1}(\Omega\CC))=:\Alg_{\EE_2}(\Omega\CC)$. This fact automatically implies the compatibility between $\mu_A^1$ and $\mu_A^2$ and the commutativity of $A$. Indeed, by composing the diagram (\ref{diag:mu1_mu2}) with the unit morphism $\eta_A: 1_\one \to A$ that are compatible with (\ref{eq:delta_a11d}) and (\ref{eq:braiding_from_2fusion_1}), we obtain two commutative diagrams:
\be
\begin{array}{c}
\xymatrix{
A\otimes^2 A \ar[rr]^\simeq \ar[dr]_{\mu_A^2} & & A\otimes^1 A \ar[dl]^{\mu_A^1} \\
& A & 
}
\end{array}
\hspace{1cm}
\begin{array}{c}
\xymatrix{
A\otimes^1 A \ar[rr]^\simeq \ar[dr]_{\mu_A^1} & & A\otimes^1 A \ar[dl]^{\mu_A^1} \\
& A & 
}
\end{array}
\ee
where the `$\simeq$' in the first diagram is defined in (\ref{eq:delta_a11d}), and the `$\simeq$' in the second diagram is precisely the braiding defined in  (\ref{eq:braiding_from_2fusion_1}). 

Moreover, we require the commutative algebra $A$ to be separable (i.e., there exists a bimodule map $\Delta: A\to A\otimes A$ such that $m\circ \Delta = \id_A$). Mathematically, this condition is needed for $\Omega\CD$ and $\Omega_m\CM$ to be separable 1-categories. Physically, this condition is a stability condition of the vacuum $A$ as explained in \cite{Kon14e}. A commutative separable algebra in $\Omega\CC$ is called a {\it condensable $\EE_2$-algebra} in $\Omega\CC$. Such an algebra is also equipped naturally with a Frobenius algebra structure (see \cite{FRS02,Kon14e}). We use the notation $A\in \Algc_{\EE_2}(\Omega\CC)$ to represent the statement that $A$ is a condensable $\EE_2$-algebra in $\Omega\CC$. Since $\SD^3$ is simple, $A$ must be simple (i.e., $\dim \hom_\CC(1_{\one_\CC}, A)=1$). A general condensable $\EE_2$-algebra in a braided fusion 1-category is a direct sum of simple ones. 

\item Anyons in $\Omega\CD$ are deconfined particles, which are necessarily endowed with a 2-dimensional action of the trivial particle $1_{\one_\CD}=A$ in $\Omega\CD$. Mathematically, it amounts to say that anyons in $\Omega\CD$ are $\EE_2$-$A$-modules (or local $A$-modules) in $\Omega\CC$. We explain this fact in a few steps from the perspective of Section\,\ref{sec:E2=E1+E1}. 

\begin{itemize}
\item[-] For $A,B\in \Algc_{\EE_2}(\Omega\CC)$, both categories $\LMod_A(\Omega\CC)$ and $\RMod_A(\Omega\CC)$ are monoidal with the monoidal structures defined as follows (see Remark\,\ref{rem:two_monoidal_structures} for another monoidal structure). 
\be \label{eq:pic-tpover_A-tpover_B}
\begin{array}{c}
\begin{tikzpicture}[scale=0.9]
\fill[teal!20] (-3,0) rectangle (3,4) ;
\draw[fill=white] (-0.05,1.45) rectangle (0.05,1.55) node[midway,above] {$x$} ;
\draw[fill=white] (-0.05,2.45) rectangle (0.05,2.55) node[midway,below] {$y$} ;
\node at (-2.5,3.5) {$\Omega\CC$} ;
\draw [dashed,->] (-1,1.5) -- (-0.2,1.5) node[midway,above] {$\mu_x^\rightarrow$}; 
\draw [dashed,->] (1,1.5) -- (0.2,1.5) node[midway,below] {$\mu_x^\leftarrow$}; 
\draw [dashed,->] (-1,2.5) -- (-0.2,2.5) node[midway,above] {$\nu_y^\rightarrow$}; 
\draw [dashed,->] (1,2.5) -- (0.2,2.5) node[midway,below] {$\nu_y^\leftarrow$}; 
\draw [dashed,->] (0,3.5) -- (0,2.7) node[midway,right] {$\nu_y^\downarrow$}; 
\draw [dashed,->] (0,0.5) -- (0,1.3) node[midway,left] {$\mu_x^\uparrow$}; 
\node at (-1.2,1.5) {$A$} ;
\node at (1.2,1.5) {$A$} ;
\node at (-1.2,2.5) {$B$} ;
\node at (1.2,2.5) {$B$} ;
\node at (0,3.7) {$B$} ;
\node at (0,0.3) {$A$} ;

\draw[->] (-3,0) -- (-2.5,0) node [very near end, above] {$x^1$} ;
\draw[->] (-3,0) -- (-3,0.5) node [very near end, left] {$x^2$} ;
\end{tikzpicture}
\end{array}
\ee
\bnu

\item The tensor unit in $\LMod_A(\Omega\CC)$ is $A$ and the tensor product is given by $\otimes_A^1$, which needs an explanation. Note that the left $A$-action $\mu_A^\uparrow: A\otimes^2 x \to x$ induces two horizontal $A$-actions on $x$ (denoted by $\mu_A^\leftarrow$ and $\mu_A^\rightarrow$) by diabatic moves as illustrated in the picture in (\ref{eq:pic-tpover_A-tpover_B}). These two horizontal $A$-actions induces a tensor product $x\otimes_A^1 x'$ defined by the coequalizer of $x\otimes^1 A \otimes^1 x' \rightrightarrows x\otimes^1 x'$ for $x,x'\in \LMod_A(\Omega\CC)$. Note that $x\otimes_A^1 x'$ has a structure of left $A$-module induced from that of $x$.

\item The tensor unit in $\RMod_B(\Omega\CC)$ is $B$ and the tensor product is given by $\otimes_B^1$, which needs an explanation. Note that the right $B$-action $\nu_y^\downarrow: B\otimes^2 y \to y$ induces two horizontal $B$-actions on $y$ (denoted by $\nu_y^\leftarrow$ and $\nu_y^\rightarrow$) by diabatic moves as illustrated in the picture in (\ref{eq:pic-tpover_A-tpover_B}). These two horizontal $B$-actions induces a tensor product $y\otimes_B^1 y'$ defined by the coequalizer of $y\otimes^1 B \otimes^1 y' \rightrightarrows y\otimes^1 y'$ for $y,y'\in \RMod_B(\Omega\CC)$. Note that $y\otimes_B^1 y'$ has a structure of right $A$-module induced from that of $y'$. 

\enu
As a consequence, the category $\BMod_{A|B}(\Omega\CC)$ inherits a natural monoidal structure due to the following equivalence of 2-categories: 
\be \label{eq:BMod=LMod-tensor-RMod_1}
\BMod_{A|B}(\Omega\CC) \simeq \LMod_A(\Omega\CC) \boxtimes_{\Omega\CC} \RMod_B(\Omega\CC). 
\ee
More explicitly, the tensor product in $\BMod_{A|B}(\Omega\CC)$, denoted by $\otimes_{A|B}^1$, is the one induced from the tensor product $\otimes^1_A \boxtimes_{\Omega\CC} \otimes^1_B$ in $\LMod_A(\Omega\CC) \boxtimes_{\Omega\CC} \RMod_B(\Omega\CC)$. The tensor unit in $\BMod_{A|B}(\Omega\CC)$ is $A\otimes^2 B$, which is induced from the tensor unit $A\boxtimes_{\Omega\CC} B$ in $\LMod_A(\Omega\CC) \boxtimes_{\Omega\CC} \RMod_B(\Omega\CC)$. The physical meaning of this tensor product is illustrated in a special case in (\ref{eq:pic-AB-tensor-x}). 

\item[-] We are only interested in the case $B=A$. In this case, $\Mod_A^{\EE_1}(\Omega\CC)=\BMod_{A|A}(\Omega\CC)$ is equipped with the monoidal structure with the tensor product $\otimes_{A|A}^1$ and the tensor unit $A\otimes^2 A$. Note that the $\EE_2$-algebra structure on $A\in \Algc_{\EE_2}(\Omega\CC)$ endows $A$ with a structure of an $\EE_1$-algebra in $\Mod_A^{\EE_1}(\Omega\CC)$. In particular, the multiplication is given by $\mu_A^1: A\otimes^1 A \to A$, and the unit morphism is given by $\mu_A^2: A\otimes^2 A \to A$, which is an algebra homomorphism. Namely, $A\in \Algc_{\EE_1}(\Mod_A^{\EE_1}(\Omega\CC))$. 
Note that, in this case, the tensor unit $A\otimes^2 A$ is not simple, and $A$ is a direct summand of the tensor unit. By the structure theorem of a multi-fusion 1-category, $\Mod_A^{\EE_1}(\Mod_A^{\EE_1}(\Omega\CC))$ is a full subcategory of $\Mod_A^{\EE_1}(\Omega\CC)$. 

\item[-] For $x\in \Mod_A^{\EE_1}(\Omega\CC)$, the left $(A\otimes^2 A)$-action on $x$ is illustrated below.  
\be \label{eq:pic-AB-tensor-x}
(A\otimes^2 A)  \otimes_{A|B}^1 \, (x) = 
\begin{array}{c}
\begin{tikzpicture}[scale=0.8]
\fill[teal!20] (-2,0) rectangle (2,3) ;

\node at (-1.5,2.5) {$\Omega\CC$} ;
\node at (0,2.5) {$A\otimes_B^1 B$} ;
\node at (0.5,2) {$\otimes_B^2$} ;
\node at (0.5,1.5) {$x$} ;
\node at (0.5, 1) {$\otimes_A^2$} ;
\node at (0,0.5) {$A\otimes_A^1 A$} ;

\draw[->] (-2,0) -- (-1.5,0) node [very near end, above] {$x^1$} ;
\draw[->] (-2,0) -- (-2,0.5) node [very near end, left] {$x^2$} ;
\end{tikzpicture}
\end{array}
\xrightarrow{\simeq} \,\, x
\ee
The right $(A\otimes^2 A)$-actions on $x$ is similar. Note that $x\in \Mod_A^{\EE_1}(\Mod_A^{\EE_1}(\Omega\CC))$ if and only if the left (resp. right) $(A\otimes^2 A)$-actions on $x$ factor through $A=A\otimes_A^2 A$. 
This condition is equivalent to the conditions that $\mu_x^\rightarrow=\nu_x^\rightarrow$ (resp. $\mu_x^\leftarrow=\nu_x^\leftarrow$). These conditions are further equivalent to the commutativity of the following diagram (where we set $\mu_x^\downarrow=\nu_x^\downarrow$): 
\be \label{diag:local-module}
\begin{array}{c}
\xymatrix{
& x\otimes^2 A \ar[dr]^\simeq \ar[d]^{\mu_x^\downarrow} &  \\
A\otimes^1 x \ar[ur]^{\simeq} \ar[r]^{\mu_x^\to} & x &  x\otimes^1 A \ar[ld]^\simeq \ar[l]_{\mu_x^\leftarrow} \\
& A\otimes^2 x \ar[lu]^\simeq   \ar[u]^{\mu_x^\uparrow} & 
}
\end{array}
\ee
where all `$\simeq$' are defined by (\ref{eq:delta_a11d}) or its inverse. The composition of adjacent two such $\simeq$'s defines a braiding as in (\ref{eq:braiding_from_2fusion_1}) and (\ref{eq:braiding_from_2fusion_2}). 
Such a module $x$, i.e. equipped with a 2-dimensional $A$-actions, is called an $\EE_2$-$A$-module. As a consequence, we have provided a physical proof of the following equivalence of 1-categories: 
\be \label{def:Mod_E2}
\Mod_A^{\EE_2}(\Omega\CC) \simeq \Mod_A^{\EE_1}(\Mod_A^{\EE_1}(\Omega\CC)),
\ee
which is a special case of a more general result (see Corollary\,\ref{cor:Em_module=mE1}). 
An $\EE_2$-$A$-module in $\Omega\CC$ is precisely a local $A$-module in the usual sense
\cite{KO02,FFRS06,Kon14e}). It is clear that $A$ itself is an $\EE_2$-$A$-module. 
\end{itemize}

\item Note that $\Mod_A^{\EE_1}(\Omega\CC)$ has another natural monoidal structure with the tensor product $\otimes_A^2$ and the tensor unit $A$. The category $\Mod_A^{\EE_1}(\Mod_A^{\EE_1}(\Omega\CC))$ is naturally equipped with two tensor products $\otimes_A^2$ and $\otimes_A^1$ that defines an $\EE_2$-monoidal structure on $\Mod_A^{\EE_1}(\Mod_A^{\EE_1}(\Omega\CC))$. 

\item If the commutativity of the diagram in (\ref{diag:local-module}) does not hold, this non-commutativity causes a strong interference effect around the $x$-particle that confines the particle. Such confined particles are necessarily confined to the gapped domain wall $\SM^2$ between $\SC^3$ and $\SD^3$. As a consequence, particles on the wall $\SM^2=(\CM,m)$ form a fusion 1-category $\Omega_m\CM=\RMod_A(\Omega\CC)$.  

\enu

\begin{rem} \label{rem:def_E2_Mod}
Note that the notion of an $\EE_2$-$A$-module for an $\EE_2$-algebra $A$ in $\Omega\CC$ is defined by four actions $\mu_x^\to, \mu_x^\downarrow, \mu_x^\leftarrow, \mu_x^\uparrow$ in four directions such that they are mutually commuting in the sense that the following diagram:
\be \label{diag:commutative_left_left} 
\begin{array}{c}
\xymatrix{ A \otimes^2 (A\otimes^1 x) \ar[rr]^\simeq \ar[d]_{1_A\otimes^2 \mu_A^\rightarrow} & & A \otimes^1 (A\otimes^2 x) \ar[d]^{\mu_A^\uparrow} \\ A\otimes^2 x \ar[rd]_{\mu_x^\uparrow} & & A\otimes^1 x \ar[ld]^{\mu_x^\rightarrow} \\ & x & }
\end{array}
\ee
and other similar diagrams commute, and are isomorphic in the sense of (\ref{diag:local-module}). In other words, there is essentially only one 2-dimensional $A$-action on $x$. 
This notion immediately generalizes to the notion of an $\EE_k$-$B$-module for an $\EE_k$-algebra $B$ in an $\EE_k$-monoidal category (see Definition\,\ref{def:Ek_Mod} and Corollary\,\ref{cor:Em_module=mE1}). 
\end{rem}

\begin{rem} \label{rem:two_monoidal_structures}
Actually, $\BMod_{A|A}(\Omega\CC)$ is equipped with the second monoidal structure with the usual tensor product $\otimes_A^2$ and the tensor unit $A$. These two monoidal structures are not equivalent. They can be obtained from two different fusion processes, respectively, as illustrated below. 
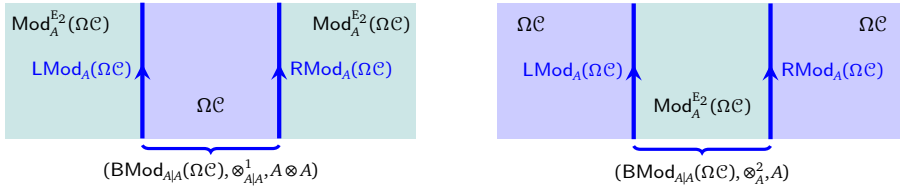
\[
\begin{array}{c}
\begin{tikzpicture}[scale=0.9]
\fill[teal!20] (-3,0) rectangle (3,2) ;
\fill[blue!20] (-1,0) rectangle (1,2) ;
\draw[blue,ultra thick,->-] (-1,0) -- (-1,2) node[midway,left] {\scriptsize $\LMod_A(\Omega\CC)$} ;
\draw[blue,ultra thick,->-] (1,0) -- (1,2) node[midway,right] {\scriptsize $\RMod_A(\Omega\CC)$} ;
\node at (-2.2,1.7) {\scriptsize $\Mod_A^{\EE_2}(\Omega\CC)$} ;
\node at (2.2,1.7) {\scriptsize $\Mod_A^{\EE_2}(\Omega\CC)$} ;
\node at (0,0.5) {\footnotesize $\Omega\CC$} ;
\draw[blue,decorate,decoration=brace,very thick] (1,-0.1)--(-1,-0.1) ;
\node at (0,-0.5) {\scriptsize $(\BMod_{A|A}(\Omega\CC), \otimes^1_{A|A},A\otimes A)$} ;
\end{tikzpicture}
\end{array}
\quad\quad
\begin{array}{c}
\begin{tikzpicture}[scale=0.9]
\fill[blue!20] (-3,0) rectangle (3,2) ;
\fill[teal!20] (-1,0) rectangle (1,2) ;
\draw[blue,ultra thick,->-] (-1,0) -- (-1,2) node[midway,left] {\scriptsize $\LMod_A(\Omega\CC)$} ;
\draw[blue,ultra thick,->-] (1,0) -- (1,2) node[midway,right] {\scriptsize $\RMod_A(\Omega\CC)$} ;
\node at (-2.5,1.7) {\footnotesize $\Omega\CC$} ;
\node at (2.5,1.7) {\footnotesize $\Omega\CC$} ;
\node at (0,0.5) {\scriptsize $\Mod_A^{\EE_2}(\Omega\CC)$} ;
\draw[blue,decorate,decoration=brace,very thick] (1,-0.1)--(-1,-0.1) ;
\node at (0,-0.5) {\scriptsize $(\BMod_{A|A}(\Omega\CC), \otimes^2_A,A)$} ;
\end{tikzpicture}
\end{array}
\]
Note that the first monoidal structure is Morita equivalent to the fusion 1-category $\Mod_A^{\EE_2}(\Omega\CC)$ and the second one is Morita equivalent to the fusion 1-category $\Omega\CC$. Therefore, these two monoidal structures are not even Morita equivalent when $\dim A >1$. We want to emphasize that the existence of multiple higher monoidal structures on a single higher category is a ubiquitous and important feature of higher categories and higher condensation theory. 
\end{rem}

In summary, we obtain the following categorical description of an anyon condensation. 
\begin{pthm}[\cite{Kon14e}]  \label{pthm:anyon-condensation}
If an anyon condensation in a 2+1D anomaly-free simple topological order $\SC^3$ produces an anomaly-free simple condensed phase 
$\SD^3$ and a gapped domain wall $\SM^2$ (see Figure \ref{fig:2d-condensation} and the picture below).
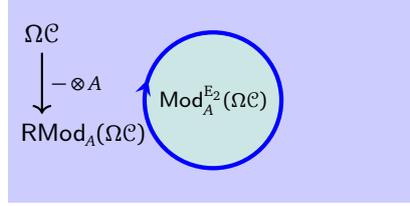
\[
\begin{tikzpicture}[scale=0.9]
\fill[blue!20] (-3,0) rectangle (3,3) ;
\fill[teal!20] (0,1.5) circle (1);
\draw[thick,->] (-2.5,2.2) -- (-2.5,1.3) node [midway, right] {\footnotesize $-\otimes A$} ;
\node at (-2.5,2.5) {$\Omega\CC$} ;
\node at (0,1.5) {\footnotesize $\Mod_A^{\EE_2}(\Omega\CC)$} ;
\node at (-1.9,1) {\small $\RMod_A(\Omega\CC)$} ;
\draw[blue, ultra thick,->-] (1,1.5) arc (360:0:1) ;
\end{tikzpicture}
\]
\bnu
\item The vacuum particle in $\Omega\SD$ can be identified with a simple condensable $\EE_2$-algebra $A$ in $\Omega\CC$, which is a composite boson\footnote{It is really a boson condensation \cite{BSS03,BS09} instead of an `anyon condensation', a term which is wildly used in literature.} (see Remark\,\ref{rem:boson_condensation}). Moreover, $\Omega\CD$ consists of all deconfined particles and can be identified with the category $\Mod_A^{\EE_2}(\Omega\CC)$ of $\EE_2$-$A$-modules (or equivalently, local $A$-modules) in $\CC$, i.e., $\Omega\CD = \Mod_A^{\EE_2}(\Omega\CC)$. 

\item Particles on the wall $\SM^2=(\CM,m)$ include all confined and deconfined particles, and can be identified with the fusion 1-category $\Omega_m\CM=\RMod_A(\Omega\CC)$.

\item Anyons in $\Omega\CC$ move to the wall according to the central functor\footnote{Being a central functor means that $-\otimes A$ factors through a braided monoidal functor $g: \Omega\CC \to \RMod_A(\Omega\CC)$, i.e. it coincides with the composed functor 
$\Omega\CC \xrightarrow{g} \FZ_1(\RMod_A(\Omega\CC)) \xrightarrow{\forget} \RMod_A(\Omega\CC)$, where $\forget$ is the forgetful functor.} $- \otimes A \colon \Omega\CC \to \RMod_A(\Omega\CC)$ defined by $x \mapsto x \otimes A$ for all $x\in \Omega\CC$. 
 
\item Anyons in $\Omega\CD$ move to the wall according to the embedding $\Mod_A^{\EE_2}(\Omega\CC) \hookrightarrow \RMod_A(\Omega\CC)$, then can move out to the $\SD^3$-side freely. 

\enu
If $A$ is also Lagrangian, i.e. $(\dim A)^2=\dim (\Omega\CC)$ \cite{KR09,DMNO13}, we have $\Omega\CD=\Mod_A^{\EE_2}(\Omega\CC)\simeq \Vect$ \cite{DMNO13}\footnote{This fact follows immediately from the formula (\ref{eq:quantum_dim}) of quantum dimensions.}. In this case, $\SM^2$ is a gapped boundary of $\SC^3$ and consists of all confined particles and deconfined particles. 
\end{pthm}

\begin{rem} \label{rem:anyon_condensation_is_not_reversable}
Different from the condensation of 1-codimensional defects (recall Theorem$^{\mathrm{ph}}$\,\ref{pthm:reversed_condensation_1d}), such defined anyon condensation process can not be reversed mathematically. More precisely, there is no condensable $\EE_2$-algebra in $\Omega\CD$ such that its condensation in $\SD^3$ reproduces $\SC^3$. The simplest way to see this fact is through the following formula of quantum dimensions: 
\be \label{eq:quantum_dim}
\dim\ \Mod_A^{\EE_2}(\Omega\CC) = \frac{\dim \Omega\CC}{(\dim A)^2}. 
\ee
Since $\dim A > 1$ if $A\neq 1_{\one_\CC}$, a non-trivial particle condensation in $\Omega\CC$ always reduces the quantum dimension. Therefore, there is no way to find a condensable $\EE_2$-algebra in $\Omega\CD$ to reproduce $\Omega\CC$.  On the other hand, the phase transition can always run in the reversed direction. This shortcoming can be fixed by taking all 1-codimensional topological defects in $\SC^3$ into account as in Section\,\ref{sec:condense_1-codim_defect}. 
\end{rem}

\begin{rem} \label{rem:boson_condensation}
When we derive the properties of the condensable $\EE_2$-algebra $A$ in $\CC$, we have ignored some properties that have played important roles in \cite{Kon14e}. 
\bnu
\item We do not emphasize the unitary structure and the Frobenius algebra structures (as in Remark\,\ref{rem:coalgebra_structure}), which allows us to obtain a lot of graphic constructions of ingredients of the condensed phases, such as the screening operator \cite[Eq.\ (20)]{Kon14e}. The reason is that the mathematical theory of unitary higher categories (see \cite{FHJF+24} for a recent progress) and that of Frobenius algebras in higher monoidal categories are still underdeveloped. 

\item Another important property of $A$ is that it should preserves the twist (or the topological spin) \cite[Eq.\ (12)]{Kon14e}. As a consequence, $A$ must have a trivial twist, i.e., $\theta_A=\id_A$. It further means that $A$ is a composite boson. This, of course, is a fundamental physical result. In 2+1D, however, a simple condensable $\EE_2$-algebra $A$ is automatically equipped with a structure of a simple special symmetric Frobenius algebra \cite{FRS02,Kon14e}, which automatically implies that $\theta_A=\id_A$. In higher dimensions, it is not clear to us how to generalize the notion of a twist to higher categories. We believe that this is a very important question because it allows us to define or characterize the statistical properties of higher dimensional topological defects. We hope to come back to this point in the future. 
\enu
\end{rem}

\begin{rem} \label{rem:recent_references_anyon_condensation}
We give a brief remark on later developments. After the categorical description of an anyon condensation was fully established in \cite{Kon14e} (see Theorem$^{\mathrm{ph}}$\,\ref{pthm:anyon-condensation}), some further discussions and explanations of this categorical description appeared in literature \cite{ERB14,HW15a,NHKSB16,NHKSB16a} (see \cite{Bur18} for a physical but not very categorical review). The anyon condensation theory for 2+1D topological orders was used and generalized to 2+1D SPT/SET orders \cite{HW14,LKW16a,LKW16,JR17,BJLP19}. Since the anyon condensation theory summarized in Theorem$^{\mathrm{ph}}$\,\ref{pthm:anyon-condensation} is model independent, it is very important to construct lattice models to realize a given anyon condensation. The microscopic or physical constructions of anyon condensations appeared first in the tensor network approach \cite{DIHVS17,GIP17,XZZ20}. 
The lattice model constructions of anyon condensations appeared only recently in \cite{HHHW22,ZHWHW23,CGHP23,LB24,ZW25,ZW25a}. We do expect more progress in this direction in the coming future. Recently, anyon condensation theory has been used systematically in the study of topological phase transitions in 1+1D \cite{CJKYZ20,CW23a,LY23,CJW25,CAW24}. It was also used recently to study many different physical systems \cite{CGS23,ZVW25,CGS24,Tak25,CGSO25,HJSWZ25}. 
\end{rem}



\subsubsection{Examples} \label{sec:example_2codim_3D}

\begin{expl} \label{expl_toric_code_2d_condensation}
Take $\SC^3$ to be the 2+1D $\Zb_2$ topological order, which is realizable by the 2+1D toric code model \cite{Kit03}. In this case, the category of particles in $\SC^3$ is $\Omega\CC=\FZ_1(\Rep(\Zb_2))$, which is the Drinfeld center of $\Rep(\Zb_2)$. It contains four simple anyons $\one, e, m, f$ with the fusion rule 
\[
e\otimes e \simeq m\otimes m \simeq f\otimes f \simeq \one, \quad \quad f\simeq e\otimes m \simeq m \otimes e.
\]
The double braiding of $e$ with $m$ is $-1$. The self-braiding of $e$ and $m$ are trivial, and the self-braiding of $f$ is $-1$. There are two nontrivial condensable $\EE_2$-algebra $A_e=\one \oplus e$ and $A_m=\one\oplus m$, both of which are Lagrangian.  
\bnu
\item The multiplication of $A_e$ is given by
\[
A_e \otimes A_e = \one \otimes \one \oplus e \otimes e \oplus \one \otimes e \oplus e \otimes \one = (\one \oplus \one) \oplus (e \oplus e) \xrightarrow{\begin{psmallmatrix}1 & 1\end{psmallmatrix} \oplus \begin{psmallmatrix}1 & 1\end{psmallmatrix}} \one \oplus e = A_e ,
\]
and the unit of $A_e$ is $\one \xrightarrow{1 \oplus 0} \one \oplus e = A_e$. It is separable because the following morphism is an $A_e$-$A_e$-bimodule map:
\[
A_e = \one \oplus e \xrightarrow{\begin{psmallmatrix}1/2 \\ 1/2\end{psmallmatrix} \oplus \begin{psmallmatrix}1/2 \\ 1/2\end{psmallmatrix}} (\one \oplus \one) \oplus (e \oplus e) = \one \otimes \one \oplus e \otimes e \oplus \one \otimes e \oplus e \otimes \one = A_e \otimes A_e .
\]
Since $\dim \FZ_1(\Rep(\Zb_2))=4$ and $\dim A_e =2$, $A_e$ is Lagrangian, i.e. $\Mod_{A_e}^{\EE_2}(\Omega\CC)\simeq \vect$. Condensing $A_e$ produces the trivial 2+1D topological order and a gapped 1+1D boundary of $\SC^3$, which is called the rough boundary \cite{BK98}. There are two simple anyons on the rough boundary, which correspond to two simple right $A_e$-modules: one is $A_e$ itself, and the other one is the free module $M_e \coloneqq m \otimes A_e = f \otimes A_e$. The fusion rule is given by $M_e \otimes_{A_e} M_e \simeq A_e$ and the associator is trivial. Hence $\RMod_{A_e}(\FZ_1(\Rep(\Zb_2))) \simeq \Rep(\Zb_2)$. The right module $M_e$ is not local because the double braiding of $e$ and $m$ is $-1$. Thus $A_e$ itself is the only simple local $A_e$-module.
\item Similarly, the multiplication of $A_m$ is given by
\[
A_m \otimes A_m = \one \otimes \one \oplus m \otimes m \oplus \one \otimes m \oplus m \otimes \one = (\one \oplus \one) \oplus (m \oplus m) \xrightarrow{\begin{psmallmatrix}1 & 1\end{psmallmatrix} \oplus \begin{psmallmatrix}1 & 1\end{psmallmatrix}} \one \oplus m = A_m ,
\]
and the unit of $A_m$ is $\one \xrightarrow{1 \oplus 0} \one \oplus m = A_m$. Again $A_m$ is Lagrangian. Condensing $A_m$ gives the trivial phase and also produces a gapped boundary of the 2+1D $\Zb_2$ topological order, which is called the smooth boundary. There are two simple anyons on the smooth boundary, which correspond to two simple right $A_m$-modules: one is $A_m$ itself, and the other one is the free module $M_m \coloneqq e \otimes A_m = f \otimes A_m$. The fusion rule is given by $M_m \otimes_{A_m} M_m \simeq A_m$ and the associator is trivial. Hence $\RMod_{A_m}(\FZ_1(\Rep(\Zb_2))) \simeq \vect_{\Zb_2}$. The right module $A_m$ is local but $M_m$ is not.
\enu
\end{expl}

\begin{expl} \label{expl_double_semion_2d_condensation}
Consider the 2+1D twisted $\Zb_2$ topological order $\SG\ST_{(\Zb_2,\omega)}^3$, 
where $\omega \in Z^3(\Zb_2;U(1))$ represents the nontrivial cohomology class in $H^3(\Zb_2;U(1)) \simeq \Zb_2$.  More explicitly we take $\omega(1,1,1) = -1$ and $\omega(a,b,c) = 1$ if one of $a,b,c$ is $0$. Note that $\SG\ST_{(\Zb_2,\omega)}^3$ can be realized by the double semion model. In this case, the category of anyons is given by the Drinfeld center of $\vect_{\Zb_2}^\omega$, i.e., $\Omega\CG\CT_{(\Zb_2,\omega)} = \FZ_1(\vect_{\Zb_2}^\omega)$. There are four simple anyons $\one,s,\bar s,s \bar s$ in $\FZ_1(\vect_{\Zb_2}^\omega)$ with the following fusion rules:
\[
s \otimes s \simeq \bar s \otimes \bar s \simeq \one , \quad \quad s \bar s \simeq s \otimes \bar s \simeq \bar s \otimes s .
\]
The double braiding of $s$ and $\bar s$ is trivial. The self-braiding of $s$ is $\mathrm i$, the self-braiding of $\bar s$ is $-\mathrm i$, and the self-braiding of $s \bar s$ is $1$.

There is only one nontrivial condensable $\EE_2$-algebra $A = \one \oplus s \bar s$. The multiplication is given by
\[
A \otimes A = \one \otimes \one \oplus s \bar s \otimes s \bar s \oplus \one \otimes s \bar s \oplus s \bar s \otimes \one = (\one \oplus \one) \oplus (s \bar s \oplus s \bar s) \xrightarrow{\begin{psmallmatrix}1 & 1\end{psmallmatrix} \oplus \begin{psmallmatrix}1 & 1\end{psmallmatrix}} \one \oplus s \bar s = A .
\]
Condensing $A$ produces a gapped boundary of $\SG\ST_{(\Zb_2,\omega)}^3$. There are two simple anyons on this boundary corresponding to two simple right $A$-modules: one is $A$ it self, and the other is the free module $M_s \coloneqq s \otimes A \simeq \bar s \otimes A$. The fusion rule is given by $M_s \otimes_A M_s \simeq A$ but the associator is non-trivial. We have $\RMod_A(\FZ_1(\vect_{\Zb_2}^\omega)) \simeq \vect_{\Zb_2}^\omega$. The right module $M_s$ is not a local module because the double braiding of $s$ and $s \bar s$ is $\mathrm i$.
\end{expl}

\begin{expl}
A 2+1D topological order $\SC^3$ is called \emph{abelian} if for every simple anyon $x$ there is an anyon $x^*$ satisfying $x \otimes x^* \simeq \one \simeq x^* \otimes x$. Thus the category $\Omega\CC$ of anyons is \emph{pointed} in the sense that every simple object is invertible under the tensor product. In other words, the simple anyons form a finite abelian group under the fusion, denoted by $G$. Then for every $x,y \in G$ we define $q(x) \in U(1)$ to be the self-braiding of $x$ and $b(x,y) \in U(1)$ to be the double braiding of $x$ with $y$. Then we have
\[
b(x,y) = \frac{q(x y)}{q(x) q(y)}
\]
and $b \colon G \times G \to U(1)$ is a bi-character, i.e., $b(x y,z) = b(x,z) b(y,z)$ for all $x,y,z \in G$. We say that $q \colon G \to \mathrm U(1)$ is a \emph{qudratic form} and $(G,q)$ is a \emph{metric group}. The pointed braided fusion category $\Omega\CC$ is completely determined by the metric group $(G,q)$ consisting of simple anyons and self-braidings.

The condensable $\EE_2$-algebras in $\Omega\CC$ are classified by subgroups $H \leq G$ such that $q(x) = 1$ for all $x \in H$. Such a subgroup is called an \emph{isotropic subgroup}. We denote the corresponding algebra by $A_H$. The multiplication of $A_H$ is given by
\[
A_H \otimes A_H = \bigoplus_{g,h \in H} g \otimes h \xrightarrow{\delta_{gh,k} \cdot \phi(g,h)} \bigoplus_{k \in H} k = A_H
\]
for some 2-cochain $\phi \in C^2(G;U(1))$ such that $\mathrm d \phi$ coincides with the associator of simple anyons in $H$. Indeed, the associator of simple anyons in $H$ must be a coboundary by the isotropic condition. All simple right $A_H$-modules are free modules, i.e., of the form $x \otimes A_H$ for some $x \in G$. Two simple modules $x \otimes A_H$ and $y \otimes A_H$ are isomorphic if and only if $x$ and $y$ lie in the same coset of $H$ in $G$, i.e., $x^{-1} y \in H$. The fusion rule simple right $A_H$-modules is given by the multiplication of the quotient group $G/H$. Therefore, the fusion category $\RMod_{A_H}(\Omega\CC)$ is a pointed fusion category with the group of simple objects being $G/H$. The associator can be found in \cite[Theorem A.1.4]{DS18}. A simple right $A_H$-module $x \otimes A_H$ is local if and only if $x$ has trivial double braiding with all $y \in H$. We define $H^\perp \coloneqq \{x \in G \mid b(x,y) = 1 , \, \forall y \in H\}$. Then the local $A_H$-modules form a pointed modular tensor category with the group of simple objects being $H^\perp/H$, and the self-braiding is the restriction of $q$ on $H^\perp/H$. Hence $A_H$ is Lagrangian if and only if $H^\perp = H$. A subgroup $H \leq G$ is called \emph{Lagrangian} if $H = H^\perp$. Lagrangian algebras in $\Omega\CC$ one-to-one correspond to Lagrangian subgroups of $G$.

Both the 2+1D $\Zb_2$ topological order and twisted $\Zb_2$ topological order are abelian topological orders (see Example \ref{expl_toric_code_2d_condensation} and Example \ref{expl_double_semion_2d_condensation}).
\bnu
\item The $\Zb_2$ topological order corresponds to $G = \Zb_2 \times \Zb_2$ and $q(a,b) = (-1)^{ab}$. There are two nontrivial isotropic subgroup generated by $(1,0) = e$ and $(0,1) = m$, and both of them are Lagrangian.
\item The twisted $\Zb_2$ topological order corresponds to $G = \Zb_2 \times \Zb_2$ and $q(a,b) = \mathrm i^{a^2-b^2}$. The only nontrivial isotropic subgroup is generated by $(1,1) = s \bar s$ and it is Lagrangian.
\enu
\end{expl}

\begin{expl} \label{expl:Davydov}
Let $G$ be a finite group. Anyons in the 2+1D $G$-gauge theory $\SG\ST_G^3$ (realizable by the Kitaev's quantum double model \cite{Kit03}) form a non-degenerate braided fusion 1-category $\FZ_1(\vect_G) \simeq \FZ_1(\Rep(G))$, which is the Drinfeld center of $\vect_G$ or $\Rep(G)$. An object in $\FZ_1(\vect_G)$ is a finite-dimensional $G$-graded vector space $V = \bigoplus_{g \in G} V_g$ equipped with a $G$-action $\rho \colon G \to \mathrm{GL}(V)$ such that $\rho(g)(V_h) \subseteq V_{ghg^{-1}}$ for all $g,h \in G$. The tensor product in $\FZ_1(\vect_G)$ is the tensor product of $G$-graded vector spaces with the diagonal $G$-action, and the tensor unit is $1_e \in \vect_G$ with the trivial $G$-action. The braiding $c_{X,Y}: X\otimes Y \to Y\otimes X$ is defined by $c_{X,Y}(x\otimes y) = g(y) \otimes x$ for $x\in X_g, y\in Y$ \cite{Dav10a}. The condensable $\EE_2$-algebras in $\FZ_1(\vect_G)$ are completely classified in \cite{Dav10a}. We list some examples here.
\bnu
\item Let $\Fun(G)$ be the space of $\Cb$-valued functions on $G$. We equipped $\Fun(G)$ with the left translation $G$-action:
\[
(g \triangleright f)(h) \coloneqq f(g^{-1} h) , \quad g,h \in G , \, f \in \Fun(G).
\]
We can also equip $\Fun(G)$ with the right translation $G$-action $(g \triangleright f)(h) \coloneqq f(hg)$, but these two $G$-representations are isomorphic. Then we equip $\Fun(G)$ with the $G$-grading concentrating on the trivial degree, i.e.,
\[
\Fun(G)_g = \begin{cases}\Fun(G) , & g = e , \\ 0 , & g \neq e .\end{cases}
\]
Then $\Fun(G)$ can be viewed as an object in $\FZ_1(\vect_G)$. It is also a condensable $\EE_2$-algebra in $\FZ_1(\vect_G)$ with the point-wise multiplication:
\[
(f \cdot f')(g) \coloneqq f(g) f'(g) , \quad f,f' \in \Fun(G) , \, g \in G .
\]
Let $\delta_g \in \Fun(G)$ be the delta function on $g \in G$. Then $\{\delta_g\}_{g \in G}$ is a basis of $\Fun(G)$ and the multiplication is given by $\delta_g \cdot \delta_h = \delta_{g,h} \delta_g$. The simple right $\Fun(G)$-modules are $\{M_g\}_{g \in G}$ where $M_g$ is the same as $\Fun(G)$ as $G$-representations but equipped with the different $G$-grading:
\[
\delta_h \in (M_g)_{hgh^{-1}} , \quad g,h \in G .
\]
In particular, $M_e$ is the regular $\Fun(G)$-modules. The fusion rule is $M_g \otimes_{\Fun(G)} M_h \simeq M_{gh}$ and the fusion category $\RMod_{\Fun(G)}(\FZ_1(\vect_G))$ is equivalent to $\vect_G$. The only local $\Fun(G)$-module is $\Fun(G)$ itself. Thus $\Fun(G) \in \FZ_1(\vect_G)$ is a Lagrangian algebra.
\item For any subgroup $H \leq G$, the space $\Fun(G/H)$ of $\Cb$-valued functions on the coset space $G/H$ equipped with the left translation $G$-action, the trivial $G$-grading and the point-wise multiplication is a condensable $\EE_2$-algebra. The modular tensor category of local $\Fun(G/H)$-modules is equivalent to $\FZ_1(\vect_H)$, and the fusion category of right $\Fun(G/H)$-modules is equivalent to the $\EE_1$-centralizer of $\vect_H$ in $\vect_G$. Two algebras $\Fun(G/H)$ and $\Fun(G/K)$ are isomorphic if and only if two subgroups $H$ and $K$ are conjugate to each other in $G$.
\item Let $\Cb[G]$ be group algebra of $G$. It has a basis $\{\tau_g\}_{g \in G}$ with the multiplication defined by $\tau_g \cdot \tau_h = \tau_{gh}$ for all $g,h \in G$. We equip $\Cb[G]$ with the canonical $G$-grading that $\tau_g \in \Cb[G]_g$ and the conjugation $G$-action that $g \triangleright \tau_h \coloneqq \tau_{ghg^{-1}}$. Then $\Cb[G]$ is a condensable $\EE_2$-algebra in $\FZ_1(\vect_G)$. Note that the multiplication of $\Cb[G]$ is not commutative in $\vect$ when $G$ is non-abelian, but is commutative in $\FZ_1(\vect_G)$ because the self-braiding of $\Cb[G]$ in $\FZ_1(\vect_G)$ is nontrivial due to the conjugation $G$-action. Every simple right $\Cb[G]$-module is isomorphic to $V \otimes \Cb[G]$ for some irreducible $G$-representation $V$, viewed as an object in $\FZ_1(\vect_G)$ with the $G$-grading concentrating on the trivial degree. The fusion category $\RMod_{\Cb[G]}(\FZ_1(\vect_G))$ is equivalent to $\Rep(G)$. The only local $\Cb[G]$-module is $\Cb[G]$ itself. Thus $\Cb[G] \in \FZ_1(\vect_G)$ is a Lagrangian algebra.
\item For any $\alpha \in Z^2(G;U(1))$, the twisted group algebra $\Cb[G,\alpha]$ is generated by $\{\tau_g\}_{g \in G}$ subject to the relation that $\tau_g \cdot \tau_h = \alpha(g,h) \tau_{gh}$ for all $g,h \in G$. We equip $\Cb[G,\alpha]$ with the canonical $G$-grading and the conjugation $G$-action:
\[
g \triangleright \tau_h = \frac{\alpha(g,h)}{\alpha(ghg^{-1},g)} \tau_{ghg^{-1}} .
\]
Then $\Cb[G,\alpha] \in \FZ_1(\vect_G)$ is a Lagrangian algebra and we still have $\RMod_{\Cb[G,\alpha]}(\FZ_1(\vect_G)) \simeq \Rep(G)$.
\item The Lagrangian algebras in $\FZ_1(\vect_G)$ are classified by pairs $(H,\alpha)$, where $H \leq G$ is a subgroup and $\alpha \in Z^2(H;U(1))$ is a 2-cocycle. We denote the corresponding algebra by $A_{H,\alpha}$. It is the induced representation of the twisted group algebra:
\[
A_{H,\alpha} = \Ind_H^G \Cb[H,\alpha] .
\]
Two Lagrangian algebras $A_{H,\alpha}$ and $A_{K,\beta}$ are isomorphic if and only if there exists an element $g \in G$ such that $gHg^{-1} = K$ and $[\alpha] = [g^* \beta] \in H^2(H;U(1))$, where $g^* \colon Z^2(K;U(1)) \to Z^2(H;U(1))$ is the pullback morphism induced by the conjugation of $g$. 

For example, when $G = S_3$, there are 4 subgroups up to conjugation: $\{e\}$, $\Zb_2$, $\Zb_3$ and $S_3$. For each subgroup $K$ the cohomology group $H^2(K,U(1))$ is trivial. Therefore, there are 4 Lagrangian algebras in $\FZ_1(\rep(S_3)) \simeq \FZ_1(\vect_{S_3})$ \cite[Section 4.2]{Dav10a}. See also \cite[Section 3.8]{CCW16} for the condensation of these Lagrangian algebras.

\item When $G$ is abelian, we can give a more explicit construction of the Lagrangian algebras $A_{H,\alpha}$. In this case $\FZ_1(\vect_G)$ is pointed, and the corresponding metric group is $(G \times \hat G,\chi)$, where $\hat G \coloneqq \Hom(G;U(1))$ is the dual group of $G$ and $\chi(g,\phi) \coloneqq \phi(g)$. Given a 2-cocycle $\alpha \in Z^2(G;U(1))$, its anti-symmetrization $\alpha_a$ is defined by
\[
\alpha_a(g,h) \coloneqq \frac{\alpha(g,h)}{\alpha(h,g)} , \quad g,h \in G .
\]
The cocycle condition implies that $\alpha_a \colon G \times G \to U(1)$ is a bi-character. We denote the space of anti-symmetric bi-characters on $G$ by $\mathrm{Alt}^2(G;U(1))$. Then $\alpha \mapsto \alpha_a$ defines a group homomorphism $Z^2(G;U(1)) \to \mathrm{Alt}^2(G;U(1))$. Moreover, if $\alpha$ is a coboundary, then $\alpha_a$ is trivial. Hence we get a group homomorphism $H^2(G;U(1)) \to \mathrm{Alt}^2(G;U(1))$. Indeed, this is an isomorphism because every symmetric 2-cocycle must be a coboundary. The Lagrangian subgroups of $(G \times \hat G,q)$ are parametrized by $(H,[\alpha])$, where $H \leq G$ is a subgroup and $[\alpha] \in H^2(H;U(1))$. The corresponding Lagrangian subgroup of $(G \times \hat G,\chi)$ is
\[
L_{H,[\alpha]} \coloneqq \{(h,\phi) \in G \times \hat G \mid h \in H , \, \phi \vert_H = \alpha_a(h,-)\} .
\]
This gives all Lagrangian algebras in $\FZ_1(\vect_G)$.
\enu
\end{expl}

\begin{expl}[\cite{DS17}]
We consider 2+1D twist finite gauge theory $\SG\ST_{(G,\omega)}^3$. Its category of anyons is given by $\Omega\CG\CT_{(G,\omega)} = \FZ_1(\vect_G^\omega)$. The classification of simple condensable $\EE_2$-algebras and Lagrangian algebras in $\FZ_1(\vect_G^\omega)$ are studied in \cite{DS17}. More precisely, the Lagrangian algebras in $\FZ_1(\vect_G^\omega)$ are classified by the pairs $(H,[\alpha])$ up to conjugation, where $H \leq G$ is a subgroup such that $[\omega] \vert_H \in H^3(H,U(1))$ is trivial and $[\alpha] \in H^2(H,U(1))$.
\end{expl}

\begin{expl}[\cite{BS09,CJKYZ20}] \label{expl:double_Ising_to_toric_code}
A very classical example of anyon condensation is the condensation from double Ising to toric code. It was first studied by Bais and Slingerland in \cite{BS09}. The complete details of this example had been worked out in \cite{CJKYZ20}. We briefly summarize the result below. The object 
$$
A=1\boxtimes 1 \oplus \psi \boxtimes \psi\in \Omega\ising\boxtimes \Omega\ising^\rev \simeq \FZ_1(\Omega\ising)
$$
has a canonical structure of a condensable $\EE_2$-algebra in $\FZ_1(\Omega\ising)$. It can be explicitly defined (see for example \cite[Example\ 7]{CJKYZ20}). There are six simple right $A$-modules in $\Omega\ising$: 
\begin{align}
1:=A, \quad e:=(\sigma\boxtimes\sigma), \quad &m:=(\sigma\boxtimes\sigma)^{\mathrm{tw}}, 
\quad f:=(\psi\boxtimes \one) \otimes A, \nn
\chi_+:=(\one\boxtimes\sigma)\otimes A, \quad\quad &\chi_-=(\sigma\boxtimes \one) \otimes A, \label{eq:1emfX}
\end{align}
among which $1,e,m,f$ are $\EE_2$-$A$-modules (or local $A$-modules) and $\chi_\pm$ are non-local $A$-modules. Moreover, the fusion and braiding structures on $1,e,m,f$ are the same as those on the 4 simple anyons in the 2+1D toric model, i.e. $\Mod_A^{\EE_2}(\FZ_1(\Omega\ising)) \simeq \FZ_1(\Rep(\Zb_2))$. 
\end{expl}

\begin{expl} \label{expl_VOA_algebra}
Let $V$ be a rational vertex operator algebra (VOA). Then the category $\Mod_V$ of $V$-modules is a modular tensor category \cite{Hua08}. The simple condensable $\EE_1$-algebras in $\Mod_V$ are open-string VOAs over $V$ \cite{HK04}, and the simple condensable $\EE_2$-algebras in $\Mod_V$ are VOA extensions of $V$ \cite{HKL15}. If $V \hookrightarrow W$ is an extension of VOAs, then $\Mod_W^{\EE_2}(\Mod_V) \simeq \Mod_W$ as modular tensor categories. More generally, if $V_L,V_R$ are rational VOAs, the simple condensable $\EE_2$-algebras in $\Mod_{V_L \otimes \bar{V}_R} \simeq \Mod_{V_L} \boxtimes \overline{\Mod_{V_R}}$ are full field algebras over $V_L \otimes \bar{V}_R$ \cite{Kon07}, and the Lagrangian algebras are precisely modular invariant CFTs over $V_L \otimes \bar{V}_R$ \cite{KR09}. When $V_L = V_R = V$, there is a well-known Lagrangian algebra
\[
\bigoplus_{m \in \Irr(\Mod_V)} m \boxtimes m^* \in \Mod_V \boxtimes \overline{\Mod_V} \simeq \Mod_{V \otimes \bar V} ,
\]
and this modular invariant CFT is also called the ``charge conjugation construction'' or the ``Cardy case''. Under the canonical equivalence $\Mod_V \boxtimes \overline{\Mod_V} \simeq \FZ_1(\Mod_V)$, this Lagrangian algebra is also the full center \cite{FFRS06,KR09,Dav10} of the tensor unit $\one = V \in \Mod_V$. In this case, Lagrangian algebras in $\FZ_1(\Mod_V)$ one-to-one correspond to indecomposable $\Mod_V$-modules in $2\vect$ \cite{KR08}. 
\end{expl}

\begin{expl}
Let $k$ be a non-negative integer and $\CC_k \coloneqq \CC(\mathfrak{sl}_2,k)$ be the category of integrable modules over $\widehat{\mathfrak{sl}_2}$ of level $k$. It is also equivalent to the semisimple part of the category of finite-dimensional representations over the quantum group $U_q(\mathfrak{sl}_2)$ with $q = \exp(\frac{\mathrm i \pi}{k+2})$. Then $\CC_k$ is a modular tensor category and it has $k+1$ simple objects, denoted by $V_0,\ldots,V_k$.

The indecomposable finite semisimple modules over $\CC_k$ are one-to-one corresponding to simply laced Dynkin diagrams of type $A,D,E$ with Coxeter number $k+2$ \cite{Ost03}, and the simple objects of the module correspond to the vertices of the Dynkin diagram. Moreover, each module of type $A_n,D_{2n},E_6,E_8$ is equivalent to the category of right modules over a unique simple condensable $\EE_2$-algebra in $\CC_k$, and this is also a complete classification of simple condensable $\EE_2$-algebras in $\CC_k$ \cite{KO02}.
\bit
\item Type $A_n$ for $n = k+1$. The corresponding module is $\CC_k$ itself as a regular $\CC_k$-module. It is equivalent to the category of right modules over the tensor unit $\one = V_0$.
\item Type $D_n$ for $2n = k+4$. In this case $k$ must be even. Then $V_0 \oplus V_k \in \CC_k$ is a simple condensable $\EE_1$-algebra. The corresponding module is $\RMod_{V_0 \oplus V_k}(\CC_k)$. The algebra $V_0 \oplus V_k$ is commutative if and only if $k$ is divided by $4$ (i.e., $n$ is even).
\item Type $E_6$. In this case $k = 10$. There is a VOA extension $(\widehat{\mathfrak{sl}_2})_{10} \hookrightarrow (\widehat{\mathfrak{sp}_4})_1$, which gives a simple condensable $\EE_1$-algebra in $\CC_{10}$ with the underlying object $V_0 \oplus V_6$. The corresponding module over $\CC_{10}$ is $\RMod_{V_0 \oplus V_6}(\CC_{10})$.
\item Type $E_7$. In this case $k = 16$. There is no VOA extension of $(\widehat{\mathfrak{sl}_2})_{16}$, but an VOA extension $(\widehat{\mathfrak{sl}_2})_{16} \otimes (\widehat{\mathfrak{sl}_3})_6 \hookrightarrow (\widehat{E_8})_1$. The corresponding module over $\CC_{16}$ can be constructed from this VOA extension. See \cite[Proof of Theorem 6]{Ost03} for more details.
\item Type $E_8$. In this case $k = 28$. There is a VOA extension $(\widehat{\mathfrak{sl}_2})_{28} \hookrightarrow (\widehat{G_2})_1$, which gives a simple condensable $\EE_1$-algebra in $\CC_{10}$ with the underlying object $V_0 \oplus V_{10} \oplus V_{18} \oplus V_{28}$. The corresponding module over $\CC_{28}$ is $\RMod_{V_0 \oplus V_{10} \oplus V_{18} \oplus V_{28}}(\CC_{28})$.
\eit
Equivalently, this also gives the complete classification of Lagrangian algebras in $\FZ_1(\CC_k)$. As explained in Examle \ref{expl_VOA_algebra}, these Lagrangian algebras are modular invariant CFTs over $(\widehat{\mathfrak{sl}_2})_k \otimes \overline{(\widehat{\mathfrak{sl}_2})}_{k}$. The partition functions of these CFTs, or the underlying objects of these Lagrangian algebras, can be found in \cite[Chapter 17]{DFMS97} for example.
\end{expl}

\begin{rem}
Some examples of anyon condensations from para-fermions to finite gauge theories can be found in \cite{LY23}. More examples in lower rank braided fusion 1-categories can be found in \cite{Kik23, Kik23a, KKH24a, Kik24}. 
\end{rem}

\subsubsection{Geometric approach} \label{sec:2codim_3D_geometric_approach}

Now we take a geometric approach towards to the anyon condensation theory in (2+1)D. The physical or geometric intuitions behind it naturally generalize to higher dimensions. 

\medskip
The setup is the same one given in Figure\,\ref{fig:2d-condensation}. Recall that the anyon condensation is determined by a condensable $\EE_2$-algebra $A$ in $\Omega \CC$, and we have $\Omega \CD \simeq \Mod^{\EE_2}_A(\Omega \CC)$, $\Omega_m\CM \simeq \RMod_A(\Omega \CC)$. We show that the algebra $A$ can be constructed geometrically. We start from recalling a known result. 
\begin{lem}[\cite{Zhe17}] \label{lem:shrink_Cphase}
Let $\Omega \CC,\Omega \CD$ be non-degenerate braided fusion 1-categories and $\Omega_m\CM$ a closed milti-fusion $\Omega \CC$-$\Omega \CD$-bimodule. We have canonical monoidal equivalences
\begin{align} 
\Omega_m\CM \boxtimes_{\Omega \CD} \Omega_m\CM^{\rm rev} &\stackrel{\simeq}{\longrightarrow} \mathrm{Fun}_{\Omega \CC}(\Omega_m\CM,\Omega_m\CM), \quad x \boxtimes_{\Omega\CD} y \mapsto x \odot - \odot y \label{eq:factorization_formula_1} \\
\Omega_m\CM \boxtimes_{\Omega \CC} \Omega_m\CM^{\rm rev} &\stackrel{\simeq}{\longrightarrow} \mathrm{Fun}_{\Omega \CD}(\Omega_m\CM,\Omega_m\CM), \quad x \boxtimes_{\Omega \CC} y \mapsto x \odot - \odot y. \label{eq:factorization_formula_2}
\end{align}
\end{lem}

This Lemma says that if we squeeze $\Omega_m\CM$ and $\Omega_m\CM^{\rm rev}$ along $\Omega \CC$ in the following physical configuration, 
\be \label{eq:shrink_Cphase_1}
\begin{array}{c}
\begin{tikzpicture}[scale=0.7]
\fill[teal!20](-2,0)--(0,0)--(0,2)--(-2,2)--cycle node[black,opacity=1] at(-1.5,1){\scriptsize $\Omega \CD$};
\fill[blue!20](-2,0)--(-2,-0.5)--(3,-0.5)--(3,0)--(1,0)--(1,2)--(3,2)--(3,2.5)--(-2,2.5)--(-2,2)--(0,2)--(0,0)--cycle node[black,opacity=1] at(0.5,1){\scriptsize $\Omega \CC$};
\draw[->-, blue, ultra thick](0,2)--(0,0) ; 
\draw[->-, blue, ultra thick](0,0)--(-2,0) node[black,opacity=1] at(-1,-0.3){\scriptsize $\Omega_m\CM$};
\fill[teal!20](3,0)--(1,0)--(1,2)--(3,2)--cycle node[black,opacity=1] at(2.5,1){\scriptsize $\Omega\CD$};

\draw[->-, blue, ultra thick](3,0)--(1,0) node[black,opacity=1] at(2.2,-0.3){\scriptsize $\Omega_m\CM$};
\draw[->-, blue, ultra thick](1,0)--(1,2) ; 
\draw[->-, blue, ultra thick](1,2)--(3,2) ; 
\draw[->-, blue, ultra thick](-2,2)--(0,2) ; 
\end{tikzpicture}
\end{array}
\stackrel{\mathrm{Squeeze}}{\rightsquigarrow}
\begin{array}{c}
\begin{tikzpicture}[scale=0.7]
\fill[blue!20](-2,2.5)--(2,2.5)--(2,2)--(-2,2)--cycle;
\fill[blue!20](-2,-0.5)--(2,-0.5)--(2,0)--(-2,0)--cycle;
\fill[teal!20](-2,0)--(0,0)--(0,2)--(-2,2)--cycle node[black,opacity=1] at(-1.5,1){\scriptsize $\Omega \CD$};
\draw[->-,blue, very thick](-2,0)--(0,0) node[black,opacity=1] at(-1,-0.3){\scriptsize $\Omega_m\CM^{\rm rev}$};
\fill[teal!20](2,0)--(0,0)--(0,2)--(2,2)--cycle node[black,opacity=1] at(1.5,1){\scriptsize $\Omega \CD$};
\draw[->-, blue, ultra thick](0,0)--(2,0) node[black,opacity=1] at(1,-0.3){\scriptsize $\Omega_m\CM^{\rm rev}$};
\draw[->-, blue, ultra  thick](0,2)--(2,2) node[black,opacity=1] at(1,2.3){\scriptsize $\Omega_m\CM$};
\draw[->-, blue, ultra thick](-2,2)--(0,2) node[black,opacity=1] at(-1,2.3){\scriptsize $\Omega_m\CM$};
\draw[->-, blue, ultra thick](0,0)--(0,2) node[black,opacity=1] at(0.3,0.6){\scriptsize $ \mathrm{Fun}_{\Omega \CD}(\Omega_m\CM,\Omega_m\CM)$};
\end{tikzpicture}
\end{array}
\quad = \quad
\begin{array}{c}
\begin{tikzpicture}[scale=0.7]
\fill[blue!20](-2,2.5)--(2,2.5)--(2,2)--(-2,2)--cycle;
\fill[blue!20](-2,-0.5)--(2,-0.5)--(2,0)--(-2,0)--cycle;
\fill[teal!20](-2,0)--(0,0)--(0,2)--(-2,2)--cycle node[black,opacity=1] at(-1.5,1){\scriptsize $\Omega \CD$};
\draw[->-,blue,ultra thick](-2,0)--(0,0) node[black,opacity=1] at(-1,-0.3){\scriptsize $\Omega_m\CM^{\rm rev}$};
\fill[teal!20](2,0)--(0,0)--(0,2)--(2,2)--cycle ; 
\draw[->-,blue,ultra thick](0,0)--(2,0) node[black,opacity=1] at(-1,-0.3){\scriptsize $\Omega_m\CM^{\rm rev}$};
\draw[->-,blue,ultra thick](0,2)--(2,2) node[black,opacity=1] at(1,2.3){\scriptsize $\Omega_m\CM$};
\draw[->-,blue,ultra thick](-2,2)--(0,2) node[black,opacity=1] at(-1,2.3){\scriptsize $\Omega_m\CM$};
\draw[->-,blue,ultra thick](0,0)--(0,0.7);
\fill[black] (0,0.6) circle (0.07) node[right,black,opacity=1] at(0,0.6){\scriptsize $\Omega_m\CM$};
\draw[->-,blue,ultra thick](0,1.3)--(0,2);
\fill[black] (0,1.4) circle (0.07) node[right,black,opacity=1] at(0,1.4){\scriptsize $(\Omega_m\CM)^{\rm op}$};
\draw[->-,black,dashed,ultra thick](0,0.6)--(0,1.4) ; 
\end{tikzpicture}
\end{array}
\ee
we obtain a domain wall whose particles form the fusion 1-category $\Fun_{\Omega \CD}(\Omega_m\CM,\Omega_m\CM)$.  The ``$=$'' is due to the equivalence $\mathrm{Fun}_{\Omega \CD}(\Omega_m\CM,\Omega_m\CM) \simeq \Omega_m\CM^\op \boxtimes_{\Omega\CD} \CM$, which maps $\id_{\Omega_m\CM}$ to $\int^{\Omega \CD}_{x \in \CM} x^R\boxtimes_{\Omega\CD} x$. The notation $\int^{\Omega \CD}_{x \in \CM}$ is called an `$\Omega\CD$-module end' and is explained in Example\,\ref{expl:idM=int_mRm} in Appendix\,\ref{Appendix^EMcal}. The domain wall $\Fun_{\Omega \CD}(\Omega_m\CM,\Omega_m\CM)$ without specifying a particle on it can be understood as the particle is trivial. Then we obtain the following physically equivalent configurations. 
\be \label{eq:shrink_Cphase_2}
\begin{array}{c}
\begin{tikzpicture}[scale=0.7]
\fill[blue!20](-2,2.5)--(2,2.5)--(2,2)--(-2,2)--cycle;
\fill[blue!20](-2,-0.5)--(2,-0.5)--(2,0)--(-2,0)--cycle;
\fill[teal!20](-2,0)--(0,0)--(0,2)--(-2,2)--cycle node[black,opacity=1] at(-1.5,1.4){\scriptsize $\Omega \CD$};
\draw[->-,blue,ultra thick](-2,0)--(0,0) node[black,opacity=1] at(-1,-0.3){\scriptsize $\Omega_m\CM^{\rm rev}$};
\fill[teal!20](2,0)--(0,0)--(0,2)--(2,2)--cycle node[black,opacity=1] at(1.5,1.4){\scriptsize $\Omega \CD$};
\draw[->-,blue,ultra thick](0,0)--(2,0) node[black,opacity=1] at(1,-0.3){\scriptsize $\Omega_m\CM^{\rm rev}$};
\draw[->-,blue,ultra thick](0,2)--(2,2) node[black,opacity=1] at(1,2.3){\scriptsize $\Omega_m\CM$};
\draw[->-,blue,ultra thick](-2,2)--(0,2) node[black,opacity=1] at(-1,2.3){\scriptsize $\Omega_m\CM$};
\draw[->-,blue,ultra thick](0,0)--(0,2) node[black,opacity=1] at(0,0.6){\scriptsize $\mathrm{Fun}_{\Omega \CD}(\Omega_m\CM,\Omega_m\CM)$};
\end{tikzpicture}
\end{array}
= \int^{\Omega \CD}_{x \in \CM}
\begin{array}{c}
\begin{tikzpicture}[scale=0.7]
\fill[blue!20](-2,2.5)--(2,2.5)--(2,2)--(-2,2)--cycle;
\fill[blue!20](-2,-0.5)--(2,-0.5)--(2,0)--(-2,0)--cycle;
\fill[teal!20](-2,0)--(0,0)--(0,2)--(-2,2)--cycle node[black,opacity=1] at(-1.5,1){\scriptsize $\Omega \CD$};
\draw[->-,blue,ultra thick](-2,0)--(0,0) node[black,opacity=1] at(-1,-0.3){\scriptsize $\Omega_m\CM^{\rm rev}$};
\fill[teal!20](2,0)--(0,0)--(0,2)--(2,2)--cycle node[black,opacity=1] at(1.5,1){\scriptsize $\Omega \CD$};
\draw[->-,blue,ultra thick](0,0)--(2,0) node[black,opacity=1] at(1,-0.3){\scriptsize $\Omega_m\CM^{\rm rev}$};
\draw[->-,blue,ultra thick](0,2)--(2,2) node[black,opacity=1] at(1,2.3){\scriptsize $\Omega_m\CM$};
\draw[->-,blue,ultra thick](-2,2)--(0,2) node[black,opacity=1] at(-1,2.3){\scriptsize $\Omega_m\CM$};
\draw[->-,blue,ultra thick](0,0)--(0,0.7);
\draw[->-,blue,ultra thick](0,1.3)--(0,2);
\draw[->-,black,dashed,ultra thick](0,0.7)--(0,1.3) node[right,black,opacity=1] at(0,1){\scriptsize $\Omega \CD$};
\fill[red] (0,0.7) circle (0.07);
\fill[red] (0,1.3) circle (0.07);
\node[black,opacity=1] at (-0.3,0.7){\scriptsize $x$};
\node[black,opacity=1] at (-0.3,1.3){\scriptsize $x^R$};
\end{tikzpicture}
\end{array} 
= \int^{\Omega \CD}_{x \in \CM}
\begin{array}{c}
\begin{tikzpicture}[scale=0.7]
\fill[blue!20](-2,2.5)--(2,2.5)--(2,2)--(-2,2)--cycle;
\fill[blue!20](-2,-0.5)--(2,-0.5)--(2,0)--(-2,0)--cycle;
\fill[teal!20](-2,0)--(2,0)--(2,2)--(-2,2)--cycle node[black,opacity=1] at(-1.2,1){\scriptsize $\Omega \CD$};
\draw[->-,blue,ultra thick](-2,0)--(0,0) node[black,opacity=1] at(-1.2,-0.3){\scriptsize $\Omega_m\CM^{\rm rev}$};
\draw[->-,blue,ultra thick](-2,2)--(0,2) node[black,opacity=1] at(-1.2,2.3){\scriptsize $\Omega_m\CM$};
\draw[->-,blue,ultra thick](0,0)--(2,0);
\draw[->-,blue,ultra thick](0,2)--(2,2);
\fill[red] (0,0) circle (0.07);
\fill[red] (0,2) circle (0.07);
\node[black,opacity=1] at (0,0.3){\scriptsize $x$};
\node[black,opacity=1] at (0,1.7){\scriptsize $x^R$};
\end{tikzpicture}
\end{array}
\ee

\medskip
Now we show that the condensable $\EE_2$-algebra $A$ can be constructed geometrically as a $\SD^3$-bubble in $\SC^3$ as illustrated below. 
\be \label{eq:A=bubble_object}
A =
\begin{array}{c}
\begin{tikzpicture}[scale=0.7]
\fill[teal!20](0.5,0.5)--(1.5,0.5)--(1.5,1.5)--(0.5,1.5)--cycle;
\fill[blue!20,even odd rule](-0.5,0)--(2.5,0)--(2.5,2)--(-0.5,2)--cycle (0.5,0.5)--(1.5,0.5)--(1.5,1.5)--(0.5,1.5)--cycle;
\draw[->-,blue,ultra thick](0.5,0.5)--(0.5,1.5); 
\draw[blue,ultra thick](0.5,0.5)--(1.5,0.5);
\draw[blue,ultra thick](1.5,0.5)--(1.5,1.5);
\draw[blue,ultra thick](1.5,1.5)--(0.5,1.5);
\node[black,opacity=1] at(2.1,0.2) {\scriptsize $\Omega \CC$};
\node[black,opacity=1] at(1,1) {\scriptsize $\Omega \CD$};
\end{tikzpicture}
\end{array}
\ee
Viewed from far away, the bubble shrink to a particle in $\Omega\CC$, it was shown in \cite{AKZ17} that this particle is precisely $A$ as an object in $\Omega\CC$. We recall this result and give a new proof. 
\begin{plem}[\cite{AKZ17}] \label{pthm:bubble=internal_hom}
For $d\in \Omega\CD$ and $y\in \Omega_m\CM$, the following bubble defines a particle in $\Omega\CC$ given by $[1_m, d\odot y]_{\Omega \CC} \in \Omega\CC$. 
$$
[1_m, d\odot y]_{\Omega \CC} \quad = \quad
\begin{array}{c}
\begin{tikzpicture}[scale=0.8]
\fill[teal!20](0.5,0.5)--(1.5,0.5)--(1.5,1.5)--(0.5,1.5)--cycle;
\fill[blue!20,even odd rule](0,0)--(2,0)--(2,2)--(0,2)--cycle (0.5,0.5)--(1.5,0.5)--(1.5,1.5)--(0.5,1.5)--cycle;
\draw[->-,blue,ultra thick](0.5,0.5)--(0.5,1.5); 
\draw[blue,ultra thick](0.5,0.5)--(1.5,0.5);
\draw[blue,ultra thick](1.5,0.5)--(1.5,1.5);
\draw[blue,ultra thick](1.5,1.5)--(0.5,1.5);
\fill[red] (1.5,1) circle (0.07) node[right,black,opacity=1] at(1.5,1) {\scriptsize $y$};
\fill[red] (1,1) circle (0.07) ;
\node at (1.2,1) {\scriptsize $d$} ;
\node[black,opacity = 1] at (0.4,1.8) {\scriptsize $\Omega \CC$};
\end{tikzpicture}
\end{array}
$$
\end{plem}
\begin{proof}
First squeeze the bubble vertically to a line then horizontally to a point. We obtain
$$
\begin{array}{c}
\begin{tikzpicture}[scale=0.8]
\fill[teal!20](0.5,0.5)--(1.5,0.5)--(1.5,1.5)--(0.5,1.5)--cycle;
\fill[blue!20,even odd rule](0,0)--(2,0)--(2,2)--(0,2)--cycle (0.5,0.5)--(1.5,0.5)--(1.5,1.5)--(0.5,1.5)--cycle;
\draw[->-,blue,ultra thick](0.5,0.5)--(0.5,1.5); 
\draw[blue,ultra thick](0.5,0.5)--(1.5,0.5);
\draw[blue,ultra thick](1.5,0.5)--(1.5,1.5);
\draw[blue,ultra thick](1.5,1.5)--(0.5,1.5);
\fill[red] (1.5,1) circle (0.07) node[right,black,opacity=1] at(1.5,1) {\scriptsize $y$};
\fill[red] (1,1) circle (0.07) ;
\node at (1.2,1) {\scriptsize $d$} ;
\node[black,opacity = 1] at (0.4,1.8) {\scriptsize $\Omega \CC$};
\end{tikzpicture}
\end{array}
=
\begin{array}{c}
\begin{tikzpicture}[scale=0.8]
\fill[blue!20,even odd rule](0,0)--(3,0)--(3,2)--(0,2)--cycle;
\draw[->-,black,ultra thick] (0.5,1)--(2.5,1) node[midway,below] {\scriptsize $\mathrm{Fun}_{\Omega \CC}(\Omega_m\CM,\Omega_m\CM)$}; 
\fill[red] (2.5,1) circle (0.07) node[above,black] {\scriptsize $d\odot y$};
\fill[black] (0.5,1) circle (0.07) node[above] {\scriptsize $1_m$};
\node[black,opacity = 1] at (0.4,1.8) {\scriptsize $\Omega \CC$};
\end{tikzpicture}
\end{array}
=
\begin{array}{c}
\begin{tikzpicture}[scale=0.8]
\fill[blue!20,even odd rule](0,0)--(2,0)--(2,2)--(0,2)--cycle; 
\fill[red] (1,1) circle (0.07) node[right,black] {\scriptsize $[1_m,d\odot y]_{\Omega \CC}$};
\node[black,opacity = 1] at (0.4,1.8) {\scriptsize $\Omega \CC$};
\end{tikzpicture}
\end{array}
$$
The second ``$=$'' is due to the monoidal equivalence $(\Omega_m\CM)^{\rm op} \boxtimes_{\CE} \Omega_m\CM \simeq \mathrm{Fun}_{\CE}(\Omega_m\CM,\Omega_m\CM) \simeq \Omega \CC$, which maps $1_m \boxtimes_{\CE} y$ to 
\[
[-,1_m]^R_{\CE} \odot y \mapsto \int^{\CE}_{x \in \Omega_m\CM}[x, [x,1_m]_{\CE}^R \odot y]_{\Omega \CD} \simeq [\int^{x \in \Omega_m\CM}_{\CE}[x, 1_m] \odot x, y]_{\Omega \CC} \simeq [1_m,y]_{\Omega \CC}.
\]
\end{proof}

The identity (\ref{eq:A=bubble_object}) is only an identity of objects. The algebraic structure on the bubble has not been constructed. We do that now. The multiplication morphisms $\mu_A: A\otimes A\to A$ can be defined geometrically by fusing two such bubbles into one bubble as follows. 
\be \label{eq:geo_define_mu_A}
A \otimes A= 
\begin{array}{c}
\begin{tikzpicture}[scale=0.8]
\fill[teal!20](0.5,0.5)--(1.5,0.5)--(1.5,1.5)--(0.5,1.5)--cycle;
\fill[teal!20](2,0.5)--(3,0.5)--(3,1.5)--(2,1.5)--cycle;
\fill[blue!20,even odd rule](0,0)--(3.5,0)--(3.5,2)--(0,2)--cycle (0.5,0.5)--(1.5,0.5)--(1.5,1.5)--(0.5,1.5)--cycle (2,0.5)--(3,0.5)--(3,1.5)--(2,1.5)--cycle;
\draw[->-,blue,ultra thick](0.5,0.5)--(0.5,1.5); 
\draw[blue,ultra thick](0.5,0.5)--(1.5,0.5);
\draw[blue,ultra thick](1.5,0.5)--(1.5,1.5);
\draw[blue,ultra thick](1.5,1.5)--(0.5,1.5);
\draw[->-,blue,ultra thick](2,0.5)--(2,1.5); 
\draw[blue,ultra thick](2,0.5)--(3,0.5);
\draw[blue,ultra thick](3,0.5)--(3,1.5);
\draw[blue,ultra thick](3,1.5)--(2,1.5);
\node[black,opacity=1] at(1,1) {\scriptsize $\Omega \CD$};
\node[black,opacity=1] at(2.5,1) {\scriptsize $\Omega \CD$};
\node[black,opacity=1] at(3.2,0.2) {\scriptsize $\Omega \CC$};
\end{tikzpicture}
\end{array}
=
\int^{\Omega \CC}_{x \in \Omega_m\CM} 
\begin{array}{c}
\begin{tikzpicture}[scale=0.8]
\fill[teal!20](0.5,0.5)--(3,0.5)--(3,1.5)--(0.5,1.5)--cycle;
\fill[blue!20,even odd rule](0,0)--(3.5,0)--(3.5,2)--(0,2)--cycle (0.5,0.5)--(3,0.5)--(3,1.5)--(0.5,1.5)--cycle;
\draw[->-,blue,ultra thick](0.5,0.5)--(0.5,1.5); 
\draw[blue,ultra thick](0.5,0.5)--(3,0.5);
\draw[blue,ultra thick](3,0.5)--(3,1.5);
\draw[blue,ultra thick](3,1.5)--(0.5,1.5);
\node[black,opacity=1] at(1.75,1) {\scriptsize $\Omega \CD$};
\fill[red] (1.75,0.5) circle (0.07);
\node[black,opacity=1] at (1.75,0.3){\scriptsize $x$};
\fill[red] (1.75,1.5) circle (0.07);
\node[black,opacity=1] at (1.75,1.75){\scriptsize $x^R$};
\node[black,opacity=1] at(3.2,0.2) {\scriptsize $\Omega \CC$};
\end{tikzpicture}
\end{array}
\xrightarrow{\mu_A}
\begin{array}{c}
\begin{tikzpicture}[scale=0.8]
\fill[teal!20](0.5,0.5)--(1.5,0.5)--(1.5,1.5)--(0.5,1.5)--cycle;
\fill[blue!20,even odd rule](0,0)--(2,0)--(2,2)--(0,2)--cycle (0.5,0.5)--(1.5,0.5)--(1.5,1.5)--(0.5,1.5)--cycle;
\draw[->-,blue,ultra thick](0.5,0.5)--(0.5,1.5); 
\draw[blue,ultra thick](0.5,0.5)--(1.5,0.5);
\draw[blue,ultra thick](1.5,0.5)--(1.5,1.5);
\draw[blue,ultra thick](1.5,1.5)--(0.5,1.5);
\node[black,opacity=1] at(1,1) {\scriptsize $\Omega \CD$};
\node[black,opacity=1] at(1.7,0.2) {\scriptsize $\Omega \CC$};
\end{tikzpicture}
\end{array}
\ee
where the arrow is defined by the canonical projection $\int^{\Omega \CC}_{x \in \Omega_m\CM}  x^R \boxtimes_{\Omega \CD} x \to 1_m \boxtimes_{\Omega \CD} 1_m$. In the rest of this subsubsection, we show that this algebraic structure on $A$ coincides with that of 
$[1_m,1_m]_{\Omega \CC}$. 
\begin{thm} \label{thm:geo_A=internal_hom}
The algebraic structure on $A$ defined in (\ref{eq:geo_define_mu_A}) coincides with that on $[1_m ,1_m]_{\Omega\CC}$. 
\end{thm}
\begin{proof}
We only sketch the idea of the proof and leave the remaining details to Appendix\,\ref{sec:proof_A=internal_hom}. 
The algebraic structure on the internal hom $[1_m ,1_m]_{\Omega\CC}$ follows from the universal property of the internal hom. This universal construction, however, is not physical or geometrical. In order to find a proof, it is enough to find a geometric construction of the algebraic structure on $[1_m ,1_m]_{\Omega\CC}$ and compare it with (\ref{eq:geo_define_mu_A}). 

The idea is that we rewrite the internal hom as follows: 
$$
[1_m ,1_m]_{\Omega\CC} = [1_m ,-]_{\Omega\CC}(1_m) = L_2^R(1_m), 
$$
where $L_2^R=[1_m ,-]_{\Omega\CC}$ is precisely the right adjoint of the functor $L_2:=-\odot 1_m: \Omega\CC \to \Omega_m\CM$ by the definition of internal hom (see Definition\,\ref{def:internal_hom}). Note that $L$ has a clear physical meaning, i.e., moving bulk anyons to the wall, and is called bulk-to-wall map. Its right adjoint functor $L^R$ also has a clear physical meaning as a physical process of creating a bubble from the wall $\SM^2$ to the bulk of $\SC^3$  
as illustrated below. 
\begin{align*}
\begin{array}{c}
\begin{tikzpicture}[scale=0.8]
\fill[blue!20](-2,0)--(0,0)--(0,2)--(-2,2)--cycle node[black,opacity=1] at(-1,1) {\scriptsize $\Omega \CC$};
\fill[teal!20](2,0)--(0,0)--(0,2)--(2,2)--cycle node[black,opacity=1] at(1,1) {\scriptsize $\Omega \CD$};
\draw[->-, blue,ultra thick] (0,0)--(0,1);
\draw[->-, blue,ultra thick] (0,1)--(0,2);
\fill[red] (0,0.9) circle (0.07) node[right,black,opacity=1] at(0,0.9) {\scriptsize $y$};
\end{tikzpicture}
\end{array}
&=
\begin{array}{c}
\begin{tikzpicture}[scale=0.8]
\fill[blue!20](-2,0)--(0,0)--(0,0.5)--(-0.8,0.5)--(-0.8,1.5)--(0,1.5)--(0,2)--(-2,2)--cycle node[black,opacity=1] at(-1.5,1) {\scriptsize $\Omega \CC$};
\fill[teal!20](2,0)--(0,0)--(0,0.5)--(-0.8,0.5)--(-0.8,1.5)--(0,1.5)--(0,2)--(2,2)--cycle node[black,opacity=1] at(1,1) {\scriptsize $\Omega \CD$}; 
\draw[blue,ultra thick](0,0.5)--(-0.8,0.5)--(-0.8,1.5)--(0,1.5);
\fill[red] (-0.8,1) circle (0.07) node[right,black,opacity=1] at(-0.75,1) {\scriptsize $y$};
\draw[->-,blue,ultra thick] (0,0)--(0,0.5);
\draw[->-,blue,ultra thick] (0,1.5)--(0,2);
\end{tikzpicture}
\end{array}
\\
\mapsto \int^{\Omega \CC}_{x \in \Omega_m\CM}
\begin{array}{c}
\begin{tikzpicture}[scale=0.8]
\fill[blue!20](-2,0)--(0,0)--(0,0.5)--(-0.8,0.5)--(-0.8,1.5)--(0,1.5)--(0,2)--(-2,2)--cycle node[black,opacity=1] at(-1.5,1) {\scriptsize $\Omega \CC$};
\fill[teal!20](2,0)--(0,0)--(0,0.5)--(-0.8,0.5)--(-0.8,1.5)--(0,1.5)--(0,2)--(2,2)--cycle node[black,opacity=1] at(1,1) {\scriptsize $\Omega \CD$}; 
\draw[blue,ultra thick](0,0.5)--(-0.8,0.5)--(-0.8,1.5)--(0,1.5);
\fill[red] (-0.8,1) circle (0.07) node[right,black,opacity=1] at(-0.75,1) {\scriptsize $y$};
\draw[->-,blue,ultra thick] (0,0)--(0,0.5);
\draw[->-,blue,ultra thick] (0,1.5)--(0,2);
\fill[black] (-0.4,0.5) circle (0.07);
\fill[black] (-0.4,1.5) circle (0.07);
\node[black,opacity=1] at (-0.4,0.3){\scriptsize $x$};
\node[black,opacity=1] at (-0.4,1.75){\scriptsize $x^R$};
\end{tikzpicture}
\end{array}
& \simeq
\begin{array}{c}
\begin{tikzpicture}[scale=0.8]
\fill[blue!20](-2,0)--(0,0)--(0,0.5)--(-0.4,0.5)--(-0.4,1.5)--(0,1.5)--(0,2)--(-2,2)--cycle node[black,opacity=1] at(-1.7,1.8) {\scriptsize $\Omega \CC$};
\fill[teal!20](2,0)--(0,0)--(0,0.5)--(-0.4,0.5)--(-0.4,1.5)--(0,1.5)--(0,2)--(2,2)--cycle node[black,opacity=1] at(1,1) {\scriptsize $\Omega \CD$}; 
\fill[teal!20](-0.6,0.5)--(-1.6,0.5)--(-1.6,1.5)--(-0.6,1.5)--cycle;
\draw[blue,ultra thick](0,0.5)--(-0.4,0.5)--(-0.4,1.5)--(0,1.5);
\draw[blue,ultra thick](-0.6,0.5)--(-1.6,0.5)--(-1.6,1.5)--(-0.6,1.5);
\draw[->-,blue,ultra thick] (0,0)--(0,0.5);
\draw[->-,blue,ultra thick] (0,1.5)--(0,2);
\draw[->-,blue,ultra thick] (-0.6,1.5)--(-0.6,0.5);
\fill[red] (-1.6,1) circle (0.07) node[right,black,opacity=1] at(-1.5,1) {\scriptsize $y$};
\end{tikzpicture}
\end{array}
\end{align*}

The algebraic structure on $[1_m,1_m]_{\Omega\CC}=L^R(1_m)$ is defined in terms of the defining data of the adjoint pair $L_2$ and $L_2^R$ as follows.  
\be \label{eq:multiplication_via_LLR}
L_2^R(1_m) \otimes L_2^R(1_m) \to L_2^RL_2(L_2^R(1_m) \otimes L_2^R(1_m))  \simeq L_2^R(L_2L_2^R(1_m) \otimes L_2L_2^R(1_m)) 
\to 
L_2^R(1_m).  
\ee
It remains to show that the morphism (\ref{eq:multiplication_via_LLR}) coincides with the one defined in (\ref{eq:geo_define_mu_A}). We postpone it to Appendix\,\ref{sec:proof_A=internal_hom}. 
\end{proof}

\begin{rem}
It is worth mentioning that the algebra we obtained by creating $\SD^3$-bubbles (taking internal homs) is a condensable algebra in $\Omega \CC$. This fact does not rely on the fact that $\SD^3$ is obtained from $\SC^3$ via a particle condensation, so it applies to any $\SD^3$ that are Morita equivalent to $\SC^3$. However, condensing this algebra does not reproduce the phase $\SD^3$ in general. For example, in the same setting as Figure\,\ref{fig:2d-condensation}, we create a $\SC^3$-bubble in $\SD^3$, which is the trivial particle in $\Omega\CD$, i.e.,
$$
\begin{array}{c}
\begin{tikzpicture}[scale=0.8]
\fill[blue!20](0.5,0.5)--(1.5,0.5)--(1.5,1.5)--(0.5,1.5)--cycle;
\fill[teal!20,even odd rule](0,0)--(2,0)--(2,2)--(0,2)--cycle (0.5,0.5)--(1.5,0.5)--(1.5,1.5)--(0.5,1.5)--cycle;
\draw[->-,blue,ultra thick](0.5,1.5)--(0.5,0.5); 
\draw[blue,ultra thick](0.5,0.5)--(1.5,0.5);
\draw[blue,ultra thick](1.5,0.5)--(1.5,1.5);
\draw[blue,ultra thick](1.5,1.5)--(0.5,1.5);
\node[black,opacity=1] at(1.6,0.2) {\scriptsize $\Omega\CD$};
\node[black,opacity=1] at(1,1) {\scriptsize $\Omega\CC$};
\end{tikzpicture}
\end{array}
\quad = \quad [A,A]_{\Omega\CD} = A = 1_{\one_\CD}. 
$$
Clearly, condensing this algebra (vacuum in $\Omega \CD$) does not produce $\SC^3$ as the condensed phase.
\end{rem}

\begin{rem}
Rolling up the domain wall $\SM^2$ (to the $\SD^3$-side) over a full circle creates a bubble in $\SC^3$, which turns out to be a condensable $\EE_2$-algebra in $\Omega\CC$. This physical process naturally generalizes to higher dimension cases, in which one can roll up potentially higher codimensional defects over higher dimensional spheres or other manifolds. We use this intuition without a proof in later sections. To write down a proof explicitly is a non-trivial task. However, we expect that the proof should be a higher dimensional analogue of the one in this subsubsection. 
\end{rem}

Similar to the physical process of condensing 1-codimensional topological defects described in Section\,\ref{sec:geometric_intuition_1codim} and \ref{sec:general_theory_1codim_nd}, we can provide a description of the physical process of a 2-codimensional defect condensation, which naturally generalizes to higher dimensional topological orders.
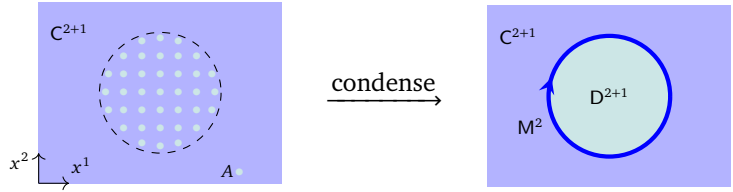
\[
\begin{array}{c}
\begin{tikzpicture}[scale=0.8]
\fill[blue!30] (-2,0) rectangle (2,3) ;
\node at (-1.5,2.5) {\scriptsize $\SC^{2+1}$} ;
\node at (0,0.9) {\scriptsize \textcolor{teal!20}{$\bullet$}} ;
\node at (0,1.2) {\scriptsize \textcolor{teal!20}{$\bullet$}} ;
\node at (0,1.5) {\scriptsize \textcolor{teal!20}{$\bullet$}} ;
\node at (0,1.8) {\scriptsize \textcolor{teal!20}{$\bullet$}} ;
\node at (0,2.1) {\scriptsize \textcolor{teal!20}{$\bullet$}} ;
\node at (0,2.4) {\scriptsize \textcolor{teal!20}{$\bullet$}} ;
\node at (0,0.6) {\scriptsize \textcolor{teal!20}{$\bullet$}} ;
\node at (0.3,1.5) {\scriptsize \textcolor{teal!20}{$\bullet$}} ;
\node at (-0.3,1.5) {\scriptsize \textcolor{teal!20}{$\bullet$}} ;
\node at (0.6,1.5) {\scriptsize \textcolor{teal!20}{$\bullet$}} ;
\node at (-0.6,1.5) {\scriptsize \textcolor{teal!20}{$\bullet$}} ;
\node at (-0.6,1.8) {\scriptsize \textcolor{teal!20}{$\bullet$}} ;
\node at (-0.6,2.1) {\scriptsize \textcolor{teal!20}{$\bullet$}} ;
\node at (-0.6,1.2) {\scriptsize \textcolor{teal!20}{$\bullet$}} ;
\node at (-0.6,0.9) {\scriptsize \textcolor{teal!20}{$\bullet$}} ;
\node at (-0.9,1.5) {\scriptsize \textcolor{teal!20}{$\bullet$}} ;
\node at (0.9,1.5) {\scriptsize \textcolor{teal!20}{$\bullet$}} ;
\node at (0.3,1.2) {\scriptsize \textcolor{teal!20}{$\bullet$}} ;
\node at (-0.3,1.2) {\scriptsize \textcolor{teal!20}{$\bullet$}} ;
\node at (0.3,1.8) {\scriptsize \textcolor{teal!20}{$\bullet$}} ;
\node at (-0.3,1.8) {\scriptsize \textcolor{teal!20}{$\bullet$}} ;
\node at (-0.85,1.8) {\scriptsize \textcolor{teal!20}{$\bullet$}} ;
\node at (-0.85,1.2) {\scriptsize \textcolor{teal!20}{$\bullet$}} ;
\node at (-0.3,2.1) {\scriptsize \textcolor{teal!20}{$\bullet$}} ;
\node at (-0.3,2.35) {\scriptsize \textcolor{teal!20}{$\bullet$}} ;
\node at (-0.3,0.9) {\scriptsize \textcolor{teal!20}{$\bullet$}} ;
\node at (-0.3,0.65) {\scriptsize \textcolor{teal!20}{$\bullet$}} ;
\node at (0.6,1.8) {\scriptsize \textcolor{teal!20}{$\bullet$}} ;
\node at (0.6,2.1) {\scriptsize \textcolor{teal!20}{$\bullet$}} ;
\node at (0.6,1.2) {\scriptsize \textcolor{teal!20}{$\bullet$}} ;
\node at (0.6,0.9) {\scriptsize \textcolor{teal!20}{$\bullet$}} ;
\node at (0.3,2.1) {\scriptsize \textcolor{teal!20}{$\bullet$}} ;
\node at (0.3,2.35) {\scriptsize \textcolor{teal!20}{$\bullet$}} ;
\node at (0.3,0.9) {\scriptsize \textcolor{teal!20}{$\bullet$}} ;
\node at (0.3,0.65) {\scriptsize \textcolor{teal!20}{$\bullet$}} ;
\node at (0.85,1.8) {\scriptsize \textcolor{teal!20}{$\bullet$}} ;
\node at (0.85,1.2) {\scriptsize \textcolor{teal!20}{$\bullet$}} ;

\node at (1.2,0.2) {\scriptsize $A\,$\textcolor{teal!20}{$\bullet$}} ;

\draw[dashed] (1,1.5) arc (360:0:1) ;
\draw[->] (-2,0) -- (-1.5,0) ; 
\node at (-1.3,0.3) {\scriptsize $x^1$};
\draw[->] (-2,0) -- (-2,0.5) node [near end,left] {\scriptsize $x^2$};
\end{tikzpicture}
\end{array}
\quad\xrightarrow{\mbox{condense}} \quad
\begin{array}{c}
\begin{tikzpicture}[scale=0.8]
\fill[blue!30] (-2,0) rectangle (2,3) ;
\fill[teal!20] (0,1.5) circle (1);
\node at (-1.5,2.5) {\scriptsize $\SC^{2+1}$} ;
\node at (0,1.5) {\scriptsize $\SD^{2+1}$} ;
\node at (-1.3,1) {\scriptsize $\SM^2$} ;
\draw[blue, ultra thick,->-] (1,1.5) arc (360:0:1) ;
\end{tikzpicture}
\end{array}
\]
\bnu

\item[(1)] First proliferating the $\CD^3$-phase in a disk-like region inside the $\SC^3$-phase; 

\item[(2)] then shrinking the $\SC^3$-phase within the region and merging two $\CD^3$-disks into a larger $\CD^3$-disk but with extra particles on the gapped boundary according to (\ref{eq:shrink_Cphase_1}) and (\ref{eq:shrink_Cphase_2}); 

\item[(3)] then annihilating the extra particles on the boundary according to   (\ref{eq:geo_define_mu_A}). 

\enu

\begin{rem}
Notice that the process of merging two $\CD^3$-disks into a larger $\CD^3$-disk can split into merging vertically then merging horizontally. This leads us to a general theory of the condensation of $k$-codimensional defects (see Section\,\ref{sec:condense_k-codim_defect}). 
\end{rem}

\subsection{Condensations of 2-codimensional defects in \texorpdfstring{$n$}{n}+1D} 
\label{sec:2codim_condensation_n+1D}

\begin{figure}[htbp]
\[
\begin{array}{c}
\begin{tikzpicture}[scale=1]
\fill[blue!30] (-2,0) rectangle (2,3) ;
\node at (-1.5,2.5) {$\SC^{n+1}$} ;
\node at (0,0.9) {\scriptsize \textcolor{teal!20}{$\bullet$}} ;
\node at (0,1.2) {\scriptsize \textcolor{teal!20}{$\bullet$}} ;
\node at (0,1.5) {\scriptsize \textcolor{teal!20}{$\bullet$}} ;
\node at (0,1.8) {\scriptsize \textcolor{teal!20}{$\bullet$}} ;
\node at (0,2.1) {\scriptsize \textcolor{teal!20}{$\bullet$}} ;
\node at (0,2.4) {\scriptsize \textcolor{teal!20}{$\bullet$}} ;
\node at (0,0.6) {\scriptsize \textcolor{teal!20}{$\bullet$}} ;
\node at (0.3,1.5) {\scriptsize \textcolor{teal!20}{$\bullet$}} ;
\node at (-0.3,1.5) {\scriptsize \textcolor{teal!20}{$\bullet$}} ;
\node at (0.6,1.5) {\scriptsize \textcolor{teal!20}{$\bullet$}} ;
\node at (-0.6,1.5) {\scriptsize \textcolor{teal!20}{$\bullet$}} ;
\node at (-0.6,1.8) {\scriptsize \textcolor{teal!20}{$\bullet$}} ;
\node at (-0.6,2.1) {\scriptsize \textcolor{teal!20}{$\bullet$}} ;
\node at (-0.6,1.2) {\scriptsize \textcolor{teal!20}{$\bullet$}} ;
\node at (-0.6,0.9) {\scriptsize \textcolor{teal!20}{$\bullet$}} ;
\node at (-0.9,1.5) {\scriptsize \textcolor{teal!20}{$\bullet$}} ;
\node at (0.9,1.5) {\scriptsize \textcolor{teal!20}{$\bullet$}} ;
\node at (0.3,1.2) {\scriptsize \textcolor{teal!20}{$\bullet$}} ;
\node at (-0.3,1.2) {\scriptsize \textcolor{teal!20}{$\bullet$}} ;
\node at (0.3,1.8) {\scriptsize \textcolor{teal!20}{$\bullet$}} ;
\node at (-0.3,1.8) {\scriptsize \textcolor{teal!20}{$\bullet$}} ;
\node at (-0.85,1.8) {\scriptsize \textcolor{teal!20}{$\bullet$}} ;
\node at (-0.85,1.2) {\scriptsize \textcolor{teal!20}{$\bullet$}} ;
\node at (-0.3,2.1) {\scriptsize \textcolor{teal!20}{$\bullet$}} ;
\node at (-0.3,2.35) {\scriptsize \textcolor{teal!20}{$\bullet$}} ;
\node at (-0.3,0.9) {\scriptsize \textcolor{teal!20}{$\bullet$}} ;
\node at (-0.3,0.65) {\scriptsize \textcolor{teal!20}{$\bullet$}} ;
\node at (0.6,1.8) {\scriptsize \textcolor{teal!20}{$\bullet$}} ;
\node at (0.6,2.1) {\scriptsize \textcolor{teal!20}{$\bullet$}} ;
\node at (0.6,1.2) {\scriptsize \textcolor{teal!20}{$\bullet$}} ;
\node at (0.6,0.9) {\scriptsize \textcolor{teal!20}{$\bullet$}} ;
\node at (0.3,2.1) {\scriptsize \textcolor{teal!20}{$\bullet$}} ;
\node at (0.3,2.35) {\scriptsize \textcolor{teal!20}{$\bullet$}} ;
\node at (0.3,0.9) {\scriptsize \textcolor{teal!20}{$\bullet$}} ;
\node at (0.3,0.65) {\scriptsize \textcolor{teal!20}{$\bullet$}} ;
\node at (0.85,1.8) {\scriptsize \textcolor{teal!20}{$\bullet$}} ;
\node at (0.85,1.2) {\scriptsize \textcolor{teal!20}{$\bullet$}} ;

\node at (1.2,0.2) { $A\,$\textcolor{teal!20}{$\bullet$}} ;

\draw[dashed] (1,1.5) arc (360:0:1) ;
\draw[->] (-2,0) -- (-1.5,0) ; 
\node at (-1.5,0.3) {\scriptsize $x^1$};
\draw[->] (-2,0) -- (-2,0.5) node [near end,left] {\scriptsize $x^2$};
\end{tikzpicture}
\end{array}
\quad\xrightarrow{\mbox{condense}} \quad
\begin{array}{c}
\begin{tikzpicture}
\fill[blue!30] (-2,0) rectangle (2,3) ;
\fill[teal!20] (0,1.5) circle (1);
\draw[->] (-2,0) -- (-1.5,0) ; 
\node at (-1.5,0.3) {\scriptsize $x^1$};
\draw[->] (-2,0) -- (-2,0.5) node [near end,left] {\scriptsize $x^2$};
\node at (-1.5,2.5) {$\SC^{n+1}$} ;
\node at (0,1.5) {$\SD^{n+1}$} ;
\node at (-1.3,1) {$\SM^n$} ;
\draw[blue, ultra thick,->-] (1,1.5) arc (360:0:1) ;
\end{tikzpicture}
\end{array}
\]
\caption{a condensation of 2-codimensional topological defect in $\SC^{n+1}$}
\label{fig:2codim-condensation_nd}
\end{figure}
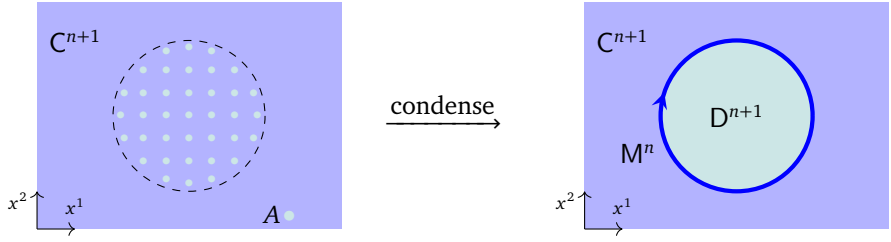

In this subsection, we discuss the theory of the condensations of 2-codimensional topological defects in $n+$1D anomaly-free simple topological orders. The setup of the question remains the same as the 2+1D cases. More precisely, consider a condensation of 2-codimensional topological defect occurring in an $n+$1D anomaly-free simple topological order $\SC^{n+1}$. It creates a new simple topological order $\SD^{n+1}$, which is also anomaly-free, and a gapped domain wall $\SM^n$ as depicted in Figure\,\ref{fig:2codim-condensation_nd}.

\subsubsection{General theory} \label{sec:2codim_nd_general_theory}
The bootstrap analysis in $n+$1D is possible and is similar to the 2+1D cases. However, it becomes much harder in higher dimensions due to the complexity of the higher coherence relations. Instead, we use the geometric intuition of anyon condensation in an anomaly-free 2+1D simple topological order, which generalizes to that of condensations of 2-codimensional topological defects in an $n+$1D anomaly-free simple topological order tautologically. 

\begin{defn}
For a braided multi-fusion $n$-category $\CA$ and $A\in \Algc_{\EE_2}(\CA)$, roughly speaking, an $\EE_2$-$A$-module in $\CA$ is an object $x\in \CA$ equipped with a 2-dimensional actions, i.e., four $A$-actions $\mu_x^\to, \mu_x^\downarrow, \mu_x^\leftarrow, \mu_x^\uparrow$ in four directions such that the diagram (\ref{diag:local-module}) is commutative. Recall (\ref{def:Mod_E2}), one can take 
$$
\Mod_A^{\EE_2}(\CA) \simeq \Mod_A^{\EE_1}(\Mod_A^{\EE_1}(\CA))
$$ 
as a working definition of an $\EE_2$-$A$-module and the category $\Mod_A^{\EE_2}(\CA)$ of $\EE_2$-$A$-modules. 
\end{defn}

Now we are ready to state the result about the condensed phase and the gapped domain wall for a condensation of a 2-codimensional topological defect in $\SC^{n+1}$. 
\begin{pthm}  \label{pthm:2codim_condensation_nd}
If an $n+$1D anomaly-free simple topological order $\SD^{n+1}$ and a gapped domain wall $\SM^n$ are obtained from an anomaly-free simple topological order $\SC^{n+1}$ via a condensation of 2-codimensional topological defect (see Figure \ref{fig:2codim-condensation_nd} and the picture below).
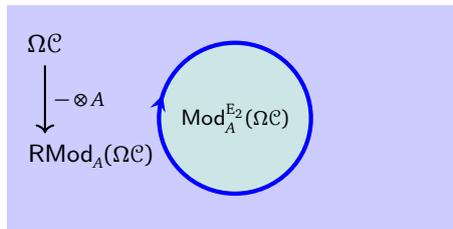
\[
\begin{tikzpicture}
\fill[blue!20] (-3,0) rectangle (3,3) ;
\fill[teal!20] (0,1.5) circle (1);
\node at (-2.5,2.5) {$\Omega\CC$} ;
\draw[thick,->] (-2.5,2.2) -- (-2.5,1.3) node [midway, right] {\footnotesize $-\otimes A$} ;
\node at (0,1.5) {\footnotesize $\Mod_A^{\EE_2}(\Omega\CC)$} ;
\node at (-1.9,1) {\small $\RMod_A(\Omega\CC)$} ;
\draw[blue, ultra thick,->-] (1,1.5) arc (360:0:1) ;
\end{tikzpicture}
\]
\bnu
\item The trivial 2-codimensional topological defect $1_{\one_\CD}$ in $\Omega\SD$ can be identified with a simple condensable $\EE_2$-algebra $A$ in $\Omega\CC$. Moreover, $\Omega\CD$ consists of all deconfined topological defects of codimension 2 and higher, and it can be identified with the category $\Mod_A^{\EE_2}(\Omega\CC)$ of $\EE_2$-$A$-modules (or equivalently, local $A$-modules) in $\CC$, i.e., $\Omega\CD = \Mod_A^{\EE_2}(\Omega\CC)$. 

\item Topological defects on the wall $\SM^n$ include all confined and deconfined 1-codimensional defects, and can be identified with the fusion 1-category $\RMod_A(\Omega\CC)$.

\item Defects in $\Omega\CC$ move to the wall according to the central functor\footnote{Being a central functor means that $-\otimes A$ factors through a braided monoidal functor $g: \Omega\CC \to \FZ_1(\RMod_A(\Omega\CC))$, i.e., it coincides with the composed functor 
$\Omega\CC \xrightarrow{g} \FZ_1(\RMod_A(\Omega\CC)) \xrightarrow{\forget} \RMod_A(\Omega\CC)$, where $\forget$ is the forgetful functor.} $L_2:=- \otimes A \colon \Omega\CC \to \RMod_A(\Omega\CC)$ defined by $x \mapsto x \otimes A$ for all $x\in \Omega\CC$. 
 
\item Defects in $\Omega\CD$ move to the wall according to the canonical forgetful functor\footnote{The ``$\EE_2$-module  condition" for a right $A$-module is a property for $n=2$ but a structure plus properties for $n>2$. This functor $R_2$ is fully faithful for $n=2$ but not fully faithful for $n>2$.} $R_2: \Mod_A^{\EE_2}(\Omega\CC) \to \RMod_A(\Omega\CC)$. 

\enu
The condensable $\EE_2$-algebra $A$ in $\Omega\CC$ is called Lagrangian if $\Mod_A^{\EE_2}(\Omega\CC)\simeq \Vect$. In this case, $\SM^n$ becomes a gapped boundary of $\SC^{n+1}$ and consists of all confined and deconfined defects. 
\end{pthm}

\begin{rem}
If a topological defect is confined (resp. deconfined), then its condensation descendants is also confined (resp. deconfined). This results seems physically natural. A rigorous proof of the $n=2$ case was given in \cite{DX24}. We will provide a mathematical proof when we develop the theory of separable higher algebras elsewhere. 
\end{rem}

\begin{pthm}
By the folding trick and the boundary-bulk relation, we obtain a braided equivalence: 
$$
\Omega\CC \boxtimes \Mod_A^{\EE_2}(\Omega\CC)^\rev \simeq \FZ_1(\RMod_A(\Omega\CC)). 
$$
This result is equivalent to the following braided equivalence: 
\be
\Mod_A^{\EE_2}(\Omega\CC)^\rev \simeq \FZ_2(\Omega\CC, \FZ_1(\RMod_A(\Omega\CC))), 
\ee
which is ready to be generalized to higher algebras and higher representations (see Theorem\,\ref{cor:EkMod_Zk_centralizer}). 
\end{pthm}

Let $\SC^{n+1}$ and $\SD^{n+1}$ be any two $n+$1D (potentially anomalous and composite) topological orders connected by a gapped domain wall $\SM^n=(\CM,m)$, where $\CM$ is a separable $n$-category and a $\CC$-$\CD$-bimodule and $m$ is an object in $\CM$ represents a distinguished gapped wall condition. 
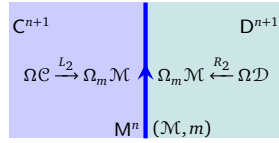
\[
\begin{array}{c}
\begin{tikzpicture}[scale=0.6]
\fill[blue!20] (-3,0) rectangle (3,3) ;
\fill[teal!20] (0,0) rectangle (3,3) ;
\draw[blue, ultra thick,->-] (0,0) -- (0,3) ; 
\node at (-1.5,1.5) {\scriptsize $\Omega\CC \xrightarrow{L_2} \Omega_m\CM$} ;
\node at (1.5,1.5) {\scriptsize $\Omega_m\CM \xleftarrow{R_2} \Omega\CD $} ;
\node at (-0.4,0.2) {\scriptsize  $\SM^n$} ;
\node at (0.8,0.2) {\scriptsize  $(\CM,m)$} ;
\node at (-2.5,2.5) {\scriptsize $\SC^{n+1}$} ;
\node at (2.5,2.5) {\scriptsize $\SD^{n+1}$} ;
\end{tikzpicture}
\end{array}
\]
There is a canonical functor $L_2: \Omega\CC \to \Omega_m\CM$ from the category $\Omega\CC$ of topological defects of codimension 2 and higher in $\SC^{n+1}$ to that of topological defects on $\SM^n$, which is defined simply by moving topological defects in $\SC^{n+1}$ onto $\SM^n$. Similarly, there is also a canonical functor $R_2: \Omega\CD \to \Omega_m\CM$. It is clear that these two functors are both monoidal (i.e. preserving the fusion product). We denote their right adjoint functor by $L_2^R$ and $R_2^R$, respectively.

By rolling up the right half (i.e. $\SM^n$ and $\SD^{n+1}$) in the spatial dimension, we obtain a solid cylinder $\mathrm{D}^1 \times \Rb^{n-1}$ with the interior filled with the $\SD^{n+1}$-phase and the boundary $S^1 \times \Rb^{n-1}$ cylinder decorated by $\SM^n$. Similarly, we can roll up the left half to obtain a solid cylinder in $\SD^{n+1}$-phase as illustrasted below.
\be \label{eq:rolling_up}
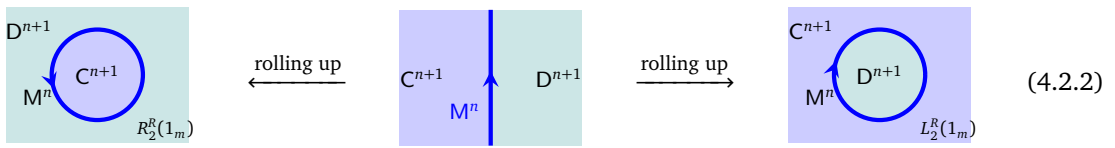
\begin{array}{c}
\begin{tikzpicture}[scale=0.6]
\fill[teal!20] (-2,0) rectangle (2,3) ;
\fill[blue!20] (0,1.5) circle (1);
\node at (-1.5,2.5) {\footnotesize $\SD^{n+1}$} ;
\node at (0,1.5) {\footnotesize $\SC^{n+1}$} ;
\node at (-1.3,1) {\footnotesize $\SM^n$} ;
\draw [blue, ultra thick,->-] (0,1.5) circle [radius=1] ;
\node at (1.5, 0.3) {\scriptsize $R_2^R(1_m)$} ;
\end{tikzpicture}
\end{array}
\quad \xleftarrow{\mbox{\footnotesize rolling up}}  \quad
\begin{array}{c}
\begin{tikzpicture}[scale=0.6]
\fill[blue!20] (-2,0) rectangle (2,3) ;
\fill[teal!20] (0,0) rectangle (2,3) ;
\draw[blue, ultra thick, ->-] (0,0) -- (0,3) node[near start, left] {\footnotesize $\SM^n$};
\node at (-1.5,1.5) {\footnotesize $\SC^{n+1}$} ;
\node at (1.5,1.5) {\footnotesize $\SD^{n+1}$} ;
\end{tikzpicture}
\end{array}
\quad \xrightarrow{\mbox{\footnotesize rolling up}} \quad 
\begin{array}{c}
\begin{tikzpicture}[scale=0.6]
\fill[blue!20] (-2,0) rectangle (2,3) ;
\fill[teal!20] (0,1.5) circle (1);
\node at (-1.5,2.5) {\footnotesize $\SC^{n+1}$} ;
\node at (0,1.5) {\footnotesize $\SD^{n+1}$} ;
\node at (-1.3,1) {\footnotesize $\SM^n$} ;
\draw[blue, ultra thick,->-] (1,1.5) arc (360:0:1) ;
\node at (1.5, 0.3) {\scriptsize $L_2^R(1_m)$} ;
\end{tikzpicture}
\end{array}
\ee
Viewed from far away, these two solid cylinders become two 2-codimensional topological defects in $\SC^{n+1}$ and $\SD^{n+1}$, respectively. The next result tells us how to compute these defects. 
\begin{pthm} \label{pthm:internal_hom_2codim_nd}
Such obtained 2-codimensional topological defects in $\SC^{n+1}$ and $\SD^{n+1}$ are given by 
\be \label{eq:inter_hom_2codim_nd_1}
L_2^R(1_m) = [1_m, 1_m]_{\Omega\CC} \in \Omega\CC, \quad\quad 
R_2^R(1_m) = [1_m, 1_m]_{\Omega\CD} \in \Omega\CD, 
\ee
both of which are condensable $\EE_2$-algebra in $\Omega\CC$ and $\Omega\CD$, respectively. Moreover, all  condensable $\EE_2$-algebra in $\Omega\CC$ can be obtained in this way by properly choosing $\SD^{n+1}$ and $\SM^n$. The Morita equivalence among gapped domain walls naturally defines the so-called 2-Morita equivalence relation among condensable $\EE_2$-algebras (see Definition\,\ref{def:ModEk_kMorita_eq}). 
\end{pthm}
\pf
We sketch a proof. Note that $\Omega\CC$ acts on $\Omega_m\CM$ in the obvious way. We denote the action functor by $\odot: \Omega\CC \times \Omega_m\CM \to \Omega_m\CM$. Therefore, the internal hom is well defined \cite[Proposition\ 3.27]{KZ22b}. Moreover, the action $\odot$ preserves the fusion product, i.e. $\odot$ is monoidal. By the universal property of the center, this action factors through the canonical action $\FZ_1(\Omega_m\CM) \times \Omega_m\CM \to \Omega_m\CM$. Equivalently, 2-codimensional topological defects in $\SC^{n+1}$ can be half-braided with 1-codimensional topological defects in $\SM^n$. This half-braiding structure upgrades the action $\odot$ to an action $\odot: \Algc_{\EE_1}(\Omega\CC)\times \Algc_{\EE_1}(\Omega_m\CM) \to \Algc_{\EE_1}(\Omega_m\CM)$. As a consequence, the internal hom $[1_m, 1_m]_{\Omega\CC} \in \Algc_{\EE_1}(\Algc_{\EE_1}(\Omega\CC))= \Algc_{\EE_2}(\Omega\CC)$.

For $A\in \Algc_{\EE_2}(\Omega\CC)$, let $\SD^{n+1}$ and $\SM^n$ be the condensed phase and gapped domain wall obtained by condensing $A$ (as in Theorem$^{\mathrm{ph}}$\,\ref{pthm:2codim_condensation_nd}), i.e., $\Omega_m\CM=\RMod_A(\Omega\CC)$ and $\Omega\CD = \Mod_A^{\EE_2}(\Omega\CC)$. In this case, $L_2=-\otimes A$ and $L_2^R=[A,-]$, where $[A,-]$ is the forgetful functor $\forget: \RMod_A(\Omega\CC) \to \Omega\CC$. Therefore, in this case, we obtain
\be \label{eq:inter_hom_2codim_nd_2}
L_2^R(1_m) = [1_m, 1_m]_{\Omega\CC} = A \in \Omega\CC; 
\ee
Therefore, all condensable $\EE_2$-algebras in $\Omega\CC$ arise in this way. 
\epf

In general, condensing $L_2^R(1_m)$ in $\Omega\CC$ does not give $\Omega\CD$, and condensing $R_2^R(1_m)$ in $\Omega\CD$ does not give $\Omega\CC$. 
\bnu
\item Let $\SC^3$ be the $\Zb_2$ topological order and $\SD^3$ be the double semion topological order and $\SM^2$ be any gapped domain wall. Then $L_2^R(1_m)=1_{\one_\CC}$ and $R_2^R(1_m)=1_{\one_\CD}$ are both trivial. 

\item When $\SC^{n+1}$ and $\SD^{n+1}$ are both anomaly-free and simple, if $\SD^{n+1}$ and $\SM^n$ are obtained from $\SC^{n+1}$ by condensing a non-trivial simple condensable $\EE_2$-algebra $A \in \Omega\CC$, we expect that this condensation is not reversible in general. For example, for $n\geq 3$ and a non-trivial finite group $G$, let $\SC^{n+1}=\SG\ST_G^{n+1}$, $\SM^n=(\CM,m)=(n\vect_G, \one)$ and $\SD^{n+1}=\mathbf{1}^{n+1}$. In this case, the internal hom $[1_m,1_m]_{\Omega\CC}$ is a Lagrangian algebra in $\Omega\CC=\FZ_1(n\vect_G)$, condensing which produces $\mathbf{1}^{n+1}$ as the condensed phase. However, the internal hom $[1_m,1_m]_{\Omega\CD}=\one_{\Omega\CD}$ is the trivial algebra in $\Omega\CD=(n-1)\vect$, condensing which does not reproduce $\SC^{n+1}$. However, for $n\geq 3$, there are some interesting exceptions (see for example Theorem\,\ref{thm:2codim_condensations_in_trivial_phase} and Remark\,\ref{rem:2codim_cond_4D_special}). 
\enu
This phenomenon is very different from the condensation of 1-codimensional topological defects (recall Theorem$^{\mathrm{ph}}$\,\ref{pthm:1codim_condensation_nd_II}). The reason is that all 1-codimensional topological defects are missing from $\Omega\CC$. If we use $A\in \Omega\CC$ to define a condensed 1-codimensional topological defects, denoted by $\Sigma A$ in $\CC$, then the condensation of $\Sigma A$ in $\CC$ is reversable. We discuss this phenomenon in Section\,\ref{sec:3D_two-step_condensation}.

\medskip
A special case is very important to us. If $\SD^{n+1}=\mathbf{1}^{n+1}$ is obtained from $\SC^{n+1}$ by condensing $A\in \Algc_{\EE_2}(\Omega\CC)$, then $A$ is a Lagrangian algebra in $\Omega\CC$, $\SM^n$ is a gapped boundary of $\SC^{n+1}$ (i.e. $\SC^{n+1}$ is non-chiral). Therefore, we obtain a map:  
$$
\{ \mbox{Lagrangian algebras in $\Omega\CC$} \} 
\xrightarrow{\phi}  \{ \mbox{gapped boundaries of $\SC^{n+1}$} \}. 
$$ 
In this case, $\SC^{n+1}=\SZ(\SM)^{n+1}$ catches all the information of the gravitational anomaly of the gapped boundary $\SM^n$ \cite{KW14,KWZ15}. The non-degenerate braided fusion $(n-1)$-category $\Omega\CC$, as the center of $\Omega_m\CM$, is maximal (or universal) in a precise sense. The internal hom $[1_m, 1_m]_{\Omega\CC}$ is again maximal by the universal property of the internal hom. Therefore, it is natural to expect that it can recover $\SM^n$. Moreover, we expect that all gapped boundaries are the consequences of the condensation of 2-codimensional topological defects. We summarize this discussion as the following result. 

\begin{pthm} \label{conj:Lagrangian=boundary}
When $\SC^{n+1}$ is anomaly-free, non-chiral and simple, $\phi$ is bijective and its inverse $\phi^{-1}$ is defined by 
\be
\SM^n = (\CM, m) \mapsto L_2^R(1_m) = [1_m, 1_m]_{\Omega\CC}
=
\begin{array}{c}
\begin{tikzpicture}[scale=0.6]
\fill[blue!20] (-2,0) rectangle (2,3) ;
\fill[white] (0,1.5) circle (1);
\node at (-1.5,2.5) {\footnotesize $\SC^{n+1}$} ;
\node at (0,1.5) {\footnotesize $\mathbf{1}^{n+1}$} ;
\node at (-1.3,1) {\footnotesize $\SM^n$} ;
\draw[blue, ultra thick,->-] (1,1.5) arc (360:0:1) ;
\node at (1.5, 0.3) {\scriptsize $L_2^R(1_m)$} ;
\end{tikzpicture}
\end{array}
\ee 
Two Lagrangian $\EE_2$-algebras $A$ and $B$ are called 2-Morita equivalence if the associated gapped boundaries (or the fusion $n$-categories $\RMod_A(\Omega\CC)$ and $\RMod_B(\Omega\CC)$) are Morita equivalent (see Definition\,\ref{def:ModEk_kMorita_eq}). We identify the Morita classes $[A]=[\RMod_A(\Omega\CC)]$. 
\end{pthm}

\begin{rem}
In a general situation, two arbitrary anomaly-free simple $n+$1D topological orders $\SC^{n+1}$ and $\SD^{n+1}$ connected by a gapped domain wall. The information of $\SM^n$ can be encoded in the internal hom $[1_m, 1_m]_{\Omega\CC\boxtimes\Omega\CD^\rev}$, which is a Lagrangian algebra in $\FZ_1(\Omega_m\CM)$. While the following two condensable $\EE_2$-algebras: 
$$
[1_m, 1_m]_{\Omega\CC} = 
[1_m,1_m]_{\Omega\CC\boxtimes\Omega\CD^\rev} \cap \Omega\CC 
\quad \mbox{and} \quad
[1_m, 1_m]_{\Omega\CD} = 
[1_m,1_m]_{\Omega\CC\boxtimes\Omega\CD^\rev} \cap \Omega\CD^\rev
$$
encode only some incomplete information of the domain wall $\SM^n$. 
\end{rem}

\begin{rem}
There are infinitely many Lagrangian algebras or gapped boundaries in a non-chiral $n+$1D topological order $\SC^{n+1}$ for $n>2$. For example, stacking a gapped boundary of $\SC^{n+1}$ with an anomaly-free $n$D topological order gives another gapped boundary. The Lagrangian algebras in $\mathbf{1}^{n+1}$ associated to an anomaly-free $n$D topological order is given more explicitly in Section\,\ref{sec:general_example_2-codim} (1) (see for example Remark\,\ref{rem:rolling_up_cylinder}). 
\end{rem}

Since the rolling-up process plays a very important role in our theory, we would like to introduce some useful notations that are inspired by the theory of factorization homology (see for example \cite{Lur17,AF15,AFT17,AFT16,AFR18,AKZ17,AF20}). Note that, in the first picture in (\ref{eq:rolling_up}), one can view $\Omega\CC$ as observables living in the interior of the solid cylinder $\mathrm{D}^2 \times \Rb^{n-2}$ (in spatial dimension), i.e., 
$$
\Obs(\mathring{\mathrm{D}}^2 \times \Rb^{n-2}) = \Omega\CD. 
$$
Since $\Omega\CC$ is an $\EE_2$-algebra in $n\vect$. It makes perfect sense to integrate it over the 2-manifold $\mathring{\mathrm{D}}^2$. Similarly, $\Omega_m\CM$ describes the observable living on the boundary of the solid cylinder, i.e., 
$$
\Obs(\partial \mathrm{D}^2 \times \Rb^{n-2}) = \Omega_m\CM. 
$$
Since $\Omega_m\CM$ is monoidal, i.e. an $\EE_1$-algebra in $n\vect$, it makes perfect sense to integrate it along the 1-manifold $\partial \mathrm{D}^2$. By integrating all the obversables on the solid cylinder, we obtain 
\begin{align*}
\Obs(\mathrm{D}^2 \times \Rb^{n-2}) &= \int_{\mathrm{D}^2 \times \Rb^{n-2}} (
\Obs(\mathring{\mathrm{D}}^2 \times \Rb^{n-2}), \Obs(\partial \mathrm{D}^2 \times \Rb^{n-2})) \nn
&=\int_{\mathrm{D}^2 \times \Rb^{n-2}} ( \Omega\CD |_{\mathring{\mathrm{D}}^2 \times \Rb^{n-2}}, \,\,\, \Omega_m\CM |_{\partial \mathrm{D}^2 \times \Rb^{n-2}}) 
= (\Omega\CC, L_2^R(1_m)). 
\end{align*}

However, from the physical point of view, choosing $\Omega\CD$ and $\Omega_m\CM$ instead of the complete data $\CD$ and $(\CM,m)$ is based on certain conventions of throwing away certain topological defects. Although it is possible to make the convention or rule explicitly, it is much more convenient to simply set:
$$
\Obs(\mathring{\mathrm{D}}^2 \times \Rb^{n-2}) := \CD,
\quad\quad
\Obs(\partial \mathrm{D}^2 \times \Rb^{n-2}) := (\CM, m). 
$$
Then we have 
\be \label{eq:integral=internal_hom}
\Obs(\mathrm{D}^2 \times \Rb^{n-2}) =\int_{\mathrm{D}^2 \times \Rb^{n-2}} ( \CD |_{\mathring{\mathrm{D}}^2 \times \Rb^{n-2}}, \,\,\, (\CM,m) |_{\partial \mathrm{D}^2 \times \Rb^{n-2}}) 
= (\CC, \one_\CC, L_2^R(1_m))
=\begin{array}{c}
\begin{tikzpicture}[scale=0.5]
\fill[blue!20] (-2,0) rectangle (2,3) ;
\fill[teal!20] (0,1.5) circle (1);
\node at (-1.5,2.5) {\scriptsize $\SC^{n+1}$} ;
\node at (0,1.5) {\scriptsize $\SD^{n+1}$} ;
\node at (-1.3,1) {\scriptsize $\SM^n$} ;
\draw[blue, ultra thick,->-] (1,1.5) arc (360:0:1) ;
\node at (1.5, 0.3) {\scriptsize $L_2^R(1_m)$} ;
\end{tikzpicture}
\end{array}
\ee
where the triple $(\CC, \one_\CC, L_2^R(1_m))$, i.e., a separable $n$-category $\CC$, together with a distinguished 0-morphism $\one_\CC\in \CC$ and a distinguished 1-morphism $L_2^R(1_m)\in \Omega_{\one_\CC}\CC$, can be viewed as an $\EE_{-1}$-algebra object in $(n+1)\vect$ \cite{KWZ15}. 

\begin{rem}
Note that there are extra consistency in our convention in (\ref{eq:integral=internal_hom}). More precisely, $\CD \in \Algc_{\EE_1}((n+1)\vect)$ and $(\CM,m)\in \Algc_{\EE_0}((n+1)\vect)$. Integrating them over a 2-manifold and a 1-manifold, respectively, amount to reduce the level of commutativity by 2 and 1, respectively. As a consequence, both cases give $\EE_{-1}$-algebras after the integration. More generally, one can view 
$$
(\CA,a_1,a_2, a_3, \cdots, a_k) \in \Algc_{\EE_{-k+1}}((n+1)\vect),
$$ 
where $\CA$ is a separable $n$-category, $a_1\in \CA$ is a 0-morphism, $a_2 \in \Omega_{a_1}\CA$ is a 1-morphism, $a_3 \in \Omega_{a_2}(\Omega_{a_1}\CA)$ is a 2-morphism, $a_k \in \Omega_{a_{k-1}}(\cdots \Omega_{a_2}(\Omega_{a_1}\CA)\cdots)$ is a $(k-1)$-morphism \cite{KWZ15}. This notation is quite useful when we study the integration of $\CD$ and $(\CM,m)$ over $\mathrm{D}^k \times \Rb^{n-k}$ in next section. We suspect that our theory might be related to the so-called $\beta$-version of factorization homology \cite{AFT17,AFT16,AFR18}, our understanding of which is too limited to say anything precise.
\end{rem}

\begin{rem} \label{rem:FH} 
The notations introduced in (\ref{eq:integral=internal_hom}) are used later. Although we introduced the integral notation heuristically, it turns out that this integral has a mathematical foundation, which is called factorization homology \cite{Lur17,AF15} (see \cite{AF20} for a recent review). We hope to clarify this point in future publications. 
\end{rem}

Consider two $n+$1D anomaly-free topological orders $\SC^{n+1}, \SD^{n+1}, \SE^{n+1}$ and two gapped domain walls $\SM^n, \SN^n$ (as illustrated in the left picture in Eq.\,(\ref{pic:2wall_rolling_up})) such that 
\be \label{pic:2wall_rolling_up}
\begin{array}{c}
\begin{tikzpicture}[scale=0.6]
\fill[blue!20] (-2,0) rectangle (2,3) ;
\fill[teal!20] (0,0) rectangle (2,3) ;
\draw[blue, ultra thick, ->-] (0,0) -- (0,3) node[near start, left] {\footnotesize $\SM^n$};
\node at (-1,1.5) {\footnotesize $\SC^{n+1}$} ;
\node at (1,1.5) {\footnotesize $\SD^{n+1}$} ;
\fill[red!20] (2,0) rectangle (4,3) ;
\draw[purple, ultra thick, ->-] (2,0) -- (2,3) node[near start, right] {\footnotesize $\SN^n$};
\node at (3.3,1.5) {\footnotesize $\SE^{n+1}$} ;
\end{tikzpicture}
\end{array}
\quad \xrightarrow{\mbox{\footnotesize rolling up}} \quad 
\begin{array}{c}
\begin{tikzpicture}[scale=0.6]
\fill[blue!20] (-2,0) rectangle (2,3) ;
\fill[teal!20] (0,1.5) circle (1);
\fill[red!20] (0,1.5) circle (0.5);
\node at (-1.5,2.5) {\footnotesize $\SC^{n+1}$} ;
\draw[blue, ultra thick,->-] (1,1.5) arc (360:0:1) ;
\draw[purple, ultra thick,->-] (0.5,1.5) arc (360:0:0.5) ;
\node at (1.5, 0.3) {\scriptsize $L_2^R(L_2^R(1_n))$} ;
\end{tikzpicture}
\end{array}
\ee
\bnu
\item[-] $\SD^{n+1}$ and $\SM^n$ are obtained from $\SC^{n+1}$ by condensing the 2-codimensional topological defect $A \in \Algc_{\EE_2}(\Omega\CC)$, i.e., $\Omega\CD = \Mod_A^{\EE_2}(\Omega\CC)$ and $\Omega_m\CM=\RMod_A(\Omega\CC)$; 
\item[-] $\SE^{n+1}$ and $\SN^n$ are obtained from $\SD^{n+1}$ by condensing the 2-codimensional topological defect $B\in \Algc_{\EE_2}(\Omega\CD)$, i.e., $\Omega\CE=\Mod_B^{\EE_2}(\Omega\CD)$ and 
$\Omega_n\CN=\RMod_B(\Omega\CD)$. 
\enu
The left bulk-to-wall functors associated to $\SM^n$ and $\SN^n$ are $L_2 = -\otimes A$ or $L_2=-\otimes_A B$, respectively. By rolling up two domain walls as in (\ref{pic:2wall_rolling_up}),  we obtain a condensable $\EE_2$-algebra in $\Omega\CC$ given by $L_2^R(L_2^R(B))=B$, where $B\in \Algc_{\EE_2}(\Mod_A^{\EE_2}(\Omega\CC))$. By condensing $B\in \Omega\CC$, we obtain $\SE^{n+1}$ as the condensed phase and a gapped domain wall $\SM^n \boxtimes_{\SD^{n+1}} \SN^n$, i.e. 
\be \label{eq:composing_two_condensations}
\Omega\CE=\Mod_B^{\EE_2}(\Omega\CC), \quad\quad\quad
\Omega_{m\boxtimes_\CD n}(\CM\boxtimes_\CD\CN) \simeq \Omega_m\CM \boxtimes_{\Omega\CD} \Omega_n\CN = \RMod_B(\Omega\CC). 
\ee

\begin{rem}
Above discussion has an immediate consequence Corollary$^{\mathrm{ph}}$\,\ref{pcor:4D_Morita_mutual_2codim_condensation} as we will show later. The `$\simeq$' in 
(\ref{eq:composing_two_condensations}) is a special case of a general principle of vanishing spatial fusion anomaly \cite{KZ20,KYZ21} for 1-codimensional topological defects. We will  discuss it elsewhere. 
\end{rem}

\subsubsection{General construction} 
\label{sec:general_example_2-codim}

In this subsubsection, we provide some general constructions of the condensations of topological defects of codimension 2. 


\medskip
(1) $\SC^{n+1}=\SD^{n+1}=\mathbf{1}^{n+1}$: In this case, the gapped domain wall $\SM^n$ is precisely an anomaly-free $n$D topological order $\SM^n=(\CM,m)$, where $\CM$ is an indecomposable separable $n$-category and $m$ is a distinguished object in $\CM$ that labels the anomaly-free $n$D topological order. The indecomposable multi-fusion $(n-1)$-category $\Omega_m\CM$ is the category of topological defects in the $n$D topological order labeled by $m$. All objects in $\CM$ label anomaly-free $n$D topological orders that are all Morita equivalent. Mathematically, $\Omega_m\CM$ and $\Omega_{m'}\CM$ are Morita equivalent for $m,m'\in \CM$, and we have
$\CM=\Sigma\Omega_m\CM$. If we further assume that $m$ is simple, then $\Omega_m\CM$ is a non-degenerate fusion $(n-1)$-category and $\Omega_m^2\CM$ is a non-degenerate braided fusion $(n-2)$-category, i.e., $\FZ_2(\Omega_m^2\CM)\simeq (n-2)\vect$. Moreover, we have
$$
\Omega_m\CM = \Sigma \Omega_m^2\CM = \RMod_{\Omega_m^2\CM}((n-1)\vect),
$$
which provides a coordinate system for $\Omega_m\CM$ that are very useful for computation. 

\medskip
Recall that all 2-codimensional topological defects in $\mathbf{1}^{n+1}$ form the trivial braided fusion $(n-1)$-category $(n-1)\vect$. In this case, note that the first level of the left bulk-to-wall functor $L_1: \CC \to \CM$ is defined by $-\odot m$, where $\odot$ denotes the natural $\CC$-action (of 1-codimensional topological defects) on $\CM$. We also have the second level of left bulk-to-wall map that maps 2-codimensional topological defects in $\mathbf{1}^{n+1}$ to $\SM^n$ defined as follows: 
\begin{align*}
(n-1)\vect \quad &\xrightarrow{L_2=-\boxtimes \Omega_m^2\CM} \quad \Omega_m\CM=\RMod_{\Omega_m^2\CM}((n-1)\vect) \\
a \quad\quad\quad\quad &\quad\quad \mapsto \quad\quad\quad a \odot 1_m = a \odot \Omega_m^2\CM.  
\end{align*}
Its physical meaing is illustrated below. 
$$
\begin{array}{c}
\begin{tikzpicture}[scale=0.8]
\draw[blue!20,fill=blue!20] (0,0)--(0,2)--(1,3)--(1,1)--cycle ;
\node at (3,1.5) {\footnotesize $\Omega_m\CM=\RMod_{\Omega_m^2\CM}((n-1)\vect)$} ;
\node at (-1.5,2.5) {\footnotesize $\mathbf{1}^{n+1}$} ;
\node at (3.5,2.5) {\footnotesize $\mathbf{1}^{n+1}$} ;
\node at (-2,0.8) {\scriptsize $(n-1)\vect \xrightarrow{L_2 = -\otimes 1_m} \Omega_m\CM$} ;
\node at (2.5,0.6) {\scriptsize $1_m = \Omega_m^2\CM \in \RMod_{\Omega_m^2\CM}((n-1)\vect)$} ; 
\end{tikzpicture}
\end{array}
$$
The right adjoint $L_2^R$ of $L_2$ is precisely the forgetful functor $\forget: \RMod_{\Omega_m\CM}((n-1)\vect) \to (n-1)\vect$. Therefore, we obtain $L_2^R(1_m) = \Omega_m^2\CM \in \Algc_{\EE_2}((n-1)\vect)$. Moreover, by Theorem$^{\mathrm{ph}}$\,\ref{conj:Lagrangian=boundary}, $L_2^R(1_m)$ is a Lagrangian algebra in $(n-1)\vect$, i.e. $\Mod_{\Omega_m\CM}^{\EE_2}((n-1)\vect) \simeq (n-1)\vect$. This result can be reformulated as a mathematical result. 

\begin{prop} \label{cor:condense_nDTO_trivial_phase_II}
A non-degenerate braided fusion $n$-category $\CA$ is automatically a Lagrangian algebra in $(n+1)\vect$ (recall Definition\,\ref{def:non-degenerate_Lagrangian}), i.e. 
\[
\Mod_\CA^{\EE_2}((n+1)\vect) \simeq (n+1)\vect. 
\]
\end{prop}

\begin{rem}
When $n=2$, the above result was proved in mathematical literature \cite{DN21,JF22,DX24}. Our results provide a physical meaning to this $n=2$ mathematical result and naturally generalize it to all dimensions. 
\end{rem}

\begin{rem} \label{rem:rolling_up_cylinder}
Recall Proposition\,\ref{cor:condense_nDTO_trivial_phase} and Remark\,\ref{rem:condensing_afTO}, Proposition\,\ref{cor:condense_nDTO_trivial_phase_II} is the second mathematical reformulation of the fact that condensing an anomaly-free $n$D topological order in the trivial $n+$1D topological order $\mathbf{1}^{n+1}$ reproduces $\mathbf{1}^{n+1}$. More precisely, 
\bnu
\item When an anomaly-free $n$D simple topological order $\SM^n$ is viewed as a 1-codimensional topological defect in $\mathbf{1}^{n+1}$, we can condense it to obtain $\mathbf{1}^{n+1}$ as the condensed phase and a gapped domain wall,  which is nothing but $\SM^n$ characterized by the pair $(\CM,m)$. Mathematically, it amounts to condense the simple condensable $\EE_1$-algebra $\Omega_m\CM$ in $n\vect$. 

\item When we rollup an anomaly-free $n$D topological order $\SM^n$ in one direction, it becomes a cylinder $S^1\times \Rb^{n-2}$ in spatial $n$-dimensions $\Rb^n$. By shrinking the size of $S^1$, it becomes a 2-codimensional topological defects in $\mathbf{1}^{n+1}$. 
$$
\begin{array}{c}
\begin{tikzpicture}[scale=0.8]
\draw[blue!20,fill=blue!20] (0,0)--(0,2)--(1,3)--(1,1)--cycle ;
\node at (0.7,2.1) {\footnotesize $\SM^n$} ; 
\node at (0.5,1.5) {\scriptsize $\Omega_m\CM$} ; 
\node at (-1,2.5) {\footnotesize $\mathbf{1}^{n+1}$} ;
\node at (2.5,2.5) {\footnotesize $\mathbf{1}^{n+1}$} ;
\end{tikzpicture}
\end{array}
\quad \xrightarrow{\mbox{\scriptsize rolling up}} \quad 
\begin{array}{c}
\begin{tikzpicture}
\draw[blue!20] (1,0)--(1,2) ;
\fill[blue!25,opacity=0.7] (0,0)--(0,2) .. controls (0,2.3) and (1,2.3) .. (1,2)--(1,0) .. controls (1,0.3) and (0,0.3) .. cycle ;
\draw[blue!50] (0,2) .. controls (0,2.3) and (1,2.3) .. (1,2) ;
\draw[dashed,blue!50] (0,0) .. controls (0,0.3) and (1,0.3) .. (1,0) ;
\node at (0.5,2) {\footnotesize $\SM^n$} ;
\fill[blue!25,opacity=0.7] (0,0)--(0,2) .. controls (0,1.7) and (1,1.7) .. (1,2)--(1,0) .. controls (1,-0.3) and (0,-0.3) .. cycle ;
\draw[blue!50] (0,2) .. controls (0,1.7) and (1,1.7) .. (1,2) ;
\draw[blue!50] (0,0) .. controls (0,-0.3) and (1,-0.3) .. (1,0) ;
\end{tikzpicture}
\end{array}
= \Omega_m^2\CM \in (n-1)\vect. 
$$
As an object in $(n-1)\vect$, it is precisely the non-degenerate braided fusion $(n-2)$-category $\Omega_m^2\CM$, which is automatically a Lagrangian $\EE_2$-algebra in $(n-1)\vect$. 
By condensing this 2-codimensional topological defects, we obtain $\mathbf{1}^{n+1}$ as the condensed phase.

\enu
Note that the second reformulation is possible because $\Omega_m^2\CM$ gives a complete characterization of all topological defects $\SM^n$ up to condensation descendants. 
\end{rem}


(2) $\SC^{n+1}=\mathbf{1}^{n+1}\neq \SD^{n+1}$: 
We have seen that condensing a Lagrangian algebra $\CA$ in $(n-1)\vect$, i.e. a non-degenerate braided fusion $(n-2)$-category, produces the trivial phase $\mathbf{1}^{n+1}$ as the condensed phase. If $\CA$ is not Lagrangian, i.e., a braided fusion $(n-2)$-category, it is precisely a simple condensable $\EE_2$-algebra in $(n-1)\vect$. By condensing $\CA$ in $(n-1)\vect$, we obtain a condensed phase $\SD^{n+1}$ and a gapped domain wall $\SM^n=(\CM,m)$ such that 
\be \label{eq:D_OmegaM_2codim}
\Omega\CD \simeq \Mod_\CA^{\EE_2}((n-1)\vect) \quad \mbox{and} \quad \Omega_m\CM\simeq \RMod_\CA((n-1)\vect) = \Sigma\CA. 
\ee
Note that we have $\CD=\Sigma\Omega\CD$ and 
$\CM=\Sigma\Omega_m\CM=\Sigma^2\CA$. 
By the boundary-bulk relation, we obtain
\begin{align}
\Mod_\CA^{\EE_2}((n-1)\vect) &\simeq \FZ_1(\RMod_\CA((n-1)\vect))^\op,  \nn
\CD =\Sigma\Mod_\CA^{\EE_2}((n-1)\vect) &\simeq \FZ_0((\CM,m)) = \Fun(\Sigma^2\CA,\Sigma^2\CA) \simeq \BMod_{\Sigma\CA|\Sigma\CA}(n\vect). 
\end{align}
Moreover, we have 
$L^R(1_m) = L^R(\CA) = [\CA,\CA]_{(n-1)\vect} = \CA$, or equivalently, 
\be \label{eq:integral=internal_hom_2}
\int_{\mathrm{D}^2 \times \Rb^{n-2}} ( \CD |_{\mathring{\mathrm{D}}^2 \times \Rb^{n-2}}, \,\,\, (\CM,m) |_{\partial \mathrm{D}^2 \times \Rb^{n-2}}) 
= (n\vect, \one, \CA). 
\ee

One can see immediately that $\SD^{n+1}$ can be obtained from $\SC^{n+1}$ by condensing 1-codimensional topological defect $\Sigma\CA$, which is a condensable $\EE_1$-algebra in $(n+2)\vect$. We restudy this phenomenon in details in Section\,\ref{sec:condense_k-codim_defect}. We summarize above results as a mathematical result. 
\begin{thm} \label{thm:2codim_condensations_in_trivial_phase}
For $m\geq 1$, a simple condensable $\EE_2$-algebra in $(m+1)\vect$ is precisely a braided fusion $m$-category $\CA$. We have 
\begin{align} \label{eq:E2_Z1_LMod}
\Mod_\CA^{\EE_2}((m+1)\vect)^\op &\simeq \FZ_1(\RMod_\CA((m+1)\vect))\simeq\FZ_1(\Sigma\CA), \\
\Sigma\Mod_\CA^{\EE_2}((m+1)\vect)^\op &\simeq \Fun(\Sigma^2\CA,\Sigma^2\CA) \simeq \BMod_{\Sigma\CA|\Sigma\CA}((m+2)\vect). 
\end{align}
\end{thm}

\begin{expl} \label{expl:condense_nRep(G)_from_1_to_GT}
Let $\CA=n\Rep(G)$ viewed as a condensable $\EE_2$-algebra in $(n+1)\vect$, which is the category of topological defects of codimension 2 and higher in $\mathbf{1}^{n+3}$. By condensing $n\Rep(G)$ in $(n+1)\vect$, we obtain 
$$
\Mod_{n\Rep(G)}^{\EE_2}((n+1)\vect) \simeq \FZ_1((n+1)\Rep(G)). 
$$
Therefore, we obtain $\SG\ST_G^{n+3}$ as the condensed phase. 
\end{expl}


\begin{rem} \label{rem:2codim_cond_4D_special}
The $n=3$ case of (\ref{eq:E2_Z1_LMod}) was first proved in \cite[Theorem\,4.11]{DN21}. In this case, D\'{e}coppet proved that every fusion 2-category $\CB$ is Morita equivalent to a connected one, i.e., $\Sigma\CA$ for a braided fusion 1-category $\CA$ \cite{Dec23b}. As a consequence, every non-chiral 3+1D topological order can be obtained by condensing a simple condensable $\EE_2$-algebra in $2\vect$, or equivalently, by condensing a 2-codimensional topological defect in $\mathbf{1}^4$. However, it is not true that all gapped boundaries of a non-chiral 4D topological order can be obtained in this way\footnote{All gapped boundaries of a non-chiral 4D topological order can be obtained from $\mathbf{1}^4$ by condensing a 2-codimensional topological defect followed by a condensation on the boundary. This combined process was called a two-step condensation in \cite{Kon14e}.}. In general, a simple gapped boundary $\SM^n$ of a simple non-chiral $n+$1D topological order $\SD^{n+1}$ can be obtained from $\SC^{n+1}=\mathbf{1}^{n+1}$ via the condensation of a topological defect of codimension 2 if and only if the fusion $(n-1)$-category $\Omega_m\CM$ is connected as a separable 2-category, i.e., $\Omega_m\CM \simeq \Sigma\CA$ for a braided fusion $(n-2)$-category. Moreover, we do not expect that D\'{e}coppet's result generalizes to fusion $k$-categories for $k\geq 3$. The $k=3$ case is special because 3+1D topological orders are very rare \cite{LKW18,LW19,JF20a}\footnote{By topological Wick rotation \cite{KZ18b,KZ20,KZ21}, this fact also suggests that 2+1D rational CFT's are rare comparing to 1+1D CFT's.}. 
\end{rem}

\begin{rem}
Similar to Section\,\ref{sec:general_example_1-codim}, one can also discuss general examples of the condensation of 2-codimensional topological defects. We postpone it to Section\,\ref{sec:general_example_k-codim} as a special case of the condensations of $k$-codimensional topological defects. 
\end{rem}

(3) $\SC^{n+1}$ is non-trivial and non-chiral: The following result provides a physical way to construct Lagrangian $\EE_2$-algebras in $\Omega\CC$ from a single one. It is an immediate consequence of Theorem$^{\mathrm{ph}}$\,\ref{pthm:recover_all_bdy_from_one}. 
\begin{pthm} \label{pthm:3-step_process}
Given an anomaly-free non-chiral simple $n+$1D topological order $\SC^{n+1}$ and an anomaly-free $n$D topological order $\SN^n$. If $A$ is a Lagrangian $\EE_2$-algebra in $\Omega\CC$, then $A\boxtimes \Omega\CN \in \Omega\CC \boxtimes (n-1)\vect \simeq \Omega\CC$ is also a Lagrangian $\EE_2$-algebra in $\Omega\CC$. Moreover, we can obtain all Lagrangian $\EE_2$-algebras in $\Omega\CC$ from a single one by the following three-step process: 
\bnu

\item[(1)] First, stacking an anomaly-free $n$D topological order $\SN^n$ to a given gapped boundary $\SX^n$ of $\SC^{n+1}$, we obtain a new gapped boundary $\SX^n \boxtimes \SN^n$ of $\SC^{n+1}$; 

\item[(2)] Secondly, condense $\SX^n \boxtimes \SN^n$ to a new but Morita equivalent gapped boundary $\SY^n$; 

\item[(3)] Thirdly, rolling up the boundary $\SY^n$ to a 2-codimensional topological defect $A_\SY$ in $\SC^{n+1}$, it gives a Lagrangian $\EE_2$-algebra in $\Omega\CC$. 

\enu  
All Lagrangian $\EE_2$-algebras in $\Omega\CC$ arise in this way from a single gapped boundary $\SX^n$. 
\end{pthm}

\begin{rem}
One has to treat the roll-up process with caution. It is possible to switch the order the second step and the third step, i.e., obtaining $A_Y$ from $A_{\SX\boxtimes \SN}$ by a condensation restricted on the 2-codimensional topological defect. 
It is because the roll-up process indeed maps a gapped domain wall between $\SX^n \boxtimes \SN^n$ and $\SY^n$ to a gapped domain wall between $A_{\SX\boxtimes \SN}$ and $A_\SY$. However, such obtained 
condensations on $A_{\SX\boxtimes \SN}$ belongs to a very limited sub-family of all possible condensations. A generic condensations on $A_{\SX\boxtimes \SN}$ produces a condensation descendant, which is not necessarily an algebra. Also notice that the roll-up process sometimes maps a gapless domain wall between two gapped boundaries to a gappable domain wall between two 2-codimensional topological defects. For example, by rolling up a 1+1D anomaly-free rational CFT, we obtain a gapless 0+1D defect in $\mathbf{1}^3$ that is gappable. 
\end{rem}

We recall a well-known result.
\begin{thm}[\cite{KR08,DMNO13}]
For a fusion 1-category $\CA$, there is a one-to-one correspondence between Lagrangian $\EE_2$-algebras in $\FZ_1(\CA)$ and indecomposable left $\CA$-modules, or equivalently, the Morita classes of indecomposable condensable $\EE_1$-algebras in $\CA$. Moreover, given an indecomposable condensable $\EE_1$-algebra $A$ in $\CA$, the associated Lagrangian $\EE_2$-algebra in $\FZ_1(\CA)$ is the full center of $A$. 
\end{thm}

\begin{expl}
On the one hand, Lagrangian $\EE_2$-algebras in $\FZ_1(\Rep(G))$ are classified by a pair $(H,\omega)$, where $H$ is a subgroup of $G$ and $\omega \in Z^2(H,\Cb^\times)$. On other other hand, left indecomposable $\Rep(G)$-modules are also classified by the same pairs $(H,\omega)$. 
\end{expl}

We have a mathematical reformulation of Corollary$^{\mathrm{ph}}$\,\ref{pcor:4D_gapped boundary_classification} in terms of Lagrangian $\EE_2$-algebras as shown below. It is a generalization of above result for the $n=1$ case and the $n=2$ case \cite{JFR23,DX24} to all $n>2$. 
\begin{thm} \label{pcor:4D_Lagrangian_1to1_modules} 
Let $\CT$ be a non-degenerate fusion $n$-category $\CT$, i.e., $\CT=\Sigma\Omega\CT$ and $\Omega\CT$ is a non-degenerate braided fusion $(n-1)$-category. 
For an indecomposable multi-fusion $n$-category $\CB$, there are one-to-one correspondences among the following three sets:
\begin{itemize}

\item[(1)] Lagrangian $\EE_2$-algebras in $\FZ_1(\CB)$ (up to isomorphisms) in the 2-Morita class $[\CB \boxtimes \CT]$ (recall Theorem$^{\mathrm{ph}}$\,\ref{conj:Lagrangian=boundary}); 

\item[(2)] left indecomposable $\CB\boxtimes \CT$-modules modulo the equivalence relation: $\CM \sim \CN$ if and only if $\CN \simeq \CM\boxtimes \CV$ for $\CV \in (n+1)\vect^\times$. 

\item[(3)] indecomposable condensable $\EE_1$-algebras in $\CB\boxtimes \CT$ modulo the equivalence relation: $A \sim B$ if $B$ is Morita equivalent to $A\boxtimes L \in (\CB\boxtimes \CT)\boxtimes n\vect \simeq \CB\boxtimes \CT$ for a Lagrangian $\EE_1$-algebra $L$ in $n\vect$ (i.e., $L$ is a non-degenerate multi-fusion $(n-1)$-category).


\end{itemize}
\end{thm}

\begin{rem}
Similar to $n=1$ case, we expect that, for an indecomposable condensable $\EE_1$-algebra $A$ in $\CB\boxtimes \CT$, the associated Lagrangian $\EE_2$-algebra in $\FZ_1(\CB)$ is given by the full center of $A$. The notion of the full center of an algebra in a fusion $n$-category can be defined by the universal property generalizing that of the full center of an algebra in fusion 1-categories \cite{Dav10, KYZ21}. We will introduce this notion as a special case of the $\EE_k$-center of an $\EE_k$-algebra in an $\EE_k$-monoidal higher category in \cite{KZZZ25}. 
\end{rem}


Now the following result provide a physical meaning to the classification result in Corollary$^{\mathrm{ph}}$\, \ref{cor:A_connected_classification_bdy_Z(A)}. 
\begin{pthm} \label{pthm:iterate_E2_condensation}
For $n\geq 2$ and a braided fusion $(n-1)$-category $\CA$, a condensable $\EE_2$-algebra in $\FZ_1(\Sigma\CA)^\op \simeq \Mod_\CA^{\EE_2}(n\vect)$ (recall (\ref{eq:E2_Z1_LMod})) is precisely given by a braided multi-fusion $(n-1)$-category $\CB$ equipped with a braided monoidal functor $\CA\to \CB$. The condensable $\EE_2$-algebra $\CB$ is simple if and only if $\CB$ is fusion, and it is Lagrangian if and only if $\CB$ is non-degenerate. The 2-Morita class of the Lagrangian $\EE_2$-algebra is determined by the Morita class of $\Sigma\CB$. 

Moreover, by condensing $\CB$ in $\FZ_1(\Sigma\CA)^\op$, where $\CB$ is viewed as a condensable defect of codimension 2 in the anomaly-free topological order $\SC^{n+1}$ such that $\CC=\FZ_0(\Sigma^2\CA)^\op\simeq \Mod_{\Sigma\CA}^{\EE_1}((n+1)\vect)$ and $\Omega\CC = \FZ_1(\Sigma\CA)^\op$ as illustrated below, 
$$
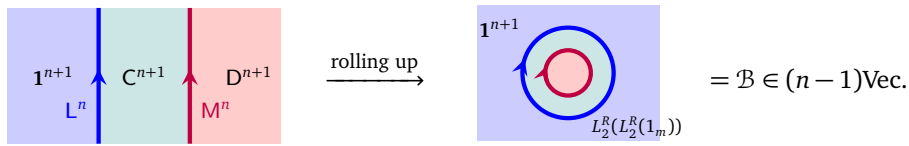
\begin{array}{c}
\begin{tikzpicture}[scale=0.6]
\fill[blue!20] (-2,0) rectangle (2,3) ;
\fill[teal!20] (0,0) rectangle (2,3) ;
\draw[blue, ultra thick, ->-] (0,0) -- (0,3) node[near start, left] {\footnotesize $\SL^n$};
\node at (-1,1.5) {\footnotesize $\mathbf{1}^{n+1}$} ;
\node at (1,1.5) {\footnotesize $\SC^{n+1}$} ;
\fill[red!20] (2,0) rectangle (4,3) ;
\draw[purple, ultra thick, ->-] (2,0) -- (2,3) node[near start, right] {\footnotesize $\SM^n$};
\node at (3.3,1.5) {\footnotesize $\SD^{n+1}$} ;
\end{tikzpicture}
\end{array}
\quad \xrightarrow{\mbox{\footnotesize rolling up}} \quad 
\begin{array}{c}
\begin{tikzpicture}[scale=0.6]
\fill[blue!20] (-2,0) rectangle (2,3) ;
\fill[teal!20] (0,1.5) circle (1);
\fill[red!20] (0,1.5) circle (0.5);
\node at (-1.5,2.5) {\footnotesize $\mathbf{1}^{n+1}$} ;
\draw[blue, ultra thick,->-] (1,1.5) arc (360:0:1) ;
\draw[purple, ultra thick,->-] (0.5,1.5) arc (360:0:0.5) ;
\node at (1.5, 0.3) {\scriptsize $L_2^R(L_2^R(1_m))$} ;
\end{tikzpicture}
\end{array}
=\CB \in (n-1)\vect. 
$$
we obtain a condensed phase $\SD^{n+1}$ and a gapped domain wall $\SM^n=(\CM,m)$ such that 
\begin{align}
&\CD\simeq \Mod_{\Sigma\CB}^{\EE_1}(\Mod_{\Sigma\CA}^{\EE_2}((n+1)\vect)), 
\quad\quad (\CM,m) = (\BMod_{\Sigma\CA|\Sigma\CB}((n+1)\vect)), \Sigma\CB), 
\nonumber \\
&\Omega\CD \simeq \Mod_\CB^{\EE_2}(\Mod_\CA^{\EE_2}(n\vect)) \simeq \Mod_\CB^{\EE_2}(n\vect) \simeq \FZ_1(\Sigma\CB)^\op, \nonumber \\
&\Omega_m\CM = \Fun_{\Sigma\CA|\Sigma\CB}(\Sigma\CB,\Sigma\CB) \simeq \RMod_\CB(\Mod_\CA^{\EE_2}(n\vect)). \label{eq:RMod_B_Mod_AE2}
\end{align}
Moreover, set $\SL^n=(\CL,l) = (\Sigma^2\CA, \Sigma\CA)$. We have 
\begin{align}
\Omega_l\CL \boxtimes_{\Omega\CC} \Omega_m\CM &= \Sigma\CA \boxtimes_{\Mod_\CA^{\EE_2}(n\vect)} \RMod_\CB(\Mod_\CA^{\EE_2}(n\vect)) \simeq \Sigma\CB. \label{eq:LCM_SigmaA} \\
\CL \boxtimes_{\CC} \CM &= \Sigma^2\CA \boxtimes_{\Sigma\Mod_\CA^{\EE_2}(n\vect)} \Sigma(\RMod_\CB(\Mod_\CA^{\EE_2}(n\vect))) \simeq \Sigma^2\CB. \nonumber
\end{align}
\end{pthm}

\begin{rem}
The $n=2$ case of Theorem$^{\mathrm{ph}}$\,\ref{pthm:iterate_E2_condensation} was obtained in \cite{JFR23,DX24,Xu24}. When $\CA=\Rep(G)$, this braided functor necessarily $\Rep(G) \to \CB$ factors through $\Rep(G) \xrightarrow{\forget} \Rep(H) \hookrightarrow \CB$ \cite{DMNO13}. This result also says that non-minimal non-degenerate extensions of $\Rep(H)$ also have physical meanings. 
\end{rem}


\begin{rem}
Note that Theorem$^{\mathrm{ph}}$\,\ref{pthm:iterate_E2_condensation} of a 2-codimensional defect condensation is an analogue of Theorem$^{\mathrm{ph}}$\,\ref{pthm:construct_cond_E1_algebras} of a 1-codimensional defect condensation, and can be generalized further to higher codimensional defect condensations (see Theorem$^{\mathrm{ph}}$\,\ref{pthm:iterate_Ek_condensation}). 
\end{rem}

\subsubsection{Witt equivalence}

\begin{pprop} 
If two anomaly-free $n+$1D topological order $\SC^{n+1}$ and $\SD^{n+1}$ are Morita equivalent (i.e., admiting a gapped anomaly-free domain wall), then there exists the third anomaly-free $n+$1D topological order $\SA^{n+1}$ such that $\SC^{n+1}$ and $\SD^{n+1}$ can be obtained from $\SA^{n+1}$ via two different condensations of 2-codimensional topological defects.
\end{pprop}
\pf
It follows from above Lemma by chossing $\SA^{n+1}=\SC^{n+1} \boxtimes \SZ(\SP)^{n+1} = \SD^{n+1} \boxtimes \SZ(\SQ)^{n+1}$. Note that $A_\CP=[\one_\CP, \one_\CP]_{\FZ_1(\CP)}$ is a Lagrangian $\EE_2$-algebra in $\FZ_1(\CP)$, and $A_\CQ=[\one_\CQ, \one_\CQ]_{\FZ_1(\CQ)}$ is a Lagrangian $\EE_2$-algebra in $\FZ_1(\CQ)$. By condensing the condensable $\EE_2$-algebra $A_1=1_{\one_\CC} \boxtimes A_\CP \in \Omega\CC \boxtimes \FZ_1(\CP)$, we obtain $\SC^{n+1}$ as the condensed phase. By condensing the condensable $\EE_2$-algebra $A_2=1_{\one_\CD} \boxtimes A_\CQ \in \Omega\CD \boxtimes \FZ_1(\CQ)$, we obtain $\SD^{n+1}$ as the condensed phase. 
\epf

\begin{rem}
When $n=2$, it was proved in \cite{DMNO13} there exists an anomaly-free $n+$1D topological order $\SB^{n+1}$
such that $\SB^{n+1}$ can be obtained from both $\SC^{n+1}$ and $\SD^{n+1}$ via condensations of 2-codimensional topological defects. It is natural to ask if it is true for $n>2$ (see a related question in Remark\,\ref{rem:anisotropic}). 
\end{rem}

The next result was first obtained in \cite{Xu24}. 
\begin{pcor} \label{pcor:4D_Morita_mutual_2codim_condensation}
Two Morita equivalent anomaly-free topological order $\SC^{3+1}$ and $\SD^{3+1}$ can be obtained from each other by condensing 2-codimensional topological defects. 
\end{pcor}
\pf
This is an immediate consequence of Proposition$^\mathrm{ph}$\,\ref{plem:2way_Witt_eq_phases}, Remark\,\ref{rem:2codim_cond_4D_special} and discussion around (\ref{pic:2wall_rolling_up}). 
\epf

\begin{rem}
Similar to Remark\,\ref{rem:2codim_cond_4D_special}, not every gapped domain wall between $\SC^4$ and $\SD^4$ can be obtained from condensing a 2-codimensional topological defect in either $\SC^4$ or $\SD^4$. 
\end{rem}


The following notion is a direct generalization of that for 1-categories introduced in \cite{DMNO13}.
\begin{defn} \label{def:Witt_equivalence}
Two non-degenerate braided fusion $n$-categories $\CA$ and $\CB$ are called {\it Witt equivalent} if there exist two fusion $n$-categories $\CP$ and $\CQ$ and a braided equivalence: 
$$
\CA \boxtimes \FZ_1(\CP) \simeq \CB \boxtimes \FZ_1(\CQ). 
$$
\end{defn}

The physical result Proposition$^{\mathrm{ph}}$\,\ref{plem:2way_Witt_eq_phases} immediately implies the following mathematical result. 
\begin{cor}
Two non-degenerate braided fusion $n$-categories $\CA$ and $\CB$ are Witt equivalent if and only if there exists a fusion $n$-category $\CM$ and braided equivalence: 
$$
\CA \boxtimes \CB^\rev \simeq \FZ_1(\CM). 
$$
\end{cor}

\begin{cor} \label{cor:A-B_from_C}
Two non-degenerate braided fusion $n$-categories $\CA$ and $\CB$ are Witt equivalent if and only if there exists a third non-degenerate braided fusion $n$-category $\CC$ and $A,B\in \Algc_{\EE_2}(\CC)$ such that 
$$
\CA \simeq \Mod_A^{\EE_2}(\CC) \quad\quad \mbox{and} \quad\quad 
\CB \simeq \Mod_B^{\EE_2}(\CC). 
$$
\end{cor}

\begin{prop}
The notion of Witt equivalence is clearly an equivalence relation. We denote the Witt equivalence class of $\CC$ by $[\CC]$. The Witt equivalence classes of non-degenerate braided fusion $n$-categories clearly form a group, called the Witt group and denoted by $\mathrm{Witt}^n$, with the multiplication defined by $[\CC] \star [\CD] := [\CC \boxtimes \CD]$, the unit defined by $n\vect$ and the inverse given by $[\CC]^{-1}=[\CC^\op]$. 
\end{prop}

\begin{rem}
It was known that the map $[\CC] \mapsto \Sigma^2\CC$ defines a group isomorphism from the Witt group $\mathrm{Witt}^n$ to the group $(n+3)\vect^\times$ of invertible separable $(n+2)$-categories \cite{JF22,KZ24}. 
\end{rem}

We generalize the notion of a completely anisotropic non-degenerate braided fusion 1-category \cite{DMNO13} to higher categories. 
\begin{defn}
A non-degenerate braided fusion $n$-category $\CB$ is called {\it completely anisotropic} if every condensable $\EE_2$-algebra in $\CB$ is isomorphic to $\one_\CB \boxtimes A \in \CB \boxtimes n\vect \simeq \CB$, where $A$ is a condensable $\EE_2$-algebra in $n\vect$. 
\end{defn}

\begin{rem} \label{rem:anisotropic}
An obvious question is if there exists a unique completely anisotropic non-degenerate braided fusion $n$-category in each Witt equivalence class as in the $n=1$ case \cite{DMNO13}. We hope to explore this question in the future. 
\end{rem}

\subsubsection{Examples in 3+1D} \label{sec:example_2codim_4D}

We have already seen the condensations of 2-codimensional topological defects in $\mathbf{1}^4$. In this subsubsection, we provide some examples in 3+1D $\Zb_2$-gauge theory.

\begin{expl} \label{expl:Ae+A1+A2_4D_TC}
Consider 3+1D $\Zb_2$ gauge theory $\CG\CT_{\Zb_2}^4$. Then $\Omega\CG\CT_{\Zb_2}^4$ is 
a nondegenerate braided fusion 2-category and we have $\Omega\CG\CT_{\Zb_2}^4\simeq \FZ_1(2\Rep(\Zb_2)) \simeq \FZ_1(2\vect_{\Zb_2})$. This braided fusion 2-category was studied in \cite{KTZ20} and via the 3+1D toric code model in \cite{KTZ20a}. We use the notation in \cite{KTZ20a,ZLZHKT23}. There are 4 simple objects $\one,\one_c,m,m_c$ in $\Omega\CG\CT_{\Zb_2}^4$ with the fusion rule
\[
m \otimes m \simeq \one , \quad \one_c \otimes \one_c \simeq \one_c \oplus \one_c , \quad m \otimes \one_c \simeq \one_c \otimes m \simeq m_c .
\]
The hom categories are depicted in the following diagram:
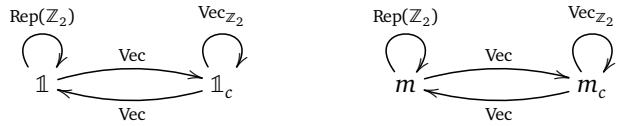
\[
\xymatrix{
\one \ar@(ul,ur)[]^{\Rep(\Zb_2)}  \ar@/^/[rr]^{\vect} & & \one_c \ar@(ul,ur)[]^{\vect_{\Zb_2}} \ar@/^/[ll]^{\vect}
& & m \ar@(ul,ur)[]^{\Rep(\Zb_2)}  \ar@/^/[rr]^{\vect} & & m_c \ar@(ul,ur)[]^{\vect_{\Zb_2}} \ar@/^/[ll]^{\vect}
}
\]
We also denote the simple objects in the hom categories in the following diagram:
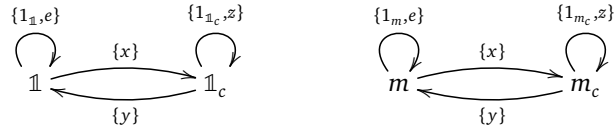
\[
\xymatrix{
\one \ar@(ul,ur)[]^{\{1_\one,e\}}  \ar@/^/[rr]^{\{x\}} & & \one_c \ar@(ul,ur)[]^{\{1_{\one_c},z\}} \ar@/^/[ll]^{\{y\}}
& & m \ar@(ul,ur)[]^{\{1_m,e\}}  \ar@/^/[rr]^{\{x\}} & & m_c \ar@(ul,ur)[]^{\{1_{m_c},z\}} \ar@/^/[ll]^{\{y\}}
}
\]
The composition rules are given by $e \circ e = 1_\one, z \circ z = 1_{\one_c}, x \circ y = 1_{\one_c} \oplus z$ and $y \circ x = 1_\one \oplus e$.

We review three examples of Lagrangian $\EE_2$-algebras in $\Omega\CG\CT_{\Zb_2}^4$ constructed in \cite{ZLZHKT23}.
\bnu
\item $A_e=\one_c$: Its multiplication morphism is given by 
\[
\one_c \otimes \one_c = \one_c  \oplus \one_c \xrightarrow{1_{\one_c} \oplus 0} \one_c.
\]
and the unit $x \colon \one \to \one_c$, together with the identity 2-associators, 2-unitors and 2-commutators is a Lagrangian algebra. There are two simple modules $\one_c$ and $m_c$, and only $\one_c$ itself is a local module. Condensing $\one_c$ leads to the rough boundary of the 3+1D toric code model. The topological defects on the boundary form a fusion 2-category $2\vect_{\Zb_2}$. In the light of Theorem$^{\mathrm{ph}}$\,\ref{pthm:iterate_E2_condensation}, $\Omega\CG\CT_{\Zb_2}^4$ is equipped with another coordinate system:
$$
\Omega\CG\CT_{\Zb_2}^4 \simeq \FZ_1(\Sigma\Rep(G)) \simeq \Mod_{\Rep(G)}^{\EE_2}(2\vect)^\op. 
$$
In this coordinate system, the Lagrangian $\EE_2$-algebra $A_e$ is the one associated to the forgetful functor $\Rep(\Zb_2) \to \vect$. 

\item $A_m=\one \oplus m$: The algebra $A_m$ has the multiplication 1-morphism defined component-wise by
\[
\begin{array}{c}
\begin{tikzcd}
\one \otimes \one \ar[d,"1_\one"] \\
\one
\end{tikzcd}
\end{array}
\begin{array}{c}
\begin{tikzcd}
\one \otimes m \ar[d,"1_{m}"] \\
m
\end{tikzcd}
\end{array}
\begin{array}{c}
\begin{tikzcd}
m \otimes \one \ar[d,"1_{m}"] \\
m
\end{tikzcd}
\end{array}
\begin{array}{c}
\begin{tikzcd}
m \otimes m \ar[d,"1_{\one}"] \\
\one
\end{tikzcd}
\end{array}
\]
and the unit 1-morphism $\one \xrightarrow{1 \oplus 0} \one \oplus m$. The 2-associator and 2-unitors are identity 2-morphisms. The 2-commutator is trivial on all components except $\beta_{m,m}=\pm 1$. Two different choices of the 2-commutator define two commutative algebra structures, but they are equivalent to each other. There are two simple $A_m$-modules $A_m$ and $\one_c \otimes A_1$, and only $A_1$ itself is a local module. Condensing $A_m$ leads to the smooth boundary of the 3+1D toric code model. The topological defects on the boundary form a fusion 2-category $2\Rep(\Zb_2)$. In the light of 
Theorem$^{\mathrm{ph}}$\,\ref{pthm:iterate_E2_condensation},

this Lagrangian $\EE_2$-algebra is the one associated to the trivial minimal modular extension $\Rep(\Zb_2) \to \FZ_1(\Rep(\Zb_2))$. 

\item $A_m^{\mathrm{tw}}=(1\oplus m)^{\mathrm{tw}}$: There is another Lagrangian algebra structure on $\one \oplus m$, denoted by $A_m^{\mathrm{tw}}$. It has the same multiplication 1-morphism and unit 1-morphism as those of $A_m$. The 2-associator has only one nontrivial component:
\[
\xymatrix{\begin{array}{c}
\begin{tikzpicture}[scale = 0.4]
\draw[black,thick](0,0)--(1,1)--(1,2);
\draw[black,thick](0,0)--(-1,1)--(-1,1.5)--(-1.5,2);
\draw[black,thick](-1,1.5)--(-0.5,2);
\draw[black,thick](0,0)--(0,-1);
\node[black] at (-1.5,2.3) {\scriptsize $m$};
\node[black] at (-0.5,2.3) {\scriptsize $m$};
\node[black] at (1,2.3) {\scriptsize $m$};
\end{tikzpicture}
\end{array}
\ar@{=>}[rr]^{\alpha_{m,m,m}=-1} 
&&
\begin{array}{c}
\begin{tikzpicture}[scale = 0.4]
\draw[black,thick](0,0)--(1,1)--(1,1.5);
\draw[black,thick](0,0)--(-1,1)--(-1,2);
\draw[black,thick](1,1.5)--(1.5,2);
\draw[black,thick](1,1.5)--(0.5,2);
\draw[black,thick](0,0)--(0,-1);
\node[black] at (1.5,2.3) {\scriptsize $m$};
\node[black] at (0.5,2.3) {\scriptsize $m$};
\node[black] at (-1,2.3) {\scriptsize $m$};
\end{tikzpicture}
\end{array}}
\]
The 2-commutator $\beta$ is trivial on all components except $\beta_{m,m}=\pm i$. Two different choices of the 2-commutator define two commutative algebra structures, but they are equivalent to each other. Condensing $A_m^{\mathrm{tw}}$ leads to a `twisted' smooth boundary. The topological defects on this boundary still form $2\Rep(\Zb_2)$, but the bulk-to-boundary map is different. In the light of Theorem$^{\mathrm{ph}}$\,\ref{pthm:iterate_E2_condensation}, this Lagrangian $\EE_2$-algebra is the one associated to the non-trivial minimal modular extension of $\Rep(\Zb_2)$ given by the double semion MTC. 
\enu
Recall that there are infinitely many condensable $\EE_2$-algebras in $\Omega\CG\CT_{\Zb_2}^4=\FZ_1(2\Rep(\Zb_2))\simeq \Mod_{\Rep(\Zb_2)}^{\EE_2}(2\vect)$, each of which is given by a braided fusion 1-category $\CQ$ equipped with a braided monoidal functor $\Rep(\Zb_2) \to \CQ$. Moreover, such a condensable $\EE_2$-algebra is Lagrangian if and only if $\CQ$ is non-degenerate. By \cite{DMNO13}, this braided functor is either (1) a braided embedding $\Rep(\Zb_2) \hookrightarrow \CQ$ or (2) factors as $\Rep(\Zb_2) \to \vect \hookrightarrow \CQ$. The case (1) covers $A_m$ and $A_m^{\mathrm{tw}}$; and the case (2) are all of the form $A_e\boxtimes \CQ \in \FZ_1(2\Rep(\Zb_2))\boxtimes 2\vect \simeq \FZ_1(2\Rep(\Zb_2))$. 
\end{expl}

\begin{expl} \label{expl:GT_G4_GtoH}
Consider 3+1D finite gauge theory $\SG\ST_G^4$. The 2-category of topological defects of codimension 2 and higher is $\FZ_1(2\vect_G)$ \cite{KTZ20}. For a subgroup $H\leq G$, the forgetful functor $\forget: \Rep(G) \to \Rep(H)$ defines a condensable $\EE_2$-algebra in $2\Rep(G)$ thus in $\FZ_1(2\vect_G)$. 
By condensing the algebra $\Rep(H)$ in $\FZ_1(2\vect_G)$, we obtain $\SG\ST_H^4$ as the condensed phase, i.e., 
\[
\Mod_{\Rep(H)}^{\EE_2}(\FZ_1(2\vect_G)) \simeq \FZ_1(2\vect_H). 
\]
The proof of this fact is given in Example\,\ref{expl:GT_G4_GtoH_II}. 
\end{expl}


\begin{expl} \label{expl:E2_algebra_Z12VecG}
By \cite{DX24}, the condensable $\EE_2$-algebras in $\FZ_1(2\vect_G)$ are $G$-crossed braided multi-fusion 1-categories. A $G$-crossed braided multi-fusion 1-category \cite{Tur00,Kir01,Mueg04,DGNO10} consists of
\bit
\item a $G$-graded multi-fusion 1-category $\CB = \bigoplus_{g \in G} \CB_g$;
\item a monoidal $G$-action on $\CB$ such that $g \odot \CB_h \subseteq \CB_{ghg^{-1}}$;
\item a $G$-crossed braiding: $c_{x,y} \colon x \otimes y \to (g \odot y) \otimes x$ for $x \in \CC_g , \, y \in \CB$.
\eit
These data also satisfy some coherence condition. The half-braiding of $\CB \in \FZ_1(2\vect_G)$ is induced from the $G$-action. In particular, the trivial component $\CB_e$ is a braided multi-fusion 1-category equipped with a braided monoidal $G$-action, that is, a $G$-equivariant braided multi-fusion category. Similar to Example \ref{expl_2RepG_algebra}, a $G$-crossed braided multi-fusion 1-category is a simple condensable $\EE_2$-algebra in $\FZ_1(2\vect_G)$ if and only if $G$ acts on the simple direct summands of the tensor unit transitively. It follows that the simple condensable $\EE_2$-algebras in $\FZ_1(2\vect_G)$ are classified by pairs $(H,\CB_0)$, where $H \leq G$ is a subgroup defined up to conjugations and $\CB_0$ is an $H$-crossed braided fusion category. Given such a pair $(H,\CB_0)$, the corresponding $\EE_2$-algebra is the induced category $\CB \coloneqq \Ind^G_H \Cb_0 \coloneqq \Fun_H(G,\CB_0)$, and it is Lagrangian if and only if the $H$-grading on $\CB_0$ is faithful and the trivial component of $\CB_0$ is a nondegenerate braided fusion 1-category \cite{Xu24a,Wen24a}. We list some examples here.
\bnu
\item The category $\Fun(G,\vect)$ with the pointwise tensor product and left translation $G$-action is a braided multi-fusion category equipped with a $G$-action, which can be viewed as a $G$-crossed braided fusion category with only the trivial component. It is a Lagrangian algebra in $\FZ_1(\vect_G)$ and $\RMod_{\Fun(G,\vect)}(\FZ_1(2\vect_G)) \simeq \vect_G$. When $G = \Zb_2$, this is the Lagrangian algebra $\one_c$ in the 3+1D toric code model (see Example \ref{expl:Ae+A1+A2_4D_TC}).
\item More generally, for every subgroup $H \leq G$, the category $\Fun(G/H,\vect)$ is a condensable $\EE_2$-algebra in $\FZ_1(2\vect_G)$ and $\Mod_{\Fun(G/H,\vect)}^{\EE_2}(\FZ_1(2\vect_G)) \simeq \FZ_1(2\vect_H)$.
\item For every 3-cocycle $\omega \in Z^3(G,U(1))$, there is a (unique up to unique isomorphism) $G$-crossed braided fusion 1-category structure on $\vect_G^\omega$ with the obvious $G$-grading and the conjugation $G$-action \cite{DGNO10}. Then $\vect_G^\omega$ is a Lagrangian algebra in $\FZ_1(\vect_G)$ and $\RMod_{\vect_G^\omega}(\FZ_1(2\vect_G)) \simeq 2\rep(G)$. When $G = \Zb_2$, we have $H^3(\Zb_2,U(1)) \simeq \Zb_2$ and the two Lagrangian algebras $\vect_{\Zb_2}^\omega$ are the two Lagrangian algebra structure on $\one \oplus m$ in the 3+1D toric code model (see Example \ref{expl:Ae+A1+A2_4D_TC}).
\item In general, for a faithfully graded nondegenerate $G$-crossed braided fusion category $\CB$, we have \newline $\RMod_\CB(\FZ_1(2\vect_G)) \simeq \Sigma (\CB_e)^G$ \cite{Wen24a}. 
\item For every subgroup $H \leq G$ and 3-cocycle $\omega \in Z^3(H,U(1))$, the induced category $\vect_H^\omega \coloneqq \Fun_H(G,\vect_H^\omega)$ is a Lagrangian algebra in $\FZ_1(2\vect_G)$.
\enu
\end{expl}

\begin{expl} \label{expl:E2_algebra_Z12RepG}
There are also two coordinate systems on $\FZ_1(2\rep(G))$.
\bit
\item The condensable $\EE_2$-algebras in $\FZ_1(2\rep(G)) \simeq \FZ_1(\LMod_{\vect_G}(2\vect))$ are also $G$-crossed braided multi-fusion categories, on which the half-braidings are equivalent to the $G$-gradings.
\item By \cite{JFR23,DX24}, the condensable $\EE_2$-algebras in $\FZ_1(\Sigma \Rep(G)) \simeq \Mod_{\Rep(G)}^{\EE_2}(2\vect)$ are braided multi-fusion categories $\CM$ equipped with braided monoidal functor $\Rep(G) \to \CM$, which are simple if and only if $\CM$ is a braided fusion category. By \cite[Lemma 3.5 and Remark 3.23]{DMNO13}, such a braided functor factors throught $\Rep(G) \to \Rep(H) \to \CM$ for some subgroup $H \leq G$ such that $\Rep(H) \to \CM$ is a braided embedding.
\eit
These two coordinate systems are related by the equivariantization and de-equivariantization. In particular, there is a one-to-one correspondence between $G$-crossed braided fusion categories and braided fusion categories containing $\Rep(G)$ \cite[Theorem 4.44]{DGNO10}.

Let $\Rep(G) \to \CM$ be a simple condensable $\EE_2$-algebra in $\FZ_1(\Sigma \Rep(G))$. We also denote it by $\CM$ for simplicity. According to \cite{DX24} (see also Theorem$^{\mathrm{ph}}$\,\ref{pthm:iterate_E2_condensation}), by condensing $\CM$ as 2-codimensional topological defect in $\SG\ST_G^4$, we obtain a condensed phase $\SD^4$ and a gapped domain wall $\SM^3$ such that  
$$
\Omega\CD \simeq \Mod_\CM^{\EE_2}(\Mod_{\Rep(G)}^{\EE_2}(2\vect)) \simeq \Mod_\CM^{\EE_2}(2\vect) \simeq \FZ_1(\Sigma \CM), 
\quad\quad
\Omega_m\CM \simeq \RMod_\CM(\Mod_{\Rep(G)}^{\EE_2}(2\vect)). 
$$ 
It follows that the Lagrangian $\EE_2$-algebras in $\FZ_1(\Sigma \Rep(G))$ are nondegenerate braided fusion categories $\CM$ equipped with braided functors $\Rep(G) \to \CM$. 
We list some examples here.
\bnu
\item For every $\omega \in Z^3(G,U(1))$, there is a braided embedding $\Rep(G) \to \FZ_1(\vect_G^\omega)$, which defines a Lagrangian $\EE_2$-algebra in $\FZ_1(\Sigma \Rep(G))$. As a consequence, the condensed phase $\SD^4=\mathbf{1}^4$ is trivial and the gapped domain wall is such that 
$$
\Omega_m\CM \simeq \RMod_{\FZ_1(\vect_G^\omega)}(\FZ_1(\Sigma \Rep(G))) \simeq 2\Rep(G).
$$ 
When $G = \Zb_2$, we have $H^3(\Zb_2,U(1)) \simeq \Zb_2$ and the two minimal nondegenerate extensions correspond to the two different Lagrangian $\EE_2$-algebra structures on $\one \oplus m$ in the 3+1D toric code model (see Example \ref{expl:Ae+A1+A2_4D_TC}).

\item When $\CM$ is a nondegenerate extension of $\Rep(G)$, the 2-category $\RMod_\CM(\FZ_1(\Sigma \Rep(G)))$ is connected and equivalent to $\Sigma \FZ_2(\Rep(G) \to \CM)$.
\item More generally, for every subgroup $H \leq G$ and $\omega \in Z^3(H,U(1))$, the composite functor $\Rep(G) \to \Rep(H) \to \FZ_1(\vect_H^\omega)$ is a Lagrangian algebra in $\FZ_1(\Sigma \Rep(G))$.
\item In particular, when $H$ is the trivial subgroup, the forgetful functor $\Rep(G) \to \vect$ is a Lagrangian algebra in $\FZ_1(\Sigma \Rep(G))$, and $\RMod_\CM(\FZ_1(\Sigma \Rep(G)))$ is equivalent to $2\vect_G$. When $G = \Zb_2$, this is the Lagrangian algebra $\one_c$ in the 3+1D toric code model (see Example\ \ref{expl:Ae+A1+A2_4D_TC}).
\enu
\end{expl}

\begin{expl}[\cite{Xu24a,Wen24a}]
Let $\pi \in Z^4(G,U(1))$. For the 3+1D $\pi$-twisted $G$ gauge theory, the condensable $\EE_2$-algebras in $\FZ_1(2\vect_G^\pi)$ are precisely $\pi$-twisted $G$-crossed braided multi-fusion 1-categories $\CA$ \cite[Lemma\,5.8]{Xu24a}. The simple condensable $\EE_2$-algebras are given in Theorem\,5.11, and Lagrangian $\EE_2$-algebras are given in Corollary\,5.12 in \cite{Xu24a}.
When $\CA$ is a braided fusion 1-category with a $G$-action, then the 2-category of right $\CA$-modules in $\FZ_1(2\vect_G^\pi)$ was computed in Theorem 5.17 in \cite{Xu24a}, and 
the 2-category of $\EE_2$-$\CA$-modules (or local $\CA$-modules) was computed in Theorem\,5.20 in \cite{Xu24a}. 
\end{expl}

\subsubsection{Examples in higher dimensions}

\begin{expl} \label{expl:G_crossed_BF_nCat}
Consider finite gauge theory $\SG\ST_G^{n+2}$ for a finite group $G$. We have $\Omega\CG\CT_G^{n+2}$ has two coordinate systems: 
$$
\Omega\CG\CT_G^{n+2} \simeq \FZ_1((n+1)\vect_G), \quad\quad
\Omega\CG\CT_G^{n+2} \simeq \FZ_1((n+1)\Rep(G)) \simeq \Mod_{n\Rep(G)}^{\EE_2}((n+1)\vect), 
$$ 
where the last $\simeq$ is due to (\ref{eq:E2_Z1_LMod}). We discuss condensable $\EE_1$-algebras in $\Omega\CG\CT_G^{n+2}$ in both coordinates. 
\bnu
\item[(1)] A condensable $\EE_2$-algebras in $\FZ_1((n+1)\vect_G)$ is precisely a $G$-crossed braided multi-fusion $n$-category.
Inspired by the definition of a $G$-crossed braided multi-fusion 1-category \cite{Tur00,Kir01,Mueg04,DGNO10}, we expect a $G$-crossed braided multi-fusion $n$-category to consist of
\bit
\item a $G$-graded multi-fusion 1-category $\CA = \bigoplus_{g \in G} \CA_g$;
\item a monoidal $G$-action on $\CA$ such that $g \odot \CA_h \subseteq \CA_{ghg^{-1}}$;
\item a $G$-crossed braiding: $c_{x,y} \colon x \otimes y \to (g \odot y) \otimes x$ for $x \in \CC_g , \, y \in \CA$. 
\eit
satisfying some (higher) coherence conditions. Higher coherence properties are always difficult in high category theory. We try to avoid higher coherence in this work. Instead, we give a convenient or almost physical way to define this notion. A $G$-crossed braided fusion $n$-category $\CA$ can be understood as a monoidal $(n+1)$-category $\mathrm{B}_G\CA$ consists of $|G|$ objects labeled by the elements in $G$ with the monoidal structure defined by the multiplication of $G$ and $\hom_{\mathrm{B}_G\CA}(g,h)=\CA_{hg^{-1}}$ (recall Example\,\ref{expl:E1_algebras_2vectG}). When $n=1$, this construction can be found in \cite{Cui19,DR18,JPR22}. 
We can understand this construction physically. If we view the condensation completion of $\mathrm{B}_G\CA$, denoted by $\Sigma_G\CA$, as the category of topological defects of an anomalous $n+$1D topological order $\SX^{n+1}$, then the object $g\in G=\mathrm{Ob}(\mathrm{B}_G\CA)$ can be viewed as an invertible 1-codimensional defect in $\SX^{n+1}$, and $\hom_{\mathrm{B}_G\CA}(g,h)$ can be viewed as the $n$-category of 2-codimensional defects living on the domain wall between two 1-codimensional defect $g, h\in G$. 

Now we show that $\Sigma_G \CA \simeq \RMod_\CA((n+1)\vect_G)$ as $(n+1)\vect_G$-modules. Denote the simple objects in $(n+1)\vect_G$ by $\{\delta_g\}_{g \in G}$. There is a $(n+1)$-functor $F: \mathrm B_G \CA \to \RMod_\CA((n+1)\vect_G)$ defined by
\bit
\item an object $g \in G$ is mapped to the free right $\CA$-module $\delta_{g^{-1}} \otimes \CA$;
\item the functors between hom $n$-categories are defined by the canonical equivalences:
\[
\Hom_{\mathrm B_G \CA}(g,h) = \CA_{hg^{-1}} \simeq \Hom_{2\vect_G}(\delta_{g^{-1}},\delta_{h^{-1}} \otimes \CA) \simeq \Hom_{\RMod_\CA(2\vect_G)}(\delta_{g^{-1}} \otimes \CA,\delta_{h^{-1}} \otimes \CA) .
\]
\eit
Moreover, this is a $G$-module $(n+1)$-functor. Thus it induces a $G$-module $(n+1)$-functor $\Sigma_G \CA \to \RMod_\CA((n+1)\vect_G)$. Note that every simple module $x \in \RMod_\CA((n+1)\vect_G)$ can be condensed from the free module $x \otimes \CA$. Thus the $G$-module $(n+1)$-functor $\Sigma_G \CA \to \RMod_\CA((n+1)\vect_G)$ is an equivalence.


\item[(2)] A condensable $\EE_2$-algebra in $\FZ_1((n+1)\Rep(G))$ is precisely given by a braided monoidal functor $n\Rep(G) \to \CB$ for a braided multi-fusion $n$-category $\CB$. It is simple if $\CB$ is fusion. 
\enu
These two ways of defining condensable $\EE_2$-algebras in $\Omega\CG\CT_G^{n+2}$ are related again by equivariantization $\CA \mapsto \CA^G$ and de-equivariantization $\CB \mapsto \CB_G$ (recall Example\,\ref{expl:E1-algebra_in_n+1RepG}). In particular, $\CB_G:=\RMod_{[\one,\one]}(\CB)$, where $\one$ is the tensor unit of $n\vect$ and $[\one,\one]$ is the internal hom algebra in $n\Rep(G)$ for the $n\Rep(G)$-module $n\vect$. In other words, the notion of a $G$-crossed braided fusion $n$-category can also be understood as the de-equivariantization of a braided monoidal functor $n\Rep(G) \to \CB$.

Recall Theorem$^{\mathrm{ph}}$\ref{pthm:iterate_E2_condensation}, by condensing the condensable $\EE_2$-algebra $\CB$ in $\FZ_1((n+1)\Rep(G))$, we obtain a condensed phase $\SD^{n+2}$ such that $\Omega\CD \simeq \FZ_1(\Sigma\CB)$. 
\end{expl}

\newpage


\section{Condensations of higher codimensional topological defects} \label{sec:condense_k-codim_defect}
In this section, we give the general theory of the condensations of $k$-codimensional topological defects in an $n+$1D topological order. We first study the relation between particle condensations and string condensations in 2+1D, then we generalize it to higher dimensions. 

\subsection{Particle condensations in 2+1D via a two-step process} \label{sec:3D_two-step_condensation}
In 2+1D, 1-codimensional defects are strings and 2-codimensional defects are particles. In Section \ref{sec:1codim_nd} and \ref{sec:1codim_2d}, we have studied the string condensation in 2+1D; in Section \ref{sec:2codim_2d}, we have studied the particle (or anyon) condensation in 2+1D. In this subsection, we study their relation, which can be generalized to higher dimensions.

\subsubsection{General theory} \label{sec:2d_two_step}
Consider a 2+1D anomaly-free simple topological orders $\SC^3$. In this case, there is a natural coordinate system of $\CC$, i.e., $\CC=\Sigma\Omega\CC=\RMod_{\Omega\CC}(2\vect)$. 

\medskip
Consider a particle condensation in $\SC^3$ defined by a simple condensable $\EE_2$-algebra $A$ in $\Omega\CC$. It produces a condensed 2+1D topological order $\SD^3$ and a gapped domain wall $\SM^2=(\CM,m)$ such that $\Omega\CD\simeq \Mod_A^{\EE_2}(\Omega\CC)$ and $\Omega_m\CM=\RMod_A(\Omega\CC)$. Now we show that the same phase transition can be realized in a different way, which splits into two steps. 
\bnu

\item In the first step, we condense the particle $A$ along a line in space as illustrated below. It produces a condensed string denoted by $\Sigma A$. 
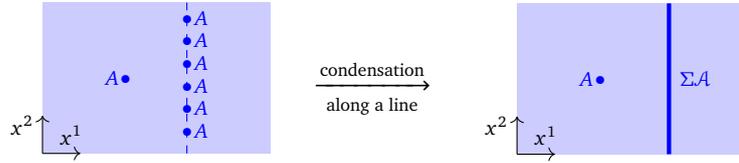
$$
\begin{array}{c}
\begin{tikzpicture}[scale=1]
\fill[blue!20] (0,0) rectangle (3,2) ;
\draw[->] (0,0) -- (0.5,0) node [near end,above] {\footnotesize $x^1$};
\draw[->] (0,0) -- (0,0.5) node [near end,left] {\footnotesize $x^2$};
\node at (1,1) {\footnotesize \textcolor{blue}{$A\, \bullet$}} ;
\node at (2,0.3) {\footnotesize \textcolor{blue}{$\bullet\, A$}} ;
\node at (2,0.6) {\footnotesize \textcolor{blue}{$\bullet\, A$}} ;
\node at (2,0.9) {\footnotesize \textcolor{blue}{$\bullet\, A$}} ;
\node at (2,1.2) {\footnotesize \textcolor{blue}{$\bullet\, A$}} ;
\node at (2,1.5) {\footnotesize \textcolor{blue}{$\bullet\, A$}} ;
\node at (2,1.8) {\footnotesize \textcolor{blue}{$\bullet\, A$}} ;
\draw[blue, dashed] (1.9,0) -- (1.9,2) ;
\end{tikzpicture}
\end{array}
\quad
\xrightarrow[\text{\scriptsize along a line}]{\text{\scriptsize condensation}}
\quad
\begin{array}{c}
\begin{tikzpicture}[scale=1]
\fill[blue!20] (0,0) rectangle (3,2) ;
\draw[->] (0,0) -- (0.5,0) node [near end,above] {\footnotesize $x^1$};
\draw[->] (0,0) -- (0,0.5) node [near end,left] {\footnotesize $x^2$};
\node at (1,1) {\footnotesize \textcolor{blue}{$A\, \bullet$}} ;
\draw[blue, ultra thick] (2,0) -- (2,2) node [midway,right] {\footnotesize $\Sigma\CA$}; 
\end{tikzpicture}
\end{array}
$$
In this step, one can forget about the $\EE_2$-algebra structure on $A$ but only remember its $\EE_1$-algebra structure in the $x^2$-direction. Recall Section\,\ref{sec:1d_cond_alg} \cite{Kon14e}, condensing $A$ on a line produces a condensed string, particles on which form the fusion 1-category $\Mod_A^{\EE_1}(\CC)$, i.e. 
$$
\hom_\CC(\Sigma A, \Sigma A) \simeq \Mod_A^{\EE_1}(\CC). 
$$
However, this result does not tell us what $\Sigma A$ is as an object in $\CC$. Now we show that $\Sigma A$ can be identified with $\RMod_A(\Omega\CC)$ as an object in $\CC=\RMod_{\Omega\CC}(2\vect)$. 
\bnu
\item First, we have already shown in Theorem$^{\mathrm{ph}}$\,\ref{thm:condensable_E1Alg_af_C} that $\RMod_A(\Omega\CC) \in \Algc_{\EE_1}(\RMod_{\Omega\CC}(2\vect)) \simeq \LMod_{\Omega\CC}(\Algc_{\EE_1}(2\vect))$ because the functor $-\otimes A: \Omega\CC \to \RMod_A(\Omega\CC)$ is a central functor. 


\item It remains to show that $\Sigma A$ can be identified with $\RMod_A(\Omega\CC)$ as an object in $\CC = \RMod_{\Omega\CC}(2\vect)$. One can see this directly from the following picture: 
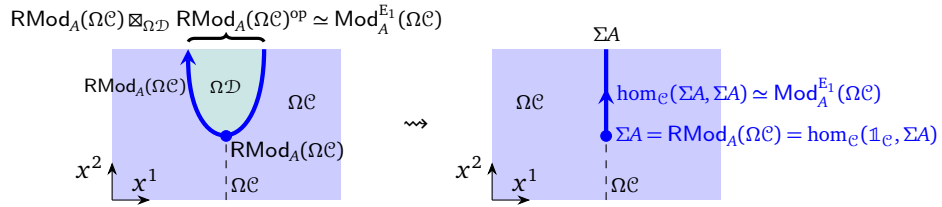
\be \label{eq:SigmaA_shrink_D}
\begin{array}{c}
\begin{tikzpicture}[>=Stealth]
\fill[blue!20] (-4,0) rectangle (-1,2) ;
\fill[teal!20] (-3,2) .. controls (-3,0.5) and (-2,0.5) .. (-2,2)--(-3,2)..controls (-2.5,2) .. (-2,2)--cycle;
\node at (-2.5,1.5) {\scriptsize $\Omega\CD$} ;
\draw[blue,ultra thick,-stealth] (-2,2) .. controls (-2,0.5) and (-3,0.5) .. (-3,2) ;
\draw[dashed] (-2.5,0) -- (-2.5,0.8) ;
\draw[decorate,decoration=brace,very thick] (-3,2.1)--(-2,2.1) ;
\node at (-3.7,1.5) {\scriptsize $\RMod_A(\Omega\CC)$} ;
\filldraw [blue] (-2.5,0.85) circle [radius=2pt] ;
\node at (-1.7,0.65) {\footnotesize $\RMod_A(\Omega\CC)$} ;
\node at (-2.5,2.4) {\footnotesize $\RMod_A(\Omega\CC) \boxtimes_{\Omega\CD} \RMod_A(\Omega\CC)^\op \simeq \Mod_A^{\EE_1}(\Omega\CC)$};
\draw[->] (-4,0) -- (-4,0.5) node [very near end, left] {$x^2$} ;
\draw[->] (-4,0) -- (-3.5,0) node [very near end, above] {$x^1$} ;
\node at (-1.5,1.3) {\footnotesize $\Omega\CC$} ;
\node at (-2.25,0.2) {\footnotesize $\Omega\CC$} ;
\node at (0,1) {$\rightsquigarrow$};
\fill[blue!20] (1,0) rectangle (4,2) ;
\draw[dashed] (2.5,0) -- (2.5,0.85) ;
\draw[blue,ultra thick,->-] (2.5,0.85) -- (2.5,2) node[midway, right] {\footnotesize $\hom_\CC(\Sigma A, \Sigma A)\simeq \Mod_A^{\EE_1}(\Omega\CC)$} ;
\filldraw [blue] (2.5,0.85) circle [radius=2pt] node[right]  {\footnotesize $\Sigma A=\RMod_A(\Omega\CC)=\hom_\CC(\one_\CC,\Sigma A)$} ;
\node at (2.5,2.2) {\footnotesize $\Sigma A$} ;
\draw[->] (1,0) -- (1.5,0) node [very near end, above] {$x^1$} ;
\draw[->] (1,0) -- (1,0.5) node [very near end, left] {$x^2$} ;
\node at (2.75,0.2) {\footnotesize $\Omega\CC$} ;
\node at (1.5,1.3) {\footnotesize $\Omega\CC$} ;
\end{tikzpicture}
\end{array}
\ee
where the green region shrink horizontally to the condensed string $\Sigma A$, and  the wall between two colored region has particles which form a fusion 1-category given by $\RMod_A(\Omega\CC)$. We recover the multi-fusion 1-category of particles on the condensed string by the following equivalences:  
$$
\RMod_A(\Omega\CC) \boxtimes_{\Omega\CD} \RMod_A(\Omega\CC)^\op \simeq 
\RMod_A(\Omega\CC) \boxtimes_{\Mod_A^{\EE_2}(\Omega\CC)} \LMod_A(\Omega\CC) \simeq \Mod_A^{\EE_1}(\Omega\CC),
$$
where we have used the natural equivalence $\RMod_A(\Omega\CC) \simeq \LMod_A(\Omega\CC)^\rev$ defined by $x\mapsto x^R=x^\ast$ in the first `$\simeq$' and the second `$\simeq$' is a mathematical fact (see for example \cite{KZ18}). 
Remember $\hom_\CC(x,y), \forall x,y\in \CC$ are physical observables in spacetime, but the label of a string can not be observed directly in spacetime. However, one can identify an object $x\in \CC$ in the coordinate system $\CC=\RMod_{\Omega\CC}(2\vect)$ with $\hom_{\CC}(\Omega\CC, x)$ because $\hom_{\RMod_{\Omega\CC}(2\vect)}(\Omega\CC, x) \simeq x$ in $\RMod_{\Omega\CC}(2\vect)$. From the physical configuration depicted in (\ref{eq:SigmaA_shrink_D}), we immediately obtain 
$$
\Sigma A \simeq \hom_{\CC}(\Omega\CC, \Sigma A) \simeq \RMod_A(\Omega\CC) \in \RMod_{\Omega\CC}(2\vect) =\CC, 
$$ 
because $\hom_{\CC}(\Omega\CC, \Sigma A)$ is precisely the observables living the domain wall between the trivial 1-codimensional topological defect $\one_\CC=\Omega\CC$ and $\Sigma A$. 
\enu
From now on, for simplicity, we set $\Sigma A:=\RMod_A(\Omega\CC) \in \RMod_{\Omega\CC}(2\vect) =\CC$ and, at the same time, $\Sigma A$ is also the abstract label of the condensed defect if we do not choose any coordinate system.

\item In the second step, we condense $\Sigma A$. We have already shown in Theorem$^{\mathrm{ph}}$\,\ref{thm:condensable_E1Alg_af_C} that $\Sigma A$ is a condensable $\EE_1$-algebra in $\CC$. This fact can also be viewed in another way. Since the $\Sigma A$-string can be obtained by shrinking the green region horizontally as depicted in (\ref{eq:SigmaA_shrink_D}), by our theory of the condensations of 1-codimensional topological defects (Theorem$^{\mathrm{ph}}$\,\ref{pthm:1codim_condensation_nd} and \ref{pthm:1codim_condensation_nd_II}), the $\Sigma A$-string has a canonical structure of a condensable $\EE_1$-algebra in $\CC$.
The algebraic structure on $\Sigma A$ is defined in (\ref{pic:condensable_E1_algebra_2}). In this argument, the $\EE_2$-algebra structure on $A$ is implicitly used. Moreover, this algebraic structure on $\Sigma A$ coincides with the monoidal structure on $\RMod_A(\Omega\CC)$ (recall (\ref{pic:OmegaM=E1_algebra})). Therefore, we can further condensing the $\Sigma A$-string thus produces a new phase. It is physically obvious that this new phase is precisely $\SD^3$, In other words, we obtain
\be
\CD \simeq \Mod_{\Sigma A}^{\EE_1}(\CC). \label{eq:D_from_SigmaA}
\ee
It immediately implies the following equivalences:
\begin{align} 
\Mod_{\Sigma A}^{\EE_1}(\CC) &\simeq \Sigma \Mod_A^{\EE_2}(\Omega\CC), \label{eq:E2_Mod_Omega_E1_Mod} \\
\Omega\Mod_{\Sigma A}^{\EE_1}(\CC) &\simeq \Mod_A^{\EE_2}(\Omega\CC). \label{eq:E2_Mod_Omega_E1_Mod_2}
\end{align}
where the first equivalence is a monoidal equivalence and the second one is a braided monoidal equivalence. Note that (\ref{eq:E2_Mod_Omega_E1_Mod_2}) is an immediate consequence of (\ref{eq:E2_Mod_Omega_E1_Mod}) and is proved directly in Theorem\,\ref{pthm:OmegaSigmaA-AOmega}. 
The gapped domain wall $\SM^2$ between $\SC^3$ and $\SD^3$ can be described the pair 
$$
(\RMod_{\Sigma A}(\CC), \Sigma A).
$$ 
The multi-fusion 1-category of particles on this domain wall is given by  
$$
\hom_{\RMod_{\Sigma A}(\CC)}(\Sigma A, \Sigma A)  \simeq \Fun_{\RMod_A(\Omega\CC)^\rev}(\RMod_A(\Omega\CC), \RMod_A(\Omega\CC))
\simeq \RMod_A(\Omega\CC),
$$
where the first `$\simeq$' is due to the fact that a functor intertwining the $\Sigma A$-action automatically intertwines the $\Omega\CC$-action, which is defined via the monoidal functor $- \otimes A: \Omega\CC \to \RMod_A(\Omega\CC)$. 
\enu
Note that the monoidal equivalences in (\ref{eq:E2_Mod_Omega_E1_Mod}) summarize precisely the relation between the direct condensation of a 2-codimensional topological defect $A$ and that of the 1-codimensional topological defect $\Sigma A$.

\medskip
All of above arguments automatically generalize to higher codimensional defects except a key point, which deserves to be reinvestigated. We see the key point here is to argue that $\Sigma A$ is automatically a condensable $\EE_1$-algebra in $\CC$ if $A$ is a condensable $\EE_2$-algebra in $\Omega\CC$. We have used the physical or geometric intuition established in Theorem$^{\mathrm{ph}}$\,\ref{pthm:1codim_condensation_nd} and \ref{pthm:1codim_condensation_nd_II} (in particular, (\ref{pic:condensable_E1_algebra_2})). 
Unfortunately, this physical or geometric intuition stops to make sense for higher codimensional topological defects because $k$-codimensional topological defects for $k>2$ do not give a complete mathematical characterization of a topological order. Whenever we want to apply the geometric intuition of a topological order $\SC^{n+1}$ we should automatically include all topological defects of codimension 1 or higher if  $\SC^{n+1}$ is anomalous; or include all topological defect of codimension (at least) 2 or higher if $\SC^{n+1}$ is anomaly-free. 
Therefore, in order to generalize this key point to higher codimensional defects, we need more fundamental arguments that are ready to be generalized. 

Now we explain mathematically why $\Sigma A$ is a condensable $\EE_1$-algebra in $\CC$. First, recall that $\otimes^2$ represents the fusion product in the vertical direction (see Figure\,\ref{fig:braiding_E2} and \ref{fig:deconfined_particles_E2_module}), and $(\Omega\CC,\otimes^2)$ is monoidal. Therefore, it makes sense to talk about the 1-category $\Alg_{\EE_1}(\Omega\CC)$ based on the monoidal structure $(\Omega\CC,\otimes^2)$. It turns out that the second tensor product $\otimes^1$ on $\Omega\CC$ (i.e., the horizontal fusion) endow both triples 
$$
(\Alg_{\EE_1}(\Omega\CC), \otimes^1, 1_{\one_\CC})
\quad\quad \mbox{and} \quad\quad
(\RMod_{\Omega\CC}(2\vect), \boxtimes_{\Omega\CC}, \Omega\CC)
$$ 
with the structure of a monoidal 1-category and a monoidal 2-category, respectively. We explain this fact below. 
\bnu
\item The tensor product in $\Alg_{\EE_1}(\Omega\CC)$ is given by the tensor product $\otimes^1$ in $\Omega\CC$. For $A,B\in \Alg_{\EE_1}(\Omega\CC)$, $A\otimes^1 B$ is a well-defined $\EE_1$-algebra in $\Omega\CC$ with the algebraic structure defined as follows: 
\[
\begin{array}{c}
\begin{tikzpicture}[scale=0.8]
\fill[teal!20] (-1,0.3) rectangle (2,2.7) ;
\draw [dashed] (0,0.3) -- (0,2.7) ; 
\draw [dashed] (1,0.3) -- (1,2.7) ; 
\draw[fill=white] (-0.05,0.95) rectangle (0.05,1.05) node[midway,left] {$A$} ;
\draw[fill=white] (-0.05,1.95) rectangle (0.05,2.05) node[midway,left] {$A$} ;
\draw[fill=white] (0.95,0.95) rectangle (1.05,1.05) node[midway,right] {$B$} ;
\draw[fill=white] (0.95,1.95) rectangle (1.05,2.05) node[midway,right] {$B$} ;
\node at (-0.5,2.5) {\scriptsize $\Omega\CC$} ;
\draw[->] (-1,0.3) -- (-0.5,0.3) node [very near end, above] {$x^1$} ;
\draw[->] (-1,0.3) -- (-1,0.8) node [very near end, left] {$x^2$} ;
\end{tikzpicture}
\end{array}
\quad\quad
\begin{array}{c}
(A\otimes^1 B) \otimes^2 (A\otimes^1 B) \xrightarrow{\delta_{A,A,B,B}} 
(A\otimes^2 A) \otimes^1 (B\otimes^2 B) 
\xrightarrow{\mu_A \mu_B} A \otimes^1 B. 
\end{array}
\]
\item A right $\Omega\CC$-module $X$ with a right $\Omega\CC$-action $\odot: X \times \Omega\CC \to X$ is automatically a left $\Omega\CC$-module with the left $\Omega\CC$-action defined by $a \odot x := x\odot a$ for $a\in \Omega\CC, x\in X$. This is a well-defined left $\Omega\CC$-module due to the existence of the braiding in $\Omega\CC$ as shown below: 
$$
a \odot (b \odot x) := (x\odot b) \odot a \simeq x \odot (b\otimes^1 a) \xrightarrow{1_x c_{b,a}^1} x \odot (a\otimes^1 b) =(a\otimes^1 b) \odot x. 
$$
where the braiding $c_{b,a}^1$ is defined in (\ref{eq:braiding_from_2fusion_1}). 
\enu
Note that there is a functor 
\be \label{eq:LMod_functor}
\Sigma(-)=\RMod_{-}(\Omega\CC): \Alg_{\EE_1}(\Omega\CC) \to \RMod_{\Omega\CC}(2\vect)=\CC
\ee
defined by 
$$
\xymatrix{
A \ar@{|->}[r] \ar[d]_f &  \RMod_A(\Omega\CC) \ar[d]^{- \otimes_A \, {}_fB} \\
B \ar@{|->}[r] & \RMod_B(\Omega\CC)  
}
$$
where ${}_fB$ is the $A$-$B$-bimodule with the right $A$-module structure induced from the algebra map $f: A \to B$. One can show that $\RMod_-(\Omega\CC): \Alg_{\EE_1}(\Omega\CC) \to \RMod_{\Omega\CC}(2\vect)$ is a lax-monoidal functor with two defining 1-morphisms: 
$$
1_{\one_\CC} \mapsto \RMod_{1_{\one_\CC}}(\Omega\CC) \simeq \Omega\CC; \quad\quad 
\RMod_A(\Omega\CC) \boxtimes_{\Omega\CC} \RMod_B(\Omega\CC) \xrightarrow{\tilde{F}} \RMod_{A\otimes^1 B}(\Omega\CC). 
$$
We give a construction of $\tilde{F}$. The functor $\RMod_A(\Omega\CC) \boxtimes \RMod_B(\Omega\CC) \xrightarrow{F} \RMod_{A\otimes B}(\Omega\CC)$ defined by $m\boxtimes n \to m\otimes^1 n$, where the $A\otimes^1B$-action on $m\otimes^1n$ is defined as follows: 
\[
\begin{array}{c}
\begin{tikzpicture}[scale=0.8]
\fill[teal!20] (-1,0.3) rectangle (2,2.7) ;
\draw [dashed] (0,0.3) -- (0,2.7) ; 
\draw [dashed] (1,0.3) -- (1,2.7) ; 
\draw[fill=white] (-0.05,0.95) rectangle (0.05,1.05) node[midway,left] {$m$} ;
\draw[fill=white] (-0.05,1.95) rectangle (0.05,2.05) node[midway,left] {$A$} ;
\draw[fill=white] (0.95,0.95) rectangle (1.05,1.05) node[midway,right] {$n$} ;
\draw[fill=white] (0.95,1.95) rectangle (1.05,2.05) node[midway,right] {$B$} ;
\draw[fill=teal!20] (0.45,1.45) rectangle (0.55,1.55) node[right] {\scriptsize $a\in \Omega\CC$} ;
\draw[->] (-1,0.3) -- (-0.5,0.3) node [very near end, above] {$x^1$} ;
\draw[->] (-1,0.3) -- (-1,0.8) node [very near end, left] {$x^2$} ;
\end{tikzpicture}
\end{array}
\quad\quad
\begin{array}{c}
(m\otimes^1 n) \otimes^2 (A\otimes^1 B) \xrightarrow{\delta_{m,n,A,B}} 
(m\otimes^2 A) \otimes^1 (n\otimes^2 B) 
\xrightarrow{\mu_m \mu_n} m \otimes^1 n. 
\end{array}
\]
From above picture, it is clear that this functor $F$ is $\Omega\CC$-balanced, i.e., $(m\otimes^1 a) \otimes^1 n \simeq m\otimes^1 (a\otimes^1 n)$. Therefore, it induces a functor $\tilde{F}: \RMod_A(\Omega\CC) \boxtimes_{\Omega\CC} \RMod_B(\Omega\CC) \to \RMod_{A\otimes B}(\Omega\CC)$. 
This functor $\tilde{F}$ endows the functor $\Sigma(-):\Alg_{\EE_1}(\Omega\CC) \to \RMod_{\Omega\CC}(2\vect)$ with the structure of a lax-monoidal functor.

Since a lax-monoidal functor maps $\EE_1$-algebras to $\EE_1$-algebras, it means that 
the functor $\RMod_-(\Omega\CC)$ can be lifted to a functor 
\begin{align}
\RMod_-(\Omega\CC): \Alg_{\EE_2}(\Omega\CC)=\Alg_{\EE_1}(\Alg_{\EE_1}(\Omega\CC)) &\to \Alg_{\EE_1}(\RMod_{\Omega\CC}(2\vect)), \label{eq:LMod_functor_2} \\
A &\mapsto \Sigma A = \RMod_A(\Omega\CC). \nonumber
\end{align}
Since the separability is defined by 1-morphisms and identities in $\Alg_{\EE_1}(\Omega\CC)$, they are mapped to 1-morphisms and identities in $\RMod_{\Omega\CC}(2\vect)$. As a consequence, we obtain a restriction of the functor $\RMod_-(\Omega\CC)$ on the subcategory $\Algc_{\EE_2}(\Omega\CC)$: 
\begin{align}
\Sigma(-)=\RMod_-(\Omega\CC): \Algc_{\EE_2}(\Omega\CC) &\to \Algc_{\EE_1}(\RMod_{\Omega\CC}(2\vect))=\Algc_{\EE_1}(\CC). \label{eq:LMod_functor_3} \\
A &\mapsto \Sigma A = \RMod_A(\Omega\CC). \nonumber
\end{align}
We summarize above result as a mathematical theorem. 
\begin{prop} \label{prop:E2Alg_mapsto_E1Alg}
For a braided fusion 1-category $\CB$, there is a well-defined functor 
$$
\Sigma(-): \Algc_{\EE_2}(\CB) \to \Algc_{\EE_1}(\Sigma\CB)
\quad\quad \mbox{defined by} \quad\quad 
A\mapsto \Sigma A = \RMod_A(\CB).
$$ 
When $\CB$ is non-degenerate, this functor maps Lagrangian $\EE_2$-algebras in $\CB$ to Lagrangian $\EE_1$-algebras in $\Sigma\CB$ bijectively. 
\end{prop}

\begin{rem}
Proposition\,\ref{prop:E2Alg_mapsto_E1Alg} provides a powerful way to construct or classify Lagrangian $\EE_2$-algebras via the classification of Lagrangian $\EE_1$-algebras, which can be obtained from Theorem$^{\mathrm{ph}}$\,\ref{pthm:construct_cond_E1_algebras}. For example, the classification of the Lagrangian algebras in 2+1D finite gauge theories, originally obtained by Davydov in \cite{Dav10a}, can be easily recovered in Example\,\ref{expl:3D_finite_gauge_1codim_condensation}. As we show later, it is a powerful tool to study Lagrangian $\EE_2$-algebras in braided fusion higher categories. 
\end{rem}

\begin{rem}
The functor $\Sigma: \Algc_{\EE_2}(\CB) \to \Algc_{\EE_1}(\Sigma\CB)$ is not surjective in general. This fact is related to the fact that two Witt equivalent non-degenerate braided fusion categories cannot be obtained from each other via anyon condensations in general; however, they can be obtained from each via condensations of 1-codimensional topological defects. 
\end{rem}

When $\SC^3$ is anomalous, the condensation of a 2-codimension topological defect $A \in \Algc_{\EE_2}(\Omega\CC)$ is defined by first condensing it on a line to produce a 1-codimensional topological defect $\Sigma A$; then condensing $\Sigma A$. In this case, the equivalence (\ref{eq:E2_Mod_Omega_E1_Mod}) is not true in general because $\Omega\CC$ does not have the complete information of $\CC$ and some 1-codimensional topological defects in $\CC$ are not condensation descendants of $\Omega\CC$. However, the equivalence (\ref{eq:E2_Mod_Omega_E1_Mod_2}), i.e.,
$\Omega\Mod_{\Sigma A}^{\EE_1}(\CC) \simeq \Mod_A^{\EE_2}(\Omega\CC)$ remains correct because this equivalence is simply a reformuation of the natural physical intuition that all 2-codimensional topological defects in the condensed (also anomalous) phase $\SD^3$ can only come from $\Omega\CC$ as deconfined 2-codimensional topological defects, i.e. as $\EE_2$-$A$-modules in $\Omega\CC$. Moreover, $\SC^3$ and $\SD^3$ should share the same gravitational anomaly, i.e., $\SZ(\SC)^4=\SZ(\SD)^4$. We summarize and reformulate above results as mathematical results.

\begin{thm} \label{pthm:OmegaSigmaA-AOmega}
Let $\CC$ be a fusion $2$-category, $A\in\Algc_{\EE_2}(\Omega\CC)$ and $\Sigma A=\RMod_A(\Omega\CC)$. Then we have $\Sigma A\in \Algc_{\EE_1}(\CC)$ and a natural braided equivalence: 
\be
\Omega\Mod_{\Sigma A}^{\EE_1}(\CC) \simeq \Mod_A^{\EE_2}(\Omega\CC). 
\ee
\end{thm}
\pf
We have already obtained this result through physical arguments. Now we sketch a mathematical proof. Note that $\Omega\Mod_{\Sigma A}^{\EE_1}(\CC) \simeq \Omega\Mod_{\Sigma A}^{\EE_{1}}(\Sigma\Omega\CC)$. A 1-morphism $\Sigma A \to \Sigma A$ in $\Mod_{\Sigma A}^{\EE_{1}}(\Sigma\Omega\CC)$ is a functor from $\Sigma A$ to $\Sigma A$ intertwining the right $\Omega\CC$-action and the two-side $\Sigma A$-actions in $x^1$-dimension. Note that a functor from $\Sigma A$ to $\Sigma A$ intertwining the right $\Omega\CC$-action is precise the functor $-\otimes_A^2 m$ for an $A$-$A$-bimodule $m$ in $\Omega\CC$, where the superscript `$2$' represents in the $x^2$-dimension. Therefore, a 1-morphism $\Sigma A \to \Sigma A$ in $\Mod_{\Sigma A}^{\EE_{1}}(\Sigma\Omega\CC)$ is precisely a functor $-\otimes_A^2 m$ intertwining the two-side $\Sigma A$-actions in $x^1$-direction. 
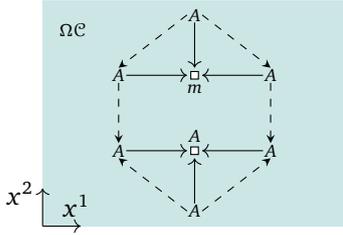
\[
\begin{array}{c}
\begin{tikzpicture}[scale=1]
\fill[teal!20] (-2,0) rectangle (2,3) ;
\draw[fill=white] (-0.05,1.95) rectangle (0.05,2.05) node[midway,below] {\scriptsize $m$} ;
\draw[fill=white] (-0.05,0.95) rectangle (0.05,1.05) node[midway,above] {\scriptsize $A$} ;
\node at (-1.6,2.6) {\scriptsize $\Omega\CC$} ;
\draw[->] (-2,0) -- (-1.5,0) node [very near end, above] {$x^1$} ;
\draw[->] (-2,0) -- (-2,0.5) node [very near end, left] {$x^2$} ;

\node at (0,2.8) {\scriptsize $A$} ; 
\node at (0,0.2) {\scriptsize $A$} ;
\node at (-1,2) {\scriptsize $A$} ;
\node at (-1,1) {\scriptsize $A$} ;
\node at (1,2) {\scriptsize $A$} ;
\node at (1,1) {\scriptsize $A$} ;

\draw[->] (0,2.7) -- (0,2.1) ; 
\draw[->] (0,0.3) -- (0,0.9) ; 
\draw[->] (-0.9,2) -- (-0.1,2) ; 
\draw[->] (-0.9,1) -- (-0.1,1) ; 
\draw[->] (0.9,2) -- (0.1,2) ; 
\draw[->] (0.9,1) -- (0.1,1) ; 

\draw[dashed,-stealth] (-0.1,2.8) -- (-1,2.1) ;
\draw[dashed,-stealth] (0.1,2.8) -- (1,2.1) ;
\draw[dashed,-stealth] (-0.1,0.2) -- (-1,0.9) ;
\draw[dashed,-stealth] (0.1,0.2) -- (1,0.9) ;

\draw[dashed,-stealth] (-1,1.9) -- (-1,1.1) ;
\draw[dashed,-stealth] (1,1.9) -- (1,1.1) ;

\end{tikzpicture}
\end{array}
\quad\quad
\begin{array}{l}
\mbox{$m$ is equipped with horizontal two-side $A$-actions} \\
\mbox{induced from the vertical right $A$-action $m\otimes^2 A \to m$.} \\
\mbox{These three $A$-actions on $m$ are illustrated as the three arrows} \\
\mbox{from $A$ to $m$ in the top triangles, which mean that the top two}\\
\mbox{triangles in diagram (\ref{diag:local-module}) (for $x=m$) is commutative.}\\
\mbox{It remains to show that the horizontal $A$-actions are also} \\
\mbox{compatible with the vertical left $A$-action $A\otimes^2 m \to m$.}
\end{array}
\]
The condition that $-\otimes_A^2 m$ intertwines the horizontal $\Sigma A$-actions says that the horizontal $A$-actions on $m$ can be equivalently defined by the horizontal $A$-actions on the $A$-factor in $A\otimes_A^2 m$. Therefore, the horizontal $A$-actions are compatible with the vertical left $A$-action $A\otimes^2 (A\otimes_A^2 m) \to m$ because $A$ itself is an $\EE_2$-$A$-module (recall the defining condition of an $\EE_2$-$A$-module in the commutative diagram (\ref{diag:local-module})). This compatibility is illustrated in the bottom triangles in the above picture. 
\epf

\begin{thm} \label{thm:ModE1SASB=SModE2AB}
For a non-degenerate braided fusion 1-category $\CB$, $A\in\Algc_{\EE_2}(\CB)$ and $\Sigma A=\RMod_A(\CB)$, we have a natural monoidal equivalence: 
\be
\Mod_{\Sigma A}^{\EE_1}(\Sigma\CB) \simeq \Sigma\Mod_A^{\EE_2}(\CB).
\ee
\end{thm}

\begin{rem} 
When $A$ is a condensable $\EE_1$-algebra in a fusion 2-fusion $\CC$, $\Sigma A \in \Sigma'\CC=\LMod_\CC(3\vect)$.  We have 
\be
\Omega_{\Sigma A}\Sigma'\CC = \Fun_{\CC}(\RMod_A(\CC), \RMod_A(\CC)) \simeq \Mod_A^{\EE_1}(\CC). 
\ee
When $A$ is a condensable $\EE_2$-algebra in braided fusion 2-category $\CB$, $\Sigma A \in \Algc_{\EE_1}(\Sigma\CB)$. We can repeat the argument to obtain $\Sigma^2 A :=\RMod_{\Sigma A}(\Sigma\CB) \in \Sigma'\Sigma\CB$. Then we obtain 
\be \label{eq:Omega2-E2Mod}
\Omega_{\Sigma^2 A}^2(\Sigma'\Sigma\CB) = \Omega \Mod_{\Sigma A}^{\EE_1}(\Sigma\CB) \simeq \Mod_A^{\EE_2}(\CB). 
\ee
The canonical equivalence (\ref{eq:Omega2-E2Mod}) provides us a convenient way to understand, compute or even define alternatively the category $\Mod_A^{\EE_2}(\CB)$ (recall Section\,\ref{sec:anyon_cond_algebra}).
\end{rem}

\begin{rem}
We have assumed the topological order $\SC^3$ to be anomaly-free and simple. If $\SC^3$ is not anomaly-free, then, in order to define a phase transition by condensing a 2-codimensional topological defect $A \in \Omega\CC$, we have to first condense $A$ along a line to create a condensed string $\Sigma A$, then condense $\Sigma A$ along the remaining transversal direction. We discuss such a condensation as a special case of more general condensations in Section\,\ref{sec:condense_k-codim_defects}. 
\end{rem}

\subsubsection{Examples} \label{sec:example_12codim_3D}

\begin{expl} \label{expl:toric_code_SigmaA}
Let us consider the 2+1D $\Zb_2$ topological order $\TC^3$, whose 2-category of topological defects is $\CT\CC$. In this case, $\Omega\CT\CC = \FZ_1(\Rep(\Zb_2))$ is a non-degenerate braided fusion 1-category with four simple objects $1,e,m,f$, where $1,e$ generate the sub-fusion category $\Rep(\Zb_2)$. By \cite{KZ22a}, there are six simple strings $\one, \vartheta, \mathrm{ss}, \mathrm{sr}, \mathrm{rs}, \mathrm{rr}$ in $\TC^2$ with the fusion rules given in (\ref{table:fusion_rule_1d_domain_wall_toric_code}). 
\bnu

\item $A_0=1$ is the trivial condensable $\EE_2$-algebra in $\Omega\CT\CC$. If we condense $A_0$ along a line, we obtain a condensed string $\one$.\footnote{Note that a condensable $\EE_1$-algebra that are Morita equivalent to $1$ (e.g., $[x,x]=x\otimes x^\ast$ for $x\in \Omega\CT\CC$) produces the `essentially same' condensed string $\one$. In general, when $A$ is only a condensable $\EE_1$-algebra, by choosing a unit morphism $\one \to \Sigma A$ different from the canonical one, the condensed defect $\Sigma A$ might be further condensed if $A$ is Morita equivalent to a condensable $\EE_2$-algebra.} 

\item $A_e=1 \oplus e$ is a condensable $\EE_2$-algebra in $\Omega\CT\CC$. If we condense $A_e$ along a line, we obtain the condensed string $\mathrm{rr}=\Sigma A_e\in \CC$. 
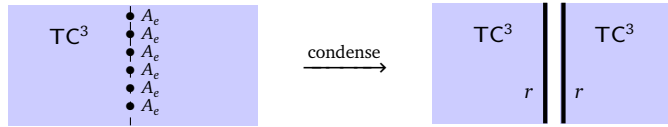
$$
\begin{array}{c}
\begin{tikzpicture}[scale=0.8]
\fill[blue!20] (-2,0) rectangle (2.1,2) ;
\draw[dashed] (0,0)--(0,2) ;
\node at (-1,1.5) {\footnotesize $\ST\SC^3$} ;
\node at (0,0.3) {\footnotesize $\bullet$}; 
\node at (0,0.6) {\footnotesize $\bullet$}; 
\node at (0,0.9) {\footnotesize $\bullet$}; 
\node at (0,1.2) {\footnotesize $\bullet$}; 
\node at (0,1.5) {\footnotesize $\bullet$}; 
\node at (0,1.8) {\footnotesize $\bullet$}; 
\node at (0.35,1.2) {\scriptsize $A_e$};
\node at (0.35,0.9) {\scriptsize $A_e$};
\node at (0.35,1.5) {\scriptsize $A_e$};
\node at (0.35,0.3) {\scriptsize $A_e$};
\node at (0.35,1.8) {\scriptsize $A_e$};
\node at (0.35,0.6) {\scriptsize $A_e$};
\end{tikzpicture}
\end{array}
\quad\xrightarrow{\mbox{\scriptsize condense}} \quad 
\begin{array}{c}
\begin{tikzpicture}[scale=0.8]
\fill[blue!20] (-2,0) rectangle (-0.1,2) ;
\fill[blue!20] (0.1,0) rectangle (2,2) ;
\draw[ultra thick] (-0.15,0)--(-0.15,2) node [near start, left] {\footnotesize $r$} ;
\draw[ultra thick] (0.15,0)--(0.15,2) node [near start, right] {\footnotesize $r$} ;
\node at (-1,1.5) {\footnotesize $\ST\SC^3$} ;
\node at (1,1.5) {\footnotesize $\ST\SC^3$} ;
\end{tikzpicture}
\end{array}
$$
As we have shown in Example\,\ref{expl:1codim_cond_TC_1}, this string $\mathrm{rr}$ is a condensable $\EE_1$-algebra that can be further condensed to create the rough boundary of $\TC^3$.

\item $A_m=1\oplus m$ is a condensable $\EE_2$-algebra in $\Omega\CT\CC$. If we condense $A_m$ along a line, we obtain a condensed string $\mathrm{ss}$. 
$$
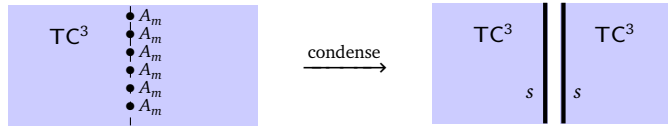
\begin{array}{c}
\begin{tikzpicture}[scale=0.8]
\fill[blue!20] (-2,0) rectangle (2.1,2) ;
\draw[dashed] (0,0)--(0,2) ;
\node at (-1,1.5) {\footnotesize $\ST\SC^3$} ;
\node at (0,0.3) {\footnotesize $\bullet$}; 
\node at (0,0.6) {\footnotesize $\bullet$}; 
\node at (0,0.9) {\footnotesize $\bullet$}; 
\node at (0,1.2) {\footnotesize $\bullet$}; 
\node at (0,1.5) {\footnotesize $\bullet$}; 
\node at (0,1.8) {\footnotesize $\bullet$}; 
\node at (0.35,1.2) {\scriptsize $A_m$};
\node at (0.35,0.9) {\scriptsize $A_m$};
\node at (0.35,1.5) {\scriptsize $A_m$};
\node at (0.35,0.3) {\scriptsize $A_m$};
\node at (0.35,1.8) {\scriptsize $A_m$};
\node at (0.35,0.6) {\scriptsize $A_m$};
\end{tikzpicture}
\end{array}
\quad\xrightarrow{\mbox{\scriptsize condense}} \quad 
\begin{array}{c}
\begin{tikzpicture}[scale=0.8]
\fill[blue!20] (-2,0) rectangle (-0.1,2) ;
\fill[blue!20] (0.1,0) rectangle (2,2) ;
\draw[ultra thick] (-0.15,0)--(-0.15,2) node [near start, left] {\footnotesize $s$} ;
\draw[ultra thick] (0.15,0)--(0.15,2) node [near start, right] {\footnotesize $s$} ;
\node at (-1,1.5) {\footnotesize $\ST\SC^3$} ;
\node at (1,1.5) {\footnotesize $\ST\SC^3$} ;
\end{tikzpicture}
\end{array}
$$
As we have shown in Example\,\ref{expl:1codim_cond_TC_1}, this string $\mathrm{ss}$ has a canonical structure of a condensable $\EE_1$-algebra in $\CC$ that can be further condensed to create the smooth boundary of $\TC^3$. 

\item $A_f=1\oplus f$ is a condensable $\EE_1$-algebra in $\Omega\CT\CC$. If we condense $A_f$ along a line, we obtain $\Sigma A_f =\vartheta$. This follows from the fact that the condensable $\EE_1$- algebra, as an internal hom, that creates the e-m-duality wall $\vartheta$ from the trivial wall is precisely $A_f=1\oplus f$ \cite{Bom10,KK12}. 
$$
\begin{array}{c}
\begin{tikzpicture}[scale=0.8]
\fill[blue!20] (-2,0) rectangle (2.1,2) ;
\draw[dashed] (0,0)--(0,2) ;
\node at (-1,1.5) {\footnotesize $\ST\SC^3$} ;
\node at (0,0.3) {\footnotesize $\bullet$}; 
\node at (0,0.6) {\footnotesize $\bullet$}; 
\node at (0,0.9) {\footnotesize $\bullet$}; 
\node at (0,1.2) {\footnotesize $\bullet$}; 
\node at (0,1.5) {\footnotesize $\bullet$}; 
\node at (0,1.8) {\footnotesize $\bullet$}; 
\node at (0.35,1.2) {\scriptsize $A_f$};
\node at (0.35,0.9) {\scriptsize $A_f$};
\node at (0.35,1.5) {\scriptsize $A_f$};
\node at (0.35,0.3) {\scriptsize $A_f$};
\node at (0.35,1.8) {\scriptsize $A_f$};
\node at (0.35,0.6) {\scriptsize $A_f$};
\end{tikzpicture}
\end{array}
\quad\xrightarrow{\mbox{\scriptsize condense}} \quad 
\begin{array}{c}
\begin{tikzpicture}[scale=0.8]
\fill[blue!20] (-2,0) rectangle (2,2) ;
\draw[dashed] (0,0)--(0,2) node [near start, right] {\footnotesize $\vartheta$} ;
\node at (-1,1.5) {\footnotesize $\ST\SC^3$} ;
\node at (1,1.5) {\footnotesize $\ST\SC^3$} ;
\end{tikzpicture}
\end{array}
$$
In this case, $\vartheta$ is not a condensable $\EE_1$-algebra in $\CT\CC$ and can not be further condensed. 

\item $A_4=1\oplus e\oplus m\oplus f$ is a condensable $\EE_1$-algebra in $\Omega\CT\CC$ isomorphic to the tensor product algebra $A_e\otimes A_m$. If we condense $A_4$ along a line, we obtain $\Sigma A_4=\mathrm{rs}$ because the internal hom algebra associated to $\mathrm{rs}$ is precisely $A_4$ \cite{JZW+19,CJKYZ20}. Note that $\Sigma A_4$ cannot be further condensed. 
$$
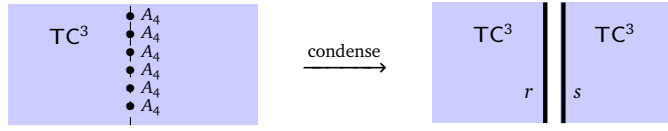
\begin{array}{c}
\begin{tikzpicture}[scale=0.8]
\fill[blue!20] (-2,0) rectangle (2.1,2) ;
\draw[dashed] (0,0)--(0,2) ;
\node at (-1,1.5) {\footnotesize $\ST\SC^3$} ;
\node at (0,0.3) {\footnotesize $\bullet$}; 
\node at (0,0.6) {\footnotesize $\bullet$}; 
\node at (0,0.9) {\footnotesize $\bullet$}; 
\node at (0,1.2) {\footnotesize $\bullet$}; 
\node at (0,1.5) {\footnotesize $\bullet$}; 
\node at (0,1.8) {\footnotesize $\bullet$}; 
\node at (0.35,1.2) {\scriptsize $A_4$};
\node at (0.35,0.9) {\scriptsize $A_4$};
\node at (0.35,1.5) {\scriptsize $A_4$};
\node at (0.35,0.3) {\scriptsize $A_4$};
\node at (0.35,1.8) {\scriptsize $A_4$};
\node at (0.35,0.6) {\scriptsize $A_4$};
\end{tikzpicture}
\end{array}
\quad\xrightarrow{\mbox{\scriptsize condense}} \quad 
\begin{array}{c}
\begin{tikzpicture}[scale=0.8]
\fill[blue!20] (-2,0) rectangle (-0.1,2) ;
\fill[blue!20] (0.1,0) rectangle (2,2) ;
\draw[ultra thick] (-0.15,0)--(-0.15,2) node [near start, left] {\footnotesize $r$} ;
\draw[ultra thick] (0.15,0)--(0.15,2) node [near start, right] {\footnotesize $s$} ;
\node at (-1,1.5) {\footnotesize $\ST\SC^3$} ;
\node at (1,1.5) {\footnotesize $\ST\SC^3$} ;
\end{tikzpicture}
\end{array}
$$

\item Note that $A_4^\op$ is also a condensable $\EE_1$-algebra in $\Omega\CT\CC$ isomorphic to the tensor product algebra $A_m\otimes A_e$. Condensing it produce the condensed string $\mathrm{sr}=\Sigma A_4^\op$, which cannot be further condensed. 
$$
\begin{array}{c}
\begin{tikzpicture}[scale=0.8]
\fill[blue!20] (-2,0) rectangle (2.1,2) ;
\draw[dashed] (0,0)--(0,2) ;
\node at (-1,1.5) {\footnotesize $\ST\SC^3$} ;
\node at (0,0.3) {\footnotesize $\bullet$}; 
\node at (0,0.6) {\footnotesize $\bullet$}; 
\node at (0,0.9) {\footnotesize $\bullet$}; 
\node at (0,1.2) {\footnotesize $\bullet$}; 
\node at (0,1.5) {\footnotesize $\bullet$}; 
\node at (0,1.8) {\footnotesize $\bullet$}; 
\node at (0.35,1.2) {\tiny $A_4^\op$};
\node at (0.35,1.8) {\tiny $A_4^\op$};
\node at (0.35,0.6) {\tiny $A_4^\op$};
\end{tikzpicture}
\end{array}
\quad\xrightarrow{\mbox{\scriptsize condense}} \quad 
\begin{array}{c}
\begin{tikzpicture}[scale=0.8]
\fill[blue!20] (-2,0) rectangle (-0.1,2) ;
\fill[blue!20] (0.1,0) rectangle (2,2) ;
\draw[ultra thick] (-0.15,0)--(-0.15,2) node [near start, left] {\footnotesize $s$} ;
\draw[ultra thick] (0.15,0)--(0.15,2) node [near start, right] {\footnotesize $r$} ;
\node at (-1,1.5) {\footnotesize $\ST\SC^3$} ;
\node at (1,1.5) {\footnotesize $\ST\SC^3$} ;
\end{tikzpicture}
\end{array}
$$

\enu
\end{expl}

\begin{expl} \label{expl:TC-to-DIsing}
We consider 2+1D double Ising topological order. 
\bnu

\item The only Lagrangian $\EE_2$-algebra in $\Omega(\ising\boxtimes\ising^\op)=\Omega\ising \boxtimes \Omega\ising^\rev$ is $A_0=1\boxtimes 1 \oplus \psi\boxtimes\psi \oplus \sigma\boxtimes \sigma$. By condensing it in an open 2-disk, we obtain the trivial phase $\mathbf{1}^3$. This condensation can be achieved in two steps. First, we condense it along a line, we obtain a condensed string $\Sigma A_0$, which can be identified with $\Omega\ising \simeq \RMod_{A_0}(\Omega\ising \boxtimes \Omega\ising^\rev) \in \Sigma\Omega(\ising\boxtimes\ising^\op)$, or, coordinate independently, as 1-codimensional defect obtained by stacking of two layers of the unique gapped boundary of the double Ising topological order. 
$$
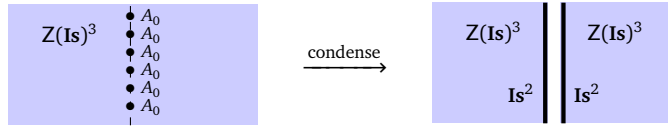
\begin{array}{c}
\begin{tikzpicture}[scale=0.8]
\fill[blue!20] (-2,0) rectangle (2.1,2) ;
\draw[dashed] (0,0)--(0,2) ;
\node at (-1,1.5) {\footnotesize $\SZ(\mathbf{Is})^3$} ;
\node at (0,0.3) {\footnotesize $\bullet$}; 
\node at (0,0.6) {\footnotesize $\bullet$}; 
\node at (0,0.9) {\footnotesize $\bullet$}; 
\node at (0,1.2) {\footnotesize $\bullet$}; 
\node at (0,1.5) {\footnotesize $\bullet$}; 
\node at (0,1.8) {\footnotesize $\bullet$}; 
\node at (0.35,1.2) {\scriptsize $A_0$};
\node at (0.35,0.9) {\scriptsize $A_0$};
\node at (0.35,1.5) {\scriptsize $A_0$};
\node at (0.35,0.3) {\scriptsize $A_0$};
\node at (0.35,1.8) {\scriptsize $A_0$};
\node at (0.35,0.6) {\scriptsize $A_0$};
\end{tikzpicture}
\end{array}
\quad\xrightarrow{\mbox{\scriptsize condense}} \quad 
\begin{array}{c}
\begin{tikzpicture}[scale=0.8]
\fill[blue!20] (-2,0) rectangle (-0.1,2) ;
\fill[blue!20] (0.1,0) rectangle (2,2) ;
\draw[ultra thick] (-0.15,0)--(-0.15,2) node [near start, left] {\footnotesize $\mathbf{Is}^2$} ;
\draw[ultra thick] (0.15,0)--(0.15,2) node [near start, right] {\footnotesize $\mathbf{Is}^2$} ;
\node at (-1,1.5) {\footnotesize $\SZ(\mathbf{Is})^3$} ;
\node at (1,1.5) {\footnotesize $\SZ(\mathbf{Is})^3$} ;
\end{tikzpicture}
\end{array}
$$
Condensing $\Sigma A_0$ produces the trivial phase $\mathbf{1}^3$ and the gapped boundary $\SM^2=\mathbf{Is}^2$, i.e., 
$$
\Mod_{\Sigma A_0}^{\EE_1}(\Sigma\Omega(\ising\boxtimes\ising^\op)) \simeq 2\vect,
\quad\quad
(\CM, m) = (\RMod_{\Sigma A_0}(\Sigma\Omega(\ising\boxtimes\ising^\op)), \Sigma A_0)
\simeq (\ising, \one). 
$$

\item The anyon condensation from double Ising to toric code was explained in Example\,\ref{expl:double_Ising_to_toric_code}. Now we reconstruct this phase transition in two steps. Recall that $A=1\boxtimes 1 \oplus \psi\boxtimes\psi \in \Omega\ising \boxtimes \Omega\ising^\rev \simeq \FZ_1(\Omega\ising)$ is a condensable $\EE_2$-algebra in $\FZ_1(\Omega\ising)$. We have $\Sigma A=\RMod_A(\FZ_1(\Omega\ising))$. By further condensing $\Sigma A$ in $\Sigma\FZ_1(\Omega\ising)=\ising \boxtimes \ising^\rev$, according to (\ref{eq:D_cond_OmM}), we obtain the $\Zb_2$ topological order as the condensed phase, i.e., 
$$
\Mod_{\Sigma A}^{\EE_1}(\Sigma\FZ_1(\Omega\ising)) \simeq \Sigma\FZ_1(\Rep(\Zb_2))=\CT\CC. 
$$
Note that $\CK=\RMod_A(\FZ_1(\Omega\ising))^\op=1\oplus \vartheta$ (recall (\ref{eq:K=1+vartheta})) is also a condensable $\EE_1$-algebra in $\CT\CC$, which is not the delooping of any condensable $\EE_2$-algebra in $\Omega\CT\CC$. Recall that condensing $\CK$ in $\ST\SC^3$ reproduces the 3D double Ising topological order, i.e., $\Mod_\CK^{\EE_1}(\CT\CC) \simeq \ising\boxtimes \ising^\op$. We explain in Example\,\ref{expl_Ising_G_crossed} that condensing $\CK$ in $\CT\CC^3$ can be understood as the gauging of the $e$-$m$-duality in $\ST\SC^3$. 

\enu
\end{expl}

\begin{expl} \label{expl:3D_Gauge_condense_E2Alg}
All the examples of anyon condensations given in Section\,\ref{sec:example_2codim_3D} can be redefined by first condensing an anyon $A$ along a line to give a condensed string $\Sigma A$ then condensing $\Sigma A$. We leave them as exercises. We explain only a special case that echo with previous examples (see Example\,\ref{expl:3D_finite_gauge_1codim_condensation}). 
\begin{itemize}

\item Let $G$ be a finite group and $H\leq G$ a subgroup. Let $A=\mathrm{Fun}(G/H)$. Note that $A\in \Algc_{\EE_2}(\FZ_1(\Rep(G)))$. The central functor $-\otimes A: \FZ_1(\Rep(G)) \to \RMod_A(\FZ_1(\Rep(G)))$ defines a condensable $\EE_1$-algebra $\Sigma A$ in $\CG\CT_G^3=\Sigma\FZ_1(\Rep(G))$; or equivalently, the monoidal functor 
$$
\Rep(G) \to \Rep(G) \boxtimes_{\FZ_1(\Rep(G))} \RMod_A(\FZ_1(\Rep(G))) \simeq \RMod_A(\Rep(G)) = \Rep(H)
$$ 
defines a condensable $\EE_1$-algebra $\Sigma A=\Rep(H)^\op$ in $\CG\CT_G^3=\BMod_{\Rep(G)|\Rep(G)}(2\vect)^\op$. By condensing $\Sigma A$ in $\CG\CT_G^3$, we obtain the condensed phase $\SD^3=\SG\ST_H^3$ with $\CD=\BMod_{\Rep(H)|\Rep(H)}(2\vect)^\op$. This coincides with the well-known result that the anyon condensation of the $\EE_2$-algebra $A$ in $\FZ_1(\Rep(G))$ produces the condensed phase $\SD^3=\SG\ST_H^3$ (recall Example\,\ref{expl:Davydov}). 

\end{itemize}
\end{expl}

\subsection{Condensation of higher codimensional topological defects} \label{sec:condense_k-codim_defects}

\subsubsection{General theory} \label{sec:main_theorem}
Consider a (potentially anomalous) $n+$1D simple topological order $\SC^{n+1}$. The category of all topological defects $\CC$ is a fusion $n$-category. 
By definition, for $k\geq 1$, the category $\Omega^{k-1}\CC$ of topological defects of codimension $k$ or higher is an $\EE_k$-fusion $(n-k+1)$-category, and $\Sigma\Omega^{k-1}\CC$ is an indecomposable separable $(n-k+2)$-category containing $1_{\one_\CC}^{k-2}$ and is the full subcategory of $\Omega^{k-2}\CC$.

\medskip
We denote the composition of $k$-morphisms in $\mathrm{B}\CC$ by $\otimes^k$. For $1\leq i\leq k$, similar to the discussion in Section\,\ref{sec:E2=E1+E1}, $\otimes^i$ also represents the fusion product of two $k$-codimensional topological defects along the $x^i$-direction. For $1\leq i,j\leq k$, if we want to discuss the fusion among $k$-codimensional topological defects along the $x^i$-direction and the $x^j$-direction, we often use the following picture in the $(x^i, x^j)$-plane: 
\be \label{pic:xi_xj_plane}
\begin{array}{c}
\begin{tikzpicture}[scale=0.8]
\fill[teal!20] (-1,0) rectangle (3,3) ;
\draw [dashed] (1,0) -- (1,3) ; 
\draw [dashed] (2,0) -- (2,3) ; 
\draw[fill=white] (0.95,0.95) rectangle (1.05,1.05) node[midway,left] {$a$} ;
\draw[fill=white] (0.95,1.95) rectangle (1.05,2.05) node[midway,left] {$b$} ;
\draw[fill=white] (1.95,0.95) rectangle (2.05,1.05) node[midway,right] {$c$} ;
\draw[fill=white] (1.95,1.95) rectangle (2.05,2.05) node[midway,right] {$d$} ;
\node at (-0.5,2.5) {\scriptsize $\Omega^{k-1}\CC$} ;
\draw[->] (-1,0) -- (-0.5,0) node [very near end, above] {$x^i$} ;
\draw[->] (-1,0) -- (-1,0.5) node [very near end, left] {$x^j$} ;
\end{tikzpicture}
\end{array}
\ee
where $a,b,c,d\in \Omega^{k-1}\CC$ and other directions of these $k$-codimensional defects $a,b,c,d$ are assumed to lie in the normal direction of above plane. By restricting to the $(x^i, x^j)$-plane, we see that these $k$-codimensional topological defects form is equipped with the following isomorphism
\be
(a\otimes^i c) \otimes^j (b\otimes^i d) \xrightarrow[\simeq]{\delta_{a,b,c,d}^{(i,j)}}
(a\otimes^j b) \otimes^i (c\otimes^j d) 
\ee
that defines the $\EE_2$-monoidal structure. These fusion products $\otimes^i$ for $1\leq i \leq k$ define the $k$-dimensional multiplication of $k$-codimensional topological defects and, therefore, define the $\EE_k$-monoidal structure on $\Omega^{k-1}\CC$.

\begin{lem} \label{lem:Sigma_functor_for_algebras}
Let $\CB$ be an $\EE_k$-multi-fusion $n$-category. The assignment $A \mapsto \Sigma A=\RMod_A(\CB)$ defines a functor:
\begin{align*}
\Sigma(-):=\RMod_-(\CB): \Algc_{\EE_k}(\CB) &\to \Algc_{\EE_{k-1}}(\Sigma\CB).
\end{align*} 
\end{lem}
\pf
Since $\CB$ is $\EE_k$-monoidal, it has fusion products $\otimes^i$ in $i$-th independent direction for $1\leq i \leq k$. We have shown in Section\,\ref{sec:2d_two_step} that $\Sigma(-): \Algc_{\EE_1}(\CB) \to \Sigma\CB$, defined as follows:  
$$
\xymatrix{
A \ar@{|->}[r] \ar[d]_f &  \RMod_A(\CB) \ar[d]^{-\otimes_A \, {}_fB} \\
B \ar@{|->}[r] & \RMod_B(\CB)  
}
$$
is a well-defined functor. We have used the $k$-th monoidal structure $\otimes^k$ on $\CB$ to define the objects in the category $\Algc_{\EE_1}(\CB)$. For $1\leq i <k$, each tensor product $\otimes^i$ endows $\Algc_{\EE_1}(\CB)$ with a monoidal structure denoted by $(\Algc_{\EE_1}(\CB), \otimes^i)$, and also endows $\RMod_A(\CB)$ with a monoidal structure denoted by $(\RMod_A(\CB), \otimes^i)$. Restricting to the $(x^i, x^k)$-plane, we have already shown in Section\,\ref{sec:2d_two_step} that $\Sigma: (\Algc_{\EE_1}(\CB), \otimes^i) \to (\Sigma\CB, \otimes^i)$ is a lax-monoidal functor, which maps algebras in $(\Algc_{\EE_1}(\CB), \otimes^i)$ to algebras in $(\Sigma\CB, \otimes^i)$. Since $\Sigma$ preserves the monoidal structure in all $i$-th directions for $1\leq i <k$, it preserves the $\EE_{k-1}$-monoidal structures on $\Algc_{\EE_1}(\CB)$ and $\Sigma\CB$. 
\epf


Now we are ready to state one of the main results of this work, which is a natural consequence of all the previous discussion in this work. 
\begin{pthm} \label{thm:main_k_codim}
In a (potentially anomalous) $n+$1D simple topological order $\SC^{n+1}$, a $k$-codimensional topological defect $A \in \Omega^{k-1}\CC$ is fully condensable, if it is equipped with the structure of a condensable $\EE_k$-algebra in $\Omega^{k-1}\CC$ (i.e., $A\in \Algc_{\EE_k}(\Omega^{k-1}\CC)$). By condensing it, we mean a $k$-step process that produces a condensed new phase $\SD^{n+1}$. 
\bnu

\item In the first step, we condense the defect $A$ along one of the transversal direction $x^k$. It means that we first proliferate the defect $A$ along the transversal direction $x^k$, then introduce interactions among these defect, or equivalently, introduce condensation maps defined by the multiplication map $A\otimes A\to A$ of the $\EE_k$-algebra in the $x^k$-direction. This produces a $(k-1)$-codimensional topological defect $\Sigma A \in \Sigma\Omega^{k-1}\CC \hookrightarrow \Omega^{k-2}\CC$. 
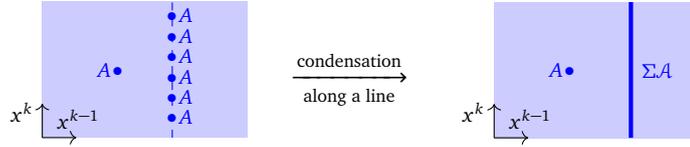
$$
\begin{array}{c}
\begin{tikzpicture}[scale=0.9]
\fill[blue!20] (0,0) rectangle (3,2) ;
\draw[->] (0,0) -- (0.5,0) node [near end,above] {\footnotesize $\quad x^{k-1}$};
\draw[->] (0,0) -- (0,0.5) node [near end,left] {\footnotesize $x^k$};
\node at (1,1) {\footnotesize \textcolor{blue}{$A\, \bullet$}} ;
\node at (2,0.3) {\footnotesize \textcolor{blue}{$\bullet\, A$}} ;
\node at (2,0.6) {\footnotesize \textcolor{blue}{$\bullet\, A$}} ;
\node at (2,0.9) {\footnotesize \textcolor{blue}{$\bullet\, A$}} ;
\node at (2,1.2) {\footnotesize \textcolor{blue}{$\bullet\, A$}} ;
\node at (2,1.5) {\footnotesize \textcolor{blue}{$\bullet\, A$}} ;
\node at (2,1.8) {\footnotesize \textcolor{blue}{$\bullet\, A$}} ;
\draw[blue, dashed] (1.9,0) -- (1.9,2) ;
\end{tikzpicture}
\end{array}
\quad
\xrightarrow[\text{\scriptsize along a line}]{\text{\scriptsize condensation}}
\quad
\begin{array}{c}
\begin{tikzpicture}[scale=0.9]
\fill[blue!20] (0,0) rectangle (3,2) ;
\draw[->] (0,0) -- (0.5,0) node [near end,above] {\footnotesize $\quad x^{k-1}$};
\draw[->] (0,0) -- (0,0.5) node [near end,left] {\footnotesize $x^k$};
\node at (1,1) {\footnotesize \textcolor{blue}{$A\, \bullet$}} ;
\draw[blue, ultra thick] (2,0) -- (2,2) node [midway,right] {\footnotesize $\Sigma\CA$}; 
\end{tikzpicture}
\end{array}
$$
If we use the coordinate system $\Sigma\Omega^{k-1}\CC=\RMod_{\Omega^{k-1}\CC}((n-k+2)\vect)$, the object $\Sigma A$ can be defined as follows: 
\be
\Sigma A:=\RMod_A(\Omega^{k-1}\CC) \in \RMod_{\Omega^{k-1}\CC}((n-k+2)\vect) \hookrightarrow \Omega^{k-2}\CC. 
\ee

\item It turns out that $\Sigma A$ has a natural structure of a condensable $\EE_{k-1}$-algebra in $\Sigma\Omega^{k-1}\CC$ or in $\Omega^{k-2}\CC$. In the second step, we condense $\Sigma A$ along one of the remaining transversal direction $x^{k-1}$ to produce a $(k-2)$-codimensional topological defect $\Sigma^2 A := \RMod_{\Sigma A}(\Omega^{k-2}\CC) \in \Sigma\Omega^{k-2}\CC \hookrightarrow \Omega^{k-3}\CC$.
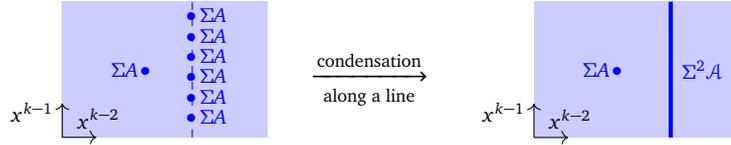
$$
\begin{array}{c}
\begin{tikzpicture}[scale=0.9]
\fill[blue!20] (0,0) rectangle (3,2) ;
\draw[->] (0,0) -- (0.5,0) node [near end,above] {\footnotesize $\quad x^{k-2}$};
\draw[->] (0,0) -- (0,0.5) node [near end,left] {\footnotesize $x^{k-1}$};
\node at (1,1) {\footnotesize \textcolor{blue}{$\Sigma A\, \bullet$}} ;
\node at (2.1,0.3) {\footnotesize \textcolor{blue}{$\bullet\, \Sigma A$}} ;
\node at (2.1,0.6) {\footnotesize \textcolor{blue}{$\bullet\, \Sigma A$}} ;
\node at (2.1,0.9) {\footnotesize \textcolor{blue}{$\bullet\, \Sigma A$}} ;
\node at (2.1,1.2) {\footnotesize \textcolor{blue}{$\bullet\, \Sigma A$}} ;
\node at (2.1,1.5) {\footnotesize \textcolor{blue}{$\bullet\, \Sigma A$}} ;
\node at (2.1,1.8) {\footnotesize \textcolor{blue}{$\bullet\, \Sigma A$}} ;
\draw[blue, dashed] (1.9,0) -- (1.9,2) ;
\end{tikzpicture}
\end{array}
\quad
\xrightarrow[\text{\scriptsize along a line}]{\text{\scriptsize condensation}}
\quad
\begin{array}{c}
\begin{tikzpicture}[scale=0.9]
\fill[blue!20] (0,0) rectangle (3,2) ;
\draw[->] (0,0) -- (0.5,0) node [near end,above] {\footnotesize $\quad x^{k-2}$};
\draw[->] (0,0) -- (0,0.5) node [near end,left] {\footnotesize $x^{k-1}$};
\node at (1,1) {\footnotesize \textcolor{blue}{$\Sigma A\, \bullet$}} ;
\draw[blue, ultra thick] (2,0) -- (2,2) node [midway,right] {\footnotesize $\Sigma^2\CA$}; 
\end{tikzpicture}
\end{array}
$$


\item Repeat above process until $\Sigma^{k-1}A$, which is a condensable $\EE_1$-algebra in $\CC$. In the $k$-th step, we condense $\Sigma^{k-1}A$ in $\CC$ along the only remaining transversal direction $x^1$. We obtain a new phase $\SD^{n+1}$ and a gapped domain wall $\SM^n=(\CM,m)$ 
such that 
\be \label{pic:C-M-D_II}
\begin{array}{c}
\begin{tikzpicture}[scale=1.7]
\fill[blue!20] (1,0) rectangle (3,1.2) ;
\draw[->] (1,0) -- (1.3,0) node [near end,above] {\footnotesize $\quad x^1$};
\draw[->] (1,0) -- (1,0.3) node [near end,left] {\footnotesize $x^2$};
\node at (1.2,0.6) {\scriptsize $\SC^{n+1}$} ;
\node at (2,-0.1) {\scriptsize $\SM^n=(\CM,m)$} ;
\fill[teal!20] (2.02,0) rectangle (3,1.2) ;
\draw[blue, ultra thick,->-] (2,0) -- (2,1.2) node [midway,left] {\scriptsize $\SM^n$}; 
\node at (2.8,0.6) {\scriptsize $\SD^{n+1}$} ;
\node at (1.5,1) {\scriptsize $\CC \xrightarrow{L_1} \CM$} ;
\node at (2.5,1) {\scriptsize $\CM \xleftarrow{R_1} \CD$} ;
\end{tikzpicture}
\end{array}
\ee
\be \label{eq:CD_CM_1}
\CD \simeq \Mod_{\Sigma^{k-1}A}^{\EE_1}(\CC) \quad \mbox{and} \quad \CM=\RMod_{\Sigma^{k-1}A}(\CC) =\Sigma^k A, \quad m=\Sigma^{k-1}A. 
\ee

\item For $1\leq i \leq k$, $i$-codimensional topological defects in $\Omega^{i-1}\CC$ move onto the wall according to the functor 
$$
L_i :=  -\otimes \Sigma^{k-i}A: \Omega^{i-1}\CC \to \Omega_m^{i-1}\CM;
$$ 
and those in $\SD^{n+1}$ move onto the wall according to the functor 
$$
R_i  := \Sigma^{k-i}A \otimes_{\Sigma^{k-i}A} -: \Omega^{i-1}\CD \to \Omega_m^{i-1}\CM.
$$ 
The functors $L_1$ and $R_1$ is illustrated in (\ref{pic:C-M-D_II}).

\item $\SC^{n+1}$ and $\SD^{n+1}$ share the same gravitational anomaly. In other words, they are gapped boundaries of the same anomaly-free $n+$2D topological order as illustrated below,
$$
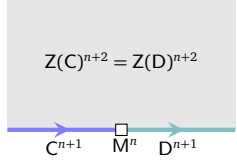
\begin{array}{c}
\begin{tikzpicture}[scale=1.5]
\fill[gray!20] (0,0) rectangle (2,1.2) ;
\node at (1,0.6) {\scriptsize $\SZ(\SC)^{n+2}=\SZ(\SD)^{n+2}$} ;
\draw[blue!50, ultra thick,->-] (0,0) -- (1,0) ; 
\draw[teal!50, ultra thick,->-] (1,0) -- (2,0) ; 
\draw[fill=white] (0.95,-0.05) rectangle (1.05,0.05) node[midway,below,scale=1] {\scriptsize $\hspace{1mm}\SM^n$} ;
\node at (0.5,-0.15) {\scriptsize $\SC^{n+1}$} ;
\node at (1.5,-0.15) {\scriptsize $\SD^{n+1}$} ;
\end{tikzpicture}
\end{array}
$$
where $\SZ(\SC)^{n+2}$ and $\SZ(\SD)^{n+2}$ denote the bulk of $\SC^{n+1}$ and $\SD^{n+1}$, respectively. Mathematically, it implies the following results. 
\begin{align}
\FZ_1(\CC) &\simeq \FZ_1(\Mod_{\Sigma^{k-1}A}^{\EE_1}(\CC)), \\
\FZ_0(\RMod_{\Sigma^{k-1}A}(\CC)) &\simeq \Mod_{\Sigma^{k-1}A}^{\EE_1}(\CC)^\rev \boxtimes_{\FZ_1(\CC)} \CC, \\
\CD^\rev \simeq \Mod_{\Sigma^{k-1}A}^{\EE_1}(\CC)^\rev 
&\simeq \FZ_1(\CC, \FZ_0(\RMod_{\Sigma^{k-1}A}(\CC))).
\end{align}

\item If $\SC^{n+1}$ is anomaly-free, then one can alternatively terminate the process at $\Sigma^{k-2} A$, which is a condensable $\EE_2$-algebra in $\Omega\CC$. More precisely, we can condense $\Sigma^{k-2}A \in \Algc_{\EE_2}(\Omega\CC)$ directly to obtain the same new phase $\SD^{n+1}$ and the same gapped domain wall such that 
\[
\begin{array}{c}
\begin{tikzpicture}[scale=0.8]
\fill[blue!30] (-2,0) rectangle (2,3) ;
\node at (-1.5,2.5) {\scriptsize $\SC^{n+1}$} ;
\node at (0,0.9) {\scriptsize \textcolor{teal!20}{$\bullet$}} ;
\node at (0,1.2) {\scriptsize \textcolor{teal!20}{$\bullet$}} ;
\node at (0,1.5) {\scriptsize \textcolor{teal!20}{$\bullet$}} ;
\node at (0,1.8) {\scriptsize \textcolor{teal!20}{$\bullet$}} ;
\node at (0,2.1) {\scriptsize \textcolor{teal!20}{$\bullet$}} ;
\node at (0,2.4) {\scriptsize \textcolor{teal!20}{$\bullet$}} ;
\node at (0,0.6) {\scriptsize \textcolor{teal!20}{$\bullet$}} ;
\node at (0.3,1.5) {\scriptsize \textcolor{teal!20}{$\bullet$}} ;
\node at (-0.3,1.5) {\scriptsize \textcolor{teal!20}{$\bullet$}} ;
\node at (0.6,1.5) {\scriptsize \textcolor{teal!20}{$\bullet$}} ;
\node at (-0.6,1.5) {\scriptsize \textcolor{teal!20}{$\bullet$}} ;
\node at (-0.6,1.8) {\scriptsize \textcolor{teal!20}{$\bullet$}} ;
\node at (-0.6,2.1) {\scriptsize \textcolor{teal!20}{$\bullet$}} ;
\node at (-0.6,1.2) {\scriptsize \textcolor{teal!20}{$\bullet$}} ;
\node at (-0.6,0.9) {\scriptsize \textcolor{teal!20}{$\bullet$}} ;
\node at (-0.9,1.5) {\scriptsize \textcolor{teal!20}{$\bullet$}} ;
\node at (0.9,1.5) {\scriptsize \textcolor{teal!20}{$\bullet$}} ;
\node at (0.3,1.2) {\scriptsize \textcolor{teal!20}{$\bullet$}} ;
\node at (-0.3,1.2) {\scriptsize \textcolor{teal!20}{$\bullet$}} ;
\node at (0.3,1.8) {\scriptsize \textcolor{teal!20}{$\bullet$}} ;
\node at (-0.3,1.8) {\scriptsize \textcolor{teal!20}{$\bullet$}} ;
\node at (-0.85,1.8) {\scriptsize \textcolor{teal!20}{$\bullet$}} ;
\node at (-0.85,1.2) {\scriptsize \textcolor{teal!20}{$\bullet$}} ;
\node at (-0.3,2.1) {\scriptsize \textcolor{teal!20}{$\bullet$}} ;
\node at (-0.3,2.35) {\scriptsize \textcolor{teal!20}{$\bullet$}} ;
\node at (-0.3,0.9) {\scriptsize \textcolor{teal!20}{$\bullet$}} ;
\node at (-0.3,0.65) {\scriptsize \textcolor{teal!20}{$\bullet$}} ;
\node at (0.6,1.8) {\scriptsize \textcolor{teal!20}{$\bullet$}} ;
\node at (0.6,2.1) {\scriptsize \textcolor{teal!20}{$\bullet$}} ;
\node at (0.6,1.2) {\scriptsize \textcolor{teal!20}{$\bullet$}} ;
\node at (0.6,0.9) {\scriptsize \textcolor{teal!20}{$\bullet$}} ;
\node at (0.3,2.1) {\scriptsize \textcolor{teal!20}{$\bullet$}} ;
\node at (0.3,2.35) {\scriptsize \textcolor{teal!20}{$\bullet$}} ;
\node at (0.3,0.9) {\scriptsize \textcolor{teal!20}{$\bullet$}} ;
\node at (0.3,0.65) {\scriptsize \textcolor{teal!20}{$\bullet$}} ;
\node at (0.85,1.8) {\scriptsize \textcolor{teal!20}{$\bullet$}} ;
\node at (0.85,1.2) {\scriptsize \textcolor{teal!20}{$\bullet$}} ;

\node at (1.2,0.2) {\scriptsize $\Sigma^{k-2}A\,$\textcolor{teal!20}{$\bullet$}} ;

\draw[dashed] (1,1.5) arc (360:0:1) ;
\draw[->] (-2,0) -- (-1.5,0) ; 
\node at (-1.3,0.3) {\scriptsize $x^1$};
\draw[->] (-2,0) -- (-2,0.5) node [near end,left] {\scriptsize $x^2$};
\end{tikzpicture}
\end{array}
\quad\xrightarrow{\mbox{\footnotesize 2-codim condensation}} \quad
\begin{array}{c}
\begin{tikzpicture}[scale=0.8]
\fill[blue!30] (-2,0) rectangle (2,3) ;
\fill[teal!20] (0,1.5) circle (1);
\node at (-1.5,2.5) {\scriptsize $\SC^{n+1}$} ;
\node at (0,1.5) {\scriptsize $\SD^{n+1}$} ;
\node at (-0.9,0.5) {\scriptsize $\SM^n$} ;
\draw[blue, ultra thick,->-] (1,1.5) arc (360:0:1) ;
\end{tikzpicture}
\end{array}
\]
\be \label{eq:CD_CM_2}
\Omega\CD \simeq \Mod_{\Sigma^{k-2}A}^{\EE_2}(\Omega\CC) \quad \mbox{and} \quad 
\Omega_m\CM = \RMod_{\Sigma^{k-2}A}(\Omega\CC). 
\ee
The consistency of two approaches immediately implies the following natural equivalences: 
\be
\Sigma\Mod_{\Sigma^{k-2}A}^{\EE_2}(\Omega\CC) \simeq \Mod_{\Sigma^{k-1}A}^{\EE_1}(\CC) \quad \mbox{and} \quad 
\Sigma\RMod_{\Sigma^{k-2}A}(\Omega\CC) \simeq \RMod_{\Sigma^{k-1}A}(\CC),
\ee
where the first equivalence is a monoidal equivalence. 
\enu
\end{pthm}

\begin{rem}
When the $k$-codimensional topological defect $A \in \Omega^{k-1}\CC$ is only a condensable $\EE_i$-algebra in $\Omega^{k-1}\CC$ for $1\leq i < k$, then condensing $A$ can only produce lower codimensional defects $\Sigma^j A$ in $\SC^{n+1}$ for $i\leq j<k$ instead of a new phase $\SD^{n+1}$. In this sense, $A$ can be called `$i$-condensable'. When $i=k$, the defect $A$ is called {\it fully condensable}. 
\end{rem}

We first provide a physical or working definition of an $\EE_m$-module over an $\EE_m$-algebra. 
\begin{defn} \label{def:Ek_Mod}
Given $\CA\in\Alg_{\EE_m}((n+1)\vect)$ and $A\in \Alg_{\EE_m}(\CA)$. 
For $m=1$, an $\EE_1$-module over $A$ in $\CA$ is just an $A$-$A$-bimodule in $\CA$. For $m=2$, an $\EE_2$-module over $A$ in $\CA$ is defined in Section\,\ref{sec:anyon_cond_algebra}. For $m\geq 3$, an $\EE_m$-module over $A$ (or an $\EE_m$-$A$-module) in $\CA$ is an object $M$ in $\CA$ equipped with two-side $A$-actions in the $x^i$-dimension for all $1\leq i \leq m$ such that $M$ is an $\EE_2$-module over $A$ in each $(x^i,x^j)$-plane for $1\leq i<j\leq m$ (recall Remark\,\ref{rem:def_E2_Mod}). 
\end{defn}

We denote the category of $\EE_m$-module over $A$ by $\Mod_A^{\EE_m}(\CA)$. When $m=0$, we set $\Mod_A^{\EE_0}(\CA):=(\CA,A)$. For $m\geq 2$, we would like to provide an alternative way to understand $\Mod_A^{\EE_m}(\CA)$. For an $\EE_m$-multi-fusion $n$-category $\CA$ and $A,B\in \Algc_{\EE_m}(\CA)$, there is a natural equivalence of $(n+1)$-categories (recall Eq.\, (\ref{eq:BMod=LMod-tensor-RMod_1})): 
$$
\BMod_{A|B}(\CA) \simeq \LMod_A(\CA) \boxtimes_\CA \RMod_B(\CA). 
$$
The $\BMod_{A|B}(\CA)$ inherits a natural structure of an $\EE_{m-1}$-multi-fusion $n$-category from those on $\LMod_A(\CA)$ and $\RMod_A(\CA)$. More explicitly, it is equipped with a tensor unit $A\otimes B$ and a tensor product $\otimes_{A|B}^i$ in the $x^i$-dimension induced from $\otimes^i_A \boxtimes_\CA \otimes^i_B$ in $\LMod_A(\CA) \boxtimes_\CA \RMod_B(\CA)$ for all $1\leq i \leq m-1$. The tensor product $\otimes_A^i$ in $\LMod_A(\CA)$ is induced from the two-side $A$-actions on modules $A\otimes^i M \to M$ and $M\otimes^i A \to M$, both of which are induced from the left $A$-action $A\otimes^m M \to M$ in the $x^m$-dimension via diabatically moving $A$. A special case of $m=2$ is illustrated in the picture in (\ref{eq:pic-tpover_A-tpover_B}). The tensor products $\otimes_B^i$ are similar.

When $A=B$, we obtain $A\in \Algc_{\EE_{m-1}}(\Mod_A^{\EE_1}(\CA))$, and $A$ is a direct summand of the tensor unit $A\otimes A$. Note that, at the same time, $\Mod_A^{\EE_1}(\CA)$ has another natural structure of $\EE_1$-multi-fusion $n$-category with the tensor product $\otimes_A^m$ (defined in the $x^m$-dimension) and the tensor unit $A$.

If we repeat this process $j$-times, the category $\Mod_A^{\EE_1}(\cdots \Mod_A^{\EE_1}(\CA))$ has two different (higher) monoidal structures: 
\bnu
\item[-] an $\EE_{m-j}$-monoidal structure with the tensor unit $A\otimes A$ and a tensor product in each $x^i$-direction for $i=1,\cdots, m-j$ such that they are all compatible; 
\item[-] an $\EE_j$-monoidal structure with a tensor unit $A$ and a tensor product $\otimes_A^i$ in the $x^i$-dimension for $i=m-j+1, \cdots, m$ such that they are all compatible. 
\enu
Therefore, we obtain a generalization of (\ref{def:Mod_E2}) and an alternative way to define $\Mod_A^{\EE_m}(\CA)$ as summarized in the following Proposition. 
\begin{prop} \label{cor:Em_module=mE1}
We have an $\EE_m$-monoidal equivalence 
$$
\Mod_A^{\EE_m}(\CA) \simeq \Mod_A^{\EE_{m-1}}(\Mod_A^{\EE_1}(\CA)) 
\simeq \Mod_A^{\EE_1}(\cdots \Mod_A^{\EE_1}(\CA)).
$$ 
\end{prop}

\begin{rem}
It is possible to give a more compact mathematical definition of $\Mod_A^{\EE_m}(\CA)$ as an iterated looping of the category of $\EE_0$-modules over an $\EE_0$-algebra. We will do that in \cite{KZZZ25}.  
\end{rem}

When $\SC^{n+1}$ is anomalous, the condensable $\EE_k$-algebra $A$ in $\Omega^{k-1}\CC$ does not directly determines the phase transition without going through the condensed defect $\Sigma^{k-1}A$. However, it determines some partial data of the condensed phase directly.  
\bnu
\item Since all (deconfined) topological defects of codimension $k$ or higher in the condensed phase $\SD^{n-1}$ necessarily come from those in $\Omega^{k-1}\CC$, and they are necessarily the $\EE_k$-modules over $A$, we must have the following $\EE_k$-monoidal equivalence: 
\be \label{eq:EkMod_OmegaE1Mod}
\Mod_A^{\EE_k}(\Omega^{k-1}\CC) \simeq \Omega^{k-1}\Mod_{\Sigma^{k-1}A}^{\EE_1}(\CC) = \Omega^{k-1}\Sigma^k A. 
\ee

\item Topological defects living on the gapped domain wall $\SM^n$ are confined topological defects. Those $k$-codimensional topological defect on $\SM^n$, i.e., objects in $\Omega_m^{k-1}\CM$, are necessarily originated from those in $\Omega^{k-1}\CC$. Recall that a deconfined $k$-codimensional defect need support an $A$-action in all $k$-dimensions (with two mutually-opposite directions in each dimension). If we drop the $A$-action in one single direction, the associated topological defect is not deconfined. On the other hand, the minimal requirement for a topological defect to survive on the wall $\SM^n$ is that it should support an $A$-action at least from one single direction (see further discussion in Remark\,\ref{rem:different_level_confinement}). Therefore, those defects in $\Omega_m^{k-1}\CM$ are necessarily right $A$-modules in $\Omega^{k-1}\CC$, i.e., 
\be \label{eq:RMod=OmegaRModSigma}
\RMod_A(\Omega^{k-1}\CC) \simeq \Omega^{k-1} \RMod_{\Sigma^{k-1}A}(\CC) = \Omega^{k-1}\Sigma^k A. 
\ee
As a consequence, we also have 
\be \label{eq:RMod=OmegaRModSigma_2}
\Omega_m^i\CM = \Omega^i \Sigma^k A = \Omega^i \RMod_{\Sigma^{k-1}A}(\CC) 
\simeq
\RMod_{\Sigma^{k-1-i}A}(\Omega^i\CC), 
\quad \mbox{for $0\leq i \leq k-1$.}
\ee
\enu

\begin{rem} \label{rem:different_level_confinement}
The condition for an object in $\Omega^{k-1}\CC$ to live on the wall $\SM^n$ is that it supports an $A$-action from at least a single direction. It can be equipped with compatible $A$-actions from other directions. When it does not support compatible $A$-actions in all $2k$ directions in all $k$-dimensions, say the compatibility of $A$-action in one of the two directions in the $x^k$-dimension is dropped, even though one can still define an $A$-action in this direction using the other $A$-action in the $x^k$-dimension via a braiding in $(x^i, x^k)$-plane, this $A$-action depends on the paths of the braidings. The incompatibility of different paths causes a strong interference that triggers the confinement. At the same time, one can also see that $(k-1)$-codimensional defects in $\Omega^{k-1}_m\CM$ can have different levels of confinement. More precisely, assume the domain wall $\SM^n$ is defined by the hyperplane defined by $x^k=x_0^k$ in space. Then all $(k-1)$-codimensional defects in $\Omega^{k-1}_m\CM$ are confined in the $x^k$-dimension. For $1\leq i <k$, if a $(k-1)$-codimensional defect in $\Omega^{k-1}_m\CM$ supports an two-side $A$-action in the $x^i$-dimension that are compatible with the $A$-action in one of the $x^k$-direction, then it is deconfined in the $x^i$-dimension. 
\end{rem}

We summarize and reformulate above results as mathematical results. 
\begin{thm} \label{thm:Ek_A=OEk-1_SigmaA}
Let $\CC$ be a fusion $n$-category and $A\in\Algc_{\EE_k}(\Omega^{k-1}\CC)$. For $1\leq i< k$, we set $\Sigma^0 A:=A$ and define inductively
$\Sigma^i A=\RMod_{\Sigma^{i-1}A}(\Omega^{k-i}\CC)$.  
Then $\Sigma^i A\in \Algc_{\EE_{k-i}}(\Omega^{k-i-1}\CC)$, and there is a natural $\EE_{k-i}$-monoidal equivalence: 
\be \label{eq:ModEk=OmegaModEk-1}
\Mod_{\Sigma^i A}^{\EE_{k-i}}(\Omega^{k-i-1}\CC) \simeq \Omega^{k-i-1}\Mod_{\Sigma^{k-1} A}^{\EE_1}(\CC). 
\ee
\end{thm}
\pf
We obtain this result through physical arguments. Now we sketch a mathematical proof of (\ref{eq:ModEk=OmegaModEk-1}) parallel to that of Theorem\,\ref{pthm:OmegaSigmaA-AOmega}. 
It is enough to prove the following $\EE_k$-monoidal equivalence: 
\be
\Mod_A^{\EE_k}(\Omega^{k-1}\CC) \simeq \Omega\Mod_{\Sigma A}^{\EE_{k-1}}(\Omega^{k-2}\CC). 
\ee
Note that $\Omega\Mod_{\Sigma A}^{\EE_{k-1}}(\Omega^{k-2}\CC) \simeq \Omega\Mod_{\Sigma A}^{\EE_{k-1}}(\Sigma\Omega^{k-1}\CC)$. A 1-morphism $\Sigma A \to \Sigma A$ in $\Mod_{\Sigma A}^{\EE_{k-1}}(\Sigma\Omega^{k-1}\CC)$ is a functor from $\Sigma A$ to $\Sigma A$ intertwining the right $\Omega^{k-1}\CC$-action and the $(k-1)$-dimensional $\Sigma A$-action. Since a functor from $\Sigma A$ to $\Sigma A$ intertwining the right $\Omega^{k-1}\CC$-action is necessarily equivalent to the functor $-\otimes_A^k m$ for an $A$-$A$-bimodule $m$ in $\Omega^{k-1}\CC$, a 1-morphism $\Sigma A \to \Sigma A$ in $\Mod_{\Sigma A}^{\EE_{k-1}}(\Sigma\Omega^{k-1}\CC)$ is precisely a functor $-\otimes_A^k m$ intertwining the $\Sigma A$-actions in the $x^i$-dimension for all $1\leq i<k$, a condition which upgrades $m$ to an $\EE_2$-$A$-module in $\Omega^{k-1}\CC$ in each $(x^i,x^k)$-plane as shown in the proof of Theorem\,\ref{pthm:OmegaSigmaA-AOmega}. Hence, $m$ is an $\EE_k$-$A$-module in $\Omega^{k-1}\CC$. 
\epf

If $\CC=\Sigma^{k-1}\Omega^{k-1}\CC$, i.e., topological defects outside $\Omega^{k-1}\CC$ are all condensation descendants of those in $\Omega^{k-1}\CC$, then we have the following stronger results. 
\bnu
\item It is natural to ask if all deconfined topological defects outside $\Omega^{k-1}\CD$ are also condensation descendants of those in $\Omega^{k-1}\CD$. It turns out that this is not correct because, for $i<k-1$, 
$\Omega^i\CD$ is not connected in general even though $\Omega^i\CC$ is connected. We give an example of this fact. Consider $\SC^5=\mathbf{1}^5$ and $A=\Rep(G)\in 2\vect=\Omega^2\CC$, then $\Sigma A = 2\Rep(G) \in 3\vect = \Omega\CC$ and $\Sigma^2 A = 3\Rep(G)$. In this case, we have $\Mod_{\Sigma A}^{\EE_2}(3\vect) \simeq \FZ_1(3\Rep(G))$, which is not indecomposable as a separable 3-category \cite{KTZ20}. Therefore, $\Mod_{\Sigma A}^{\EE_2}(3\vect) \neq \Sigma \Mod_A^{\EE_3}(2\vect)$. Moreover, for $i=0$, it was known that every fusion 2-category is Morita equivalent to a connected one (i.e., $\Sigma\CB$ for a braided fusion 1-category $\CB$) \cite{Dec23b}.


\item In this case, all confined topological defects on the gapped domain wall are necessarily condensation descendants of those in $\Omega_m^{k-1}(\CM)$. This fact will be explained mathematically in \cite{KZZZ25}. Therefore, we obtain the following natural equivalence: 
\be \label{eq:Sigma_distribute_2}
\Sigma^{k-1}\RMod_A(\Omega^{k-1}\CC) \simeq \RMod_{\Sigma^{k-1}A}(\CC).
\ee
Note that 
(\ref{eq:RMod=OmegaRModSigma}) is an immediate consequence of 
(\ref{eq:Sigma_distribute_2}), respectively. A special case of (\ref{eq:Sigma_distribute_2}) will be useful later. For a subgroup $H$ of a finite group $G$, we have
\be \label{eq:nRep(G)-nRep(H)}
\RMod_{n\Rep(H)}((n+1)\Rep(G)) \simeq \Sigma^n(\RMod_{\mathrm{Fun}(G/H)}(\Rep(G))) \simeq \Sigma^n(\Rep(H)) \simeq n\Rep(H). 
\ee
\enu

We reformulate (\ref{eq:EkMod_OmegaE1Mod}), (\ref{eq:RMod=OmegaRModSigma}) and 
(\ref{eq:Sigma_distribute_2}) as mathematical results below. 
\begin{thm} \label{thm:Ek-fusion_nCat_EkAlg_A_condense}
Let $\CB$ be an $\EE_k$-fusion $n$-category and $A\in\Algc_{\EE_k}(\CB)$. We set $\Sigma^0 A=A$ and, for $0\leq i< k$, we define inductively $\Sigma^i A=\RMod_{\Sigma^{i-1}A}(\Sigma^{i-1}\CB)$. We have a natural $\EE_{k-i}$-monoidal equivalence: 
\be \label{eq:Eki_from_Ek}
\Mod_{\Sigma^i A}^{\EE_{k-i}}(\Sigma^i\CB) \simeq \Omega^{k-i}\Mod_{\Sigma^k A}^{\EE_0}(\Sigma^k\CB),
\ee
which is proved in the proof of Theorem\,\ref{thm:Ek_A=OEk-1_SigmaA}. Moreover, we have a natural $\EE_{k-1-i}$-monoidal equivalence and an $\EE_{k-1}$-monoidal equivalence, respectively, given as follows:  
\be
\Sigma^i\RMod_A(\CB) \simeq \RMod_{\Sigma^i A}(\Sigma^i\CB), \quad\quad
\RMod_A(\CB) \simeq \Omega^i\RMod_{\Sigma^i A}(\Sigma^i\CB). 
\ee
\end{thm}

\begin{expl} \label{expl:Ek-module_over_nRep(G)}
The $\EE_k$-monoidal equivalence \eqref{eq:Eki_from_Ek} provides a convenient way to understand and compute the category $\Mod_A^{\EE_k}(\CB)$. It can also serves as a working definition of $\Mod_A^{\EE_k}(\CB)$. We give an example. Let $A=n\Rep(G)$, which can be viewed as a condensable $\EE_k$-algebra in $(n+1)\vect$ for all $k\geq 1$. We have the following results.
\bnu

\item $\Mod_{n\Rep(G)}^{\EE_1}((n+1)\vect)  \simeq \CG\CT_G^{n+2}$. 

\item $\Mod_{n\Rep(G)}^{\EE_2}((n+1)\vect) \simeq \Omega\Mod_{(n+1)\Rep(G)}^{\EE_1}((n+2)\vect) \simeq \FZ_1((n+1)\Rep(G)) \simeq \Omega\CG\CT_G^{n+2}$. 

\item For $k>2$, we have
\be
\Mod_{n\Rep(G)}^{\EE_k}((n+1)\vect) \simeq (n+1) \Rep(G)
\ee
which follows from
$$
\Mod_{n\Rep(G)}^{\EE_k}((n+1)\vect) \simeq
\Omega^{k-1}(\Mod_{(n+k-1)\Rep(G)}^{\EE_1}((n+k)\vect)) \simeq \Omega^{k-2}
\FZ_1((n+k-1)\Rep(G)). 
$$
\enu
\end{expl}


The following result generalizes the equivalence (\ref{eq:E1Mod_Z1Z0}). 
\begin{thm} \label{cor:EkMod_Zk_centralizer}
Let $\CB$ be an $\EE_k$-fusion $n$-category and $A\in\Algc_{\EE_k}(\CB)$. We have 
\be 
\Mod_A^{\EE_k}(\CB)^\rev \simeq \FZ_k(\CB, \FZ_{k-1}(\RMod_A(\CB))). 
\ee
When $\CB=n\vect$, we obtain $\Mod_A^{\EE_k}(n\vect)^\rev \simeq \FZ_{k-1}(\Sigma A)$ (recall Example\,\ref{expl:Z1Z0_centralizer}). 
\end{thm}
\pf
It follows from the the following equivalences: 
\begin{align*} 
\Mod_A^{\EE_k}(\CB)^\rev &\simeq \Omega^{k-1}(\Mod_{\Sigma^{k-1}A}^{\EE_1}(\Sigma^{k-1}\CB)^\rev) \\
&\simeq \Omega^{k-1} \FZ_1(\Sigma^{k-1}\CB, \FZ_0(\RMod_{\Sigma^{k-1}A}(\Sigma^{k-1}\CB))) \\
&\simeq \FZ_k(\CB, \Omega^{k-1}\FZ_0(\Sigma^{k-1}\RMod_A(\CB))) \\
&\simeq \FZ_k(\CB, \FZ_{k-1}(\RMod_A(\CB))),
\end{align*}
where the first `$\simeq$' is due to \eqref{eq:Eki_from_Ek}, 
the second `$\simeq$' is due to \eqref{eq:E1Mod_Z1Z0}, the third `$\simeq$' is due to \eqref{eq:FZm-FZm-1} and the 4th `$\simeq$' is due to \eqref{eq:Zm_center}. 
\epf

\begin{rem}
The special cases of $n=2$ and $k=1,2,3$ of Theorem\,\ref{cor:EkMod_Zk_centralizer} are also obtained independently in \cite{Xu24}, which appeared on arXiv on the same day as ours. These results generalize similar results for 1-categories obtained much earlier in \cite{Sch01}. 
\end{rem}

\begin{expl}
When $\CA=(n-1)\Rep(G)$ and $\CB=n\vect$, we have 
\begin{align*}
\Mod_{(n-1)\Rep(G)}^{\EE_1}(n\vect)^\rev &\simeq \FZ_{0}(n\Rep(G)), \\
\Mod_{(n-1)\Rep(G)}^{\EE_2}(n\vect)^\rev &\simeq \FZ_{1}(n\Rep(G)),
\end{align*}
which are also the consequences of (\ref{eq:condense_nDTO_trivial_phase_2}) and (\ref{eq:E2_Z1_LMod}), respectively. For $k\geq 3$, we obtain the following equivalences: 
\be \label{eq:Zk-1_nRepG}
 \Mod_{(n-1)\Rep(G)}^{\EE_k}(n\vect)^\rev \simeq \FZ_{k-1}(n\Rep(G)) \simeq n\Rep(G), 
\ee
where the second `$\simeq$' is due to Example\,\ref{expl:Ek-module_over_nRep(G)} and was rigorously proved in \cite[Proposition\,3.11]{KTZ20} for $n=2$ and $k=3$. The other cases of this result is new. 
\end{expl}

\begin{rem}
Note that (\ref{eq:Zk-1_nRepG}) shows certain stability of higher centers that might hold for more general $\EE_m$-fusion $n$-categories. Actually, for an $\EE_{k+1}$-fusion $n$-category $\CE$, we have
$$
\CE = \Omega\Sigma\CE \simeq \FZ_k(\CE),
$$ 
if and only if the canonical $\EE_k$-monoidal functor $\Sigma\CE \to \FZ_{k-1}(\Sigma\CE)$ is fully faithful. This condition clearly holds for $\CE=n\Rep(G)$ and $k\geq 2$. 
\end{rem}

\subsubsection{General constructions} \label{sec:general_example_k-codim}

Similar to Section\,\ref{sec:general_example_1-codim}, in this subsubsection, we provide some general examples or constructions of the condensations of $k$-codimensional topological defects in $\SC^{n+1}$. We would like to begin with a discussion on how to construction condensable $\EE_k$-algebras.


\medskip
For an $n+$1D topological order $\SC^{n+1}$, some condensable $\EE_k$-algebras in $\Omega^{k-1}\CC$ can be constructed from the condensable $\EE_{k+1}$-algebras in $\Omega^k\CC$ by delooping. In general, there are additional condensable $\EE_k$-algebra in $\Omega^{k-1}\CC$. In particular, those condensable $\EE_k$-algebras in the sub-category $\Sigma\Omega^k\CC$ of $\Omega^{k-1}\CC$ can all be constructed physically.  More explicitly, consider $\SD^{n+1}$ and a gapped (potentially anomalous) domain wall $\SM^n$ between $\SC^{n+1}$ and $\SD^{n+1}$. For $k+1\leq n$, we illustrate the $(k+1)$-th left bulk-to-wall map $L_{k+1}$ as follow. 
$$
\begin{array}{c}
\begin{tikzpicture}[scale=0.8]
\draw[blue!20,fill=blue!20] (0,0)--(0,2)--(1,3)--(1,1)--cycle ;
\node at (0.7,1.5) {\footnotesize $\Omega_m^k\CM$}; 
\node at (-1.5,2.5) {\footnotesize $\SC^{n+1}$} ;
\node at (3.5,2.5) {\footnotesize $\SD^{n+1}$} ;
\node at (-2,1.7) {\scriptsize $\Omega^k\CC \times \Omega_m^k\CM \xrightarrow{\odot} \Omega_m^k\CM$} ;
\node at (-2,0.8) {\scriptsize $\Omega^k\CC \xrightarrow{L_{k+1} = -\odot 1_m^k} \Omega_m^k\CM$} ;
\end{tikzpicture}
\end{array}
$$
The category $\Omega^k\CM$ is $\EE_k$-monoidal. It is clear that $\Omega^k\CM \in \LMod_{\Omega^k\CC}((n-k+1)\vect)$. Moreover, the left $\Omega^k\CC$-action compatible with the fusion products $\otimes^i$ in $\Omega^k\CM$ for $i=1,\cdots, k$ because the $\Omega^k\CC$ is acting from the $x^{k+1}$-dimension that are orthogonal to the $x^1, \dots, x^k$-dimensions. In other words, $\Omega^k\CM \in \LMod_{\Omega^k\CC}(\Algc_{\EE_k}(n-k+1)\vect)$. 
Similar to the proof of (\ref{eq:AlgRMod=LModAlg}), by running the arguments in each $(x^i,x^k)$-plane (for $i<k$), we must have 
$$
\Omega_m^k\CM \in \Algc_{\EE_k}(\RMod_{\Omega^k\CC}((n-k+1)\vect))=\LMod_{\Omega^k\CC}(\Algc_{\EE_k}(n-k+1)\vect). 
$$
Similar to the discussion in Section\,\ref{sec:general_example_1-codim}, all condensable $\EE_k$-algebras in $\Sigma\Omega^k\CC$ should arise in this way. We summarize this results as a mathematical theorem. 

\begin{thm}
For an $\EE_k$-fusion $n$-category $\CB$, we have a natural equivalence: 
$$
\Algc_{\EE_{k-1}}(\RMod_\CB((n+1)\vect) \simeq \LMod_\CB(\Algc_{\EE_{k-1}}((n+1)\vect). 
$$
\end{thm}

\begin{rem}
Note that the action $\odot: \Omega^k\CC \times \Omega_m^k\CM \to \Omega_m^k\CM$ also induces an internal hom $[1_m^k,1_m^k]_{\Omega^k\CC}$, which is automatically a condensable $\EE_{k+1}$-algebra in $\Omega^k\CC$, i.e., 
\be \label{eq:internal_hom_of_Ek_is_Ek+1}
[1_m^k,1_m^k]_{\Omega^k\CC} \in \Algc_{\EE_{k+1}}(\Omega^k\CC). 
\ee
We give a brief explanation of (\ref{eq:internal_hom_of_Ek_is_Ek+1}) in Appendix\,\ref{Appendix:Zm-center_Em-algebra}. 
In general, it is not true that condensing $[1_m^k,1_m^k]_{\Omega^k\CC}$ can recover the gapped domain wall $\SM^n$. 
In general, $\RMod_{[1_m^k,1_m^k]_{\Omega^k\CC}}(\Omega^k\CC)$ is only a sub-fusion category of $\Omega_m^k\CM$. However, if $\CC=\Sigma^k\Omega^k\CC$ and $\SM^n$ is obtained from the condensation of $A\in \Algc_{\EE_{k+1}}(\Omega^k\CC)$, i.e., $\CM=\RMod_{\Sigma^k A}(\CC)$, then we have $\Omega^k\CM \simeq \RMod_{A}(\Omega^k\CC)$. As a consequence, we must have $A=[1_m^k,1_m^k]_{\Omega^k\CC}$. 
\end{rem}

\begin{expl} \label{expl:Ek-monoidal-functor}
The basic idea behind all explicit constructions of condensable $\EE_k$-algebras in an $\EE_k$-fusion $n$-category $\CB$ is to find a nice coordinate system for $\CB$. 
\bnu

\item[(1)] Consider an $\EE_k$-fusion $(n-1)$-category $\CA$. Recall Theorem\,\ref{cor:EkMod_Zk_centralizer}, the $\EE_k$-fusion $n$-category $\FZ_{k-1}(\Sigma\CA)^\op$ is equipped with a coordinate system $\FZ_{k-1}(\Sigma\CA)^\op=\Mod_{\CA}^{\EE_k}(n\vect)$. In this case, a condensable $\EE_k$-algebra in $\Mod_{\CA}^{\EE_k}(n\vect)$ is precisely an $\EE_k$-multi-fusion $(n-1)$-category $\CP$ equipped with an $\EE_k$-monoidal functor $\phi: \CA \to \CP$. Moreover, we have 
$$
\Mod_\CP^{\EE_k}(\FZ_{k-1}(\Sigma\CA)^\op) \simeq \Mod_\CP^{\EE_k}(\Mod_{\CA}^{\EE_k}(n\vect)) \simeq \Mod_\CP^{\EE_k}(n\vect) 
\simeq \FZ_{k-1}(\Sigma\CP)^\op
$$

Examples of such $\CP$ can be constructed. Let $\CX \in \LMod_\CA(\Algc_{\EE_{k-1}}(n\vect))$. The $\EE_{k-1}$-monoidal $\CA$-action $\CA \boxtimes \CX \to \CX$ induces a canonical $\EE_k$-monoidal functor $\CA \to \FZ_{k-1}(\CX)$, which defines a condensable $\EE_k$-algebra in $\CB$. 
For example, one can choose $\CX=\CA$. 
\bnu

\item Let $\CA=n\Rep(G)$ and $\CB=\Mod_{n\Rep(G)}^{\EE_k}((n+1)\vect)$ (recall Example\,\ref{expl:Ek-module_over_nRep(G)}). 
The canonical $\EE_k$-monoidal functors \cite{KZ24}: 
$$
n\Rep(G) \xrightarrow{\forget} n\Rep(H), \quad\quad n\Rep(G) \to \FZ_{k-1}(n\Rep(H)),
$$
where $H\leq G$ is a subgroup, define condensable $\EE_k$-algebras in $\CB$.

\enu

\item[(2)] Assume $\CB =\Sigma\FZ_k(\CA)$ for an $\EE_k$-fusion $(n-1)$-category $\CA$. In this case, we have 
\be \label{eq:RModZk=ModEk} 
\Sigma\FZ_k(\CA) = \RMod_{\FZ_k(\CA)}(n\vect) \simeq \Sigma\Omega\Mod_{\CA}^{\EE_k}(n\vect) \simeq \Sigma\Omega\FZ_{k-1}(\Sigma\CA). 
\ee
where we have used the mathematical fact $\FZ_k(\CA) \simeq \Fun_\CA^{\EE_k}(\CA,\CA) \simeq \Omega(\Mod_{\CA}^{\EE_k}(n\vect))$ followed directly from the universal property of the $\EE_k$-center $\FZ_k(-)$ (see \cite{Lur17,Fra12}). 
Therefore, this $\CB$ is a full subcategory of $\Mod_{\CA}^{\EE_k}(n\vect)$. More precisely, a condensable $\EE_k$-algebra in $\CB=\Sigma\FZ_k(\CA)$ is precisely an $\EE_k$-fusion $(n-1)$-category $\CP$ lying in $\Sigma\Omega\Mod_{\CA}^{\EE_k}(n\vect)$ and equipped with an $\EE_k$-monoidal functor $\phi: \CA \to \CP$. 
\enu
\end{expl}

Now we provide some general constructions of $k$-codimensional defect condensations. 

\medskip
(1) $\SC^{n+1}=\mathbf{1}^{n+1}$: In this case, we have $\CC=n\vect$ and $\Omega^{k-1}\CC=(n-k+1)\vect$ for $k\leq n$. A simple condensable $\EE_k$-algebra in $(n-k+1)\vect$ is precisely an $\EE_k$-fusion $(n-k)$-category $\CA$. Following Theorem$^{\mathrm{ph}}$\,\ref{thm:main_k_codim}, we first condense the $k$-codimensional defect $\CA$ along a transversal direction of the defect $\CA$ and obtain a $(k-1)$-codimensional defect $\Sigma\CA$, then we repeat the process to obtain a2-codimensional defect $\Sigma^{k-2}\CA$ and a 1-codimensional defect $\Sigma^{k-1}\CA$. By condensing $\Sigma^{k-1}\CA$ along the remaining transversal direction, we obtain a condensed phase $\SD^{n+1}$ and a gapped domain wall $\SM^n$ such that $\SD^{n+1} = \overline{\SZ(\SM)^{n+1}}$,  
$$
\CD \simeq \Mod_{\Sigma^{k-1}\CA}^{\EE_1}(n\vect) \simeq \FZ_0(\Sigma^k\CA)^\op, \quad\quad
(\CM, m) = (\Sigma^k\CA, \Sigma^{k-1}\CA). 
$$
Moreover, we have, for $1\leq i \leq k-1$, 
\begin{align*}
\Omega^i\CD &\simeq \Mod_{\Sigma^{k-i-1}\CA}^{\EE_{i+1}}((n-i)\vect) \simeq \Omega^i\FZ_0(\Sigma^k\CA)^\op\simeq \FZ_i(\Sigma^{k-i}\CA)^\op, \quad\quad
\Omega_m^i\CM \simeq \Sigma^{k-i}\CA.  
\end{align*}
For $i=k-1$, we recover $\Mod_\CA^{\EE_k}((n-k+1)\vect) \simeq \FZ_{k-1}(\Sigma\CA)^\op$ obtained previously as a special case in Theorem\,\ref{cor:EkMod_Zk_centralizer}.

\medskip
(2) Now we consider a condensation of $k$-codimensional defect in $\SD^{n+1}$ obtained above. A simple condensable $\EE_k$-algebra $B$ in $\Omega^{k-1}\CD$ in the coordinate system $\Omega^{k-1}\CD\simeq \Mod_\CA^{\EE_k}((n-k+1)\vect)$ is precisely given by an $\EE_k$-monoidal functor $\CA \to \CB$ for an $\EE_k$-fusion $(n-k)$-category $\CB$. Now we condense $B$ viewed as a $k$-codimensional defect in $\SD^{n+1}$ according to the $k$-step process described in Theorem$^{\mathrm{ph}}$\,\ref{thm:main_k_codim}. First, we condensed the $k$-codimensional defect $B$ along one of its transversal direction. We obtain a $(k-1)$-codimensional defect $\Sigma B$ defined by an $\EE_{k-1}$-monoidal functor $\Sigma\CA \to \Sigma\CB$. We repeat the process. We obtain a 2-codimensional defect $\Sigma^{k-2}B$ defined by the $\EE_2$-monoidal functor $\Sigma^{k-2}\CA \to \Sigma^{k-2}\CB$ and a 1-codimensional defect $\Sigma^{k-1}B$ defined by the $\EE_1$-monoidal functor $\Sigma^{k-1}\CA \to \Sigma^{k-1}\CB$. By further condensing such obtained 1-codimensional defect, we obtain a new condensed phase $\SE^{n+1}$ and a new gapped domain wall $\SN^n$ as illustrated below: 
$$
\begin{array}{c}
\begin{tikzpicture}[scale=0.6]
\fill[blue!20] (-2,0) rectangle (2,3) ;
\fill[teal!20] (0,0) rectangle (2,3) ;
\draw[blue, ultra thick, ->-] (0,0) -- (0,3) node[near start, left] {\footnotesize $\SM^n$};
\node at (-1,1.5) {\footnotesize $\mathbf{1}^{n+1}$} ;
\node at (1,1.5) {\footnotesize $\SD^{n+1}$} ;
\fill[red!20] (2,0) rectangle (4,3) ;
\draw[purple, ultra thick, ->-] (2,0) -- (2,3) node[near start, right] {\footnotesize $\SN^n$};
\node at (3.3,1.5) {\footnotesize $\SE^{n+1}$} ;
\end{tikzpicture}
\end{array}
\quad \xrightarrow{\mbox{\footnotesize rolling up}} \quad 
\begin{array}{c}
\begin{tikzpicture}[scale=0.6]
\fill[blue!20] (-2,0) rectangle (2,3) ;
\fill[teal!20] (0,1.5) circle (1);
\fill[red!20] (0,1.5) circle (0.5);
\node at (-1.5,2.5) {\footnotesize $\mathbf{1}^{n+1}$} ;
\draw[blue, ultra thick,->-] (1,1.5) arc (360:0:1) ;
\draw[purple, ultra thick,->-] (0.5,1.5) arc (360:0:0.5) ;
\node at (1.5, 0.3) {\scriptsize $L_2^R(L_2^R(1_n))$} ;
\end{tikzpicture}
\end{array}
=\Sigma^{k-2}\CB \in (n-1)\vect. 
$$
such that 
\begin{align*}
\CE &\simeq \Mod_{\Sigma^{k-1}B}^{\EE_1}(\CD) \simeq \Mod_{\Sigma^{k-1}\CB}^{\EE_1}(n\vect) \simeq \FZ_0(\Sigma^k\CB)^\op, \\
(\CN,n) &\simeq (\RMod_{\Sigma^{k-1}B}(\CD), \Sigma^{k-1}B)  
\simeq (\BMod_{\Sigma^{k-1}\CA|\Sigma^{k-1}\CB}(n\vect), \Sigma^{k-1}\CB). 
\end{align*}
Moreover, we have, for $1\leq i \leq k-1$, 
\begin{align*}
\Omega^i\CE &\simeq \Mod_{\Sigma^{k-i-1}B}^{\EE_{i+1}}(\Omega^i\CD) 
\simeq \Mod_{\Sigma^{k-i-1}B}^{\EE_{i+1}}(\Mod_{\Sigma^{k-i-1}\CA}^{\EE_{i+1}}((n-i)\vect))
\simeq \Mod_{\Sigma^{k-i-1}B}^{\EE_{i+1}}((n-i)\vect) \simeq \FZ_i(\Sigma^{k-i}\CB)^\op.
\end{align*}
Moreover, we have the following (monoidal) equivalences: 
\begin{align*}
\CM \boxtimes_\CD \CN &\simeq \RMod_{\Sigma^{k-1}\CB}(n\vect) \simeq \Sigma^k\CB, \\
\Omega_m\CM \boxtimes_{\Omega\CD} \Omega_n\CN &\simeq
\Sigma^{k-1}\CA \boxtimes_{\FZ_1(\Sigma^{k-1}\CA)} \Fun_{\Sigma^{k-1}\CA|\Sigma^{k-1}\CB}(\Sigma^{k-1}\CB,\Sigma^{k-1}\CB) \\
&\simeq \Fun_{(n-1)\vect|\Sigma^{k-1}\CB}(\Sigma^{k-1}\CB,\Sigma^{k-1}\CB) \simeq \Sigma^{k-1}\CB \simeq \Omega_{m\boxtimes_\CD n}(\CM\boxtimes_\CD\CN).
\end{align*}
Note that $\SE^{n+1}=\mathbf{1}^{n+1}$ if and only if $\Sigma^{k-1}\CB$ is non-degenerate. Moreover, by \cite[Corollary*\ 2.10]{KZ21}, we obtain $\SE^{n+1}=\mathbf{1}^{n+1}$ if and only if $\Sigma^k\CB$ is invertible as a separable $n$-category. This motivates us to introduce the following notion \cite{KZZZ25}.
\begin{defn}
An $\EE_k$-fusion $m$-category $\CB$ is called non-degenerate if $\Sigma^k\CB$ is invertible separable $(m+k)$-category. 
\end{defn}

\begin{lem}
If an $\EE_k$-fusion $m$-category $\CB$ is non-degenerate, then we have 
$\Mod_\CB^{\EE_k}((m+1)\vect) \simeq m\vect$, which further implies that $\FZ_k(\CB) \simeq m\vect$. 
\end{lem}
\pf
Note the first condition (i.e., $\CB$ being non-degenerate) means $\FZ_0(\Sigma^k\CB)$ is trivial; and the second condition means $\FZ_{k-1}(\Sigma\CB)\simeq \Omega^{k-1}\FZ_0(\Sigma^k\CB)$ is trivial; and the third condition means $\FZ_k(\CB)\simeq \Omega^k\FZ_0(\Sigma^k\CB)$ is trivial.
\epf

We summarize above results as a theorem that generalizes Theorem$^{\mathrm{ph}}$\,\ref{pthm:construct_cond_E1_algebras} and Theorem$^{\mathrm{ph}}$\,\ref{pthm:iterate_E2_condensation}. 
\begin{pthm} \label{pthm:iterate_Ek_condensation}
For an $\EE_k$-fusion $m$-category $\CA$, a simple condensable $\EE_k$-algebra $B$ in $\FZ_{k-1}(\Sigma\CA)^\op$ in the coordinate system (recall Theorem\,\ref{cor:EkMod_Zk_centralizer})
$$
\FZ_{k-1}(\Sigma\CA)^\op \simeq \Mod_\CA^{\EE_k}((m+1)\vect)
$$ 
is precisely defined by an $\EE_k$-monoidal functor $\CA\to \CB$ for an $\EE_k$-multi-fusion $m$-category $\CB$. The $\EE_k$-algebra $B$ is Lagrangian if and only if $\CB$ is Lagrangian. 
Moreover, we have 
$$
\Mod_B^{\EE_k}(\FZ_{k-1}(\Sigma\CA)^\op) \simeq \Mod_\CB^{\EE_k}((m+1)\vect) 
\simeq \FZ_{k-1}(\Sigma\CB)^\op. 
$$

This result is precisely the $k$-codimensional layer of a physical condensation from the initial phase $\SC^{m+k+1}$, which is anomaly-free, simple and non-chiral, such that 
$$
\CC = \Mod_{\Sigma^{k-1}\CA}^{\EE_1}((m+k)\vect) \simeq \FZ_0(\Sigma^k\CA)^\op, \quad\quad
\Omega^{k-1}\CC \simeq \Omega^{k-1} \FZ_0(\Sigma^k\CA)^\op \simeq \FZ_{k-1}(\Sigma\CA)^\op
$$
to the condensed phase $\SD^{m+k+1}$ such that 
$$
\CD=\Mod_{\Sigma^{k-1}\CB}^{\EE_1}(\CC) \simeq \Mod_{\Sigma^{k-1}\CB}^{\EE_1}((m+k)\vect), \quad\quad
\Omega^{k-1}\CD \simeq \Mod_\CB^{\EE_k}((m+1)\vect) \simeq \FZ_{k-1}(\Sigma\CB)^\op, 
$$ 
where $\Sigma^{k-1}\CB$ is equipped with an $\EE_1$-monoidal functor $\Sigma^{k-1} \CA \to \Sigma^{k-1}\CB$ and should be viewed as a condensable $\EE_1$-algebra in $\CC$. Moreover, the gapped domain wall $\SM^{m+k}$ produced by this condensation is given by 
\begin{align*}
&(\CM,m) = (\BMod_{\Sigma^{k-1}\CA|\Sigma^{k-1}\CB}(\Sigma^{k-1}\CB,\Sigma^{k-1}\CB), \Sigma^{k-1}\CB), \\
&\Omega_m^{k-1}\CM = \Omega^{k-2}\Fun_{\Sigma^{k-1}\CA|\Sigma^{k-1}\CB}(\Sigma^{k-1}\CB,\Sigma^{k-1}\CB) \simeq \RMod_\CB(\Mod_\CA^{\EE_k}((m+1)\vect)), 
\end{align*}
which generalizes the $k=2$ case in (\ref{eq:RMod_B_Mod_AE2}). 
\end{pthm}

\begin{expl}
Recall (\ref{eq:Zk-1_nRepG}), for $k\geq 3$, we have following equivalences: 
\be 
 \Mod_{n\Rep(G)}^{\EE_k}((n+1)\vect) \simeq \FZ_{k-1}((n+1)\Rep(G))^\op \simeq (n+1)\Rep(G)^\op = (n+1)\Rep(G), 
\ee
In this coordinate system of $(n+1)\Rep(G)$, a condensable $\EE_k$-algebra $B$ in $(n+1)\Rep(G)$ is precisely given by an $\EE_k$-monoidal functor $n\Rep(G) \to \CB$ for an $\EE_k$-multi-fusion $n$-category $\CB$. We have
$$
\Mod_B^{\EE_k}((n+1)\Rep(G)) \simeq \Mod_\CB^{\EE_k}((n+1)\vect) 
\simeq \FZ_{k-1}(\Sigma\CB)^\op. 
$$
For example, let $\CB=\Rep(H)$ for a subgroup $H\leq G$. Let $B$ be defined by the forgetful functor $\forget: \Rep(G) \to \Rep(H)$. Then for $k\geq 3$, we obtain 
$$
\Mod_B^{\EE_k}((n+1)\Rep(G)) \simeq \Mod_{n\Rep(H)}^{\EE_k}((n+1)\vect) 
\simeq \FZ_{k-1}((n+1)\Rep(H))^\op \simeq (n+1)\Rep(H),
$$
where we have used (\ref{eq:Zk-1_nRepG}) in the last `$\simeq$'. 
\end{expl}

\subsubsection{Concrete examples} \label{sec:example_kcodim}

\begin{expl}
Consider the 3+1D $\Zb_2$-gauge theory $\SG\ST_{\Zb_2}^4$ (recall the notations in Example\,\ref{expl:Ae+A1+A2_4D_TC}).
\bnu
\item Condensing the $e$-particles: 
\bnu
\item We first condensing the $e$-particle along a line, we obtain the $\one_c$-string. Mathematically, it amounts to condense the condensable $\EE_3$-algebra $A_e^0=1_{\one} \oplus e$ in $\Omega^2\CT\CC=\Rep(\Zb_2)$ along a line and we obtain $\one_c = \Sigma A_e^0 = \RMod_{A_e^0}(\Rep(\Zb_2)) \simeq \vect$ in $2\Rep(\Zb_2)=\RMod_{\Rep(\Zb_2)}(2\vect)$. 

\item Then we condense the $\one_c$-strings along one of the remaining transversal direction. We obtain a condensed defect $\Sigma^2 A_e^0=\mathrm{rr}$, where $\mathrm{rr}$ denotes the 1-codimensional topological defect obtained by stacking two rough boundaries. It is clear that $\mathrm{rr}$ is condensable $\EE_1$-algebra in $\CG\CT_{\Zb_2}^4$. By further condensing $\Sigma^2 A_e^0$, we obtain the trivial phase $\mathbf{1}^4$ as the condensed phase and the rough boundary as the domain wall.

\item Alternatively, we can condense the $\one_c$-string as a condensable $\EE_2$-algebra in $\Omega\CG\CT_{\Zb_2}^4$, we obtain the trivial phase $\mathbf{1}^4$ as the condensed phase and the rough boundary, on which topological defects form the fusion 2-category $2\vect_{\Zb_2}$.

\enu

\item Condensing the $A_m=(1\oplus m)$-string: 
\bnu
\item We first condense $A_m$ along one of the transversal direction. It produces a membrane $\Sigma A_m = \mathrm{ss}$, where $\mathrm{ss}$ denotes the 1-codimensional defect obtained by stacking two smooth boundaries. By further condensing $\mathrm{ss}$, we obtain $\mathbf{1}^4$ as the condensed phase and the smooth boundary, on which topological defects form the fusion 2-category $2\Rep(\Zb_2)$. 

\item Alternatively, we can condense $A_m$ directly in the remaining two transversal directions to obtain the smooth boundary. 
\enu

\item Condensing the $A_m^{\mathrm{tw}}=(1\oplus m)$-string:
\bnu
\item We first condense $A_m^{\mathrm{tw}}$ along one of the transversal direction. It produces a membrane $\Sigma A_m^{\mathrm{tw}} = \mathrm{tt}$, where $\mathrm{tt}$ denotes the 1-codimensional defect obtained by stacking two twist smooth boundaries. By further condensing $\mathrm{tt}$, we obtain $\mathbf{1}^4$ as the condensed phase and the twist smooth boundary, on which topological defects form the fusion 2-category $2\Rep(\Zb_2)$. 

\item Alternatively, we can condense $A_m^{\mathrm{tw}}$ directly in the remaining two transversal directions to obtain the twist smooth boundary. 
\enu
\enu
\end{expl}

\begin{expl} \label{expl:GT_G4_GtoH_II}
Consider the $3+$1D $G$-gauge theory $\SG\ST_G^4$. As fusion 3-categories, we have the following two different coordinate systems:  
$$
\CG\CT_G^4 = \RMod_{\FZ_1(2\Rep(G))}(3\vect), \quad\quad \CG\CT_G^4=\BMod_{2\Rep(G)|2\Rep(G)}(3\vect)^\op.
$$ 
Its particles form a symmetric fusion 1-category $\Omega^2\CG\CT_G^4=\Rep(G)$, and its strings form a braided fusion 2-category $\Omega\CG\CT_G^4 = \FZ_1(2\Rep(G))$. 

Let $H\leq G$ be a subgroup of $G$. We have $A=\mathrm{Fun}(G/H) \in \Algc_{\EE_3}(\Rep(G))$. By condensing it along one of the transversal directions, we obtain 
$$
\Sigma A:=\RMod_A(\Rep(G))\simeq \Rep(H) \in 2\Rep(G) \hookrightarrow \FZ_1(2\Rep(G)).
$$ 
By condensing $\Sigma A$ along one of the remaining transversal directions, we obtain  
$$
\Sigma^2 A = \RMod_{\Sigma A}(\FZ_1(2\Rep(G))) \in \Algc_{\EE_1}(\Sigma\FZ_1(2\Rep(G))). 
$$
Now we translate $\Sigma^2 A$ into the second coordinate system of $\CG\CT_G^4$ as follows:  
\be \label{eq:M_A=MCC_A}
2\Rep(G) \to 2\Rep(G) \boxtimes_{\FZ_1(2\Rep(G))} \RMod_{\Sigma A}(\FZ_1(2\Rep(G))) \simeq \RMod_{\Sigma A}(2\Rep(G)) \simeq 2\Rep(H), 
\ee
where the first `$\simeq$' a natural generalization of results in 1-categories (see for example \cite{KZ18,XY25} and also Remark\,\ref{rem:M_A=MCC_A}) and the second `$\simeq$' is due to (\ref{eq:nRep(G)-nRep(H)}). 
By condensing $\Sigma^2 A=2\Rep(H)^\op$ in $\CG\CT_G^4$, we obtain $\SG\ST_H^4$ as the condensed phase because
$$
\Mod_{\Sigma^2 A}^{\EE_1}(\CG\CT_G^4) \simeq \BMod_{2\Rep(H)|2\Rep(H)}(3\vect)^\op \simeq \CG\CT_H^4. 
$$
As a consequence, it also implies that we should have 
$$
\Mod_{\Sigma A}^{\EE_2}(\FZ_1(2\vect_G)) \simeq \FZ_1(2\vect_H). 
$$
This provides a proof of Example\,\ref{expl:GT_G4_GtoH}. 
The gapped domain wall $\SM^3=(\CM,m)$ is given by 
$$
(\CM,m) \simeq (\RMod_{\Sigma^2 A}(\CG\CT_G^4), \Sigma^2 A) 
\simeq (\BMod_{2\Rep(H)|2\Rep(G)}(3\vect), 2\Rep(H)). 
$$
We see that $\Omega_m\CM=\Fun_{2\Rep(H)|2\Rep(G)}(2\Rep(H),2\Rep(H))$. When $A=\Fun(G)$, the condensed phase is trivial and $\SM^2$ is a gapped boundary of $\SG\ST_G^4$ such that the boundary defects form the fusion 2-category $\Omega_m\CM\simeq 2\vect_G$. 
\end{expl}

\begin{rem} \label{rem:M_A=MCC_A}
In Eq.\,(\ref{eq:M_A=MCC_A}), we have used a natural equivalence without giving a proof. It is actually a special case of a more general result. For $\CA\in \Algc_{\EE_k}(n\vect)$, $\CM \in \RMod_\CA(\Algc_{\EE_{k-1}}(n\vect))$ and $A\in \Algc_{\EE_k}(\CA)$, we have natural $\EE_{k-1}$-monoidal equivalences: 
$$
\RMod_A(\CM) \simeq \RMod_A(\CM\boxtimes_\CA\CA) \simeq \CM \boxtimes_\CA \RMod_A(\CA). 
$$ 
When $n=2$, the proof above results, which is quite standard (see for example \cite{KZ18,XY25}), is based on the colimits theory for 1-categories. It is natural to expect that we need a fully developed colimit theory for separable higher categories for a proof of the $(n>2)$-cases. 
\end{rem}

\begin{expl} \label{expl:GT_Gn+1_GtoH}
Consider the $n+$2D $G$-gauge theory $\SG\ST_G^{n+2}$. As fusion $(n+1)$-categories, we have two coordinate systems: 
\be \label{eq:two_coordinate_systems_GT_Gn+2}
\CG\CT_G^{n+2} = \RMod_{\FZ_1(n\Rep(G))}((n+1)\vect), \quad\quad \CG\CT_G^{n+2}=\BMod_{n\Rep(G)|n\Rep(G)}((n+1)\vect)^\op.
\ee 
Its particles form a symmetric fusion 1-category $\Omega^n\CG\CT_G^{n+2}=\Rep(G)$. In general, we have
$$
\Omega^{n-k}\CG\CT_G^{n+2} = (k+1)\Rep(G) \quad \mbox{for $0\leq k< n-1$}
$$
and $\Omega\CG\CT_G^{n+2}=\FZ_1(n\Rep(G))$.

Let $H\leq G$ be a subgroup of $G$. The commutative algebra $A=\mathrm{Fun}(G/H)$ of all $\Cb$-valued functions on $G$ is an $\EE_{n+1}$-algebra in $\Omega^{n}\CG\CT_G^{n+2}=\Rep(G)$. As a composed particle, $\mathrm{Fun}(G/H)$ can be condensed. More precisely, by condensing the $\mathrm{Fun}(G/H)$-particles, we mean the following procedures. 
\bnu
\item We first condensing the $A$-particle along a line, we obtain a string 
$$
\Sigma A =\RMod_A(\Rep(G)) \simeq \Rep(H) \in \Algc_{\EE_n}(2\Rep(G)). 
$$ 

\item We further condense the $\Sigma A$-string along one of the remaining transversal directions, we obtain 
$$
\Sigma^2 A = \RMod_{\Sigma A}(2\Rep(G)) \simeq \RMod_{\Sigma A} (2\vect) \simeq 2\Rep(H). 
$$

\item We repeat the procedure until $\Sigma^{n-1}A = (n-1)\Rep(H) \in n\Rep(G) \hookrightarrow \FZ_1(n\Rep(G))$. By condensing $\Sigma^{n-1}A$ along one of the remaining two transversal directions, we obtain 
$$
\Sigma^n A = \RMod_{\Sigma^{n-1} A}(\FZ_1(n\Rep(G))) \in \Algc_{\EE_1}(\Sigma\FZ_1(n\Rep(G))). 
$$
Now we translate $\Sigma^n A$ into an $\EE_1$-algebra in the second coordinate system of $\CG\CT_G^{n+2}$ in (\ref{eq:two_coordinate_systems_GT_Gn+2}).  It is defined by the following monoidal functor (recall Remark\,\ref{rem:M_A=MCC_A}): 
$$
n\Rep(G) \to n\Rep(G) \boxtimes_{\FZ_1(n\Rep(G))} \RMod_{\Sigma^{n-1} A}(\FZ_1(n\Rep(G))) \simeq \RMod_{\Sigma^{n-1} A}(n\Rep(G)) \simeq n\Rep(H). 
$$

\item By condensing $\Sigma^n A=n\Rep(H)$, we obtain $\SG\ST_H^{n+2}$ as the condensed phase because
$$
\SG\ST_H^{n+2} \simeq \Mod_{n\Rep(H)^\op}^{\EE_1}((n+1)\vect) \simeq \Mod_{n\Rep(H)^\op}^{\EE_1}(\CG\CT_G^{n+2}).
$$
\enu
In this example, we provides the precise mathematical theory behind the folklore that breaking the $G$-gauge ``symmetry'' in $\SG\ST_G^{n+2}$ to a subgroup $H$ gives the $H$-gauge theory $\SG\ST_H^{n+2}$. 
\end{expl}


\subsection{Gauging a \texorpdfstring{$G$}{G}-symmetry} \label{sec:gauging_G-symmetry}

Recall that $\Omega\CT\CC^3$ has a $\Zb_2$-symmetry defined by the non-trivial braided auto-equivalence of $\FZ_1(\Rep(\Zb_2))$, which is also called the $e$-$m$-duality. It was known that the double Ising modular tensor category $\FZ_1(\Omega\ising)$ can be obtained from the toric code $\Omega\CT\CC$ via an equivariantization of this $\Zb_2$-symmetry \cite{DGNO10}. In physics, this equivariantization is viewed as a process of  `gauging' this $e$-$m$-duality in $\Omega\CT\CC$. In this subsection, we provide a mathematical foundation of this alleged `gauging' process by explaining that it is a natural consequence of the condensation of 1-codimensional topological defects or the gauging of a generalized 0-form symmetry.  Mathematically, it amounts to establish the a previously unknown relation between $G$-crossed braided extensions (or minimal non-degenerate extensions) and 1-codimensional defect condensations. 

\subsubsection{Gauging a \texorpdfstring{$G$}{G}-symmetry in a 2+1D SET order}

Let $G$ be a finite group. We use 2+1D $G$-SET order to mean a symmetry enriched topological order with an onsite $G$-symmetry (or a generalized 0-form symmetry). There are several different approaches towards the mathematical characterization of a 2+1D $G$-SET order \cite{BBCW19,LKW16a,KLWZZ20,KZ22b,LYW24}.  

\begin{figure}
\centering
\begin{tikzpicture}
\fill[gray!20] (0,0) rectangle (4,3) ;
\draw[very thick,dashed] (2,1)--(2,3) ;
\fill (1,1.8) circle (0.07) node[above] {$x \in \CC$} ;
\fill (3,1.8) circle (0.07) node[above] {$\rho_g(x)$} ;
\draw[-latex] (1.2,1.8)--(2.8,1.8) ;
\draw[thick,fill=white] (1.9,0.9) rectangle (2.1,1.1) node[midway,below] {$y \in \CC_g$} ;
\end{tikzpicture}
\caption{The dashed line is the invertible domain wall of the topological order such that moving an anyon across the domain wall realizes the braided equivalence $\rho_g \colon \CC \to \CC$. The topological defects living on the end of this invertible domain wall are called the $g$-defects, and they form the category $\CC_g$. Equivalently, moving $x \in \CC$ around $y \in \CC_g$ yields the auto-equivalence $\rho_g$. One can think of that $y$ is attached with a flux $g$.}
\label{fig_g_defect}
\end{figure}

\medskip
The first approach towards 2+1D $G$-SET orders is based on the idea of equipping a topological order with a $G$-action. Let $\CC$ be a nondegenerate braided fusion category  equipped with a 2-group homomorphism (monoidal functor) $\rho \colon G \to \Aut^{\EE_2}(\CC)$, where $\Aut^{\EE_2}(\CC)$ is the 2-group of braided auto-equivalences of $\CC$. It turns out that this $G$-action is not enough to characterize the $G$-SET order. In \cite{BBCW19}, Barkeshli, Bonderson, Cheng and Wang showed that a $G$-SET order is a $G$-crossed extension of $\CC$, i.e., a $G$-crossed braided fusion category
\[
\CC_G^\times = \bigoplus_{g \in G} \CC_g ,
\]
where $\CC_e = \CC$ and the category $\CC_g\neq 0$ consists of the $g$-defects living on the end of an invertible 1+1D domain wall realizing the braided auto-equivalence $\rho_g$ as illustrated in Figure \ref{fig_g_defect}. The topological defects living on the invertible 1+1D domain wall form the multi-fusion 1-category $\Fun_\CC(\CC_g, \CC_g)^\op \simeq \CC$ because the wall is invertible. As a consequence, $\CC_g$ is an invertible $\CC$-$\CC$-bimodule, i.e., $\CC_g \in \mathrm{Pic}(\CC)$, where $\Pic(\CC)$ is the 3-group of invertible $\CC$-modules, invertible $\CC$-module functors and invertible $\CC$-module natural transformations. According to \cite{BBCW19}, the nondegenerate braided fusion category of the particle-like topological defects in the gauged topological order is the $G$-equivariantization $(\CC_G^\times)^G$ of $\CC_G^\times$, which consists of `$G$-invariant' symmetry defects.

\begin{rem}
According to \cite{ENO10}, a $G$-crossed extension of $\CC$ is equivalent to lifting a group homomorphism $g \mapsto \CC_g$ to a 3-group homomorphism (i.e., a monoidal 2-functor) $\bar \rho \colon G \to \Pic(\CC)$. This lifting can be achieved in two steps. 
\begin{itemize}
\item To lift it to a 2-group homomorphism $G \to \Pi_2(\Pic(\CC))$ (i.e., a monoidal 1-functor), where $\Pi_2(-)$ denotes the underlying 2-group, we need to endow the 1-functor with a monoidal structure, which is defined by $\CC$-module equivalences
\be \label{eq_monoidal_structure_tensor_product}
\CC_g \boxtimes_\CC \CC_h \simeq \CC_{gh} , \quad \forall g,h \in G
\ee
satisfying the axioms of a monoidal structure. These equivalences are equivalent to the tensor product functor $\otimes \colon \CC_g \times \CC_h \to \CC_{gh}$ of $\CC_G^\times$, because $\otimes$ is $\CC_e$-balanced. This lifting exists if and only if certain obstruction class $O_3 \in H^3(G;\pi_2(\Pic(\CC))) \simeq H^3(G;\pi_1(\CC^\times))$ vanishes. When $O_3$ vanishes, all the liftings form a torsor over $H^2(G;\pi_1(\CC^\times))$.
\item To lift it to a 3-group homomorphism $G \to \Pic(\CC)$, we need to specify some $\CC$-module natural isomophisms as the monoidal structure, which are precisely the associators and $G$-crossed braidings of the associated $G$-crossed extension $\CC_G^\times$. This lifting exists if and only if certain obstruction class $O_4 \in H^4(G;\pi_3(\Pic(\CC))) \simeq H^3(G;\bk^\times)$ vanishes. When $O_4$ vanishes, all the liftings form a torsor over $H^3(G;\bk^\times)$.
\end{itemize}
Since there is a canonical 3-group homomorphism $\partial \colon \Pic(\CC) \to \Aut^{\EE_2}(\CC)$ \cite{ENO10,DN13}, which restricts to a 2-group equivalence $\Pi_2(\Pic(\CC)) \to \Aut^{\EE_2}(\CC)$ when $\CC$ is nondegenerate, a $G$-crossed extension of $\CC$ (i.e., a 3-group homomorphism $\bar \rho \colon G \to \Pic(\CC)$) automatically defines a $G$-action on $\CC$ by the composed functor
\[
G \xrightarrow{\bar \rho} \Pic(\CC) \xrightarrow{\partial} \Aut^{\EE_2}(\CC) .
\]
Hence, the 3-group homomorphism $\bar \rho \colon G \to \Pic(\CC)$ lifts the the 2-group homomorphism $\rho \colon G \to \Aut^{\EE_2}(\CC)$.
\end{rem}

The second approach towards 2+1D $G$-SET orders, due to Lan, Kong and Wen \cite{LKW16a,LKW17}, is based on the categorical description of $G$-invariant particle-like topological defects. Such defects can be fused and braided thus form a braided fusion 1-category $\CD$. Symmetry charges $\Rep(G)$ form a full subcategory of $\CD$ that is equivalent to $\FZ_2(\CD)$, i.e. $\Rep(G) \to \FZ_2(\CD)$. Gauging the $G$ symmetry amounts to a minimal nondegenerate extension of $\CD$, i.e., a nondegenerate braided fusion 1-category $\CM$ equipped with a braided embedding $\CD \hookrightarrow \CM$, which induces an equivalence $\Rep(G) \hookrightarrow \FZ_2(\CD,\CM)$. A minimal nondegenerate extension of $\CD$ provides a complete categorical description of a 2+1D (bosonic\footnote{Replacing the bosonic symmetry charges $\Rep(G)$ by fermionic symmetry charges $\Rep(G,z)$, where $z$ is the fermion parity, automatically gives a categorical description of a fermionic $(G,z)$-SET order \cite{LKW16a,LKW17}.}) $G$-SET order. 

\begin{rem}
There are two additional approaches towards SET orders based on boundary-bulk relation \cite{KLWZZ20} and topological Wick rotations \cite{KZ22b}. We do not need them here. 
\end{rem}

For a bosonic finite symmetry $G$, the minimal nondegenerate extension approach is equivalent to the $G$-crossed extension approach. More precisely, by \cite[Theorem 4.44]{DGNO10}, the following two constructions are mutually inverse of each other: 
\begin{itemize}
\item given a $G$-crossed braided fusion category $\CK$, its equivariantization $\CK^G$ is a braided fusion category containing $\Rep(G)$; 
\item given a braided fusion category $\CL$ containing $\Rep(G)$, its de-equivariantization $\CL_G$ is a $G$-crossed braided fusion category. 
\end{itemize}
Moreover, $\CK^G$ is nondegenerate if and only if $\CK_e$ is nondegenerate and each $\CK_g$ is nonzero by \cite[Theorem 4.56]{DGNO10}.  Therefore, the notion of a $G$-crossed extension is equivalent to that of a minimal nondegenerate extension.
For 2+1D $G$-SET order, we have $\CD\simeq \CC^G$ and $\CM=(\CC_G^\times)^G$. We summarize the relation between these categories in the following diagram:
\[
\xymatrix@C=4cm{
\vect \ar@<-.7ex>[d]_{(-)^G}  \ar@{^{(}->}[r]^-{\text{inclusion}} & \CC \ar@<-.7ex>[d]_{(-)^G} \ar@{^{(}->}[r]^-{\text{$G$-crossed extension}} 
& \CC_G^\times \ar@<-.7ex>[d]_{(-)^G} \\
\Rep(G) \ar@<-.7ex>[u]_{(-)_G} \ar@{^{(}->}[r]_-{\text{inclusion}} & \CD \ar@<-.7ex>[u]_{(-)_G} \ar@{^{(}->}[r]_-{\text{minimal nondegenerate extension}} 
& \CM \ar@<-.7ex>[u]_{(-)_G}
}
\]

\medskip
The third approach towards $n$+1D SET orders, preceded by earlier works \cite{DKR11,TW24,KLWZZ20a} on a similar formulation of SPT orders and some online talks in 2022-2023 by the first author of this work \cite{Kon22Talk}, was explicitly proposed by Lan, Yue and Wang in \cite{LYW24}. An $n$+1D SET order can be described by a linear monoidal $n$-functor
\be \label{eq:phi_TB}
\phi \colon \CT \to \CB ,
\ee
where the fusion $n$-category $\CT$ describes the (non-invertible) $n$-from symmetries and the fusion $n$-category $\CB$ is the category of topological defects of a topological phase $\SB^{n+1}$. To gauge the symmetry amounts to choose a $\CT$-module $\CK$ and the gauged theory is described by the indecomposable multi-fusion $n$-category $\Fun_\CB(\CB \boxtimes_\CT \CK,\CB \boxtimes_\CT \CK)$. When $\CK \simeq \RMod_A(\CT)$ for a condensable $\EE_1$-algebra $A \in \CT$ (see also \cite[Remark 4.5]{LYW24}), the gauged theory is described by $\BMod_{\phi(A)|\phi(A)}(\CB)$. In other words, this gauging is a special case of 1-codimensional defect condensations and the same as the condensation of $\phi(A) \in \Algc_{\EE_1}(\CB)$. In the rest of this subsubsection, we established a previously unknown connection between this gauging and that defined by a $G$-crossed extension (or that defined by a minimal nondegenerate extension). 


In the setting of the 2+1D $G$-SET order in this subsubsection. Recall that $\Sigma \CC$ is a fusion 2-category describing all topological defects in a 2+1D anomaly-free topological order. Since $\mathrm{Pic}(\CC) \simeq (\Sigma \CC)^\times$, a 3-group homomorphism $\phi \colon G \to \mathrm{Pic}(\CC)$ is equivalent to a linear monoidal 2-functor $\phi \colon 2\vect_G \to \Sigma \CC$. Therefore, the $G$-SET order can be equivalently described by a linear monoidal 2-functor $\phi\colon 2\vect_G \to \Sigma \CC$. One can recover the $G$-crossed extension $\CC_G^\times$ of $\CC$ canonically as follows. Note that $\vect_G \in 2\vect_G$ is a condensable $\EE_1$-algebra with the multiplication induced by the group multiplication of $G$. 

\begin{lem}
The monoidal structure of $\phi$ endows its image $\phi(\vect_G)$ with a structure of  a condensable $\EE_1$-algebra in $\Sigma\CC$, which is precisely the $G$-crossed braided fusion category $\CC_G^\times$. 
\end{lem}
\begin{proof}
Recall that, in the coordinate system $\Sigma\CC=\RMod_\CC(2\vect)$, the condensable $\EE_1$-algebras in $\Sigma \CC$ are multi-fusion $\CC$-modules, i.e., multi-fusion categories $\CP$ equipped with a braided functor $\CC \to \FZ_1(\CP)$ \cite[Definition-Proposition 3.2]{BJS21} (see also \cite[Lemma 2.2.6]{Dec24}) or Theorem$^{\mathrm{ph}}$\,\ref{thm:condensable_E1Alg_af_C}. In this coordinate system, the $G$-crossed extension $\CC_G^\times$ of $\CC$ is equipped with a canonical braided functor $\CC \to \FZ_1(\CC_G^\times)$, which is induced by the $G$-crossed braiding on $\CC_G^\times$. It endows $\CC_G^\times$ with a condensable $\EE_1$-algebra structure in $\Sigma\CC$. 

On the other hand, as an object,
\[
\phi(\vect_G) \simeq \bigoplus_{g \in G} \phi(g) = \bigoplus_{g \in G} \CC_g = \CC_G^\times .
\]
The multiplication of $\phi(\vect_G)$ is induced by the group multiplication of $G$ and the monoidal structure of $\phi$:
\[
\CC_g \boxtimes_\CC \CC_h = \phi(g) \boxtimes_\CC \phi(h) \xrightarrow{\phi^2_{g,h}} \phi(gh) , \quad g,h \in G ,
\]
where $\phi^2$ is part of the monoidal structure of $\phi$. According to the discussion below \eqref{eq_monoidal_structure_tensor_product}, this monoidal structure is exactly the tensor product of $\CC_G^\times$. Similarly, the higher morphisms of the monoidal structure of $\phi$ induce the associator and $G$-crossed braiding of $\CC_G^\times$. Therefore, the condensable algebra $\phi(\vect_G) \in \Sigma \CC$ is exactly the $G$-crossed braided fusion category $\CC_G^\times$.
\end{proof}

\begin{prop}
Condensing the condensable $\EE_1$-algebra $\CC_G^\times$ in $\Sigma \CC$ produces the condensed phase given by $\Sigma \CM$ (where $\CM=(\CC_G^\times)^G$), i.e., we have a canonical monoidal equivalence 
\be \label{eq:BMod_CGtimes=SigmaM}
\BMod_{\CC_G^\times|\CC_G^\times}(\Sigma \CC) \simeq \Sigma \CM.
\ee
\end{prop}
\begin{proof}
Let $A \coloneqq \Fun(G) \in \Rep(G) \subset \CM$. We have $\CC_G^\times\simeq \CM_G = \RMod_A(\CM)$. Moreover, we have $\CC \simeq \Mod_A^{\EE_2}(\CM)$ \cite{Kir01,Mueg04} (see also \cite[Proposition 4.55]{DGNO10}). Since $\CM$ is nondegenerate, by \cite[Corollary 3.30]{DMNO13}, we obtain a canonical braided equivalence:
\[
\FZ_1(\CC_G^\times) \simeq \CC \boxtimes \overline{\CM}, 
\]
which implies that $\FZ_2(\CC,\FZ_1(\CC_G^\times)) \simeq \overline{\CM}$. 
Using (\ref{eq:E1Mod_Z1Z0}) and (\ref{eq:FZm-FZm-1}), we obtain a canonical equivalence of fusion 2-categories $\Mod_{\CC_G^\times}^{\EE_1}(\Sigma \CC)^\op \simeq \FZ_1(\Sigma\CC, \FZ_0(\RMod_{\CC_G^\times}(\Sigma\CC))) 
\simeq \Sigma\FZ_2(\CC,\FZ_1(\CC_G^\times)) \simeq \Sigma \overline{\CM}$. 
\end{proof}

\begin{rem}
Above proposition simply says that gauging the $G$-symmetry in a 2+1D $G$-SET order defined by the equivariantization of a $G$-crossed braided fusion 1-category $\CC_G^\times$ is equivalent to gauging the symmetry $2\vect_G \to \Sigma\CC$ in the setting of (\ref{eq:phi_TB}) for $\CT = 2\vect_G$, $\CK = 2\vect$ and $A = \vect_G \in \Algc_{\EE_1}(2\vect_G)$. 
\end{rem}

\begin{rem}
A generalized $0$-from $G$-symmetry in an $n+$1D topological order $\SC^{n+1}$ can be defined by a monoidal functor $n\vect_G \to \CC$. By \cite{LYW24}, gauging it amounts to the gauging process in the case $\CT=n\vect_G$, $\CK=n\vect$ and $A=(n-1)\vect_G$, where the $\CT$-action on $\CK$ is defined by the forgetful functor $\forget: n\vect_G \to n\vect \simeq \Fun(n\vect, n\vect)$. 
\end{rem}

\begin{rem}
Recently, we were informed by Nils Carqueville that the relation between $G$-crossed extensions and orbiford TQFT's was also independently established in \cite{CH25,HPRW25}. 
\end{rem}

\subsubsection{General constructions and examples}

\begin{expl}
A $G$-SPT (symmetry protected topological) order is a $G$-SET order with a trivial topological order. Thus 2+1D $G$-SPT orders are classified by the $G$-crossed extensions of $\vect$, which are $\vect_G^\omega$ for $\omega \in H^3(G;\Cb^\times)$ \cite[Proposition 4.61]{DGNO10}. The physical meaning of $\vect_G^\omega$ is that, the only simple $g$-defect is the flux $g$ for every $g \in G$, and the associators of the fusion of $G$-fluxes are defined by $\omega$.

The equivariantization gives $\vect^G=\Rep(G)$ and $(\vect_G^\omega)^G \simeq \FZ_1(\vect_G^\omega)$. Note that $(\vect_G^\omega)^G$ for $\omega\in H^3(G,\Cb^\times)$ are precisely all the minimal nondegenerate extensions of $\Rep(G)$. 

The $G$-crossed extensions of $\vect$, i.e., $\vect_G^\omega$, are condensable $\EE_1$-algebras in $2\vect=\Sigma\vect$. By condensing $\vect_G^\omega$ in the trivial 2+1D topological order, we obtain a non-chiral topological order:
$$
\Mod_{\vect_G^\omega}^{\EE_1}(2\vect) \simeq \Sigma\FZ_1(\vect_G^\omega) \simeq \FZ_0(\Sigma\vect_G^\omega) \simeq \Sigma(\vect_G^\omega)^G.
$$

Note that the forgetful functor $\forget: 2\vect_G \to 2\vect$ is a monoidal functor. We can twist its monoidal structure by defining the hexagonator by a non-trivial 3-cocycle $\omega \in Z^3(G,\Cb^\times)$. We denote the resulting monoidal functor by $\forget^\omega: 2\vect_G \to 2\vect$. It is easy to check that $\forget^\omega$ maps $\vect_G \in \Algc_{\EE_1}(2\vect_G)$ to $\vect_G^\omega \in \Algc_{\EE_1}(2\vect)$. Therefore, gauging the $G$-symmetry in the $G$-crossed extension $\vect_G^\omega$ of $\vect$ is equivalent to gauging the $2\vect_G$-symmetry defined by  the monoidal functor $\forget^\omega: 2\vect_G \to 2\vect$. 
\end{expl}

\begin{expl} \label{expl_G_defects_boundary}
Suppose $\CA$ is a fusion 1-category and $\CC = \FZ_1(\CA)$. Physically this means that $\CC$ is the category of anyons in a 2+1D topological order that has a gapped boundary and the boundary particles form a fusion 1-category $\CA$. Note that there is an equivalence of 3-groups 
$$
\mathrm{BrPic}(\CA) = \BMod_{\CA|\CA}(2\vect)^\times \simeq \RMod_{\FZ_1(\CA)}(2\vect)^\times \simeq \mathrm{Pic}(\FZ_1(\CA)),
$$ 
where $\mathrm{BrPic}(\CA)$ is the 3-group of invertible $\CA$-$\CA$-bimodules \cite{ENO10}. More explicitly, this equivalence sends an invertible $\CA$-$\CA$-bimodule $\CX$ to its `center' $\FZ_\CA(\CM) \simeq \Fun_{\CA|\CA}(\CA,\CX)$ as defined in \cite{GNN09}. The physical meaning of this equivalence was explained in \cite[Figure 4]{KLWZZ20}. Then a $G$-crossed extension of $\FZ_1(\CA)$ is equivalent to a 3-group homomorphism $G \to \mathrm{BrPic}(\CA)$, which by \cite{ENO10} is equivalent to a $G$-graded extension of $\CA$:
\[
\bigoplus_{g \in G} \CA_g , \quad \CA_e = \CA .
\]
The category $\CA_g$ consists of the $g$-defects living on the junction between the invertible domain wall realizing the equivalence $\rho_g$ and the boundary (see Figure \ref{fig_g_defect_boundary}). The corresponding $G$-crossed extension of $\FZ_1(\CA)$ is
\[
\FZ_\CA(\bigoplus_{g \in G} \CA_g) \simeq \Fun_{\CA|\CA}\biggl(\CA,\bigoplus_{g \in G} \CA_g \biggr) \simeq \bigoplus_{g \in G} \FZ_\CA(\CA_g) \simeq \FZ_1(\CA, \bigoplus_{g \in G} \CA_g),
\]
where $\FZ_1(\CA, \bigoplus_{g \in G} \CA_g)$ is the $\EE_1$-centralizer (recall Definition\,\ref{def:centralizer+center}). Its $G$-crossed braided structure was defined in \cite[Theorem 3.3]{GNN09}. Moreover, by \cite[Theorem 3.5]{GNN09}, the equivariantization of the $G$-crossed braided fusion category $\FZ_1(\CA, \bigoplus_{g \in G} \CA_g)$ is precisely the Drinfeld center $\FZ_1(\bigoplus_{g \in G} \CA_g)$, i.e., 
$$
\FZ_1(\CA, \bigoplus_{g \in G} \CA_g)^G \simeq \FZ_1(\bigoplus_{g \in G} \CA_g). 
$$ 
Physically, this means that the topological defects on the boundary after gauging the symmetry are all symmetry defects on the boundary. We summarize various relations in the following diagram. 
\[
\xymatrix@C=4cm{
& \CA \ar@{^{(}->}[r]^-{\text{$G$-graded extension}} 
\ar@{|->}[d]_{\FZ_1(\CA,-)}
& \bigoplus_{g\in G} \CA_g \ar@{|->}[d]^{\FZ_1(\CA,-)}  
\\
\vect \ar@<-.7ex>[d]_{(-)^G}  \ar@{^{(}->}[r]^-{\text{inclusion}} & \FZ_1(\CA) \ar@<-.7ex>[d]_{(-)^G} \ar@{^{(}->}[r]^-{\text{$G$-crossed extension}} 
& \FZ_1(\CA, \bigoplus_{g\in G}\CA_g) \ar@<-.7ex>[d]_{(-)^G} \\
\Rep(G) \ar@<-.7ex>[u]_{(-)_G} \ar@{^{(}->}[r]_-{\text{inclusion}} & (\FZ_1(\CA))^G \ar@<-.7ex>[u]_{(-)_G} \ar@{^{(}->}[r]_-{\text{minimal nondegenerate extension}} 
& \FZ_1(\bigoplus_{g\in G}\CA_g) \ar@<-.7ex>[u]_{(-)_G}
}
\]

The 3-group homomorphism $G \to \mathrm{BrPic}(\CA)$ extends to a monoidal 2-functor 
$$
\phi: 2\vect_G \to \BMod_{\CA|\CA}(2\vect)^\op \simeq \Sigma\FZ_1(\CA).
$$ 
It maps $\vect_G \in \Algc_{\EE_1}(2\vect_G)$ to a condensable $\EE_1$-algebra $\phi(\vect_G)$ in $\Sigma\FZ_1(\CA)$, which is precisely $\FZ_1(\CA, \bigoplus_g \CA_g)$. Using the coordinate transformation between $\BMod_{\CA|\CA}(2\vect)^\op$ and $\Sigma\FZ_1(\CA)$ (recall Theorem$^{\mathrm{ph}}$\,\ref{pthm:construct_cond_E1_algebras}), 
$$
\begin{array}{c}
\begin{tikzpicture}[scale=1]
\fill[blue!20] (-2,0) rectangle (0,2) ;
\fill[teal!20] (0,0) rectangle (2,2) ;
\draw[->-,ultra thick] (-2,0)--(0,0) node[midway,below] {\scriptsize $\CA$} ;
\draw[->-,ultra thick] (-2,2)--(-2,0) node[midway,left] {\scriptsize $\CA$} ;
\draw[->-,ultra thick] (0,0)--(2,0) node[midway,below] {\scriptsize $\oplus_g\CA_g$} ;

\draw[blue,->-,ultra thick] (0,2)--(0,0) ; 
\draw[fill=white] (-0.07,-0.07) rectangle (0.07,0.07) ; 
%
%
\node at (-1,1.5) {\scriptsize $\FZ_1(\CA)$} ;
\node at (1,1.5) {\scriptsize $\FZ_1(\oplus_g\CA_g)$} ;
\node at (0,-0.25) {\scriptsize $\oplus_g\CA_g$} ;
\node at (0,2.2) {\scriptsize $\FZ_1(\CA,\oplus_g\CA_g)$} ;
\node at (-2,2.2) {\scriptsize $\CA$} ;

\draw[decorate,decoration=brace,very thick] (-2,2.4)--(0,2.4) node[midway, above] {\scriptsize $\oplus_g \CA_g$};

\end{tikzpicture}
\end{array}
\quad\quad
\begin{array}{l}
\mbox{\small Two definitions of the same algebra in $\CC$}:  \\
\mbox{\small (1) the monoidal functor $\CA \to \bigoplus_g \CA_g$} \\
\mbox{\small \quad\,\, defines $(\bigoplus_g \CA_g)^\op \in \Algc_{\EE_1}(\BMod_{\CA|\CA}(2\vect)^\op)$}; \\
\mbox{\small (2) the central functor $\FZ_1(\CA) \to \Fun_{\CA|\oplus_g \CA_g}(\oplus_g \CA_g,\oplus_g \CA_g)^\op$} \\
\mbox{\small \quad\,\, defines $\FZ_1(\CA,\oplus_g\CA_g)^\op \in \Algc_{\EE_1}(\RMod_{\FZ_1(\CA)}(2\vect))$},
\end{array}
$$
where we have used the fact that $\FZ_1(\CA,\oplus_g\CA_g) \simeq \Fun_{\CA|\oplus_g \CA_g}(\oplus_g \CA_g,\oplus_g \CA_g)$ as fusion 1-categories. In other words, $\phi(\vect_G) = (\bigoplus_{g\in G} \CA_g)^\op$ in $\Algc_{\EE_1}(\BMod_{\CA|\CA}(2\vect)^\op)$. By condensing $\phi(\vect_G)$ in $\BMod_{\CA|\CA}(2\vect)^\op$ or $\Sigma\FZ_1(\CA)$, we obtain the condensed topological order determined by $\FZ_1(\bigoplus_{g\in G} \CA_g)$. 
\end{expl}

\begin{figure}
\centering
\begin{tikzpicture}
\fill[gray!20] (0,0) rectangle (4,2) ;
\draw[very thick] (0,0)--(4,0) ;
\draw[very thick,dashed] (2,0)--(2,2) ;
\fill (1,0.9) circle (0.07) node[above] {\small $x \in \FZ_1(\CA)$} ;
\fill (3,0.9) circle (0.07) node[above] {\small $\rho_g(x)$} ;
\draw[-latex] (1.2,0.9)--(2.8,0.9) ;
\draw[thick,fill=white] (1.9,-0.1) rectangle (2.1,0.1) node[midway,below] {\small $a \in \CA_g$} ;
\end{tikzpicture}
\caption{The invertible domain wall realizing the braided equivalence $\rho_g \colon \CC \to \CC$ can end on the boundary. The topological defects living on the junction are called the $g$-defects on the boundary, and they form the category $\CA_g$.}
\label{fig_g_defect_boundary}
\end{figure}

\begin{expl} \label{expl_Ising_G_crossed}
This example explains the well-known statement that the double Ising topological order can be obtained from the toric code model by gauging a 0-form $\Zb_2$-symmetry called the $e$-$m$-duality (see also \cite[Section X.G.2]{BBCW19}) as a special case of Example\,\ref{expl_G_defects_boundary}.

There is a $\Zb_2$-action on the toric code model given by exchanging $e$ and $m$. This gives a 2-group homomorphism $\Zb_2 \to \Aut^{\EE_2}(\FZ_1(\Rep(\Zb_2)))$. An $\Zb_2$-SET order lifting this symmetry action, by Example \ref{expl_G_defects_boundary}, is equivalent to a $\Zb_2$-graded extension of $\Rep(\Zb_2)$:
\[
\CA_0 \oplus \CA_1 , \quad \CA_0 = \Rep(\Zb_2) .
\]
By considering the 2-group homomorphism $\Zb_2 \to \Aut^{\EE_2}(\FZ_1(\Rep(\Zb_2))) \simeq \Pi_2(\mathrm{BrPic}(\Rep(\Zb_2)))$, it is not hard to see that $\CA_1 \simeq \vect$. Physically, $\CA_1$ consists of the topological defects living on the junction between the $e$-$m$-exchange domain wall and the rough boundary of the toric code model. Thus we can also check that $\CA_1 \simeq \vect$ in the concrete lattice model (see for example \cite[Section 4.4.3]{KZ22a}).

An $\Zb_2$-graded extension $\Rep(\Zb_2) \oplus \vect$ of $\Rep(\Zb_2)$ is the same as an Ising type fusion 1-category, that is, a fusion 1-category with three simple objects $\one,\psi,\sigma$ and the fusion rule
\[
\psi \otimes \psi \simeq \one , \quad \psi \otimes \sigma = \sigma = \sigma \otimes \psi , \quad \sigma \otimes \sigma = \one \oplus \psi .
\]
By \cite{TY98} (see also \cite[Section B.1]{DGNO10} or \cite[Example 9.4]{ENO10}), there are two Ising type fusion 1-categories up to equivalence, denoted by $\Ising_\pm$. They differ in the definition of the associators\footnote{Note that $\Ising_+=\Omega\ising$ as fusion 1-categories.}. According to the discussion in Example \ref{expl_G_defects_boundary}, the cateogry of particle-like topological defects in the gauged theory is $\FZ_1(\Ising_\pm)$. It also means that there are two 3-group homomorphisms $\phi_\pm: G \to \mathrm{BrPic}(\Rep(\Zb_2))$. Figure\,\ref{fig_toric_to_Ising2} depicts the physical relation between the $\Zb_2$ and double Ising topological orders. 

\begin{figure}
\centering
\begin{tikzpicture}[scale=1]
\fill[blue!20] (-2,0) rectangle (0,2) ;
\fill[teal!20] (0,0) rectangle (2,2) ;
\draw[->-,ultra thick] (-2,0)--(0,0) node[midway,below] {\scriptsize $\Rep(\Zb_2)$} ;
\draw[->-,ultra thick] (-2,2)--(-2,0) node[midway,left] {\scriptsize $\Rep(\Zb_2)$} ;
\draw[->-,ultra thick] (0,0)--(2,0) node[midway,below] {\scriptsize $\Ising_\pm$} ;

\draw[blue,->-,ultra thick] (0,2)--(0,0) ; 
\draw[fill=white] (-0.07,-0.07) rectangle (0.07,0.07) ; 
%
%
\node at (-1,1.5) {\scriptsize $\FZ_1(\Rep(\Zb_2))$} ;
\node at (1,1.5) {\scriptsize $\FZ_1(\Ising_\pm)$} ;
\node at (0,-0.25) {\scriptsize $\Ising_\pm$} ;
\node at (0,2.2) {\scriptsize $\FZ_1(\Rep(\Zb_2),\Ising_\pm)$} ;
\node at (-2,2.2) {\scriptsize $\Rep(\Zb_2)$} ;

\draw[decorate,decoration=brace,very thick] (-2,2.4)--(0,2.4) node[midway, above] {\scriptsize $\Ising_\pm$};

\end{tikzpicture}
\caption{The physical relation between the $\Zb_2$ and double Ising topological orders.}
\label{fig_toric_to_Ising2}
\end{figure}

By Example \ref{expl_G_defects_boundary}, we know that the $\Zb_2$-crossed extensions of $\FZ_1(\Rep(\Zb_2))$ are given by 
$$
\FZ_1(\Rep(\Zb_2), \Ising_\pm) \simeq \Fun_{\Rep(\Zb_2)|\Ising_\pm}(\Ising_\pm, \Ising_\pm), 
$$
which can be computed directly. On the other hand, it can also be determined directly by the fact that it is precisely the de-equivariantization of $\FZ_1(\Ising_\pm)$. This de-equivariantization can be determined as follows. By \cite[Section B.3]{DGNO10}, there are 8 Ising type braided fusion categories $\Ising_\zeta$ up to equivalence, labeled by complex numbers $\zeta$ satisfying $\zeta^8 = -1$, and both of them are nondegenerate. The underlying fusion category of $\Ising_\zeta$ is determined by the sign $\pm$ in the equation $\zeta^2 + \zeta^{-2} = \pm \sqrt 2$. Therefore, there are braided equivalences
$\Ising_\zeta \boxtimes \overline{\Ising_\zeta} \simeq \FZ_1(\Ising_\pm)$
for some choices of $\zeta$. After taking the equivariantization, we obtain two minimal nondegenerate extensions:
\[
\FZ_1(\Rep(\Zb_2))^{\Zb_2} \hookrightarrow \FZ_1(\Ising_\pm) = \Ising_\zeta \boxtimes \overline{\Ising_\zeta}
\]
The simple objects of $\FZ_1(\rep(\Zb_2))^{\Zb_2}$ in $\Ising_\zeta \boxtimes \overline{\Ising_\zeta}$ are 
\[
\one \boxtimes \one, \quad \one \boxtimes \psi, \quad \psi \boxtimes \one, \quad \psi \boxtimes \psi, \quad \sigma \boxtimes \sigma .
\]
More precisely, we have
\[
\langle \one \boxtimes \one , \psi \boxtimes \psi \rangle = \langle \one \rangle^{\Zb_2} \simeq \rep(\Zb_2) , \quad \langle \one \boxtimes \psi , \psi \boxtimes \one \rangle = \langle f \rangle^{\Zb_2} , \quad \langle \sigma \boxtimes \sigma \rangle = \langle e , m \rangle^{\Zb_2} ,
\]
where $\langle x_i \rangle$ denotes the full subcategory consisting of the direct sums of the simple objects $x_i$. Moreover, notice that $\FZ_2(\FZ_1(\rep(\Zb_2))^{\Zb_2}) =\langle \one \boxtimes \one , \psi \boxtimes \psi \rangle \simeq \Rep(\Zb_2)$. Then the de-equivariantization of $\FZ_1(\Ising_\pm)$ is equivalent to take the modules over the condensable $\EE_2$-algebra
\[
A \coloneqq \one \boxtimes \one \oplus \psi \boxtimes \psi = \mathrm{Fun}(\Zb_2) \in \Rep(\Zb_2) \hookrightarrow \Ising_\zeta \boxtimes \Ising_\zeta .
\]
Recall that this algebra defines the well-known anyon condensation from the double Ising topological order to the $\Zb_2$ topological order. We obtain two $\Zb_2$-crossed extensions of $\FZ_1(\Rep(\Zb_2))$:
\be \label{eq:Z2-crossed_extensions_TC_mtc}
\FZ_1(\Rep(\Zb_2), \Ising_\pm) \simeq (\FZ_1(\Ising_\pm))_G \simeq \RMod_A(\Ising_\zeta \boxtimes \overline{\Ising_\zeta}) \simeq \langle \one,e,m,f \rangle \oplus \langle \chi_+, \chi_- \rangle,
\ee
which was shown in \cite[Section 3.2]{CJKYZ20}\footnote{Indeed, \cite[Section 3.2]{CJKYZ20} only consider the case $\zeta = \exp(-\pi \mathrm i/8)$, but there is no essential difference for other choices of $\zeta$.} to consist of 4 simple local $A$-modules $\one,e,m,f$ generating the trivial graded component (which is equivalent to $\FZ_1(\Rep(\Zb_2))$), and 2 simple non-local $A$-modules $\chi_\pm$ generating the nontrivial graded component.

The two 3-group homomorphisms $G \to \mathrm{BrPic}(\Rep(\Zb_2))$ naturally induce two monoidal 2-functors 
$$
\phi_\pm: 2\vect_{\Zb_2} \to \BMod_{\Rep(\Zb_2)|\Rep(\Zb_2)}(2\vect)^\op \simeq \Sigma\FZ_1(\Rep(\Zb_2)).
$$ 
The image of $\phi_\pm(\vect_{\Zb_2})$ are condensable $\EE_1$-algebras in $\CT\CC$. In the coordinate system $\Sigma\FZ_1(\Rep(\Zb_2))$, they are precisely the two $\Zb_2$-crossed extensions of $\FZ_1(\Rep(\Zb_2))$, explicitly computed in (\ref{eq:Z2-crossed_extensions_TC_mtc}). In the coordinate system  
$\BMod_{\Rep(\Zb_2)|\Rep(\Zb_2)}(2\vect)^\op$, they are precisely given by $\Ising_\pm$ as illustrated in Figure\,\ref{fig_toric_to_Ising2}. By condensing $\phi_\pm(\vect_{\Zb_2})$, we obtain the fusion 2-category $\Sigma\FZ_1(\Ising_\pm)$ associated to the condensed phases. Therefore, gauging the $e$-$m$-duality (i.e., a 0-form symmetry) in the 2+1D $\Zb_2$ topological order is equivalent to gauging the $2\vect_{\Zb_2}$-symmetry in $\CT\CC=\Sigma\FZ_1(\Rep(\Zb_2))$. 
\end{expl}

\subsubsection{Generalizations, examples and remarks} \label{sec:general_gauging}

In this subsubsection, we revisit an example of 1-form symmetry with 't Hooft anomaly and provide an example of gauging a 0-form symmetry in 4D. We remark on further generalizations of the notions of a symmetry and gauging. 

\begin{expl}
Let $A$ be a finite abelian group. Its delooping $\mathrm B A$ is a 2-group, describing a 1-form symmetry. Let $\CC$ be the modular tensor category of anyons in a 2+1D topological order. Then a 1-form symmetry $\mathrm B A$ on this topological order is a monoidal 2-functor $\phi \colon \mathrm B A \to \Sigma \CC$ or $\phi \colon 2\vect_{\mathrm B A} \simeq 2\rep(\hat A) \to \Sigma \CC$. A monoidal 2-functor $\phi \colon \mathrm B A \to \Sigma \CC$ is the same as a braided functor $\phi \colon A \to \CC$, which maps each $a \in A$ to an abelian anyon (invertible object in $\CC$). Moreover, since the braiding on $A$ is trivial, the anyons $\phi(a)$ have trivial double braiding with each other. Therefore, $E \coloneqq \bigoplus_{a \in A} \phi(a) \in \CC$ is a condensable $\EE_2$-algebra.

Gauging this 1-form symmetry is the same as condensing the algebra $\phi(\vect_{\mathrm B A}) \in \Sigma \CC$. Here $\vect_{\mathrm B A}$ is the condensation completion (or Karoubi completion) of $\mathrm B \bk[A]$. In other words, $\vect_{\mathrm B A}$ can be obtained by condensing $\bk[A]$ from the tensor unit $\vect \in 2\vect_{\mathrm B A}$. Thus $\phi(\vect_{\mathrm B A}) \in \Sigma \CC$ can be obtained by condensing $\phi(\bk[A]) = \bigoplus_{a \in A} \phi(a) = E$ from the tensor unit of $\Sigma \CC$. In other words, $\phi(\vect_{\mathrm B A}) = \Sigma E$, which is the 1+1D domain wall obtained by condensing the condensable $\EE_2$-algebra $E$ along a line. Hence gauging this 1-form symmetry is the same as the 2-codimensional condensation (i.e., anyon condensation) of $E$.

In literature, a subgroup $A$ of abelian anyons with nontrivial braiding is also called a 1-form symmetry with 't Hooft anomaly (see for example \cite{GKSW15}). In this case, these abelian anyons span a fusion subcategory of $\CC$, and its associator and braiding determine a class in the abelian cohomology (i.e., $\EE_2$ Eilenberg-MacLane cohomology) $\pi \in H^4_{\EE_2}(A;\Cb^\times)$. This fusion subcategory is denoted by $\vect_A^\pi$. On the other hand, $2\vect_{\mathrm B A}$ can also be twisted by $\pi \in H^4(\mathrm B A;\Cb^\times) \simeq H^4_{\EE_2}(A;\Cb^\times)$. Then this anomalous 1-form symmetry is a monoidal 2-functor $\phi \colon 2\vect_{\mathrm B A}^\omega \simeq \Sigma \vect_A^\omega \to \Sigma \CC$.

When $\pi$ is nontrivial, $\vect_{\mathrm B A}$ is not a condensable algebra in $2\vect_{\mathrm B A}^\pi$ because there is no fiber 2-functor $2\vect_{\mathrm B A}^\pi \to 2\vect$. So the $\mathrm B A$-symmetry cannot be gauged. We may say that the nontrivial class $\pi \in H^4_{\EE_2}(A;\Cb^\times)$ characterizes the 't Hooft anomaly of this 1-form symmetry.
\end{expl}

\begin{expl}
The codimension-2 topological defects in the 3+1D toric code model form a $\EE_2$-fusion 2-category $\CC \coloneqq \FZ_1(2\rep(\Zb_2))$ \cite{KTZ20a}. There is an invertible domain wall in the 3+1D toric code model corresponding to the nontrivial 2+1D $\Zb_2$ SPT order \cite[Theorem$^{\text{ph}}$ 3.34 and Figure 5]{KLWZZ20}. Also, by \cite[Theorem$^{\text{ph}}$ 3.37]{KLWZZ20}, this invertible domain wall in the coordinate system 
$$
\Sigma \CC \simeq \BMod_{2\rep(\Zb_2)|2\rep(\Zb_2)}(3\vect)
$$ 
is $2\rep(\Zb_2,\omega) = \Sigma \vect_{\Zb_2}^\omega$, where $\omega \in H^3(\Zb_2;\Cb^\times) \simeq \Zb_2$ is the nontrivial class.

This invertible domain wall has order $2$. It is natural to ask whether this invertible domain wall can be promoted to $\Zb_2$-symmetry. If such a $\Zb_2$-symmetry exists, it is a monoidal 3-functor $\phi \colon 3\vect_{\Zb_2} \to \Sigma \CC$ which maps the nontrivial simple object to the above invertible domain wall. In the coordinate system $\Sigma \CC \simeq \BMod_{2\rep(\Zb_2)|2\rep(\Zb_2)}(3\vect)$, it maps the tensor unit of $3\vect_{\Zb_2}$ to $2\rep(\Zb_2)$ and the nontrivial simple object to $2\rep(\Zb_2,\omega)$. Denote $\CP_0 \coloneqq 2\rep(\Zb_2)$ and $\CP_1 \coloneqq 2\rep(\Zb_2,\omega)$. The monoidal structure of $\phi$ is equivalent to a $\Zb_2$-graded fusion 2-category structure on $\CP = \CP_0 \oplus \CP_1$.

By the obstruction theory (similar to \cite{ENO10}), such $\Zb_2$-graded fusion 2-categories $\CP$ form a torsor over $H^4(\Zb_2;\Cb^\times) = 0$, hence are unique. It is also known that there is a 2-group $\CG$ defined by $\pi_1(\CG) = \pi_2(\CG) = \Zb_2$ and the Postnikov class (associator) $\omega \in H^3(\Zb_2;\Zb_2) \simeq H^3(\Zb_2;\Cb^\times) \simeq \Zb_2$ such that $\CP = 2\rep(\CG)$ (see for example \cite[Example 3.36]{HZ23}).

By gauging this $\Zb_2$-symmetry, we obtain a 3+1D topological order with the fusion 3-category of topological defects being
\[
\BMod_{\CP|\CP}(\BMod_{2\rep(\Zb_2)|2\rep(\Zb_2)}(3\vect)) \simeq \BMod_{\CP|\CP}(3\vect) \simeq \Sigma \FZ_1(\CP) .
\]
By \cite[Proposition 4.2.1]{DY25}, $\CP = 2\rep(\CG)$ is Morita equivalent to $2\vect_{\Zb_2 \times \Zb_2}^\pi$, where the 4-cocycle $\pi \in H^4(\Zb_2 \times \Zb_2;\Cb^\times)$ is defined by
\[
\pi((a_1,b_1),(a_2,b_2),(a_3,b_3),(a_4,b_4)) \coloneqq \omega(a_2,a_3,a_4)^{b_1} .
\]
Then $\FZ_1(\CP)$ is braided equivalent to $\FZ_1(2\vect_{\Zb_2 \times \Zb_2}^\pi)$. Hence the gauged theroy is $\SG\ST_{(\Zb_2\times \Zb_2,\pi)}^4$, i.e., the 3+1D $\pi$-twisted $\Zb_2 \times \Zb_2$ gauge theory.
\end{expl}

\begin{rem}
It seems that the framework of formulating an SET order as a monoidal functor also works for continuous group $G$ symmetry, but a monoidal $n$-functor $G \to \CC^\times$ should be understood as a pointed continuous map $\mathrm B G \to \mathrm B \CC^\times$, where $\mathrm B$ denotes the classifying space. For example, consider the 2+1D trivial phase equipped with a compact group $G$ symmetry. The space of invertible topological defects in the 2+1D trivial phase is $\CC^\times \simeq \mathrm B^2 \Cb^\times \simeq \mathrm B^3 \Zb = K(\Zb,3)$. The set of homotopy classes of pointed maps $\mathrm B G \to \mathrm B \CC^\times \simeq K(\Zb,4)$ is $H^4(\mathrm B G,\Zb)$, which exactly classifies the 2+1D Chern-Simons $G$-gauge theory \cite{DW90}. In other words, the 2+1D SPT orders with a compact group $G$ symmetry should be classified by $H^4(\mathrm B G,\Zb)$, and their gauged theories are the Chern-Simons theories. However, the categorical description of the gauging (or condensation) process in this case is not clear.
\end{rem}

\begin{rem}
Let $G$ be a finite group. Given an $n$+1D $G$-SET order described by a monoidal $n$-functor $\phi \colon n\vect_G \to \CC$, its partition functions can be constructed as follows. A spacetime manifold is an $(n+1)$-dimensional closed manifold $M$ equipped with a principal $G$-bundle, or equivalently, a continuous map $\gamma \colon M \to \mathrm B G$. If we fix a triangulation on $M$, then the map $\gamma$ is equivalent to a flat $G$-connection, that is, an assignment to each 1-simplex of $M$ a group element such that the holonomy on each 2-simplex vanishes. The monoidal $n$-functor $\phi$ is equivalent to a continuous map $\mathrm B \phi \colon \mathrm B G \to \mathrm B \CC^\times$. The composition of these two maps $(\mathrm B \phi) \circ \gamma \colon M \to \mathrm B \CC^\times$ assigns
\bit
\item to each 1-simplex $e$ of $M$ the codimension-1 domain wall $\phi(\gamma(e)) \in \CC$;
\item to each 2-simplex of $M$ the codimension-2 topological defect corresponding to the monoidal structure of $\phi$;
\item to each $(n+1)$-simplex of $M$ a codimension-$(n+1)$ topological defect (a number in $\Cb^\times$).
\eit
This is a $\CC$-colored graph on $M$, and the monoidal structure of $\phi$ ensures that this graph is invariant under the Pachner moves. Therefore, this graph can be evaluated to a number using the graph calculus in $\CC$ because $M$ is closed. This number is defined to be the partition function $Z(M,\gamma)$. For the gauged theory, the partition function need to sum over all possible $G$-bundles on $M$:
\[
Z_{\text{gauged}}(M) \coloneqq \frac{1}{\lvert G \rvert} \sum_{\gamma} Z(M,\gamma) .
\]
This partition function is the same as a graph in $M$ whose 1-simplices are labeled by the algebra $A \coloneqq \phi((n-1)\vect_G) \in \CC$, 2-simplices are labeled by the multiplication of $A$, and higher simplices are labeled by the higher associativity constraints of $A$.


This construction generalizes the Dijkgraaf-Witten theory \cite{DW90} and the decorated domain wall construction of SPT wave functions \cite{CLV14}, and should be a special case of the orbifold construction of defect TQFT \cite{CRS19}.
\end{rem}

\begin{rem} \label{rem:further_generalization_symmetry}
Further generalizations of the notion of a symmetry are possible. 
\bnu
\item[(1)] According to the mathematical theory of gapped/gapless quantum liquids \cite{KZ18,KZ20,KZ21,KZ22b,KZ22}, the topological skeleton $\SX_{\mathrm{top}}^{n+1}$ of an $n$D gapped/gapless quantum liquid $\SX^n$ is determined by a generalized quiche $(\SB^{n+1}, \SQ^n)$ such that the fusion $n$-category $\CQ$ is equipped with a braided monoidal functor $\varphi: \Omega\CB^\op \to \FZ_1(\CQ)$. The pair becomes an ordinary quiche when $\varphi$ is an equivalence.

The notion of a generalized symmetry can be defined as a morphism between pairs $(\CA,\CP) \to (\CB,\CQ)$, or equivalently, a morphism between two `quiches' $(\SA^{n+1}, \SP^n) \to (\SB^{n+1},\SQ^n)$ by a `generalized sandwich construction' as depicted below: 
\be \label{pic:enriched functor}
\begin{array}{c}
\begin{tikzpicture}[scale=1.5]
\fill[gray!20] (-2,0) rectangle (2,0.8) ;
\node at (-1,0.4) {\footnotesize $\SZ(\SB)^{n+2}$} ;
\node at (1,0.4) {\scriptsize $\SZ(\SA)^{n+2}$} ;
\draw[black, ultra thick,->-] (-2,0) -- (0,0) node[midway,below,scale=1] {\scriptsize $\SD^{n+1}$} ; 
\draw[black, ultra thick,->-] (0,0) -- (2,0) node[midway,below,scale=1] {\scriptsize $\SC^{n+1}$} ; 
\draw[black, ultra thick,->-] (0,0) -- (0,0.8) node[midway,right] {\scriptsize $\SF^{n+1}$} ; 
\draw[black, ultra thick,->-] (2,0) -- (2,0.8) node[midway,right] {\scriptsize $\SA^{n+1}$} ; 
\draw[fill=white] (-0.05,-0.05) rectangle (0.05,0.05) node[midway,below,scale=1] {\scriptsize $\SG^n$} ;
\draw[fill=white] (1.95,-0.05) rectangle (2.05,0.05) node[midway,below,scale=1] {\scriptsize $\hspace{1mm}\SP^n$} ;
\draw[decorate,decoration=brace,very thick] (2,-0.3)--(0,-0.3) ;
\node at (1,-0.5) {\scriptsize $\SG^n\boxtimes_{\SC} \SP^n =\SQ^n$} ;
\draw[decorate,decoration=brace,very thick] (0,0.85)--(2,0.85) ;
\node at (1,1.05) {\scriptsize $\SF^n\boxtimes_{\SZ(\SA)} \SA^n =\SB^n$} ;
\end{tikzpicture}
\end{array}
\ee
where the generalized `quiche' $(\SF^{n+1}, \SG^n)$ defines the morphism, and $\SC^{n+1}$ (resp. $\SD^{n+1}$) is the unique topological order that is Morita equivalent to $\SA^{n+1}$ (resp. $\SB^{n+1}$) via $\SP^n$ (resp. $\SQ^n$). 

When $\SA^{n+1}=\SB^{n+1}=\mathbf{1}^{n+1}$, by \cite{KZ18,KZ24}, this generalized symmetry recovers the symmetry defined by a monoidal functor $\CG: \CP \to \CQ$ as proposed in \cite{LYW24}. When $n=1$, this `generalized sandwich construction' provides a physical realization of an enriched monoidal 1-functor \cite{KYZZ24}. For general $n$, we expect this `generalized sandwich construction' to provide a physical realization of an enriched monoidal higher functor. 

\item[(2)] A new type of `topological symmetries', generalizing that of `quiche' \cite{FMT24}, was proposed in \cite{Kon22Talk} as an application of topological Wick rotation.
This new type of a `topological symmetry' consists of a topological order, a gapped boundary (codimension 1), a gapped corner (a defect of codimension 2 living on the boundary), a higher corner (a defect of codimension $>2$ living on the boundary), etc., largely generalizing the notion of a `quiche' (i.e., a topological order + a gapped boundary). Maybe it can be called a ``higher quiche''. A symmetry can be defined as a morphism between two generalized `topological symmetries' via a `generalized sandwich construction', i.e., via a topless higher dimensional cubic generalizing the topless square (with three edges labeled by $\SF^{n+1}$, $\SC^{n+1}$, $\SA^{n+1}$) in (\ref{pic:enriched functor}). This leads to an interesting relation between theories in two different dimensions (i.e., a generalized topological holography) \cite{Kon22Talk}. 
\enu
The theory of gauging these further-generalized symmetries is automatic or tautologically followed from our theory of defect condensations in topological orders because all the ingredients in the `generalized sandwich construction' are all (potentially anomalous) topological orders. In the light of this discussion, it is beneficial to simply define a symmetry by a topological order or a quiche or a higher quiche. In particular, this generalized gauging of symmetry allows us to obtain an SPT/SET/SSB order from a topological order via gauging and vice versa. 
\end{rem}

\newpage


\section{Generalizations and Applications} \label{sec:app}
In this work, we have developed a general theory of the condensations of topological defects in higher dimensions. In this section, we show that this theory leads to many new interesting questions, generalizations, applications and outlooks. We want to emphasize that this work is not an end of condensation theory but a beginning of a much richer theory in both mathematics and physics. For example, the natural questions in condensations lead us to a general higher Morita theory; the physical intuition of integrating local observables to a global leads us to a more general theory of factorization homology; the generalization to gapless liquid-like defects leads us a more complete theory of condensation. In this section, we can only scratch the surfaces of these directions but leave the thorough development to future publications. 

\subsection{Higher Morita theory} \label{sec:Higher_Morita_Theory}
In this subsection, we study some questions that naturally arise from the physics of condensations. The most basic question is when two condensations produce the same condensed phase or the same condensed defect.  

\subsubsection{Higher Morita equivalences}

Given a topological order $\SC^{n+1}$ and two condensations defined by $A,B\in \Algc_{\EE_1}(\CC)$, it is natural to ask when they produces the same condensed phase $\SD^{n+1}$. We formulate the question in more mathematical terms.   


\begin{defn} \label{def:ModE1_RMod_eq_MN}
Let $\CB\in \Algc_{\EE_1}((n+1)\vect)$ and $A,B\in \Algc_{\EE_1}(\CB)$. 
\begin{itemize}

\item For $\CM \in \Mod_\CB^{\EE_1}((n+1)\vect)$, $A$ and $B$ are called {\it $\Mod^{\EE_1}(\CM)$-equivalent} if $\Mod_A^{\EE_1}(\CM) \simeq \Mod_B^{\EE_1}(\CM)$ in $\Algc_{\EE_1}((n+1)\vect)$. When $\CM=\CB$, we abbreviate this equivalence to $\Mod^{\EE_1}$-equivalence. 

\item For $\CN \in \LMod_\CB((n+1)\vect)$, $A$ and $B$ are called  {\it $\LMod(\CN)$-equivalent} if $\LMod_A(\CN) \simeq \LMod_B(\CN)$ in $(n+1)\vect$. When $\CN=\CB$, we abbreviate this equivalence to $\LMod$-equivalence, which is just the usual Morita equivalence. The definition of a $\RMod(\CP)$-equivalence for $\CP \in \RMod_\CB((n+1)\vect)$ is similar. 

\end{itemize}
\end{defn}

\begin{expl}
Recall that Morita equivalent topological orders can be obtained from each other via condensations of 1-codimensional topological defects. Therefore, for fusion $n$-category $\CC$ with a trivial $\EE_1$-center, there is a one-to-one correspondence between the $\Mod^{\EE_1}$-equivalence classes of simple condensable $\EE_1$-algebras in $\CC$ and the isomorphic classes of non-degenerate braided fusion $(n-1)$-categories within the Witt class of $[\Omega\CC]$. 
\end{expl}

\begin{expl}
Recall Theorem$^{\mathrm{ph}}$\,\ref{pthm:construct_cond_E1_algebras}, for a fixed $\CP$, all the condensable $\EE_1$-algebras $\Fun_{\CP^\rev}(\CX,\CX)$ in $\CC=\BMod_{\CB|\CB}(n\vect)$ for $\CX \in \BMod_{\CB|\CP}(n\vect)$ are all $\Mod^{\EE_1}$-equivalent. 
\end{expl}

\begin{prop} \label{prop:LModeq_implies_ModE1eq}
If $A$ and $B$ are $\LMod$-equivalence, then they are $\Mod^{\EE_1}$-equivalent. 
\end{prop}
\pf
When $A$ and $B$ are $\LMod$-equivalent, set $\CM=\LMod_A(\CB) \simeq \LMod_B(\CB)$. We have $\Mod_A^{\EE_1}(\CB) \simeq \Fun_\CB(\CM,\CM)^\rev \simeq \Mod_B^{\EE_1}(\CB)$. 
\epf

\begin{rem} \label{rem:ModE1_not=_LMod}
Conversely, $\Mod^{\EE_1}$-equivalence does not imply $\LMod$-equivalence. We give an example. For $\CB=\Rep(\Zb_2)$ and $A=\one, B= \one \oplus e \in \Algc_{\EE_1}(\Rep(\Zb_2))$, we have 
$$
\LMod_A(\Rep(\Zb_2)) \simeq \Rep(\Zb_2), \quad \quad \LMod_B(\Rep(\Zb_2)) \simeq \vect.
$$ 
On the other hand, we have $\Mod_A^{\EE_1}(\Rep(\Zb_2)) \simeq \vect_{\Zb_2} \simeq \Fun_{\Rep(\Zb_2)}(\vect, \vect) \simeq \Mod_B^{\EE_1}(\Rep(\Zb_2))$. 
\end{rem}



We generalize the equivalences in Definition\,\ref{def:ModE1_RMod_eq_MN} to $\EE_k$-algebras. 
\begin{defn} \label{def:ModEk_kMorita_eq}
Let $\CB\in \Algc_{\EE_k}((n+1)\vect)$ and $A,B\in \Algc_{\EE_k}(\CB)$. 
\bnu
\item For $\CM \in \Mod_\CB^{\EE_k}((n+1)\vect)$, $A$ and $B$ are called {\it $\Mod^{\EE_k}(\CM)$-equivalent} if 
$$
\Mod_A^{\EE_k}(\CM) \simeq \Mod_B^{\EE_k}(\CM) \quad \in \Algc_{\EE_k}((n+1)\vect). 
$$ 
When $\CM=\CB$, we abbreviate `$\Mod^{\EE_k}(\CB)$-equivalent' to `$\Mod^{\EE_k}$-equivalent'.

\item For $l<k$ and $\CN \in \LMod_\CB(\Algc_{\EE_l}((n+1)\vect))$, $A$ and $B$ are called  {\it $\LMod(\CN)$-$\EE_l$-equivalent} if 
$$
\LMod_A(\CN) \simeq \LMod_B(\CN) \quad \in \Algc_{\EE_l}((n+1)\vect).
$$ 
When $\CN=\CB\in \LMod_\CB(\Algc_{\EE_{k-1}}((n+1)\vect))$, we abbreviate the $\LMod(\CB)$-$\EE_{k-1}$-equivalence to $\LMod$-$\EE_{k-1}$-equivalence. 

\item Recall that $\Sigma\CB=\RMod_\CB((n+1)\vect) \in \Algc_{\EE_{k-1}}((n+2)\vect)$ and $\Sigma A := \RMod_A(\CB) \in \Algc_{\EE_{k-1}}(\Sigma\CB)$. 
For $k\geq p> 1$, we define $\Sigma^pA$ iteratively as $\Sigma^p A:=\RMod_{\Sigma^{p-1}A}(\Sigma^{p-1}\CB) \in \Algc_{\EE_{k-p}}(\Sigma^p\CB)$. Then $A$ and $B$ are called $k$-Morita equivalent if $\Sigma^k A \simeq \Sigma^k B$ in $\Sigma^k\CB$. 
\enu
\end{defn}

\begin{expl}  \label{def:k_Morita_eq}
We give an example of $k$-Morita equivalence when $\CB=n\vect$. We spell out the $k$-Morita equivalence in this case more explicitly. 
\begin{itemize}
\item Two $\EE_k$-multi-fusion $n$-categories $\CA$ and $\CB$ are called {\it $k$-Morita equivalent} if $\Sigma^k \CA \simeq \Sigma^k\CB$ as separable $(n+k)$-categories. 
\end{itemize}
This $k$-Morita equivalence generalizes the usual Morita equivalence and Witt equivalence. 
\bnu
\item When $k=1$, $1$-Morita equivalence is precisely the usual Morita equivalence. 
\item When $k=2$ and both $\CA$ and $\CB$ are nondegenerate, the $2$-Morita equivalence is precisely the usual Witt equivalence \cite{JF22,KZ22b}. The advantage of the notion of 2-Morita equivalence is that it also works for (not nondegenerate) braided fusion categories. 
\enu
The $k$-Morita equivalence is clearly a well-defined equivalence relation. We denote the $k$-Morita equivalence class associated to $\CA$ by $[\CA]_k$. 
\end{expl}

\begin{expl}
Let $\CB$ be a non-degenerate braided fusion $n$-category. Then all Lagrangian $\EE_2$-algebras in $\CB$ are $\Mod^{\EE_2}$-equivalent by definition. It is an interesting question to work out when two condensable $\EE_2$-algebra in $\CB$ are $\Mod^{\EE_2}$-equivalent. When $n=1$, an interesting answer to this question can be found in \cite{FFRS06}. A work devoted to the study of this question for $n=1$ appeared recently \cite{XY25}. 
\end{expl}

\begin{rem}
When $\CB$ is non-degenerate, the functor $\Sigma(-): \Algc_{\EE_2}(\CB) \to \Algc_{\EE_1}(\Sigma\CB)$ induces an injection from the set of $\Mod^{\EE_2}$-equivalence classes of condensable $\EE_2$-algebras in $\Omega\CB$ to the set of $\Mod^{\EE_1}$-equivalence classes of condensable $\EE_1$-algebras in $\CB$.
\end{rem}

\begin{rem}
There is a natural higher category called Morita category, in which many notions and problems of higher Morita theory can be formulated and studied. It consists of of $\EE_m$-algebras as 0-morphisms, $\EE_{m-1}$-algebras over $\EE_m$-algebras as 1-morphisms, $\EE_{m-2}$-algebras over $\EE_{m-1}$-algebras as 2-morphisms, so on and so forth. We will discuss this Morita category elsewhere \cite{KZZZ25}. 
\end{rem}

We expect the following result to be true. We will provide a proof of this result, together with a more complete theory of $k$-Morita equivalence, elsewhere \cite{KZZZ25}.
\begin{prop} \label{prop:k-Morita_eq}
For $\CA \in \Algc_{\EE_k}((n+1)\vect)$ and $A\in \Algc_{\EE_k}(\CA)$, $\CA$ and $\Mod_A^{\EE_k}(\CA)$ are $k$-Morita equivalent. 
\end{prop}

\begin{prop}
For $\CA,\CB\in \Algc_{\EE_k}((n+1)\vect)$, if $\CA$ and $\CB$ are $k$-Morita equivalent, then $\FZ_k(\CA) \simeq \FZ_k(\CB)$ in $\Algc_{\EE_{k+1}}((n+1)\vect)$.
\end{prop}
\pf
We have $\EE_{k+1}$-monoidal equivalences: $\FZ_k(\CA) \simeq \Omega^k\FZ_0(\Sigma^k \CA) \simeq \Omega^k\FZ_0(\Sigma^k \CB) \simeq \FZ_k(\CB)$.  
\epf

\begin{rem}
There is a more general notion of the $\EE_k$-center of an $\EE_k$-algebra in an $\EE_k$-monoidal $n$-category $\CC$. It is necessarily living in the category $\FZ_k(\CC)$. Hence, one can expect a generalization of above proposition in this setting. We will develop this theory in \cite{KZZZ25}. 
\end{rem}

The following Lemma is a result in \cite{Lur17,Fra12} reformulated in the world of separable $n$-categories. It is briefly explained in Appendix\,\ref{Appendix:Zm-center_Em-algebra}. 
\begin{lem} \label{lem:francis}
For $\CA, \CB \in \Algc_{\EE_k}((n+1)\vect)$ and an $\EE_k$-monoidal functor $F: \CA \to \CB$, we have 
$$
\FZ_k(F)=\FZ_k(\CA,\CB) \simeq \hom_{\Mod_\CA^{\EE_k}((n+1)\vect)}(\CA, \CB) \quad \in \Algc_{\EE_{k}}((n+1)\vect). 
$$
When $F=\id_\CA$, we obtain $\FZ_k(\CA) \simeq \hom_{\Mod_\CA^{\EE_k}((n+1)\vect)}(\CA, \CA) \in \Algc_{\EE_{k+1}}((n+1)\vect)$, which is also a consequence of Theorem\,\ref{cor:EkMod_Zk_centralizer}. 
\end{lem}


\begin{prop}
For $\CA, \CB \in \Algc_{\EE_k}((n+1)\vect)$, if $\CA$ and $\CB$ are $\Mod^{\EE_k}$-equivalent, then $\FZ_k(\CA) \simeq \FZ_k(\CB)$. 
\end{prop}

\begin{prop}
For $\CB\in \Algc_{\EE_k}((n+1)\vect)$ and $A,B\in \Algc_{\EE_k}(\CB)$, if $A,B\in \Algc_{\EE_k}(\CB)$ are $k$-Morita equivalent, then they are $\Mod^{\EE_k}$-equivalent. 
\end{prop}
\pf
It follows from $\Mod_A^{\EE_k}(\CB) \simeq \Omega^{k-1}\Mod_{\Sigma^{k-1}A}^{\EE_1}(\Sigma^{k-1}\CB) \simeq \Omega^{k-1}\Mod_{\Sigma^{k-1}B}^{\EE_1}(\Sigma^{k-1}\CB) \simeq \Mod_B^{\EE_k}(\CB)$, where the second `$\simeq$' is a consequence of the assumption $\Sigma^k A \simeq \Sigma^k B$ in $\Sigma^k \CB$ and Proposition\,\ref{prop:LModeq_implies_ModE1eq}. 
\epf


\subsubsection{Enveloping algebras} \label{sec:enveloping_algebra}

In order to study the $\Mod^{\EE_k}$-equivalence among $\EE_k$-algebras, it is beneficial to introduce an $\Mod^{\EE_k}$-invariant, which is called the enveloping algebra $U_A$ of an $\EE_k$-algebra $A \in \Algc_{\EE_k}(\CC)$ for an $\EE_k$-fusion $n$-category $\CC$ such that $\Mod_A^{\EE_k}(\CC) \simeq \LMod_{U_A}(\CC)$. This notion allows us to reduce the problem of studying $\Mod^{\EE_k}$-equivalence to that of studying usual Morita equivalence. 

\medskip
Let us start from the $k=1$ case. Let us first recall an $\EE_1$-algebra $A$ in $\vect$ is just an ordinary $\Cb$-algebra. An $A$-$A$-bimodule is the same as a left $A\otimes_\Cb A^\op$-module. Therefore, $U_A=A\otimes_\Cb A^\op$. However, this construction does not generalize to $A\in \Algc_{\EE_1}(\CC)$ naively because $\CC$ is only monoidal without braidings. 

In order to find the proper definition, we first look at the physical meaning behind the data $A$ and $\CC$. When $k=1$, the fusion $n$-category $\CC$ is the category of topological defects of an $n+$1D topological order, which has potentially non-trivial gravitational anomaly $\SZ(\SC)^{n+2}$ as illustrated in Figure\,\ref{fig:U_C+U_A} (a), where $x\in \Mod_A^{\EE_1}(\CC)$ and only 1-spatial-dimension (the horizontal direction) is shown and the remaining $(n-1)$-spatial-dimensions are orthogonal to the paper. Since the remaining dimensions do not appear in our discussion. Without lose of generality, we can simply assume $n=1$. Since 1-codimensional topological defects in $\SZ(\SC)^{n+2}$ change the boundary condition, we only consider 2-codimensional topological defects in $\SZ(\SC)^{n+2}$. They form a braided fusion $n$-category given by the Drinfeld center $\FZ_1(\CC)$ of $\CC$ by boundary-bulk relation \cite{KWZ15,KWZ17}.

The shaded region is $\mathrm{D}^1 \times \Rb$, where $\mathrm{D}^1$ is an oriented closed 1-disk and its oriented boundary is a 0-sphere, i.e. $\partial \mathrm{D}^1 = \partial \mathrm{D}^1_+ \cup \partial \mathrm{D}^1_- = S^0 = \{ +, - \}$. Notice that the macroscopic observables in the interior of the shaded region (i.e. $\mathring{D}^1 \times \Rb$) is given by $\FZ_1(\CC)$. We denote this fact by $\Obs(\mathring{D}^1 \times \Rb)=\FZ_1(\CC)$. Similarly, we have $\Obs(\partial \mathrm{D}_+^1 \times \Rb)=\CC$ and $\Obs(\partial \mathrm{D}_-^1 \times \Rb)=\CC^\rev$, which should be viewed as observables living in an open neighborhood of $\partial\mathrm{D}_\pm \times \Rb$. One can summarize all the observables living in the shaded region $\mathrm{D}^1 \times \Rb$ by integrating them over the whole region \cite{AKZ17}. More explicitly, we have 
\begin{align}
\Obs(\mathrm{D}^1 \times \Rb) &=\int_{\mathrm{D}^1 \times \Rb} (\Obs(\mathring{D}^1 \times \Rb), \Obs(\partial \mathrm{D}_+^1 \times \Rb), \Obs(\partial \mathrm{D}_-^1 \times \Rb)  \nn
&= \int_{\mathrm{D}^1 \times \Rb} (\FZ_1(\CC)|_{\mathring{D}^1 \times \Rb}, \CC|_{\partial \mathrm{D}^1_+ \times \Rb}, \CC^\rev|_{\partial \mathrm{D}^1_- \times \Rb}) := \CC \boxtimes_{\FZ_1(\CC)} \CC^\rev, \label{eq:Obs_fan}
\end{align}
where $\boxtimes_{\FZ_1(\CC)}$ is the relative tensor product over $\FZ_1(\CC)$. One way to see why $\CC \boxtimes_{\FZ_1(\CC)} \CC^\rev$ is qualified to be called the `global observable' is to show that $\Obs(\mathring{D}^1 \times \Rb), \Obs(\partial \mathrm{D}_+ \times \Rb), \Obs(\partial \mathrm{D}_- \times \Rb)$ naturally map into the global one. Indeed, one can define these three maps explicitly and precisely below. 
\begin{align}
&\CC \to \CC \boxtimes_{\FZ_1(\CC)} \CC^\rev, \quad \FZ_1(\CC) \to \CC \boxtimes_{\FZ_1(\CC)} \FZ_1(\CC) \boxtimes_{\FZ_1(\CC)} \CC^\rev 
\simeq \CC \boxtimes_{\FZ_1(\CC)} \CC^\rev, \quad \CC^\rev \to \CC \boxtimes_{\FZ_1(\CC)} \CC^\rev \nn
&a \mapsto a\boxtimes_{\FZ_1(\CC)} \one_\CC, \quad\quad\quad\,\,\,\,
z \mapsto \one_\CC \boxtimes_{\FZ_1(\CC)} z \boxtimes_{\FZ_1(\CC)} \one_\CC, \hspace{3.6cm}
b \mapsto \one_\CC \boxtimes_{\FZ_1(\CC)} b \nonumber
\end{align}
Moreover, all three maps preserve the $\EE_1$-algebraic structure of three observables algebras. In other words, they are all monoidal functors as required by the obvious physical intuitions. 

\begin{figure}
$$
\begin{array}{c}
\begin{tikzpicture}[scale =1.2]
\fill[gray!20] (2,0) arc (0:180:2) -- cycle ;
\draw[color = black, ->-, ultra thick](-2,0)--(-0.2,0);
\draw[color = black, ->-, ultra thick](2,0)--(0.2,0);
\fill[white] (0.2,0) arc (0:180:0.2) -- cycle ;
\draw[fill=white] (-0.05,-0.05) rectangle (0.05,0.05) node[midway,below,scale=1] {\small $(\CC,x)$} ;
\node[black] at(-1.5,-0.22) {\small $(\CC,A)$};
\node[black] at(1.5,-0.22) {\small $(\CC^\rev,A)$};
\node[black] at(1,1.2) {\small $\FZ_1(\CC)$};
\draw[dashed,->] (9mm,0mm) arc [start angle=0, end angle=180, radius=9mm] node[near end,left,scale=1] {\small $\mathrm{D}^1$};
\draw[dashed,->] (0,2) -- (0,0.2) node[very near end,right] {\small $\Rb$};
\end{tikzpicture} 
\\
(a)
\end{array}
\quad\quad\quad\quad
\begin{array}{c}
\begin{tikzpicture}[scale =1.2]
\fill[gray!20] (2,0) arc (0:180:2) -- cycle ;
\draw[color = black, ->-, ultra thick](-2,0)--(-0.2,0);
\draw[color = black, ->-, ultra thick](2,0)--(0.2,0);
\fill[white] (0.2,0) arc (0:180:0.2) -- cycle ;
\draw[fill=white] (-0.05,-0.05) rectangle (0.05,0.05) node[midway,below,scale=1] {\small $\CM$} ;
\node[black] at(-1.5,-0.25) {\small $\CC$};
\node[black] at(1.5,-0.25) {\small $\CC^\rev$};
\node[black] at(1,1.2) {\small $\FZ_1(\CC)$};
\draw[dashed,->] (9mm,0mm) arc [start angle=0, end angle=180, radius=9mm] node[near end,left,scale=1] {\small $\mathrm{D}^1$};
\draw[color=black, ->-, ultra thick] (0,2) -- (0,0.2) node[very near start,right] {\footnotesize $\Fun_{\CC|\CC}(\CM,\CM)$};
\end{tikzpicture}
\\
(b)
\end{array}
$$
\caption{Physical intuition of the enveloping algebras of $\CC$ and $A\in \Algc_{\EE_1}(\CC)$}
\label{fig:U_C+U_A}
\end{figure}

Therefore, it is clear that the two-side $A$-action on $x$ is the same as the left $A\boxtimes A$-action on $x$ through the following three logical steps: 
\bnu
\item $A \boxtimes A$ is an $\EE_1$-algebra in $\CC\boxtimes \CC^\rev$.

\item There are two canonical monoidal functors: 
\begin{align}
\CC\boxtimes \CC^\rev &\xrightarrow{\boxtimes_{\FZ_1(\CC)}} \CC\boxtimes_{\FZ_1(\CC)} \CC^\rev \xrightarrow{\simeq} \Fun(\CC,\CC)  \nn
a\boxtimes b &\quad \mapsto \,\,\,\,\, a\boxtimes_{\FZ_1(\CC)} b \quad \mapsto \,\,\,  a\otimes - \otimes b,
\label{eq:a-b-action}
\end{align}
where the relative tensor product $\CC\boxtimes_{\FZ_1(\CC)} \CC^\rev$ amounts to closing fan in Figure\,\ref{fig:U_C+U_A} and the equivalence `$\simeq$' is due to the boundary-bulk relation \cite{KWZ15,KWZ17}. Therefore, $A\boxtimes A$ is mapped to an $\EE_1$-algebra in $\Fun(\CC,\CC)$. 

\item $\Fun(\CC,\CC)$ acts on $\CC$ canonically. Therefore, $A\boxtimes A$ acts on $x\in \CC$ canonically. More precisely, by (\ref{eq:a-b-action}), this action $(A\boxtimes A) \odot x := A\otimes x \otimes A$ is precisely the correct two-side $A$-action.
\enu
Therefore, the correct definition of $U_A$ is $A\boxtimes A \in \CC\boxtimes \CC^\rev$. 

\begin{rem}
Note that the fusion product $\otimes$ in the category $\CC \boxtimes_{\FZ_1(\CC)} \CC^\rev$ is defined by 
$$
(a\boxtimes_{\FZ_1(\CC)} b) \otimes (a' \boxtimes_{\FZ_1(\CC)} b') := (a\otimes_\CC a') \boxtimes_{\FZ_1(\CC)} (b'\otimes_\CC b').
$$
Geometrically, it says that, by integrating over $\mathrm{D}^1$, the fusion product along this $\mathrm{D}^1$-direction  in $\FZ_1(\CC)$ has been integrated out, and no longer makes sense after the integration. However, the fusions along the $\Rb$-direction remains a well-defined fusion product. By ignoring the fusion products in $\CC,\FZ_1(\CC), \CC^\rev$ in the $\Rb$-direction, it makes sense to integrate over $\mathrm{D}^1$ only. Moreover, in $2\vect$, we have 
$$
\int_{\mathrm{D}^1 \times \Rb} (\FZ_1(\CC)|_{\mathring{D}^1 \times \Rb}, \CC|_{\partial \mathrm{D}^1_+ \times \Rb}, \CC^\rev|_{\partial \mathrm{D}^1_- \times \Rb}) = 
\int_{\mathrm{D}^1} (\FZ_1(\CC)|_{\mathring{D}^1}, \CC|_{\partial \mathrm{D}^1_+}, \CC^\rev|_{\partial \mathrm{D}^1_-}). 
$$
Since remembering the fusion product in the $\Rb$-direction in the right hand side causes no confusion, we agree that above identity holds in $\Alg_{\EE_1}((n+1)\vect)$ as well. We use them interchangeably. 
\end{rem}

Before we give the official definition of $U_A$, we first discuss the meaning of $\CC\boxtimes \CC^\rev$. Since $\CC$ is an $\EE_1$-algebra in $(n+1)\vect$, mathematically, an $\CC$-$\CC$-module is precisely a left $\CC\boxtimes \CC^\rev$-module. This mathematical fact has a nice physical meaning as shown in Remark\,\ref{rem:U_C}. Therefore, $\CC\boxtimes \CC^\rev$ is precisely the enveloping algebra of $\CC$. 

\begin{rem} \label{rem:U_C}
The physical meaning of a $\CC$-$\CC$-bimodule $\CM$ is a the category of all wall conditions on a gapped domain wall $\SM^n$ between $\SC^{n+1}$ and $\SC^{n+1}$ as illustrated in Figure\,\ref{fig:U_C+U_A} (b). Moreover, this wall uniquely determines a relative bulk or a relative gravitational anomaly $\SZ^{(1)}(\SM)^{n+1}$ such that the category of topological defects on it is given by $\Fun_{\CC|\CC}(\CM,\CM)$ (recall Figure\,\ref{fig:functoriality_bbr}) \cite{KWZ15,KZ24}. By definition, to say $\CM$ is a $\CC$-$\CC$-bimodule is equivalent to say there is a monoidal functor from $f: \CC\boxtimes \CC^\rev$ to $\Fun(\CM,\CM)$, which is precisely a morphism from $\SC^{n+1} \boxtimes \overline{\SC^{n+1}}$ to $\SZ(\SM)^{n+1}$. By the definition of a morphism between topological order \cite{KWZ15,KWZ17}, this morphism is precisely defined by the domain wall $\SZ^{(1)}(\SM)^{n+1}$ between $\SZ(\SC)^{n+2}$ and $\SZ(\SC)^{n+2}$ (see also \cite{KZ18,KZ24}). 
\end{rem}

We summarize above discussion in the following mathematical definition and Lemma. 
\begin{defn}
For $\CC\in \Algc_{\EE_1}(2\vect)$, we define its enveloping algebra $U_\CC$ in $(n+1)\vect$ as follows:  
\be
U_\CC:=\int_{S^0} \CC := \CC \boxtimes \CC^\rev \in \Algc_{\EE_1}((n+1)\vect).
\ee  
where $S^0$ denotes the 0-sphere consisting of two points, i.e. $S^0=\{ +,- \}$.
For $A\in \Algc_{\EE_1}(\CC)$, we define the enveloping algebra $U_A$ of $A$ as follows: 
$$
U_A := \int_{S^0} A = A \boxtimes A \in \Algc_{\EE_1}(U_\CC). 
$$
\end{defn}

\begin{rem}
One can also view the pair $(\CC,A)$ as an $\EE_1$-algebra \cite{Str12} and its enveloping algebra is also defined by a pair $U_{(\CC,A)}:=(U_\CC,U_A) = \int_{S^0} (\CC,A)$. 
\end{rem}

Since $\CC$ is naturally a left $U_\CC$-module, it makes sense to talk about a left $U_A$-module in $\CC$. More precisely, a left $U_A$-module in $\CC$ is an object $x\in \CC$ equipped with a unital left $U_A$-action $\mu_x: U_A \odot x \to x$ satisfying the usual left unit and associativity properties of a left module. Moreover, the category $\LMod_{U_A}(\CC)$ of left $U_A$-modules in $\CC$ is well-defined. The following Lemma is tautological. 
\begin{lem}
We have $\Mod_\CC^{\EE_1}((n+1)\vect) \simeq \LMod_{U_\CC}((n+1)\vect)$ and 
$$
\Mod_A^{\EE_1}(\CC) \simeq \LMod_{U_A}(\CC).
$$
\end{lem}

Since $\Algc_{\EE_k}(-)=\Algc_{\EE_1}(\Algc_{\EE_{k-1}}(-))$ and $\Mod^{\EE_k}(-)=\Mod^{\EE_1}(\Mod^{\EE_{k-1}}(-))$ and the $i$-th multiplication or $i$-th action occurs in $i$-th spatial direction that is orthogonal to the $j$-th direction for $i\neq j$, the generalization to $\EE_k$-algebra is straightforward. 
\begin{defn}
For $\CC\in \Algc_{\EE_k}((n+1)\vect)$ and $A\in \Algc_{\EE_k}(\CC)$, we define the enveloping algebra $U_{(\CC,A)}^{\EE_k}$ of $(\CC,A)$ as follows: 
$$
U_{(\CC,A)}^{\EE_k} =\int_{S^{k-1}} (\CC,A) = (U_\CC^{\EE_k}, U_A^{\EE_k}) = (\smallint_{S^{k-1}} \CC, \quad \smallint_{S^{k-1}} A). 
$$
\end{defn}

\begin{prop}
We have $\CC\in \Mod_\CC^{\EE_k}((n+1)\vect) \simeq \LMod_{U_\CC^{\EE_k}}((n+1)\vect)$ and 
$$
\Mod_A^{\EE_k}(\CC) \simeq \LMod_{U_A^{\EE_k}}(\CC).
$$
As a consequence, two $A,B \in \Algc_{\EE_k}(\CC)$ are $\Mod^{\EE_k}$-equivalent if and only if $U_A^{\EE_k}, U_B^{\EE_k} \in \Algc_{\EE_1}(U_\CC^{\EE_k})$ are $\LMod(\CC)$-equivalent. 
\end{prop}

\subsection{Theory of integrals}\label{sec:FH}
The study of dimensional reduction process in topological orders is deeply related to the mathematical theory of factorization homology as shown for 2+1D anomaly-free topological orders in \cite{AKZ17}. However, in general anomalous and higher dimensional settings, it is certainly different from the usual theory (the so-called $\alpha$-version) of factorization homology \cite{Lur17,AF20,AFT17,AFT16} (see Remark\,\ref{rem:alpha_version_FH}). We believe that the theory of integrals emerging from this work is related to the so-called $\beta$-version of factorization homology partially developed in \cite{AFR18} (see Remark\,\ref{rem:alpha_version_FH}). We hope to come back to this point in the future. In this subsection, we only briefly discuss some results of integrating physical observables that follow directly from physical intuitions.

\medskip
Consider two $n+$1D  Morita equivalent topological orders $\SC^{n+1}$ and $\SD^{n+1}$ connected by a gapped domain wall $\SM^n$. We specify the wall condition by a pair $(\CM,m)$, where the separable $n$-category $\CM$ is the category of wall conditions and the distinguished object $m \in \CM$ specifies a single wall condition. 

\medskip
Topological defects of codimension $k$ in $\SC^{n+1}$ can be fused with a defect (of the same dimension) on the wall and become topological defects on the wall. This fusion defines an action 
\begin{align} \label{eq:odot_k}
\odot: \Omega^{k-1}\CC \times \Omega_m^{k-1}\CM &\to \Omega_m^{k-1}\CM \\
(a,x) &\mapsto a\odot x. 
\end{align}
Note that the category $\Omega^{k-1}\CC$ is $\EE_k$-monoidal and the category $\Omega_m^{k-1}\CM$ is $\EE_{k-1}$-monoidal. The functor $\odot$ is $\EE_{k-1}$-monoidal. We can define internal homs associated to this action, i.e. $\forall x,y\in \Omega_m^{k-1}\CM$, 
$$
\hom_{\Omega_m^{k-1}\CM}(a\odot x, y) \simeq \hom_{\Omega^{k-1}\CC}(a, [x,y]_{\Omega^{k-1}\CC}). 
$$
When $x=y$, the internal hom $[x,x]_{\Omega^{k-1}\CC} \in \Algc_{\EE_1}(\Omega^{k-1}(\CC))$. Moreover, we have 
$$
[1_m^{k-1}, 1_m^{k-1}]_{\Omega^{k-1}\CC} \in \Algc_{\EE_k}(\Omega^{k-1}(\CC)).
$$ 

This action functor $\odot$ naturally induces the following bulk-to-wall map (recall $1_m^0=m$): 
\be \label{eq:L_k}
L_k := -\odot 1_m^{k-1}: \Omega^{k-1}\CC \to \Omega_m^{k-1}\CM,
\ee
which is again $\EE_{k-1}$-monoidal. We illustrate two examples of $L_k$ in the following picture. 
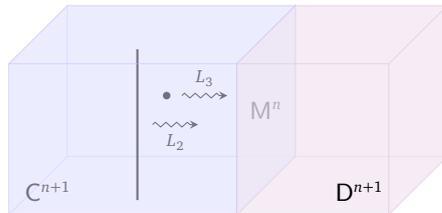
\[
\begin{array}{c}
\begin{tikzpicture}
\draw[blue!20,fill=blue!10,opacity=0.5] (-1,0,1)--(-1,2,1)--(-1,2,-1)--(-1,0,-1)--cycle ;
\draw[blue!20,fill=blue!10,opacity=0.5] (-1,0,-1)--(2,0,-1)--(2,0,1)--(-1,0,1)--cycle ;
\draw[blue!20,fill=blue!10,opacity=0.5] (-1,0,-1) rectangle (2,2,-1) ;
\node at (-0.6,0.2,0.7) {\small $\SC^{n+1}$} ; 
\draw[violet!20,fill=violet!10,opacity=0.5] (2,0,1)--(2,2,1)--(2,2,-1)--(2,0,-1)--cycle ;
\draw[violet!20,fill=violet!10,opacity=0.5] (2,0,-1)--(4,0,-1)--(4,0,1)--(2,0,1)--cycle ;
\draw[violet!20,fill=violet!10,opacity=0.5] (2,0,-1) rectangle (4,2,-1) ;
\fill (0.5,1,-0.5) circle (0.05) ;
\draw[thick] (0.5,0,0.5)--(0.5,2,0.5) ;
\draw [-stealth,
line join=round,
decorate, decoration={
    zigzag,
    segment length=4,
    amplitude=.9,post=lineto,
    post length=2pt
}] (0.7,1,0.5) -- (1.3,1,0.5) node [midway,below] {\scriptsize $L_2$} ;
\draw [-stealth,
line join=round,
decorate, decoration={
    zigzag,
    segment length=4,
    amplitude=.9,post=lineto,
    post length=2pt
}] (0.7,1,-0.5) -- (1.3,1,-0.5) node [midway,above] {\scriptsize $L_3$} ;
\node at (2,1,0) {\small $\SM^n$} ;
\draw[blue!20,fill=blue!10,opacity=0.5] (2,0,1)--(2,2,1)--(2,2,-1)--(2,0,-1)--cycle ;
\draw[blue!20,fill=blue!10,opacity=0.5] (-1,2,-1)--(2,2,-1)--(2,2,1)--(-1,2,1)--cycle ;
\draw[blue!20,fill=blue!10,opacity=0.5] (-1,0,1) rectangle (2,2,1) ;
\draw[violet!20,fill=violet!10,opacity=0.5] (4,0,1)--(4,2,1)--(4,2,-1)--(4,0,-1)--cycle ;
\draw[violet!20,fill=violet!10,opacity=0.5] (2,2,-1)--(4,2,-1)--(4,2,1)--(2,2,1)--cycle ;
\draw[violet!20,fill=violet!10,opacity=0.5] (2,0,1) rectangle (4,2,1) ;
\node at (3.5,0.2,0.7) {\small $\SD^{n+1}$} ; 
\end{tikzpicture}
\end{array}
\quad\quad
\begin{array}{c}
\mbox{bulk to wall maps}: \\
\\
\Omega^2\CC \xrightarrow{L_3=-\odot 1_m^2} \Omega_m^2\CM, \\
\Omega\CC \xrightarrow{L_2=-\odot 1_m} \Omega_m\CM, \\
\CC \xrightarrow{L_1= -\odot m} \CM. 
\end{array}
\]
Note that the right adjoint functor of $L_k$ is given by $L_k^R=[1_m^{k-1}, -]_{\Omega_m^{k-1}\CM}: \Omega_m^{k-1}\CM \to \Omega^{k-1}\CC$.

By rolling up the $\SM^n$ and $\SD^{n+1}$ along $N^k \times \Rb^{n+1-k}$, where $N^k$ is a $k$-dimensional compact manifold, we obtain a $k$-codimensional topological defect in $\SC^{n+1}$ as depicted below.  
\be \label{eq:rolling_up_M_N}
\int_{N^k \times \Rb^{n+1-k}} (\SD^{n+1} |_{\mathring{M}^k \times \Rb^{n+1-k}}, \SM^n |_{\partial N^k \times \Rb^{n+1-k}}) \quad :=
\begin{array}{c}
\begin{tikzpicture}
\draw[blue!20,fill=blue!10,opacity=0.5] (-1,0,1)--(-1,2,1)--(-1,2,-1)--(-1,0,-1)--cycle ;
\draw[blue!20,fill=blue!10,opacity=0.5] (-1,0,-1)--(2,0,-1)--(2,0,1)--(-1,0,1)--cycle ;
\draw[blue!20,fill=blue!10,opacity=0.5] (-1,0,-1) rectangle (2,2,-1) ;
\draw[red!20] (0,0)--(0,2) node[black,midway,left] {\footnotesize $\Rb^{n+1-k}$};
\draw[red!20] (1,0)--(1,2) ;
\fill[red!10,opacity=0.7] (0,0)--(0,2) .. controls (0,2.3) and (1,2.3) .. (1,2)--(1,0) .. controls (1,0.3) and (0,0.3) .. cycle ;
\draw[red!20] (0,2) .. controls (0,2.3) and (1,2.3) .. (1,2) ;
\draw[red!20] (0,0) .. controls (0,0.3) and (1,0.3) .. (1,0) ;
\node at (0.5,2) {\footnotesize $M^k$} ;
\fill[red!10,opacity=0.7] (0,0)--(0,2) .. controls (0,1.7) and (1,1.7) .. (1,2)--(1,0) .. controls (1,-0.3) and (0,-0.3) .. cycle ;
\draw[red!20] (0,2) .. controls (0,1.7) and (1,1.7) .. (1,2) ;
\draw[red!20] (0,0) .. controls (0,-0.3) and (1,-0.3) .. (1,0) ;
\draw[blue!20,fill=blue!10,opacity=0.5] (2,0,1)--(2,2,1)--(2,2,-1)--(2,0,-1)--cycle ;
\draw[blue!20,fill=blue!10,opacity=0.5] (-1,2,-1)--(2,2,-1)--(2,2,1)--(-1,2,1)--cycle ;
\draw[blue!20,fill=blue!10,opacity=0.5] (-1,0,1) rectangle (2,2,1) ;
\node at (-0.8,-0.2) {\footnotesize $\SC^{n+1}$} ; 
\end{tikzpicture}
\end{array}
\quad .
\ee

\begin{pthm} \label{pthm:integral=internal_hom_general}
Let $\SC^{n+1}$ and $\SD^{n+1}$ be two Morita equivalent $n+$1D topological orders connected by an anomaly-free gapped domain wall $\SM^n=(\CM,m)$ and $m\in \CM$. We have
\begin{align}
&\int_{\mathrm{D}^k \times \Rb^{n+1-k}} (\SD^{n+1} |_{\mathring{\mathrm{D}}^k \times \Rb^{n+1-k}}, \SM^n |_{\partial \mathrm{D}^k \times \Rb^{n+1-k}}) :=
\int_{\mathrm{D}^k \times \Rb^{n+1-k}} (\CD |_{\mathring{\mathrm{D}}^k \times \Rb^{n+1-k}}, (\CM,m) |_{\partial \mathrm{D}^k \times \Rb^{n+1-k}}) \nn
&\hspace{1cm} = (\Omega^{k-1}\CC,  [1_m^{k-1},1_m^{k-1}]_{\Omega^{k-1}\CC})  = (\CC, \one_\CC, \cdots, 1_{\one_\CC}^{k-2}, [1_m^{k-1},1_m^{k-1}]_{\Omega^{k-1}\CC}).
\label{eq:integral_Dn+1}
\end{align}
When $N^k=\mathrm{D}^k$, let $P:=\int_{\mathrm{D}^k \times \Rb^{n+1-k}} (\SD^{n+1} |_{\mathring{\mathrm{D}}^k \times \Rb^{n+1-k}}, \SM^n |_{\partial \mathrm{D}^k \times \Rb^{n-k}})$. Then higher codimensional topological defects living on the defect $P$ form a multi-fusion higher category $\Omega_P\CC$ given by 
\be
\Omega_P\CC = \hom_{\Omega^{k-1}\CC}([1_m^{k-1},1_m^{k-1}]_{\Omega^{k-1}\CC}, [1_m^{k-1},1_m^{k-1}]_{\Omega^{k-1}\CC}). 
\ee
\end{pthm}

\begin{rem} \label{rem:alpha_version_FH}
Our theory of integral is beyond the usual theory of factorization homology \cite{Lur17,AF15}, which was also called the $\alpha$-version of factorization homology (see \cite{AF20} for a review). For example, all $(n-k)$-dimensional topological defects in $\SM^n$ form the category $\Omega_m^k\CM$, which is an $\EE_k$-fusion $(n-k)$-category; and all $(n-k)$-dimensional topological defects in $\SD^{n+1}$ form the category 
$\Omega^k\CD$, which is an $\EE_{k+1}$-fusion $(n-k)$-category. By the usual theory of factorization homology \cite{Lur17,AF20}, the following factorization homology 
\[
\int_{\mathrm{D}^k \times \Rb^{n+1-k}} ( \Omega^k\CD |_{\mathring{\mathrm{D}}^k \times \Rb^{n+1-k}}, \Omega_m^k\CM |_{\partial \mathrm{D}^k \times \Rb^{n+1-k}})
\]
is well-defined in the usual sense (see also \cite{AKZ17}). There is a natural $\EE_1$-monoidal functor 
\[
\int_{\mathrm{D}^k \times \Rb^{n+1-k}} ( \Omega^k\CD |_{\mathring{\mathrm{D}}^k \times \Rb^{n+1-k}}, \Omega_m^k\CM |_{\partial \mathrm{D}^k \times \Rb^{n+1-k}}) 
\longrightarrow \hom_{\Omega^{k-1}\CC}([1_m^{k-1},1_m^{k-1}]_{\Omega^{k-1}\CC}, [1_m^{k-1},1_m^{k-1}]_{\Omega^{k-1}\CC}),
\]
which, in general, is not an equivalence. This non-equivalence is a very common phenomenon, which says that integrating local observables living on each stratum over a stratified manifold $N$ only produces parts of the global observables on $N$ because certain non-local observables become local when we shrink the size of $M$. This phenomenon was called {\it spatial fusion anomaly} appeared for topological defects of codimension 2 or higher
in \cite{KZ20,KZ21}. Interestingly, this anomaly vanishes for the fusions of 1-codimensional gapped domain walls among anomaly-free topological orders \cite{AKZ17,KZ20,KZ24,KZ21}. 

In the summer of 2025, one of us (Liang Kong) had a private communication with John Francis. It was clarified that results in Theorem$^{\mathrm{ph}}$\,\ref{pthm:integral=internal_hom_general} are compatible with the predictions from the $\beta$-version of factorization homology. It convinces us that the mathematical foundation of the theory of integrals proposed in this work should be precisely the $\beta$-version of factorization homology proposed in \cite{AFR18}. According to Francis, the mathematical foundation of the $\beta$-version of factorization homology has not yet been fully established. Some recent developments in this program can be found in \cite{AF24}.
\end{rem}

\begin{expl} \label{expl:X3}
Consider an anomaly-free simple 2+1D topological order $\SX^3$, i.e. $\Omega\CX$ is a non-degenerate braided fusion 1-category. 
\bnu
\item By rolling up a 2+1D topological order $\SX^3$ along a circle $S^1$ (see Remark\,\ref{rem:Francis}), we obtain a 2-codimensional topological defect (i.e. a string-like defect) in $\mathbf{1}^4$; 
\be \label{eq:for_Francis_remark}
\int_{\mathrm{D}^2 \times \Rb^{2}} (\mathbf{1}^4 |_{\mathring{\mathrm{D}}^2 \times \Rb^2}, \SX^3 |_{\partial \mathrm{D}^2 \times \Rb^2}) = (2\vect, \Omega\CX). 
\ee

\item By rolling up $\SX^3$ along a 2-sphere $S^2$ (see Remark\,\ref{rem:Francis}), we obtain a 3-codimensional topological defect (i.e. a particle-like defect) in $\mathbf{1}^4$. 
$$ 
\int_{\mathrm{D}^3 \times \Rb^{1}} (\mathbf{1}^4 |_{\mathring{\mathrm{D}}^3 \times \Rb^1}, \SX^3 |_{\partial \mathrm{D}^3 \times \Rb^1}) 
 = (\vect, \Omega^2\CX) \simeq (\vect, \Cb).
 $$
This result is compatible with the $\alpha$-version of factorization homology \cite{AKZ17}. This is due to a general principle that the fusions of 1-codimensional topological defects are free of the spatial fusion anomaly \cite{KZ21}. 

\item By rolling up $\SX^3$ along a 2-torus $S^1 \times S^1$, or equivalently, rolling up $\int_{\mathrm{D}^2 \times \Rb^2} \SX^3$ along $S^1$, we obtain a 3-codimensional topological defect (i.e. a particle-like defect) in $\mathbf{1}^4$. 
\begin{align*} 
\int_{\mathrm{D}^2 \times S^1 \times \Rb^1} (\mathbf{1}^4 |_{\mathring{\mathrm{D}}^2 \times S^1\times \Rb^1}, \SX^3 |_{\partial \mathrm{D}^2 \times S^1 \times \Rb^1}) 
 &= (\vect, \hom_{\Fun(\Omega\CX,\Omega\CX)}(\id_{\one_\CX}, \id_{\one_\CX}))  \\
&\simeq (\vect, \hom_{\Omega\CX}(1_{\one_\CX}, \oplus_{i\in \mathrm{Irr}(\Omega\CX)} i \otimes i^R)). 
\end{align*}
This result is compatible with the $\alpha$-version of factorization homology \cite{AKZ17} for the same reason. 
\enu

\end{expl}

\begin{rem} \label{rem:Francis}
In an earlier version of this paper, we used a `simplified' notion for the formula (\ref{eq:for_Francis_remark}) written as $\int_{S^1 \times \Rb^2} \SX^3 = (2\vect, \Omega\CX)$. This simplification ignores the data on $\mathring{\mathrm{D}}^2 \times \Rb^2$ because this missing data (i.e., the trivial topological order $\mathbf{1}^4$) is automatically implied by the anomaly-freeness of $\SX^3$. This simplification is harmless physically. Mathematically, however, without the support of the geometric intuition of ``anomaly-freeness'', this simplification is `structurally wrong' as pointed out to us by John Francis. In this work, we do not distinguish them for the simplicity of notations. But for those who are interested in the relation between our physical theory and the mathematical theory of the $\beta$-version of factorization homology, one should be aware of this subtleness. 
\end{rem}

\begin{expl}
Consider the 3+1D $\Zb_2$ topological order $\SC^4=\SG\ST_{\Zb_2}^4$ and $\SD^4=\mathbf{1}^4$. 
\bnu
\item Let $\SM^3$ be the (twist) smooth boundary of $\SG\ST_{\Zb_2}^4$, i.e. $\SM^3=(2\Rep(\Zb_2),\one_{2\Rep(\Zb_2)})$. We have 
\begin{align*}
\int_{\mathrm{D}^2 \times \Rb^{2}} (3\vect |_{\mathring{\mathrm{D}}^2 \times \Rb^2}, (2\Rep(\Zb_2),\one_{2\Rep(\Zb_2)}) |_{\partial \mathrm{D}^2 \times \Rb^2}) &= (\Omega\CC, 1_{\one_\CC} \oplus m), \\
\int_{\mathrm{D}^3 \times \Rb^{1}} (3\vect |_{\mathring{\mathrm{D}}^3 \times \Rb^1}, (2\Rep(\Zb_2),\one_{2\Rep(\Zb_2)}) |_{\partial \mathrm{D}^3 \times \Rb^1}) &= (\Omega^2\CC, 1_{\one_\CC}^2), \\
\int_{\mathrm{D}^2 \times S^1 \times \Rb^{1}} (3\vect |_{\mathring{\mathrm{D}}^2 \times S^1 \times \Rb^1}, (2\Rep(\Zb_2),\one_{2\Rep(\Zb_2)}) |_{\partial \mathrm{D}^2 \times S^1 \times \Rb^1}) &= \int_{S^1} 1_{\one_\CC} \oplus m = (\Omega^2\CC, 1_{\one_\CC}^2 \oplus 1_{\one_\CC}^2),
\end{align*}
where $1_{\one_\CC} \oplus m$ is precisely the Lagrangian algebra in $\Omega\CT\CC$ that determines the (twist) smooth boundary; the second identity follows from the fact that the left hand side amounts to a line segment of the $(1_{\one_\CC} \oplus m)$-string; the third identity follows from the fact that an $m$-loop gives the trivial particle $1_{\one_\CC}^2$ \cite{KTZ20a}. 

\item Let $\SM^3$ be the rough boundary of $\SG\ST_{\Zb_2}^4$, $\SM^3=(2\vect_{\Zb_2},\one_{2\vect_{\Zb_2}})$. We have 
\begin{align*}
\int_{\mathrm{D}^2 \times \Rb^{2}} (3\vect |_{\mathring{\mathrm{D}}^2 \times \Rb^2}, (2\vect_{\Zb_2},\one_{2\vect_{\Zb_2}}) |_{\partial \mathrm{D}^2 \times \Rb^2}) &= (\Omega\CC, \one_c), \\
\int_{\mathrm{D}^3 \times \Rb^{1}} (3\vect |_{\mathring{\mathrm{D}}^3 \times \Rb^1}, (2\vect_{\Zb_2},\one_{2\vect_{\Zb_2}}) |_{\partial \mathrm{D}^3 \times \Rb^1}) &= (\Omega^2\CC, 1_{\one_\CC}^2\oplus e), \\
\int_{\mathrm{D}^2 \times S^1 \times \Rb^{1}} (3\vect |_{\mathring{\mathrm{D}}^2 \times S^1 \times \Rb^1}, (2\vect_{\Zb_2},\one_{2\vect_{\Zb_2}}) |_{\partial \mathrm{D}^2 \times S^1 \times \Rb^1}) &= \int_{S^1} \one_c = (\Omega^2\CC, 1_{\one_\CC}^2\oplus e),
\end{align*}
where $\one_c$ is the Lagrangian algebra that determines the rough boundary; the second identity follows from the fact that the left hand side amounts to a line segment of the $\one_c$-string; the third identity follows from the fact that an $\one_c$-loop gives the particle $1_{\one_\CC}^2 \oplus e$ \cite[Eq.\ (4.6)]{KTZ20a}. 
\enu
\end{expl}

Note that the pair $(\CM,m)$ only specifies a very special wall condition. The most general wall condition $\SM^n$ can be defined by a stratified $n$-disk with each $p$-codimensional stratum labeled by a $p$-morphism in $\CM$ for $0\leq p \leq n$. In particular, we can label different $0$-codimensional strata (i.e. $n$-cells) by different $0$-morphisms in $\CM$. We would like to know how to compute the $k$-codimensional defect in $\SC^{n+1}$ obtained by rolling up $\SM^n$ with such a general wall condition along $M^i \times \Rb^{n+1-i}$ as long as all strata glue properly. At the current stage, the explicit computation in such a general setup is not available. In the remaining of this subsection, we provide some explicit examples in the case $\SC^{n+1}=\SD^{n+1}$ and $\SM^n=\SC^n$.

\begin{pprop} \label{pprop:integral_on_spheres}
Let $X^{n+1-l}$ be an $l$-codimensional topological defect in $\SC^{n+1}$. For $l\leq k<n+1$, we have a canonical action $\odot: \Omega^{k-1}\CC \times \Omega_X^{k-l}(\CC) \to \Omega_X^{k-l}(\CC)$, which is an $\EE_{k-l}$-monoidal functor. By rolling up $X$ along $S^{k-l}\times \Rb^{n+1-k}$, we obtain a $k$-codimensional topological defect $\int_{S^{k-l} \times \Rb^{n+1-k}} X$ given as follows: 
\be
\int_{S^{k-l} \times \Rb^{n+1-k}} X = (\Omega^{k-1}\CC, [1_X^{k-l}, 1_X^{k-l}]_{\Omega^{k-1}\CC}), 
\ee 
where $[1_X^{k-l}, 1_X^{k-l}]_{\Omega^{k-1}\CC} \in \Algc_{\EE_{k-l+1}}(\Omega^{k-1}\CC)$ and, when $k=l$, we have $[1_X^{k-l}, 1_X^{k-l}]_{\Omega^{k-1}\CC} =X\otimes X^R$.
\end{pprop}

\begin{expl}
In 2+1D $\Zb_2$ topological order $\TC^3$, there are six simple 1-codimensional topological defects $\one, \mathrm{ss}, \mathrm{sr}, \mathrm{rr}, \mathrm{rs}, \vartheta$ and four simple 2-codimensional topological defects $1_\one, m, e, f$. Then we have 
$$
\int_{S^1 \times \Rb^1} X = \left\{ \begin{array}{ll} (\FZ_1(\Rep(\Zb_2)), 
1_\one) & \mbox{if $X=\one, \vartheta$}; \\
(\FZ_1(\Rep(\Zb_2)), 1_\one \oplus m) & \mbox{if $X=\mathrm{ss}, \mathrm{sr}$}; \\
(\FZ_1(\Rep(\Zb_2)), 1_\one\oplus e) & \mbox{if $X=\mathrm{rr}, \mathrm{rs}$}.
\end{array} \right.
$$
\end{expl}

\begin{rem}
If we select a slightly more general wall condition, more precisely, we specify an $(n-1)$-dimensional topological defect $m_1\in \Omega_m\CM$ on the wall and an $(n-1)$-dimensional topological defect $d\in \Omega\CD$, then, by rolling up the wall along $S^1\times \Rb^{n-1}$, we obtain the following formula (recall Theorem$^{\mathrm{ph}}$\ \ref{pthm:bubble=internal_hom}).  
$$
\begin{array}{c}
\begin{tikzpicture}[scale=0.6]
\fill[blue!20] (-2,0) rectangle (2,3) ;
\fill[teal!20] (0,1.5) circle (1);
\node at (-1.5,2.5) {\footnotesize $\SC^{n+1}$} ;
\node at (0,2) {\scriptsize $\SD^{n+1}$} ;
\node at (0.3,1.4) {\scriptsize $d$} ; 
\node at (1.2,0.4) {\scriptsize $(\CM,m)$} ;
\draw [blue, ultra thick] (0,1.5) circle [radius=1] ;
\fill[black] (-1,1.5) circle (0.08) node[black] at (-1.5,1.5) {\scriptsize $m_1$};
\fill[black] (0,1.5) circle (0.08) ;
\end{tikzpicture}
\end{array}
\quad\quad
\int_{\mathrm{D}^2 \times \Rb^{n-2}} ((\CD,d) |_{\mathring{\mathrm{D}}^2 \times \Rb^{n-2}}, (\CM,m,m_1) |_{\partial \mathrm{D}^2 \times \Rb^{n-2}}) = (\Omega\CC, [1_{m}, m_1\odot d]_{\Omega\CC} ).
$$
Moreover, $[1_m,-\odot -]$ is a well-defined functor, which maps $\Omega_m\CM \times \Omega\CD \to \Omega\CC$. In other words, this roll-up process is functorial and $[1_m,-\odot -]$ also tells us how to obtain the higher codimensional topological defect in $\Omega\CC$ resulting from rolling up higher codimensional defects in $\Omega_m\CM$ and $\Omega\CD$. 
\end{rem}

\subsection{Condensations of gapless liquid-like defects} \label{sec:gapless}

In this subsection, we sketch a generalization of the condensation theory to gapless liquid-like defects. We leave a detailed study to future publications.

\subsubsection{General theory}
Let $\SC^{n+1}$ be an $n+$1D topological order. It can have gapless defects. For example, for $\SC^3$, by stacking a 2D topological defect with a 1+1D anomaly-free gapless phase (e.g., a 2D rational CFT), we obtain a gapless 1-codimensional defect in $\SC^3$. We give another example. Consider two $k$-codimensional simple topological defects $X$ and $Y$ in $\SC^{n+1}$ connected by a $(k+1)$-codimensional topological defect. By proliferating $Y$ within $X$, we can construct an $(n+1-k)$D phase transition from $X$ to $Y$ (without altering $\SC^{n+1}$). At the critical point of the phase transition, the gap (in a neighborhood of the defect) is the closed, thus we obtain a $k$-codimensional gapless defect in $\SC^{n+1}$. 

\medskip
A general gapless defect can be really wild. In general, the fusions among general defects are not necessarily well-defined. Therefore, we limit ourselves to a special class of gapless defects, which we call {\it liquid-like} defects \cite{KZ22b,KZ22}. By `liquid-like', we mean that the defect is `soft' enough so that it can be bent freely without altering the defect. Physically, it means that such a defect should be transparent to energy-momentum tensor\footnote{The `liquid-like' properties are in some sense topological properties. In physics literature, gapless boundaries or domain walls are often called non-topological. However, some gapless defect lines in 1+1D rational CFT's are also called topological. To avoid confusion, we will not use the term `a non-topological defect' for a gapless liquid-like defect.}. Such an $n$-dimensional liquid-like defect can also be viewed as an anomalous $n$D quantum phase, which was called an (anomalous) {\it quantum liquid} \cite{KZ22b}. The mathematical theory of quantum liquids was developed in \cite{KZ22b,KZ22}. In particular, a quantum liquid $\SX^{n+1}$ can be described by a pair $(\SX_{\mathrm{lqs}}, \SX_{\mathrm{top}})$, where 
\begin{itemize}
\item $\SX_{\mathrm{lqs}}$ summarizes all the dynamical data (called {\it local quantum symmetry}), a mathematical theory of which was developed in \cite{KZ22}; 
\item $\SX_{\mathrm{top}}$ summarizes all the topological (or categorical) data (called {\it topological skeleton}).
\end{itemize}
The fusion of the topological skeletons of two quantum liquids is well-defined. It was  conjectured that the fusion of local quantum symmetry can also be properly defined so that it is compatible with that of the topological skeletons \cite{KZ22b}. A possible theory to make sense of this conjecture was proposed and developed in \cite{KZ22} based on a generalization of conformal nets in 1+1D to topological/defect nets in arbitrary dimensions\footnote{If using the VOA approach for 1+1D CFT's instead of the conformal net approach, there might be some technical issues.}. We do not want to go to the details. But simply take it for granted that the fusions among quantum liquids are well-defined. As a consequence, it means that the category of all gapped and gapless liquid-like defects in a topological order is well-defined. Given an $n+$1D topological order $\SC^{n+1}$, we denote the category of all liquid-like defects by $\LC$. It is important to notice that a topological defect is always connected to a large family of gappable liquid-like defects (see Example\,\ref{expl:stacking_CFT}). Mathematically, the softness of the liquid-like defects simply means that the category of such defects is fully dualizable.

\begin{expl} \label{expl:stacking_CFT}
A 1+1D rational conformal field theory (RCFT) can be viewed as a 1-codimensional liquid-like defect in $\mathbf{1}^3$. In a general 2+1D topological order $\SC^3$, we can stack a 1-codimensional topological defect with a 1+1D RCFT to obtain a 1-codimensional liquid-like defect in $\SC^3$. For 2+1D $\Zb_2$ topological order $\TC^3$, it has two gapped boundaries: the smooth boundary and the rough boundary. In \cite{CJKYZ20}, an explicit purely boundary phase transition was constructed via lattice model, and the critical point, viewed as a gappable gapless boundary of $\TC^3$, was shown to be liquid-like. More liquid-like gapless boundaries of 2+1D $\Zb_2$ topological order $\TC^3$ have been constructed in \cite{KZ21}.  
\end{expl}

To avoid the technical issues of local quantum symmetry and to simplify the discussion, we consider a simplified category. By forgeting the data of local quantum symmetry, we obtain a new category $\LC_{\mathrm{top}}$. It was shown that this category is monoidal, fully dualizable and condensation complete \cite{KZ22b,KZ22}. Note that when $\SC^{n+1}=\mathbf{1}^{n+1}$, it was shown that $\LC_{\mathrm{top}}$ is equivalent to $\bullet/(n+2)\vect$ \cite{KZ22b,KZ22}\footnote{We ignore the unitarity here.}. We will leave a mathematically detailed study of condensations in $\LC_{\mathrm{top}}$ or $\bullet/(n+2)\vect$ elsewhere. Instead, we give a heuristic discussion of condensations in $\LC_{\mathrm{top}}$. 

We expect that the condensation theory of liquid-like defects in $\LC_{\mathrm{top}}$ is completely parallel to that of topological defects in a topological order. In particular, by condensing a $k$-codimensional liquid-like defect in $\LC_{\mathrm{top}}$, we mean a multiple-step process. 
\bnu

\item We first condense a condensable $\EE_k$-algebra $A$ in one of the transversal directions. We obtain a $(k-1)$-codimensional liquid-like defect $\Sigma A$, which is automatically a condensable $\EE_{k-1}$-algebra in $\Omega^{k-1}\LC_{\mathrm{top}}$. 

\item We repeat the process $(k-1)$-times and obtain a 1-codimensional liquid-like defect $\Sigma^{k-1}A$, which is automatically a condensable $\EE_1$-algebra in $\LC_{\mathrm{top}}$. It can be further condensed to give a new phase $\SD^{n+1}$ such that $\LD_{\mathrm{top}}\simeq\Mod_{\Sigma^{k-1}A}^{\EE_1}(\LC_{\mathrm{top}})$ and a potentially gapless wall $(\LM_{\mathrm{top}},m)$ such that $\LM_{\mathrm{top}}\simeq\RMod_{\Sigma^{k-1}A}(\LC_{\mathrm{top}})$ and $m=\Sigma^{k-1}A$. 

\item Alternatively, one can condense the condensable $\EE_2$-algebra $\Sigma^{k-2}A$ in $\Omega\LC_{\mathrm{top}}$ directly to obtain the condensed phase $\SD^{n+1}$ such that $\Omega\LD_{\mathrm{top}} \simeq \Mod_{\Sigma^{k-2}A}^{\EE_2}(\Omega\LC_{\mathrm{top}})$ and $\Omega_m\LM_{\mathrm{top}}\simeq\RMod_{\Sigma^{k-2}A}(\Omega\LC_{\mathrm{top}})$. 
\enu
Since any topological order can have a gapped or gapless liquid-like boundary, one expect that any topological order can be obtained via a condensation in the trivial phase. This theory largely generalizes the main results in this paper and will be developed in details elsewhere.

The complete mathematical theory of condensation of gapless liquid-like defects are incredibly rich. It relies on some not-yet developed mathematical theory of enriched (higher) categories and needed-to-be-further-developed theory of local quantum symmetries \cite{KZ22}. In the next subsubsection, avoiding mathematical details, we give an illustrating example in 2+1D based on the physical intuition of topological Wick rotation \cite{KZ18b,KZ20,KZ21}. 

\subsubsection{An illustrating example in 2+1D} \label{sec:gapless_condensation_3D}

\begin{figure}
$$ 
\raisebox{-2em}{\begin{tikzpicture}
\draw[thick] (-1,0)--(0.5,1.5) node[midway,left] {} ;
\draw[thick] (0,0)--(1.5,1.5) node[midway,left] {} ;
\draw[thick] (2,0)--(3.5,1.5) node[midway,left] {} ;
\draw[thick] (-1,0)--(2,0) node[midway,left] {} ;
\draw[thick] (0.5,1.5)--(3.5,1.5) node[midway,left] {} ;

\node at (-0.3,0.2) {\scriptsize $\SB^3$} ;
\node at (1,0.7) {\scriptsize $\SP^2$} ;
\node at (2.3,1) {\scriptsize $\SC^3$} ;
\end{tikzpicture}}
\xrightarrow{\mbox{\scriptsize topological Wick rotation}}
\raisebox{-2.5em}{\begin{tikzpicture}
\draw[thick] (0,0)--(0,0.8) node[midway,left] {} ;
\draw[thick] (0,0.8)--(1.5,2.3) node[midway,left] {\scriptsize $\SX_{\mathrm{top}}={}^{\Omega\CB^\rev}\CP$} ;
\draw[thick] (1.5,1.5)--(1.5,2.3) node[midway,left] {} ;
\draw[thick] (0,0)--(1.5,1.5) node[midway,left] {} ;
\draw[thick] (2,0)--(3.5,1.5) node[midway,left] {} ;
\draw[thick] (0,0)--(2,0) node[midway,left] {} ;
\draw[thick] (1.5,1.5)--(3.5,1.5) node[midway,left] {} ;
\node at (0.9,1.3) {\scriptsize $\SX^2$} ;
\node at (2.3,1) {\scriptsize $\SC^3$} ;
\end{tikzpicture}}
$$
\caption{the idea of topological Wick rotation \cite{KZ18b,KZ20,KZ21}
}
\label{fig:TWR}
\end{figure}
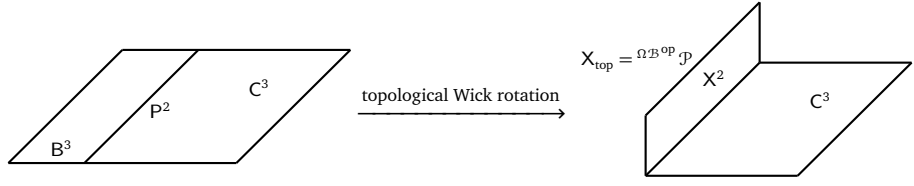

The mathematical theory of gapped/gapless boundaries of 2+1D topological orders was developed in \cite{KZ18b,KZ20,KZ21}. We do not need the complete theory here. We only focus on the topological data of gapless boundaries and the physical intuition provided by the so-called topological Wick rotation as illustrated in Figure\,\ref{fig:TWR}. More precisely, the left picture depicts in the spatial dimension a gapped domain wall $\SP^2$ between two anomaly-free 2+1D simple topological order $\SC^3$ and $\SB^3$. The 2-codimensional topological defects in $\SB^3$ (resp. $\SC^3$) form a non-degenerate braided fusion 1-category $\Omega\CB$ (resp. $\Omega\CC$). For convenience, we also assume $\SP^2=(\CP,p)$ is simple. Then the 1-codimensional topological defects in $\SP^2$ form a fusion 1-category $\Omega_p\CP$. We have a canonical braided equivalence $\Omega\CB^\rev \boxtimes \Omega\CC \simeq \FZ_1(\Omega_p\CP)$. The canonical action of $\Omega\CB^\rev$ on $\Omega_p\CP$ defines an enriched fusion category ${}^{\Omega\CB^\rev}\Omega_p\CP$ via the so-called canonical construction \cite{MP17}. It turns out that the topological skeleton $\SX_{\mathrm{top}}=\CX$ of a gapless boundary $\SX^2$ of a 2+1D topological $\SC^3$ is precisely given by an enriched fusion category $\CX={}^{\Omega\CB^\rev}\Omega_p\CP$, in which the original finite dimensional hom space $\hom_{\Omega_p\CP}(x,y)$ is replaced by an internal hom $[x,y]$ in $\Omega\CB^\op$. The physical meaning of $\Omega\CB$ is the category of modules over a rational vertex operator algebra, and that of the internal hom space $[x,y]$ is an infinite dimensional vector space of defect fields \cite{KZ18b,KZ20,KZ21}. This suggests that the topological skeleton of the gapless boundary $\SX^2$ can be obtained by rotating the 2+1D phase $\SB^3$ to the time direction as a fictional phase in spacetime as depicted in the second picture in Figure\,\ref{fig:TWR}. This fictional phase can be viewed as a bookkeeping device for the combination of $\Omega\CB^\rev$ with $\CP$ to form an $\Omega\CB^\rev$-enriched fusion category ${}^{\Omega\CB^\rev}\Omega_p\CP$ \cite{KZ18b}. Actually, it suggests something much deeper than a bookkeeping device as shown already in \cite{KZ18b} that it led a powerful formula for computing the fusion of gapless domain walls \cite[Eq.\ 5.3]{KZ18b} and a generalization of holographic duality to higher dimensions \cite{KZ20,KZ21,KZ22b}. From now on, we take it for granted that the second picture in Figure\,\ref{fig:TWR} represents a gapless boundary $\SX^2$ of $\SC^3$. 

\medskip
Similar to Section\,\ref{sec:2codim_3D_geometric_approach}, by rolling up the gapless boundary $\SX^2$, we obtain a hole with a gapless boundary (depicted as a cylinder in (\ref{eq:def-hole}) based on the geometric intuition of topological Wick rotation) or a 2-codimensional liquid-like defect in $\Omega\LC_{\mathrm{top}}$. 
\be \label{eq:def-hole}
A = \int_{S^1} \CX \quad = \int_{S^1} {}^{\Omega\CB^\rev}\CP \quad = \quad
\begin{array}{c}
\begin{tikzpicture}[scale=0.6]
\draw[fill=white] (-2,-1)--(3,-1)--(4,1)--(-1,1)--cycle ;
\draw (0,0)--(0,2) ;
\draw (2,0)--(2,2) node [very near end, right] {\scriptsize ${}^{\Omega\CB^\rev}\CP$} ;
\fill[blue!10,opacity=0.7] (0,0)--(0,2) .. controls (0,2.5) and (2,2.5) .. (2,2)--(2,0) .. controls (2,0.5) and (0,0.5) .. cycle ;
\draw (0,2) .. controls (0,2.5) and (2,2.5) .. (2,2) ;
\draw (0,0) .. controls (0,0.5) and (2,0.5) .. (2,0) ;
\fill[blue!10,opacity=0.7] (0,0)--(0,2) .. controls (0,1.5) and (2,1.5) .. (2,2)--(2,0) .. controls (2,-0.5) and (0,-0.5) .. cycle ;
\draw (0,2) .. controls (0,1.5) and (2,1.5) .. (2,2) ;
\draw (0,0) .. controls (0,-0.5) and (2,-0.5) .. (2,0) ;
\node at (-1.3,-0.7) {\scriptsize $\SC^3$} ; 
\node at (1,-0.6) {\scriptsize $\SP^2$} ;
\node at (1,1) {\scriptsize $\SB^3$} ;
\end{tikzpicture}
\end{array}
\quad\quad \in \Omega\LC. 
\ee
Note that this 2-codimensional liquid-like defect is not a particle (or anyon) in the usual sense. It should be viewed as a generalized particle (or a virtual\footnote{In physical reality, by shrinking the radius of the hole, the hole will be gapped out and becomes an ordinary particle. However, in our mathematical theory, there is no length scale. The radius of the cylinder does not play any role. One has to gap it out by hand as it was done in \cite{KZ21}. In this sense, without gapping it out, the cylinder in (\ref{eq:def-hole}) should be viewed as a virtual particle (more precisely, it lives in the category of `generalized nerves'). We will provide more details elsewhere.} particle). 
Similar to the discussion in Section\,\ref{sec:2codim_3D_geometric_approach}, this liquid-like defect is an internal hom $[\one_\CX,\one_\CX]_{\Omega\LC}$. The internal hom is again defined by the natural action $\odot: \Omega\LC_{\mathrm{top}} \times \CX \to \CX$, i.e. 
$$
\hom_\CX(a\odot x, y) \simeq \hom_{\Omega\LC_{\mathrm{top}}}(a, [x,y]_{\Omega\LC_{\mathrm{top}}}). 
$$
Therefore, $A=[\one_\CX,\one_\CX]_{\Omega\LC}$ has a canonical structure of a condensable $\EE_2$-algebra in $\LC_{\mathrm{top}}$. The geometric meaning of the algebraic structure of $A$ was explained in details in \cite[Section\ 6.2, Remark\ 6.3]{KZ21} as a natural\footnote{There is no unnatural choice involved in the geometric definition of the fusion of two cylinders depicted in (\ref{eq:def-hole}).} geometric fusion of two such `cylinders' depicted in (\ref{eq:def-hole}). From the geometric construction of this multiplication, it is clear that $A$ is an $\EE_2$-algebra in $\Omega\LC_{\mathrm{top}}$. 

\medskip
Notice that a defect in $\CX$ also defines an object in $\Omega\LC_{\mathrm{top}}$ as illustrated below: 
\be \label{eq:def-hole_xX}
\int_{S^1} (\CX,x) \quad = \quad
\begin{array}{c}
\begin{tikzpicture}[scale=0.6]
\draw[fill=white] (-2,-1)--(3,-1)--(4,1)--(-1,1)--cycle ;
\draw (0,0)--(0,2) ;
\draw (2,0)--(2,2) node [very near end, right] {\scriptsize ${}^{\Omega\CB^\rev}\CP$} ;
\fill[blue!10,opacity=0.7] (0,0)--(0,2) .. controls (0,2.5) and (2,2.5) .. (2,2)--(2,0) .. controls (2,0.5) and (0,0.5) .. cycle ;
\draw (0,2) .. controls (0,2.5) and (2,2.5) .. (2,2) ;
\draw (0,0) .. controls (0,0.5) and (2,0.5) .. (2,0) ;
\fill[blue!10,opacity=0.7] (0,0)--(0,2) .. controls (0,1.5) and (2,1.5) .. (2,2)--(2,0) .. controls (2,-0.5) and (0,-0.5) .. cycle ;
\draw (0,2) .. controls (0,1.5) and (2,1.5) .. (2,2) ;
\draw (0,0) .. controls (0,-0.5) and (2,-0.5) .. (2,0) ;
\node at (-1.3,-0.7) {\scriptsize $\SC^3$} ; 
\node at (1.5,-0.6) {\scriptsize $x\in \CX$} ;
\draw (1,-0.35) -- (1,1.65) ;
\end{tikzpicture}
\end{array}
\quad\quad \in \Omega\LC. 
\ee
The same geometric fusion of two cylinders defined in \cite[Section\ 6.2]{KZ21} defines an $A$-action on $\int_{S^1} (\CX,x)$. As a consequence, $\int_{S^1} (\CX,x)$ is only a right $A$-module (not an $\EE_2$-module) in $\Omega\LC_{\mathrm{top}}$. Geometrically, it is clear that such a right $A$-module is an $\EE_2$-$A$-module if and only if $x\in \CX$ can be moved into the interior of the hole (i.e., trivial phase $\mathbf{1}^3$). However, the only simple 2-codimensional liquid-like defect is the trivial particle. We obtain $\Mod_A^{\EE_2}(\Omega\LC_{\mathrm{top}}) \simeq \vect$. We summarize all the results in this subsubsection in the following theorem. 

\begin{pthm} \label{pthm:condensing_cylinder_3D}
The topological skeleton $\SX_{\mathrm{top}}=\CX$ of a gapless boundary $\SX^2$ of an anomaly-free 2+1D topological order $\SC^3$ can be obtained by condensing a liquid-like defect $A:=\int_{S^1}\CX$, which is an $\EE_2$-algebra in $\LC_{\mathrm{top}}$. Moreover, we have 
$$
\CX \simeq \RMod_A(\Omega\LC_{\mathrm{top}}), \quad\quad  \Mod_A^{\EE_2}(\Omega\LC_{\mathrm{top}}) \simeq \vect. 
$$
\end{pthm}

One can also condense $A$ along a line in space. As a consequence, we obtain a condensed 1-codimensional liquid-like defect $\Sigma A=\SX^2 \boxtimes \overline{\SX^2} \in \LC_{\mathrm{top}}$, which is a condensable $\EE_1$-algebra in $\LC_{\mathrm{top}}$ and can be graphically represented by the following picture. 
$$
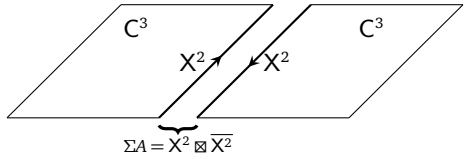
\begin{tikzpicture}
\draw[thick,->-] (-0.5,0)--(1,1.5) node[midway,left] {\small  $\SX^2$} ;
\draw[] (-2.5,0)--(-1,1.5) node[midway,left] {} ;
\draw[] (-0.5,0)--(-2.5,0) node[midway,left] {} ;
\draw[] (-1,1.5)--(1,1.5) node[midway,left] {} ;

\draw[thick,->-] (1.5,1.5)--(0,0) node[midway,right] {\small  $\SX^2$} ;
\draw[] (2,0)--(3.5,1.5) node[midway,left] {} ;
\draw[] (0,0)--(2,0) node[midway,left] {} ;
\draw[] (1.5,1.5)--(3.5,1.5) node[midway,left] {} ;
\node at (2.3,1.2) {\small $\SC^3$} ;
\node at (-0.8,1.2) {\small $\SC^3$} ;
\draw[decorate,decoration=brace,very thick] (0,-0.1)--(-0.5,-0.1) ;
\node at (-0.25,-0.33) {\scriptsize $\Sigma A = \SX^2 \boxtimes \overline{\SX^2}$} ;
\end{tikzpicture}
$$
By condensing $\Sigma A$ in $\LC_{\mathrm{top}}$, we again obtain $\mathbf{1}^3$ as the condensed phase and a gapless boundary, which is precisely $\SX^2$. 

\begin{rem} \label{rem:virtual_anyons}
When $\SC^3=\mathbf{1}^3$,  its gapless boundary $\SX^2$ is just a 2D RCFT (with liquid-like defects). By condensing the condensable $\EE_2$-algebra $A=\int_{S^1} \CX$, or equivalently, condensing $\Sigma A$, we obtain $\mathbf{1}^3$ as the condensed phase and a gapless domain wall $\SM^2=\SX^2$. This result says that condensing a 2D RCFT $\SX^2$ in $\mathbf{1}^3$ reproduces $\SX^2$ as the domain wall between the initial phase $\mathbf{1}^3$ and the condensed phase $\mathbf{1}^3$. Note that if $\SX^2$ is gapped (i.e., $\Omega\CB^\op$ is trivial), this construction is already beyond usual anyon condensation theory because we have included the condensation of `virtual particles' (in the category of `generalized nerves'). For example, if the cylinder-like defect depicted in (\ref{eq:def-hole}) and (\ref{eq:def-hole_xX}) has a non-trivial interior, then some non-trivial defect $x\in \CX$ in (\ref{eq:def-hole_xX}) can be moved into the interior and becomes an $\EE_2$-$A$-module. In this case, the condensed phase $\SD^3$ is non-trivial and we have $\Omega\hat{\CD}_{\mathrm{top}}=\Mod_A^{\EE_2}(\Omega\hat{\CC}_{\mathrm{top}})$. This special case already largely generalizes the usual anyon condensation theory even when the wall $\SX^2$ is gapped. We will provide a more complete story elsewhere. 
\end{rem}

\begin{rem} 
Although this illustrating example is about the condensation in a 2+1D topological order with a gapless liquid-like boundary, it tautologically generalizes to higher dimensional topological orders with gapless liquid-like boundaries (or domain walls) because topological Wick rotation also applies to gapless liquid-like boundaries of higher dimensional topological orders \cite{KZ20,KLWZZ20,KLWZZ20a,KZ22b}. 
\end{rem}

\begin{rem} \label{rem:infinite-type_condensation_gapless}
We have ignored the dynamical data $\SX_{\mathrm{lqs}}$ completely and have provided a `generalized condensation' only for the topological data. The complete story, including the dynamical data, is not yet known but far richer and more interesting. One possible way to complete the story is to use the mathematical theory of topological nets developed in \cite{KZ22}. Since the hom spaces in the enriched categories are internal homs in the category of modules over a vertex operator algebra, the physical realization of these hom spaces are infinite dimensional vector space of defect fields. This infinite feature already shows that the condensation theory presented in this subsection is of infinite type and beyond the main results of this work (recall Remark\,\ref{rem:infinite-type_condensation}).

However, we want to remark that the generalized condensation theory presented in this subsection seems to suggest a geometric way to understand the usual representation theory of vertex operator algebra (VOA) as a geometric condensation theory. More precisely, consider a rational VOA $V$, i.e., $\Mod_V$ is a modular tensor category. Consider a 2+1D topological order $\SC^3$ and a braided monoidal equivalence $\phi: \Mod_V \to \Omega\CC$. According to the mathematical theory of a gapless boundary of a 2+1D topological order, the triple $(V,\phi,{}^{\Omega\CC^\op}\Omega\CC)$ provides a complete mathematical characterization of the so-called canonical gapless boundary of $\SC^3$. Consider the solid cylinder in spacetime depicted in the following picture. 
\be \label{eq:pic-cylinder}
\begin{array}{c}
\begin{tikzpicture}[scale=0.6]
\draw[fill=white] (-2,-1)--(3,-1)--(4,1)--(-1,1)--cycle ;
\draw (0,0)--(0,2) ;
\draw (2,0)--(2,2) node [very near end, right] {\scriptsize $(V,\phi,{}^{\Omega\CC^\op}\Omega\CC)$} ;
\fill[blue!10,opacity=0.7] (0,0)--(0,2) .. controls (0,2.5) and (2,2.5) .. (2,2)--(2,0) .. controls (2,0.5) and (0,0.5) .. cycle ;
\draw (0,2) .. controls (0,2.5) and (2,2.5) .. (2,2) ;
\draw (0,0) .. controls (0,0.5) and (2,0.5) .. (2,0) ;
\fill[blue!10,opacity=0.7] (0,0)--(0,2) .. controls (0,1.5) and (2,1.5) .. (2,2)--(2,0) .. controls (2,-0.5) and (0,-0.5) .. cycle ;
\draw (0,2) .. controls (0,1.5) and (2,1.5) .. (2,2) ;
\draw (0,0) .. controls (0,-0.5) and (2,-0.5) .. (2,0) ;
\node at (-1.3,-0.7) {\scriptsize $\mathbf{1}^3$} ; 
\node at (1,-0.6) {\scriptsize $\SC^2$} ;
\node at (1,1) {\scriptsize $\SC^3$} ;
\node at (1,2) {\scriptsize $\SC^3$} ;
\end{tikzpicture}
\end{array}
\ee
It is a spacetime picture of the topological order $\SC^3$ and its canonical gapless boundary living in an empty 3D spacetime. The meaning of the label $\SC^2$ is explained in Example\,\ref{expl:folding}. Then the representation theory of $V$ or the fact that $\CC \simeq \Mod_V$ can be physically understood as the condensation of the solid cylinder (as an $\EE_2$-algebra) in $\mathbf{1}^3$ (see Remark\,\ref{rem:VOA}). Recall that a $V$-module in the usual sense \cite{FHL93} is precisely an $\EE_2$-module over $V$. The same picture works if we replace a rational VOA by a rational full field algebras \cite{HK07,Kon07}. 
\end{rem}

\begin{rem} \label{rem:VOA}
In order to provide a physical meaning or a physical intuition to a mathematical theory, one has to associate a mathematical object to its physical realization precisely according to physically laws. For example, in mathematical literature, it is perfectly fine to say that a VOA defines an algebra over a spherical partial operad in Huang's geometric theory of VOA \cite{Hua95a}, or, a local observable algebra on a 1+1D Riemann surface as in the theory of chiral algebra by Beilinson and Drinfeld \cite{BD04}. However,  from a physical point of view, a VOA with a non-trivial central charge can not live on a 1+1D worldsheet alone because it is physically anomalous in 1+1D even when it is a holomorphic VOA. It can only be physically realized as a 1+1D chiral gapless boundary of a 2+1D topological order. Moreover, the algebraic structure of the cylinder in (\ref{eq:pic-cylinder}) is defined by geometrically fusing the cylinder which was defined in \cite[Section\ 6.2]{KZ21}. In other words, we can depict this algebraic structure as a (3D filled) pant with two (filled) holes lying the plane labeled by $\mathbf{1}^3$ and the third (filled) hole lying above the plane. Recall that, in \cite[Section\ 6.2]{KZ21}, this fusion is defined topologically. In order to understand the algebraic structure of a VOA physically in this context, it is better to equip the surface of the solid cylinder with a complex structure\footnote{Such a solid cylinder can be viewed as a special case of manifolds with metric.}.
This yet-conjectural geometric meanings of a VOA is different from Huang's geometric theory of VOA \cite{Hua95a} and Beilinson-Drinfeld's chiral algebra \cite{BD04}. Therefore, it becomes a very interesting problem to develop the mathematical foundation of above geometric pictures. Another deeply related problem, proposed by the first author of this paper a few years ago, is to established a functorial quantum field theory (as a functor from cobordism category to a linear category) that combines the usual Reshetikhin-Turaev TQFT's with a gapless boundary.
\end{rem}

\begin{rem}
Similar to (\ref{eq:AlgRMod=LModAlg}), we expect that 1-codimensional condensations of liquid-like gapless defect are all given by the liquid-like gapless domain walls between the initial phase and the condensed phase. As explained in (\ref{pic:OmegaM=E1_algebra}), a condensable $\EE_1$-algebra in the initial phase is precisely defined by the $\EE_1$-algebra of observables living on the gapless domain wall. As a special case of this picture, we expect that a modular-invariant closed 1+1D rational CFT can be obtained by condensing any one of its 0+1D boundary CFT's, which is viewed as a condensable $\EE_1$-algebra in the trivial phase. 
\end{rem}

\subsection{Condensations as mutually commuting interactions}

We have shown that the condensation theory is essentially about higher algebras and higher representations, the mathematical theory of which is far from been developed. In this subsection, we discuss some physical or philosophical enlightenments from such a general theory of condensations.

\medskip
Physically, if we reduce the condensation to its simplest form, it is nothing but a way to select a sub-Hilbert space from a Hilbert space as emphasized in \cite{Kon14e}. For example, consider the tensor product of two spins spaces $V$'s, i.e. 
\be
V \otimes_\Cb V \simeq V_0 \oplus V_0^\perp, 
\ee
which is quantum mechanical system instead of a quantum many-body system. Introducing an interaction between these two spin spaces amounts to introducing a projector to the subspace $V_0$, which is more energy favorable than $V_0^\perp$. Although condensation is a term applying to quantum many-body systems, we can view an interaction between two spin spaces, or more generally, an interaction among a few spin spaces, as a condensation in a quantum few-body system.

When we have a chain of spin spaces $\otimes_{i\in \Zb} V$, by introducing mutually commuting local projectors (MCLP) (or condensation) among adjacent spins, we obtain a 1+1D lattice model (see Remark\,\ref{rem:finite_infinite_type_condensation}). We refer to this type of lattice models by {\it MCLP lattice models}. If a MCLP lattice model realizes a quantum phase $\SC^2$, it is reasonable to say that this lattice model realize a condensation from the trivial phase $\mathbf{1}^2$ to $\SC^2$. This point of view automatically generalizes to all dimensions. Namely, a MCLP lattice model realization of a quantum phase $\SC^{n+1}$ is precisely a physical realization of a condensation from the trivial phase $\mathbf{1}^{n+1}$ to $\SC^{n+1}$ (see Remark\,\ref{rem:finite_infinite_type_condensation}). In other words, a MCLP lattice model and a condensation are essentially the same thing.

\begin{rem} \label{rem:finite_infinite_type_condensation}
Our condensation theory is based on condensation maps, which are mutually commuting projectors. Therefore, it only applies to lattice models with mutually commuting local interactions (or condensation). On the other hand, a MCLP lattice model construction is already more general than what is covered in this work because the quantum phase $\SC^{n+1}$ can also be a non-liquid phase (see \cite{Cha05,SSWC18,SSC19,VHF16,Wen20} and references therein). A quantum liquid can be viewed as a quantum phase satisfying certain finite properties because fully dualizability condition is a finiteness condition. Therefore, when $\SC^{n+1}$ is a quantum liquid, we can view a condensation from $\mathbf{1}^{n+1}$ to $\SC^{n+1}$ as a finite-type condensation; when $\SC^{n+1}$ is a quantum non-liquid, the associated condensation is a infinite-type condensation. We hope to give a precise mathematical definition of a finite-type or infinite-type condensation in the future. If we allow condensations of infinite-type, then the trivial phase $\mathbf{1}^{n+1}$ can already be viewed as a `theory of everything'. 
\end{rem}

This unification of these terminologies or ideas leads us to many natural generalizations of previous ideas. We give some examples. 
\bnu
\item An $n+$1D MCLP lattice model can be viewed as first introducing an $n$D MCLP lattice model followed by a layer construction. In other words, higher dimensional MCLP lattice models are layer constructions of lower dimensional MCLP lattice models. 

\begin{figure}[htbp] 
\[
\begin{tikzpicture}[scale=1.5]
\draw[help lines,step=1cm] (-0.5,-0.5) grid (3.5,2.5) ;
\foreach \x in {0,...,3}
 \foreach \y in {0,...,2}{
  \draw[thick,->-=0.5,fill=blue!20] (\x-0.25,\y-0.25) rectangle (\x+0.25,\y+0.25) node[midway] {\scriptsize $\SC^{n+1}$};
  \node[left] at (\x-0.21,\y) {\tiny $\SX^n$} ;
  \node[right] at (\x+0.21,\y) {\tiny $\SX^n$} ;
  \node[above] at (\x,\y+0.21) {\tiny $\SX^n$} ;
  \node[below] at (\x,\y-0.21) {\tiny $\SX^n$} ;
  }
\end{tikzpicture}
\]
\caption{the construction of a universal MCLP lattice model}
\label{fig:universal_lattice_model}
\end{figure}
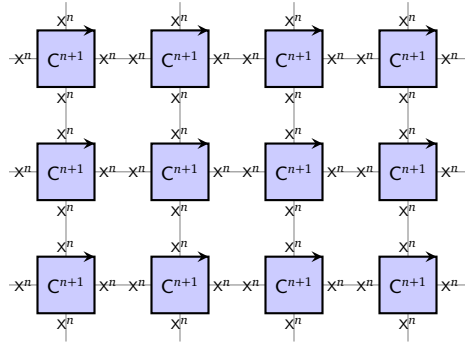

\item It also suggests that, for a quantum phase $\SC^{n+1}$, there is a universal construction of a `MCLP lattice model' that can realize $\SC^{n+1}$. The idea is illustrated in Figure\,\ref{fig:universal_lattice_model}. Each square depicts an $n+$1D quantum phase $\SC^{n+1}$ surrounded by a (potentially gapless) boundary phase $\SX^n$. Such a square should be viewed as a generalized local spin space $V$, and is clearly infinite dimensional. Note that some local degrees of freedom in $V$ are those on the ``local boundary'' $\SX^n$ and others are those in the $\SC^{n+1}$-phase. We define the interaction between two adjacent squares simply by introducing an interactions between the local degrees of freedom in $V$ that belong to the ``local boundaries" $\SX^n$ of two adjacent squares such that this interaction defines a condensation that erases the local boundary and connects the interior of two adjacent squares (see Remark\,\ref{rem:interaction}). This MCLP lattice model realizes a condensation from $\mathbf{1}^{n+1}$. 
\begin{itemize}
\item When $\SX^n$ is gapped, this MCLP lattice model realizes a condensation of `finite type' from $\mathbf{1}^{n+1}$. We expect that this condensation is a continuous phase transition.  
\item When $\SX^n$ is not-gappable gapless and liquid-like\footnote{Then we expect that its bulk $\SC^{n+1}$ is also liquid-like.}, this MCLP lattice model realizes a condensation from $\mathbf{1}^{n+1}$ that is more violent than the finite type. Therefore, it is reasonable to call such a condensation an `infinite type'. We still expect that the phase transition to be continuous. When $\SX^n$ is not liquid-like, then we expect it to be `discontinuous'\footnote{The mathematical theory of phase transitions might need some radical new ideas. For example, we expect that there are hidden structures underlying the traditional notion of a continuous or discontinuous phase transition.}. 
\end{itemize}

\item Since the notion of condensation is defined not only from $\mathbf{1}^{n+1}$ to $\SC^{n+1}$ but also from $\SB^{n+1}$ to $\SC^{n+1}$, we should also have a notion of a {\it lattice model over a background phase $\SB^{n+1}$},  which realizes a new phase $\SC^{n+1}$ and, at the same time, a condensation from $\SB^{n+1}$ to $\SC^{n+1}$. Moreover, the condensation process defined by first proliferating $\SC^{n+1}$-phase-islands within the $\SB^{n+1}$-phase then introducing proper interactions among $\SC^{n+1}$-phase-islands immediately leads us to a universal construction of a MCLP lattice model over the background phase $\SB^{n+1}$ that realizes the phase $\SC^{n+1}$ (see Remark\,\ref{rem:interaction}). An illustration of this construction is given in Figure\,\ref{fig:universal_lattice_model} with the background phase given by $\SB^{n+1}$. An example of such a MCLP lattice model over a non-trivial background phase $\SB^3$ is given in \cite{MRV23}. 
\begin{figure}[htbp] 
\[
\begin{tikzpicture}[scale=1.5]
\fill[green!10] (-0.7,-0.7) rectangle (3.7,2.7) ;
\node at (3.5,2.6) {\small $\SB^{n+1}$} ;
\draw[help lines,step=1cm] (-0.5,-0.5) grid (3.5,2.5) ;
\foreach \x in {0,...,3}
 \foreach \y in {0,...,2}{
  \draw[thick,->-=0.5,fill=blue!20] (\x-0.25,\y-0.25) rectangle (\x+0.25,\y+0.25) node[midway] {\scriptsize $\SC^{n+1}$};
  \node[left] at (\x-0.21,\y) {\tiny $\SX^n$} ;
  \node[right] at (\x+0.21,\y) {\tiny $\SX^n$} ;
  \node[above] at (\x,\y+0.21) {\tiny $\SX^n$} ;
  \node[below] at (\x,\y-0.21) {\tiny $\SX^n$} ;
  }
\end{tikzpicture}
\]
\caption{a universal MCLP lattice model with a brackground phase $\SB^{n+1}$}
\label{fig:universal_lattice_model_2}
\end{figure}
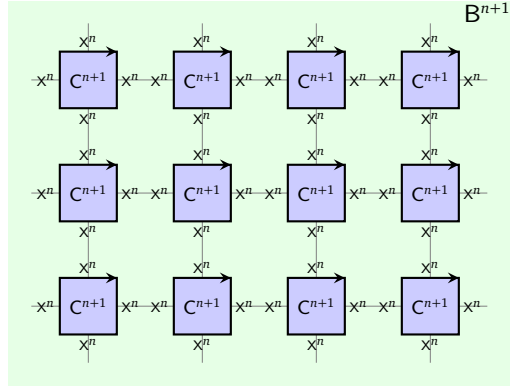
\begin{itemize}
\item When $\SX^n$ is gapped, this MCLP lattice model realizes a condensation of `finite type' from $\SB^{n+1}$. We expect this condensation to be a continuous phase transition.  
\item When $\SX^n$ is not-gappable gapless but liquid-like, this MCLP lattice model realizes a condensation from $\SB^{n+1}$ that is of `infinite type'. We still expect it to be a continuous phase transition. If $\SX^n$ is not liquid-like, we expect it to be a discontinuous phase transition. 
\end{itemize}

\enu

\begin{rem}
In a {\it lattice model over a background phase $\SB^{n+1}$}, the tensor product of local spin spaces is a tensor product in $\Omega^{n-1}\CB$ instead of the vector space tensor product $\otimes_\Cb$. Therefore, such a model is not local in the usual sense (see \cite{MRV23}). However, we believe that such generalized lattice models are very natural and valuable for theoretical studies. 
\end{rem}

\begin{rem} \label{rem:interaction}
We use the interaction between local degrees of freedom in $V$ to effectively erase the local boundaries and connect the interior of two adjacent squares in order to make sure that we are indeed talking about a lattice model in the usual sense. However, it is perhaps more interesting to view the process of erasing the local boundaries and connecting the interior of two adjacent squares as a generalization of the local interactions in a usual lattice model. For example, we can define this process by a  condensation, which can be defined categorically without referring to any local Hilbert space or Hamiltonian. As a consequence, we can generalize the notion of a lattice model by replacing local interactions among local degrees of freedom by local condensations or local phase transitions or other categorical or quantum many-body process. That amounts to say that ``the space of local degrees of freedom'' (i.e., a local Hilbert space) can be replaced by ``a more structuralized space of local degrees of freedom'', such as an ($\EE_k$-monoidal) higher category; and ``local interactions'' can be replaced by ``local condensations''. We believe that we are at the beginning stage of developing a structuralized new calculus for quantum many-body physics or QFT's. 
\end{rem}

\begin{conj}
Above discussion motivates the following two conjectures. 
\bnu

\item For each $n+$1D anomaly-free topological order $\SC^{n+1}$ Morita equivalent to $\SB^{n+1}$, there is an $n+$1D MCLP lattice model over the background phase $\SB^{n+1}$ realizing $\SC^{n+1}$ such that the local spin space $V$ is finite dimensional. 

\item For each $n+$1D anomaly-free topological order $\SC^{n+1}$ not Morita equivalent to $\SB^{n+1}$, there is an $n+$1D MCLP lattice model over the background phase $\SB^{n+1}$ realizing $\SC^{n+1}$ such that the local spin space $V$ is necessarily infinite dimensional. 
\enu
\end{conj}

\begin{rem}
2+1D Levin-Wen models \cite{LW05} or their higher dimensional analogues (see \cite{GJF19} and Remark\,5.11 in \cite{KZ22b}) are MCLP lattice model realizations of non-chiral topological orders. It is natural to expect that these models are deeply related to the universal MCLP lattice models if not the same. We hope to study their relation in the future. 
\end{rem}

\begin{rem}
The physical configuration depicted in Figure\,\ref{fig:universal_lattice_model} or \ref{fig:universal_lattice_model_2} also shows that, before we introduce the interactions, the system is a non-liquid because the ground state degeneracy scales as the size of the system. This fact also suggests that a complete mathematical theory of condensations needs a mathematical theory of non-liquid phases and that of gapless quantum liquids. We believe that it is reasonable to call this yet-conjectural complete theory of condensations as a `theory of everything' not only because it contains all interesting phases and phase transitions among them but also because it provides an interesting approach towards a theory of quantum gravity. Indeed, a theory of quantum gravity is expected to describe the early universe, which should produce a topological order or quantum liquid via finitely many condensations and vice versa. This was an untold dream that has been guiding our research for many years. 
\end{rem}

\newpage


\appendix

\section{Appendix}

\subsection{Proof of Theorem \texorpdfstring{\ref{thm:geo_A=internal_hom}}{geo}} \label{sec:proof_A=internal_hom}

It remains to show that the morphism (\ref{eq:multiplication_via_LLR}) coincides with the one defined in (\ref{eq:geo_define_mu_A}). The morphism (\ref{eq:multiplication_via_LLR}) is defined via the composition of a series of morphisms. We consider a more general condition by replacing the object $1_m$ by arbitrary object $x,y$ in $\Omega_m \CM$. The morphism we need to study is provided by the following composition
\begin{equation}\label{eq_algebraic_multi}
L_2^R(x) \otimes L_2^R(y) \to L_2^R(L_2(L_2^R(x) \otimes L_2^R(y))) \simeq L_2^R(L_2L_2^R(x) \otimes L_2L_2^R(y)) \to L_2^R(x \otimes y)
\end{equation}
We examine the geometric meaning of each morphism appearing in the above composition. It turns out that the geometric meaning of (\ref{eq_algebraic_multi}) is the fusion of two bubbles. 
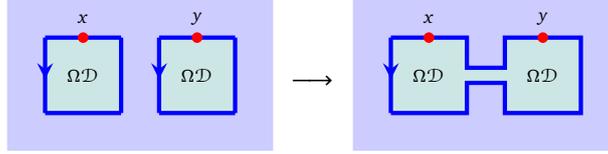
\begin{align*}
\begin{array}{c}
\begin{tikzpicture}[scale=1.0]
\fill[teal!20](0.5,0.5)--(1.5,0.5)--(1.5,1.5)--(0.5,1.5)--cycle;
\fill[teal!20](2,0.5)--(3,0.5)--(3,1.5)--(2,1.5)--cycle;
\fill[blue!20,even odd rule](0,0)--(3.5,0)--(3.5,2)--(0,2)--cycle (0.5,0.5)--(1.5,0.5)--(1.5,1.5)--(0.5,1.5)--cycle (2,0.5)--(3,0.5)--(3,1.5)--(2,1.5)--cycle;
\draw[->-,blue,ultra thick](0.5,1.5)--(0.5,0.5); 
\draw[blue,ultra thick](0.5,0.5)--(1.5,0.5);
\draw[blue,ultra thick](1.5,0.5)--(1.5,1.5);
\draw[blue,ultra thick](1.5,1.5)--(0.5,1.5);
\draw[->-,blue,ultra thick](2,1.5)--(2,0.5); 
\draw[blue,ultra thick](2,0.5)--(3,0.5);
\draw[blue,ultra thick](3,0.5)--(3,1.5);
\draw[blue,ultra thick](3,1.5)--(2,1.5);
\node[black,opacity=1] at(1,1) {\scriptsize $\Omega \CD$};
\node[black,opacity=1] at(2.5,1) {\scriptsize $\Omega \CD$};
\fill[red] (1,1.5) circle (0.07) node[black,opacity=1] at(1,1.75) {\scriptsize $x$};
\fill[red] (2.5,1.5) circle (0.07) node[black,opacity=1] at(2.5,1.75) {\scriptsize $y$};
\end{tikzpicture}
\end{array}
\longrightarrow
\begin{array}{c}
\begin{tikzpicture}[scale=1.0]
\fill[teal!20](0.5,0.5)--(1.5,0.5)--(1.5,0.9)--(2,0.9)--(2,0.5)--(3,0.5)--(3,1.5)--(2,1.5)--(2,1.1)--(1.5,1.1)--(1.5,1.5)--(0.5,1.5)--cycle;
\fill[blue!20,even odd rule](0,0)--(3.5,0)--(3.5,2)--(0,2)--cycle (0.5,0.5)--(1.5,0.5)--(1.5,0.9)--(2,0.9)--(2,0.5)--(3,0.5)--(3,1.5)--(2,1.5)--(2,1.1)--(1.5,1.1)--(1.5,1.5)--(0.5,1.5)--cycle;
\draw[->-,blue,ultra thick](0.5,1.5)--(0.5,0.5); 
\draw[blue,ultra thick](0.5,0.5)--(1.5,0.5)--(1.5,0.9)--(2,0.9)--(2,0.5)--(3,0.5)--(3,1.5)--(2,1.5)--(2,1.1)--(1.5,1.1)--(1.5,1.5)--(0.5,1.5);
\node[black,opacity=1] at(1,1) {\scriptsize $\Omega \CD$};
\node[black,opacity=1] at(2.5,1) {\scriptsize $\Omega \CD$};
\fill[red] (1,1.5) circle (0.07) node[black,opacity=1] at(1,1.75) {\scriptsize $x$};
\fill[red] (2.5,1.5) circle (0.07) node[black,opacity=1] at(2.5,1.75) {\scriptsize $y$};
\end{tikzpicture}
\end{array}
\end{align*}
As a result, by taking $x,y = 1_m$, we recover the morphism (\ref{eq:geo_define_mu_A}).

Before we start, we analyse the geometric meaning of the unit and counit of the adjunction $L_2 \dashv [1_m,-]$, which are used throughout the proof. Recall that, the bulk-to-wall map $L_2:\Omega \CC \to \Omega_m\CM$ means we move a point-like excitation $a$ in $\Omega \CC $ next to the gapped domain wall $\Omega_m\CM$, so that we can view $a$ as an excitation on $\Omega_m\CM$, and we denote it by $L_2(a)$. 
\begin{itemize}
\item Unit of the adjunction $\eta: \id_{\Omega \CC} \Rightarrow [1_m,L_2(-)] = L_2^R L_2(-) = - \otimes L_2^R (1_m)$:
\begin{align*}
\eta_a:
\begin{array}{c}
\begin{tikzpicture}[scale=0.8]
\fill[blue!20](-2,0)--(0,0)--(0,2)--(-2,2)--cycle node[black,opacity=1] at(-1,0.2) {\scriptsize $\Omega \CC$};
\fill[teal!20](2,0)--(0,0)--(0,2)--(2,2)--cycle node[black,opacity=1] at(1,0.2) {\scriptsize $\Omega \CD$};
\draw[->-,blue,ultra thick] (0,0)--(0,2);
\fill[red] (-1,0.9) circle (0.07) node[left,black,opacity=1] {\scriptsize $a$};
\end{tikzpicture}
\end{array}
\to \int^{\Omega \CD}_{x \in \Omega_m\CM}
\begin{array}{c}
\begin{tikzpicture}[scale=0.8]
\fill[blue!20](-2,0)--(0,0)--(0,0.5)--(-0.8,0.5)--(-0.8,1.5)--(0,1.5)--(0,2)--(-2,2)--cycle node[black,opacity=1] at(-1,0.2) {\scriptsize $\Omega \CC$};
\fill[teal!20](2,0)--(0,0)--(0,0.5)--(-0.8,0.5)--(-0.8,1.5)--(0,1.5)--(0,2)--(2,2)--cycle node[black,opacity=1] at(1.6,0.2) {\scriptsize $\Omega \CD$}; 
\draw[->-,blue,ultra thick] (0,0)--(0,0.5);
\draw[->-,blue,ultra thick] (0,1.5)--(0,2);
\draw[blue,ultra thick,opacity=1](0,0.5)--(-0.8,0.5)--(-0.8,1.5)--(0,1.5);
\fill[black] (-0.4,0.5) circle (0.05);
\fill[black] (-0.4,1.5) circle (0.05);
\node[black,opacity=1] at (-0.4,0.3){\scriptsize $x$};
\node[black,opacity=1] at (-0.4,1.7){\scriptsize $x^R$};
\fill[red] (-1,1) circle (0.07) node[left,black,opacity=1] {\scriptsize $a$};
\end{tikzpicture}
\end{array}
\simeq
\begin{array}{c}
\begin{tikzpicture}[scale=0.8]
\fill[blue!20](-2,0)--(0,0)--(0,2)--(-2,2)--cycle node[black,opacity=1] at(-1,0.2) {\scriptsize $\Omega \CC$};
\fill[teal!20](-0.6,0.5)--(-1.6,0.5)--(-1.6,1.5)--(-0.6,1.5)--cycle;
\fill[teal!20,even odd rule](0,0)--(2,0)--(2,2)--(0,2)--cycle (-0.6,0.5)--(-1.6,0.5)--(-1.6,1.5)--(-0.6,1.5)--cycle node[black,opacity=1] at(1,0.2) {\scriptsize $\Omega \CD$};
\draw[->-,blue,ultra thick] (0,0)--(0,2);
\draw[->-,blue,ultra thick](-0.6,1.5)--(-0.6,0.5);
\draw[blue,ultra thick](-0.6,1.5)--(-0.6,0.5); 
\draw[blue,ultra thick](-0.6,0.5)--(-1.6,0.5);
\draw[blue,ultra thick](-1.6,0.5)--(-1.6,1.5);
\draw[blue,ultra thick](-1.6,1.5)--(-0.6,1.5);
\fill[red] (-1.8,1) circle (0.07) node[left,black,opacity=1] {\scriptsize $a$};
\end{tikzpicture}
\end{array}
\end{align*}

\item Counit of the adjunction $\varepsilon:L_2 L_2^R(-) L_2([1_m,-]) \Rightarrow \id_{\Omega_m\CM}$:
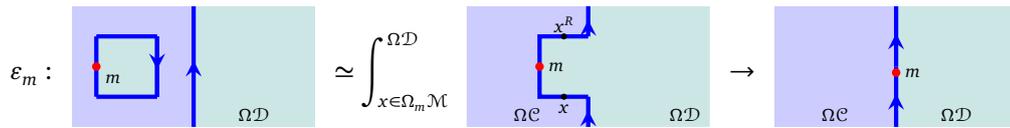
\begin{align*}
\varepsilon_m:
\begin{array}{c}
\begin{tikzpicture}[scale=0.8]
\fill[teal!20](2,0)--(0,0)--(0,2)--(2,2)--cycle node[black,opacity=1] at(1,0.2) {\scriptsize $\Omega \CD$};
\fill[teal!20](-0.6,0.5)--(-1.6,0.5)--(-1.6,1.5)--(-0.6,1.5)--cycle;
\fill[blue!20,even odd rule](0,0)--(-2,0)--(-2,2)--(0,2)--cycle (-0.6,0.5)--(-1.6,0.5)--(-1.6,1.5)--(-0.6,1.5)--cycle;
\draw[->-,blue,ultra thick] (0,0)--(0,2);
\draw[->-,blue,ultra thick](-0.6,1.5)--(-0.6,0.5);
\draw[blue,ultra thick](-0.6,1.5)--(-0.6,0.5); 
\draw[blue,ultra thick](-0.6,0.5)--(-1.6,0.5);
\draw[blue,ultra thick](-1.6,0.5)--(-1.6,1.5);
\draw[blue,ultra thick](-1.6,1.5)--(-0.6,1.5);
\fill[red] (-1.6,1) circle (0.07) node[right,black,opacity=1] at(-1.6,0.8) {\scriptsize $m$};
\end{tikzpicture}
\end{array}
\simeq
\int^{\Omega \CD}_{x \in \Omega_m\CM}
\begin{array}{c}
\begin{tikzpicture}[scale=0.8]
\fill[blue!20](-2,0)--(0,0)--(0,0.5)--(-0.8,0.5)--(-0.8,1.5)--(0,1.5)--(0,2)--(-2,2)--cycle node[black,opacity=1] at(-1,0.2) {\scriptsize $\Omega \CC$};
\fill[teal!20](2,0)--(0,0)--(0,0.5)--(-0.8,0.5)--(-0.8,1.5)--(0,1.5)--(0,2)--(2,2)--cycle node[black,opacity=1] at(1.6,0.2) {\scriptsize $\Omega \CD$}; 
\draw[->-,blue,ultra  thick] (0,0)--(0,0.5);
\draw[->-,blue,ultra thick] (0,1.5)--(0,2);
\draw[blue,ultra thick,opacity=1](0,0.5)--(-0.8,0.5)--(-0.8,1.5)--(0,1.5);
\fill[black] (-0.4,0.5) circle (0.05);
\fill[black] (-0.4,1.5) circle (0.05);
\node[black,opacity=1] at (-0.4,0.3){\scriptsize $x$};
\node[black,opacity=1] at (-0.4,1.7){\scriptsize $x^R$};
\fill[red] (-0.8,1) circle (0.07) node[right,black,opacity=1] at(-0.8,1) {\scriptsize $m$};
\end{tikzpicture}
\end{array}
\to
\begin{array}{c}
\begin{tikzpicture}[scale=0.8]
\fill[blue!20](-2,0)--(0,0)--(0,2)--(-2,2)--cycle node[black,opacity=1] at(-1,0.2) {\scriptsize $\Omega \CC$};
\fill[teal!20](2,0)--(0,0)--(0,2)--(2,2)--cycle node[black,opacity=1] at(1,0.2) {\scriptsize $\Omega \CD$};
\draw[->-,blue,ultra thick] (0,1)--(0,2);
\draw[->-,blue,ultra thick] (0,0)--(0,1);
\fill[red] (0,0.9) circle (0.07) node[right,black,opacity=1] at(0,0.9) {\scriptsize $m$};
\end{tikzpicture}
\end{array}
\end{align*}
\end{itemize}

Then we need to analyze the geometric meaning of following morphisms, whose composition gives rise to (\ref{eq_algebraic_multi}).
\begin{itemize}
\item[1.] The first morphism is a component of the unit $\eta$. It is given by
\begin{equation}\label{eq_unit}
\eta_{L_2^R(x) \otimes L_2^R(y)}: L_2^R(x) \otimes L_2^R(y) \to L_2^R(L_2(L_2^R(x) \otimes L_2^R(y)))
\end{equation}

\item[2.] The second morphism 
\begin{equation}
L_2^R(L_2(L_2^R(x) \otimes L_2^R(y))) \simeq L_2^R(L_2(L_2^R(x)) \otimes L_2(L_2^R(y)))
\end{equation}
only exploits the monoidal structure of $L_2$, hence is geometrically trivial. We omit this morphism in our geometric analysis.

\item[3.] The third morphism is given by
\[
L_2^R(\varepsilon_x \otimes \varepsilon_y): L_2^R(L_2(L_2^R(x)) \otimes L_2(L_2^R(y))) \to L_2^R(x \otimes y)
\]
We study the morphism inside $L_2^R(-)$, that is, the tensor product of two components of $\varepsilon$:
\begin{equation}\label{eq_counit}
\varepsilon_x \otimes \varepsilon_y: L_2(L_2^R(x)) \otimes L_2(L_2^R(y)) \to x \otimes y.
\end{equation}
The analysis of the morphism (\ref{eq_counit}) occupies the main space of this appendix.
\end{itemize}

The geometric meaning of the morphism (\ref{eq_unit}) is: 
$$
\eta_{L_2^R(x) \otimes L_2^R(y)}:
\begin{array}{c}
\begin{tikzpicture}[scale=0.8]
\fill[teal!20](0.5,0.5)--(1.5,0.5)--(1.5,1.5)--(0.5,1.5)--cycle;
\fill[teal!20](2,0.5)--(3,0.5)--(3,1.5)--(2,1.5)--cycle;
\fill[blue!20,even odd rule](0,0)--(3.8,0)--(3.8,2)--(0,2)--cycle (0.5,0.5)--(1.5,0.5)--(1.5,1.5)--(0.5,1.5)--cycle (2,0.5)--(3,0.5)--(3,1.5)--(2,1.5)--cycle;
\draw[->-,blue,ultra thick](0.5,1.5)--(0.5,0.5); 
\draw[blue,ultra thick](0.5,0.5)--(1.5,0.5);
\draw[blue,ultra thick](1.5,0.5)--(1.5,1.5);
\draw[blue,ultra thick](1.5,1.5)--(0.5,1.5);
\draw[->-,blue,ultra thick](2,1.5)--(2,0.5); 
\draw[blue,ultra thick](2,0.5)--(3,0.5);
\draw[blue,ultra thick](3,0.5)--(3,1.5);
\draw[blue,ultra thick](3,1.5)--(2,1.5);
\node[black,opacity=1] at(3.5,1) {\scriptsize $\Omega \CC$};
\node[black,opacity=1] at(1,1) {\scriptsize $\Omega \CD$};
\node[black,opacity=1] at(2.5,1) {\scriptsize $\Omega \CD$};
\fill[red] (1,1.5) circle (0.07) node[black,opacity=1] at(1,1.75) {\scriptsize $x$};
\fill[red] (2.5,1.5) circle (0.07) node[black,opacity=1] at(2.5,1.75) {\scriptsize $y$};
\end{tikzpicture}
\end{array}
\longrightarrow
\begin{array}{c}
\begin{tikzpicture}[scale=0.7]
\fill[teal!20](0.5,0.5)--(1.5,0.5)--(1.5,1.5)--(0.5,1.5)--cycle;
\fill[teal!20](2,0.5)--(3,0.5)--(3,1.5)--(2,1.5)--cycle;
\fill[teal!20](1,0)--(2.5,0)--(2.5,-1)--(1,-1)--cycle;
\fill[blue!20,even odd rule](0,2)--(3.8,2)--(3.8,-1.5)--(0,-1.5)--cycle  (0.5,0.5)--(1.5,0.5)--(1.5,1.5)--(0.5,1.5)--cycle  (2,0.5)--(3,0.5)--(3,1.5)--(2,1.5)--cycle  (1,0)--(2.5,0)--(2.5,-1)--(1,-1)--cycle;
\draw[->-,blue,ultra thick](0.5,1.5)--(0.5,0.5); 
\draw[blue,ultra thick](0.5,0.5)--(1.5,0.5);
\draw[blue,ultra thick](1.5,0.5)--(1.5,1.5);
\draw[blue,ultra thick](1.5,1.5)--(0.5,1.5);
\draw[->-,blue,ultra thick](2,1.5)--(2,0.5); 
\draw[blue,ultra thick](2,0.5)--(3,0.5);
\draw[blue,ultra thick](3,0.5)--(3,1.5);
\draw[blue,ultra thick](3,1.5)--(2,1.5);
\node[black,opacity=1] at(3.5,0) {\scriptsize $\Omega \CC$};
\node[black,opacity=1] at(1,1) {\scriptsize $\Omega \CD$};
\node[black,opacity=1] at(2.5,1) {\scriptsize $\Omega \CD$};
\fill[red] (1,1.5) circle (0.07) node[black,opacity=1] at(1,1.75) {\scriptsize $x$};
\fill[red] (2.5,1.5) circle (0.07) node[black,opacity=1] at(2.5,1.75) {\scriptsize $y$};
\draw[->-,blue,ultra thick](1,0)--(1,-1); 
\draw[blue,ultra thick](1,0)--(2.5,0);
\draw[blue,ultra thick](2.5,0)--(2.5,-1);
\draw[blue,ultra thick](1,-1)--(2.5,-1);
\node[black,opacity=1] at(1.75,-0.5) {\scriptsize $\Omega \CD$};
\end{tikzpicture}
\end{array} 
$$
The geometric meaning of the morphism (\ref{eq_counit}) is:
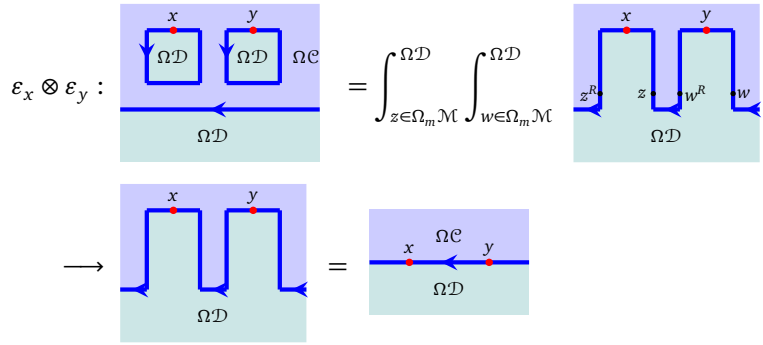
\begin{align*}
\varepsilon_x \otimes \varepsilon_y: &
\begin{array}{c}
\begin{tikzpicture}[scale=0.7]
\fill[teal!20](0.5,0.5)--(1.5,0.5)--(1.5,1.5)--(0.5,1.5)--cycle;
\fill[teal!20](2,0.5)--(3,0.5)--(3,1.5)--(2,1.5)--cycle;
\fill[blue!20,even odd rule](0,0)--(3.75,0)--(3.75,2)--(0,2)--cycle (0.5,0.5)--(1.5,0.5)--(1.5,1.5)--(0.5,1.5)--cycle (2,0.5)--(3,0.5)--(3,1.5)--(2,1.5)--cycle;
\draw[->-,blue,ultra thick](0.5,1.5)--(0.5,0.5); 
\draw[blue,ultra thick](0.5,0.5)--(1.5,0.5);
\draw[blue,ultra thick](1.5,0.5)--(1.5,1.5);
\draw[blue,ultra thick](1.5,1.5)--(0.5,1.5);
\draw[->-,blue,ultra thick](2,1.5)--(2,0.5); 
\draw[blue,ultra thick](2,0.5)--(3,0.5);
\draw[blue,ultra thick](3,0.5)--(3,1.5);
\draw[blue,ultra thick](3,1.5)--(2,1.5);
\node[black,opacity=1] at(3.5,1) {\scriptsize $\Omega \CC$};
\node[black,opacity=1] at(1,1) {\scriptsize $\Omega \CD$};
\node[black,opacity=1] at(2.5,1) {\scriptsize $\Omega \CD$};
\fill[red] (1,1.5) circle (0.07) node[black,opacity=1] at(1,1.75) {\scriptsize $x$};
\fill[red] (2.5,1.5) circle (0.07) node[black,opacity=1] at(2.5,1.75) {\scriptsize $y$};
\fill[teal!20](0,0)--(3.75,0)--(3.75,-1)--(0,-1)--cycle;
\node[black,opacity=1] at(1.75,-0.5) {\scriptsize $\Omega \CD$};
\draw[->-,blue,ultra thick](3.75,0)--(0,0);
\end{tikzpicture}
\end{array}
= \int^{\Omega \CD}_{z \in \Omega_m\CM} \int^{\Omega \CD}_{w \in \Omega_m\CM}
\begin{array}{c}
\begin{tikzpicture}[scale=0.7]
\fill[blue!20](0,0)--(0.5,0)--(0.5,1.5)--(1.5,1.5)--(1.5,0)--(2,0)--(2,1.5)--(3,1.5)--(3,0)--(3.5,0)--(3.5,2)--(0,2)--cycle;
\fill[teal!20] (0,0)--(0.5,0)--(0.5,1.5)--(1.5,1.5)--(1.5,0)--(2,0)--(2,1.5)--(3,1.5)--(3,0)--(3.5,0)--(3.5,-1)--(0,-1)--cycle;
\draw[->-,blue,ultra thick](0.5,0)--(0,0);
\draw[blue,ultra thick](0.5,0)--(0.5,1.5);
\draw[blue,ultra thick](0.5,1.5)--(1.5,1.5);
\draw[blue,ultra thick](1.5,1.5)--(1.5,0);
\draw[->-,blue,ultra thick](2,0)--(1.5,0);
\draw[blue,ultra thick](2,0)--(2,1.5);
\draw[blue,ultra thick](3,1.5)--(2,1.5);
\draw[blue,ultra thick](3,1.5)--(3,0);
\draw[->-,blue,ultra thick](3.5,0)--(3,0);
\fill[red] (1,1.5) circle (0.07) node[black,opacity=1] at(1,1.75) {\scriptsize $x$};
\fill[red] (2.5,1.5) circle (0.07) node[black,opacity=1] at(2.5,1.75) {\scriptsize $y$};
\fill[black] (0.5,0.3) circle (0.05);
\fill[black] (1.5,0.3) circle (0.05);
\fill[black] (2,0.3) circle (0.05);
\fill[black] (3,0.3) circle (0.05);
\node[black,opacity=1] at(0.3,0.3) {\scriptsize $z^R$};
\node[black,opacity=1] at(1.3,0.3) {\scriptsize $z$};
\node[black,opacity=1] at(2.3,0.3) {\scriptsize $w^R$};
\node[black,opacity=1] at(3.2,0.3) {\scriptsize $w$};
\node[black,opacity=1] at(1.75,-0.5) {\scriptsize$\Omega \CD$};
\end{tikzpicture} 
\end{array} \\
\longrightarrow &
\begin{array}{c}
\begin{tikzpicture}[scale=0.7]
\fill[blue!20](0,0)--(0.5,0)--(0.5,1.5)--(1.5,1.5)--(1.5,0)--(2,0)--(2,1.5)--(3,1.5)--(3,0)--(3.5,0)--(3.5,2)--(0,2)--cycle;
\fill[teal!20] (0,0)--(0.5,0)--(0.5,1.5)--(1.5,1.5)--(1.5,0)--(2,0)--(2,1.5)--(3,1.5)--(3,0)--(3.5,0)--(3.5,-1)--(0,-1)--cycle;
\draw[->-,blue,ultra thick](0.5,0)--(0,0);
\draw[blue,ultra thick](0.5,0)--(0.5,1.5);
\draw[blue,ultra thick](0.5,1.5)--(1.5,1.5);
\draw[blue,ultra thick](1.5,1.5)--(1.5,0);
\draw[->-,blue,ultra thick](2,0)--(1.5,0);
\draw[blue,ultra thick](2,0)--(2,1.5);
\draw[blue,ultra thick](3,1.5)--(2,1.5);
\draw[blue,ultra thick](3,1.5)--(3,0);
\draw[->-,blue,ultra thick](3.5,0)--(3,0);
\fill[red] (1,1.5) circle (0.07) node[black,opacity=1] at(1,1.75) {\scriptsize $x$};
\fill[red] (2.5,1.5) circle (0.07) node[black,opacity=1] at(2.5,1.75) {\scriptsize $y$};
\node[black,opacity=1] at(1.75,-0.5) {\scriptsize $\Omega \CD$};
\end{tikzpicture} 
\end{array}
= 
\begin{array}{c}
\begin{tikzpicture}[scale=0.7]
\fill[blue!20](0,0)--(3,0)--(3,1)--(0,1)--cycle;
\fill[teal!20](0,0)--(3,0)--(3,-1)--(0,-1)--cycle;
\draw[->-,blue,ultra thick] (3,0)--(0,0);
\node[black,opacity=1] at(1.5,-0.5) {\scriptsize $\Omega \CD$};
\node[black,opacity=1] at(1.5,0.5) {\scriptsize $\Omega \CC$};
\fill[red] (0.75,0) circle (0.07) node[black,opacity=1] at(0.75,0.25) {\scriptsize $x$};
\fill[red] (2.25,0) circle (0.07) node[black,opacity=1] at(2.25,0.25) {\scriptsize $y$};
\end{tikzpicture} 
\end{array}
\end{align*}

To proceed, let us note that there is a canonical isomorphism:

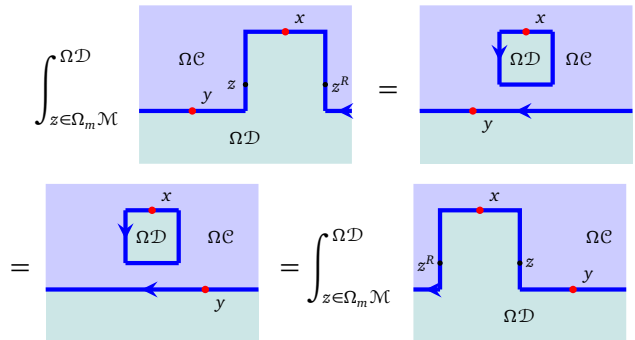
\begin{align*}
& \int^{\Omega \CD}_{z\in \Omega_m\CM}
\begin{array}{c}
\begin{tikzpicture}[scale=0.7]
\fill[teal!20](-1,0)--(1,0)--(1,1.5)--(2.5,1.5)--(2.5,0)--(3,0)--(3,-1)--(-1,-1)--cycle node[black,opacity=1] at(1,-0.5) {\scriptsize $\Omega \CD$};
\fill[blue!20](-1,0)--(1,0)--(1,1.5)--(2.5,1.5)--(2.5,0)--(3,0)--(3,2)--(-1,2)--cycle node[black,opacity=1] at(0,1) {\scriptsize $\Omega \CC$}; 
\draw[->-,blue,ultra thick] (3,0)--(2.5,0);
\draw[blue,ultra thick,opacity=1](-1,0)--(1,0);
\draw[blue,ultra thick,opacity=1](1,0)--(1,1.5)--(2.5,1.5)--(2.5,0);
\fill[black] (1,0.5) circle (0.05);
\fill[black] (2.5,0.5) circle (0.05);
\node[black,opacity=1] at (0.75,0.5){\scriptsize $z$};
\node[black,opacity=1] at (2.8,0.5){\scriptsize $z^R$};
\fill[red] (1.75,1.5) circle (0.07) node[right,black,opacity=1] at(1.75,1.75) {\scriptsize $x$};
\fill[red] (0,0) circle (0.07) node[right,black,opacity=1] at(0,0.25) {\scriptsize $y$};
\end{tikzpicture}
\end{array}
=
\begin{array}{c}
\begin{tikzpicture}[scale=0.7]
\fill[teal!20](0.5,0.5)--(1.5,0.5)--(1.5,1.5)--(0.5,1.5)--cycle;
\fill[blue!20,even odd rule](-1,0)--(3,0)--(3,2)--(-1,2)--cycle (0.5,0.5)--(1.5,0.5)--(1.5,1.5)--(0.5,1.5)--cycle;
\fill[teal!20](-1,0)--(3,0)--(3,-1)--(-1,-1)--cycle;
\draw[->-,blue,ultra thick](3,0)--(-1,0);
\draw[->-,blue,ultra thick](0.5,1.5)--(0.5,0.5); 
\draw[blue,ultra thick](0.5,0.5)--(1.5,0.5);
\draw[blue,ultra thick](1.5,0.5)--(1.5,1.5);
\draw[blue,ultra thick](1.5,1.5)--(0.5,1.5);
\node[black,opacity=1] at(2,1) {\scriptsize $\Omega \CC$};
\node[black,opacity=1] at(1,1) {\scriptsize $\Omega \CD$};
\fill[red] (1,1.5) circle (0.07) node[right,black,opacity=1] at(1,1.75) {\scriptsize $x$};
\fill[red] (0,0) circle (0.07) node[right,black,opacity=1] at(0,-0.3) {\scriptsize $y$};
\end{tikzpicture}
\end{array}
\\
= &
\begin{array}{c}
\begin{tikzpicture}[scale=0.7]
\fill[teal!20](0.5,0.5)--(1.5,0.5)--(1.5,1.5)--(0.5,1.5)--cycle;
\fill[blue!20,even odd rule](-1,0)--(3,0)--(3,2)--(-1,2)--cycle (0.5,0.5)--(1.5,0.5)--(1.5,1.5)--(0.5,1.5)--cycle;
\fill[teal!20](-1,0)--(3,0)--(3,-1)--(-1,-1)--cycle;
\draw[->-,blue,ultra thick](3,0)--(-1,0);
\draw[->-,blue,ultra thick](0.5,1.5)--(0.5,0.5); 
\draw[blue,ultra thick](0.5,0.5)--(1.5,0.5);
\draw[blue,ultra thick](1.5,0.5)--(1.5,1.5);
\draw[blue,ultra thick](1.5,1.5)--(0.5,1.5);
\node[black,opacity=1] at(2.3,1) {\scriptsize $\Omega \CC$};
\node[black,opacity=1] at(1,1) {\scriptsize $\Omega \CD$};
\fill[red] (1,1.5) circle (0.07) node[right,black,opacity=1] at(1,1.75) {\scriptsize $x$};
\fill[red] (2,0) circle (0.07) node[right,black,opacity=1] at(2,-0.3) {\scriptsize $y$};
\end{tikzpicture}
\end{array}
=
\int^{\Omega \CD}_{z\in \Omega_m\CM}
\begin{array}{c}
\begin{tikzpicture}[scale=0.7]
\fill[teal!20](1,0)--(-1,0)--(-1,1.5)--(-2.5,1.5)--(-2.5,0)--(-3,0)--(-3,-1)--(1,-1)--cycle node[black,opacity=1] at(-1,-0.5) {\scriptsize $\Omega \CD$};
\fill[blue!20](1,0)--(-1,0)--(-1,1.5)--(-2.5,1.5)--(-2.5,0)--(-3,0)--(-3,2)--(1,2)--cycle node[black,opacity=1] at(0.5,1) {\scriptsize $\Omega \CC$}; 
\draw[->-,blue,ultra thick] (-2.5,0)--(-3,0);
\draw[blue,ultra thick,opacity=1](1,0)--(-1,0);
\draw[blue,ultra thick,opacity=1](-1,0)--(-1,1.5)--(-2.5,1.5)--(-2.5,0);
\fill[black] (-1,0.5) circle (0.05);
\fill[black] (-2.5,0.5) circle (0.05);
\node[black,opacity=1] at (-0.8,0.5){\scriptsize $z$};
\node[black,opacity=1] at (-2.75,0.5){\scriptsize $z^R$};
\fill[red] (-1.75,1.5) circle (0.07) node[right,black,opacity=1] at(-1.75,1.75) {\scriptsize $x$};
\fill[red] (0,0) circle (0.07) node[right,black,opacity=1] at(0,0.25) {\scriptsize $y$};
\end{tikzpicture}
\end{array}
\end{align*}

Algebraically, this means that the object $\int^{\Omega \CD}_{z \in \Omega_m\CM} z^R \otimes x \otimes z$ is equipped with a half braiding $\beta$ whose components are $\int^{\Omega \CD}_{z \in \Omega_m\CM} z^R \otimes x \otimes z \otimes y \simeq y \otimes( \int^{\Omega \CD}_{z \in \Omega_m\CM}  z^R \otimes x \otimes z) \simeq \int^{\Omega \CD}_{z \in \Omega_m\CM} y \otimes z^R \otimes x \otimes z $.  The mathematical details of the construction of the half-braiding are postponed to the next appendix, see Example \ref{Expl_comonad} and Example \ref{Expl_half-braiding}.

Now we would like to show that, the morphism $\varepsilon_x \otimes \varepsilon_y$ is equal to the following morphism:
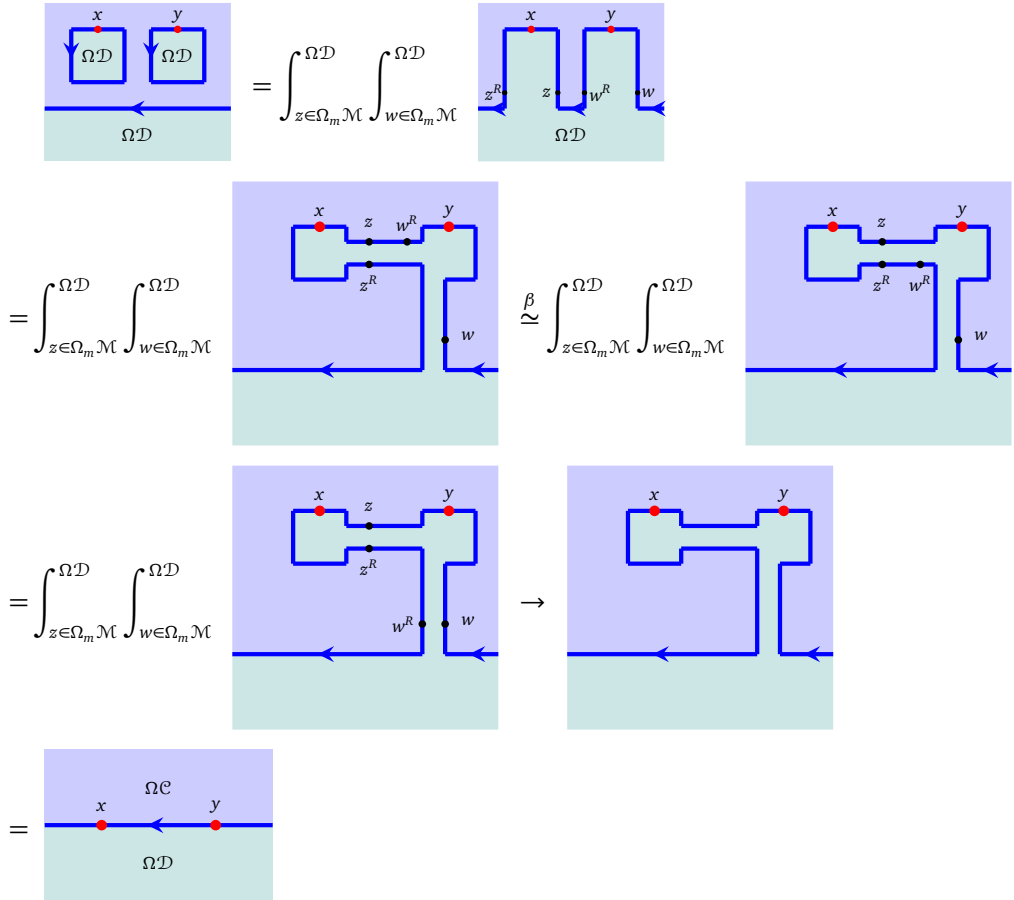
\begin{align*}
&
\begin{array}{c}
\begin{tikzpicture}[scale=0.7]
\fill[teal!20](0.5,0.5)--(1.5,0.5)--(1.5,1.5)--(0.5,1.5)--cycle;
\fill[teal!20](2,0.5)--(3,0.5)--(3,1.5)--(2,1.5)--cycle;
\fill[blue!20,even odd rule](0,0)--(3.5,0)--(3.5,2)--(0,2)--cycle (0.5,0.5)--(1.5,0.5)--(1.5,1.5)--(0.5,1.5)--cycle (2,0.5)--(3,0.5)--(3,1.5)--(2,1.5)--cycle;
\draw[->-,blue,ultra thick](0.5,1.5)--(0.5,0.5); 
\draw[blue,ultra thick](0.5,0.5)--(1.5,0.5);
\draw[blue,ultra thick](1.5,0.5)--(1.5,1.5);
\draw[blue,ultra thick](1.5,1.5)--(0.5,1.5);
\draw[->-,blue,ultra thick](2,1.5)--(2,0.5); 
\draw[blue,ultra thick](2,0.5)--(3,0.5);
\draw[blue,ultra thick](3,0.5)--(3,1.5);
\draw[blue,ultra thick](3,1.5)--(2,1.5);
\node[black,opacity=1] at(1,1) {\scriptsize $\Omega \CD$};
\node[black,opacity=1] at(2.5,1) {\scriptsize $\Omega \CD$};
\fill[red] (1,1.5) circle (0.07) node[black,opacity=1] at(1,1.75) {\scriptsize $x$};
\fill[red] (2.5,1.5) circle (0.07) node[black,opacity=1] at(2.5,1.75) {\scriptsize $y$};
\fill[teal!20](0,0)--(3.5,0)--(3.5,-1)--(0,-1)--cycle;
\node[black,opacity=1] at(1.75,-0.5) {\scriptsize $\Omega \CD$};
\draw[->-,blue,ultra thick](3.5,0)--(0,0);
\end{tikzpicture}
\end{array}
= \int^{\Omega \CD}_{z \in \Omega_m\CM} \int^{\Omega \CD}_{w \in \Omega_m\CM}
\begin{array}{c}
\begin{tikzpicture}[scale=0.7]
\fill[blue!20](0,0)--(0.5,0)--(0.5,1.5)--(1.5,1.5)--(1.5,0)--(2,0)--(2,1.5)--(3,1.5)--(3,0)--(3.5,0)--(3.5,2)--(0,2)--cycle;
\fill[teal!20] (0,0)--(0.5,0)--(0.5,1.5)--(1.5,1.5)--(1.5,0)--(2,0)--(2,1.5)--(3,1.5)--(3,0)--(3.5,0)--(3.5,-1)--(0,-1)--cycle;
\draw[->-,blue,ultra thick](0.5,0)--(0,0);
\draw[blue,ultra thick](0.5,0)--(0.5,1.5);
\draw[blue,ultra thick](0.5,1.5)--(1.5,1.5);
\draw[blue,ultra thick](1.5,1.5)--(1.5,0);
\draw[->-,blue,ultra thick](2,0)--(1.5,0);
\draw[blue,ultra thick](2,0)--(2,1.5);
\draw[blue,ultra thick](3,1.5)--(2,1.5);
\draw[blue,ultra thick](3,1.5)--(3,0);
\draw[->-,blue,ultra thick](3.5,0)--(3,0);
\fill[red] (1,1.5) circle (0.07) node[black,opacity=1] at(1,1.75) {\scriptsize $x$};
\fill[red] (2.5,1.5) circle (0.07) node[black,opacity=1] at(2.5,1.75) {\scriptsize $y$};
\fill[black] (0.5,0.3) circle (0.05);
\fill[black] (1.5,0.3) circle (0.05);
\fill[black] (2,0.3) circle (0.05);
\fill[black] (3,0.3) circle (0.05);
\node[black,opacity=1] at(0.3,0.3) {\scriptsize $z^R$};
\node[black,opacity=1] at(1.3,0.3) {\scriptsize $z$};
\node[black,opacity=1] at(2.3,0.3) {\scriptsize $w^R$};
\node[black,opacity=1] at(3.2,0.3) {\scriptsize $w$};
\node[black,opacity=1] at(1.75,-0.5) {\scriptsize $\Omega \CD$};
\end{tikzpicture} 
\end{array} \\
= & \int^{\Omega \CD}_{z \in \Omega_m\CM} \int^{\Omega \CD}_{w \in \Omega_m\CM}
\begin{array}{c}
\begin{tikzpicture}
\fill[blue!20](0,0)--(2.5,0)--(2.5,1.4)--(1.5,1.4)--(1.5,1.2)--(0.8,1.2)--(0.8,1.9)--(1.5,1.9)--(1.5,1.7)--(2.5,1.7)--(2.5,1.9)--(3.2,1.9)--(3.2,1.2)--(2.8,1.2)--(2.8,0)--(3.5,0)--(3.5,2.5)--(0,2.5)--cycle;
\fill[teal!20](0,0)--(2.5,0)--(2.5,1.4)--(1.5,1.4)--(1.5,1.2)--(0.8,1.2)--(0.8,1.9)--(1.5,1.9)--(1.5,1.7)--(2.5,1.7)--(2.5,1.9)--(3.2,1.9)--(3.2,1.2)--(2.8,1.2)--(2.8,0)--(3.5,0)--(3.5,-1)--(0,-1)--cycle;
\draw[->-,blue,ultra thick](3.5,0)--(2.8,0);
\draw[blue,ultra thick](2.8,1.2)--(2.8,0);
\draw[blue,ultra thick](3.2,1.2)--(2.8,1.2);
\draw[blue,ultra thick](3.2,1.9)--(3.2,1.2);
\draw[blue,ultra thick](2.5,1.9)--(3.2,1.9);
\draw[blue,ultra thick](2.5,1.7)--(2.5,1.9);
\draw[blue,ultra thick](1.5,1.7)--(2.5,1.7);
\draw[blue,ultra thick](1.5,1.9)--(1.5,1.7);
\draw[blue,ultra thick](0.8,1.9)--(1.5,1.9);
\draw[blue,ultra thick](1.5,1.2)--(0.8,1.2);
\draw[blue,ultra thick](1.5,1.4)--(1.5,1.2);
\draw[blue,ultra thick](2.5,1.4)--(1.5,1.4);
\draw[blue,ultra thick](2.5,0)--(2.5,1.4);
\draw[->-,blue,ultra thick](2.5,0)--(0,0);
\draw[blue,ultra thick](0.8,1.2)--(0.8,1.9);
\fill[black] (1.8,1.4) circle (0.05) node[black,opacity=1] at(1.8,1.15) {\scriptsize $z^R$};
\fill[black] (1.8,1.7) circle (0.05) node[black,opacity=1] at(1.8,1.95) {\scriptsize $z$};
\fill[black] (2.3,1.7) circle (0.05) node[black,opacity=1] at(2.3,1.95) {\scriptsize $w^R$};
\fill[black] (2.8,0.4) circle (0.05) node[black,opacity=1] at(3.1,0.45) {\scriptsize $w$};
\fill[red] (1.15,1.9) circle (0.07) node[black,opacity=1] at(1.15,2.1) {\scriptsize $x$};
\fill[red] (2.85,1.9) circle (0.07) node[black,opacity=1] at(2.85,2.1) {\scriptsize $y$};
\end{tikzpicture}
\end{array}
\stackrel{\beta}{\simeq} 
\int^{\Omega \CD}_{z \in \Omega_m\CM} \int^{\Omega \CD}_{w \in \Omega_m\CM}
\begin{array}{c}
\begin{tikzpicture}
\fill[blue!20](0,0)--(2.5,0)--(2.5,1.4)--(1.5,1.4)--(1.5,1.2)--(0.8,1.2)--(0.8,1.9)--(1.5,1.9)--(1.5,1.7)--(2.5,1.7)--(2.5,1.9)--(3.2,1.9)--(3.2,1.2)--(2.8,1.2)--(2.8,0)--(3.5,0)--(3.5,2.5)--(0,2.5)--cycle;
\fill[teal!20](0,0)--(2.5,0)--(2.5,1.4)--(1.5,1.4)--(1.5,1.2)--(0.8,1.2)--(0.8,1.9)--(1.5,1.9)--(1.5,1.7)--(2.5,1.7)--(2.5,1.9)--(3.2,1.9)--(3.2,1.2)--(2.8,1.2)--(2.8,0)--(3.5,0)--(3.5,-1)--(0,-1)--cycle;
\draw[->-,blue,ultra thick](3.5,0)--(2.8,0);
\draw[blue,ultra thick](2.8,1.2)--(2.8,0);
\draw[blue,ultra thick](3.2,1.2)--(2.8,1.2);
\draw[blue,ultra thick](3.2,1.9)--(3.2,1.2);
\draw[blue,ultra thick](2.5,1.9)--(3.2,1.9);
\draw[blue,ultra thick](2.5,1.7)--(2.5,1.9);
\draw[blue,ultra thick](1.5,1.7)--(2.5,1.7);
\draw[blue,ultra thick](1.5,1.9)--(1.5,1.7);
\draw[blue,ultra thick](0.8,1.9)--(1.5,1.9);
\draw[blue,ultra thick](1.5,1.2)--(0.8,1.2);
\draw[blue,ultra thick](1.5,1.4)--(1.5,1.2);
\draw[blue,ultra thick](2.5,1.4)--(1.5,1.4);
\draw[blue,ultra thick](2.5,0)--(2.5,1.4);
\draw[->-,blue,ultra thick](2.5,0)--(0,0);
\draw[blue,ultra thick](0.8,1.2)--(0.8,1.9);
\fill[black] (1.8,1.4) circle (0.05) node[black,opacity=1] at(1.8,1.15) {\scriptsize $z^R$};
\fill[black] (1.8,1.7) circle (0.05) node[black,opacity=1] at(1.8,1.95) {\scriptsize $z$};
\fill[black] (2.3,1.4) circle (0.05) node[black,opacity=1] at(2.3,1.15) {\scriptsize $w^R$};
\fill[black] (2.8,0.4) circle (0.05) node[black,opacity=1] at(3.1,0.45) {\scriptsize $w$};
\fill[red] (1.15,1.9) circle (0.07) node[black,opacity=1] at(1.15,2.1) {\scriptsize $x$};
\fill[red] (2.85,1.9) circle (0.07) node[black,opacity=1] at(2.85,2.1) {\scriptsize $y$};
\end{tikzpicture}
\end{array}
\\
= &
\int^{\Omega \CD}_{z \in \Omega_m\CM} \int^{\Omega \CD}_{w \in \Omega_m\CM}
\begin{array}{c}
\begin{tikzpicture}
\fill[blue!20](0,0)--(2.5,0)--(2.5,1.4)--(1.5,1.4)--(1.5,1.2)--(0.8,1.2)--(0.8,1.9)--(1.5,1.9)--(1.5,1.7)--(2.5,1.7)--(2.5,1.9)--(3.2,1.9)--(3.2,1.2)--(2.8,1.2)--(2.8,0)--(3.5,0)--(3.5,2.5)--(0,2.5)--cycle;
\fill[teal!20](0,0)--(2.5,0)--(2.5,1.4)--(1.5,1.4)--(1.5,1.2)--(0.8,1.2)--(0.8,1.9)--(1.5,1.9)--(1.5,1.7)--(2.5,1.7)--(2.5,1.9)--(3.2,1.9)--(3.2,1.2)--(2.8,1.2)--(2.8,0)--(3.5,0)--(3.5,-1)--(0,-1)--cycle;
\draw[->-,blue,ultra thick](3.5,0)--(2.8,0);
\draw[blue,ultra thick](2.8,1.2)--(2.8,0);
\draw[blue,ultra thick](3.2,1.2)--(2.8,1.2);
\draw[blue,ultra thick](3.2,1.9)--(3.2,1.2);
\draw[blue,ultra thick](2.5,1.9)--(3.2,1.9);
\draw[blue,ultra thick](2.5,1.7)--(2.5,1.9);
\draw[blue,ultra thick](1.5,1.7)--(2.5,1.7);
\draw[blue,ultra thick](1.5,1.9)--(1.5,1.7);
\draw[blue,ultra thick](0.8,1.9)--(1.5,1.9);
\draw[blue,ultra thick](1.5,1.2)--(0.8,1.2);
\draw[blue,ultra thick](1.5,1.4)--(1.5,1.2);
\draw[blue,ultra thick](2.5,1.4)--(1.5,1.4);
\draw[blue,ultra thick](2.5,0)--(2.5,1.4);
\draw[->-,blue,ultra thick](2.5,0)--(0,0);
\draw[blue,ultra thick](0.8,1.2)--(0.8,1.9);
\fill[black] (1.8,1.4) circle (0.05) node[black,opacity=1] at(1.8,1.15) {\scriptsize $z^R$};
\fill[black] (1.8,1.7) circle (0.05) node[black,opacity=1] at(1.8,1.95) {\scriptsize $z$};
\fill[black] (2.5,0.4) circle (0.05) node[black,opacity=1] at(2.25,0.4) {\scriptsize $w^R$};
\fill[black] (2.8,0.4) circle (0.05) node[black,opacity=1] at(3.1,0.45) {\scriptsize $w$};
\fill[red] (1.15,1.9) circle (0.07) node[black,opacity=1] at(1.15,2.1) {\scriptsize $x$};
\fill[red] (2.85,1.9) circle (0.07) node[black,opacity=1] at(2.85,2.1) {\scriptsize $y$};
\end{tikzpicture}
\end{array}
\to
\begin{array}{c}
\begin{tikzpicture}
\fill[blue!20](0,0)--(2.5,0)--(2.5,1.4)--(1.5,1.4)--(1.5,1.2)--(0.8,1.2)--(0.8,1.9)--(1.5,1.9)--(1.5,1.7)--(2.5,1.7)--(2.5,1.9)--(3.2,1.9)--(3.2,1.2)--(2.8,1.2)--(2.8,0)--(3.5,0)--(3.5,2.5)--(0,2.5)--cycle;
\fill[teal!20](0,0)--(2.5,0)--(2.5,1.4)--(1.5,1.4)--(1.5,1.2)--(0.8,1.2)--(0.8,1.9)--(1.5,1.9)--(1.5,1.7)--(2.5,1.7)--(2.5,1.9)--(3.2,1.9)--(3.2,1.2)--(2.8,1.2)--(2.8,0)--(3.5,0)--(3.5,-1)--(0,-1)--cycle;
\draw[->-,blue,ultra thick](3.5,0)--(2.8,0);
\draw[blue,ultra thick](2.8,1.2)--(2.8,0);
\draw[blue,ultra thick](3.2,1.2)--(2.8,1.2);
\draw[blue,ultra thick](3.2,1.9)--(3.2,1.2);
\draw[blue,ultra thick](2.5,1.9)--(3.2,1.9);
\draw[blue,ultra thick](2.5,1.7)--(2.5,1.9);
\draw[blue,ultra thick](1.5,1.7)--(2.5,1.7);
\draw[blue,ultra thick](1.5,1.9)--(1.5,1.7);
\draw[blue,ultra thick](0.8,1.9)--(1.5,1.9);
\draw[blue,ultra thick](1.5,1.2)--(0.8,1.2);
\draw[blue,ultra thick](1.5,1.4)--(1.5,1.2);
\draw[blue,ultra thick](2.5,1.4)--(1.5,1.4);
\draw[blue,ultra thick](2.5,0)--(2.5,1.4);
\draw[->-,blue,ultra thick](2.5,0)--(0,0);
\draw[blue,ultra thick](0.8,1.2)--(0.8,1.9);
\fill[red] (1.15,1.9) circle (0.07) node[black,opacity=1] at(1.15,2.1) {\scriptsize $x$};
\fill[red] (2.85,1.9) circle (0.07) node[black,opacity=1] at(2.85,2.1) {\scriptsize $y$};
\end{tikzpicture}
\end{array}
\\ 
= & 
\begin{array}{c}
\begin{tikzpicture}[scale=1.0]
\fill[blue!20](0,0)--(3,0)--(3,1)--(0,1)--cycle;
\fill[teal!20](0,0)--(3,0)--(3,-1)--(0,-1)--cycle;
\draw[->-,blue,ultra thick] (3,0)--(0,0);
\node[black,opacity=1] at(1.5,-0.5) {\scriptsize $\Omega \CD$};
\node[black,opacity=1] at(1.5,0.5) {\scriptsize $\Omega \CC$};
\fill[red] (0.75,0) circle (0.07) node[black,opacity=1] at(0.75,0.25) {\scriptsize $x$};
\fill[red] (2.25,0) circle (0.07) node[black,opacity=1] at(2.25,0.25) {\scriptsize $y$};
\end{tikzpicture} 
\end{array}
\end{align*}

Mathematically, this amounts to the commutativity of the following diagram
\[
\begin{tikzcd}[scale cd = 1]
{\int^{\Omega \CD}_{z \in \Omega_m\CM} z^R \otimes x \otimes z \otimes \int^{\Omega \CD}_{w \in \Omega_m\CM} w^R \otimes y \otimes w } \arrow[dd,"\pi_{1_m} \otimes \pi_{1_m}"] \arrow[rr]  &  & {\int^{\Omega \CD}_{w \in \Omega_m\CM} w^R \otimes (\int^{\Omega \CD}_{z \in \Omega_m\CM} z^R \otimes x \otimes z) \otimes y \otimes w} \arrow[d,"\pi_{1_m}"] \\
                          &  & {\int^{\Omega \CD}_{w \in \Omega_m\CM} w^R \otimes x \otimes y \otimes w} \arrow[d,"\pi_{1_m}"] \\
{x \otimes y} \arrow[rr,"="] &  & {x\otimes y}          
\end{tikzcd}
\]
This commutativity can be demonstrated by considering the following diagram:
$$
\begin{tikzcd}[scale cd = 0.8]
{\int^{\Omega \CD}_{z \in \Omega_m\CM} z^R \otimes x \otimes z \otimes \int^{\Omega \CD}_{w \in \Omega_m\CM} w^R \otimes y \otimes w} \arrow[d] \arrow[rr] &  & {\int^{\Omega \CD}_{w \in \Omega_m\CM} w^R \otimes \int^{\Omega \CD}_{z \in \Omega_m\CM} z^R \otimes x \otimes z \otimes y \otimes w} \arrow[d] \arrow[rd] &    \\
{\int^{\Omega \CD}_{z \in \Omega_m\CM} z^R \otimes x \otimes z \otimes 1_m \otimes y \otimes 1_m} \arrow[d] \arrow[rr] &  & {1_m \otimes \int^{\Omega \CD}_{z \in \Omega_m\CM} z^R \otimes x \otimes z \otimes y \otimes 1_m} \arrow[d]            & {\int^{\Omega \CD}_{w \in \Omega_m\CM} w^R \otimes x \otimes y \otimes w} \\
{x \otimes y} \arrow[rr]           &  & {x \otimes y} \arrow[ru]           &   
\end{tikzcd}
$$
The upper square commutes due to the definition of the upper horizontal arrow. The middle horizontal morphism is the half-braiding with the tensor unit $1_m$ and so is trivial, hence the lower square commutes. The right triangle commutes trivially.

As a result, the morphism $L_2^R(\varepsilon_x \otimes \varepsilon_y)$ can be depicted as (where we use the ``tunnels" between bubbles to indicate the way we fuse bubbles):
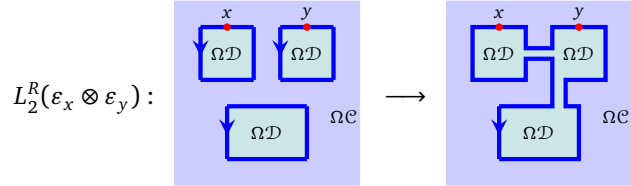
\begin{align*}
L_2^R(\varepsilon_x \otimes \varepsilon_y):
\begin{array}{c}
\begin{tikzpicture}[scale=0.7]
\fill[teal!20](0.5,0.5)--(1.5,0.5)--(1.5,1.5)--(0.5,1.5)--cycle;
\fill[teal!20](2,0.5)--(3,0.5)--(3,1.5)--(2,1.5)--cycle;
\fill[teal!20](1,0)--(2.5,0)--(2.5,-1)--(1,-1)--cycle;
\fill[blue!20,even odd rule](0,2)--(3.5,2)--(3.5,-1.5)--(0,-1.5)--cycle  (0.5,0.5)--(1.5,0.5)--(1.5,1.5)--(0.5,1.5)--cycle  (2,0.5)--(3,0.5)--(3,1.5)--(2,1.5)--cycle  (1,0)--(2.5,0)--(2.5,-1)--(1,-1)--cycle;
\draw[->-,blue,ultra thick](0.5,1.5)--(0.5,0.5); 
\draw[blue,ultra thick](0.5,0.5)--(1.5,0.5);
\draw[blue,ultra thick](1.5,0.5)--(1.5,1.5);
\draw[blue,ultra thick](1.5,1.5)--(0.5,1.5);
\draw[->-,blue,ultra thick](2,1.5)--(2,0.5); 
\draw[blue,ultra thick](2,0.5)--(3,0.5);
\draw[blue,ultra thick](3,0.5)--(3,1.5);
\draw[blue,ultra thick](3,1.5)--(2,1.5);
\node[black,opacity=1] at(3.2,-0.2) {\scriptsize $\Omega \CC$};
\node[black,opacity=1] at(1,1) {\scriptsize $\Omega \CD$};
\node[black,opacity=1] at(2.5,1) {\scriptsize $\Omega \CD$};
\fill[red] (1,1.5) circle (0.07) node[black,opacity=1] at(1,1.75) {\scriptsize $x$};
\fill[red] (2.5,1.5) circle (0.07) node[black,opacity=1] at(2.5,1.75) {\scriptsize $y$};
\draw[->-,blue,ultra thick](1,0)--(1,-1); 
\draw[blue,ultra thick](1,0)--(2.5,0);
\draw[blue,ultra thick](2.5,0)--(2.5,-1);
\draw[blue,ultra thick](1,-1)--(2.5,-1);
\node[black,opacity=1] at(1.75,-0.5) {\scriptsize $\Omega \CD$};
\end{tikzpicture}
\end{array} 
\longrightarrow
\begin{array}{c}
\begin{tikzpicture}[scale=0.7]
\fill[teal!20](1,0)--(2,0)--(2,0.9)--(1.5,0.9)--(1.5,0.5)--(0.5,0.5)--(0.5,1.5)--(1.5,1.5)--(1.5,1.1)--(2,1.1)--(2,1.5)--(3,1.5)--(3,0.5)--(2.25,0.5)--(2.25,0)--(2.5,0)--(2.5,-1)--(1,-1)--cycle;
\fill[blue!20,even odd rule](0,2)--(3.5,2)--(3.5,-1.5)--(0,-1.5)--cycle   (1,0)--(2,0)--(2,0.9)--(1.5,0.9)--(1.5,0.5)--(0.5,0.5)--(0.5,1.5)--(1.5,1.5)--(1.5,1.1)--(2,1.1)--(2,1.5)--(3,1.5)--(3,0.5)--(2.25,0.5)--(2.25,0)--(2.5,0)--(2.5,-1)--(1,-1)--cycle;
\draw[blue,ultra thick](1,0)--(2,0)--(2,0.9)--(1.5,0.9)--(1.5,0.5)--(0.5,0.5)--(0.5,1.5)--(1.5,1.5)--(1.5,1.1)--(2,1.1)--(2,1.5)--(3,1.5)--(3,0.5)--(2.25,0.5)--(2.25,0)--(2.5,0)--(2.5,-1)--(1,-1);
\draw[->-,blue,ultra thick](1,0)--(1,-1);
\node[black,opacity=1] at(3.2,-0.2) {\scriptsize $\Omega \CC$};
\node[black,opacity=1] at(1,1) {\scriptsize $\Omega \CD$};
\node[black,opacity=1] at(2.5,1) {\scriptsize $\Omega \CD$};
\fill[red] (1,1.5) circle (0.07) node[black,opacity=1] at(1,1.75) {\scriptsize $x$};
\fill[red] (2.5,1.5) circle (0.07) node[black,opacity=1] at(2.5,1.75) {\scriptsize $y$}; 
\node[black,opacity=1] at(1.75,-0.5) {\scriptsize $\Omega \CD$};
\end{tikzpicture}
\end{array} 
\end{align*}

Summarizing above analysis, we obtain that the composed morphism (\ref{eq_algebraic_multi}) can be written as the composition of the following maps:
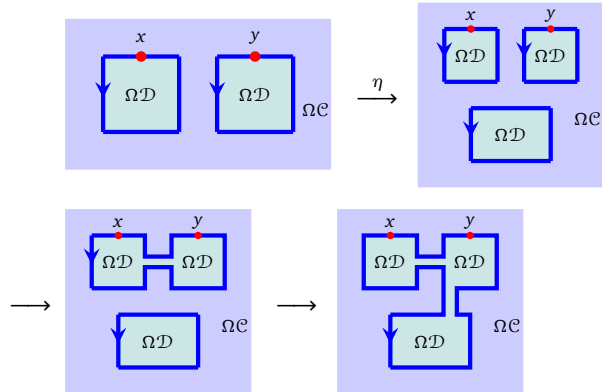
\begin{align*}
&
\begin{array}{c}
\begin{tikzpicture}[scale=1.0]
\fill[teal!20](0.5,0.5)--(1.5,0.5)--(1.5,1.5)--(0.5,1.5)--cycle;
\fill[teal!20](2,0.5)--(3,0.5)--(3,1.5)--(2,1.5)--cycle;
\fill[blue!20,even odd rule](0,0)--(3.5,0)--(3.5,2)--(0,2)--cycle (0.5,0.5)--(1.5,0.5)--(1.5,1.5)--(0.5,1.5)--cycle (2,0.5)--(3,0.5)--(3,1.5)--(2,1.5)--cycle;
\draw[->-,blue,ultra thick](0.5,1.5)--(0.5,0.5); 
\draw[blue,ultra thick](0.5,0.5)--(1.5,0.5);
\draw[blue,ultra thick](1.5,0.5)--(1.5,1.5);
\draw[blue,ultra thick](1.5,1.5)--(0.5,1.5);
\draw[->-,blue,ultra thick](2,1.5)--(2,0.5); 
\draw[blue,ultra thick](2,0.5)--(3,0.5);
\draw[blue,ultra thick](3,0.5)--(3,1.5);
\draw[blue,ultra thick](3,1.5)--(2,1.5);
\node[black,opacity=1] at(3.3,0.8) {\scriptsize $\Omega \CC$};
\node[black,opacity=1] at(1,1) {\scriptsize $\Omega \CD$};
\node[black,opacity=1] at(2.5,1) {\scriptsize $\Omega \CD$};
\fill[red] (1,1.5) circle (0.07) node[black,opacity=1] at(1,1.75) {\scriptsize $x$};
\fill[red] (2.5,1.5) circle (0.07) node[black,opacity=1] at(2.5,1.75) {\scriptsize $y$};
\end{tikzpicture}
\end{array}
\stackrel{\eta}{\longrightarrow}
\begin{array}{c}
\begin{tikzpicture}[scale=0.7]
\fill[teal!20](0.5,0.5)--(1.5,0.5)--(1.5,1.5)--(0.5,1.5)--cycle;
\fill[teal!20](2,0.5)--(3,0.5)--(3,1.5)--(2,1.5)--cycle;
\fill[teal!20](1,0)--(2.5,0)--(2.5,-1)--(1,-1)--cycle;
\fill[blue!20,even odd rule](0,2)--(3.5,2)--(3.5,-1.5)--(0,-1.5)--cycle  (0.5,0.5)--(1.5,0.5)--(1.5,1.5)--(0.5,1.5)--cycle  (2,0.5)--(3,0.5)--(3,1.5)--(2,1.5)--cycle  (1,0)--(2.5,0)--(2.5,-1)--(1,-1)--cycle;
\draw[->-,blue,ultra thick](0.5,1.5)--(0.5,0.5); 
\draw[blue,ultra thick](0.5,0.5)--(1.5,0.5);
\draw[blue,ultra thick](1.5,0.5)--(1.5,1.5);
\draw[blue,ultra thick](1.5,1.5)--(0.5,1.5);
\draw[->-,blue,ultra thick](2,1.5)--(2,0.5); 
\draw[blue,ultra thick](2,0.5)--(3,0.5);
\draw[blue,ultra thick](3,0.5)--(3,1.5);
\draw[blue,ultra thick](3,1.5)--(2,1.5);
\node[black,opacity=1] at(3.2,-0.2) {\scriptsize $\Omega \CC$};
\node[black,opacity=1] at(1,1) {\scriptsize $\Omega \CD$};
\node[black,opacity=1] at(2.5,1) {\scriptsize $\Omega \CD$};
\fill[red] (1,1.5) circle (0.07) node[black,opacity=1] at(1,1.75) {\scriptsize $x$};
\fill[red] (2.5,1.5) circle (0.07) node[black,opacity=1] at(2.5,1.75) {\scriptsize $y$};
\draw[->-,blue,ultra thick](1,0)--(1,-1); 
\draw[blue,ultra thick](1,0)--(2.5,0);
\draw[blue,ultra thick](2.5,0)--(2.5,-1);
\draw[blue,ultra thick](1,-1)--(2.5,-1);
\node[black,opacity=1] at(1.75,-0.5) {\scriptsize $\Omega \CD$};
\end{tikzpicture}
\end{array} 
\\
\longrightarrow &
\begin{array}{c}
\begin{tikzpicture}[scale=0.7]
\fill[teal!20](0.5,0.5)--(1.5,0.5)--(1.5,0.9)--(2,0.9)--(2,0.5)--(3,0.5)--(3,1.5)--(2,1.5)--(2,1.1)--(1.5,1.1)--(1.5,1.5)--(0.5,1.5)--cycle;
\fill[teal!20](1,0)--(2.5,0)--(2.5,-1)--(1,-1)--cycle;
\fill[blue!20,even odd rule](0,2)--(3.5,2)--(3.5,-1.5)--(0,-1.5)--cycle   (0.5,0.5)--(1.5,0.5)--(1.5,0.9)--(2,0.9)--(2,0.5)--(3,0.5)--(3,1.5)--(2,1.5)--(2,1.1)--(1.5,1.1)--(1.5,1.5)--(0.5,1.5)--cycle   (1,0)--(2.5,0)--(2.5,-1)--(1,-1)--cycle;
\draw[->-,blue,ultra thick](0.5,1.5)--(0.5,0.5); 
\draw[blue,ultra thick](0.5,0.5)--(1.5,0.5)--(1.5,0.9)--(2,0.9)--(2,0.5)--(3,0.5)--(3,1.5)--(2,1.5)--(2,1.1)--(1.5,1.1)--(1.5,1.5)--(0.5,1.5);
\node[black,opacity=1] at(3.2,-0.2) {\scriptsize $\Omega \CC$};
\node[black,opacity=1] at(1,1) {\scriptsize $\Omega \CD$};
\node[black,opacity=1] at(2.5,1) {\scriptsize $\Omega \CD$};
\fill[red] (1,1.5) circle (0.07) node[black,opacity=1] at(1,1.75) {\scriptsize $x$};
\fill[red] (2.5,1.5) circle (0.07) node[black,opacity=1] at(2.5,1.75) {\scriptsize $y$};
\draw[->-,blue,ultra thick](1,0)--(1,-1); 
\draw[blue,ultra thick](1,0)--(2.5,0);
\draw[blue,ultra thick](2.5,0)--(2.5,-1);
\draw[blue,ultra thick](1,-1)--(2.5,-1);
\node[black,opacity=1] at(1.75,-0.5) {\scriptsize $\Omega \CD$};
\end{tikzpicture}
\end{array} 
\longrightarrow
\begin{array}{c}
\begin{tikzpicture}[scale=0.7]
\fill[teal!20](1,0)--(2,0)--(2,0.9)--(1.5,0.9)--(1.5,0.5)--(0.5,0.5)--(0.5,1.5)--(1.5,1.5)--(1.5,1.1)--(2,1.1)--(2,1.5)--(3,1.5)--(3,0.5)--(2.25,0.5)--(2.25,0)--(2.5,0)--(2.5,-1)--(1,-1)--cycle;
\fill[blue!20,even odd rule](0,2)--(3.5,2)--(3.5,-1.5)--(0,-1.5)--cycle   (1,0)--(2,0)--(2,0.9)--(1.5,0.9)--(1.5,0.5)--(0.5,0.5)--(0.5,1.5)--(1.5,1.5)--(1.5,1.1)--(2,1.1)--(2,1.5)--(3,1.5)--(3,0.5)--(2.25,0.5)--(2.25,0)--(2.5,0)--(2.5,-1)--(1,-1)--cycle;
\draw[blue,ultra thick](1,0)--(2,0)--(2,0.9)--(1.5,0.9)--(1.5,0.5)--(0.5,0.5)--(0.5,1.5)--(1.5,1.5)--(1.5,1.1)--(2,1.1)--(2,1.5)--(3,1.5)--(3,0.5)--(2.25,0.5)--(2.25,0)--(2.5,0)--(2.5,-1)--(1,-1);
\draw[->-,blue,ultra thick](1,0)--(1,-1);
\node[black,opacity=1] at(3.2,-0.2) {\scriptsize $\Omega \CC$};
\node[black,opacity=1] at(1,1) {\scriptsize $\Omega \CD$};
\node[black,opacity=1] at(2.5,1) {\scriptsize $\Omega \CD$};
\fill[red] (1,1.5) circle (0.07) node[black,opacity=1] at(1,1.75) {\scriptsize $x$};
\fill[red] (2.5,1.5) circle (0.07) node[black,opacity=1] at(2.5,1.75) {\scriptsize $y$}; 
\node[black,opacity=1] at(1.75,-0.5) {\scriptsize $\Omega \CD$};
\end{tikzpicture}
\end{array} 
\end{align*}

By the naturality of $\eta$ we may rewrite the above map as
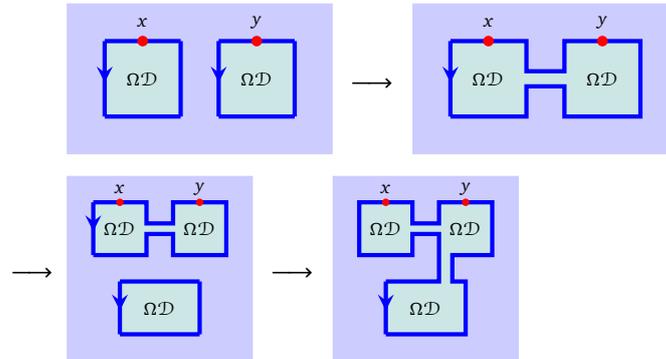
\begin{align*}
&
\begin{array}{c}
\begin{tikzpicture}[scale=1.0]
\fill[teal!20](0.5,0.5)--(1.5,0.5)--(1.5,1.5)--(0.5,1.5)--cycle;
\fill[teal!20](2,0.5)--(3,0.5)--(3,1.5)--(2,1.5)--cycle;
\fill[blue!20,even odd rule](0,0)--(3.5,0)--(3.5,2)--(0,2)--cycle (0.5,0.5)--(1.5,0.5)--(1.5,1.5)--(0.5,1.5)--cycle (2,0.5)--(3,0.5)--(3,1.5)--(2,1.5)--cycle;
\draw[->-,blue,ultra thick](0.5,1.5)--(0.5,0.5); 
\draw[blue,ultra thick](0.5,0.5)--(1.5,0.5);
\draw[blue,ultra thick](1.5,0.5)--(1.5,1.5);
\draw[blue,ultra thick](1.5,1.5)--(0.5,1.5);
\draw[->-,blue,ultra thick](2,1.5)--(2,0.5); 
\draw[blue,ultra thick](2,0.5)--(3,0.5);
\draw[blue,ultra thick](3,0.5)--(3,1.5);
\draw[blue,ultra thick](3,1.5)--(2,1.5);
\node[black,opacity=1] at(1,1) {\scriptsize $\Omega \CD$};
\node[black,opacity=1] at(2.5,1) {\scriptsize $\Omega \CD$};
\fill[red] (1,1.5) circle (0.07) node[black,opacity=1] at(1,1.75) {\scriptsize $x$};
\fill[red] (2.5,1.5) circle (0.07) node[black,opacity=1] at(2.5,1.75) {\scriptsize $y$};
\end{tikzpicture}
\end{array}
\longrightarrow
\begin{array}{c}
\begin{tikzpicture}[scale=1.0]
\fill[teal!20](0.5,0.5)--(1.5,0.5)--(1.5,0.9)--(2,0.9)--(2,0.5)--(3,0.5)--(3,1.5)--(2,1.5)--(2,1.1)--(1.5,1.1)--(1.5,1.5)--(0.5,1.5)--cycle;
\fill[blue!20,even odd rule](0,0)--(3.5,0)--(3.5,2)--(0,2)--cycle (0.5,0.5)--(1.5,0.5)--(1.5,0.9)--(2,0.9)--(2,0.5)--(3,0.5)--(3,1.5)--(2,1.5)--(2,1.1)--(1.5,1.1)--(1.5,1.5)--(0.5,1.5)--cycle;
\draw[->-,blue,ultra thick](0.5,1.5)--(0.5,0.5); 
\draw[blue,ultra thick](0.5,0.5)--(1.5,0.5)--(1.5,0.9)--(2,0.9)--(2,0.5)--(3,0.5)--(3,1.5)--(2,1.5)--(2,1.1)--(1.5,1.1)--(1.5,1.5)--(0.5,1.5);
\node[black,opacity=1] at(1,1) {\scriptsize $\Omega \CD$};
\node[black,opacity=1] at(2.5,1) {\scriptsize $\Omega \CD$};
\fill[red] (1,1.5) circle (0.07) node[black,opacity=1] at(1,1.75) {\scriptsize $x$};
\fill[red] (2.5,1.5) circle (0.07) node[black,opacity=1] at(2.5,1.75) {\scriptsize $y$};
\end{tikzpicture}
\end{array}
\\
\longrightarrow &
\begin{array}{c}
\begin{tikzpicture}[scale=0.7]
\fill[teal!20](0.5,0.5)--(1.5,0.5)--(1.5,0.9)--(2,0.9)--(2,0.5)--(3,0.5)--(3,1.5)--(2,1.5)--(2,1.1)--(1.5,1.1)--(1.5,1.5)--(0.5,1.5)--cycle;
\fill[teal!20](1,0)--(2.5,0)--(2.5,-1)--(1,-1)--cycle;
\fill[blue!20,even odd rule](0,2)--(3.5,2)--(3.5,-1.5)--(0,-1.5)--cycle   (0.5,0.5)--(1.5,0.5)--(1.5,0.9)--(2,0.9)--(2,0.5)--(3,0.5)--(3,1.5)--(2,1.5)--(2,1.1)--(1.5,1.1)--(1.5,1.5)--(0.5,1.5)--cycle   (1,0)--(2.5,0)--(2.5,-1)--(1,-1)--cycle;
\draw[->-,blue,ultra thick](0.5,1.5)--(0.5,0.5); 
\draw[blue,ultra thick](0.5,0.5)--(1.5,0.5)--(1.5,0.9)--(2,0.9)--(2,0.5)--(3,0.5)--(3,1.5)--(2,1.5)--(2,1.1)--(1.5,1.1)--(1.5,1.5)--(0.5,1.5);
\node[black,opacity=1] at(1,1) {\scriptsize $\Omega \CD$};
\node[black,opacity=1] at(2.5,1) {\scriptsize $\Omega \CD$};
\fill[red] (1,1.5) circle (0.07) node[black,opacity=1] at(1,1.75) {\scriptsize $x$};
\fill[red] (2.5,1.5) circle (0.07) node[black,opacity=1] at(2.5,1.75) {\scriptsize $y$};
\draw[->-,blue,ultra thick](1,0)--(1,-1); 
\draw[blue,ultra thick](1,0)--(2.5,0);
\draw[blue,ultra thick](2.5,0)--(2.5,-1);
\draw[blue,ultra thick](1,-1)--(2.5,-1);
\node[black,opacity=1] at(1.75,-0.5) {\scriptsize $\Omega \CD$};
\end{tikzpicture}
\end{array} 
\longrightarrow
\begin{array}{c}
\begin{tikzpicture}[scale=0.7]
\fill[teal!20](1,0)--(2,0)--(2,0.9)--(1.5,0.9)--(1.5,0.5)--(0.5,0.5)--(0.5,1.5)--(1.5,1.5)--(1.5,1.1)--(2,1.1)--(2,1.5)--(3,1.5)--(3,0.5)--(2.25,0.5)--(2.25,0)--(2.5,0)--(2.5,-1)--(1,-1)--cycle;
\fill[blue!20,even odd rule](0,2)--(3.5,2)--(3.5,-1.5)--(0,-1.5)--cycle   (1,0)--(2,0)--(2,0.9)--(1.5,0.9)--(1.5,0.5)--(0.5,0.5)--(0.5,1.5)--(1.5,1.5)--(1.5,1.1)--(2,1.1)--(2,1.5)--(3,1.5)--(3,0.5)--(2.25,0.5)--(2.25,0)--(2.5,0)--(2.5,-1)--(1,-1)--cycle;
\draw[blue,ultra thick](1,0)--(2,0)--(2,0.9)--(1.5,0.9)--(1.5,0.5)--(0.5,0.5)--(0.5,1.5)--(1.5,1.5)--(1.5,1.1)--(2,1.1)--(2,1.5)--(3,1.5)--(3,0.5)--(2.25,0.5)--(2.25,0)--(2.5,0)--(2.5,-1)--(1,-1);
\draw[->-,blue,ultra thick](1,0)--(1,-1);
\node[black,opacity=1] at(1,1) {\scriptsize $\Omega \CD$};
\node[black,opacity=1] at(2.5,1) {\scriptsize $\Omega \CD$};
\fill[red] (1,1.5) circle (0.07) node[black,opacity=1] at(1,1.75) {\scriptsize $x$};
\fill[red] (2.5,1.5) circle (0.07) node[black,opacity=1] at(2.5,1.75) {\scriptsize $y$}; 
\node[black,opacity=1] at(1.75,-0.5) {\scriptsize $\Omega \CD$};
\end{tikzpicture}
\end{array} 
\end{align*}

Note that the composition of the last two maps is just identity, by the zig-zag equation of adjunction:
$$
\id: L_2^R(x \otimes y) \to L_2^R(x \otimes y) \otimes L_2^R(1_m) = L_2^R(L_2L_2^R(x \otimes y)) \stackrel{L_2^R(\varepsilon_{x \otimes y})}{\longrightarrow} L_2^R(x \otimes y)
$$
Finally, we see that the morphism $L_2^R(x) \otimes L_2^R(y) \to L_2^R(x \otimes y)$ is given by fusion of two bubbles:
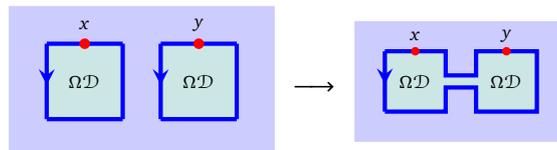
\begin{align*}
\begin{array}{c}
\begin{tikzpicture}[scale=1.0]
\fill[teal!20](0.5,0.5)--(1.5,0.5)--(1.5,1.5)--(0.5,1.5)--cycle;
\fill[teal!20](2,0.5)--(3,0.5)--(3,1.5)--(2,1.5)--cycle;
\fill[blue!20,even odd rule](0,0)--(3.5,0)--(3.5,2)--(0,2)--cycle (0.5,0.5)--(1.5,0.5)--(1.5,1.5)--(0.5,1.5)--cycle (2,0.5)--(3,0.5)--(3,1.5)--(2,1.5)--cycle;
\draw[->-,blue,ultra thick](0.5,1.5)--(0.5,0.5); 
\draw[blue,ultra thick](0.5,0.5)--(1.5,0.5);
\draw[blue,ultra thick](1.5,0.5)--(1.5,1.5);
\draw[blue,ultra thick](1.5,1.5)--(0.5,1.5);
\draw[->-,blue,ultra thick](2,1.5)--(2,0.5); 
\draw[blue,ultra thick](2,0.5)--(3,0.5);
\draw[blue,ultra thick](3,0.5)--(3,1.5);
\draw[blue,ultra thick](3,1.5)--(2,1.5);
\node[black,opacity=1] at(1,1) {\scriptsize $\Omega \CD$};
\node[black,opacity=1] at(2.5,1) {\scriptsize $\Omega \CD$};
\fill[red] (1,1.5) circle (0.07) node[black,opacity=1] at(1,1.75) {\scriptsize $x$};
\fill[red] (2.5,1.5) circle (0.07) node[black,opacity=1] at(2.5,1.75) {\scriptsize $y$};
\end{tikzpicture}
\end{array}
\longrightarrow
\begin{array}{c}
\begin{tikzpicture}[scale=0.8]
\fill[teal!20](0.5,0.5)--(1.5,0.5)--(1.5,0.9)--(2,0.9)--(2,0.5)--(3,0.5)--(3,1.5)--(2,1.5)--(2,1.1)--(1.5,1.1)--(1.5,1.5)--(0.5,1.5)--cycle;
\fill[blue!20,even odd rule](0,0)--(3.5,0)--(3.5,2)--(0,2)--cycle (0.5,0.5)--(1.5,0.5)--(1.5,0.9)--(2,0.9)--(2,0.5)--(3,0.5)--(3,1.5)--(2,1.5)--(2,1.1)--(1.5,1.1)--(1.5,1.5)--(0.5,1.5)--cycle;
\draw[->-,blue,ultra thick](0.5,1.5)--(0.5,0.5); 
\draw[blue,ultra thick](0.5,0.5)--(1.5,0.5)--(1.5,0.9)--(2,0.9)--(2,0.5)--(3,0.5)--(3,1.5)--(2,1.5)--(2,1.1)--(1.5,1.1)--(1.5,1.5)--(0.5,1.5);
\node[black,opacity=1] at(1,1) {\scriptsize $\Omega \CD$};
\node[black,opacity=1] at(2.5,1) {\scriptsize $\Omega \CD$};
\fill[red] (1,1.5) circle (0.07) node[black,opacity=1] at(1,1.75) {\scriptsize $x$};
\fill[red] (2.5,1.5) circle (0.07) node[black,opacity=1] at(2.5,1.75) {\scriptsize $y$};
\end{tikzpicture}
\end{array}
\end{align*}
This completes the proof of Theorem \ref{thm:geo_A=internal_hom}.

\subsection{Module Eilenberg-Watts calculus}\label{Appendix^EMcal}

In this section, we develop a relative version of classical Eilenberg-Watts calculus. It is different from the one developed in \cite{FSS20}. All categories we consider in this appendix are finite 1-categories, i.e. linear categories that are equivalent to the category of finite dimensional modules over some finite dimensional algebra $A$.

\begin{conv}[Convention for duality]
Our convention for left/right duality is different from that in \cite{DR18} \cite{Lur17}. For $\CC$ a monoidal category and $x \in \CC$ an object, a left dual of $x$ consists of a triple $(x^L, \mathrm{ev}_x, \mathrm{coev}_x)$ where $x^L \in \CC$ is an object, $\mathrm{coev}_x: \one_\CC \to x \otimes x^L$ is the coevaluation map and $\mathrm{ev}_x: x^L \otimes x \to \one_\CC$ is the evaluation map. The (co)evaluation maps are required to satisfy the zig-zag equation. A right dual of $x$ is a left dual in $\CC^{\rm rev}$.
\end{conv}

\begin{notation}
Let $\CC$ be a finite tensor category and $\CM$ is a left $\CC$-module, whose module structure is denoted by $\odot: \CC \times \CM \to \CM$. Rigidity of $\CC$ induces two right $\CC$-module structures over $\CM^{\rm op}$, defined in the following way:
\begin{itemize}
\item The module structure 
$$
\odot^L: \CM^{\rm op} \times \CC \to \CM^{\rm op}, \quad (m,c) \mapsto c^L \odot m.
$$
For simplicity, we denote the right $\CC$-module structure by $\CM^{\mathrm{op}|L}$.
\item The module structure 
$$
\odot^R: \CM^{\rm op} \times \CC \to \CM^{\rm op}, \quad (m,c) \mapsto c^R \odot m.
$$
For simplicity, we denote the right $\CC$-module structure by $\CM^{\mathrm{op}|R}$.
\end{itemize}
\end{notation}

\begin{notation}
Let $\CC$ be a finite tensor category and $\CM$, $\CN$ be finite left $\CC$-modules. We use $\Fun_{\CC}^L(\CM,\CN)$ to denote the finite category of left exact $\CC$-module functors from $\CM$ to $\CN$. Similarly, we use $\Fun_{\CC}^R(\CM,\CN)$ to denote the finite category of right exact $\CC$-module functors from $\CM$ to $\CN$. For two $\CC$-module functors $(F,\eta_F)$, $(G,\eta_G)$ in $\Fun_\CC(\CM,\CN)$, we write $\mathrm{Nat}_\CC(F,G)$ for the vector space of $\CC$-module natural transformations from $F$ to $G$.
\end{notation}

\begin{defn}
Let $\CC$ be a finite tensor category and $\CM$ a left $\CC$-module. We equip $\CM^{\rm op}$ with the right $\CC$-module structure $\CM^{\mathrm{op}|L}$. Let $(F,e):\CM^{\mathrm{op}|L} \times \CM  \to \CD$ be a balanced $\CC$-module functor with $e$ being the balancing natural isomorphism. The {\it $\CC$-module end of $F$} is a pair $(\int^{\CC}_{x \in \CM} F(x,x),\pi)$ with $\int^{\CC}_{x \in \CM} F(x,x) \in \CD$ an object and $\pi: \Delta_{\int^{\CC}_{x \in \CM} F(x,x)} \stackrel{\bullet \bullet}{\to} F$ is a dinarutal transformation such that the following diagram is commutative:
$$
\begin{tikzcd}
\int^{\CC}_{x \in \CM} F(x,x) \ar[r,"\pi_m"] \ar[d,"\pi_{c^L \odot m}"'] 
 & F(m,m) \ar[d,"\mathrm{coev}_c"]  \\
F(c^L \odot m ,c^L \odot m) \ar[r,"{e_{m,c,c^L \odot m}}"]
 & F(m,c \odot c^L \odot m)
\end{tikzcd}
$$
and the pair $(\int^{\CC}_{x \in \CM} F(x,x),\pi)$ is terminal among all such pairs.
\end{defn}

This is a dual notion defined as follows:

\begin{defn}
Let $\CC$ be a finite tensor category and $\CM$ a left $\CC$-module. We equip $\CM^{\rm op}$ with the right $\CC$-module structure $\CM^{\mathrm{op}|R}$. Let $(F,e):\CM^{\mathrm{op}|R} \times \CM  \to \CD$ be a balanced $\CC$-module functor with $e$ being the balancing natural isomorphism. The {\it $\CC$-module coend of $F$} is a pair $(\int_{\CC}^{x \in \CM} F(x,x),\pi)$ with $\int_{\CC}^{x \in \CM} F(x,x) \in \CD$ an object and $\pi:F \stackrel{\bullet \bullet}{\to} \Delta_{\int_{\CC}^{x \in \CM} F(x,x)} $ a dinatural transformation, such that the following diagram is commutative:
$$
\begin{tikzcd}
{F(m,(c \otimes c^R) \odot m)} \arrow[d,"\mathrm{ev}_{c^R}"] \arrow[r] & {F(c^R \odot m,c^R \odot m)} \arrow[d] \\
{F(m,m)} \arrow[r]           & {\int_{\CC}^{x \in \CM} F(x,x)}          
\end{tikzcd}
$$
and the pair $(\int_{\CC}^{x \in \CM} F(x,x),\pi)$ is initial among all such pairs.
\end{defn}

\begin{rem}
The $\CC$-module end maybe written as the following equalizer:
$$\int^{\CC}_{m \in \CM} F(m,m) \simeq \begin{tikzcd}
\mathrm{Eq}(\int_{m \in \CM} F(m,m)
\ar[r,shift left = 0.5mm,"f"]
\ar[r,shift right = 0.5mm,"g"']
& \int_{m \in \CM} \int_{c \in \CC} F(m, c \odot c^L \odot m)
\end{tikzcd}
$$
Componentwisely, $f$ and $g$ can be written respectively as (we choose the component labelled by $d \in \CC$ and $n \in \CM$)
$$
f_{d,n}= \int_{m \in \CM} F(m,m) \stackrel{\pi_n}{\longrightarrow} F(n,n) \to F(n, d \otimes d^L \odot n) 
$$ 
and
$$
g_{d,n}= \int_{m \in \CM} F(m,m) \stackrel{\pi_{d^L \odot n}}{\longrightarrow} F(d^L \odot n, d^L \odot n) \simeq F(n,d^L \otimes d \odot n). 
$$
where we use the balancing structure of $F$ in the last step. Dually the $\CC$-module coend may be written as a coequalizer of two coends.
\end{rem}

The following lemma generalizes the familiar formula for usual ends:
\begin{lem}
Let $\CM,\CN$ be left $\CC$-modules where $\CC$ is a finite tensor category. Let $(F,\eta^F),(G,\eta^G) \in \Fun_{\CC}(\CM,\CN)$ be $\CC$-module functors. Then $\int^{\CC}_{m \in \CM} \hom_{\CN}(F(m),G(m)) \simeq \mathrm{Nat}_{\CC}(F,G) $. Note that the functor $\hom_{\CN}(F(-),G(-)): \CM^{\mathrm{op}|L} \times \CM \to \vect$ is equipped with a canonical balancing structure.
\end{lem}
\begin{proof}
Take $\alpha \in \mathrm{Nat}_{\CC}(F,G)$ and $\pi_m(\alpha) = \alpha_m$ be its component at $m$. Then the commutativity of the following diagram
$$
\begin{tikzcd}
\mathrm{Nat}_{\CC}(F,G) \ar[r,"\pi_m"] \ar[d,"\pi_{c^L \odot m}"'] 
 & \hom_{\CN}(F(m),G(m)) \ar[d]  \\
\hom_{\CN}(F(c^L \odot m),G(c^L \odot m)) \ar[r,"{e_{m,c,c^L \odot m}}"]
 & \hom_{\CN}(F(m),G(c \odot c^L \odot m)) 
\end{tikzcd}
$$
is equivalent to be commutativity of the following diagram
$$
\begin{tikzcd}
c^L \odot F(m) \ar[r,"c^L \odot \alpha_m"] \ar[d,"\eta^F_{c^L,m}"]
 & c^L \odot G(m) \ar[d,"\eta^G_{c^L,m}"] \\
F(c^L \odot m) \ar[r,"\alpha_{c^L \odot m}"]
 & G(c^L \odot m)
\end{tikzcd}
$$
\end{proof}

The following lemma is a simple corollary of the enriched Yoneda Lemma, which will be useful in our proof of Lemma \ref{lem_yoneda}.

\begin{lem}
Let $\CM,\CN$ be left $\CC$-modules and $(K,\eta^K) \in \Fun_{\CC}(\CM,\CN)$ is a $\CC$-module functor. There is a canonical isomorphism of vector spaces for each $y \in \CM$, $z \in \CN$:
$$
\mathrm{Nat}_{\CC}([y,-],[z,K(-)]) \simeq \hom_{\CN}(z,K(y)).
$$
Similarly we have
$$
\mathrm{Nat}_{\CC}([-,y],[K(-),z]) \simeq \hom_{\CN}(K(y),z).
$$
\end{lem}

\begin{lem}[Yoneda and co-Yoneda Lemma via (co)ends]\label{lem_yoneda}
Let $(G,\eta^G) \in \Fun_{\CC}(\CM,\CN)$, then there is a canonical isomorphism of $\CC$-module functors:
\begin{equation}
G(-) \simeq \int^{x \in \CM}_{\CC} [x,-] \odot G(x), \quad G(-) \simeq \int_{x \in \CM}^{\CC} [-,x]^R \odot G(x) 
\end{equation}
\end{lem}

\begin{proof}
$$
\begin{aligned}
\hom(\int^{x \in \CM}_{\CC}[x,y] \odot G(x),z) & \simeq \int_{x \in \CM}^{\CC} \hom_{\CN}([x,y] \odot G(x),z) \\
& \simeq \int_{x \in \CM}^{\CC} \hom_{\CC}([x,y],[G(x),z]) \\
& \simeq \mathrm{Nat}_{\CC}([-,y], [G(-),z]) \\
& \simeq \hom_{\CN}(G(y),z).
\end{aligned}
$$
\end{proof}

\begin{expl} \label{expl:1xRx}
Let us regard $\CC$ as a regular left $\CC$-module. In this case the internal hom is easily computed as
$$
[x,y] = y \otimes x^L
$$
Applying the Yoneda lemma \ref{lem_yoneda} to the identity functor $\id: \CC \to \CC$ (which is equipped with a canonical $\CC$-module functor structure), we have
\be \label{eq:factorize_1}
y \simeq \int^\CC_{x \in \CC} [y,x]^R \otimes x \simeq \int^\CC_{x \in \CC}  y \otimes x^R \otimes x, \quad\quad \one_\CC \simeq \int^\CC_{x \in \CC} x^R \otimes x.
\ee
\end{expl}

\begin{thm}[Generalized Eilenberg-Watts calculus]\label{thm_gen_eilen_watts}
Let $\CC$ be a finite tensor category. We use $\Fun^{R}_{\CC}(\CM,\CN)$ to denote the category of right exact $\CC$-module functors from $\CM$ to $\CN$ and $\Fun^{L}_{\CC}(\CM,\CN)$ the category of left exact ones. Then there are pairs of adjoint equivalences:
$$
\begin{aligned}
& \Phi^l: \CM^{\mathrm{op}|R} \boxtimes_{\CC} \CN \to \Fun^L_{\CC}(\CM,\CN) , \quad x \boxtimes_{\CC} y \mapsto [x,-] \odot y \\
& \Psi^l: \Fun^L_{\CC}(\CM,\CN) \to \CM^{\mathrm{op}|R} \boxtimes_{\CC} \CN, \quad F \mapsto \int^{m \in \CM}_{\CC} m \boxtimes_{\CC} F(m).
\end{aligned}
$$
and
$$
\begin{aligned}
& \Phi^r:\CM^{\mathrm{op}|L} \boxtimes_{\CC} \CN \to \Fun^R_{\CC}(\CM,\CN), \quad x \boxtimes_{\CC} y \mapsto [-,x]^R \odot y \\
& \Psi^r: \Fun^R_{\CC}(\CM,\CN) \to \CM^{\mathrm{op}|L} \boxtimes_{\CC} \CN, \quad F \mapsto \int^{\CC}_{m \in \CM} m \boxtimes_{\CC} F(m).
\end{aligned}
$$
\end{thm}

\begin{expl} \label{expl:idM=int_mRm}
Let $\CC$ be a braided fusion category, $\CM$ be a multi-fusion category and $L \colon \CC \to \CM$ be a central functor. $L$ equips $\CM$ with a structure of left $\CC$-module in a manifest way. According to Theorem \ref{thm_gen_eilen_watts}, there is an equivalence $\Psi^r: \Fun_\CC(\CM,\CM) \simeq \CM^{\mathrm{op}|L} \boxtimes_\CC \CM$, which sends a functor $F$ to $\int_{m \in \CM}^\CC m \boxtimes F(m)$. In particular, $\Psi^r$ sends the identity functor $\id_\CM$ to $\int_{m \in \CM}^\CC m \boxtimes m$. Taking the equivalence of right $\CC$-modules
\[
\begin{aligned}
\CM^{\mathrm{op}|L} & \simeq \CM^{\rm rev}, \\
m & \mapsto m^R
\end{aligned}
\]
into consideration, we obtain a composed equivalence:
\[
\Fun_\CC(\CM,\CM) \simeq \CM^{\mathrm{op}|L} \boxtimes_\CC \CM \simeq \CM^{\rm rev} \boxtimes_\CC \CM, 
\]
which sends the identity functor $\Id_\CM$ to:
\begin{equation}
\id_\CM \mapsto \int_{m \in \CM}^\CC m \boxtimes_\CC m \mapsto \int_{m \in \CM}^\CC m^R \boxtimes_\CC m.
\end{equation}
This image of $\id_\CM$ is used in Section\,\ref{sec:2codim_3D_geometric_approach} and Section\,\ref{sec:proof_A=internal_hom}.
\end{expl}

To give a proof of Theorem\,\ref{thm_gen_eilen_watts}, we first summarize the following results from \cite{KZ18}, Corollary 2.2.6 and Proposition 2.2.7.

\begin{thm}\label{thm_cite}
Let $\CM$ and $\CN$ be left $\CC$-modules. $x,x^\prime \in \CM$ and $y,y^\prime \in \CN$.
\begin{itemize}
\item[(1)] There is a natural isomorphism 
$$
\hom_{\CM^{\mathrm{op}|L} \boxtimes_{\CC} \CN}(x \boxtimes_{\CC} y, x^{\prime} \boxtimes_{\CC} y^{\prime}) \simeq \hom_{\CC}(1,[y,y^{\prime}] \otimes [x^{\prime},x])
$$
\item[(2)] The formula $x \boxtimes_{\CC} y \mapsto y \boxtimes_{\CC} x$ determines an equivalence $\CM^{\mathrm{op}|R} \boxtimes_{\CC} \CN \simeq (\CN^{\mathrm{op}|L} \boxtimes_{\CC} \CM)^{\rm op}$

\item[(3)] The functor $x \boxtimes_{\CC} y \mapsto [x,-] \odot y$ determines an equivalence between $\CM^{\mathrm{op}|R} \boxtimes_{\CC} \CN$ and $\Fun_{\CC}^L(\CM,\CN)$; similarly, the functor $x \boxtimes_\CC y \mapsto [-,x]^R \odot y$ determines an equivalence between $\CM^{\mathrm{op}|L} \boxtimes_\CC \CN \simeq \Fun_\CC^R(\CM,\CN)$.
\end{itemize}
\end{thm}
\begin{proof}[Proof of Theorem \ref{thm_gen_eilen_watts}]
According to Theorem \ref{thm_cite} (3), $\Phi^l$ is an equivalence. It suffices to show that $\Psi^l$ is the left adjoint functor of $\Phi^l$. This follows from the following computation:
$$
\begin{aligned}
& \mathrm{Nat}_{\CC}(F,[x,-] \odot y)  \simeq \int_{m \in \CM}^\CC \hom_{\CN}(F(m),[x,m] \odot y) \\
\simeq & \int_{m \in \CM}^\CC \hom_{\CC}(1,[x,m] \otimes [F(m),y]) \simeq \int_{m \in \CM}^\CC \hom_{\CN^{\mathrm{op}|L} \boxtimes_{\CC} \CM}(y \boxtimes_{\CC} x, F(m) \boxtimes_{\CC} m)  \\
\simeq & \int_{m \in \CM}^\CC \hom_{(\CN^{\mathrm{op}|L} \boxtimes_{\CC} \CM)^{\mathrm{op}}}(F(m) \boxtimes_{\CC} m, y \boxtimes_{\CC} x) \\
\simeq & \hom_{\CM^{\mathrm{op}|R} \boxtimes_{\CC} \CN}(\int^{m \in \CM}_\CC m \boxtimes_{\CC} F(m), x \boxtimes_{\CC} y).
\end{aligned}
$$
The first isomorphism follows from Lemma \ref{lem_yoneda}; the second isomorphism is routine; the third isomorphism follows from Theorem \ref{thm_cite} (1); the fourth isomorphism is trivial and the last isomorphism follows from Theorem \ref{thm_cite} (2).

In a similar manner, we can show that $\Psi^r$ is right adjoint to $\Phi^r$. 
\end{proof}

\begin{expl}
Let $\CC$ and $\CD$ be Morita equivalent finite tensor categories and $_{\CC} \CM_{\CD}$ be the invertible module. It is well-known that there is a canonical equivalence of monoidal categories:
$$
u: \CD^{\rev} \to \Fun_{\CC}(\CM,\CM), \quad d \mapsto - \odot d.
$$
Now we are able to write down the quasi-inverse(right adjoint) of $u$:
\begin{equation}
u^R:\Fun_{\CC}(\CM,\CM) \to \CD^{\rev}, \quad F \mapsto \int_{m \in \CM}^\CC [m,F(m)]_{\CD^{\rev}}
\end{equation}
which follows from the following easy calculation:
\[
\begin{aligned}
\hom_{\CD}(d,\int_{m \in \CM}^\CC [m,F(m)]_{\CD^{\rev}}) &\simeq \int_{m \in \CM}^\CC \hom_{\CD}(d,[m,F(m)]_{\CD^{\rev}}) \\
& \simeq \mathrm{Nat}_{\CC}(- \odot d, F(-)).
\end{aligned}
\]
\end{expl}

\begin{expl}\label{Expl_comonad}
Let $\CC,\CD$ be non-degenerate braided fusion categories, $\CM$ be a closed multi-fusion $\CC$-$\CD$-bimodule. Let $F:\CC \to \CM$ be the central functor, whose central structure is witnessed by the braided tensor functor $\tilde{F}:\CC \to \mathfrak{Z}_1(\CM)$. Given $m \in \CM$, we define $T(m)$ to be the $\CC$-module end
$$
 T(m): =  \int^{\CC}_{x \in \CM} x^R \otimes m \otimes x
$$
Note that the central structure of $F$ is indispensable for the definition. When $m$ varies, the assignments $m \mapsto T(m)$ can be extended to a functor $T \colon \CM \to\CM$ which actually defines a comonad over $\CM$. Let us look at the Eilenberg-Moore category (i.e. the category of comodules) of this comonad. The Eilenberg-Moore category of this comonad consists of pairs $(m,\delta_m)$ where $m \in \CM$ and $\delta_m$ is a half braiding of $m$ subject to the extra property that $(m,\delta_m)$ should lie in the double centralizer of the essential image of $\tilde{F}: \CC \to \mathfrak{Z}_1(\CM)$, where $\tilde{F}$ is the lift of $F$. As a result, the Eilenberg-Moore category of $T$ is nothing but $\CD$.
\end{expl}

\begin{expl}\label{Expl_half-braiding}
The $\CC$-module end $ \int^{\CC}_{x \in \CM} x^R \otimes m \otimes x \in \CM$ defined in Example \ref{Expl_comonad} is equipped with a canonical half-braiding $\gamma$:
$$
\gamma_y \colon y \otimes \int^{\CC}_{x \in \CM} x^R \otimes m \otimes x  = \int^{\CC}_{x \in \CM} y \otimes x^R \otimes m \otimes x  \simeq \int^{\CC}_{x \in \CM} x^R \otimes m \otimes x \otimes y.
$$
$\gamma_y$ is induced by the universal property of $\CC$-module end. Let $\kappa_a$ be the morphism defined by the following composition
$$
\kappa_a \colon \int^{\CC}_{x \in \CM} x^R \otimes m \otimes x \otimes y \stackrel{\pi_{a \otimes y^L}}{\longrightarrow} (a \otimes y^L)^R \otimes m \otimes a \otimes y^L \otimes y \xrightarrow{\id \otimes \mathrm{coev}_y} y \otimes a^R \otimes m \otimes a.
$$
for each $a \in \CM$. All these $\kappa_a$ can be organized into a single morphism
$$
\int^{\CC}_{x \in \CM} x^R \otimes m \otimes x \otimes y \to \int^\CC_{a \in \CM} y \otimes a^R \otimes m \otimes a =  y \otimes \int^\CC_{a \in \CM} a^R \otimes m \otimes a.
$$
In a similar manner, one can construct a morphism
\[
\gamma_y \colon y \otimes \int^\CC_{a \in \CM} a^R \otimes m \otimes a \to \int^{\CC}_{x \in \CM} x^R \otimes m \otimes x \otimes y.
\]
The two morphisms are inverse to each other, giving rise to the desired half-braiding.
\end{expl}

\subsection{\texorpdfstring{$\EE_k$}{Ek}-center of an \texorpdfstring{$\EE_k$}{Ek}-algebra} \label{Appendix:Zm-center_Em-algebra}

In this subsection, we give brief explanations (instead of proofs) of the results in Lemma\,\ref{lem:francis} \cite{Lur17,Fra12} and \eqref{eq:internal_hom_of_Ek_is_Ek+1} that can enhance our understanding. 

\medskip
We explain Lemma\,\ref{lem:francis} by first recalling Definition\,\ref{def:centralizer+center}. For $\CA, \CB\in \Algc_{\EE_k}((n+1)\vect)$ and an $\EE_k$-monoidal functor $F: \CA \to \CB$, the $\EE_k$-centralizer of $F$ is the universal $\EE_k$-multi-fusion $n$-category $\FZ_k(F)$ equipped with a unital action $G: \FZ_k(F) \boxtimes \CA \to \CB$, i.e. 
an $\EE_k$-monoidal functor exhibiting the outer triangle of the following diagram commutative (up to equivalences) as illustrated in the following commutative diagram. 
\be \label{diag:universal_property_center_2}
\begin{array}{c}
\xymatrix{
& \FZ_k(F) \boxtimes \CA \ar@/^2pc/[ddr]^G   &    \\
& \CX \boxtimes \CA \ar[dr]^{H} \ar[u]^{\exists ! \,f \boxtimes \id_\CA}   & \\
\CA \ar@/^2pc/[uur]^{\one_{\FZ_k(F)}\boxtimes \id_\CA} \ar[ur]^{\one_\CA \boxtimes \id_\CM}  \ar[rr]^{F} & & \CB \, 
}
\end{array}
\ee
\bnu

\item For $\CA, \CO\in \Algc_{\EE_k}((n+1)\vect)$, an $\EE_k$-monoidal functor $u_\CO: \CA \to \CO$ endows $\CO$ with a structure of $\EE_k$-$A$-module. Moreover, $\CO$ and $u_\CO$ are an object and a 1-morphism in $\Alg_{\EE_k}^c(\Mod_\CA^{\EE_k}((n+1)\vect))$, respectively, and $u_\CO$ is precisely the unit of $\CO$ as an $\EE_k$-algebra in $\Mod_\CA^{\EE_k}((n+1)\vect)$. 

\item An $\EE_k$-algebra homomorphism $g: \CO \to \CB$ in $\Mod_\CA^{\EE_k}((n+1)\vect)$ is equivalent to an $\EE_k$-monoidal functor $g: \CO \to \CB$ such that the composed functor $(\CA \xrightarrow{u_\CO}  \CO \xrightarrow{g} \CB)$ is isomorphic to $u_\CB$, a condition which means that $g$ preserves the units. 

\item The canonical $(n+1)\vect$-action on $\Mod_\CA^{\EE_k}((n+1)\vect)$ can be lifted to an action on the level of $\EE_k$-algebras as illustrated below. 
\be \label{diag:forget_functor_from_algc}
\begin{array}{c}
\xymatrix{
\Algc_{\EE_k}((n+1)\vect) \times \Algc_{\EE_k}(\Mod_\CA^{\EE_k}((n+1)\vect)) \ar[r]^-\boxtimes \ar[d]_{\forget \times \forget} &   \Algc_{\EE_k}(\Mod_\CA^{\EE_k}((n+1)\vect)) \ar[d]^\forget \\
(n+1)\vect \times \Mod_\CA^{\EE_k}((n+1)\vect) \ar[r]^-\boxtimes & 
\Mod_\CA^{\EE_k}((n+1)\vect)
}
\end{array}
\ee

Therefore, we can simplify the universal property of $\FZ_k(F)$ to that of an internal hom: 
$$
\xymatrix{
& \FZ_k(F) \boxtimes \CA \ar[dr]^G   &    \\
\CX \boxtimes \CA \ar[rr]^{H} \ar[ur]^{\exists ! \,f \boxtimes \id_\CA}   & & \CB\, ,
}
$$
i.e., the pair $(\FZ_k(F),G)$ precisely defines the internal hom $[\CA,\CB]$ in $\Algc_{\EE_k}(\Mod_\CA^{\EE_k}((n+1)\vect))$. Then Lemma\,\ref{lem:francis} simply says that the forgetful functors in (\ref{diag:forget_functor_from_algc}) creates the internal hom. 
\enu

\medskip
Now we explain (\ref{eq:internal_hom_of_Ek_is_Ek+1}). Consider $\CA \in \Algc_{\EE_{k+1}}((n+1)\vect)$ and $\CB \in \LMod_\CA(\Algc_{\EE_k}((n+1)\vect)$. By definition, there is an left $\CA$-action on $\CB$ given by an $\EE_k$-monoidal functor 
$\CA \times \CB \xrightarrow{\odot} \CB$ such that $(a,b) \mapsto a\odot b$ for $a\in \CA, b\in \CB$. It is helpful to think of this action $\odot$ to act in the $x^{k+1}$-direction that is transversal to a (potentially anomalous) topological order associated to $\CB$. It turns out that this $\EE_k$-monoidal action can be lifted to an action functor $\Algc_{\EE_k}(\CA) \times \Algc_{\EE_k}(\CB) \xrightarrow{\odot} \Algc_{\EE_k}(\CB)$. Since $\one_\CB \in \Algc_{\EE_k}(\CB)$, we obtain $[\one_\CB,\one_\CB] \in \Algc_{\EE_1}(\Algc_{\EE_k}(\CA)) \simeq \Alg_{\EE_{k+1}}(\CA)$. 

\newpage

\bibliography{Top}

\end{document}